\documentclass[12pt,a4]{report}
\usepackage{amsmath}
\usepackage{amssymb}
\usepackage{mathtools}
\usepackage{tensor}
\usepackage{amsfonts}
\usepackage[hmargin={26mm,26mm},vmargin={38mm,30mm}]{geometry}
\usepackage{hyperref}
\usepackage{verbatim}
\usepackage[shortlabels]{enumitem}
\usepackage{slashed}
\usepackage{braket}
\usepackage{xcolor}
\usepackage{caption}
\usepackage{tcolorbox}
\usepackage{cite}
\usepackage{tikz}
\usetikzlibrary{arrows.meta,calc,decorations.markings,math,arrows.meta,shapes.misc,decorations.pathreplacing,decorations.pathmorphing}
\usepackage{nicefrac}
\usepackage{bbm}
\usepackage{url}
\usepackage{dsfont}
\usetikzlibrary{arrows.meta,calc,decorations.markings,math,arrows.meta}
\usepackage{mdframed}
\usepackage{sectsty}

\usepackage{fancyhdr}
\usepackage{titletoc,tocloft}
\newcommand{\nn}{\nonumber}

\def \ifempty#1{\def\temp{#1} \ifx\temp\empty }
\def \checkslot#1{ \ifempty{#1} \,\cdot\, \else #1 \fi }
\newcommand{\angl}[2]{\left\langle #1, #2 \right\rangle}
\newcommand{\trip}[3]{\left[\checkslot{#1},\checkslot{#2},\checkslot{#3}\right]}
\newcommand{\trips}[3]{\left[\checkslot{#1},\checkslot{#2};\checkslot{#3}\right]}
\newcommand{\bareps}{\bar{\epsilon}}

\newcommand{\deltahat}{\hat{\delta}}

\newcommand{\BetaijBeta}[2]{\left( \bar{\beta}\Gamma_+ \Gamma_{#1 #2} \beta \right)}

\newcommand{\BetaIijBeta}[3]{\left( \bar{\beta}\Gamma_+ \Gamma^{#1} \Gamma_{#2 #3} \beta \right)}

\newcommand{\EpsijBeta}[2]{\left( \bar{\epsilon}_- \Gamma_+ \Gamma_{#1 #2} \beta \right)}

\newcommand{\om}{R}

\newcommand{\ii}{\mathrm{i}}
\newcommand{\n}{n}
\renewcommand{\k}{k}
\newcommand{\Npt}{N}
\newcommand{\s}{s}

\newcommand{\tr}{\mathrm{tr}}

\newcommand{\I}{\mathcal{I}}

\newcommand{\sixd}{\text{6d}}

\newcommand{\sOp}{\mathcal{O}}
\newcommand{\fOp}{\Phi}
\newcommand{\fields}{\phi}
\newcommand{\tHooft}{\Upsilon}

\newcommand{\bigO}{\mathcal{O}}

\newcommand{\vex}{\vec{x}}
\newcommand{\vey}{\vec{y}}

\newcommand{\su}{\frak{su}}

\tikzset{cross/.style={cross out, draw=black, fill=none, minimum size=2*(#1-\pgflinewidth), inner sep=0pt, outer sep=0pt}, cross/.default={2pt}}
\tikzset{
  on each segment/.style={
    decorate,
    decoration={
      show path construction,
      moveto code={},
      lineto code={
        \path [#1]
        (\tikzinputsegmentfirst) -- (\tikzinputsegmentlast);
      },
      curveto code={
        \path [#1] (\tikzinputsegmentfirst)
        .. controls
        (\tikzinputsegmentsupporta) and (\tikzinputsegmentsupportb)
        ..
        (\tikzinputsegmentlast);
      },
      closepath code={
        \path [#1]
        (\tikzinputsegmentfirst) -- (\tikzinputsegmentlast);
      },
    },
  },
  mid arrow/.style={postaction={decorate,decoration={
        markings,
        mark=at position .5 with {\arrow[#1]{stealth}}
      }}},
}
\tikzset{snake it/.style={decorate, decoration=snake}}

\newcommand{\und}[1]{\underline{#1}}

\numberwithin{equation}{section}

\renewcommand{\baselinestretch}{1.3}

\newcommand{\defineauthorcolor}[2]{%
  \colorlet{author#1}{#2}
  \expandafter\def\csname authoredby#1\endcsname{
    \renewcommand{\cftpartfont}{\large\bfseries\color{author#1}}
    \renewcommand{\cftchapfont}{\bfseries\color{author#1}}
    \renewcommand{\cftsecfont}{\color{author#1}}
    \renewcommand{\cftsubsecfont}{\color{author#1}}}
}
\makeatletter
\newcommand{\authoredby}[1]{\addtocontents{toc}{\protect\@nameuse{authoredby#1}}}%
\makeatother

\definecolor{Icolor}{RGB}{ 207, 72, 62}
\definecolor{IIcolor}{RGB}{119, 37, 196}

\defineauthorcolor{I}{Icolor}
\defineauthorcolor{II}{IIcolor}
\defineauthorcolor{final}{black}

\definecolor{boxback}{RGB}{204,226,227}

\begin{document}

\begin{center}

\includegraphics[width=40mm]{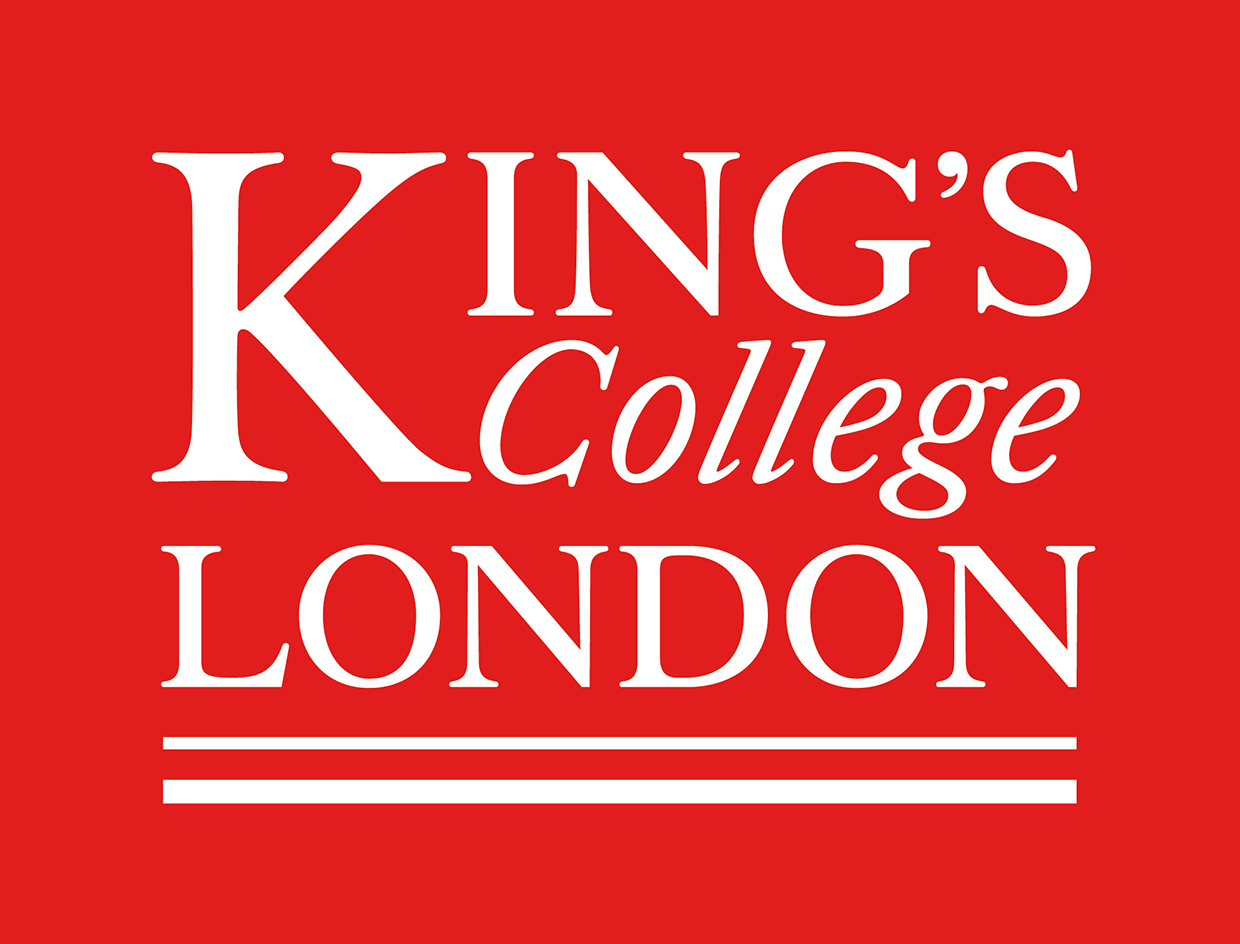}

\vspace{15mm}
{
\huge\bf
Non-Lorentzian Supersymmetric\\[0.5em]
Models and M-Theory Branes
}

 \vspace{20mm}

 {\Large\bf Rishi Talwar Mouland}
     
\vspace{1cm}

\textit{
Department of Mathematics,\\
King's College London,\\
The Strand, WC2R 2LS,\\
United Kingdom
}

\vfill 

{
A thesis submitted for the degree of \\
Doctor of Philosophy in Applied Mathematics\\
and Theoretical Physics\\[1em]
June 2021
}

\end{center}


 
\thispagestyle{empty}

\chapter*{Abstract}

The overarching theme of this thesis is the study of field theories generically without Lorentz symmetry, but possessing an inhomogeneous scaling symmetry. A number of aspects of such models are explored, including the addition of supersymmetry, as well as their application in the construction of more conventional conformal field theories.

In Part I, we describe a scaling technique by which non-Lorentzian theories with an inhomogeneous scaling symmetry are obtained from Lorentzian supersymmetric models, while crucially retaining all supersymmetry of the parent theory. The dynamics of the resulting theories are generically constrained to the moduli space of some BPS soliton. We explore this scaling technique and the resulting reduction to moduli space superconformal quantum mechanics for some examples, including several relevant to the branes of M-theory. 

In Part II, we build from the ground up the theory of models in five dimensions with an exotic $SU(1,3)$ spacetime symmetry, which includes an inhomogeneous scaling. After deriving and solving the corresponding Ward-Takahashi identities for correlators, we demonstrate how such models naturally describe any six-dimensional conformal field theory on a conformal compactification of Minkowski space. In doing so, we derive necessary conditions for a generic $SU(1,3)$ theory to admit such a six-dimensional interpretation, and also explore a degenerate limit of the construction that recovers the standard Discrete Lightcone Quantisation picture. Finally, we study in detail an explicit $SU(1,3)$ model found through the technique of Part I, which is conjectured to provide a Lagrangian description of the non-Abelian $(2,0)$ theory. The theory is shown to realise six-dimensional physics through the inclusion of singular points carrying non-zero instanton charge, which are encoded by instanton operators in the path integral. We further explore the constrained dynamics of the theory, which are shown to describe the propagation of Yang-Mills instanton-particles between such points.


\chapter*{Related publications}
Much of the material featured in this thesis also features in publications by the author, partly in collaboration.\\

\noindent The results described in Part I first appeared in the following publications,

\begin{enumerate}
  \item \textbf{N. Lambert and R. Mouland},\vspace{-3mm}
  
  \textsl{Non-Lorentzian RG Flows and Supersymmetry},\vspace{-3mm}
  
  \textsl{JHEP} \textbf{06} (2019) 130, [arXiv:\texttt{1904.05071}]
  \item \textbf{R. Mouland},\vspace{-3mm}
  
  \textsl{Supersymmetric Soliton $\sigma$-models from Non-Lorentzian Field Theories},\vspace{-3mm}
  
  \textsl{JHEP} \textbf{04} (2020) 129, [arXiv:\texttt{1911.11504}]
\end{enumerate}

\noindent Many of the results of Part II first appeared in the following publications,
\begin{enumerate}
  \setcounter{enumi}{2}
  \item \textbf{N. Lambert, A. Lipstein, R. Mouland and P. Richmond},\vspace{-3mm}
  
  \textsl{Bosonic Symmetries of $(2,0)$ DLCQ Field Theories},\vspace{-3mm}
  
  \textsl{JHEP} \textbf{01} (2020) 166, [arXiv:\texttt{1912.02638}]
  \item \textbf{N. Lambert, A. Lipstein, R. Mouland and P. Richmond},\vspace{-3mm}
  
  \textsl{Five-Dimensional Non-Lorentzian CFTs and their Relation to Six-Dimensions},\vspace{-3mm}
  
  \textsl{JHEP} \textbf{03} (2021) 053, [arXiv:\texttt{2012.00626}]
  \item \textbf{N. Lambert, A. Lipstein, R. Mouland and P. Richmond},\vspace{-3mm}
  
  \textsl{Instanton Worldlines in Five-Dimensional $\Omega$-deformed Gauge Theory},\vspace{-3mm}
  
  \textsl{Accepted for publication in JHEP}, [arXiv:\texttt{2105.02008}]
\end{enumerate}
Finally, some additional results in Part II are the author's sole work, and are due to appear in forthcoming publications.

\chapter*{Acknowledgements}
\lhead{\textsl{\nouppercase{Acknowledgements}}}

The work compiled in this thesis is the culmination of almost four years of research. A lot has happened in this time, and so now as I look back, I'm very glad to have the opportunity to thank the many people that have supported me along the way.\\

I am really very grateful to be part of the theory group at King's, which provided a welcoming and supportive environment as I made my first few steps into the world of research. I am particularly thankful to many faculty members and postdoctoral students for making the group an encouraging setting in which to begin presenting my own work in the early days. I'd like to especially thank Nadav Drukker, Dionysios Anninos and Peter West for their consistent support both in and outside of research.

But where would I be without my PhD pals? I'd like to first thank Malte, whose unrelenting good humour has kept me from ever taking things too seriously. The team we formed from the very beginning along with Maxime and Julius has many times propelled me out of ruts of frustration, for which I will always be grateful. Thanks also go to Manya, the most dependable friend, and to Tristan, who's a real pain. And last but very much not least is my good pal Michaella, who has never failed to make anything a whole lot more fun. Thank you for always giving me a very good reason to come into the office. \\

I'd like next to thank my collaborators Paul Richmond and Arthur Lipstein, alongside whom much of Part II of this thesis was produced. What began as a small project in late 2019 blossomed into a productive and very enjoyable collaboration. I am grateful to both of them not only for our work together, but also for their invaluable career advice and support as I navigate the road ahead.\\

And now, the big guns. I am indebted to my PhD supervisor, Neil Lambert, for his constant support and guidance over the last four years. After helping greatly in the early days as I got up to speed in the research world, in recent years our collaboration has thrived, becoming a great source of enjoyment and motivation. I can only hope that my future collaborations are as good fun.

I would be remiss to not also thank all of my family and friends who have supported me over the last few years. I could start reeling off all their names but it would get silly quite fast; and let's be honest, none of them will ever read this. But I am especially grateful to my parents, Myra and David, and sister Satya, for providing a comforting home to which I can always return.

Finally, I cannot thank Sally enough for her complete and unwavering encouragement and support throughout the course of this PhD. Even a bad day at the office has always ended with at least a laugh or two, while an especially good day has been been celebrated in style. And I must of course also thank Saffy, without whose constant cries for attention and food\footnote{She is a cat}, this PhD would have been much too easy.\\[3em]

\noindent\textbf{Added August 2021:} I would also like to thank my examiners, Kostas Skenderis and Jelle Hartong, for all their work in assessing this thesis.

\newpage
\lhead{\textsl{\nouppercase{\leftmark}}}
\renewcommand{\baselinestretch}{1.2}\normalsize
\tableofcontents
\renewcommand{\baselinestretch}{1.3}\normalsize

\chapter*{Introduction}
\addcontentsline{toc}{part}{Introduction}
\chaptermark{Introduction}
\lhead{\textsl{\nouppercase{Introduction}}}


If one were to try to describe theoretical physics in a sentence---perhaps to feature in a pithy thesis introduction---they might call it the task of pairing models with phenomena. Whether the phenomenon is at home, in the laboratory, or in a lofty Theory of Everything, this is what it ultimately boils down to. And so it is in many ways remarkable that across this breadth of topics lies a ubiquitous tool: symmetry. 

On one hand, one may begin with an ordained symmetry. This could come from some observation made of a phenomenon, or some consistency condition on the system you aim to describe. From here, the task is to find a theory that exhibits this symmetry. Indeed, in high energy theory one often recasts this as a classification problem, seeking some generic structure for whole classes of such theories. Such a perspective has for instance proven hugely fruitful in the study of supersymmetric gauge theories in diverse dimensions.

Conversely, given a theory with a given symmetry, it is natural---and often extremely powerful---to ask how these symmetries constrain the theory's predictions. Indeed, in the growing zoo of interesting non-Lagrangian theories, it is often \textit{only} through such approaches that we can hope to extract any information. Nowhere has such a perspective proven more profound than in conformal field theory, where the once rudimentary implications of conformal symmetry on correlators have evolved into strikingly general constraints on CFT data. 

This thesis approaches symmetries from both of these perspectives, for the particular topic of field theories with an inhomogeneous scaling symmetry, and often some degree of supersymmetry. These theories forgo Lorentz symmetry, often held sacred in quantum field theory, but in doing so offer exciting new opportunities both in their construction and application. \\


The study of quantum field theory, and in particular supersymmetric gauge theories, is inextricably tied to the study of branes in string and M-theory. Indeed, it is through the very definition of D-branes as the anchors of open strings that we know to describe their low energy dynamics with super-Yang-Mills theories. From here, one enters a playground of brane constructions, leading to profound and powerful results on the perturbative and non-perturbative properties of such theories.

M-theory is not so kind. Its extended objects---the M2-- and M5--branes---shy away from such immediate formulation, and instead we are forced to follow a more pragmatic approach, guided by symmetry. Thankfully, the decoupling limit from M-theory implies a rich superconformal structure on the branes, providing a strong constraint on the possible field theories describing low energy dynamics. Nonetheless, the path to constructing such models---particularly for an arbitrary number of branes---is far from simple, and in the case of the M5-brane, far from complete. \\


An overarching motivation for much of the work in this thesis is the use of non-Lorentzian supersymmetric Lagrangians towards a useful action principle for the $(2,0)$ theory of M5-branes. However, at every step along the way, the focus is really on abstraction, and generalisation.

In contrast to the amount known about super-Poincar\'{e} and superconformal field theories, less is known about supersymmetric theories with more exotic spacetime symmetries. This work provides foundational results in the construction and study of some such models, which may apply far more broadly than to just M-theory. Further, this thesis provides fundamental results on theories in five dimensions with an exotic $SU(1,3)$ symmetry group---which includes a inhomogeneous scaling---regardless of any supersymmetry. These results evoke the basic construction of conformal field theory, and lay the groundwork for a more general modern treatment of field theories with some (not necessarily homogeneous) scaling symmetry.

\begin{center}
\begin{tcolorbox}[width=0.9\textwidth,
                  boxrule=0pt,toprule=0pt,
                  colback=boxback,
                  arc=10pt,
                  adjusted title={\centering \Large\bf The structure of this thesis }
                  ]
\vspace{0.5em}
The remainder of this Introduction consists of a brief outline of the results of this thesis, followed by two reviews of introductory material providing useful context for these results.\vspace{0.5em}

Then, the main body is split into two Parts. While each Part is self-contained, the fundamental result of the former is instrumental to much of the latter. Each Part begins with an introductory chapter describing its key results, and concludes with a more in-depth discussion of these results, their implications, and potential future directions.\vspace{0.5em}

There is finally a number of Appendices, which provide more detailed calculations and formulae to support a number of results in the main body.\vspace{0.5em}

\end{tcolorbox}
\end{center}
The starting point of Part I is the construction of a technique in which a classical RG flow is induced upon a supersymmetric Lorentzian field theory. Under some mild assumptions, this flow is shown to produce at its fixed point a supersymmetric theory generically without Lorentz symmetry, but with an inhomogeneous---or Lifshitz---scaling symmetry\footnote{Here and throughout, we abuse notation slightly and refer to a scaling symmetry under which each coordinate may scale differently as either inhomogeneous or Lifshitz, interchangeably. We do not necessarily mean that the symmetries include the entire Lifshitz \textit{group}, although in Chapter \ref{chap: DLCQ} we will see an example of a theory with Schr\"odinger symmetry, which includes the Lifshitz group}. This technique is demonstrated explicitly for the $\mathcal{N}=2$ super-Yang-Mills theory in five-dimensions, relevant for M5-branes, as well as for the Chern-Simon-matter theories relevant to M2-branes.

A striking general feature of these fixed point models is a reduction of their dynamics to the moduli space of some BPS soliton of the parent theory. This phenomenon is explored explicitly first for an instructive model in $(1+1)$-dimensions, with dynamics reducing to the moduli space of half-BPS kinks. We then return to the five-dimensional Yang-Mills-like model, whose dynamics are shown to reduce to a superconformal quantum mechanics on the moduli space of instantons. In particular, we arrive at an explicit Lagrangian realisation of the DLCQ proposal for the M5-brane. 


In Part II, we change gears somewhat, and first build from the ground up the theory of five-dimensional field theories with a curious $SU(1,3)$ geometric symmetry. In addition to their Lifshitz scaling symmetry, such theories have a number of exotic features such as symmetry under a non-Abelian translation subgroup. In analogy with the standard treatment of conformal field theories, we investigate the consequence of this symmetry on correlation functions, determining its constraining power to lie somewhere between that of the five-dimensional Poincar\'{e} and conformal groups.

Next we explain how such models arise naturally in the context of more conventional conformal field theory. Although these $SU(1,3)$ theories can be studied in their own right, we demonstrate that there exists a conformal compactification of six-dimensional Minkowski space on which the modes of \textit{any} six-dimensional conformal field theory are described by such an $SU(1,3)$ theory. Owing to the relative power of the six-dimensional conformal group, we derive further necessary conditions on the correlation functions of an $SU(1,3)$ theory to admit such a six-dimensional interpretation. We additionally demonstrate how the classic DLCQ picture is found in a particular degenerate limit of our construction.

With this framework built, we study an explicit $SU(1,3)$ gauge theory with a high degree of supersymmetry, which is conjectured to provide a Lagrangian description of the non-Abelian $(2,0)$ theory on non-compact six-dimensional Minkowski space. We first outline how this model was derived, through an application of the technique of Part I to a particular holographic M5-brane setup. We verify explicitly that the model enjoys a classical $SU(1,3)$ spacetime symmetry when fields are regular over $\mathbb{R}^5$, thus describing the zero modes of the compactification from six dimensions. However, we find that the theory indeed describes higher Kaluza-Klein modes too, when the configuration space is extended to allow singular behaviour of the gauge field strength at isolated points. The resulting non-trivial gauge bundles on the punctured space are naturally encoded in the path integral by disorder operators called instanton operators. Then, local operators inserted at such singular points are identified precisely with non-zero Kaluza-Klein modes of the conformal compactification, thus making concrete the six-dimensional interpretation of the theory.

Like the theories in Part I, the dynamics of this $SU(1,3)$ Lagrangian model for M5-branes are constrained to a moduli space, defined by a particular deformation of the familiar anti-instanton equation. We finally explore a broad family of solutions to this equation, which are shown to describe the propagation of anti-instantons between the singular points.


Before we get going, however, we present a couple of topic reviews, providing useful context for the results of this thesis.

\section*{\makebox[4.62mm][l]{$\boldsymbol\alpha$} M-theory and its branes}
\addcontentsline{toc}{section}{{$\alpha$}\hspace{3.0mm} M-theory and its branes}


There are many ways in which one might define M-theory, depending on taste. To pay respect to its origin, one may define it as the strong coupling limit of Type IIA string theory \cite{Townsend:1995kk,Witten:1995ex}. In this picture, one identifies threshold bound states of $N$ D0-branes in Type IIA string theory as the $N^\text{th}$ Kaluza-Klein modes of the supergraviton of eleven-dimensional supergravity on a circle of radius $R_{11}=\ell_s g_s$. Then, as $g_s\to\infty$ and the $D0$ states become light, the string theory unfurls into a new eleven-dimensional theory, whose low-energy limit is precisely eleven-dimensional supergravity. This is what we call M-theory. One can also play a similar game with the $E_8\times E_8$ heterotic theory, which is identified as M-theory compactified on an interval of length $\pi R_{11}$, and thus bounded by a pair of end-of-world branes \cite{Horava:1995qa,Horava:1996ma}. There are a multitude of review articles \cite{Townsend:1996xj,Schwarz:1996bh,Vafa:1997pm,Sen:1998kr,Obers:1998fb} and books \cite{Becker:2007zj,Polchinski:1998rr} that discuss the intricate web of M-theory and superstring dualities.

Crucial to our understanding of string theory are the half--BPS D-branes, whose stability is due to their electric or magnetic charge under some form field of ten-dimensional supergravity. So it is natural to ask, what are the stable supersymmetric branes of eleven-dimensional supergravity? Here, we have just one form field---a 3-form---and thus there are just two stable branes present in M-theory, both of which are half-BPS: the electrically charged M2-brane, and the magnetically charged M5-brane. 

With the M-theory interpretation of the D0-brane already in hand, we can continue to recover all of the D$p$-branes of IIA string theory from an M-theory perspective. For instance, a D2-brane is reinterpreted as an M2-brane lying \textit{transverse} to the M-theory circle, while a D4-brane is an M5-brane \textit{wrapping} the circle. Indeed, as we shall soon see, these identifications correspond to interesting field theory relationships, with the theories lying on M-branes identified as the strong coupling limit of the Yang-Mills theories of D-branes in string theory.   \\

Of course, the idea of M-theory as the strong coupling limit of IIA string theory may be viewed not as a definition but as merely a \textit{requirement} of a theory that warrants a more fundamental construction. The search for such a definition has in turn provided a driving force behind a number of striking and fruitful directions in modern high energy theory. Many of the resulting constructions leverage some manifestation of gauge/gravity duality to conjecture fundamental definitions of M-theory in certain background geometries. Let us now review some of these ideas, as they provide useful context for a number of results in this thesis.

The first such construction that came along, dubbed \textit{Matrix theory}, is at its core a modern avatar of an old idea due to Weinberg known as the \textit{infinite momentum frame} \cite{Weinberg:1966jm}. The broad idea here is that by performing a very large boost in a Lorentzian theory, the resulting dynamics is dominated by processes with strictly positive momentum in the opposite direction. By reformulating this idea in terms of the older still lightcone quantisation of Dirac \cite{Dirac:1949cp,Susskind:1967rg,Bardakci:1969dv,Chang:1968bh}, and regulating the limit by way of a null compactification \cite{Maskawa:1975ky}, one arrives at Discrete Lightcone Quantisation (DLCQ). Weinberg's original argument becomes the mantra that upon null compactification, all states in the theory with negative null momentum decouple. Additionally, the zero mode becomes non-dynamical, and so can \textit{in principle} be integrated out, resulting in a Fock space containing only states of strictly positive null Kaluza-Klein momentum. Despite a slew of conceptual difficulties in dealing with the zero mode---sometimes dubbed the \textit{zero mode problem} \cite{Nakanishi:1976vf}---from here began a long and fruitful application of these ideas for their original purpose: the simplification of computation in perturbative quantum field theory. Our interest here, however, is rather different. Following this DLCQ mantra, the central idea of Matrix theory is that upon null compactification, M-theory is wholly described by the dynamics of states with strictly positive null Kaluza-Klein momentum \cite{Banks:1996vh,Susskind:1997cw}. As we've seen, these are the bound states of $N$ D0-branes in Type IIA string theory. Then, one arrives at the striking proposal that the $N^\text{th}$ sector of the DLCQ of M-theory is precisely described by the worldvolume action of $N$ D0-branes; an $N\times N$ matrix quantum mechanics. Excitingly, this construction then admits a decompactification limit, where the large $N$ limit of this matrix model is taken to \textit{define} M-theory on eleven non-compact dimensions. More relevant, however, for this thesis is the related DLCQ description of the M5-brane, which is detailed below. 

Not long afterwards came another realisation of gauge/gravity duality in the form of the \textit{AdS/CFT correspondence}. In its original and simplest form, a complete equivalence is proposed in the large $N$ limit between the non-Abelian gauge theory on a stack of $N$ branes in string/M-theory, and the string/M-theory living on the branes' asymptotically anti-de-Sitter near horizon geometry. The original basic examples \cite{Maldacena:1997re,Aharony:1999ti} defined a correspondence for the following superconformal branes:\vspace{-2mm}
\begin{itemize}
  \item $N$ D3-branes $\quad\longleftrightarrow\quad$ Type IIB string theory on $\text{AdS}_5\times S^5$\vspace{-3mm}
  \item $N$ M2-branes $\quad\longleftrightarrow\quad$ M-theory on $\text{AdS}_4\times S^7$\vspace{-3mm}
  \item $N$ M5-branes $\quad\longleftrightarrow\quad$ M-theory on $\text{AdS}_7\times S^4$\vspace{-2mm}
\end{itemize}
In particular, the latter two dualities can be viewed as \textit{definitions} of M-theory, at least on the respective geometries. For these examples as well as a plethora of extensions and generalisations, the AdS/CFT correspondence has been checked in more and more sophisticated ways over the decades since its inception. It stands now as an almost ubiquitous tool of high energy theory, both guiding our hand towards a more complete theory of quantum gravity, while conversely offering a toolbox for computations in strongly coupled quantum field theories.   \\

The remainder of this review is concerned with the study of the superconformal field theories living on stacks of M-theory branes. An understanding of such models provides us not only with a formulation of the fundamental constituents of M-theory, but with a description of M-theory as a whole via the AdS/CFT correspondence. Further, the M2-- and M5--brane models are just two of a much broader zoo of superconformal field theories, whose rich structure has motivated a surge of interest in recent years. Our particular focus will be the journey towards useful Lagrangians for these theories, and the roadblocks along the way.

\subsection*{M2-branes}

M-theory includes BPS branes coupled electrically to the 3-form of eleven-dimensional supergravity. Symmetry alone tells us that the field theory describing a stack of $N$ M2-branes is some three-dimensional maximally supersymmetric ($\mathcal{N}=8$) superconformal field theory (SCFT), which should have eight scalar fields and an $SO(8)$ R-symmetry corresponding to directions transverse to the branes.

 Following the IIA/M-theory dictionary, this SCFT arises from $N$ D2-branes in the limit that the Type IIA coupling becomes large, which corresponds to the gauge theory coupling also becoming large. These D2-branes are in turn described by three-dimensional maximally supersymmetric Yang-Mills (MSYM) with gauge group $U(N)$, with strong coupling realised in the deep IR. Thus, three-dimensional MSYM is predicted to admit an IR fixed point, at which its R-symmetry is enhanced $SO(7)\to SO(8)$. Indeed, this enhancement is seen directly in the case of a single brane, where the Abelian gauge field is dualised to provide an eight scalar field that becomes non-compact at strong coupling \cite{Townsend:1995af,Schmidhuber:1996fy,Bergshoeff:1996tu} .\\
 
 In trying to write down a Lagrangian for such a theory, the trouble begins when we want to describe multiple branes. For D-branes this goes smoothly, starting with a $U(1)$ super-Yang-Mills theory describing a lone brane and generalising to the corresponding non-Abelian $U(N)$ theory for $N$ coincident branes. Once again, M-theory is not so kind. Indeed, there are a number of curious features this $\mathcal{N}=8$ SCFT must satisfy which for a long time made a Lagrangian description seem out of reach. For instance, aside from not containing any continuous parameters, one expects the number of degrees of freedom to grow as $N^{3/2}$ at large $N$ \cite{Klebanov:1996un}.
 
However, despite these conceptual roadblocks, one finds salvation in the form of Chern-Simons-matter theories. The first such models applied to M2-branes are the Bagger-Lambert-Gustavsson (BLG) theories \cite{Bagger:2007jr,Gustavsson:2007vu}, whose Lagrangians realise the full $\mathcal{N}=8$ supersymmetry as well as $SO(8)$ R-symmetry manifestly. Their construction in the language of 3-algebras, however, leads to restrictions on the allowed gauge algebra, as well as subtle dependence on the choice of global gauge group. The upshot is that these models describe either two or three M2-branes, but no more.\\

To achieve a description of an arbitrary number of branes, some concessions must be made---and manifest symmetries lost. The generalisation of the BLG theory to an arbitrary number of branes was made in the form of the ABJM Chern-Simons-matter theory \cite{Aharony:2008ug}. The theory has gauge group\footnote{One may more generally consider the so-called ABJ theories, with gauge group $U(N)\times  U(M)$ \cite{Aharony:2008gk}} $U(N)\times U(N)$, and a single discrete coupling $g=1/k$ with $k\in\mathbb{N}$. It describes $N$ coincident M2-branes probing a $\mathbb{C}^4/\mathbb{Z}_k$ transverse geometry, with $\mathbb{Z}_k$ acting as a $k$-fold overall phase shift on the four complex coordinates. For generic $k$, such a theory should have only $\mathcal{N}=6$ superconformal symmetry, and $SU(4)\times U(1)$ R-symmetry; this is indeed the case for the ABJM model.

It is interesting and will later prove useful to understand the ABJM model and its symmetries from a holographic perspective; indeed, it was in the context of the AdS/CFT correspondence that the theory was originally derived. The near horizon geometry of $N$ M2-branes on $\mathbb{C}^4=\mathbb{R}^8$ is $\text{AdS}_4\times S^7$, with $N$ units of magnetic flux on the $S^7$. To understand how this geometry is changed when we replace $\mathbb{C}^4$ with the orbifold $\mathbb{C}^4/\mathbb{Z}_k$, first view the $S^7$ as a (non-trivial) circle fibration over complex projective space, $S^1 \hookrightarrow S^7 \rightarrow \mathbb{CP}^3$. One finds that the $\mathbb{Z}_k$ quotient acts at each point on $\mathbb{CP}^3$ as a $k$-fold phase shift over the $S^1$ fibre, defining an orbifold $S^7/\mathbb{Z}_k$. Then, the ABJM theory at level $k$ is dual to M-theory on $\text{AdS}_4\times S^7/\mathbb{Z}_k$. At generic $k$, this geometry preserves only three-quarters of the supersymmetry, the $\mathcal{N}=6$ supersymmetry of the ABJM model. Additionally, the spacetime symmetry group is a product of $SO(2,3)$, the isometry group of $\text{AdS}_4$ or equivalently the conformal group in three-dimensions, along with the subgroup $SU(4)\times U(1)\subset SO(8)$ of the total isometries of $S^7$ which preserve the $\mathbb{Z}_k$ orbifold. In particular, $SU(4)$ is the isometry group of $\mathbb{CP}^3$, while the $U(1)$ factor translates along the orbifolded $S^1$ fibre. These are indeed the spacetime and R-symmetry groups, respectively, of the ABJM model.\\

Although the theory has no continuous parameters, it still makes sense to think about strong ($k\sim 1$) and weak ($k>>1$) coupling regimes. In particular, in taking $k$ large to approach weak coupling, we return to where we started: a stack of D2-branes \cite{Mukhi:2008ux}. In the large $k$ limit on the gravity side, the $S^1$ fibre shrinks to zero radius and the bulk M-theory is described by Type IIA string theory on $\text{AdS}_4\times \mathbb{CP}^3$. 

The situation at strong coupling is more subtle, with a full appreciation of the quantum theory requiring some new tools. Our focus once again will be on symmetries. In particular, for $k=1$, there is no orbifold and we seem to have an issue: the theory describes M2-branes on flat space $\mathbb{C}^4=\mathbb{R}^8$, and so \textit{should} realise the full $SO(8)$ R-symmetry group as well as the whole $\mathcal{N}=8$ superconformal symmetries. To realise these symmetries, one must extend the configuration space to allow for singular behaviour of the gauge field strength $F$ at isolated points, around which one finds non-vanishing (and quantised) magnetic charge. Such configurations are naturally encoded in the path integral in terms of \textit{monopole operators}, using which one can construct additional Noether currents and demonstrate the enhancement of the theory's supersymmetry and R-symmetry.\\

The construction of three-dimensional superconformal field theories as Chern-Simons-matter theories marked a new direction in the study of brane dynamics, prompting sustained attention in the following years. Despite posing new challenges, these theories have proven invaluable in the study of M2-branes and their holographic dual, with for instance localisation techniques proving successful in recovering the expected $N^{3/2}$ scaling in degrees of freedom at large $N$ \cite{Drukker:2010nc}. 

A comprehensive review of the origins, construction and analysis of models of multiple M2-branes can be found in \cite{Bagger:2012jb}.

\subsection*{M5-branes}


M-theory also includes BPS branes coupled magnetically to the 3-form of eleven-dimensional supergravity. The theory on a stack of $N$ M5-branes must then be some interacting six-dimensional field theory with $(2,0)$ superconformal symmetry, which should have five scalar fields and an $SO(5)$ R-symmetry corresponding to directions transverse to the branes.

Let us again appeal to the IIA/M-theory dictionary. This tells us that if we wrap our stack of M5-branes on the M-theory circle of radius $R_{11}$, we arrive at $N$ D4-branes in Type IIA string theory. Now, these branes are described once again by a super-Yang-Mills theory; this time, it's five-dimensional maximal ($\mathcal{N}=2$) super-Yang-Mills (MSYM) with gauge group $U(N)$ and coupling $g_\text{YM}=4\pi^2 R_{11}$. In contrast to the M2-brane case, the D4-branes approach strong coupling in the deep UV; indeed, five-dimensional MSYM makes sense at best as an effective field theory, since it is power-counting non-renormalisable. Hence, the theory on $N$ M5-branes, also known as the non-Abelian $(2,0)$ theory, must provide a UV completion of five-dimensional MSYM. Further, at its UV fixed point the theory must grow an extra dimension, with its spacetime symmetry being enhanced as $SO(1,4)\to SO(1,5)$, and indeed to the full six-dimensional conformal group $SO(2,6)$.\\

The task of constructing a Lagrangian for the non-Abelian $(2,0)$ theory is daunting. There are a number of very good reasons why we shouldn't expect any old conventional field theory to cut it. Chief amongst them, perhaps, is that \textit{any} interacting six-dimensional Lagrangian is power-counting non-renormalisable, or else has a Hamiltonian unbounded from below. Despite this and a number of other concerns, progress has been made in a variety of directions towards a useful action principle for the $(2,0)$ theory. Two of these proposed descriptions are particularly relevant to the results of Part II, and both rely heavily on the realisation of the eleventh dimension of M-theory from the perspective of D4-branes in Type IIA. The Kaluza-Klein modes required for the lift of D4-branes to M5-branes are precisely described by D0-branes within the D4-brane worldvolume, or rather \textit{D0-D4 bound states}. From the perspective of the Yang-Mills theory on the D4-branes, these D0-branes are particle-like topological solitons called \textit{Yang-Mills instantons}.

From here comes fairly naturally a construction that does for the M5-brane what Matrix theory did for M-theory as a whole. In this so-called DLCQ proposal for the M5-brane \cite{Aharony:1997th,Aharony:1997an}, one considers $N$ M5-branes compactified on a null direction. As we saw for Matrix theory, the DLCQ mantra then stipulates that the theory is wholly captured by the states of the theory with Kaluza-Klein momentum\footnote{This sign is of course a matter of convention. By considering compactification on the opposite null direction, one arrives at $n<0$} $n>0$; these are states of $n$ instantons in the $U(N)$ super-Yang-Mills on the D4-branes. Indeed, one finds that finite energy requires slow motion of these particle-like instantons, which by the standard geodesic approximation \cite{Manton:1981mp} is effectively described by a quantum mechanics, with target space given precisely by the \textit{moduli space} of $n$ instantons of $SU(N)$, denoted $\mathcal{M}_{N,n}$. Thus, the M5-brane counterpart to Matrix theory claims that we can describe the $n^\text{th}$ sector of the DLCQ of the $U(N)$ $(2,0)$ theory precisely by quantum mechanics on $\mathcal{M}_{N,n}$. However, questions over precisely how to deal with the zero mode sector of a field theory DLCQ---and indeed concerns that any attempt to do so will render the construction ill-defined \cite{Hellerman:1997yu}---have left this description of the $(2,0)$ theory somewhat ambiguous. Further, note that as with Matrix theory, this picture admits a decompactification limit, with the $U(N)$ $(2,0)$ on non-compact space recovered only in the $n\to \infty$ limit of this construction.

Another perspective, however, is that we're making a mountain out of a molehill, and that the answer was right in front of us all along. In Type IIA string theory, the careful inclusion of D0-brane bound states in the spectrum \textit{defines} an eleven dimensional theory. So surely the theory on $N$ D4-branes---namely five-dimensional MSYM---with the careful inclusion of instanton states simply defines a six-dimensional theory; namely, the non-Abelian $(2,0)$ theory with gauge group $U(N)$ \cite{Rozali:1997cb,Berkooz:1997cq,Seiberg:1997ax}. While this interpretation of the $(2,0)$ theory is conceptually concrete---in these sense it is as true as M-theory itself---it turns out that the use of ``simply'' is a little optimistic. The initial problem here is that five-dimensional MSYM is non-renormalisable, and so although we can probe the IR for evidence that instanton are indeed Kaluza-Klein states \cite{Lee:1999xb,Dorey:2001ym}, it's hard to say anything at all as we go to higher energies where new UV degrees of freedom are required. More recently, however, this picture has been turned on its head, with evidence to suggest that five-dimensional MSYM is, in its own weird way, actually UV complete when its configuration space is taken to include instantons in an appropriate fashion \cite{Lambert:2010iw,Douglas:2010iu,Papageorgakis:2014dma}. One can take this a step further and formally construct operators in five-dimensional MSYM that carry Kaluza-Klein momentum on the M-theory circle. In detail, one dresses a standard local operator with disorder operators $\I_n(x)$ called \textit{instanton operators}\cite{Lambert:2014jna}, which prescribe in the path integral a fixed instanton flux $n$ on a small $S^4$ surrounding the point $x$. In this way, one proposes that five-dimensional MSYM with such operators included is in fact a six-dimensional theory, albeit with one spatial dimension wrapped on a circle of finite radius which only unfurls in the strongly-coupled limit.

In many ways, the punchline of Part II of this thesis is a hybrid of these two ideas: the DLCQ of M5-branes, and the interpretation of a five-dimensional Yang-Mills-like theory as a six-dimensional field theory.


\section*{\makebox[4.62mm][l]{$\boldsymbol\beta$} Topological solitons}\label{sec: topological solitons}
\addcontentsline{toc}{section}{{$\beta$}\hspace{3.1mm} Topological solitons}

Central to a number of results in this thesis is the notion of a soliton. Such objects have appeared in physics, in one way or another, for almost 200 years. One can try to define solitons in some broad sense in terms of their stubbornness: they are objects in some physical theory or observation that are robust to perturbation. 

Our arena is field theory, and we can hone this definition significantly. Suppose we have a theory defined on some manifold $M$, and a field $\phi$ valued in some other manifold $N$, so that $\phi:M\to N$. Most of the time we'll take $M,N$ to just be $\mathbb{R}^n$ for some $n$. Suppose further that $M$ has boundary $\partial M\neq \emptyset$, and that we impose some boundary conditions on $\phi$; we stipulate that $\phi$ restricted to $\partial M$ is valued in some submanifold $N_\text{vac.}\subset N$, sometimes called the \textit{vacuum manifold}. Such a boundary condition often arises from the condition that the configuration $\phi$ has finite energy.

So, what configurations of $\phi$ are allowed? Well, it is certainly the case that any $\phi$ satisfying the boundary condition on $\partial M$ defines a map $\phi|_{\partial M}:\partial M \to N_\text{vac.}$. We can then classify such maps in the language of \textit{homotopy theory}, with $\phi|_{\partial M}$ belonging to some homotopy class of maps $\partial M \to N_\text{vac.}$. Indeed, the space of allowed configurations $\phi$ is stratified into subspaces, each categorised by the homotopy class of $\phi|_{\partial M}$. By the very definition of these homotopy classes, these subspaces are then disjoint; we can't perturb our way from one to another. For this reason, we sometimes refer to these subspaces as different \textit{topological sectors} of the theory. 

So whenever we talk about a configuration $\phi$, we should specify which topological sector we sit in. Then, we can do what we always do: seek solutions to the classical equations of motion. Except if we sit in a non-trivial sector, we're not even allowed to set $\phi=0$, and hence any solutions we find---known as \textit{topological solitons}---are necessarily not continuously connected to the trivial vacuum $\phi=0$. We have come full circle: no matter how we let our configuration evolve, we will never get to $\phi=0$. Indeed, we will never leave our topological sector. It is in this sense that these solitons are \textit{robust}.\\

The remainder of this review grounds this general theory through a couple of examples, both of which are important to the results of the thesis.

We will first explore kinks in $\sigma$-models in $(1+1)$-dimensions, where the vacuum manifold is simply a pair of disjoint points in $\mathbb{R}$. Here, we will encounter perhaps the simplest examples of almost all of the universal players in the study of solitons, such as topological invariants and moduli. Further, it is in this particularly straightforward setting that we realise the simplest example of the limiting technique of Part I, as discussed in Section \ref{sec: a simple kink model}.

We then provide the basics of instantons in four-dimensional Yang-Mills theory, which in many ways are the most fundamental topological solitons of gauge theory in any dimension. In particular, their existence as particle-like states in five-dimensional Yang-Mills-like theories will prove crucial in the full quantum analysis of the $SU(1,3)$ gauge theories discussed in Part II.


\subsection*{Kinks in $\sigma$ models}
\newcounter{BetaEquations}
\stepcounter{BetaEquations}


While topological solitons come in all shapes and sizes, it is instructive to first consider perhaps the simplest field theory setting in which they are encountered. We consider a $(1+1)$-dimensional $\sigma$-model with $d$-dimensional pseudo-Riemannian target manifold $(M,g)$. This is a theory of scalar fields $\phi^i(t,x)$, $i=1,\dots, d$ that are interpreted as the coordinates of points in $M$. Thus, such a model describes the dynamics of some 1-dimensional extended object propagating in the ambient space $M$; in other words, a string. Indeed, such models recover precisely the standard formulation of the bosonic string in conformal gauge, coupled to background graviton $g$. 

\subsubsection*{Topological sectors}

For our purposes, we simply take $(M,g)=(\mathbb{R},\delta)$, and so we have a single scalar field $\phi$. Such $\sigma$-models are defined up to a generic function $W(\phi)$ on $M$ which controls the dynamics. We then have action
\begin{align}\tag{$\beta.\theBetaEquations$}
  S=\int dt\, dx\left(\frac{1}{2}\left(\phi_t^2 - \phi_x^2\right) - \frac{1}{2}W'(\phi)^2\right)\ ,
\end{align}\stepcounter{BetaEquations}%
where $\phi_t=\partial_t \phi$ and $\phi_x=\partial_x \phi$. Thus, we have simply the action of a single scalar field, propagating in a potential $V(\phi)=\frac{1}{2}W'(\phi)^2$, with equation of motion $\phi_{tt}-\phi_{xx}+V'(\phi)=0$. We can also consider the corresponding energy $E$ of some configuration at some time $t$, given by
\begin{align}\tag{$\beta.\theBetaEquations$}
  E= \int dx \left(\frac{1}{2}\left(\phi_t^2 + \phi_x^2\right) + \frac{1}{2}W'(\phi)^2\right)\ .
  \label{test}
\end{align}\stepcounter{BetaEquations}%
It is immediate that $V(\phi)>0$, and indeed we can shift $S$ by a constant to ensure that there exists at least some $\phi$ for which $V(\phi)=0$ is attained. But as is familiar from spontaneous symmetry breaking, let us suppose that $W'(\phi)$ and hence $V(\phi)$ has zeroes at a generic number of isolated points $p\in\mathcal{P}\subset \mathbb{R}$. Then, a necessary condition for some configuration $\phi$ to have finite energy is that at each time $t$, we have $\phi_\pm:=\lim_{x\to\pm\infty}\phi\in\mathcal{P}$. Crucially, whatever the values $\phi_\pm$ are, they are \textit{robust} under time evolution: it would require infinite energy to pop $\phi_+$ from one element of $\mathcal{P}$ to another, and similarly for $\phi_-$. In this sense, we find our first topological solitons, known as (scalar) \textit{kinks}. The configuration space of finite energy solution is then split into distinct \textbf{topological sectors}, defined by the asymptotic values of $\phi$. In the language of before, the \textbf{vacuum manifold} is $\mathcal{P}\subset \mathbb{R}$, and the finite energy configurations are categorised by the maps $\{-\infty,\infty\}\to \mathcal{P}$. We note in particular that if $\phi_+\neq \phi_-$, the trivial vacuum $\phi=0$ is forbidden, and we must work harder to determine what classical solutions in the sector look like. We often call such sectors \textbf{non-trivial}.

\subsubsection*{The Bogomol'nyi equation}

Up until now, we have not used the somewhat unconventional form of $V(\phi)$ in terms of the function $W(\phi)$. The motivation for this notation is manifest when we consider the classic \textbf{Bogomol'nyi trick} applied to our $\sigma$-model. The idea here is to look for finite energy, \textit{static} configurations which minimise $E$. After some manipulation, we can write
\begin{align}\tag{$\beta.\theBetaEquations$}
  E = \mp \left(W(\phi_+)-W(\phi_-)\right) + \frac{1}{2} \int dx\, \big(\phi_x \pm W'(\phi)\big)\big(\phi_x \pm W'(\phi)\big)\ .
\end{align}\stepcounter{BetaEquations}%
Thus, we find $E\ge |W(\phi_+)-W(\phi_-)|$. Importantly, $|W(\phi_+)-W(\phi_-)|$ depends only on $\phi_\pm$, which are fixed depending on which topological sector we lie in. In other words, it is a \textbf{topological invariant}.

Then, in each sector, we achieve the minimal energy precisely if:
\begin{align}\tag{$\beta.\theBetaEquations$}
  \left\{\begin{aligned}
  \phi_x - W'(\phi) = 0 \quad &\text{if } W(\phi_+)-W(\phi_-)\ge 0\quad\text{(kink)}\\
  \phi_x + W'(\phi) = 0 \quad &\text{if } W(\phi_+)-W(\phi_-)\le 0\quad\text{(anti-kink)}\ .
  \end{aligned}\right.
\end{align}\stepcounter{BetaEquations}%
In particular, in a topologically-trivial sector we have simply $\phi(x) = \phi_+=\phi_-$. Furthermore, it is easy to show that any solution to the first-order equation $\phi_x\pm W'(\phi)=0$ satisfies the second-order equation of motion $\phi_{xx} = V'(\phi) = W'(\phi)W''(\phi)$. And thus we learn, remarkably, that the static, minimal energy classical solutions must solve a first-order equation, known as the \textbf{Bogomol'nyi equation}.

\subsubsection*{The kink moduli space}

To arrive at a few final, universal soliton buzzwords---including the all-important moduli space---it is helpful to take a particular $W(\phi)=a^2 \phi - \frac{1}{3}\phi^3$ for some $a>0$, corresponding to the `double dip' potential $V(\phi) = \frac{1}{2}\left(\phi^2 - a^2\right)^2$. Then, we have precisely $\mathcal{P}=\{-a, a\}$, and thus configurations $\phi$ with finite energy are categorised by functions $\{-\infty,\infty\}\to \{-a,a\}$, of which there are precisely four.

We can then consider the solutions to the Bogomol'nyi equations in each of these four topological sectors. Clearly, if $\phi_+=\phi_-=\pm a$, then we have simply $\phi(x) = \pm a$ everywhere. It is the other two sectors that are more interesting. Take $\phi_+=-\phi_-=a$. Then, we find $W(\phi_+)-W(\phi_-)=\frac{4}{3}a^3>0$, and so we require $\phi_x = W'(\phi) = a^2 - \phi^2$. This is solved generally by $\phi(x)=a\tanh\left[a(x-y)\right]$, for any $y\in\mathbb{R}$. Thus, we have a simple, explicit form for the general static, minimal energy solution in this topological sector. It is easily checked that $\phi(x)$ solves the classical equations of motion. And crucially: we do indeed have that $\lim_{x\to \pm\infty}\phi(x) = \pm a$, for any $y$! The profile of these solutions is fixed, and is shown in Figure \ref{fig: kink}.
\begin{center}
\begin{minipage}{0.8\textwidth}
\centering
\captionof{figure}{The profile of a static kink, with $a=2$ and $y=1$}\label{fig: kink}
\includegraphics[width=100mm]{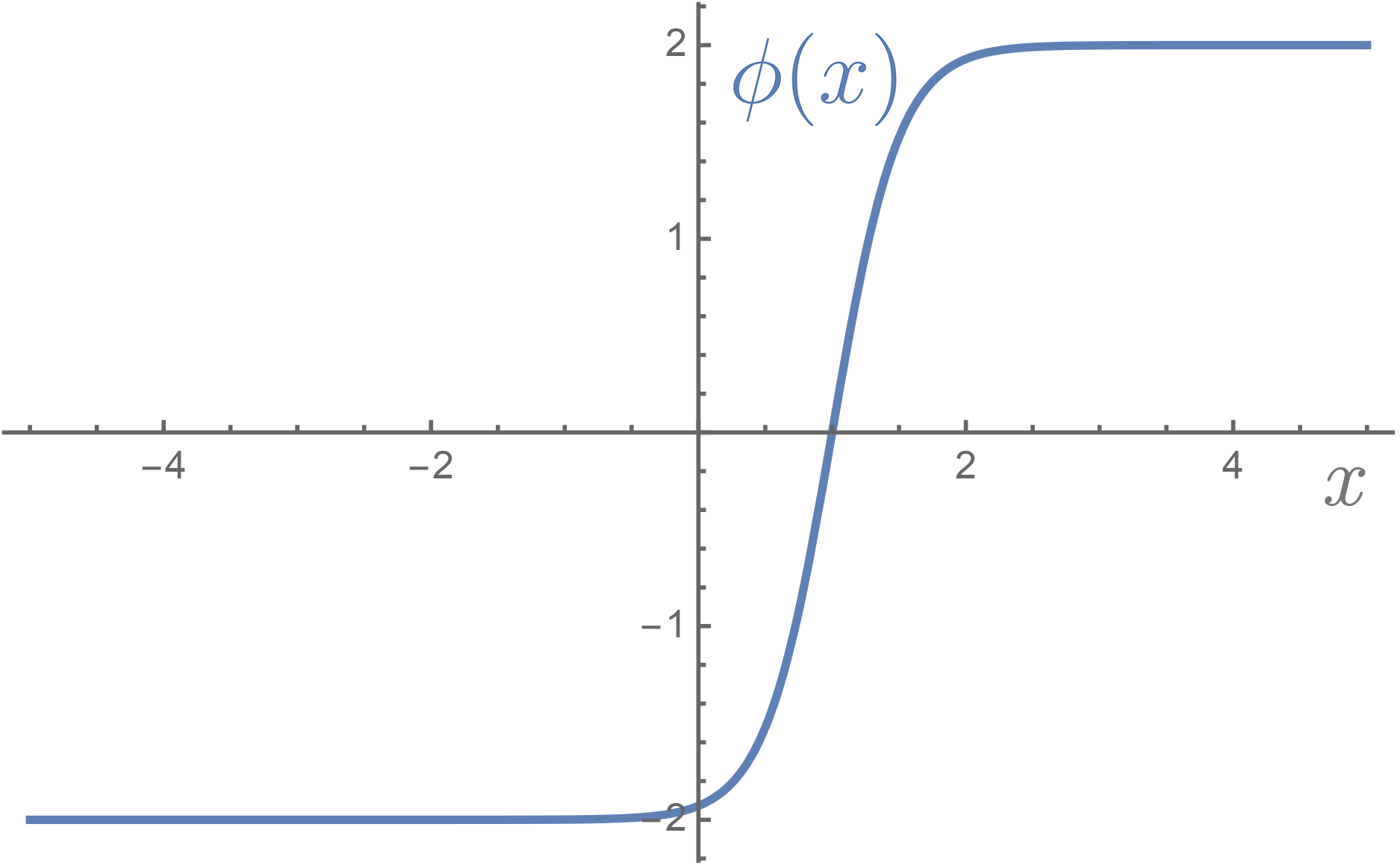}	
\end{minipage}
\end{center}
We see that $\phi$ stays near $\phi=\phi_-=-a$, but then as we approach $x=y$, the values suddenly flips up to near $\phi=\phi_+=+a$. Indeed, there is a region over which this transition largely occurs, with characteristic width $a^{-1}$. The lion's share of the kinks total energy comes from this region. In this sense, we can think of the kink as a lump of energy that is roughly localised around $x=y$.

Let us now drape some further jargon over this solution. Unsurprisingly, the solution to the Bogomol'nyi equation is not unique. But it is nonetheless very constrained: the full space of solutions in the sector is parameterised by a single real number $y$. We call $y$ a \textbf{modulus}. We can view $y$ as a coordinate on $\mathbb{R}$, in this way identifying $\mathbb{R}$ as the \textbf{moduli space}. 

With this static solution in hand, it is natural to ask: what about a moving configuration? If we want the resulting $\phi(t,x)$ to still be a classical solution, we can always Lorentz boost the static solution we've got, since the equations of motion are Lorentz-invariant. But another, simpler thing to do would be to simply allow the modulus $y$ to vary with time, and so arrive at $\phi(t,x)=a\tanh\left[a(x-y(t))\right]$. Such a solution will generically \textit{not} solve the equations of motion. However, so long as $y(t)$ is bounded for all time, $\phi(t,x)$ at any fixed time always lives in the same topological sector.\\

It is interesting to ask how useful such a form for $\phi(t,x)$ is if we wish to study dynamics in the $\phi_+=-\phi_-=a$ sector of the theory. After all, we have not claimed that under time evolution from some initial configuration, $\phi$ will remain in its Bogomol'nyi form; indeed this is \textit{not} the case in general. However, following a much more general principle known as the geodesic approximation \cite{Manton:1981mp}, if we assume that time evolution is slow\footnote{In this case, slow compared to the scale defined by $a$}, then one finds that the equations of motion are approximately solved by $\phi(t,x)=a\tanh\left[a(x-y(t))\right]$ provided $y(t)$ follows geodesic motion on the moduli space. The metric on the moduli space that defines this geodesic motion follows from the action $S$. The main point here is that if we start on the moduli space then we \textit{remain}, at least to first approximation, on the moduli space. We arrive at the striking revelation that we can study the small-velocity dynamics of the theory in some non-trivial sector by instead studying a quantum mechanical model on the soliton moduli space! This notion, which applies much more generally to models with solitons, laid the foundation for the successful study of the complicated interactions of slow-moving solitons, such as monopoles in Yang-Mills theories \cite{Manton:1981mp,Manton:1985hs,Gibbons:1995yw,Lee:1996kz}. \\

The notion of moduli, and their moduli spaces, is a very general concept that lies at the very heart of the study of all topological solitons. By virtue of the geodesic approximation, by understanding the moduli space of solutions, and the metric on it, we open a window to studying highly non-trivial dynamics in theories that seem otherwise impenetrable. It is therefore worth noting that, unfortunately, the ease with which we have studied this simple kink model is very rare indeed. More generally, even the question of the \textit{dimension} of the moduli space (as a topological manifold) is rather non-trivial, and has deep connections to the study of differential operators and their indices. To go a step further and find the global structure and metric on such moduli spaces is in many cases seemingly intractable. However, for the solitons of Yang-Mills theories most relevant to this thesis, the combined application of deep mathematics, physical intuition and even the properties of D-branes in string theory have lead to great steps forward in our understanding of these moduli spaces and their metrics.


\subsection*{Instantons and friends in Yang-Mills theory}


Before getting on with it, we have one last topic to briefly address: Yang-Mills instantons. These topological solitons play a central role throughout this thesis, and are indeed in many ways the most fundamental soliton of non-Abelian gauge theory.

\subsubsection*{A general outlook}

So, our arena now is gauge theory. Before jumping into instantons specifically, it is interesting to consider more broadly what kind of solitons we might find in a gauge theory. Let us first make a comment about the dimensionality---or rather codimensionality---of a soliton. Consider the $\sigma$-model of the previous section, and suppose we add a second spatial dimension $z\in[0,1]$. Such a model could for instance arise in the study of a long, narrow strip of some single-layer material. Then, we \textit{still} have our topological soliton: the kink (and anti-kink), given simply by $\tilde{\phi}(t,x,z)=\phi(t,x)$. In other words, we extend the solution trivially, making it constant in the $z$ direction. The centre of the resulting soliton is not a point but rather the line $(y,z)$ for all $z\in [0,1]$. What \textit{hasn't} changed is the codimension of the kink, which is still one (the $x$ direction). It is straightforward to see how this generalises: suppose we find a soliton in a $d$-dimensional theory that extends over $n$ dimensions. Such a soliton has codimension $(d-n)$, and so if we embed it in some theory of dimension $d'=d+\Delta$, we will still have a codimension $(d-n)$ soliton, which now extends over $(n+\Delta)$ directions. Hence, the number of dimensions over which some type of soliton extends depends on the total dimension of spacetime.

With this in mind, the important question to ask is: what codimensions $n$ are allowed for topological solitons in a generic gauge theory? Requiring that such solitons have finite energy when the number of spatial dimensions is also $n$ and thus they are particle-like, a beautiful argument due to Derrick \cite{Derrick:1964ww,Dunajski:2010zz} shows that, remarkably, we may only have $n=1,2,3$ or $4$. In fact, we have already at least in essence realised the $n=1$ case, which is more properly realised as a gauged form of the kinks of the previous section. In other contexts, especially theories in higher dimensions, these kinks are also known as \textit{domain walls}.

On the other end, the codimension $n=4$ solitons are realised by \textit{Yang-Mills instantons}. These solitons, which we explore next, exist in pure Yang-Mills theory. Our focus will be on their basic features, as well as on some of the early, most accessible constructions of explicit solutions to their Bogomol'nyi equations.

Finally, from instantons one can consider various compactifications to find Yang-Mills solitons with $n=2$ (monopoles) and $n=1$ (vortices, or lumps). 

\subsubsection*{The topology of $SU(2)$ instantons}

So let us now consider pure Yang-Mills theory in four flat Euclidean dimensions.
Following our previous discussion, by adding additional dimensions the point-like instantons in four dimensions become $d$-dimensional extended objects in $4+d$ dimensional Yang-Mills, and so in particular in a five-dimensional Lorentzian theory, instantons are realised as particles.

In the study of branes, one most often considers Yang-Mills theories with gauge group $U(N)$, with $N$ corresponding to the number of coincident branes. So long as one is wary of the global gauge group, we can decouple the Abelian factor and consider gauge group $SU(N)$. In fact, as is standard in the study of instantons, we will specialise to $SU(2)$. One can then construct instantons for $SU(N)$ gauge theory by considering embeddings of $SU(2)$ into $SU(N)$, although we will not explore these details here.

We have then the action,
\begin{align}\tag{$\beta.\theBetaEquations$}
  S= \int_{\mathbb{R}^4} \text{tr}\left(F\wedge \star F\right) = \frac{1}{2}\int_{\mathbb{R}^4} \text{tr}\left(F_{ij} F_{ij}\right) \ge 0\ ,
\end{align}\stepcounter{BetaEquations}%
where we have fixed the coupling. Here, $\star$ denotes the Hodge dual\footnote{We use the convention $\left(\star \alpha\right)_{i_p+1\dots i_4}=(1/p!)\,\varepsilon_{j_1\dots j_p i_{p+1}\dots j_4}\alpha_{j_1\dots j_p}$ for $p$-form $\alpha_p$. Then, $\alpha_p\wedge \star \beta_p = (1/p!)\alpha_{i_1\dots i_p}\beta_{i_1\dots j_p}d^4 x$} with respect to the Euclidean metric on $\mathbb{R}^4$, which satisfies $\star^2=(-1)^p$ on $p$-forms.

The (Hermitian) field strength $F$ is defined locally as $F=dA-iA\wedge A$ in terms of a (Hermitian) gauge field $A$. The corresponding Bianchi identity is then $DF:= dF-iA\wedge F+iF\wedge A = 0$. Varying $S$ with respect to $A$, we find equation of motion $D\star F=0$. We've also written $S$ in component notation, with respect to Cartesian coordinates $x^i$ on $\mathbb{R}^4$. Since we're using the Euclidean metric, we allow the indices $i$ to be up or down freely. Finally, the trace is just a matrix trace, and we make no particular choice of basis for $\frak{su}(2)$.\\

Following our study of kinks, what can we say at this stage about the configuration space, and in particular its topological sectors? Let us sharpen this question slightly by restricting ourselves to configurations with finite action. Such configurations must necessarily have vanishing $F$ at infinity. More formally, this implies that we can extend the theory to live on the one-point compactification $S^4=\mathbb{R}^4\cup \{\infty\}$. Then, a field configuration is some $SU(2)$-principal bundle $P\to S^4$, with connection $A$. Such bundles can be classified by their \textit{second Chern class}, which is in turn measured by
\begin{align}\tag{$\beta.\theBetaEquations$}
  n=\frac{1}{8\pi^2}\int_{S^4} \text{tr}\left(F\wedge F \right)\in\mathbb{Z}~,
\end{align}\stepcounter{BetaEquations}%
often known in the physics literature as the \textit{instanton number}. Thus, in analogy with the four topological sectors of the kink model, here we find a topological sector for every $n\in\mathbb{Z}$. Once we are in such a sector, no continuous perturbation of the gauge field $A$ will get us to another.

We have argued that at a topological level, the theory of finite action configurations on $\mathbb{R}^4$ is equivalent to the theory defined on the one-point compactification $S^4=\mathbb{R}^4\cup \{\infty\}$. It is quite non-trivial but nonetheless true that this equivalence continues to hold at the level of the action, its equations of motion, and their solutions \cite{Uhlenbeck:1982zm}. Further details of this argument, which relies heavily on the conformally-flat metric on $S^4$ and the conformal invariance of the equations motion, can be found for instance in \cite{Dunajski:2010zz}. \\

If we wish to do some physics, and in particular couple the theory to some matter, we should understand what this all means for the gauge field $A$. The important thing to note here is that $\text{tr}\left(F\wedge F\right)$ is closed\footnote{Of course, all four-forms in a four-dimensional theory are closed. However, $\text{tr}\left(F\wedge F\right)$ is still closed once we add additional dimensions, as can be seen by noting it can still be written locally as $d\omega_3(A)$}, and so locally we can write $\text{tr}\left(F\wedge F\right)=d\omega_3(A)$ where $\omega_3(A)=\text{tr}\left(A\wedge dA-\frac{2i}{3}A\wedge A\wedge A\right)$ is the Chern-Simons three-form. Now, if $A$ and hence $\omega_3(A)$ are defined globally over $S^4$, then $\text{tr}\left(F\wedge F\right)$ is exact, and so the fact that $S^4$ is closed implies that we have simply $n=0$. We see therefore that we can only have non-zero $n$ when $\omega_3(A)$ is \textit{not} globally defined, and $\text{tr}\left(F\wedge F\right)$ represents a non-trivial element of the de Rahm cohomology group $H^4(S^4)$.

We can understand a little better how the integer $n$ arises by considering how to construct the bundle $P$ for generic $n$. Indeed, the construction of bundles over spheres is made relatively simple by use of the \textit{clutching construction}, which tells us that it is sufficient to cover $S^4$ with just two patches $U_N,U_S$---topologically \textit{hemispheres}---which have a small overlap $\Sigma\cong S^3$, often called the equator of $S^4$. Then, we can always find well-behaved gauge fields $A_N,A_S$ on $U_N,U_S$, respectively. The bundle is sealed together on the equator $\Sigma$ by a single transition function $t:\Sigma\to SU(2)$, which on $\Sigma$ relates the two gauge fields as $A_N=t^{-1} A_S t + i\,t^{-1}dt$. 

So we are now lead to consider what form this transition function $t$ can take. Since $\Sigma$ is homotopic to $S^3$, the maps $t:\Sigma\to SU(2)$ are classified by the homotopy group $\pi_3(SU(2))=\mathbb{Z}$. Indeed, the particular homotopy class of some $t$, called it's degree $\deg(t)$, can be computed by
\begin{align}\tag{$\beta.\theBetaEquations$}
  \deg(t) = \frac{1}{24\pi^2}\int_\Sigma t^{-1} dt\wedge t^{-1} dt\wedge t^{-1} dt \in \mathbb{Z}\ .
\end{align}\stepcounter{BetaEquations}Now, we only really care about the bundle $P$ up to different choices of local trivialisation on the $U_N,U_S$, i.e. up to gauge transformations enacted in each patch by some $g_{N/S}:U_{N/S}\to SU(2)$. Under such transformations, the transition function changes as $t\to g_N t g_S^{-1}$. It is non-trivial but nonetheless true that the degree of maps $S^3\to SU(2)$ is additive with respect to $SU(2)$ group multiplication, i.e. for two maps $f_{1,2}:S^3\to SU(2)$, we have $\deg(f_1 f_2) = \deg (f_1) + \deg(f_2)$. Crucially, we must be able to continue $g_{N}$ and $g_S$ smoothly throughout the contractible patches $U_N$ and $U_S$, respectively, and so we have $\deg(g_N|_\Sigma)=\deg(g_S|_\Sigma)=0$. Thus, we learn that $\deg(t)$ is unchanged; it is a \textit{topological invariant}. Note that in the case that $\deg(t)=0$, we can choose $g_N=\mathds{1}_2$ and $g_S$ such that $g_S|_\Sigma=t$. Then, $t$ becomes simply the identity map and we have successfully sealed up the bundle with trivial transition function: the bundle $P$ is trivialisable. If instead $\deg(t)\neq 0$, this is never possible, and we have a truly non-trivial bundle.

We can now finally relate the instanton number $n$ to the degree $\deg(t)$ of the transition function $t$. We calculate,
\begin{align}
  n &= \frac{1}{8\pi^2}\left(\int_{U_N}d\omega_3(A_N) + \int_{U_S}d\omega_3(A_S) \right)	\nn\\
  &= \frac{1}{8\pi^2}\int_\Sigma \Big(\omega_3(A_N) - \omega_3(A_S)\Big)		\nn\\
  &= \frac{1}{24\pi^2}\int_\Sigma t^{-1} dt\wedge t^{-1} dt\wedge t^{-1} dt = \deg(t)\ .
  \tag{$\beta.\theBetaEquations$}\label{eq: n from deg}
\end{align}\stepcounter{BetaEquations}%
Hence, the instanton number $n$ corresponds precisely to the number of times the transition function `winds' around $SU(2)$ as we go around the equator $\Sigma$.\\

It is finally useful to connect a little more closely with the physics literature; after all, physicists generally don't want to have to worry about defining fields only in patches! So let us review some of the ways in which this issue is skirted in many physical applications. In the above construction, with our view of $\mathbb{R}^4\cup\{\infty\}\cong S^4$ as the union of a northern and southern hemisphere $U_N,U_S\cong B^4$, respectively, let us identify the north pole with the origin, and the south pole as the point at infinity. Now, we can always move the equator southwards, so that it becomes more like an Antarctic Circle. In fact, let us continue to move it until it is an infinitesimally small $S^3$ encircling the south pole. Further, the clutching construction tells us we may always choose $A_S=0$ while retaining full generality for the bundle $P$. Then, we can view our theory living just on the patch $U_N=S^4\setminus \{\infty\}=\mathbb{R}^4$; we're back to where we started. But crucially, we've fixed the boundary conditions for $A_N$ as we approach $\partial \mathbb{R}^4=\Sigma\cong S^3$, where we must find the asymptotics $A_N\sim t^{-1} A_S t + i\,t^{-1}dt=i\,t^{-1}dt$. We have recovered the standard statement, that $A=A_N$ must approach pure gauge as we go to spatial infinity, with the instanton number $n$ given by the degree of $t$ as it winds around the $S^3$ at infinity. In the context of the first and simplest explicit form for $A$ in the $n=\pm 1$ sectors, known as the BPST instanton \cite{Belavin:1975fg}, this choice of patching is known as \textit{regular gauge}, for obvious reasons: the gauge field $A$ is regular throughout $\mathbb{R}^4$. We will look at this solution and others in more detail below.

However, let us consider the converse set-up: suppose we had shrunk the equator around the north pole---i.e. the origin---instead. We can once again set $A_N=0$, and so as we approach the origin in $U_S$, we must have $A_S\sim i\,t\, dt^{-1}$. This of course still gives rise to instanton number $n=\deg(t)$, as we have simply chosen a different covering of the same bundle. Indeed, we can see it directly by noting that although now $\frac{1}{8\pi^2}\int_\Sigma \omega_3(A_S) = \deg(t^{-1})=-\deg(t)$, this minus sign is cancelled by the opposite relative orientation of $U_S$ and $\Sigma$, i.e. $n = -\frac{1}{8\pi^2}\int_\Sigma \omega_3(A_S)$. Note further that if $\deg(t)\neq 0$, then $t$ winds non-trivially around the small $S^3$ surrounding the origin, and so $A_S$ necessarily does not have a finite limit as we approach the origin. Nonetheless, we can still do physics on $\mathbb{R}^4$, so long as we allow that $A=A_S$ is in fact singular at the origin. For this reason, this patching is known in the context of the BPST instanton as \textit{singular gauge}. From the perspective of physics on $\mathbb{R}^4$, such configurations are related to their regular gauge counterparts by gauge transformations that are regular only in the annulus $\mathbb{R}^4\setminus\{0\}$, which have precisely the effect of moving the equator from around the origin to around infinity.

One may shy away from singular gauge at first---who wants singularities---but in fact when we later come to construct the BPST instanton and in particular its multi-centred generalisation, we will find that it is \textit{only} when we allow for singular gauges that we may hope to write down simple and elegant forms for the gauge field $A$ of such configurations. 

\subsubsection*{The return of Bogomol'nyi}

We now proceed, once again guided by the simple kink example. There, we sought configurations in each topological sector which additionally minimised the energy. The equivalent condition here in our Euclidean theory is to look for configurations that minimise the action $S$. Then, we can once again apply the Bogomol'nyi trick, writing
\begin{align}
  S= \int_{\mathbb{R}^4} \text{tr}\left(F\wedge \star F\right) &= \mp \int_{\mathbb{R}^4}\text{tr}\left(F\wedge F\right) + \frac{1}{2} \int_{\mathbb{R}^4}\text{tr}\,\Big(\left(F\pm \star F\right)\wedge \star \left(F\pm \star F\right)\Big)		\nn\\
  &= \mp 8\pi^2 n + \frac{1}{2} \int_{\mathbb{R}^4}\text{tr}\,\Big(\left(F\pm \star F\right)\wedge \star \left(F\pm \star F\right)\Big)	\ ,
  \tag{$\beta.\theBetaEquations$} 
\end{align}\stepcounter{BetaEquations}%
where following our previous discussion, we've assumed that $A$ approaches pure gauge at infinity. Then, for either sign the second term here is positive definite, and hence $S$ is bounded by $S\ge 8\pi^2 |n| $. In each sector, this bound is saturated precisely if
\begin{align}\tag{$\beta.\theBetaEquations$}\label{eq: intro instanton equations}
  \left\{\begin{aligned}
  F-\star F=0 \quad &\text{if } n\ge 0\quad\text{(instanton)}\\
  F+\star F=0 \quad &\text{if } n\le 0\quad\text{(anti-instanton)}\ .
  \end{aligned}\right.
\end{align}\stepcounter{BetaEquations}%
In particular, $n=0$ requires a flat connection, $F=0$. The equation $F=\pm \star F$ is the instanton's Bogomol'nyi equation, sometimes simply called the (anti-)instanton equation.

Once again, we find that the action is minimised by solutions to a first-order equation for $A$, which in particular also provide solutions for the equations of motion since if $F=\pm \star F$, then $D\star F = \pm DF=0$ by the Bianchi identity.

\subsubsection*{Instanton moduli space, and the BPST solution}

We finally want to get a feel for what solutions to the instanton equations $F=\pm \star F$ look like. Let us first suppose we've found the general solution for the gauge field $A(m^A)$, depending on a number of parameters $m^A$. These parameters, the moduli or collective coordinates, are entirely analogous to the single real parameter $y$ that parameterised our kink solutions. They provide coordinates on a manifold: the instanton moduli space. By linearising the instanton equation and appealing to deep index theorem arguments, we find that the dimension of this moduli space is precisely $8|n|$, which is generalised to $4N|n|$ in the $SU(N)$ theory.

One can then go much further, to learn about the hyper-K\"ahler geometry of this moduli space, the metric on it, and even in principle construct \textit{every} solution. The art of doing all this---known as the ADHM construction---can be formulated in a number of ways, from the depths of twistor theory to the playground of D-braneology. We will here forgo these details, as they are not especially relevant to this thesis, and instead focus on the simpler albeit more constrained solutions that came along first.

Before proceeding, we need to define some very useful notation, that will pop up again and again throughout this thesis. The study of instantons in four Euclidean dimensions is made all the more elegant by noting a useful isomorphism of algebras. On one hand, we are in the business of investigating a 2-form $F=\frac{1}{2}F_{ij}dx^i\wedge dx^j$. The space of 2-forms at each point---i.e. anti-symmetric $4\times 4$ matrices---is isomorphic to the Lie algebra $\frak{so}(4)=\frak{su}(2)_+\oplus \frak{su}(2)_-$. In particular, the space of self-dual matrices form $\frak{su}(2)_+$, while the anti-self-dual matrices make up $\frak{su}(2)_-$. On the other hand, $iF$ is also valued in the gauge Lie algebra $\frak{su}(2)$, spanned by the Pauli matrices $\sigma^\alpha$, $\alpha=1,2,3$. So, everything's $\frak{su}(2)$; it is precisely this observation that makes it possible to write down simple non-trivial solutions to the instanton equation.

To efficiently leverage this fact, let us introduce the self-dual and anti-self-dual 't Hooft symbols $\eta^\alpha_{ij}$ and $\bar{\eta}^\alpha_{ij}$, respectively. These are defined by
\begin{align}
  \eta^1_{ij}=\left(\begin{smallmatrix}
	0 & 0 & 0 & 1 \\[0.3em]
	0 & 0 & 1 & 0 \\[0.3em]
	0 & -1 & 0 & 0 \\[0.3em]
	-1 & 0 & 0 & 0 
\end{smallmatrix}\right) && 
 \eta^2_{ij}=\left(\begin{smallmatrix}
	0 & 0 & -1 & 0 \\[0.3em]
	0 & 0 & 0 & 1 \\[0.3em]
	1 & 0 & 0 & 0 \\[0.3em]
	0 & -1 & 0 & 0 
\end{smallmatrix}\right) &&
 \eta^3_{ij}=\left(\begin{smallmatrix}
	0 & 1 & 0 & 0 \\[0.3em]
	-1 & 0 & 0 & 0 \\[0.3em]
	0 & 0 & 0 & 1 \\[0.3em]
	0 & 0 & -1 & 0 
\end{smallmatrix}\right)		\phantom{\ .}\nn\\[0.5em]
  \bar{\eta}^1_{ij}=\left(\begin{smallmatrix}
	0 & 0 & 0 & -1 \\[0.3em]
	0 & 0 & 1 & 0 \\[0.3em]
	0 & -1 & 0 & 0 \\[0.3em]
	1 & 0 & 0 & 0
\end{smallmatrix}\right) && 
 \bar{\eta}^2_{ij}=\left(\begin{smallmatrix}
	0 & 0 & -1 & 0 \\[0.3em]
	0 & 0 & 0 & -1 \\[0.3em]
	1 & 0 & 0 & 0 \\[0.3em]
	0 & 1 & 0 & 0 
\end{smallmatrix}\right) &&
 \bar{\eta}^3_{ij}=\left(\begin{smallmatrix}
	0 & 1 & 0 & 0 \\[0.3em]
	-1 & 0 & 0 & 0 \\[0.3em]
	0 & 0 & 0 & -1 \\[0.3em]
	0 & 0 & 1 & 0 
\end{smallmatrix}\right)\ .
\tag{$\beta.\theBetaEquations$}
\end{align}\stepcounter{BetaEquations}%
Then, the $\eta^\alpha_{ij}$ form a basis for $\frak{su}(2)_+$, and the $\bar{\eta}^\alpha_{ij}$ for $\frak{su}(2)_-$. Indeed, from here the $\eta^\alpha_{ij}$, $\bar{\eta}^\alpha_{ij}$ form natural tools in the construction of chiral spinor representations of $\frak{so}(4)$. For further details, and a number of useful algebraic relations satisfied by the $\eta,\bar{\eta}$, see for instance \cite{Vandoren:2008xg}. Our interest in them will however be a little different. For us, their crucial property is their intertwining of the gauge index $\alpha$ with spacetime indices $i,j$.\\

To get our bearings, let us simply write down the first and simplest solution to the anti-instanton equation $F=-\star F$. Named after its discoverers as the BPST instanton \cite{Belavin:1975fg}, it is given by
\begin{align}\tag{$\beta.\theBetaEquations$}
  A_i(x) = \frac{(x - y)^j}{(x-y)^2+\rho^2}\bar{\eta}^\alpha_{ij} \sigma^\alpha\ .
  \label{eq: BPST regular gauge}
\end{align}
\stepcounter{BetaEquations}%
Here, we have five moduli, made up of real number $\rho$ and point $y^i$ in $\mathbb{R}^4$. There is a lot we can now say about this solution, to make some more concrete sense of the topological treatment in the previous section. For starters, $A$ is clearly regular everywhere in $\mathbb{R}^4$. Further, as we approach $|x|\to\infty$, we do indeed find that $A\to i\, t^{-1}\,dt$ for a map $t:S^3_\infty\to SU(2)$ that winds non-trivially; it has $\deg(t)=-1$. As such, we find that the solution has instanton number $n=-1$. This is indeed consistent with our derivation of the Bogomol'nyi equations, where we found that $F=-\star F$ should correspond to $n\le 0$. We call this solution the single anti-instanton. Note, to flip all the signs, and get a solution with $F=\star F$ and $n=+1$, i.e. an instanton rather than anti-instanton, we simply swap $\bar\eta\to\eta$.

Next, let's think about the profile of this solution. After all, we've repeatedly referred to instantons as codimension 4 solitons, and as such, this solution $A$ should have a centre: a single point in $\mathbb{R}^4$ where it is in some sense localised. To assess this, it's useful to note the corresponding field strength,
\begin{align}\tag{$\beta.\theBetaEquations$}
  F_{ij} = -\frac{2\rho^2}{\left(\left(x-y\right)^2+\rho^2\right)^2}\bar{\eta}^\alpha_{ij} \sigma^\alpha \quad \implies \quad \frac{1}{2}\text{tr}\left(F_{ij}F_{ij}\right) = \frac{48\rho^4}{\left(\left(x-y\right)^2+\rho^2\right)^4}\ .
\end{align}\stepcounter{BetaEquations}%
We see therefore that the Lagrangian density $\mathcal{L}$ achieves a global maximum at $x=y$ and falls monotonically as we move away radially; for this reason, we can think of this point as the instanton's centre. Further, the modulus $\rho$ sets a characteristic width of the peak of $\mathcal{L}$, and so we call it the instanton's size.

Now, we said that the dimension of the moduli space at $n=\pm 1$ should be $8$, but at first blush it looks like we only have $5$ moduli. The additional $3$ are realised by noting that we've made a particular choice of the solution's embedding into the gauge group. Just as we've chosen the Pauli matrices $\sigma^\alpha$, we can just as easily choose $g\sigma^\alpha g^{-1}$ for any $g\in SU(2)$. Thus, we find our extra $3=\dim(SU(2))$ moduli.

It is important to distinguish such a change to $A$---which affects its value everywhere including asymptotically---from a gauge transformation, which should die off at infinity. Indeed, the gauge invariance of the instanton equation ensures we can always get more solutions through gauge transformations. The moduli are defined therefore as directions in the space of solutions that lie transverse to gauge orbits.\\ 

So we've learnt, perhaps remarkably, that we can write down the general solution for the $n=\pm 1$ instanton\footnote{More concretely, we should also think about the global structure of the moduli space, which for instance has a conical singularity as we approach $\rho\to 0$. We omit these details here.}. Our single (anti-)instanton is specified by three pieces of information: a size, a position, and a gauge orientation. One can show, in fact, that this interpretation of the moduli persists for the $8m$-dimensional moduli space of an $|n|=m$ (anti-)instanton. In particular, there are solutions that look approximately like $m$ well-separated (anti-)instantons, each with size, position and gauge orientation. It is when these instantons approach one another, however, that things become much more subtle, and an appreciation of the full moduli space of solutions is attained only through the ADHM construction.

\subsubsection*{Solutions with generic $n$}

Nonetheless, we can still construct quite simple explicit solutions for $A$ with generic $n$, although they will not cover the entire moduli space at general $n$. Let us again focus on $n<0$, and consider the ansatz \cite{Corrigan:1976wk,Wilczek:1976qp}
\begin{align}\tag{$\beta.\theBetaEquations$}
  A_i = -\frac{1}{2}\eta^\alpha_{ij} \sigma^\alpha\partial_j \log\phi\ .
  \label{eq: tHooft ansatz}
\end{align}\stepcounter{BetaEquations}%
Remarkably, then, we have $F=-\star F$ precisely if the function $\phi$ is harmonic: $\partial_i \partial_i \phi$. The opposite result, with $n>0$ and $F=+\star F$, is found by swapping $\eta^\alpha_{ij}\to \bar{\eta}^\alpha_{ij}$.

Now we can see why the 't Hooft symbol is so powerful: we have gone from a highly non-trivial and daunting set of non-linear partial differential equations $F=-\star F$ to a single, linear one. And not just any PDE, but one that we're pretty good at solving. Consider the solution
\begin{align}\tag{$\beta.\theBetaEquations$}
  \phi=1+\frac{\rho^2}{(x-y)^2}\ ,
\end{align}\stepcounter{BetaEquations}%
giving rise to gauge field
\begin{align}\tag{$\beta.\theBetaEquations$}
  A_i = \frac{\rho^2(x - y)^j}{(x-y)^2\left((x-y)^2+\rho^2\right)}\eta^\alpha_{ij} \sigma^\alpha\ .
  \label{eq: BPST singular gauge}
\end{align}\stepcounter{BetaEquations}%
Let us now ask the natural question: what is the instanton number of this solution? In the language of the previous section, what topological sector does it lie in? The important thing to notice is that, in contrast to our previous solution (\ref{eq: BPST regular gauge}), $A$ is singular at the point $x=y$. Thus, a priori, this solution only gives the gauge field on the annulus-shaped patch $\mathbb{R}^4\setminus \{y\}$. However, as we approach $x\to y$, we have $A\sim i\,t\, dt^{-1}$ with $t$ of non-trivial winding: $\deg(t)=+1$. Thus, as we've seen, we can simply choose $A=0$ on a small 4-ball surrounding $y$, which is patched together with the annulus by transition function $t$. Hence, we can write
\begin{align}\tag{$\beta.\theBetaEquations$}
  n=\frac{1}{8\pi^2}\int_{\mathbb{R}^4} \text{tr}\left(F\wedge F \right)= \frac{1}{8\pi^2} \int_{S^3_\infty}\omega_3(A) - \frac{1}{8\pi^2} \int_{S^3_y}\omega_3(A)~,
\end{align}\stepcounter{BetaEquations}%
where $S^3_y$ is a small 3-sphere around the point $x=y$. By explicit computation, we find that the contribution from $S^3_\infty$ vanishes. Indeed, $A$ vanishes sufficiently quickly as $|x|\to\infty$ that $A$ actually gives a well-defined gauge field over the hemisphere $S^4\setminus\{y\}$ where $S^4=\mathbb{R}^4\cup \{\infty\}$, as can be seen by an appropriate stereographic projection. Conversely, following (\ref{eq: n from deg}), we see the contribution from $S^3_y$ is $\deg(t)=+1$, and hence $n=-1$.

Now, the field strength and resulting Lagrangian density is peaked at $x=y$, and thus we have another solution that we might call an anti-instanton, of size $\rho$, centred at $x=y$! We have, of course, preempted this in the previous section. It turns out that our new, singular solution (\ref{eq: BPST singular gauge}) and our old one (\ref{eq: BPST regular gauge}) are both connections on the \textit{same} $SU(2)$-principal bundle, just arising from different choices of open covering. For the regular solution, the `equator' $\Sigma$ has been blown out to infinity; for the singular solution, it has been shrunk down to the point $x=y$. For this reason, the solution (\ref{eq: BPST singular gauge}) is known as the BPST instanton in \textit{singular gauge}, which is indeed related to (\ref{eq: BPST regular gauge}) by a (necessarily singular) gauge transformation.

Finally, note that just as with the BPST instanton in singular gauge, we are free to adorn our singular solution with 3 parameters-worth of gauge orientation, taking the total number of moduli up to $8$.\\

Now, we can keep going and now consider the solution
\begin{align}\tag{$\beta.\theBetaEquations$}
  \phi = 1+ \sum_{A=1}^N \frac{\rho_A^2}{(x-y_A)^2}\ ,
  \label{eq: intro t Hooft solution}
\end{align}\stepcounter{BetaEquations}%
as first considered by 't Hooft \cite{tHooft:unpublished}. The resulting gauge field $A$ is now singular not at one but at $N$ points $y_A$. We find that $A$ then provides a connection on $\mathbb{R}^4\setminus\{y_A\}$ (and indeed on $S^4\setminus\{y_A\}$). Further, $A$ tends precisely to pure gauge with unit winding near each of these removed points, and thus we are once again able to set $A=0$ in the small 4-ball patches surrounding them. We find then
\begin{align}\tag{$\beta.\theBetaEquations$}
  n=\frac{1}{8\pi^2}\int_{\mathbb{R}^4} \text{tr}\left(F\wedge F \right)= \frac{1}{8\pi^2} \int_{S^3_\infty}\omega_3(A)- \frac{1}{8\pi^2} \sum_{A=1}^N \int_{S^3_{y_A}}\omega_3(A) = -N\ ,
\end{align}\stepcounter{BetaEquations}%
where the contribution from $S^3_\infty$ vanishes. Thus, this solution beautifully describes $N$ anti-instantons at points $y_A$, each with size $\rho_A$. We can again do a little better, and add an overall gauge orientation to provide a solution with $5N+3$ moduli. In particular, for $N\ge 2$ we fall short of the full $8N$ coordinates of the whole moduli space.\\

We finally note that we can generalise one step even further, without too much grief. The resulting JNR solution \cite{Jackiw:1976fs} corresponds to the ansatz (\ref{eq: tHooft ansatz}) with choice
\begin{align}\tag{$\beta.\theBetaEquations$}
  \phi = \sum_{A=1}^{N+1} \frac{\rho_A^2}{(x-y_A)^2}\ ,
\end{align}\stepcounter{BetaEquations}%
for the harmonic function $\phi$. At first this solution looks very similar to the 't Hooft form (\ref{eq: intro t Hooft solution}). However its interpretation is rather different. First note that the resulting $A$ is singular at $(N+1)$ points, and we so may expect $n=-(N+1)$. However, in contrast to the 't Hooft solution, the contribution to $n$ from $S^3_\infty$ no longer vanishes, but instead gives $+1$. Thus, in total, we find $n=1-(N+1)= -N$. Topologically, the JNR solution gives a connection on the region $\mathbb{R}^4\setminus \{y_A\}$ that does \textit{not} extend to one on $S^4\setminus \{y_A\}$. Instead, we must deal with the small 4-ball about infinity in the same way as we do the points $y_A$, setting $A=0$ on a small 4-ball surrounding it which is now non-trivially patched onto $\mathbb{R}^4\setminus \{y_A\}$. With suitable care taken, we find the JNR solution has $5N+7$ moduli, some of which are redundant when $N=1,2$ and hence $5N+7> 8N$. The 't Hooft solution in recovered in the limit that one of the $(N+1)$ singular points is sent to infinity, along with its size.  \\

Let us finally make a brief comment about gauges. We argued in the previous section that for \textit{any} $SU(2)$-principal bundle over $S^4$, we could simply cover it in a pair of patches $S^4\setminus\{\infty\}=\mathbb{R}$ along with a small 4-ball around infinity in which $A=0$. Thus, we can always write down a gauge field $A$ that is defined and regular throughout $\mathbb{R}^4$, which certainly sounds like a nice prospect (for physics anyway). We saw precisely how this worked for the single BPST instanton, which took nice forms (\ref{eq: BPST regular gauge}) and (\ref{eq: BPST singular gauge}) in regular and singular gauge, respectively. We then generalised the singular gauge solution to now be singular at $N$ (or $N+1$) points, and have instanton number $n=-N$. But we \textit{know} that we must be able to choose a different open covering---or equivalently perform a singular gauge transformation---to get to a solution that is regular throughout $\mathbb{R}^4$. This can indeed be done \cite{Giambiagi:1977yg}, but for general $N$ the resulting regular $A$ is must too complicated to be practically useful. Thus, in many physical applications, it is the singular solutions above that find the most utility.


\authoredby{I}
\partfont{\color{Icolor}}
\part{Supersymmetric non-Lorentzian models}

\chapterfont{\color{Icolor}}
\chapter*{Introduction}
\addcontentsline{toc}{chapter}{Introduction}
\lhead{\textsl{\nouppercase{Part I, Introduction}}}


The study of string theory and M-theory relies on our understanding of the dynamics of strings and branes, which has in turn motivated the longstanding study of supersymmetric field theories. While one usually considers Lorentz-invariant theories, there are many scenarios in which non-Lorentzian theories can arise, such as when a fixed frame of reference or Lorentz-violating background are chosen. In many cases, the result is a field theory with spacetime symmetries given by some contraction of the Lorentz group, such as the Galilean or Carrollian symmetry groups \cite{Duval:2014uoa,Leblanc:1992wu,Bandos:2008fr,Bagchi:2015qcw}. Additionally, of interest particularly in condensed matter systems are theories that include an inhomogeneous, or Lifshitz, scaling symmetry, for example in \cite{Orlando:2009az,Gomes:2014tua,Chapman:2015wha,Xue:2010ih}. Note, there is also considerable literature on the AdS duals of such non-Lorentzian theories (for example see \cite{Balasubramanian:2008dm,Son:2008ye,Barbon:2008bg,Goldberger:2008vg,Herzog:2008wg,Adams:2008wt,Maldacena:2008wh,Donos:2010tu,Taylor:2015glc}).

Another well established topic in high energy theory concerns classical vacua in topologically non-trivial sectors of a theory's configuration space: solitons. In particular, one is often interested in solitons that saturate some BPS bound, as in supersymmetric theories these describe non-perturbative vacua preserving some amount of supersymmetry. Central to the study of the low-energy scattering of solitons was the work of Manton \cite{Manton:1981mp} on BPS monopoles, where it was first argued that the dynamics of slowly moving solitons can be captured by geodesic motion on the moduli space of static solutions. Later work explored many different avenues, including applying similar methods to Yang-Mills solitons of other codimension \cite{Ruback:1988ba,Ward:1985ij}, and most relevant  to this work, the study of a supersymmetric extension to this approximation \cite{Gauntlett:1992yj,Cederwall:1995bc,Gauntlett:1995fu,Moss:1998jf} to determine the supersymmetric effective theory for both the bosonic and fermionic soliton zero modes.\\


The results of Part I of the thesis tie together these two rich areas of physics, and in doing so recover and extend a number of classic results in a new and more general context.


\subsection*{Outline of results}


Let us now outline the results of Part I of this thesis. The broad motivation of Chapter \ref{chap: construction from Lorentzian models} is to develop a way to construct supersymmetric models without Lorentz symmetry---but with an inhomogeneous, or Lifshitz, scaling symmetry---by starting with their better understood Lorentzian cousins. We first consider a rescaling of fields and coordinates in some generic Lagrangian supersymmetric theory, parameterised by some real parameter $\eta$. For $\eta\neq 0$, the rescaling is invertible and thus not especially interesting. However, the limit $\eta\to 0$ is ill-defined, with the action going as $S=\eta^{-1}S_{-1}+\bigO(\eta^0)$ for some $S_{-1}$. Thus, in the $\eta\to 0$ limit we localise onto the classical minima of $S_{-1}$.

If we further specialise to $S_{-1}$ schematically of the form $S_{-1}=\int \Omega^2$ for some function $\Omega$ of fields in the theory, then we see that as we approach $\eta\to 0$, the theory localises to the solutions of $\Omega=0$. Thus, we are lead to propose a fixed point action at $\eta=0$ with a Lagrange multiplier field imposing this constraint. Indeed, in the examples we are most interested in, $\Omega=0$ is a Bogomol'nyi-type equation for some topological soliton of the original theory, and thus at the fixed point, dynamics are entirely constrained to the corresponding moduli space. Our key result is that, under some further mild assumptions, this fixed point action retains all supersymmetry of the original theory. A similar process has been discussed for supergravity in \cite{Bergshoeff:2015uaa}, although those authors pursue a different treatment of divergent terms in the action.

We next explore a number of examples of this procedure, all relevant to the branes of M-theory. We first consider the $\mathcal{N}=2$ super-Yang-Mills theory in five dimensions, and impose upon it a particular rescaling. Applying the above technique, we arrive at a new theory at the fixed point $\eta\to 0$, with a number of interesting features. For one, while the theory necessarily retains the 16 real supercharges of the initial Lorentzian theory, it is shown that an additional 8 real conformal supercharges emerge in the limit. Further, the Lagrange multiplier imposes that dynamics are constrained to instanton moduli space. In fact, this model is known to describe a null compactification of the $(2,0)$ theory of M5-branes, with its equations of motion first found in \cite{Lambert:2010wm}, their solutions explored in \cite{Lambert:2011gb}, and indeed the action found here first proposed in \cite{Lambert:2018lgt}. Thus, we arrive at an explicit realisation of the DLCQ proposal of the M5-brane, beautifully recovering the restriction of dynamics to instanton moduli space.

We also consider an analogous scaling limit of the $\mathcal{N}=8$ Chern-Simons-matter theories modelling M2-branes, with an aim to arrive at non-Lorentzian models known to be U-dual to the null-compactified M5-brane model above \cite{Kucharski:2017jwv}. Despite a number of additional subtleties requiring corresponding adjustments, the technique is shown to indeed reproduce the theory of \cite{Kucharski:2017jwv} at its fixed point. We go further, and take a comparable limit of the $\mathcal{N}=6$ ABJM/ABJ models \cite{Aharony:2008ug,Aharony:2008gk}, arriving again at a non-Lorentzian theory retaining all of the initial supersymmetry. In each of these cases, the resulting theory's dynamics is constrained to the moduli space of some variant of the Hitchin system \cite{Hitchin:1986vp}.\\

Chapter \ref{chap: reduction to soliton quantum mechanics} then takes the logical next step: the explicit reduction of dynamics to superconformal quantum mechanics on the moduli space $\Omega=0$ for a few insightful examples.

First, we apply the technique of Chapter \ref{chap: construction from Lorentzian models} to perhaps the simplest model yet: an $\mathcal{N}=(1,1)$ $\sigma$-model in $(1+1)$-dimensions with general target space, on which we allow generic scalar potential $W$. This model is a broad generalisation of the simple kink model explored in Section $\beta$, and as such admits topological solitons called kinks whenever $W'$ has more than one isolated zero. We consider a rescaling which, as $\eta\to 0$, restricts to a sector of the theory in which velocities are small relative to spatial variations. The dynamics of the resulting fixed point action are localised to kink moduli space, and hence the procedure provides a explicit realisation of---and indeed supersymmetric extension to---the exact Manton approximation of slow soliton motion \cite{Manton:1981mp}. Further, it is in this context that are able to probe the quantum aspects of the reduction to moduli space, showing that the 1-loop determinant arising from integrating out the Lagrange multiplier precisely matches that which arises in a saddlepoint approximation around $\eta=0$ in the initial theory.

Finally, we return to the five-dimensional Yang-Mills-like theory of Section \ref{sec: M5s}. We show how the theory with gauge group $SU(N)$ is in principle reduced to an $\mathcal{N}=(4,4)$ (16 real supercharges) superconformal quantum mechanics on ADHM moduli space, with the final 8 original supercharges reducing to fermionic shift symmetries. The theory is also generically coupled to a number of time-dependent parameters corresponding to fluctuating zero modes of fields in the initial theory. We then perform this reduction explicitly, for the single instanton BPST solution \cite{Belavin:1975fg} in the $SU(2)$ theory, arriving at a superconformal quantum mechanics on the moduli space $\mathbb{R}^4\times \mathbb{R}^4/\mathbb{Z}_2$.


\chapter{Construction from Lorentzian models}\label{chap: construction from Lorentzian models}
\lhead{\textsl{\nouppercase{Part I, \leftmark}}}


In this first chapter, we detail an abstract scaling procedure one can perform on generic supersymmetric theories to arrive at a new theory, preserving all supersymmetry and possessing a Lifshitz scaling symmetry. We then explore this procedure for a number of example relevant to the branes of M-theory.


\section{A supersymmetric scaling limit}\label{sec: a supersymmetric scaling limit}


Let $S$ be a supersymmetric action  and let $\{\Phi^\alpha\}$ be the set of all fields, both bosonic and fermionic, each taking values in some vector space $\mathcal{V}_\alpha$. For brevity and clarity, take the $\Phi^\alpha$ real, and let $(\,\,.\,\, ,\,\, .\,\,)$ be some real inner product on $\mathcal{V}=\oplus_\alpha\mathcal{V}_\alpha$. \\

Now suppose further that we introduce a continuous parameter $\eta$ and rescale all the fields and coordinates by some power of $\eta$; $\Phi^\alpha\to \eta^{\lambda_\alpha}\Phi^\alpha$, $x^\mu \to \eta^{\lambda_\mu}x^\mu$. The action may now be written
\begin{align}\label{Sexpand}
  S= \sum_{\lambda} \eta^\lambda S_\lambda\ ,
\end{align}
where each of the $S_\lambda$ is independent of $\eta$. In this paper we will restrict our attention to cases where, after a suitable choice of $\eta$ and scaling weights, the action just contains three terms:
\begin{align}
  S= \eta^{-1}S_{-1}+S_0 +\eta S_1\ ,
\end{align} 
although our results trivially extend to cases where there are an arbitrary number of 
terms with positive powers of $\eta$. For a fixed $\eta\ne 0$ nothing has really changed and the dynamics is equivalent to the undeformed case. Our aim here is to try to make sense of the theory in the limit that $\eta\to 0$. 

We also extract the $\eta$ dependance of the supersymmetry variations $\delta$ and we allow for the supersymmetry parameter  $\epsilon$ to also scale. In general these can be expanded in a similar expansion as (\ref{Sexpand}) but here we only consider cases where
\begin{align}
  \delta =  \eta^{-1} \delta_{-1} + \delta_0+\eta\delta_1\ .
\end{align}
Again our results also apply if there is an arbitrary number of terms with postive powers of $\eta$. 
Then, the fact that the action is supersymmetric, {\it i.e.} $\delta S=0$, can now be written as a tower of invariance equations for each $\lambda$:
\begin{align}
 \sum_{\lambda'} \delta_{\lambda'} S_{\lambda-\lambda'} = 0\ .
\end{align}
In particular we will have the invariance conditions:
\begin{align}\label{eq: InvCon}
  \delta_{-1} S_{-1} &= 0\ ,\nonumber\\
  \delta_{-1}S_0 + \delta_0 S_{-1} &=0\ ,\nonumber\\
  \delta_{-1}S_1 + \delta_0 S_0 + \delta_1 S_{-1} &= 0\ .
\end{align}
We introduce one final piece of notation. Just as the action has been split into a series in $\eta$, so can the equations of motion arising from the variation of each of the $\Phi^\alpha$. We let $E^{(\lambda)}_A$ denote the equation of motion for $\Phi^\alpha$ at level $\lambda$, {\it i.e.} under \textit{general} infinitesimal variations of the fields,
\begin{align}
  \delta S_{\lambda} = \int d^d x\,\, \left( \delta \Phi^\alpha , E^{(\lambda)}_\alpha\right)\ .
\end{align}
To analyse the limit  $\eta\to 0$  we consider the case where $S_{-1}$ is of the form
\begin{align}
  S_{-1} = \int d^d x\,\, \left[\frac{1}{2} \kappa^{ab} \left(\Omega_a, \Omega_b \right) \right]\ .
  \label{eq: simple S_{-1}}
\end{align}
Here  the $\Omega_a$ are generically composite fields made up of the $\Phi^\alpha$, and $\kappa^{ab}$ is symmetric and non-degenerate. The $\lambda=-2$ invariance condition implies that
\begin{align}\label{eq: S-2}
 \int d^d x\,\,\kappa^{ab}\left(\delta_{-1}\Phi^\alpha\frac{\delta \Omega_a}{\delta\Phi^\alpha},\Omega_b \right) =0\ .
\end{align}
To begin with we restrict our attention to the cases where $\Omega_a$ is purely bosonic and $\delta_{-1}\Phi^\alpha$ is only non-zero for fermionic fields. Thus (\ref{eq: S-2}) will be trivially satisfied as ${\delta \Omega_a}/{\delta\Phi^\alpha}=0$ for fermionic choices of $\alpha$. Our analysis can be extended to other cases (indeed we will study one such extension below). 

In the physically relevant cases where $\kappa^{ab}$ and hence $S_{-1}$ is positive definite, there is a straightforward interpretation of the theory in the limit $\eta\to 0$. We see that in this limit, we localise onto the classical minima of $S_{-1}$, while disregarding $S_\lambda$ for $\lambda > 0$. This is however only strictly true if there exist configurations satisfying $\Omega_a=0$. In contrast, in all examples we are interested in, $S_{-1}$ will be subject to some generalisation of the Bogomol'nyi trick \cite{Bogomolny:1975de}, and thus will take the schematic form $S_{-1} = A + \int \tilde{\Omega}^2$ for some\footnote{In each example we consider, this Bogomol'nyi bound corresponds to a particle-like soliton, and thus $A$ as defined here is proportional to a divergent integral over the time direction. A more rigorous treatment in these cases involves identical manipulations not of the action but of the Lagrangian $L=\int_\text{space} \mathcal{L}\,d^{d-1}x $.} $A\ge 0$ and composite field $\tilde{\Omega}$. In particular, the value of $A$ corresponds to which topological sector of the configuration space we lie in, in which $\tilde{\Omega}=0$ is attainable. Thus, in sectors with $A\neq 0$, we should take the intermediate step\footnote{An alternative perspective is that we should not shift $S_{-1}$, and instead take $A$ to be absorbed into the path integral measure. Such a treatment changes our choice of the $\Omega_a$, and thus results in adjustments to later steps in the procedure. This is explored in detail in an example in Section \ref{subsec: 1-param family}.} of shifting $S_{-1}$ by $-A$ so as to ensure that $S_{-1}=0$ is attainable within the corresponding topological sector. Shifting by $A$, or equivalently shifting the Lagrangian by an exact term, does not affect the supersymmetry of the action nor the on-shell dynamics.

So let us now assume that such a shift, if required, has been done and so classical minima of $S_{-1}$ simply correspond to $\Omega_a=0$. Thus  in the limit $\eta\to0$  we expect that the dynamics is captured by the simple action
\begin{align}
  \tilde S  = S_0+ \int d^d x \,\, \kappa^{ab}\left(  \Omega_a,G_b \right)\ ,
\end{align} 
where $G_a$ are new fields that impose the constraints $\Omega_a=0$. 
As we now show, under mild assumptions,  we can suitably modify the supersymmetry transformation $\delta \to \tilde \delta=\delta_0+\delta'$ so as to ensure that $\tilde \delta\tilde S=0$. \\

To establish this we observe that for such a form of $S_{-1}$, the $\lambda=-1$ invariance equation implies that
\begin{align}
  \delta_{-1}S_0 + \delta_0 S_{-1} = \int d^d x\,\, \left[ \left( \delta_{-1}\Phi^\alpha, E^{(0)}_\alpha \right) + \kappa^{ab}\left( \delta_0 \Omega_a, \Omega_b \right)   \right] = 0\ .
  \label{eq: -1 invariance equation}
\end{align}
We emphasise that this equation is satisfied purely algebraically and off-shell. If $\delta_{-1}\Phi^\alpha$ is only non-zero for fermionic fields and $\Omega_a$ is bosonic we see that $ E^{(0)}_\alpha$ and $\delta_0\Omega_a$ are fermionic. Therefore we can find functions $\Sigma^\alpha_a $ such that 
\begin{align} \delta_0\Omega_a &= - \Sigma_a^{\alpha} E^{(0)}_\alpha \ ,
\end{align}
and hence
\begin{align}
  \delta_{-1} \Phi^\alpha &= \kappa^{ab}{}^*\Sigma^\alpha_a  \Omega_b  \ ,\label{eq: Sigma}
\end{align}
where ${}^*\Sigma_b^{\alpha}$ is the adjoint map to $\Sigma_b^{\alpha}$ with respect to the $(\ ,\ )$ inner-product.
 With these relations in hand, we turn our attention to the $\lambda=0$ invariance equation,
\begin{align}
  \delta_0 S_0 + \delta_{-1} S_1 + \delta_1 S_{-1} = 0\ .
\end{align}
This means that $S_0$ is not invariant under the $\delta_0$ supersymmetry but rather 
\begin{align}
\delta_0S_0 &=  - \delta_{-1} S_1 - \delta_1 S_{-1} \nonumber\\
& = -\int d^d x\,\,  \kappa^{ab}\left({}^*\Sigma_b^{\alpha} \Omega_a,  E^{(1)}_\alpha\right) -\int d^d x\,\, \kappa^{ab}\left(\Omega_a, \delta_1  \Omega_b \right)\nonumber\\
& = -\int d^d x\,\,  \kappa^{ab}\left(\Omega_a,  \Sigma_b^{\alpha} E^{(1)}_\alpha\right) -\int d^d x\,\, \kappa^{ab}\left(\Omega_a, \delta_1 \Omega_b \right)\ .
\end{align}
Our task now is to find the  corrected supersymmetry $\delta'$ to ensure that $\tilde \delta \tilde S=0$.  Since the   terms  in $\delta_0S_0$ are all proportional to $\Omega_a$ we can cancel them by taking
\begin{align}
\delta' G_a = \Sigma^\alpha_a E^{(1)}_\alpha + \delta_1\Omega_a\ .
\end{align}
This leaves us with
\begin{align}
\tilde\delta \tilde S & = \delta'S_0 + \int d^d x \,\, \kappa^{ab}\left( (\delta_0+\delta')\Omega_a,G_b \right)\nonumber\\
& =  \int d^d x \,\,(\delta' \Phi^\alpha, E_\alpha^{(0)})- \int d^d x \,\, \kappa^{ab}\left( \Sigma_a^\alpha E^{(0)}_\alpha,G_b \right) +   \int d^d x \,\, \kappa^{ab}\left(  \delta'\Omega_a,G_b \right)\ .
\end{align}
We can now set $\tilde\delta \tilde S=0$ by taking
\begin{align}
\delta' \Phi^\alpha &= {}^*\Sigma^\alpha_a G_b\kappa^{ab}\ ,
\end{align}
provided that we can also have $ \delta' \Omega_a  =0$. Namely we require that
\begin{align}
\delta' \Omega_a = \delta' \Phi^\alpha \frac{\partial \Omega_a}{\partial\Phi^\alpha} = {}^*\Sigma^\alpha_a G_b\kappa^{ab}\frac{\partial \Omega_a}{\partial\Phi^\alpha}=0\ .
\end{align}
Since $G_a$ is an independent field that doesn't appear in $\Omega_a$ or $ {}^*\Sigma^\alpha_a$ this implies that we need
\begin{align}
{}^*\Sigma^\alpha_b  \frac{\partial \Omega_a}{\partial\Phi^\alpha}=0\ .\label{eq: condition}
\end{align}
When $\Omega_a$ is only made of bosonic fields  $\Sigma^\alpha_a$ is only non-zero for fermionic choices of the $\alpha$-index but ${\partial \Omega_a}/{\partial\Phi^\alpha}$ is only non-zero for bosonic choices of $\alpha$ and hence (\ref{eq: condition}) holds. 

Thus in summary, and in slightly more generality, we find that if the scaling leads to a divergent term in the action of the form (\ref{eq: simple S_{-1}}) then, assuming that we can construct the map $\Sigma^\alpha_a$ defined in (\ref{eq: Sigma}), we can construct a supersymmetric action $\tilde S$  where 
\begin{align}
 \tilde S  &= S_0+ \int d^d x \,\, \kappa^{ab}\left(  \Omega_a,G_b \right)\ ,\nonumber\\
 \tilde\delta \Phi^\alpha & = \delta \Phi^\alpha +{}^*\Sigma^\alpha_a G_b\kappa^{ab}\ ,\nonumber\\
 \tilde \delta G_a & = \Sigma^\alpha_a E^{(1)}_\alpha + \delta_1\Omega_a\ ,
 \label{eq: final fixed point action and variations}
\end{align}
provided that (\ref{eq: condition}) holds. In particular we have argued that this can always be done in the case that $S_{-1}$ is bosonic and $\delta_{-1}$ is fermionic. 
 
Lastly we point out that   we started off with a scaling  that was not a symmetry of the original action $S$, otherwise we would have $S_0=S$, $\delta_0=\delta $ and hence $\tilde S=S$, $\tilde \delta=\delta$, meaning that the whole process was trivial. However this scaling is a symmetry of $\tilde S$. In particular $S_0$ is scale invariant by construction and, in order to have appeared in $S_{-1}$, $\Omega_a$ must  scale  as $\Omega_a\to \eta^{-(\lambda_0+\lambda_1\dots+\lambda_{d-1}+1)/2}\Omega_a$. Thus if we let $G_a\to \eta^{-(\lambda_0+\lambda_1\ldots+\lambda_{d-1}-1)/2}G_a $ then $\tilde S$ will be  invariant under the scale transformation. 

In other words, by introducing a scaling we have produced an inhomogeneous RG flow resulting in a fixed point theory in the limit $\eta\to 0$, which is invariant under the scaling.

It is interesting to note that, with the action (\ref{eq: final fixed point action and variations}), one could consider integrating out not the Lagrange multiplier $G_{ij}$---resulting in a quantum mechanics on the moduli space $\Omega_a=0$---but instead some other field(s) of the theory. In doing so, one may expect to uncover a dual description of the $\Omega_a=0$ quantum mechanics.

\subsection*{A more general prescription}\label{subsec: more general prescription}

The crucial quality of the initial theory that allowed for this procedure to work was the quite simple form (\ref{eq: simple S_{-1}}) of $S_{-1}$. A related condition is that  $\delta_{-1}\Phi^\alpha\neq 0$ only for  fermionic $\Phi^\alpha$. We can adapt this prescription for a more general situation as we now discuss, although the details are better left to the explicit examples below.

We saw that the scaling led to an RG flow with $\tilde S$ the fixed point action. The flow is rather trivial in that all the fields simply obey their naive scaling behaviour. However one could allow for field redefinitions and mixings along the flow. Thus we could consider making the field redefinition 
\begin{align}
\Phi^\alpha \to \Phi^\alpha - \eta^{-1}\chi^\alpha\ ,
\end{align}
in the rescaled theory, 
where $\chi^\alpha$ is some function of the fields\footnote{Again one could allow for more general powers of $\eta$ but we leave that to future work. Here we assume the same  expansion in terms of $\eta^{-1}$, $\eta^0$ and $\eta$ that we considered above.}. This has two key effects. Firstly it leads to a shift in $\delta_{-1}\Phi^\alpha$:
\begin{align}
\delta_{-1}\Phi^\alpha \to \delta_{-1}\Phi^\alpha - \delta_0\chi^\alpha\ ,
\end{align} 
and secondly it will affect the form of $S_{-1}$:
\begin{align}
S_{-1} \to S_{-1}- \int d^d x \,\, \left(\chi^\alpha,E^{(0)}_\alpha\right) +   \frac 12 \int d^d x \,\, \left(\chi^\alpha\chi^\beta,\frac{\partial E^{(1)}_\alpha}{\partial\Phi^\beta}\right)\ ,
\end{align} 
as well as  shifting other terms around. In general this will introduce  $S_{-2}$ and $\delta_{-2}$ and even more divergent terms. These would in turn lead to additional terms in the invariance conditions (\ref{eq: InvCon}) and invalidate the previous discussion. However we will see   below that there are  cases where a suitable choice of field redefinition   maps the theory back to the situation studied   above where $S_{-1}$ takes the form of (\ref{eq: simple S_{-1}})  and $\delta_{-1}\Phi^\alpha\ne 0$ only for fermions, without introducing higher divergences.

This provides a procedure for cases the where $\delta_{-1}\Phi^\alpha\ne 0$ for bosons. Namely one first looks for a field definition such that $\delta_{-1}\Phi^\alpha= 0$ for the new bosonic fields and then determines the resulting form of $S_{-1}$. If it is of the form (\ref{eq: simple S_{-1}}), with no further divergences, then one can proceed as in the previous discussion.


\section{Five-dimensional super-Yang-Mills}\label{sec: M5s}


The simplest application of the above construction starts with five-dimensional maximally supersymmetric Yang-Mills, with action
\begin{align}
S = \frac{1}{g^2} \text{tr}\int d^5 x\, &\left(  - \frac{1}{4} F_{\mu\nu} F^{\mu\nu} -\frac{1}{2} D_{\mu }X^I D^{\mu } X^I + \frac{1}{4}[X^I,X^J][X^I,X^J]\right.  \nonumber\\ 
&\left. +\frac{i}{2}\bar\Psi\Gamma^\mu D_\mu\Psi -\frac{1}{2}\bar\Psi\Gamma_5 \Gamma^I[X^I,\Psi]\right)\ .
\end{align}
Our spacetime is $(1+4)$-dimensional Minkowski space with metric $\eta_{\mu\nu}=\text{diag}(-,+,+,+,+)$, and coordinates $x^\mu$ with $\mu,\nu,\dots=0,1,2,3,4$. We take gauge group $SU(N)$, and have gauge field $A$ with corresponding field strength $F=dA-iA\wedge A$. We further have five scalar fields $X^I$, where $I,J,\dots=6,7,8,9,10$, transforming in the adjoint of $SU(N)$. Finally, we have a single fermion $\Psi$, which is a real 32-component spinor of $\text{Spin}(1,10)$ also transforming in the adjoint of $SU(N)$. The matrices $\{\Gamma_0,\Gamma_1,\dots, \Gamma_{10}\}$ form a real $32 \times 32$ representation of the $(1+10)$-dimensional Clifford algebra with signature $(-,+,\dots, +)$. Additionally, the fermions satisfy $\Gamma_{012345}\Psi=-\Psi$. 

This action has $\mathcal{N}=2$ supersymmetry, corresponding to 16 real supercharges. These supersymmetries are realised in this formulation by
\begin{align}
\delta X^I & = i\bar\xi \Gamma^I\Psi\ ,\nonumber\\
\delta A_\mu &= i\bar\xi\Gamma_\mu\Gamma_5\Psi\ ,\nonumber\\
\delta \Psi & = \frac12 \Gamma^{\mu\nu}\Gamma_5F_{\mu\nu}\xi + \Gamma^\mu\Gamma^ID_\mu X^I \xi  - \frac{i}{2}\Gamma^{IJ}\Gamma_5[X^I,X^J]\xi\ ,
\label{eq: YM SUSYs}
\end{align}
with real 32-component spinor $\xi$ satisfying $\Gamma_{012345}\xi=\xi$.

Next we want to consider a rescaling of the theory. In particular we take
\begin{align}\label{eq: Lifshitz}
x^0&\to \eta^{-1}  x^0\ ,\nonumber\\
x^i&\to \eta^{-1/2} x^i\ ,\nonumber\\
X^I&\to   \eta X^I\ ,\nonumber\\
\Psi_- &\to  \eta \Psi_-\ ,\nonumber\\
\Psi_+ &\to  \eta^{3/2} \Psi_+\ ,\nonumber\\ 
\xi_- &\to \eta^{-1/2}\xi_-\ ,\nonumber\\
\xi_+&\to  \eta^0 \xi_+\ ,
\end{align}
where $i,j=1,2,3,4$ and we introduce
\begin{align}
  \Gamma_\pm = \frac{1}{\sqrt{2}}\left( \Gamma_0 \pm \Gamma_5 \right)\ ,
\end{align}
and the corresponding projections 
\begin{align}
\Psi_{\pm} = \frac12(1\pm \Gamma_{05})\Psi\ ,\qquad  \xi_{\pm} = \frac12(1\pm \Gamma_{05})\xi\ ,
\end{align}
and so $\Gamma_\pm \Psi = \Gamma_\pm \Psi_\mp$ and $\Gamma_\pm \xi = \Gamma_\pm \xi_\mp$. Note, in general if some spinor $\chi$ has $\Gamma_{012345}\chi = a\chi$ and $\Gamma_{05}\chi=b\chi$ for some $a,b\in\{\pm 1\}$, then $\Gamma_{1234}\chi=ab\chi$ and hence $\Gamma_{ij}\chi$ is (anti-)self-dual on its $\{i,j\}$ indices: $\Gamma_{ij} \chi = -\frac{1}{2}ab\,\varepsilon_{ijkl} \Gamma_{kl} \chi$. \\

Following the discussion above this scaling leads to 
\begin{align}
  S = \eta^{-1} S_{-1} + S_0 + \eta^1 S_1\ ,
\end{align}
where  (we also further rescale  all the negative chirality spinors by a factor of $2^{1/4}$ and positive chirality ones by $2^{-1/4}$ so as to agree with \cite{Lambert:2018lgt})\begin{align}
  S_{-1} &= \frac{1}{g^2} \text{tr}\int d^4 x\, dx^0 \left( -\frac{1}{4} F_{ij} F_{ij} \right) \ , \nonumber\\ 
  S_0 &=  \frac{1}{g^2} \text{tr}\int d^4 x\, dx^0 \Bigg( \frac{1}{2} F_{0i} F_{0i} - \frac{1}{2} \left( D_i X^I \right) \left( D_i X^I \right) - \frac{i}{2} \bar{\Psi} \Gamma_+ D_0 \Psi \nonumber\\
  &\hspace{36mm}  +\frac{i}{2} \bar{\Psi} \Gamma_i D_i \Psi - \frac{1}{2} \bar{\Psi} \Gamma_+ \Gamma^I [X^I,\Psi] \Bigg)\ ,\nonumber\\
  S_{1} &= \frac{1}{g^2} \text{tr}\int d^4 x\, dx^0 \Bigg( \frac{1}{2} \left( D_0 X^I \right) \left( D_0 X^I \right) - \frac{i}{4}\bar{\Psi} \Gamma_- D_0 \Psi \nonumber\\
  &\hspace{36mm} +\frac{1}{4} \bar{\Psi} \Gamma_- \Gamma^I [X^I, \Psi] + \frac{1}{4} [X^I, X^J][X^I, X^J] \Bigg)\ .
  \label{eq: YM scaled}
\end{align}
The  supersymmetries take the form
\begin{align}
  \delta = \eta^{-1} \delta_{-1} + \delta_0 + \eta \delta_{1}\ ,
\end{align}
with
\begin{align}
  \delta_{-1} \Psi &= \tfrac{1}{2} F_{ij} \Gamma_{ij} \Gamma_+ \xi  \ ,\\[1em]
  \delta_0 X^I &= i \bar{\xi} \Gamma^I \Psi \nonumber \ ,\\
  \delta_0 A_0 &= i \bar{\xi} \Gamma_{-+} \Psi \nonumber \ ,\\
  \delta_0 A_i &= i\bar{\xi} \Gamma_i \Gamma_+ \Psi 
   \nn\ ,\\
  \delta_0 \Psi &= F_{0i} \Gamma_i \Gamma_{-+}\xi - \tfrac{1}{4} F_{ij} \Gamma_{ij} \Gamma_- \xi + \left( D_0 X^I \right) \Gamma^I \Gamma_+ \xi \nonumber \\
  &\hspace{8mm}+ \left( D_i X^I \right) \Gamma_i \Gamma^I \xi - \tfrac{i}{2} [X^I, X^J] \Gamma^{IJ} \Gamma_+ \xi \nonumber \ ,\\[1em]
  \delta_1 A_i &= -\tfrac{i}{2} \bar{\xi} \Gamma_i \Gamma_- \Psi
  \nonumber \ ,\\
  \delta_1 \Psi &= \tfrac{1}{2} \left( D_0 X^I \right) \Gamma^I \Gamma_- \xi + \tfrac{i}{4} [X^I, X^J] \Gamma^{IJ} \Gamma_- \xi\ .
  \label{eq: YM scaled SUSYs}
\end{align}
Following the discussion in Section \ref{sec: a supersymmetric scaling limit}, we want to make the constraint $S_{-1}=0$ as weak as possible, so we shift $S_{-1}$ by a topological piece proportional to $\varepsilon^{ijkl}{\rm tr}(F_{ij}   F_{kl})$   to obtain
\begin{align}
 S_{-1} &= -\frac{1}{2g^2} \text{tr}\int d^4 x\, dx^0 \left( F^+_{ij} F^+_{ij} \right)\ ,
\end{align}
where $F^+_{ij}=\frac12 \left(F_{ij}+\frac{1}{2}\varepsilon_{ijkl}F_{kl}\right)$.
In terms of the notation above we have $a\to[ij]$, $\kappa_{ab}\to\kappa_{ij,kl}=-\delta_{ik}\delta_{jl}$ and $\Omega_{ij} = \frac{1}{g}F_{ij}^+$. We then find
\begin{align}
\delta_0\Omega_{ij} & =\frac{i}{2g} \bar\xi \Gamma_+\Gamma_{ij}\Gamma_kD_k\Psi = \frac{g}{2} \bar\xi \Gamma_+\Gamma_{ij} E^{(0)}_{\Psi} \ ,\nonumber\\
\delta_{-1} \Psi &= \frac{1}{2} F_{ij}^+ \Gamma_{ij} \Gamma_+ \xi\ = \frac{g}{2} \Omega_{ij} \Gamma_{ij} \Gamma_+ \xi\ .
\end{align}
Hence, if we take $\Sigma_{ij}^\Psi  =  -\tfrac{g}{2} \bar\xi  \Gamma_+ \Gamma_{ij}$, with $\Sigma^\alpha_{ij}=0$ for all $\alpha\neq\Psi$, we indeed have
\begin{align}
 \delta_0 \Omega_{ij} &= -\Sigma_{ij}^\Psi E_\Psi^{(0)}\ ,\nonumber\\
 \delta_{-1} \Psi &= \kappa^{ij,kl} \, {}^*\Sigma_{ij}^\Psi \Omega_{kl} = -{}^*\Sigma_{ij}^\Psi \Omega_{ij} \ ,
\end{align}
where here ${}^*$ is simply the Dirac conjugate.
So, we are safe to proceed with the procedure. The end result is that, in the limit $\eta\to 0$, the theory is described by the following action 
\begin{align}
  S = \frac{1}{g^2} \text{tr}\int d^4 x\, dx^0 \Bigg(& \frac{1}{2} F_{0i} F_{0i} + \frac{1}{2} F_{ij} G_{ij} - \frac{1}{2} \left( D_i X^I \right) \left( D_i X^I \right) \nonumber\\
  &\quad - \frac{i}{2} \bar{\Psi} \Gamma_+ D_0 \Psi  +\frac{i}{2} \bar{\Psi} \Gamma_i D_i \Psi - \frac{1}{2} \bar{\Psi} \Gamma_+ \Gamma^I [X^I,\Psi] \Bigg)\ ,
  \label{eq: final M5 action}
\end{align}
where the new Lagrange multiplier is an self-dual spatial 2-form $G_{ij}$ transforming in the adjoint of the gauge group. The supersymmetry variations are
\begin{align}
  \delta X^I &= i \bar{\xi} \Gamma^I \Psi\ , \nonumber \\
  \delta A_0 &= i \bar{\xi} \Gamma_{-+} \Psi \ ,\nonumber \\
  \delta A_i &= i\bar{\xi} \Gamma_i \Gamma_+ \Psi \quad \ ,\nonumber \\
  \delta \Psi &= F_{0i} \Gamma_i \Gamma_{-+}\xi - \tfrac{1}{4} F_{ij} \Gamma_{ij} \Gamma_- \xi + \left( D_0 X^I \right) \Gamma^I \Gamma_+ \xi \nonumber \\
  &\hspace{8mm}+ \left( D_i X^I \right) \Gamma_i \Gamma^I \xi - \tfrac{i}{2} [X^I, X^J] \Gamma^{IJ} \Gamma_+ \xi - \tfrac{1}{4} G_{ij} \Gamma_{ij} \Gamma_+ \xi \ ,\nonumber\\
  \delta G_{ij} &= -\tfrac{i}{2}\bar{\xi}\Gamma_- \Gamma_k \Gamma_{ij} D_k \Psi - \tfrac{i}{2}\bar{\xi}\Gamma_+ \Gamma_{ij} \Gamma_- D_0 \Psi + \tfrac{1}{2} \bar{\xi}\Gamma_+ \Gamma_{ij} \Gamma_- \Gamma^I [X^I, \Psi]\ .
\end{align}
The equations of motion of this theory were first derived as a special solution to a set of equations defining a representation of the $(2,0)$ tensor multiplet \cite{Lambert:2010wm}, implying that the theory describes $N$ M5-branes compactified on a null circle of radius $R_+=g^2/4\pi^2$ with $x^0$ identified as the opposite null direction. In particular, the theory's dynamics are constrained to instanton moduli space ($F^+_{ij}=0$), imposed in this context by the Lagrange multiplier $G_{ij}$. In this way, we directly recover a Lagrangian realisation of the DLCQ prescription for the M5-brane \cite{Aharony:1997th,Aharony:1997an}.

The bosonic sector the resulting quantum mechanics on instanton moduli space was later explored \cite{Lambert:2011gb}, while later work first proposed the action (\ref{eq: final M5 action}) giving rise to these equations of motion \cite{Lambert:2018lgt}.

We will later show in Part II that this theory arises in a particular degenerate limit of a more general construction. This will indeed make manifest the interpretation of the theory as describing null-compactified M5-branes.\\

It is easy to check that (\ref{eq: final M5 action}) has the Lifshitz scaling symmetry
(\ref{eq: Lifshitz}) provided that $G_{ij}\to \eta^2 G_{ij}$. However one can also see that it has an additional superconformal symmetry  which does not have an analogue in the original theory:
\begin{align}
  \delta X^I &= i \bar{\epsilon} \Gamma^I \Psi \ ,\nonumber \\
  \delta A_0 &= i \bar{\epsilon} \Gamma_{-+} \Psi \ ,\nonumber \\
  \delta A_i &= i\bar{\epsilon} \Gamma_i \Gamma_+ \Psi \ ,\quad \nonumber \\
  \delta \Psi &= F_{0i} \Gamma_i \Gamma_{-+}\epsilon  - \tfrac{1}{4} F_{ij} \Gamma_{ij} \Gamma_- \epsilon  + \left( D_0 X^I \right) \Gamma^I \Gamma_+ \epsilon + \left( D_i X^I \right) \Gamma_i \Gamma^I \epsilon  \nonumber \\
  &\hspace{8mm}- \tfrac{i}{2} [X^I, X^J] \Gamma^{IJ} \Gamma_+ \epsilon - \tfrac{1}{4} G_{ij} \Gamma_{ij} \Gamma_+ \epsilon -4X^I\Gamma^I\zeta_+\ ,\nonumber\\
  \delta G_{ij} &= -\tfrac{i}{2}\bar{\epsilon}\Gamma_- \Gamma_k \Gamma_{ij} D_k \Psi - \tfrac{i}{2}\bar{\epsilon}\Gamma_+ \Gamma_{ij} \Gamma_- D_0 \Psi \nn\\
  &\hspace{8mm}+ \tfrac{1}{2} \bar{\epsilon}\Gamma_+ \Gamma_{ij} \Gamma_- \Gamma^I [X^I, \Psi] - 3i\bar\zeta_+\Gamma_{-}\Gamma_{ij}\Psi\ ,
  \label{eq: 5d fixed point theory SUSYs}
\end{align}
where now  $\epsilon=\xi +  x^0\Gamma_-\zeta_+ +x^i\Gamma_i\zeta_+$. Thus there are 16 supersymmetries parameterised by a constant $\xi$ and an additional 8 conformal supersymmetries parameterised by a constant $\zeta_+$. These are indeed consistent with the 32 super(conformal) symmetries of the M5-brane theory reduced on a null direction $x^+$ with the restriction that all fields and supersymmetries are independent of $x^+$ (which explains why there is no $\zeta_-$ superconformal symmetry as the resulting $\epsilon$ would be linear in $x^+$). \\

Note then that under (\ref{eq: 5d fixed point theory SUSYs}) the action transforms as 
\begin{align}
  \delta S = \frac{1}{g^2}\text{tr}\int d^5 x\,\, \partial_i \Big[\,\, &\bareps_- \left( -i F_{0i} \Psi_+ - i \left( D_i X^I \right)\Gamma^I \Psi_+ \right.\nonumber\\
  &\qquad  \left. + i \left( D_0 X^I \right) \Gamma_+ \Gamma^I \Gamma_i \Psi_- + \tfrac{1}{2} \left[ X^I, X^J \right] \Gamma_+ \Gamma^{IJ} \Gamma_i \Psi_- \right) \nonumber\\[0.8em]
  &+\bareps_+ \left( i F_{0j} \Gamma_{ij} \Psi_- - i \left( D_j X^I \right) \Gamma^I \Gamma_{ij} \Psi_- - \tfrac{i}{4} F_{jk} \Gamma_- \Gamma_i \Gamma_{jk} \Psi_+ \right) \nonumber\\[0.8em]
  &+  4i X^I \bar{\zeta}_+ \Gamma^I \Gamma_i \Psi_- \Big]\ ,
\end{align}
where we've neglecting boundary terms at temporal infinity. Thus, for suitable boundary conditions we have $\delta S=0$.\\

Evident from its construction as a scaling limit, $S$ enjoys a Lifshitz scaling symmetry. We introduce the notation $[\Phi]=\left( a,b \right)$ for an object $\Phi$, where $a$ is the mass dimension, and $b$ the Lifshitz scaling dimension. Then,
\begin{align}
  \left[ x^0 \right] = - \left[ A_0 \right] & = \left( -1 , -1 \right)\ , & \left[ \Psi_- \right] & = \left( \tfrac{3}{2} , 1 \right)\ , \nonumber\\
  \left[ x^i \right] = - \left[ A_i \right]  & = \left( -1 , -\tfrac{1}{2} \right)\ , & \left[ \Psi_+ \right] & = \left( \tfrac{3}{2} , \tfrac{3}{2} \right)\ , \nonumber\\
  \left[ X^I \right] & = \left( 1 , 1 \right)\ , & \left[ \xi_- \right] & = \left( -\tfrac{1}{2} ,-\tfrac{1}{2} \right)\ , \nonumber\\
  \left[ G_{ij} \right] & = \left( 2 , 2 \right)\ , & \left[ \xi_+ \right] & = \left( -\tfrac{1}{2} , 0 \right)\ , \nonumber\\
  \left[ 1/g^2 \right]& = \left( 1 , 0 \right)\ ,  & \left[ \zeta_+ \right] & = \left( \tfrac{1}{2} , \tfrac{1}{2} \right)\ .
  \label{eq: scaling weights}
\end{align}

\subsection{A 1-parameter family of alternatives}\label{subsec: 1-param family}

We could ask what would happen had we not shifted $S_{-1}$ by a topological piece. Indeed, there is a 1-parameter family of ways in which we can write $S_{-1}$, each related by a shift by some multiple of $\varepsilon^{ijkl}{\rm tr}(F_{ij} F_{kl})$. It's easily seen that the form we chose is the unique choice such that the integrand (\ref{eq: -1 invariance equation}) vanishes identically. Otherwise, it is equal to a boundary term, thanks to the Bianchi identity for $F_{ij}$.\\

What this  boils down to is the fact that, for other choices of the boundary term,  we must also relax the condition that $G_{ij}$ be self-dual and allow it to be a general spatial 2-form. So let us add to $S_{-1}$ some generic multiple of $\varepsilon^{ijkl}{\rm tr}(F_{ij} F_{kl})$, and follow the procedure through. One finds that  $\delta G_{ij}$  is then shifted by
\begin{align}
  \delta G_{ij} \to \delta G_{ij} + i\alpha  \bar{\epsilon}\Gamma_-  \Gamma_{ijk} D_k \Psi = \delta G_{ij} + \tfrac{i\alpha}{2} ( \underbrace{\bar{\epsilon}\Gamma_-  \Gamma_{ij}\Gamma_{k} D_k \Psi}_{\text{anti-self-dual}} + \underbrace{\bar{\epsilon}\Gamma_-  \Gamma_{k}\Gamma_{ij} D_k \Psi}_{\text{self-dual}} )\ ,
\end{align}
for some $\alpha\in\mathbb{R}$. For any $\alpha\neq 0$, $\delta G_{ij}$ has both non-zero self-dual and anti-self-dual parts, and we are forced to regard $G_{ij}$ as a generic 2-form. Upon integrating out such a $G_{ij}$, we have the flat connection condition $F_{ij}=0$.

Thus, the theory described by the action (\ref{eq: final M5 action}) can be seen as a special case, where $\alpha=0$, $\delta G^-_{ij}=0$, and we are safely able to regard $G_{ij}$ as self-dual. As we've already mentioned, integrating out $G_{ij}$ puts us onto instanton moduli space, with instanton number $k\le 0$. Of course in the trivial sector ($k=0$), the condition is simply $F_{ij}=0$. Thus, we haven't lost anything by going to this special case, $\alpha=0$. Notably, however, we see that only the anti-instanton sectors (as opposed to instanton) are accessible. Of course the reverse would be the case if we'd rescaled the different chirality fermions in the opposite way, corresponding to a null compactification in the opposite null direction \cite{Lambert:2010wm}.

\section{Chern-Simons-matter theories}\label{sec: M2s}

We now turn our attention to similar limits of worldvolume theories for multiple M2-branes. In \cite{Kucharski:2017jwv}, a non-Lorentzian variant of the ${\cal N}=8$ BLG theory was constructed  from the $(2,0)$ system of \cite{Lambert:2010wm,Lambert:2016xbs}. It was further explained that this system is U-dual to the DLCQ description of M5-branes. As we have just seen, the latter can be obtained via a scaling limit of five-dimensional maximally supersymmetric Yang-Mills. Thus   we expect that a similar scaling limit of the ${\cal N}=8$  theory will obtain the theory of \cite{Kucharski:2017jwv}.  

We first look at the ${\cal N}=8$  theory, where it turns out there is essentially a unique way in which we must scale the fields and coordinates in order to arrive at the theory of \cite{Lambert:2018lgt}. In section \ref{subsec: ABJM}, we apply the same scaling to the 
ABJM/ABJ action, and thus derive the associated  non-Lorentzian fixed-point theory, which nonetheless still has manifest $\mathcal{N}=6$ supersymmetry.

\subsection{${\cal N}=8$}\label{subsec: BLG}

Our starting point is the regular ${\cal N}=8$  theory \cite{Bagger:2007jr,Gustavsson:2007vu}. The dynamical fields take values in a three-algebra $\mathcal{V}$ with invariant inner product $\langle \, \cdot \, , \, \cdot \, \rangle$ and totally anti-symmetric product
\begin{align}
  [\,\cdot\, , \,\cdot\, , \,\cdot\, ]: \mathcal{V} \otimes \mathcal{V} \otimes \mathcal{V} \to \mathcal{V}\ ,
\end{align}
which acts on itself as a derivation,
\begin{align}
  [U,V,[X,Y,Z]] = [[U,V,X],Y,Z] + [X,[U,V,Y],Z] + [X,Y,[U,V,Z]]\ .
\end{align}
Additionally, the three-algebra generates a Lie-algebra $\mathcal{G}$ by the analogue of the adjoint map, $X\to \varphi_{U,V}(X) = [U,V,X]$ for any $U,V\in \mathcal{V}$. This naturally induces an invariant inner product $( \, \cdot \, , \, \cdot \, )$ on $\mathcal{G}$, which satisfies
\begin{align}
  (T,\varphi_{U,V}) = \langle T(U), V \rangle \ ,
\end{align}
for any $T\in\mathcal{G}$ and $U,V\in \mathcal{V}$.

The action is
\begin{align}
  S = \int d^3 x\,\, \bigg( -\frac{1}{2} \angl{ D_\mu X^{I}}{ D^\mu X^{I} } - \frac{1}{12} \angl{\trip{X^{I}}{X^{J}}{X^{K}}}{\trip{X^{I}}{X^{J}}{X^{K}}} +\frac{i}{2} \angl{\bar{\Psi}}{\Gamma^\mu D_\mu \Psi}\nonumber\\
   + \frac{i}{4} \angl{\bar{\Psi}}{\Gamma^{{I}{J}}\trip{X^{I}}{X^{J}}{\Psi}} + \frac{1}{2} \epsilon^{\mu\nu\lambda}\left( \left( A_\mu, \partial_\nu A_\lambda \right) - \frac{2}{3}\left( A_\mu , A_\nu A_\lambda \right) \right)\bigg)\ ,
   \label{eq: plain BLG action}
\end{align}
where in this section $\mu,\nu,\lambda=0,1,2$, and we use the convention $\epsilon^{012}=-1$. The scalars $X^{I}$, ${I}=3,4,\dots,10$ and 32-component real spinors $\Psi$ take values in $\mathcal{V}$, while the gauge field $A_\mu$ takes values in $\mathcal{G}$. We take the field strength to be $F_{\mu\nu}=-[D_\mu,D_\nu] = \partial_\mu A_\nu - \partial_\nu A_\mu - [A_\mu,A_\nu]$. The spinors additionally satisfy $\Gamma_{012}\Psi = -\Psi$. We have
\begin{align}
  D_\mu X^{I} = \partial_\mu X^{I} - A_\mu \left( X^{I} \right)\ ,
\end{align}
and similarly for $\Psi$. The theory possesses 16 supercharges corresponding to rigid supersymmetry, and an additional 16 corresponding to superconformal symmetry. In particular, we have $\delta S=0$, with
\begin{align}
  \delta X^{I} &= i \bar{\epsilon} \Gamma^{I} \Psi\ ,\nonumber\\
  \delta \Psi &= \left( D_\mu X^{I} \right) \Gamma^\mu \Gamma^{I} \epsilon - \tfrac{1}{6} \trip{X^{I}}{X^{J}}{X^{K}} \Gamma^{{I}{J}{K}} \epsilon  -\tfrac{1}{3}X^I \Gamma^I \Gamma^\mu \partial_\mu\epsilon \ , \nonumber\\
  \delta A_\mu ( \,\cdot\, ) &= i\bar{\epsilon} \Gamma_\mu \Gamma^{I} \trip{X^{I}}{\Psi}{\,\cdot\,}\ ,
\end{align}
where $\epsilon$ takes the form $\epsilon=\xi + x^\mu \Gamma_\mu \zeta$. We additionally have $\Gamma_{012}\xi  = \xi$ and $\Gamma_{012}\zeta  = \zeta$, and so $\Gamma_{012}\epsilon = \epsilon$. 

Motivated by the theory obtained in \cite{Lambert:2018lgt}, we first introduce complex coordinates on the worldvolume $z=x^1+ix^2$. We also combine two of the off-brane scalars into a complex pair, $Z=X^3+iX^4$, while labelling the rest $X^A$ with $A=5,6,\dots,10$. Further, we split spinors into definite chiralities under $\Gamma_{034} = -2i \Gamma_0 \Gamma_{Z\bar{Z}}$, so that {\it e.g.} $\Psi_\pm := \frac{1}{2}\left( 1\pm \Gamma_{034} \right)\Psi$.

Given this form of $S$ and $\delta$, we consider the scaling
\begin{align}
  x^0 &\to \eta^{-1} x^0\ , & \xi_+ &\to \eta^{-1/2} \xi_+ \ ,\nonumber \\
  z,\bar{z} &\to \eta^{- 1/2} z,\bar{z}\ , & \xi_- &\to \xi_-\ ,\nonumber\\
  Z,\bar{Z} &\to Z,\bar{Z}\ , & \zeta_+ &\to \eta^{1/2} \zeta_+\ ,\nonumber\\
  X^A &\to \eta^{1/2} X^A \ ,& \zeta_- &\to \eta  \zeta_-\ ,\nonumber\\
  \Psi_+ &\to \eta^{1/2} \Psi_+\ ,\nonumber\\
  \Psi_- &\to \eta \Psi_-\ .
  \label{eq: BLG scaling}
\end{align}
The scaling of $\xi$ and $\zeta$ imply that in terms of $\epsilon_\pm$  we have
\begin{align}
  \epsilon_+ &\to \eta^{-1/2} \left( \left( 1-z\partial -\bar{z}\bar{\partial} \right)\epsilon_+ + \eta  \left( z\partial +\bar{z}\bar{\partial} \right)\epsilon_+\right)\ , \nonumber\\
  \epsilon_- &\to \epsilon_-\ .
\end{align}
After this scaling, the action is of the form $S=\eta^{-1} S_{-1} + S_0 + \eta S_1$, with
{\allowdisplaybreaks
\begin{align}
  S_{-1} &= \int d^3 x\, \bigg( -\angl{DZ}{\bar{D}\bar{Z}} - \angl{D\bar{Z}}{\bar{D}Z} \nn\\*
  & \hspace{22mm}+ \frac{1}{8} \angl{\trip{X^A}{Z}{\bar{Z}}}{\trip{X^A}{Z}{\bar{Z}}}
  + \frac{1}{4} \angl{\bar{\Psi}_+}{\Gamma_0\trip{Z}{\bar{Z}}{\Psi_+}}\bigg) \ ,\\
  S_0 &= \int d^3 x\, \bigg( \frac{1}{2} \angl{D_0 Z}{D_0 \bar{Z}} - 2 \angl{DX^A}{\bar{D}X^A} - \frac{1}{4} \angl{\trip{X^A}{X^B}{Z}}{\trip{X^A}{X^B}{\bar{Z}}} \nn\\*
  &\hspace{22mm}  -\frac{i}{2} \angl{\bar{\Psi}_+}{\Gamma_0 D_0 \Psi_+} + 2i \angl{\bar{\Psi}_+}{\left( \Gamma_z \bar{D} + \Gamma_{\bar{z}}D\right)\Psi_-} \nonumber\\*
  &\hspace{22mm} - \frac{1}{4}\angl{\bar{\Psi}_-}{\Gamma_0 \trip{Z}{\bar{Z}}{\Psi_-}}+i\angl{\bar{\Psi}_+}{\Gamma^A\Gamma_Z\trip{X^A}{Z}{\Psi_-}}   \nonumber\\*
  &\hspace{22mm} + i\angl{\bar{\Psi}_+}{\Gamma^A\Gamma_{\bar{Z}}\trip{X^A}{\bar{Z}}{\Psi_-}}+ \frac{i}{4} \angl{\bar{\Psi}_+}{\Gamma^{AB}\trip{X^A}{X^B}{\Psi_+}}\nn\\
  &\hspace{22mm}+ i \big( \left( A_0, F_{z\bar{z}} \right) + \left( A_{\bar{z}}, F_{0z} \right) + \left( A_z, F_{\bar{z}0} \right) + \left( A_0, \left[ A_z, A_{\bar{z}} \right] \right) \big) \bigg) \ ,\\
  S_{1} &= \int d^3 x\, \bigg( \frac{1}{2} \angl{D_0 X^A}{D_0 X^A} - \frac{1}{12} \angl{\trip{X^A}{X^B}{X^C}}{\trip{X^A}{X^B}{X^C}}\nonumber\\*
  &\hspace{22mm} - \frac{i}{2}\angl{\bar{\Psi}_-}{\Gamma_0 D_0 \Psi_-} + \frac{i}{4}\angl{\Bar{\Psi}_-}{\Gamma^{AB}\trip{X^A}{X^B}{\Psi_-}} \bigg)\ .
  \label{eq: BLG pre-redef}
\end{align}
}%
This action is invariant under $\delta = \eta^{-1} \delta_{-1} + \delta_0 + \eta \delta_1$, with
\newcommand{\oneminz}{\left( 1-z\partial \right)}
\newcommand{\oneminzbar}{\left( 1-\bar{z}\bar{\partial} \right)}
\newcommand{\oneminboth}{\left( 1-z\partial -\bar{z}\bar{\partial} \right)}
{\allowdisplaybreaks
\begin{align}
  \delta_{-1} \Psi_- &= 2 \left( \bar{D}Z \right) \Gamma_z \Gamma_Z \oneminzbar\epsilon_+ + 2\left( D\bar{Z} \right)\Gamma_{\bar{z}}\Gamma_{\bar{Z}}\oneminz\epsilon_+  \\*
  &\qquad +\tfrac{i}{2} \trip{X^A}{Z}{\bar{Z}}\Gamma^A \Gamma_0\oneminboth\epsilon_+\ ,\nonumber\\
  \delta_{-1} A_0 (\,\cdot\,) &= -\bar{\epsilon}_+ \Gamma_Z \trip{Z}{\Psi_+}{} + \bar{\epsilon}_+ \Gamma_{\bar{Z}} \trip{\bar{Z}}{\Psi_+}{}\nonumber\\[1em]
  \delta_0 Z &= 2i\left( \oneminz\bar{\epsilon}_+ \right) \Gamma_{\bar{Z}}\Psi_+  \ ,\\
  \delta_0 \bar{Z} &= 2i\left( \oneminzbar\bar{\epsilon}_+ \right) \Gamma_{Z}\Psi_+  \ ,\nonumber\\
  \delta_0 X^A &= i\left(\oneminboth \bar{\epsilon}_+ \right) \Gamma^A \Psi_- + i\bar{\epsilon}_- \Gamma^A \Psi_+  \ ,\nonumber\\
  \delta_0 \Psi_+ &= - \left( D_0 Z \right) \Gamma_0 \Gamma_Z \oneminzbar\epsilon_+ - \left( D_0 \bar{Z} \right) \Gamma_0 \Gamma_{\bar{Z}} \oneminz\epsilon_+ + 2 \left( DZ \right) \Gamma_{\bar{z}}\Gamma_Z \epsilon_-  \nonumber\\*
  &\hspace{4.7mm} + 2 \left( \bar{D}\bar{Z} \right) \Gamma_{z}\Gamma_{\bar{Z}} \epsilon_- + 2 \left( DX^A \right) \Gamma_{\bar{z}} \Gamma^A \oneminz\epsilon_+ + 2 \left( \bar{D} X^A \right) \Gamma_{z} \Gamma^A \oneminzbar\epsilon_+  \nonumber\\*
  &\hspace{4.7mm} -\tfrac{1}{2} \trip{X^A}{X^B}{Z}\Gamma^{AB} \Gamma_Z \oneminzbar\epsilon_+ -\tfrac{1}{2} \trip{X^A}{X^B}{\bar{Z}}\Gamma^{AB} \Gamma_{\bar{Z}} \oneminz\epsilon_+ \nonumber\\*
  &\hspace{4.7mm} - \tfrac{i}{2}\trip{X^A}{Z}{\bar{Z}}\Gamma^A \Gamma_0\epsilon_- + \tfrac{2}{3} Z\Gamma_{\bar{z}} \Gamma_Z \partial\epsilon_- + \tfrac{2}{3} \bar{Z}\Gamma_{z} \Gamma_{\bar{Z}} \bar{\partial}\epsilon_- \nonumber\\*
  &\hspace{4.7mm} + \tfrac{1}{3}Z\Gamma_Z \Gamma_0 \partial_0 \epsilon_+ + \tfrac{1}{3} \bar{Z}\Gamma_{\bar{Z}}\Gamma_0 \partial_- \epsilon_+ \ ,\nonumber\\
  \delta_0 \Psi_- &= - \left( D_0 X^A \right) \Gamma_0 \Gamma^A \oneminboth\epsilon_+ + 2 \left( D X^A \right) \Gamma_{\bar{z}}\Gamma^A \epsilon_- + 2 \left( \bar{D} X^A \right) \Gamma_{z}\Gamma^A \epsilon_- \nonumber\\
  &\hspace{4.7mm} - \left( D_0 Z \right) \Gamma_0\Gamma_Z \epsilon_- -\left( D_0 \bar{Z} \right) \Gamma_0\Gamma_{\bar{Z}} \epsilon_- -\tfrac{1}{6}\trip{X^A}{X^B}{X^C}\Gamma^{ABC}\oneminboth\epsilon_+ \nonumber\\
  &\hspace{4.7mm} -\tfrac{1}{2}\trip{X^A}{X^B}{Z}\Gamma^{AB}\Gamma_Z\epsilon_- - \tfrac{1}{2}\trip{X^A}{X^B}{\bar{Z}}\Gamma^{AB}\Gamma_{\bar{Z}}\epsilon_- + 2\bar{z}\left( \bar{D}Z \right) \Gamma_z \Gamma_Z \bar{\partial}\epsilon_+ \nonumber\\
  &\hspace{4.7mm} + 2z\left( D\bar{Z} \right) \Gamma_{\bar{z}} \Gamma_{\bar{Z}} \partial\epsilon_+ + \tfrac{i}{2} \trip{X^A}{Z}{\bar{Z}} \Gamma^A \Gamma_0 \left( z\partial + \bar{z}\bar{\partial} \right) \epsilon_+ + \tfrac{2}{3} Z\Gamma_z\Gamma_Z\bar{\partial}\epsilon_+ \nonumber\\
  &\hspace{4.7mm} + \tfrac{2}{3} \bar{Z}\Gamma_{\bar{z}}\Gamma_{\bar{Z}}\partial\epsilon_+ + \tfrac{1}{3} Z\Gamma_Z \Gamma_0 \partial_0 \epsilon_- + \tfrac{1}{3} \bar{Z} \Gamma_{\bar{Z}} \Gamma_0 \partial_0 \epsilon_- \nonumber\\
  &\hspace{4.7mm} - \tfrac{2}{3} X^A \Gamma^A \Gamma_{\bar{z}} \partial \epsilon_- - \tfrac{2}{3} X^A \Gamma^A \Gamma_{z} \bar{\partial} \epsilon_- +\tfrac{1}{3} X^A \Gamma^A \Gamma_0 \partial_0 \epsilon_+ \ ,\nonumber\\
  \delta_0 A_0(\,\cdot\,) &= \bar{\epsilon}_- \Gamma_Z \trip{Z}{\Psi_-}{} - \bar{\epsilon}_- \Gamma_{\bar{Z}} \trip{\bar{Z}}{\Psi_-}{} \nonumber\\
  &\hspace{4.7mm} + i\left( \oneminboth\bar{\epsilon}_+ \right) \Gamma_0 \Gamma^A \trip{X^A}{\Psi_-}{} + i\bar{\epsilon}_- \Gamma_0 \Gamma^A \trip{X^A}{\Psi_+}{}  \ ,\nonumber\\
  \delta_0 A_z(\,\cdot\,) &= i \left( \oneminzbar\bar{\epsilon}_+ \right) \Gamma_z \Gamma_Z \trip{Z}{\Psi_-}{} + i \bar{\epsilon}_- \Gamma_z \Gamma_{\bar{Z}} \trip{\bar{Z}}{\Psi_+}{} \nonumber\\
  &\hspace{4.7mm} + i\left( \oneminzbar\bar{\epsilon}_+ \right)\Gamma_z \Gamma^A \trip{X^A}{\Psi_+}{} \ ,\nonumber\\
  \delta_0 A_{\bar{z}}(\,\cdot\,) &= i \bar{\epsilon}_- \Gamma_{\bar{z}} \Gamma_Z \trip{Z}{\Psi_+}{} + i \left( \oneminz\bar{\epsilon}_+ \right) \Gamma_{\bar{z}} \Gamma_{\bar{Z}} \trip{\bar{Z}}{\Psi_-}{} \nonumber\\
  &\hspace{4.7mm} + i\left( \oneminz\bar{\epsilon}_+ \right)\Gamma_{\bar{z}} \Gamma^A \trip{X^A}{\Psi_+}{}  \ ,\nonumber\\[1em]
  \delta_1 Z &= 2i\bar{\epsilon}_- \Gamma_{\bar{Z}}\Psi_- + 2iz \left( \partial \bareps_+ \right) \Gamma_{\bar{Z}}\Psi_+ \ ,\\
  \delta_1 \bar{Z} &= 2i\bar{\epsilon}_- \Gamma_{Z}\Psi_- + 2i\bar{z} \left( \bar{\partial} \bareps_+ \right) \Gamma_Z\Psi_+ \ ,\nonumber\\
  \delta_1 X^A &= i \left( \left( z\partial + \bar{z}\bar{\partial} \right) \bareps_+ \right) \Gamma^A \Psi_-\ ,\nonumber\\
  \delta_1 \Psi_+ &= -\left( D_0 X^A \right) \Gamma_0 \Gamma^A \epsilon_- - \tfrac{1}{6} \trip{X^A}{X^B}{X^C}\Gamma^{ABC}\epsilon_- -\bar{z}\left( D_0 Z \right)\Gamma_0 \Gamma_Z \bar{\partial} \epsilon_+ \nonumber\\
  &\hspace{4.7mm} - z\left( D_0 \bar{Z} \right)\Gamma_0 \Gamma_{\bar{Z}} \partial \epsilon_+ + 2z \left( DX^A \right)\Gamma_{\bar{z}}\Gamma^A \partial \epsilon_+ + 2\bar{z} \left( \bar{D}X^A \right)\Gamma_{z}\Gamma^A \bar{\partial} \epsilon_+ \nonumber\\
  &\hspace{4.7mm} - \tfrac{1}{2}\bar{z}\trip{X^A}{X^B}{Z}\Gamma^{AB} \Gamma_Z \bar{\partial} \epsilon_+ - \tfrac{1}{2}z\trip{X^A}{X^B}{\bar{Z}}\Gamma^{AB} \Gamma_{\bar{Z}} \partial \epsilon_+ \nonumber\\
  &\hspace{4.7mm} - \tfrac{2}{3} X^A \Gamma^A \Gamma_{\bar{z}}\partial \epsilon_+ - \tfrac{2}{3} X^A \Gamma^A \Gamma_{z}\bar{\partial} \epsilon_+ + \tfrac{2}{3} X^A \Gamma^A \Gamma_0 \partial_0 \epsilon_- \ ,\nonumber\\
  \delta_1 \Psi_- &= - \left( D_0 X^A \right) \Gamma_0 \Gamma^A \left( z\partial + \bar{z}\bar{\partial}  \right) \epsilon_+ - \tfrac{1}{6} \trip{X^A}{X^B}{X^C} \Gamma^{ABC} \left( z\partial + \bar{z}\bar{\partial} \right)\epsilon_+ \ ,\nonumber\\
  \delta_1 A_0(\,\cdot\,) &= i \left( \left( z\partial + \bar{z}\bar{\partial} \right) \bareps_+ \right) \Gamma_0 \Gamma^A \trip{X^A}{\Psi_-}{} \ ,\nonumber\\
  \delta_1 A_z(\,\cdot\,) &= i \bar{\epsilon}_- \Gamma_z \Gamma^A \trip{X^A}{\Psi_-}{} + i\bar{z} \left( \bar{\partial} \bareps_+ \right)\Gamma_z \Gamma_Z \trip{Z}{\Psi_-}{} + i\bar{z} \left( \bar{\partial} \bareps_+ \right) \Gamma_z \Gamma^A \trip{X^A}{\Psi_+}{} \ ,\nonumber\\
  \delta_1 A_{\bar{z}}(\,\cdot\,) &= i \bar{\epsilon}_- \Gamma_{\bar{z}} \Gamma^A \trip{X^A}{\Psi_-}{} + iz \left( \partial \bareps_+ \right)\Gamma_{\bar{z}} \Gamma_{\bar{Z}} \trip{\bar{Z}}{\Psi_-}{} + iz \left( \partial \bareps_+ \right) \Gamma_{\bar{z}} \Gamma^A \trip{X^A}{\Psi_+}{}\nonumber\ ,
  \label{eq: BLG SUSYs pre-redef}
\end{align}}%
and $\epsilon=\xi + x^\mu \Gamma_\mu \zeta = \xi + \left( x^0 \Gamma_0 + z\Gamma_z + \bar{z} \Gamma_{\bar{z}} \right)\zeta$.

We see that $S_{-1}$ is not of the form (\ref{eq: simple S_{-1}}), and also that $\delta_{-1}$ acts non-trivially on the bosonic field $A_0$, as well as $\Psi_-$. However, observe that
\begin{align}
  \delta_{-1} A_0 (\,\cdot\,) - \frac{i}{2} \delta_0\trip{Z}{\bar{Z}}{} = 0\ .
\end{align}
Thus, following the discussion in Section \ref{sec: a supersymmetric scaling limit}, we consider the following field redefiniton:
\begin{align}
  A_0 \to \hat{A}_0 = A_0 + \frac{1}{\eta} \frac{i}{2} \trip{Z}{\bar{Z}}{}\ .
  \label{eq: BLG field redef}
\end{align}
We can now calculate how both the action and the supersymmetry field transformations are changed. As described in section \ref{subsec: more general prescription}, we find corrections to $S_{-1}$ arising from terms linear in $\hat{A}_0$ in $S_0$, as well as from terms quadratic in $\hat{A}_0$ in $S_1$. Similarly, $S_0$ is corrected by terms linear in $\hat{A}_0$ in $S_1$.

The resulting shift of the action is
\begin{align}
  S_{-1} &\to S_{-1} + \int d^3 x\, \Big( -\angl{\bar{Z}}{F_{z\bar{z}}(Z)} - \tfrac{1}{8} \angl{\trip{X^A}{Z}{\bar{Z}}}{\trip{X^A}{Z}{\bar{Z}}}
  - \tfrac{1}{4} \angl{\bar{\Psi}_+}{\Gamma_0\trip{Z}{\bar{Z}}{\Psi_+}}\Big)\ ,\nonumber\\
  S_0 &\to S_0 + \int d^3 x\, \Big( - \tfrac{i}{2}\angl{D_0 X^A}{\trip{X^A}{Z}{\bar{Z}}} - \tfrac{1}{4}\angl{\bar{\Psi}_-}{\Gamma_0 \trip{Z}{\bar{Z}}{\Psi_-}}\Big)\ ,
\end{align}
while the supersymmetry transformations are shifted by
\begin{align}
  \delta_{-1}\Psi_- &\to \delta_{-1}\Psi_- - \tfrac{i}{2} \trip{X^A}{Z}{\bar{Z}}\Gamma^A \Gamma_0\oneminboth\epsilon_+ \ ,\nonumber\\
  \delta_{-1}A_0(\,\cdot\,) &\to \delta_{-1}A_0(\,\cdot\,) +\bar{\epsilon}_+ \Gamma_Z \trip{Z}{\Psi_+}{} - \bar{\epsilon}_+ \Gamma_{\bar{Z}} \trip{\bar{Z}}{\Psi_+}{} \ ,\nonumber\\[1em]
  \delta_0\Psi_+ &\to \delta_0\Psi_+ - \tfrac{i}{2}\trip{X^A}{Z}{\bar{Z}}\Gamma^A \Gamma_0\epsilon_- \ ,\nonumber\\
  \delta_0\Psi_- &\to \delta_0\Psi_- - \tfrac{i}{2} \trip{X^A}{Z}{\bar{Z}} \Gamma^A \Gamma_0 \left( z\partial + \bar{z}\bar{\partial} \right) \epsilon_+\ ,\nonumber\\
  \delta_{0}A_0(\,\cdot\,) &\to \delta_{0}A_0(\,\cdot\,) + \bar{\epsilon}_- \Gamma_Z \trip{Z}{\Psi_-}{} - \bar{\epsilon}_- \Gamma_{\bar{Z}} \trip{\bar{Z}}{\Psi_-}{} \ .  
  \end{align}
In particular we find that 
\begin{align}
  S_{-1} &= \int d^3 x\, \left( -2\angl{\bar{D}Z}{D\bar{Z}}  \right)\ ,
\end{align}
and
\begin{align}
  \delta_{-1} \Psi_- &= 2 \left( \bar{D}Z \right) \Gamma_z \Gamma_Z \oneminzbar\epsilon_+ + 2\left( D\bar{Z} \right)\Gamma_{\bar{z}}\Gamma_{\bar{Z}}\oneminz\epsilon_+\ ,
\end{align}
with $\delta_{-1}$ vanishing on all other fields. 

The theory is now in the correct form to proceed. We introduce Lagrange multiplier field $H$, in the same representation of $\mathcal{G}$ as $Z$, and imposing the condition $\bar{D}Z=0$. The end result is that in the $\eta\to 0$ limit, the theory is described by the action
\begin{align}
 \tilde S &=  \int d^3 x\, \bigg( \frac{1}{2} \angl{D_0 Z}{D_0 \bar{Z}} - 2 \angl{DX^A}{\bar{D}X^A} + \angl{H}{\bar{D}Z} + \angl{\bar{H}}{D\bar{Z}} \nonumber\\
  &\hspace{10mm} - \frac{1}{4} \angl{\trip{X^A}{X^B}{Z}}{\trip{X^A}{X^B}{\bar{Z}}} - \frac{i}{2}\angl{D_0 X^A}{\trip{X^A}{Z}{\bar{Z}}}  -\frac{i}{2} \angl{\bar{\Psi}_+}{\Gamma_0 D_0 \Psi_+} \nonumber\\
  &\hspace{10mm} + 2i \angl{\bar{\Psi}_+}{\left( \Gamma_z \bar{D} + \Gamma_{\bar{z}}D\right)\Psi_-} - \frac{1}{2}\angl{\bar{\Psi}_-}{\Gamma_0 \trip{Z}{\bar{Z}}{\Psi_-}} +i\angl{\bar{\Psi}_+}{\Gamma^A\Gamma_Z\trip{X^A}{Z}{\Psi_-}}  \nonumber\\
  &\hspace{10mm} + i\angl{\bar{\Psi}_+}{\Gamma^A\Gamma_{\bar{Z}}\trip{X^A}{\bar{Z}}{\Psi_-}} + \frac{i}{4} \angl{\bar{\Psi}_+}{\Gamma^{AB}\trip{X^A}{X^B}{\Psi_+}} \nonumber\\
  &\hspace{10mm} + i \big( \left( A_0, F_{z\bar{z}} \right) + \left( A_{\bar{z}}, F_{0z} \right) + \left( A_z, F_{\bar{z}0} \right) + \left( A_0, \left[ A_z, A_{\bar{z}} \right] \right) \big) \bigg)\ ,
  \label{eq: final scaled BLG action}
\end{align}
which is invariant under
{\allowdisplaybreaks
\begin{align}
 \tilde \delta Z &= 2i\left( \oneminz\bar{\epsilon}_+ \right) \Gamma_{\bar{Z}}\Psi_+  \ ,\nonumber\\
 \tilde \delta \bar{Z} &= 2i\left( \oneminzbar\bar{\epsilon}_+ \right) \Gamma_{Z}\Psi_+  \ ,\nonumber\\
 \tilde \delta X^A &= i\left(\oneminboth \bar{\epsilon}_+ \right) \Gamma^A \Psi_- + i\bar{\epsilon}_- \Gamma^A \Psi_+  \ ,\nonumber\\
 \tilde \delta \Psi_+ &= - \left( D_0 Z \right) \Gamma_0 \Gamma_Z \oneminzbar\epsilon_+ - \left( D_0 \bar{Z} \right) \Gamma_0 \Gamma_{\bar{Z}} \oneminz\epsilon_+ + 2 \left( DZ \right) \Gamma_{\bar{z}}\Gamma_Z \epsilon_-  \nonumber\\
  &\hspace{4.7mm} + 2 \left( \bar{D}\bar{Z} \right) \Gamma_{z}\Gamma_{\bar{Z}} \epsilon_- + 2 \left( DX^A \right) \Gamma_{\bar{z}} \Gamma^A \oneminz\epsilon_+ + 2 \left( \bar{D} X^A \right) \Gamma_{z} \Gamma^A \oneminzbar\epsilon_+  \nonumber\\
  &\hspace{4.7mm} -\tfrac{1}{2} \trip{X^A}{X^B}{Z}\Gamma^{AB} \Gamma_Z \oneminzbar\epsilon_+ -\tfrac{1}{2} \trip{X^A}{X^B}{\bar{Z}}\Gamma^{AB} \Gamma_{\bar{Z}} \oneminz\epsilon_+ \nonumber\\
  &\hspace{4.7mm} - i\trip{X^A}{Z}{\bar{Z}}\Gamma^A \Gamma_0\epsilon_- + \tfrac{2}{3} Z\Gamma_{\bar{z}} \Gamma_Z \partial\epsilon_- + \tfrac{2}{3} \bar{Z}\Gamma_{z} \Gamma_{\bar{Z}} \bar{\partial}\epsilon_- \nonumber\\
  &\hspace{4.7mm} + \tfrac{1}{3}Z\Gamma_Z \Gamma_0 \partial_0 \epsilon_+ + \tfrac{1}{3} \bar{Z}\Gamma_{\bar{Z}}\Gamma_0 \partial_- \epsilon_+ \ ,\nonumber\\
 \tilde \delta \Psi_- &= - \left( D_0 X^A \right) \Gamma_0 \Gamma^A \oneminboth\epsilon_+ + 2 \left( D X^A \right) \Gamma_{\bar{z}}\Gamma^A \epsilon_- + 2 \left( \bar{D} X^A \right) \Gamma_{z}\Gamma^A \epsilon_- \nonumber\\
  &\hspace{4.7mm} - \left( D_0 Z \right) \Gamma_0\Gamma_Z \epsilon_- -\left( D_0 \bar{Z} \right) \Gamma_0\Gamma_{\bar{Z}} \epsilon_- -\tfrac{1}{6}\trip{X^A}{X^B}{X^C}\Gamma^{ABC}\oneminboth\epsilon_+ \nonumber\\
  &\hspace{4.7mm} -\tfrac{1}{2}\trip{X^A}{X^B}{Z}\Gamma^{AB}\Gamma_Z\epsilon_- - \tfrac{1}{2}\trip{X^A}{X^B}{\bar{Z}}\Gamma^{AB}\Gamma_{\bar{Z}}\epsilon_- + 2\bar{z}\left( \bar{D}Z \right) \Gamma_z \Gamma_Z \bar{\partial}\epsilon_+ \nonumber\\
  &\hspace{4.7mm} + 2z\left( D\bar{Z} \right) \Gamma_{\bar{z}} \Gamma_{\bar{Z}} \partial\epsilon_+ + \tfrac{2}{3} Z\Gamma_z\Gamma_Z\bar{\partial}\epsilon_+ + \tfrac{2}{3} \bar{Z}\Gamma_{\bar{z}}\Gamma_{\bar{Z}}\partial\epsilon_+ + \tfrac{1}{3} Z\Gamma_Z \Gamma_0 \partial_0 \epsilon_- \nonumber\\
  &\hspace{4.7mm}  + \tfrac{1}{3} \bar{Z} \Gamma_{\bar{Z}} \Gamma_0 \partial_0 \epsilon_- - \tfrac{2}{3} X^A \Gamma^A \Gamma_{\bar{z}} \partial \epsilon_- - \tfrac{2}{3} X^A \Gamma^A \Gamma_{z} \bar{\partial} \epsilon_- +\tfrac{1}{3} X^A \Gamma^A \Gamma_0 \partial_0 \epsilon_+ \nonumber\\
  &\hspace{4.7mm} - H \Gamma_{\bar{z}} \Gamma_{\bar{Z}}\oneminz \epsilon_+ - \bar{H} \Gamma_z \Gamma_Z \oneminzbar \epsilon_+ \ ,\nonumber\\
\tilde  \delta A_0(\,\cdot\,) &= 2\bar{\epsilon}_- \Gamma_Z \trip{Z}{\Psi_-}{} - 2\bar{\epsilon}_- \Gamma_{\bar{Z}} \trip{\bar{Z}}{\Psi_-}{} \nonumber\\
  &\hspace{4.7mm} + i\left( \oneminboth\bar{\epsilon}_+ \right) \Gamma_0 \Gamma^A \trip{X^A}{\Psi_-}{} + i\bar{\epsilon}_- \Gamma_0 \Gamma^A \trip{X^A}{\Psi_+}{}  \ ,\nonumber\\
\tilde  \delta A_z(\,\cdot\,) &= i \left( \oneminzbar\bar{\epsilon}_+ \right) \Gamma_z \Gamma_Z \trip{Z}{\Psi_-}{} + i \bar{\epsilon}_- \Gamma_z \Gamma_{\bar{Z}} \trip{\bar{Z}}{\Psi_+}{} \nonumber\\
  &\hspace{4.7mm} + i\left( \oneminzbar\bar{\epsilon}_+ \right)\Gamma_z \Gamma^A \trip{X^A}{\Psi_+}{} \ ,\nonumber\\
 \tilde \delta A_{\bar{z}}(\,\cdot\,) &= i \bar{\epsilon}_- \Gamma_{\bar{z}} \Gamma_Z \trip{Z}{\Psi_+}{} + i \left( \oneminz\bar{\epsilon}_+ \right) \Gamma_{\bar{z}} \Gamma_{\bar{Z}} \trip{\bar{Z}}{\Psi_-}{} \nonumber\\
  &\hspace{4.7mm} + i\left( \oneminz\bar{\epsilon}_+ \right)\Gamma_{\bar{z}} \Gamma^A \trip{X^A}{\Psi_+}{}\ ,\nonumber\\
\tilde  \delta H &= -2 \left( \oneminzbar\bareps_+ \right)\Gamma_z \Gamma_Z D_0 \Psi_- - i \left( \oneminzbar\bareps_+ \right) \Gamma_z \Gamma_Z \Gamma^{AB} \trip{X^A}{X^B}{\Psi_-}\nonumber\\
  &\hspace{4.7mm} -4i \bareps_- \Gamma_Z D\Psi_- -4i \left( \partial \bareps_- \right)\Gamma_Z \Psi_- - 4i\bar{z}\left( \bar{\partial}\bareps_+ \right)\Gamma_Z D\Psi_+ - 2i \bareps_- \Gamma_z \Gamma^A \trip{X^A}{\bar{Z}}{\Psi_-} \nonumber\\
   &\hspace{4.7mm}  - 2i \bar{z} \left( \bar{\partial}\bareps_+ \right)\Gamma_z\Gamma_Z \trip{Z}{\bar{Z}}{\Psi_-} -2i\bar{z} \left( \bar{\partial}\bareps_+ \right)\Gamma_z \Gamma^A \trip{X^A}{\bar{Z}}{\Psi_+}\ ,
\end{align}}where
\begin{align}
  \epsilon = \xi + x^\mu \Gamma_\mu \zeta = \xi + \left( x^0 \Gamma_0 + z\Gamma_z + \bar{z} \Gamma_{\bar{z}} \right)\zeta \ .
\end{align}
For $\zeta=0$, we recover the rigid supersymmetry of $S$ as found in \cite{Lambert:2018lgt}. However, in this more general construction, we see that all 32 supercharges of the original ${\cal N}=8$ theory survive in the limiting theory. As discussed in \cite{Kucharski:2017jwv}, the constraint $\bar{D}Z=0$ together with the Gauss law constraint arising from integrating out the non-dynamical $A_0$ form a three-algebra variation of the Hitchin system for $SU(2)$ gauge group.

\subsection{${\cal N}=6$}\label{subsec: ABJM}

It is natural to ask whether we can obtain similar results when applying an analogous scaling to the ABJM/ABJ theory \cite{Aharony:2008ug,Aharony:2008gk}. This is a $U(N)\times U(M)$ Chern-Simons-matter theory with action
\begin{align}
  S = \text{tr} \int & d^3 x \bigg(  - \left( D_\mu Z^I \right) \left( D^\mu Z_I \right) - \frac{8\pi^2}{3k^2}\Upsilon^{KL}_J \Upsilon_{KL}^J\nonumber\\
  &+\frac{k}{4\pi} \epsilon^{\mu\nu\lambda} \left( \left( A_\mu^L \partial_\nu A_\lambda^L - \frac{2i}{3}A_\mu^L A_\nu^L A_\lambda^L \right) - \left( A_\mu^R \partial_\nu A_\lambda^R - \frac{2i}{3}A_\mu^R A_\nu^R A_\lambda^R \right) \right)\nonumber\\
  &-i \bar{\Psi}^I \gamma^\mu D_\mu \Psi_I - \frac{2\pi i}{k}\bar{\Psi}^I [\Psi_I, Z^J;Z_J] + \frac{4\pi i}{k} \bar{\Psi}^I [\Psi_J, Z^J; Z_I]\nonumber\\
  & +\frac{\pi i}{k}\varepsilon_{IJKL} \bar{\Psi}^I [Z^K, Z^L; \Psi^J] - \frac{\pi i}{k}\varepsilon^{IJKL} \bar{\Psi}_I[Z_K, Z_L, \Psi_J] \bigg)\ ,
\end{align}
where we have used the conventions of \cite{Bagger:2008se}. In particular, we have
\begin{align}
  [Z^I, Z^J; Z_K] &= Z^I Z_K Z^J - Z^J Z_K Z^I\ ,\nonumber\\
  \Upsilon^{KL}_J &= [Z^K, Z^L; Z_J] - \tfrac{1}{2} \delta_J^K [Z^E, Z^L; Z_E] + \tfrac{1}{2} \delta_J^L [Z^E, Z^K, Z_E]\ .
\end{align}
The matter fields here are $N\times M$ complex matrices, while $A_m^L$ is a hermitian $N\times N$ matrix, and $A_m^R$  a hermitian $M\times M$ matrix. Hermitian conjugation acts to raise/lower the R-symmetry index $I=1,2,3,4$. The fields $Z^I$ and $\psi_I$ transform in the $(\mathbf{N},\bar{\mathbf{M}})$ of the $U(N)_L\times U(M)_R$ gauge symmetry, and so we have
\begin{align}
  D_\mu Z^I = \partial_\mu Z^I - i A_\mu^L Z^I + i Z^I A_\mu^R\ ,
\end{align}
and similarly for $\psi_I$. Finally, we choose conventions in which $\epsilon^{012}=-1$, and a representation for the $(1+2)$-dimensional Clifford algebra is chosen such that the $\gamma^m$, $\mu=0,1,2$ are real $2\times 2$ matrices satisfying $\gamma^0 \gamma^1 \gamma^2 = 1$. 

The theory possesses 12 supercharges corresponding to $\mathcal{N}=6$ rigid supersymmetry, and an additional 12 corresponding to superconformal symmetry. In particular, we find $\delta S=0$ where
\begin{align}
  \delta Z^I &=  i \bar{\epsilon}^{IJ} \psi_J \ ,\nonumber\\
\delta A_\mu^L &= \tfrac{2\pi}{k} \left( \bar{\epsilon}^{IJ} \gamma_\mu \Psi_I Z_J - \bar{\epsilon}_{IJ} \gamma_\mu Z^J \Psi^I \right) \ ,\nonumber\\
\delta A_\mu^R &= \tfrac{2\pi}{k} \left( \bar{\epsilon}^{IJ} \gamma_\mu Z_J \Psi_I  - \bar{\epsilon}_{IJ} \gamma_\mu \Psi^I Z^J  \right) \ ,\nonumber\\
\delta \Psi_J &= D_\mu Z^I \gamma^\mu \epsilon_{IJ} + \tfrac{2\pi}{k} \Upsilon^{KL}_J \epsilon_{KL} + \tfrac{1}{3}Z^I \gamma^\mu \partial_\mu \epsilon_{IJ}\ ,
\end{align}
where the supersymmetry parameter $\epsilon_{IJ}$ takes the form $\epsilon_{IJ}=\xi_{IJ}+x^\mu\gamma_\mu\zeta_{IJ}$ for constant $\xi_{IJ},\zeta_{IJ}$. What's more, $\epsilon_{IJ}$ is anti-symmetric in its R-symmetry indices, and satisfies the reality condition $\epsilon^{IJ}=\frac{1}{2} \varepsilon^{IJKL}\epsilon_{KL}$, which automatically ensures that $\xi_{IJ}$ and $\zeta_{IJ}$ satisfy the same condition. 

Next, we seek a scaling analogous to that performed on the ${\cal N}=8$ theory. In particular, we single out $Z^1$, and look for a scaling which, in the $\eta\to 0$ limit, will localise onto the static $\tfrac{1}{2}$-BPS state $\bar{D} Z^1 = 0$, where we have once again defined the worldvolume complex coordinate $z=x^1+ix^2$, and $D\equiv D_{z}$. Further, we split spinors into definite chirality under $i\gamma^0$, so that $\Psi^\pm_I := P_\pm \Psi_I$ and $\epsilon_{IJ}^\pm := P_\pm \epsilon_{IJ}$ with $P_\pm=\tfrac{1}{2}\left(1\pm i\gamma^0\right)$. Then, upon comparing the form of $\delta$ to that of the ${\cal N}=8$ theory, we find that $\Psi_1^-$ and $\Psi_A^+$ play a role analogous to that of $\Psi_+$ in the ${\cal N}=8$ theory, while $\Psi_1^+$ and $\Psi_A^-$ are analogous to $\Psi_-$, where here $A=2,3,4$. We find similar correspondences for the constant components of $\epsilon_{IJ}$, giving us the full scalings:
\begin{align}
  x^0 &\to \eta^{-1} x^0 \ ,& \xi_{1A}^+ , \xi_{AB}^- &\to \eta^{-1/2} \xi_{1A}^+ , \eta^{-1/2}\xi_{AB}^- \ ,\nonumber \\
  z,\bar{z} &\to \eta^{-1/2} z,\eta^{-1/2}\bar{z} \ ,& \xi_{1A}^- , \xi_{AB}^+ &\to \xi_{1A}^- , \xi_{AB}^+\ ,\nonumber\\
  Z^1 &\to  Z^1 \ ,& \zeta_{1A}^+ , \zeta_{AB}^- &\to \eta^{1/2} \zeta_{1A}^+ , \eta^{1/2}\zeta_{AB}^-\ ,\nonumber\\
  Z^A &\to \eta^{1/2} Z^A \ ,& \zeta_{1A}^- , \zeta_{AB}^+ &\to \eta \, \zeta_{1A}^- , \eta\,\zeta_{AB}^+\ ,\nonumber\\
  \Psi_1^-,\Psi_A^+ &\to \eta^{1/2} \Psi_1^-,\eta^{1/2}\Psi_A^+ \ ,\nonumber \\
  \Psi_1^+,\Psi_A^- &\to \eta \Psi_1^+,\eta\Psi_A^-\ .
  \label{eq: ABJM scaling}
\end{align}
Having performed this scaling, the action takes the form $S=\eta^{-1} S_{-1} + S_0 + \eta S_1$, with
{\allowdisplaybreaks
\begin{align}
  S_{-1} &= \text{tr} \int d^3 x \bigg[ -2 \left( DZ^1 \right) \left( \bar{D} Z_1 \right) -2 \left( \bar{D}Z^1 \right) \left( D Z_1 \right)  + \tfrac{4\pi^2}{k^2}\trips{Z^1}{Z^A}{Z_1}\trips{Z_1}{Z_A}{Z^1}\nonumber\\*
  & \hspace{23mm} - \tfrac{2\pi i}{k} \left( \bar{\Psi}^{A,+} \trips{\Psi_A^+}{Z^1}{Z_1} -\bar{\Psi}^{1,-}\trips{\Psi_1^-}{Z^1}{Z_1} \right) \bigg]\ , \nonumber\\
  S_0 &=  \text{tr}  \int  d^3 x\, \bigg[ \left( D_0 Z^1 \right) \left( D_0 Z_1 \right) - 2 \left( D Z^A \right) \left( \bar{D} Z_A \right) - 2 \left( \bar{D} Z^A \right) \left( D Z_A \right) \nonumber\\*
  &\hspace{25mm} +\tfrac{4\pi^2}{3 k^2} \Big( - \trips{Z^A}{Z^1}{Z_A}\trips{Z_B}{Z_1}{Z^B} + 4\trips{Z^A}{Z^1}{Z_B}\trips{Z_A}{Z_1}{Z^B}\nonumber\\*
  &\hspace{39mm} - \trips{Z^1}{Z^A}{Z_1}\trips{Z_B}{Z_A}{Z^B}+ 2\trips{Z^A}{Z^B}{Z_1}\trips{Z_A}{Z_B}{Z^1} \nn\\*
  &\hspace{39mm}- \trips{Z_1}{Z_A}{Z^1}\trips{Z^B}{Z^A}{Z_B} \Big) \nonumber\\*
  &\hspace{25mm}+\tfrac{ki}{2\pi} \Big( \left( A_0^L F_{z\bar{z}}^L + A_{\bar{z}}^L F_{0z}^L + A_z^L F_{\bar{z}0}^L + i A_0^L \left[ A_z^L, A_{\bar{z}}^L \right] \right) \nn\\*
  &\hspace{39mm}- \left( A_0^R F_{z\bar{z}}^R + A_{\bar{z}}^R F_{0z}^R + A_z^R F_{\bar{z}0}^R + i A_0^R \left[ A_z^R, A_{\bar{z}}^R \right] \right) \Big)\nonumber\\*
  &\hspace{25mm}+\bar{\Psi}^{1,-} D_0 \Psi_1^- - 2i\bar{\Psi}^{1,+} \gamma_{\bar{z}} D \Psi_1^- - 2i\bar{\Psi}^{1,-} \gamma_{z} \bar{D} \Psi_1^+ \nn\\
  &\hspace{25mm}-\bar{\Psi}^{A,+} D_0 \Psi_A^+ - 2i\bar{\Psi}^{A,+} \gamma_{\bar{z}} D \Psi_A^- - 2i\bar{\Psi}^{A,-} \gamma_{z} \bar{D} \Psi_A^+\nonumber\\*
  &\hspace{25mm}+\tfrac{2\pi i}{k} \Big(  \bar{\Psi}^{1,+} \trips{\Psi_1^+}{Z^1}{Z_1} - \bar{\Psi}^{1,-}\trips{\Psi_1^-}{Z^A}{Z_A} - \bar{\Psi}^{A,-}\trips{\Psi^-_A}{Z^1}{Z_1} \nonumber\\*
  &\hspace{39mm} - \bar{\Psi}^{A,+} \trips{\Psi_A^+}{Z^B}{Z_B} + 2\bar{\Psi}^{1,+}\trips{\Psi_A^+}{Z^A}{Z_1} \nonumber\\*
  &\hspace{39mm} + 2 \bar{\Psi}^{1,-}\trips{\Psi_A^-}{Z^A}{Z_1}+ 2\bar{\Psi}^{A,+} \trips{\Psi_1^+}{Z^1}{Z_A}  \nn\\*
  &\hspace{39mm}+ 2 \bar{\Psi}^{A,-}\trips{\Psi_1^-}{Z^1}{Z_A}+ 2 \bar{\Psi}^{A,+}\trips{\Psi_B^+}{Z^B}{Z_A} \Big)\nonumber\\*
  &\hspace{25mm}+\tfrac{4\pi i}{k}\left( \varepsilon_{ABC}\bar{\Psi}^{A,+}\trips{Z^B}{Z^1}{\Psi^{C,-}} - \varepsilon^{ABC} \bar{\Psi}_A^+ \trips{Z_B}{Z_1}{\Psi_C^-} \right) \bigg]\ ,\nonumber\\
  S_1 &= \text{tr} \int d^3 x \bigg[ \left( D_0 Z^A \right) \left( D_0 Z_A \right) -\bar{\Psi}^{1,+}  D_0 \Psi_1^+ + \bar{\Psi}^{A,-}D_0 \Psi_A^- \nonumber\\*
  &\hspace{23mm}+ \tfrac{4\pi^2}{3k^2} \left( 2\trips{Z^B}{Z^C}{Z_A}\trips{Z_B}{Z_C}{Z^A} - \trips{Z^B}{Z^A}{Z_B}\trips{Z_C}{Z_A}{Z^C} \right) \nonumber\\*
  & \hspace{23mm}- \tfrac{2\pi i}{k} \left( \bar{\Psi}^{1,+}\trips{\Psi_1^+}{Z^A}{Z_A} + \bar{\Psi}^{A,-}\trips{\Psi_A^-}{Z^B}{Z_B} \right) \bigg]\ .
\end{align}
}%
This satisfies $\delta S=0$, with $\delta = \eta^{-1}\delta_{-1} + \delta_0 + \eta \delta_1$, with
{\allowdisplaybreaks
\begin{align}
  \delta_{-1}\Psi_1^+ &= \tfrac{2\pi}{k} \trips{Z^1}{Z^A}{Z_1}\oneminzbar\epsilon_{1A}^+ \ ,\nn\\
  \delta_{-1}\Psi_A^- &= 2 \left( \bar{D} Z^1 \right) \gamma_z \oneminzbar \epsilon_{1A}^+ - \tfrac{2\pi}{k} \trips{Z^1}{Z^B}{Z_1} \oneminz \epsilon_{AB}^- \ ,\nonumber\\
  \delta_{-1}A_0^L &= \tfrac{2\pi i}{k} \left( - \left( \oneminz \bareps^{1A,+} \right) \Psi_A^+ Z_1 - \left( \oneminzbar \bareps^{+}_{1A} \right) Z^1 \Psi^{A,+} \right) \ ,\nonumber\\[1em]
  \delta_{0}Z^1 &= i\left( \oneminz  \bareps^{1A,+} \right)\Psi_A^+  \ ,\nn\\
  \delta_{0}Z^A &= -i\left( \oneminz \bareps^{1A,+} \right)\Psi_1^+ - i\bareps^{1A,-}\Psi_1^- + i\bareps^{AB,+} \Psi_B^+ + i\left( \oneminzbar \bareps^{AB,-} \right) \Psi_B^- \ ,\nonumber\\
  \delta_{0}\Psi_1^+ &= i\left( D_0 Z^A \right)\oneminzbar\epsilon_{1A}^+ - 2\left( DZ^A \right) \gamma_{\bar{z}} \epsilon_{1A}^- + \tfrac{i}{3}Z^A \partial_0 \epsilon_{1A}^+ -\tfrac{2}{3} Z^A \gamma_{\bar{z}}\partial \epsilon_{1A}^- \nonumber\\*
  & \hspace{4.7mm} + \tfrac{2\pi}{k} \left( \trips{Z^A}{Z^B}{Z_1}\epsilon_{AB}^+ -\trips{Z^B}{Z^A}{Z_B}\oneminzbar\epsilon_{1A}^+ + \bar{z}\trips{Z^1}{Z^A}{Z_1} \bar{\partial}\epsilon_{1A}^+ \right) \ ,\nonumber\\
  \delta_{0}\Psi_1^- &= -2\left( \bar{D}Z^A \right)\gamma_{z}\oneminzbar\epsilon_{1A}^+ +\tfrac{2\pi}{k} \left( \trips{Z^A}{Z^B}{Z_1}\oneminz\epsilon_{AB}^- + \trips{Z^1}{Z^A}{Z_1}\epsilon_{1A}^- \right) \ ,\nonumber\\
  \delta_{0}\Psi_A^+ &= -i\left( D_0 Z^1 \right) \oneminzbar \epsilon_{1A}^+ + 2\left( DZ^1 \right) \gamma_{\bar{z}} \epsilon_{1A}^- - 2\left( DZ^B \right) \gamma_{\bar{z}} \oneminz \epsilon_{AB}^- \nonumber\\*
  &\hspace{4.7mm} +\tfrac{2\pi}{k} \left( \trips{Z^B}{Z^1}{Z_B} \oneminzbar \epsilon_{1A}^+ -2\trips{Z^B}{Z^1}{Z_A}\oneminzbar\epsilon_{1B}^+ - \trips{Z^1}{Z^B}{Z_1}\epsilon_{AB}^+ \right) \nonumber\\*
  &\hspace{4.7mm} - \tfrac{i}{3} Z^1 \partial_0 \epsilon_{1A}^+ + \tfrac{2}{3} Z^1 \gamma_{\bar{z}}\partial \epsilon_{1A}^-\ ,\nonumber\\
  \delta_{0}\Psi_A^- &= i \left( D_0 Z^1 \right) \epsilon_{1A}^- - i \left( D_0 Z^B \right) \oneminz  \epsilon_{AB}^- - 2\left( \bar{D} Z^B \right) \gamma_z\epsilon_{AB}^+  \nonumber\\*
  &\hspace{4.7mm} + \tfrac{2\pi}{k} \big( \trips{Z^B}{Z^C}{Z_A}\oneminz\epsilon_{BC}^- -2\trips{Z^B}{Z^1}{Z_A}\epsilon_{1B}^- + \trips{Z^B}{Z^1}{Z_B}\epsilon_{1A}^- \nonumber\\*
  &\hspace{4.7mm} - \trips{Z^1}{Z^B}{Z_1} \oneminz\epsilon_{AB}^- -z\trips{Z^1}{Z^B}{Z_1} \partial\epsilon_{AB}^- \big)  \nonumber\\*
  &\hspace{4.7mm} +\tfrac{i}{3} Z^1 \partial_0 \epsilon_{1A}^- - \tfrac{i}{3} Z^B \partial_0 \epsilon_{AB}^- - \tfrac{2}{3}Z^B \gamma_z \bar{\partial} \epsilon_{AB}^+ + 2\bar{z} \left( \bar{D}Z^1 \right) \gamma_z \bar{\partial}\epsilon_{1A}^+ + \tfrac{2}{3} Z^1 \gamma_z \bar{\partial}\epsilon_{1A}^+ \ ,\nonumber\\
  \delta_{0}A_0^L &= \tfrac{2\pi i}{k} \big( \left( \oneminz\bareps^{1A,+} \right) \Psi_1^+ Z_A - \bareps^{1A,-} \Psi_1^- Z_A + \bareps^{1A,-} \Psi_A^- Z_1 + \bareps^{AB,+} \Psi_A^+ Z_B \nonumber\\*
  &\hspace{4.7mm} - \left( \oneminzbar\bareps^{AB,-} \right) \Psi_A^- Z_B + \left( \oneminzbar \bareps^{+}_{1A} \right) Z^A \Psi^{1,+} - \bareps^{-}_{1A} Z^A \Psi^{1,-} + \bareps^{-}_{1A} Z^1 \Psi^{A,-} \nonumber\\*
  &\hspace{4.7mm} + \bareps^{+}_{AB} Z^B\Psi^{A,+} - \left( \oneminz\bareps^{-}_{AB} \right) Z^B\Psi^{A,-} -z \left( \partial \bareps^{1A,+} \right) \Psi_A^+ Z_1 - \bar{z} \left( \bar{\partial}\bareps_{1A}^+ \right) Z^1 \Psi^{A,+} \big) \ ,\nonumber\\
  \delta_{0}A_z^L &= \tfrac{2\pi}{k} \Big( -\bareps^{1A,-}\gamma_z \Psi_A^+ Z_1 + \left( \oneminzbar\bareps^{AB,-} \right)\gamma_z \Psi_A^+ Z_B \nn\\*
  &\hspace{4.7mm}+ \left( \oneminzbar\bareps_{1A}^+ \right) \gamma_z \left( Z^1 \Psi^{A,-} - Z^A \Psi^{1,-} \right) \Big) \ ,\nonumber\\
  \delta_{0}A_{\bar{z}}^L &= \tfrac{2\pi}{k} \Big( \bareps_{1A}^- \gamma_{\bar{z}} Z^1 \Psi^{A,+} - \left( \oneminz\bareps_{AB}^- \right) \gamma_{\bar{z}}Z^B \Psi^{A,+} \nn\\*
  &\hspace{4.7mm}- \left( \oneminz \bareps^{1A,+} \right)\gamma_{\bar{z}}\left( \Psi_A^- Z_1 - \Psi_1^- Z_A \right)  \Big) \ ,\nonumber\\[1em]
  \delta_{1} Z^1 &= i \bareps^{1A,-} \Psi_A^- + iz \left( \partial \bareps^{1A,+} \right) \Psi_A^+ \ ,\nn\\
  \delta_{1} Z^A &= -iz \left( \partial \bareps^{1A,+} \right)\Psi_1^+ + i \bar{z} \left( \bar{\partial}\bareps^{AB,-} \right)\Psi_B^- \ ,\nonumber\\
  \delta_{1} \Psi_1^+ &= i\bar{z} \left( D_0 Z^A \right) \bar{\partial}\epsilon_{1A}^+ - \tfrac{2\pi}{k} \bar{z} \trips{Z^B}{Z^A}{Z_B}\bar{\partial}\epsilon_{1A}^+ \ ,\nonumber\\
  \delta_{1} \Psi_1^- &=   \tfrac{2\pi}{k} \left( - \trips{Z^B}{Z^A}{Z_B}\epsilon_{1A}^- + z \trips{Z^A}{Z^B}{Z_1} \partial \epsilon_{AB}^- \right) - \tfrac{i}{3} Z^A \partial_0 \epsilon_{1A}^- \nonumber\\*
  &\hspace{4.7mm} - i \left( D_0 Z^A \right) \epsilon_{1A}^--2\bar{z} \left( \bar{D}Z^A \right)\gamma_z \bar{\partial}\epsilon_{1A}^+ - \tfrac{2}{3}Z^A \gamma_z \bar{\partial}\epsilon_{1A}^+ \ ,\nonumber\\
  \delta_{1} \Psi_A^+ &=   \tfrac{2\pi}{k} \left( \trips{Z^B}{Z^C}{Z_A}\epsilon_{BC}^+ + \bar{z} \trips{Z^B}{Z^1}{Z_B}\bar{\partial}\epsilon_{1A}^+ - 2 z \trips{Z^B}{Z^1}{Z_A}\bar{\partial}\epsilon_{1B}^+ \right) \nonumber\\*
  &\hspace{4.7mm} + i \left( D_0 Z^B \right) \epsilon_{AB}^++ \tfrac{i}{3} Z^B \partial_0 \epsilon_{AB}^+ -i\bar{z} \left( D_0 Z^1 \right) \bar{\partial}\epsilon_{1A}^+ - \tfrac{2}{3} Z^B \gamma_{\bar{z}} \partial \epsilon_{AB}^- \ ,\nonumber\\
  \delta_{1} \Psi_A^- &= \tfrac{2\pi}{k} z \left( \trips{Z^B}{Z^C}{Z_A}\partial \epsilon_{BC}^- - \trips{Z^1}{Z^B}{Z_1} \partial \epsilon_{AB}^- \right) - iz \left( D_0 Z^B \right) \partial \epsilon_{AB}^- \ ,\nonumber\\
  \delta_{1}A_0^L &= \tfrac{2\pi i}{k} \left( z \left( \partial \bareps^{1A,+}\right) \Psi_1^+ Z_A - \bar{z} \left( \bar{\partial}\bareps^{AB,-} \right) \Psi_A^- Z_B + \bar{z} \left( \bar{\partial} \bareps_{1A}^+ \right) Z^A \Psi^{1,+} - z \left( \partial \bareps_{AB}^- \right) Z^B \Psi^{A,-} \right) \ ,\nonumber\\
  \delta_{1}A_z^L &= \tfrac{2\pi}{k} \big( \bareps^{1A,-} \gamma_z \Psi_1^+ Z_A - \bareps_{AB}^+ \gamma_z Z^B \Psi^{A,-} + \bar{z} \left( \bar{\partial} \bareps^{AB,-} \right) \gamma_z \Psi_A^+ Z_B \nonumber\\*
  &\hspace{4.7mm} + \bar{z} \left( \bar{\partial}\bareps_{1A}^+ \right) \gamma_z\left( Z^1 \Psi^{A,-} -  Z^A \Psi^{1,-} \right) \big) \ ,\nonumber\\
  \delta_{1}A_{\bar{z}}^L &= \tfrac{2\pi}{k} \big( -\bareps_{1A}^- \gamma_{\bar{z}} Z^A \Psi^{1,+} + \bareps^{AB,+} \gamma_{\bar{z}} \Psi^{-}_A Z_B  - z \left( \partial \bareps^{-}_{AB} \right) \gamma_{\bar{z}} Z^B \Psi^{A,+} \nonumber\\*
  &\hspace{4.7mm} - z \left( \partial\bareps^{1A,+} \right) \gamma_{\bar{z}} \left( \Psi^{-}_A  Z_1 - \Psi^{-}_1 Z_A \right) \big)\ ,
\end{align}
}where one obtains $\delta A_m^R$ by swapping the order of all fields in the expressions for $\delta A_m^L$. Here, $\epsilon_{IJ}$ is once again of the form
\begin{align}
  \epsilon_{IJ} = \xi_{IJ} + x^\mu\gamma_\mu \zeta_{IJ} = \xi_{IJ} + \left( x^0\gamma_0 + z\gamma_z + \bar{z} \gamma_{\bar{z}} \right) \zeta_{IJ}\ .
\end{align}
As in the ${\cal N}=8$ case, we see that neither $S_{-1}$ or $\delta_{-1}$ are in the simple form required to directly find the limiting theory. However, we note that
\begin{align}
  \delta_{-1}A_0^L + \frac{2\pi}{k} \delta_0\left( Z^1 Z_1 \right) = 0 \ , \qquad 
  \delta_{-1}A_0^R + \frac{2\pi}{k} \delta_0\left( Z_1 Z^1 \right) = 0 \ .
\end{align}
Thus, we consider the field redefiniton
\begin{align}
  A_0^L \to \hat{A}_0^L = A_0^L - \frac{1}{\eta} \frac{2\pi}{k} Z^1 Z_1 \ ,\qquad 
  A_0^R \to \hat{A}_0^R = A_0^R - \frac{1}{\eta} \frac{2\pi}{k} Z_1 Z^1 \ .
\end{align}
The shift to the action is then
\begin{align}
  S_{-1} &\to S_{-1} + \text{tr} \int d^3 x\,\, \Big( -2i Z^1 Z_1 F_{z\bar{z}}^L + 2i Z_1 Z^1 F_{z\bar{z}}^R - \tfrac{4\pi^2}{k^2}\trips{Z^1}{Z^A}{Z_1}\trips{Z_1}{Z_A}{Z^1} \nonumber\\
  & \hspace{38mm}+ \tfrac{2\pi i}{k} \left( \bar{\Psi}^{A,+} \trips{\Psi_A^+}{Z^1}{Z_1} -\bar{\Psi}^{1,-}\trips{\Psi_1^-}{Z^1}{Z_1} \right) \Big) \ ,\nonumber\\
  S_0 &\to S_0 + \text{tr} \int d^3 x\,\, \Big( -\tfrac{2\pi i}{k}\left( \left( D_0 Z_A \right) \trips{Z^A}{Z^1}{Z_1} + \left( D_0 Z^A \right) \trips{Z_A}{Z_1}{Z^1} \right) \nonumber\\
  & \hspace{38mm} +\tfrac{2\pi i}{k}  \left( \bar{\Psi}^{1,+} \trips{\Psi_1^+}{Z^1}{Z_1}  - \bar{\Psi}^{A,-}\trips{\Psi^-_A}{Z^1}{Z_1} \right) \Big)\ ,
\end{align}
while the supersymmetry transformations are shifted by
\begin{align}
  \delta_{-1} \Psi_1^+ &\to \delta_{-1}\Psi_1^+ - \tfrac{2\pi}{k} \trips{Z^1}{Z^A}{Z_1}\oneminzbar \epsilon_{1A}^+ \ ,\nonumber  \\
  \delta_{-1} \Psi_A^- &\to \delta_{-1} \Psi_A^- + \tfrac{2\pi}{k} \trips{Z^1}{Z^B}{Z_1}\oneminz \epsilon_{AB}^-\ ,\nonumber\\
  \delta_{-1} A_0^L &\to A_0^L + \tfrac{2\pi i}{k} \left( \left( \oneminz \bareps^{1A,+} \right) \Psi_A^+ Z_1 + \left( \oneminzbar \bareps^{+}_{1A} \right) Z^1 \Psi^{A,+} \right)\ ,\nonumber\\[1em]
  \delta_0 \Psi_1^+ &\to \delta_0 \Psi_1^+ - \tfrac{2\pi}{k}\bar{z}\trips{Z^1}{Z^A}{Z_1} \bar{\partial}\epsilon_{1A}^+ \ ,\nonumber \\
  \delta_0 \Psi_1^- &\to \delta_0 \Psi_1^- + \tfrac{2\pi}{k}\trips{Z^1}{Z^A}{Z_1} \epsilon_{1A}^- \ ,\nonumber\\
  \delta_0 \Psi_A^+ &\to \delta_0 \Psi_A^+ - \tfrac{2\pi}{k}\trips{Z^1}{Z^B}{Z_1} \epsilon_{AB}^+ \ ,\nonumber\\
  \delta_0 \Psi_A^- &\to \delta_0 \Psi_A^- + \tfrac{2\pi}{k}z\trips{Z^1}{Z^B}{Z_1} \partial\epsilon_{AB}^- \ ,\nonumber\\
  \delta_0 A_0^L &\to \delta_0 A_0^L + \tfrac{2\pi i}{k} \left( \bareps^{1A,-}\Psi_A^- Z_1 + \bareps_{1A}^- Z^1 \Psi^{A,-} + z \left( \partial \bareps^{1A,+} \right) \Psi_A^+ Z_1 + \bar{z} \left( \bar{\partial}\bareps_{1A}^+ \right) Z^1 \Psi^{A,+} \right)\ ,
\end{align}
where again the shift to $\delta A_0^R$ can be determined from that of $\delta A_0^L$ by simply swapping all pairs of fields. 
In particular, we now have that
\begin{align}
  S_{-1} = \text{tr} \int d^3 x\,\, \Big( -4 \left( \bar{D} Z^1 \right) \left( D Z_1 \right) \Big)\ , \qquad 
    \delta_{-1}\Psi_A^- &= 2 \left( \bar{D} Z^1 \right) \gamma_z \oneminzbar \epsilon_{1A}^+\ ,
\end{align}
with $\delta_{-1}=0$ on all other fields. This is now in the correct form to proceed. The end result is that in the $\eta\to 0$ limit, the theory is described by the action
\begin{align}
   \tilde S &=  \text{tr}  \int  d^3 x\, \bigg[ \left( D_0 Z^1 \right) \left( D_0 Z_1 \right) - 2 \left( D Z^A \right) \left( \bar{D} Z_A \right) - 2 \left( \bar{D} Z^A \right) \left( D Z_A \right)\nn\\
   &\hspace{25mm} + H \left( DZ_1 \right) + \bar{H}\left( \bar{D}Z^1 \right) \nonumber\\
  &\hspace{25mm} -\tfrac{2\pi i}{k}\left( \left( D_0 Z_A \right) \trips{Z^A}{Z^1}{Z_1} + \left( D_0 Z^A \right) \trips{Z_A}{Z_1}{Z^1} \right) \nonumber\\
  &\hspace{25mm} +\tfrac{4\pi^2}{3 k^2} \Big( - \trips{Z^A}{Z^1}{Z_A}\trips{Z_B}{Z_1}{Z^B} + 4\trips{Z^A}{Z^1}{Z_B}\trips{Z_A}{Z_1}{Z^B}\nonumber\\
  &\hspace{39mm}- \trips{Z^1}{Z^A}{Z_1}\trips{Z_B}{Z_A}{Z^B} + 2\trips{Z^A}{Z^B}{Z_1}\trips{Z_A}{Z_B}{Z^1} \nn\\
  &\hspace{39mm}- \trips{Z_1}{Z_A}{Z^1}\trips{Z^B}{Z^A}{Z_B} \Big) \nonumber\\
  &\hspace{25mm}+\tfrac{ki}{2\pi} \Big( \left( A_0^L F_{z\bar{z}}^L + A_{\bar{z}}^L F_{0z}^L + A_z^L F_{\bar{z}0}^L + i A_0^L \left[ A_z^L, A_{\bar{z}}^L \right] \right) \nn\\
  &\hspace{39mm}- \left( A_0^R F_{z\bar{z}}^R + A_{\bar{z}}^R F_{0z}^R + A_z^R F_{\bar{z}0}^R + i A_0^R \left[ A_z^R, A_{\bar{z}}^R \right] \right) \Big)\nonumber\\
  &\hspace{25mm}+\bar{\Psi}^{1,-} D_0 \Psi_1^- - 2i\bar{\Psi}^{1,+} \gamma_{\bar{z}} D \Psi_1^- - 2i\bar{\Psi}^{1,-} \gamma_{z} \bar{D} \Psi_1^+ \nn\\
  &\hspace{25mm}-\bar{\Psi}^{A,+} D_0 \Psi_A^+ - 2i\bar{\Psi}^{A,+} \gamma_{\bar{z}} D \Psi_A^- - 2i\bar{\Psi}^{A,-} \gamma_{z} \bar{D} \Psi_A^+\nonumber\\
  &\hspace{25mm}+\tfrac{2\pi i}{k} \Big(  \bar{\Psi}^{1,+} \trips{\Psi_1^+}{Z^1}{Z_1} - \bar{\Psi}^{1,-}\trips{\Psi_1^-}{Z^A}{Z_A} - \bar{\Psi}^{A,-}\trips{\Psi^-_A}{Z^1}{Z_1} \nonumber\\
  &\hspace{39mm} - \bar{\Psi}^{A,+} \trips{\Psi_A^+}{Z^B}{Z_B} + 2\bar{\Psi}^{1,+}\trips{\Psi_A^+}{Z^A}{Z_1} \nonumber\\
  &\hspace{39mm}+ 2 \bar{\Psi}^{1,-}\trips{\Psi_A^-}{Z^A}{Z_1} + 2\bar{\Psi}^{A,+} \trips{\Psi_1^+}{Z^1}{Z_A} \nn\\
  &\hspace{39mm}+ 2 \bar{\Psi}^{A,-}\trips{\Psi_1^-}{Z^1}{Z_A} + 2 \bar{\Psi}^{A,+}\trips{\Psi_B^+}{Z^B}{Z_A} \Big)\nonumber\\
  &\hspace{25mm}+\tfrac{4\pi}{k}\left( \varepsilon_{ABC}\bar{\Psi}^{A,+}\trips{Z^B}{Z^1}{\Psi^{C,-}} - \varepsilon^{ABC} \bar{\Psi}_A^+ \trips{Z_B}{Z_1}{\Psi_C^-} \right) \bigg]\ ,
\end{align}
which preserves the full 24 supersymmetries of the original ${\cal N}=6$ theory. In particular, we have $\tilde \delta \tilde S=0$, with 
{\allowdisplaybreaks
\begin{align}
    \tilde\delta Z^1 &= i\left( \oneminz  \bareps^{1A,+} \right)\Psi_A^+ \ ,\nonumber\\
 \tilde \delta Z^A &= -i\left( \oneminz \bareps^{1A,+} \right)\Psi_1^+ - i\bareps^{1A,-}\Psi_1^- + i\bareps^{AB,+} \Psi_B^+ + i\left( \oneminzbar \bareps^{AB,-} \right) \Psi_B^- \ ,\nonumber\\
  \tilde\delta \Psi_1^+ &= i\left( D_0 Z^A \right)\oneminzbar\epsilon_{1A}^+ - 2\left( DZ^A \right) \gamma_{\bar{z}} \epsilon_{1A}^- + \tfrac{i}{3}Z^A \partial_0 \epsilon_{1A}^+ -\tfrac{2}{3} Z^A \gamma_{\bar{z}}\partial \epsilon_{1A}^- \nonumber\\*
  & \hspace{4.7mm} + \tfrac{2\pi}{k} \left( \trips{Z^A}{Z^B}{Z_1}\epsilon_{AB}^+ -\trips{Z^B}{Z^A}{Z_B}\oneminzbar\epsilon_{1A}^+  \right) \ ,\nonumber\\
 \tilde \delta \Psi_1^- &= -2\left( \bar{D}Z^A \right)\gamma_{z}\oneminzbar\epsilon_{1A}^+ +\tfrac{2\pi}{k} \left( \trips{Z^A}{Z^B}{Z_1}\oneminz\epsilon_{AB}^- + 2\trips{Z^1}{Z^A}{Z_1}\epsilon_{1A}^- \right) \ ,\nonumber\\
  \tilde\delta \Psi_A^+ &= -i\left( D_0 Z^1 \right) \oneminzbar \epsilon_{1A}^+ + 2\left( DZ^1 \right) \gamma_{\bar{z}} \epsilon_{1A}^- - 2\left( DZ^B \right) \gamma_{\bar{z}} \oneminz \epsilon_{AB}^- \nonumber\\*
  &\hspace{4.7mm} +\tfrac{2\pi}{k} \left( \trips{Z^B}{Z^1}{Z_B} \oneminzbar \epsilon_{1A}^+ -2\trips{Z^B}{Z^1}{Z_A}\oneminzbar\epsilon_{1B}^+ - 2\trips{Z^1}{Z^B}{Z_1}\epsilon_{AB}^+ \right) \nonumber\\*
  &\hspace{4.7mm} - \tfrac{i}{3} Z^1 \partial_0 \epsilon_{1A}^+ + \tfrac{2}{3} Z^1 \gamma_{\bar{z}}\partial \epsilon_{1A}^-\ ,\nonumber\\
  \tilde\delta \Psi_A^- &= i \left( D_0 Z^1 \right) \epsilon_{1A}^- - i \left( D_0 Z^B \right) \oneminz  \epsilon_{AB}^- - 2\left( \bar{D} Z^B \right) \gamma_z\epsilon_{AB}^+  \nn\\*
  &\hspace{4.7mm}+ \tfrac{2\pi}{k} \big( \trips{Z^B}{Z^C}{Z_A}\oneminz\epsilon_{BC}^- -2\trips{Z^B}{Z^1}{Z_A}\epsilon_{1B}^- + \trips{Z^B}{Z^1}{Z_B}\epsilon_{1A}^- \nn\\*
  &\hspace{4.7mm}- \trips{Z^1}{Z^B}{Z_1} \oneminz\epsilon_{AB}^-  \big)  +\tfrac{i}{3} Z^1 \partial_0 \epsilon_{1A}^-  - \tfrac{i}{3} Z^B \partial_0 \epsilon_{AB}^- - \tfrac{2}{3}Z^B \gamma_z \bar{\partial} \epsilon_{AB}^+ \nn\\*
  &\hspace{4.7mm}+ 2\bar{z} \left( \bar{D}Z^1 \right) \gamma_z \bar{\partial}\epsilon_{1A}^+ + \tfrac{2}{3} Z^1 \gamma_z \bar{\partial}\epsilon_{1A}^+ -\tfrac{1}{2}H \gamma_z \oneminzbar \epsilon_{1A}^+ \ ,\nonumber\\
 \tilde \delta A_0^L &= \tfrac{2\pi i}{k} \big( \left( \oneminz\bareps^{1A,+} \right) \Psi_1^+ Z_A - \bareps^{1A,-} \Psi_1^- Z_A + 2\bareps^{1A,-} \Psi_A^- Z_1 + \bareps^{AB,+} \Psi_A^+ Z_B \nonumber\\
  &\hspace{4.7mm} - \left( \oneminzbar\bareps^{AB,-} \right) \Psi_A^- Z_B + \left( \oneminzbar \bareps^{+}_{1A} \right) Z^A \Psi^{1,+} - \bareps^{-}_{1A} Z^A \Psi^{1,-} + 2\bareps^{-}_{1A} Z^1 \Psi^{A,-} \nonumber\\
  &\hspace{4.7mm} + \bareps^{+}_{AB} Z^B\Psi^{A,+} - \left( \oneminz\bareps^{-}_{AB} \right) Z^B\Psi^{A,-} \big) \ ,\nonumber\\
 \tilde \delta A_z^L &= \tfrac{2\pi}{k} \Big( -\bareps^{1A,-}\gamma_z \Psi_A^+ Z_1 + \left( \oneminzbar\bareps^{AB,-} \right)\gamma_z \Psi_A^+ Z_B \nn\\*
 &\hspace{4.7mm}+ \left( \oneminzbar\bareps_{1A}^+ \right) \gamma_z \left( Z^1 \Psi^{A,-} - Z^A \Psi^{1,-} \right) \Big) \ ,\nonumber\\
 \tilde \delta A_{\bar{z}}^L &= \tfrac{2\pi}{k} \Big( \bareps_{1A}^- \gamma_{\bar{z}} Z^1 \Psi^{A,+} - \left( \oneminz\bareps_{AB}^- \right) \gamma_{\bar{z}}Z^B \Psi^{A,+} \nn\\*
  &\hspace{4.7mm}- \left( \oneminz \bareps^{1A,+} \right)\gamma_{\bar{z}}\left( \Psi_A^- Z_1 - \Psi_1^- Z_A \right)  \Big) \ ,\nonumber\\
 \tilde \delta H &= -2 \left( \oneminz \bareps^{1A,+}  \right)\gamma_{\bar{z}} \left( D_0 \Psi_A^- - \tfrac{2\pi i}{k}\trips{\Psi_A^-}{Z^B}{Z_B} \right) - 4i \left( \partial \bareps^{1A,-} \right)\Psi_A^- - 4i \bareps^{1A,-}\bar{D}\Psi_A^- \nonumber\\
  &\hspace{4.7mm} -4i z \left( \partial \bareps^{1A,+} \right) \bar{D} \Psi_A^+ + \tfrac{8\pi i}{k} \big( \bareps^{AB,+}\gamma_{\bar{z}}\trips{\Psi_A^-}{Z^1}{Z_B} - \bareps_{1A}^- \gamma_{\bar{z}}\trips{Z^A}{Z^1}{\Psi^{1,+}} \nonumber\\
  & \hspace{4.7mm} -z \left( \partial \bareps_{AB}^- \right) \gamma_{\bar{z}} \trips{Z^B}{Z^1}{\Psi^{A,+}} -z \left( \partial \bareps^{1A,+} \right)\gamma_{\bar{z}}\left( \trips{\Psi_A^-}{Z^1}{Z_1}-\trips{\Psi_1^-}{Z^1}{Z_A} \right)\big)\ ,
\end{align}}where once again, $\tilde\delta A_\mu^R$ is obtained from $\tilde\delta A_\mu^L$ by swapping the order of all pairs of fields, and $\epsilon_{IJ}$ is given by
\begin{align}
  \epsilon_{IJ} = \xi_{IJ} + x^\mu \gamma_\mu \zeta_{IJ} = \xi_{IJ} +\left( x^0\gamma_0 + z\gamma_z + \bar{z} \gamma_{\bar{z}} \right)\zeta_{IJ}\ ,
\end{align}
with $\xi_{IJ}$, $\zeta_{IJ}$, and hence $\epsilon_{IJ}$ satisfying the reality condition $\epsilon^{IJ}=\tfrac{1}{2} \varepsilon^{IJKL}\epsilon_{KL}$. We have introduced the $N\times M$ complex matrix $H$ transforming in the $\left( \mathbf{N}, \bar{\mathbf{M}} \right)$ of $U(N)_L\times U(M)_R$, which acts a a Lagrange multiplier imposing $\bar{D}Z^1=0$. Its Hermitian conjugate, transforming in the $\left(\bar{\mathbf{N}}, \mathbf{M} \right)$, is denoted $\bar{H}$.

As with the ${\cal N}=8$ case above  there are two constraints. The first comes from integrating out $\bar H$ which simply implies
\begin{align}
\bar D Z^1 = 0\ .
\end{align}
However there is also the Gauss law constraint that comes from integrating out $A^{L/R}_0$:
\begin{align}
F^L_{z\bar z}    &= \frac{ 2\pi^2 i}{k^2}(Z^A[Z_A,Z_1;Z^1]- [Z^A,Z^1;Z_1]Z_A)\ ,\nonumber\\
 F^R_{z\bar z}  &= -\frac{ 2\pi^2 i}{k^2}([Z_A,Z_1;Z^1]Z^A- Z_A[Z^A,Z^1;Z_1])\ ,
\end{align}
where for simplicity  we have set the fermions to zero and assumed static configurations with $D_0Z^1=0$. These constraints then form a bi-fundamental version of the Hitchin system.


\chapter{Reduction to soliton quantum mechanics}\label{chap: reduction to soliton quantum mechanics}


A generic feature of the fixed point actions obtained in the previous Chapter is the localisation of their dynamics to the classical minima of $S_{-1}$, imposed as a constraint in the fixed point theory via a Lagrange multiplier. Indeed, in the examples considered, this constraint was the Bogolmol'nyi equation of some soliton. One can then in principle solve this equation, and reduce the fixed point theory to a supersymmetric quantum mechanics on the corresponding moduli space. In this Chapter, we explore this reduction explicitly for a number of interesting examples. 


\section{A simple kink model}\label{sec: a simple kink model}


We first present a scaling limit and the subsequent reduction to a quantum mechanical model for perhaps the simplest supersymmetric theory containing BPS solitons: the $\mathcal{N}=(1,1)$, $(1+1)$-dimensional $\sigma$-model. This is a supersymmetric extension to the bosonic kink model explored in Section $\beta$.

\subsection{The parent theory}

We consider the $(1+1)$-dimensional $\sigma$-model with $d$-dimensional Riemannian target manifold $(M,g)$, local coordinates $\phi^i$ and generic scalar potential $W(\phi)$. Letting $\Xi$ denote the 2-dimensional worldsheet with signature $(-,+)$, we have action
  \begin{align}
  S = \frac{1}{2} \int_\Xi d^2 x\,\, \bigg(& -g_{ij} \partial_\mu \phi^i \partial^\mu \phi^j - g^{ij} D_i W D_j W + i g_{ij} \bar{\psi}^i \slashed{D}\psi^j \nn\\
  &\qquad\qquad- \frac{1}{6} R_{ijkl} \bar{\psi}^i \psi^k \bar{\psi}^j \psi^l + i \left( D_i D_j W \right)\bar{\psi}^i \psi^j \bigg)\ .
\end{align}
The $\left\{ \psi^i \right\}_{i=1}^d$ are two-component Majorana spinors, with conjugate $\bar{\psi}^i= (\psi^i)^T \gamma_t$. The covariant derivative of the $\psi^i$ is defined in terms of the pullback of the Levi-Civita connection $\Gamma^i_{jk}$ in coordinate basis $\{\phi^i\}$ on $M$,
\begin{align}
  D_\mu \psi^j = \partial_\mu \psi^j + \Gamma^j_{kl}(\partial_\mu \phi^k)\psi^l\ .
\end{align}
We then find that $S$ enjoys $\mathcal{N}=(1,1)$ supersymmetry,
\begin{align}
  \delta \phi^i &= i\bar{\epsilon} \psi^i \ ,\nonumber\\
  \delta \psi^i &= -\slashed{\partial} \phi^i \epsilon + (D^i W)\epsilon - i \Gamma^i_{jk} (\bareps \psi^j)\psi^k \ ,
\end{align}
for a constant two-component Majorana spinor $\epsilon$. Note, in general one can consider further potential terms for $S$ in terms of Killing vector fields on $M$ \cite{AlvarezGaume:1983ab}, which we choose to omit.\\

Suppose now that $D_i W$ has zeros at points $p\in \mathcal{P}\subset M$, and is non-zero elsewhere. Thus, any finite energy configuration $\phi$ must satisfy $\lim_{x\to \pm\infty}\phi\in \mathcal{P}$. Further imposing that the points in $\mathcal{P}$ are isolated, we see that the space of finite energy configuration splits into topological sectors corresponding to which element of $\mathcal{P}$ $\phi$ settles at as $x\to \pm \infty$.

We briefly review the construction of classical \textit{static} solutions in each topological sector. The bosonic part of the energy functional is given by
\begin{align}
  E_\text{bos.} = \frac{1}{2} \int dx \left( g_{ij} \partial_t\phi^i \partial_t\phi^j + g_{ij} \partial_x\phi^i \partial_x\phi^j + g^{ij} D_i W D_j W \right)\ .
\end{align}
So, setting $\partial_t\phi^i=0$, we perform the standard Bogomol'nyi recasting of $E_\text{bos.}$ to arrive at
\begin{align}
  E_\text{bos.} = \frac{1}{2} \int dx &\,\, g_{ij} \left( \partial_x\phi^i \pm D^i W \right)\left( \partial_x\phi^j \pm D^j W \right) \nonumber\\
  &\mp \left( W(\phi(\infty)) - W(\phi(-\infty)) \right)\ .
  \label{eq: Bogomol'nyi argument}
\end{align}
Hence, we have
\begin{align}
  E_\text{bos.} \ge |W(\phi(\infty)) - W(\phi(-\infty))|\ ,
\end{align}
with equality precisely for configurations satisfying
\begin{align}
  \left\{\begin{aligned}
  \partial_x\phi^i - D^i W = 0 \quad &\text{if } W(\phi(\infty)) - W(\phi(-\infty))\ge 0\quad\text{(kink)}\\
  \partial_x\phi^i + D^i W = 0 \quad &\text{if } W(\phi(\infty)) - W(\phi(-\infty))\le 0\quad\text{(anti-kink)}\ .
  \end{aligned}\right.
  \label{eq: Bogomol'nyi equations}
\end{align}
It is easily verified that such kink solutions satisfy the second-order bosonic equations of motion for $\phi^i$, and are thus classical bosonic solutions. Upon solving (\ref{eq: Bogomol'nyi equations}) for the relevant topological sector, one can then use supersymmetry to generate fermionic zero modes.

These static bosonic solutions define $\tfrac{1}{2}$-BPS states of the theory. This is easily seen by decomposing spinors as $\psi^i = \psi^i_+ + \psi^i_-$, with $\psi^i_\pm = \frac{1}{2}\left( 1\pm \gamma_x \right)\psi^i$ and similarly for $\epsilon$, and noting
\begin{align}
  \delta\psi^i_+ &= \gamma_t \partial_t\phi^i \epsilon_- - \left( \partial_x\phi^i - D^i W \right) \epsilon_+ + \text{fermions} \ ,\nonumber\\
  \delta\psi^i_- &= \gamma_t \partial_t\phi^i \epsilon_+ + \left( \partial_x\phi^i + D^i W \right) \epsilon_- + \text{fermions}\ ,
\end{align}
and so depending on the sign of $W(\phi(\infty)) - W(\phi(-\infty))$, one of $\epsilon_\pm$ is broken, while the other preserved\footnote{Of course, if $W(\infty)-W(-\infty)=0$ then we have the trivial vacuum $\phi=\text{constant}\in\mathcal{P}$, and full supersymmetry is preserved.}.

\subsection{Scaling limit and fixed point action}

We now consider a rescaling of the fields and worldsheet coordinates of the theory with parameter $\eta$. In particular, we choose our scaling such that the small $\eta$ limit corresponds to slow motion of the coordinates $\phi^i$, i.e. $\partial_t \phi^i << \partial_x \phi^i$, and hence the theory at the fixed point $\eta\to 0$ will provide an effective theory for the leading order dynamics of a slow moving (i.e. low energy) kink.

We expect to recover the standard argument after Manton \cite{Manton:1981mp} that these dynamics are described by geodesic motion on the moduli space of the static solution. In this construction, the metric on the moduli space descends from the kinetic terms in the action. To be more precise, if $\delta_\alpha \phi^i$ denotes the variation of $\phi^i$ with respect to a modulus $m^\alpha$, then this metric is
\begin{align}
  \mathcal{G}_{\alpha \beta} = \int dx\,\, g_{ij} \delta_\alpha \phi^i \delta_\beta \phi^j\ .
  \label{eq: induced metric}
\end{align}
One can propose an analogous metric for the Grassmann moduli \cite{Gauntlett:1992yj}, and thus try to construct a supersymmetric quantum mechanical model that pairs the bosonic and fermionic perturbations around the static bosonic kink. We will see that precisely this form for the quantum mechanics emerges as we take $\eta\to 0$.\\

Without loss of generality, we seek an effective action describing the slow motion of a kink rather than anti-kink, and so restrict $\phi^i$ to lie in the sectors of configuration space satisfying $W(\phi(\infty)) - W(\phi(-\infty))\ge 0$. The vacua in such sectors preserve the supersymmetry $\epsilon_+$, which pairs $\phi^i$ with $\psi^i_-$. Any fluctuations above these vacua will fall into representations of this unbroken supersymmetry $\epsilon_+$.

So we seek a scaling of both the coordinates and fields such that the limit $\eta\to 0$ will provide us with an effective theory for the leading order dynamics about a static kink solution. Such a scaling should satisfy the following.
\begin{itemize}
	\item Velocities should be suppressed relative to spatial variation, i.e. $\partial_t << \partial_x$
	\item We want to describe the dynamics of the perturbations about the static solution, and so the kinetic part of the action for $\phi^i$ should appear in the scaled theory at order $\eta^0$
	\item The original static kink should remain a classical solution, implying that $\left( \partial_x \phi^i - D^i W \right)$ should scale homogeneously
	\item As we take $\eta\to 0$, $\phi^i$ should remain paired to $\psi^i_-$ under the supersymmetry $\epsilon_+$
\end{itemize}
These requirements determine an essentially unique choice of scaling, given by 
\begin{align}
  t &\to \eta^{-1} t \ ,& \psi^i_+ &\to \eta^{3/4} \psi^i_+ \ ,\nonumber\\
  x &\to \eta^{-1/2} x \ ,& \psi^i_- &\to \eta^{1/4} \psi^i_- \ ,\nonumber\\
  \phi^i &\to \eta^{-1/4} \phi^i \ ,& \epsilon_+ &\to \eta^{-1/2} \epsilon_+ \ ,\nonumber\\
  && \epsilon_- &\to  \epsilon_-\ ,
\end{align}
where we use the two degrees of scaling symmetry already present in $S$ (scaling by worldsheet `mass dimension', and local scaling diffeomorphisms of $M$) to keep $g_{ij}$ and $W$ fixed under the scaling by $\eta$. The action and supersymmetry variations then take the form
\begin{align}
  S &= \eta^{-1} S_{-1} + S_0 + \eta S_1 \ ,\nonumber\\
  \delta &= \eta^{-1} \delta_{-1} + \delta_0 + \eta \delta_1\ ,
\end{align}
with
\begin{align}
  S_{-1} &= \frac{1}{2} \int_\Xi d^2 x\,\, \Big( -g_{ij} \partial_x \phi^i \partial_x \phi^j - g^{ij} D_i W D_j W \Big) \ ,\nonumber \\
  S_0 &= \frac{1}{2} \int_\Xi d^2 x\,\, \left( g_{ij} \partial_t\phi^i \partial_t\phi^j - i g_{ij} \bar{\psi}^i_- \gamma_t D_t \psi^j_- - 2i \bar{\psi}^i_+\big( g_{ij} D_x - \left( D_i D_j W \right) \big)\psi^j_- \right) \ ,\nonumber\\
  S_1 &= \frac{1}{2} \int_\Xi d^2 x\,\, \left( -ig_{ij} \bar{\psi}^i_+ \gamma_t D_t \psi^j_+ -\tfrac{1}{3}\left( R_{ijkl}+R_{ilkj} \right)\bar{\psi}^i_+ \psi^k_- \bar{\psi}^j_+ \psi^l_-  \right)\ ,
  \label{eq: rescaled kink action}
\end{align}
and
\begin{align}
  \delta_{-1} \psi^i_+ &= - \left( \partial_x\phi^i - D^i W \right)\epsilon_+\ ,\nonumber\\[0.5em]
  \delta_0 \phi^i & = i\bareps_+ \psi^i_- \ ,\nonumber\\
  \delta_0 \psi^i_+ & = \gamma_t \partial_t\phi^i \epsilon_ - + i \Gamma^i_{jk}( \bar{\psi}^j_+ \psi^k_- ) \epsilon_+ \ ,\nonumber\\
  \delta_0 \psi^i_- & = \gamma_t \partial_t\phi^i \epsilon_+ + \left( \partial_x\phi^i + D^i W \right) \epsilon_-  \ ,\nonumber\\[0.5em]
  \delta_1 \phi^i & = i\bareps_- \psi^i_+ \ ,\nonumber\\
  \delta_1 \psi^i_- & =  i \Gamma^i_{jk}( \bar{\psi}^j_+ \psi^k_- ) \epsilon_-\ .
\end{align}
We've used that for a torsion-free connection, $\Gamma^i_{jk} ( \bareps_+ \psi^j_- ) \psi^k_- = \Gamma^i_{jk} ( \bareps_- \psi^j_+ ) \psi^k_+=0$.\\

In the $\eta\to 0$ limit, we localise onto classical minima of $S_{-1}$, i.e. solutions to $\partial_x\phi^i- D^i W=0$. By utilising a generalisation of the procedure of Chapter \ref{chap: construction from Lorentzian models} for non-constant metrics in the quadratic constraint imposed by $S_{-1}$, we propose the following action for the theory at the fixed point $\eta\to 0$,
\begin{align}
  \tilde{S} = \frac{1}{2} \int_\Xi d^2 x\,\, \Big(& g_{ij} \partial_t\phi^i \partial_t\phi^j + 2 G_i \left( \partial_x\phi^i - D^i W \right) \nonumber\\ 
  &- i g_{ij} \bar{\psi}^i_- \gamma_t D_t \psi^j_- - 2i \bar{\psi}^i_+\big( g_{ij} D_x - \left( D_i D_j W \right) \big)\psi^j_- \Big)\ ,
  \label{eq: kink fixed point action}
\end{align}
where here the target space 1-form $G_i(t,x)$ acts as a Lagrange multiplier imposing the Bogomol'nyi equation. $\tilde{S}$ then has $\mathcal{N}=(1,1)$ supersymmetry, given by
\begin{align}
  \tilde{\delta} \phi^i & = i\bareps_+ \psi^i_- \ ,\nonumber\\
  \tilde{\delta} \psi^i_+ & = \gamma_t \partial_t\phi^i \epsilon_- + i \Gamma^i_{jk}( \bar{\psi}^j_+ \psi^k_- ) \epsilon_+ +  G^i \epsilon_+ \ ,\nonumber\\
  \tilde{\delta} \psi^i_- & = \gamma_t \partial_t\phi^i \epsilon_+ + \left( \partial_x\phi^i + D^i W \right) \epsilon_- \ ,\nonumber\\
  \tilde{\delta} G_i &= i\bareps_+\left( \Gamma^j_{ik} G_j \psi^k_- + g_{ij}\gamma_t D_t \psi^i_+ + \tfrac{1}{3}\left( R_{ijkl}+R_{ilkj} \right)\psi^k_- (\bar{\psi}^j_+\psi^l_-) \right)\nonumber\\
  &\qquad + i\bareps_-\left( -g_{ij} D_x \psi^j_+ + (D_i D_j W)\psi^j_+ \right)\ .
  \label{eq: reduced sigma model SUSY}
\end{align}

\subsection{Reduction to constraint surface}\label{subsec: reduction to constraint surface}

The Lagrange multiplier $G_i$ imposes the constraint $\phi^i_x - D^i W=0$, and hence $\phi^i$ is constrained to live on the kink moduli space, where crucially the moduli $m^\alpha(t)$ are allowed to depend on time. Then, the bosonic part of the action on this constraint surface is given by
\begin{align}
  \tilde{S}_\Sigma^\text{bos.} = \frac{1}{2}\int_\Xi d^2 x\,\, g_{ij} \partial_t\phi^i \partial_t\phi^j \Big|_{\phi^i_x - D^i W=0} = \frac{1}{2} \int dt\,\, \mathcal{G}_{\alpha\beta} \dot{m}^\alpha \dot{m}^\beta\ ,
\end{align}
with $\mathcal{G}_{\alpha\beta}$ as defined in (\ref{eq: induced metric}). Thus, as expected we reproduce the standard moduli space approximation.

We also have a fermionic Lagrange multiplier, given by the now non-dynamical $\psi_+^i$. This imposes a fermionic constraint, such that the complete constraint surface can be denoted $\Sigma=\{\mathcal{C}_1^i=0,\mathcal{C}_2^i=0\}$, where
\begin{align}
  \mathcal{C}^i_1 &= \partial_x\phi^i - D^i W \ ,\nonumber\\
  \mathcal{C}^i_2 &= D_x \psi^i_- - (D^i D_j W) \psi^j_-\ .
  \label{eq: sigma model constraints}
\end{align}
We then find\footnote{We necessarily have that the full set of equations of motion from $\tilde{S}$ transform into one-another under $\tilde{\delta}$. We see here, however, that the constraints defining $\Sigma$ transform only amongst themselves}
\begin{align}
  \tilde{\delta}\mathcal{C}_1 &= i \bareps_+ \left( \mathcal{C}^i_2 - \Gamma^i_{jk} \mathcal{C}^j_1 \psi^k_- \right) \ ,\nonumber\\
  \tilde{\delta}\mathcal{C}_2 &= \left( D_t \mathcal{C}^i_1 + \tfrac{i}{2} R^i_{jkl} \mathcal{C}^i_1 ( \bar{\psi}^k_- \gamma_t \psi^l_- ) - i \Gamma^i_{jk} (\bar{\psi}^j_- \gamma_t \mathcal{C}^k_2) \right)\gamma_t \epsilon_+ \nonumber\\
  &\qquad +\left( D_x \mathcal{C}^i_1 + (D^i D_j W)\mathcal{C}^j_1 \right)\epsilon_-\ ,
\end{align}
and hence we are able to restrict to $\Sigma$ without breaking any supersymmetry. In principle, we can then solve the constraints (\ref{eq: sigma model constraints}) to determine $\phi^i$ and $\psi^i_-$ in terms of a set of both bosonic and fermionic time-dependant moduli. Upon substituting these forms for $\phi^i$ and $\psi^i_-$ back into $\tilde{S}$, we determine the supersymmetrised low energy effective action. The supersymmetry transformation rules for the moduli are determined from those of $\phi^i$ and $\psi^i_-$ (\ref{eq: reduced sigma model SUSY}). In particular, we see that the bosonic and fermionic moduli are paired under the supersymmetry $\epsilon_+$, while the `broken' supersymmetry $\epsilon_-$ lives on as a shift symmetry for its Goldstino mode $\psi^i_-$.\\

Finally, it is instructive to consider an explicit example. We take $(M,g)=(\mathbb{R},\delta)$ and $W(\phi)=\lambda\left( a^2 \phi - \tfrac{1}{3}\phi^3 \right)$. This choice describes a single scalar field in a `double dip' potential $(\partial_\phi W)^2=\lambda^2 \left( \phi^2 - a^2 \right)^2$. Finite energy configurations have $\lim_{x\to \pm\infty}\phi=\pm a$. Restricting to sectors satisfying $W(\phi(\infty)) - W(\phi(-\infty))\ge 0$ corresponds to requiring $\phi(\infty)\ge \phi(-\infty)$.

We now solve the constraints $\mathcal{C}_1^i=0,\mathcal{C}_2^i=0$ for the sector defined by $\phi(\infty)=a, \phi(-\infty)=-a$, i.e. the kink solution. These read
\begin{align}
  \partial_x\phi - \lambda \left( a^2 - \phi^2 \right) &=0 \ ,\nonumber\\
  \partial_x \psi_- + 2\lambda \phi \psi_- &= 0\ .
\end{align}
We find the general solutions
\begin{align}
  \phi(t,x) &= a \tanh \left[ \lambda a \left( x - y(t) \right) \right] \ ,\nonumber\\
  \psi_- (t,x) &= -\lambda a^2\,\text{sech}^2\left[ \lambda a \left( x-y(t) \right) \right] \chi_-(t)\ ,
\end{align}
where $y(t)\in\mathbb{R}$ is the position of the kink at time $t$, and $\chi_-(t)$ is a negative chirality spinor. The action $\tilde{S}$ evaluated on $\Sigma$ is then
\begin{align}
  \tilde{S}_\Sigma = \frac{2}{3} \lambda a^3 \int dt\,\, \Big( \left( \partial_t y \right)^2 - i \bar{\chi}_- \gamma_t \partial_t \chi_- \Big)\ ,
\end{align}
with supersymmetry
\begin{align}
  \tilde\delta y &= i \bareps_+ \chi_- \ ,\nonumber \\
  \tilde{\delta} \chi_- &= \left( \partial_t y \right) \gamma_t \epsilon_+ - 2\epsilon_-\ .
\end{align}
We indeed find that $\epsilon_+$ defines an $\mathcal{N}=1$ supersymmetry on this (free) QM model, while $\epsilon_-$ describes a trivial shift symmetry for its Goldstino mode $\chi_-$.

\subsection{$\eta\to 0$ in the quantum theory}

Our focus in analysing this $(1+1)$-dimensional model has been to develop a toy model for the conceptually comparable but technically much more difficult example of the non-Lorentzian five-dimensional model for the null-compactified M5-brane that was constructed in Section \ref{sec: M5s}. In particular, in considering the $\eta\to 0$ limit we have localised exactly to the soliton moduli space, and thus have neglected possible quantum corrections to the resulting quantum mechanical model. While the focus is this classical construction, it is worth briefly discussing how one can include quantum corrections to this procedure, so as to make closer contact with the existing literature on the quantisation of fields around solitons \cite{Weinberg:2006rq}. In particular, we shall see that by taking the $\eta\to 0$ limit in the full quantum theory, we will generically have 1-loop corrections to the final quantum-mechanical action corresponding to loops of the fluctuations transverse to the moduli space. Conversely, we will see that 1-loop corrections also arise if we instead start with the fixed point action (\ref{eq: kink fixed point action}) and integrate out the Lagrange multiplier, and that (at least in this simple example) these corrections precisely match those of the former calculation. \\

We first consider the partition function of the theory (\ref{eq: rescaled kink action}) at finite $\eta$, analytically continued to Euclidean signature,
\begin{align}
  \mathcal{Z}_\eta = \int D\phi\, D\psi\, \exp\Big\{-\left( \eta^{-1} S_{-1} + S_0 + \eta S_1 \right)\Big\}\ .
\end{align}
Then, as is familiar from localisation calculations \cite{Cremonesi:2014dva}, we can determine the $\eta\to 0$ limit of $\mathcal{Z}_\eta$ by treating $\eta$ as a semiclassical expansion parameter - sometimes referred to as an auxiliary Planck constant\footnote{It is implicit that the overall quantum parameter $\hbar$ has been set to 1}. In other words, we expand in $\eta$ around the (moduli space of) saddle points of $S_{-1}$. 

We choose boundary conditions such that we lie in a particular kink sector. In other words, we fix $\phi$ at $x=\pm\infty$, and assume $W(\phi(\infty))-W(\phi(-\infty))>0$. Then, to describe the QFT around the classical minima of such a sector, we shift $S_{-1}$ as
\begin{align}
  S_{-1} \to S_{-1} - [W(\phi(\infty))-W(\phi(-\infty))] = \frac{1}{2} \int d^2 x\,\, g_{ij} \left( \partial_x\phi^i - D^i W \right) \left( \partial_x\phi^j - D^j W \right)\ ,
\end{align}
where the overall sign difference from (\ref{eq: rescaled kink action}) arises from the Wick rotation to Euclidean signature. Now let $\phi_\text{cl}^i(x; m^\alpha(t))$ be a solution for $\phi^i$ lying on the saddle point locus of $S_{-1}$, i.e. $\partial_x\phi_\text{cl}^i - D^i W(\phi_\text{cl})=0$. Here, the functions $m^\alpha(t)$ denote the (time-dependant) moduli. We then expand,
\begin{align}
  \phi^i(t,x) = \phi_\text{cl}^i(x; m^\alpha(t)) + \eta^{1/2}\delta \phi^i\ ,
\end{align}
where the perturbations $\delta \phi^i$ lie transverse to the moduli space,
\begin{align}
  \int d^2x\,\,g_{ij} \delta \phi^i \frac{\partial \phi^j_\text{cl}}{\partial m^\alpha}=0\ .
\end{align}
Then, we find
\begin{align}
  \mathcal{Z}_\eta = \int Dm\, D(\delta \phi)\, D\psi \exp \left\{ -S_0[\phi_\text{cl}] -\frac{1}{2} \int d^2 x\,\, \delta \phi^i \left( \frac{\delta^2 S_{-1}}{\delta\phi^i \delta\phi^j}[\phi_\text{cl}] \right) \delta \phi^j + \bigO(\eta^{1/2}) \right\}\ .
\end{align}
Hence, taking $\eta\to 0$, we have
\begin{align}
  \mathcal{Z}_0 = \int Dm\, D\psi \,\mathcal{Z}_\text{1-loop}\, e^{-S_0[\phi_\text{cl}]}\ ,
\end{align}
where
\begin{align}
  \mathcal{Z}_\text{1-loop} = \int D(\delta \phi) \exp \left\{ -\frac{1}{2} \int d^2 x\,\, \delta \phi^i \left( \frac{\delta^2 S_{-1}}{\delta\phi^i \delta\phi^j}[\phi_\text{cl}] \right) \delta \phi^j \right\} = \left[ \det \left( \frac{\delta^2 S_{-1}}{\delta\phi^i \delta\phi^j}[\phi_\text{cl}] \right) \right]^{-1/2}\ .
\end{align}
Therefore, the classical moduli space action $S_0[\phi_\text{cl}]$, precisely as discussed in section \ref{subsec: reduction to constraint surface} up to a Wick rotation, receives quantum corrections from $\mathcal{Z}_\text{1-loop}$, which we have neglected in our classical analysis. In particular, setting $g_{ij}=\delta_{ij}$ for simplicity, we have
\begin{align}
  \frac{\delta^2 S_{-1}}{\delta\phi^i \delta\phi^j} [\phi_\text{cl}] = -\delta_{ij}\partial_x^2 +( \partial_k \partial_i \partial_j W )(\phi_\text{cl})\,\partial_x\phi^k_\text{cl} + (\partial_i \partial_k W)(\phi_\text{cl})(\partial_j \partial_k W)(\phi_\text{cl})\ .
\end{align}
It is natural then to ask how we should interpret the quantum field theory defined by what we called the `fixed point' action $\tilde{S}$ (\ref{eq: kink fixed point action}). We can once again consider the path integral, this time remaining in Lorentzian signature,
\begin{align}
  \tilde{\mathcal{Z}} = \int D\phi\, D\psi\, DG\, e^{i\tilde{S}[\phi,\psi]} = \int D\phi\, D\psi\, DG\,\exp\left\{ i S_0[\phi,\psi] + i\int d^2 x\, G_i \left( \partial_x\phi^i - D^i W \right)  \right\}\ .
\end{align}
Integrating out the Lagrange multiplier $G_i$, we arrive at
\begin{align}
  \tilde{\mathcal{Z}}_0 = \int D\phi\, D\psi\,\, \delta\left( \partial_x\phi^i - D^i W \right) e^{iS_0[\phi,\psi]} = \int Dm\, D\psi\, \tilde{\mathcal{Z}}_\text{1-loop} e^{iS_0[\phi_\text{cl},\psi]}\ ,
\end{align}
where, again taking $g_{ij}=\delta_{ij}$ for simplicity,
\begin{align}
  \tilde{\mathcal{Z}}_\text{1-loop} = \left[\det \Big( \delta_{ij} \partial_x - \partial_i\partial_j W (\phi_\text{cl})  \Big) \right]^{-1}\ .
\end{align}
Once again, we find that the action constrained exactly to the moduli space $S_0[\phi_\text{cl}]$ receives corrections in the form of a one-loop determinant. Indeed, it is not hard to see that we have $\mathcal{Z}_\text{1-loop}=\tilde{\mathcal{Z}}_\text{1-loop}$. Defining $\Delta_{ij}=\delta_{ij} \partial_x - \partial_i\partial_j W (\phi_\text{cl})$, we have that
\begin{align}
  \left( \Delta^\dagger \Delta \right)_{ij} &= \left( -\delta_{ik} \partial_x - \partial_i \partial_k W(\phi_\text{cl}) \right)\left(\delta_{kj} \partial_x - \partial_k \partial_j W(\phi_\text{cl}) \right)\nonumber\\
  &= -\delta_{ij}\partial_x^2 +( \partial_k \partial_i \partial_j W )(\phi_\text{cl})\,\partial_x\phi^k_\text{cl} + (\partial_i \partial_k W)(\partial_j \partial_k W) \nonumber\\
  &= \frac{\delta^2 S_{-1}}{\delta\phi^i \delta\phi^j} [\phi_\text{cl}]\ ,
\end{align}
where the adjoint here is respect to the $L^2$ inner product, and so
\begin{align}
  \mathcal{Z}_\text{1-loop} = \left[\det \left( \Delta^\dagger \Delta \right)\right]^{-1/2} = \left[\det ( \Delta^\dagger ) \det \left( \Delta \right) \right]^{-1/2} = \left[ \det \left( \Delta \right) \right]^{-1} = \tilde{\mathcal{Z}}_\text{1-loop}\ .
\end{align}
Hence we have shown, at least in this simple kink example and with flat target space, that the one-loop corrections to the moduli space action are independent of our choice to either consider the $\eta\to 0$ limit in the full quantum theory, or to just use the fixed point action $\tilde{S}$ to begin with.


\section{Five-dimensional super-Yang Mills and the M5-brane}\label{sec: Five-dimensional super-Yang Mills and the M5-brane}


We now return to the theory (\ref{eq: final M5 action}) found at a fixed point of a classical RG flow from maximal Lorentzian super-Yang Mills in five dimensions. The dynamics of the theory are then constrained to anti-instanton moduli space $F_{ij}^+=0$. The bosonic sector of the resulting quantum mechanics on this moduli space was first discussed in \cite{Lambert:2011gb}, which additionally specialised to spherically symmetric, commuting zero modes for the fields $A_0,X^I$.

Our aim is now to use the explicit action (\ref{eq: final M5 action}) to solve the full set of constraints of the theory, and thus reduce it to a superconformal quantum mechanics on instanton moduli space.

\subsection{Reduction to constraint surface}

We can reduce the theory by integrating out non-dynamical fields. Letting $E^{(\Phi)}=g^2 \frac{\delta S}{\delta \Phi}$ denote the on-shell condition corresponding to a field $\Phi$, we find
\begin{align}
  E^{\left( G_{ij} \right)} &= \tfrac{1}{2} F_{ij}^+ \ ,\nonumber\\
  E^{\left( \Psi_+ \right)} &= i \Gamma_i D_i \Psi_- \ ,\nonumber\\
  E^{\left( X^I \right)} &= D_i D_i X^I + \bar{\Psi}_- \Gamma_+ \Gamma^I \Psi_- \ ,\nonumber\\
  E^{\left( A_0 \right)} &= D_i F_{0i} + \bar{\Psi}_- \Gamma_+ \Psi_- \label{eq: SYM constraints}\ ,\\[1em]
  E^{\left( A_i \right)} &= - D_0 F_{0i} + D_j G_{ij} + i \left[ X^I, D_i X^I \right] - \bar{\Psi}_- \Gamma_i \Psi_+ - \bar{\Psi}_+ \Gamma_i \Psi_-  \ ,\nonumber\\
  E^{\left( \Psi_- \right)} &= -i\Gamma_+ D_0 \Psi_+ + i \Gamma_i D_i \Psi_+ - \Gamma_+ \Gamma^I \left[ X^I, \Psi_- \right]\ ,
\end{align}
where we have assumed suitable (i.e. Dirichlet or Neumann) boundary conditions for $X^I$ and $A_0$ at spatial infinity. Indeed, it is the time-dependent parameters describing such boundary conditions that determines the energy of these fields, and hence they will generically appear in the reduced quantum mechanics. 

The first four of the equations (\ref{eq: SYM constraints}), corresponding to the non-dynamical fields in $S$, define the constraint surface $\Sigma=\left\{E^{\left( G_{ij} \right)}, E^{\left( \Psi_+ \right)}, E^{\left( X^I \right)}, E^{\left( A_0 \right)}=0\right\}$. The $G_{ij}$ equation of motion restricts $A_i$ to the moduli space $\mathcal{I}_{k,N}$ of a degree $k$ instanton for some $k\le 0$. In particular, we have $\dim\left( \mathcal{I}_{k,N} \right)=4|k|N$ \cite{Belitsky:2000ws}. The $\Psi_+$ equations restricts $\Psi_-$ to solve the gauge covariant Dirac equation. Using index theorem techniques, we find that the moduli space of normalisable solutions is described by $8|k|N$ Grassmann parameters. This calculation follows the standard argument \cite{Belitsky:2000ws}, but crucially differs\footnote{Usually, the relevant index is proportional to $\text{tr}(\gamma_5 \gamma_\mu\gamma_\nu\gamma_\rho\gamma_\sigma)=4\varepsilon_{\mu\nu\rho\sigma}$ where the $\{\gamma_\mu\}_{\mu=1}^4$ form a basis for the 4d Euclidean Clifford algebra, and the fermionic zero modes in the background of an anti-instanton (rather than instanton) have positive chirality under $\gamma_5=\gamma_1\gamma_2\gamma_3\gamma_4$. The equivalent object in our formulation is $\text{tr}(\Gamma_{1234}\Gamma_i\Gamma_j\Gamma_k\Gamma_l)=32\varepsilon_{ijkl}$. We indeed have $\Gamma_{1234}\Psi_- = \Psi_-$, and so have non-trivial zero modes. Taking into account the additional chirality condition on $\Psi_-$ under $\Gamma_{05}$, we find $\frac{1}{2}\times \frac{32}{4}=4$ times as many zero modes.} by a factor of 4 from the standard result of $2|k|N$ for an adjoint fermion in four dimensions due to our use of 32-component spinors of $\text{Spin}(1,10)$.


The kinetic term for $\Psi_-$ in $S$ implies kinetic terms for these Grassmann parameters in the reduced theory. Hence, on-shell in the reduced theory, there are $8|k|N/2=4|k|N$ Grassmann degrees of freedom, and thus Bose-Fermi degeneracy is recovered on-shell. The remaining two equations restrict the non-dynamical $X^I$ and $A_0$ to the solution space of gauge covariant Poisson equations, and thus will be determined up a set of zero modes.\\

Our aim is now to constrain the theory to $\Sigma$ without breaking any supersymmetry. We find that under the supersymmetry (\ref{eq: 5d fixed point theory SUSYs}), the constraints transform as
\begin{align}
  \delta E^{\left( G_{ij} \right)} &= \tfrac{1}{4}\bareps_- \Gamma_+ \Gamma_{ij} E^{\left( \Psi_+ \right)}  \ ,\nonumber\\
  \delta E^{\left( \Psi_+ \right)} &= i \left( E^{\left( X^I \right)}\Gamma^I - E^{\left( A_0 \right)} \right)\epsilon_-  \ ,\nonumber\\
  \delta E^{\left( X^I \right)} &= \bareps_- \left( \Gamma^I \Gamma_i D_i E^{\left( \Psi_- \right)} + \Gamma_+ \left( \Gamma^I D_0 + i \Gamma^{IJ}\left[ X^J, \,\cdot\,  \right] \right) E^{\left( \Psi_+ \right)} - \Gamma^I \Gamma_{ij} \left[ E^{\left( G_{ij} \right)},\Psi_+ \right] \right) \nonumber\\
   & \qquad + \bareps_+ \Gamma^I \Gamma_i D_i E^{\left( \Psi_+ \right)} + 2 \bar{\zeta}_+ \Gamma^I E^{\left( \Psi_+ \right)}  \ ,\nonumber\\
  \delta E^{\left( A_0 \right)} &= -\bareps_- \left( \Gamma_i D_i E^{\left( \Psi_- \right)} - i \Gamma_+ \Gamma^I \left[ X^I, E^{\left( \Psi_+ \right)} \right] - \Gamma_{ij} \left[ E^{\left( G_{ij} \right)},\Psi_+ \right] \right) \nonumber\\
   & \qquad + \bareps_+  \Gamma_i D_i E^{\left( \Psi_+ \right)} - 3 \bar{\zeta}_+ E^{\left( \Psi_+ \right)}\ .
\end{align}
In particular, we note that $\left.\delta E^{(X^I)}\right|_\Sigma$ and $\left.\delta E^{(A_0)}\right|_\Sigma$ are generically non-zero for $\epsilon_-$ non-zero, corresponding to $\xi_- = \zeta_+=0$, and thus only the supersymmetry $\xi_+$ is unbroken by restricting to $\Sigma$.\\

We now show that this issue can be remedied such that the surface $\Sigma$ preserves the full 24 supercharges. We introduce a shifted supersymmetry $\hat{\delta}$, defined by
\begin{align}
  \hat\delta X^I &= \delta X^I -i\bar{\epsilon}_- \Gamma^I \Psi_+ + \bareps_- \Gamma_+ \Gamma^I \chi + \phi^I_{(X)} &=&\,\,  i\bar{\epsilon}_+ \Gamma^I \Psi_- + \bareps_- \Gamma_+ \Gamma^I \chi + \phi^I_{(X)} \ ,\nonumber \\
  \hat\delta A_0 &= \delta A_0 - i \bar{\epsilon}_- \Psi_+ - \bareps_-\Gamma_+ \chi + \phi_{(A)} &=&\,\,  -i \bar{\epsilon}_+ \Psi_- - \bareps_-\Gamma_+ \chi + \phi_{(A)}\ ,
  \label{eq: shifted SUSY}
\end{align}
with $\hat{\delta}=\delta$ on other fields. Here, $\chi$ is a spinorial field satisfying
\begin{align}
  D_i D_i \chi = -\Gamma_i \left[ F_{0i}, \Psi_- \right] + \Gamma_i \Gamma^I \left[ D_i X^I, \Psi_- \right]\ ,
  \label{eq: chi def}
\end{align}
while $\phi^I_{(X)},\phi_{(A)}$ are each solutions to the gauge-covariant Laplace equation, i.e. $D_i D_i \phi^I_{(X)}=0$, $ D_i D_i \phi_{(A)}=0$, which effectively allow for different boundary conditions on $\chi$. Then, it is easily seen that we have $\left.\hat{\delta} E^{(X^I)}\right|_\Sigma=0$, $\left.\hat{\delta} E^{(A_0)}\right|_\Sigma=0$ and thus we can restrict to $\Sigma$ without breaking any supersymmetry. We then calculate (neglecting boundary terms at temporal infinity)
\begin{align}
  \hat{\delta} S = &\frac{1}{g^2} \text{tr}\int d^4 x\, dx^0\,\, \Big( \left( -i\bareps_-\Psi_+ - \bareps_-\Gamma_+\chi + \phi_{(A)} \right) E^{\left( A_0 \right)} \nonumber\\
  &\hspace{30mm} + \left( -i\bareps_-\Gamma^I\Psi_+ + \bareps_-\Gamma_+\Gamma^I\chi + \phi^I_{(X)} \right)E^{\left( X^I \right)}  \Big) \nonumber\\
  +&\frac{1}{g^2}\text{tr}\int d^4 x\, dx^0\,\, \partial_i \Big[\,\, \bareps_- \left(  F_{0i} \Gamma_+ \chi - \left( D_i X^I \right)\Gamma_+\Gamma^I \chi \right.\nonumber\\
  &  \hspace{34mm}\left. + i \left( D_0 X^I \right) \Gamma_+ \Gamma^I \Gamma_i \Psi_- + \tfrac{1}{2} \left[ X^I, X^J \right] \Gamma_+ \Gamma^{IJ} \Gamma_i \Psi_- \right) \nonumber\\[0.8em]
  &\hspace{21mm}+\bareps_+ \left( i F_{0j} \Gamma_{ij} \Psi_- - i \left( D_j X^I \right) \Gamma^I \Gamma_{ij} \Psi_- - \tfrac{i}{2} E^{(G_{jk})} \Gamma_- \Gamma_i \Gamma_{jk} \Psi_+ \right) \nonumber\\[0.8em]
  &\hspace{21mm}+  4i X^I \bar{\zeta}_+ \Gamma^I \Gamma_i \Psi_- - F_{0i} \phi_{(A)} - \left( D_i X^I \right) \phi^I_{(X)} \Big]\ .
  \label{eq: delta hat S}
\end{align}
Thus, we find that for suitable boundary conditions at spatial infinity, and choices of $\phi_{(X)}^I, \phi_{(A)}$, we have $\hat{\delta}S|_\Sigma =0$.\\

To proceed, we need to write down the general solution for $A_i$ in some gauge, as described by the ADHM construction \cite{Atiyah:1978ri}. One can then seek the general solutions to the three remaining constraints in terms of this ADHM data. The quantum mechanical action for the reduced theory $S_\Sigma$ is then determined by evaluating $S$ on these solutions and performing the spatial integral. In particular, the bosonic sector of the theory will reproduce the standard $\sigma$-model with the usual moduli space metric as given by
\begin{align}
  \mathcal{G}_{AB} = \text{tr}\int d^4 x\,\,  \left( \delta_A A_i  \right)\left( \delta_B A_i  \right) \ ,
\end{align}
where the indices $A,B$ run over the $4|k|N$ moduli of $A_i$, and we have fixed the time evolution of $A_i$ to lie transverse to gauge orbits. The full $S_\Sigma$ hence must be an extension to the maximal $\mathcal{N}=(4,4)$ $\Sigma$-model on instanton moduli space $\mathcal{I}_{k,N}$ to include coupling to the zero modes of $X^I$ and $A_0$. The theory will possess in total 24 supersymmetries, made up of the regular 8 rigid supersymmetries contained in $\xi_-$, their 8 superconformal partners in $\zeta_+$, and finally an additional 8 contained in $\xi_+$ which generically act only on the fermions and zero modes.

\subsection{The single $SU(2)$ instanton}

We now focus on the particular choice of gauge group $G=SU(2)$, and on the single anti-instanton ($k=-1$) sector. In the M-theory picture, this corresponds to a single unit of momentum along the compact null direction. This particular gauge group and sector are special, as they allow us to generate \textit{all} $8|k|N=16$ Grassmann moduli of $\Psi_-$ using an infinitesimal super(conformal) perturbation from a purely bosonic solution for the constraints (\ref{eq: SYM constraints}). For higher $|k|$ and/or $N$, one must seek the remaining $8(|k|N-2)$ zero modes via other means. 

\subsubsection{Solving the constraints}

We solve the constraints (\ref{eq: SYM constraints}) defining the constraint surface $\Sigma$, first finding a purely bosonic solution. The general solution for the instanton equation in the $k=-1$ sector is
\begin{align}
  A_i(t,x) = h \hat{A_i} h^{-1} - i \left( \partial_i h \right) h^{-1}\ ,
\end{align}
where $\hat{A}_i$ is the $k=-1$ instanton in singular gauge \cite{Belavin:1975fg},
\begin{align}
  \hat{A_i}(t,x) = \frac{\rho^2}{\left( x-y \right)^2 \left( \left( x-y \right)^2 + \rho^2 \right)} \eta_{ij}^\alpha \left( x-y \right)_j g\sigma^\alpha g^{-1} \ ,
\end{align}
for Pauli matrices $\sigma^\alpha$, $\alpha=1,2,3$, and $\eta^\alpha_{ij}$ the `t Hooft symbols as discussed in the Introduction.
Here, we have \textit{time-dependant} moduli $y_i(t)\in\mathbb{R}^4$, $\rho(t)\in\mathbb{R}$, and $g(t)\in SU(2)$, where we write $t=x^0$. Additionally, we have allowed our solution to move freely along gauge orbits over time, as signified by the gauge parameter $h:\mathbb{R}^{1,4}\to SU(2)$. We now choose the time evolution of $h$ such that the dynamical degrees of freedom of $A_i$ lie transverse to gauge orbits, and as such the resulting model describes only these physical degrees of freedom. One can show \cite{Vandoren:2008xg,Brown:1977bj} that this is ensured precisely by requiring that at any fixed time,
\begin{align}
  \text{tr}\int_{t=t_0}d^4 x\,\, \left( D_i\Lambda \right)\partial_0 A_i = 0\ ,
  \label{eq: transverse to gauge orbits}
\end{align}
for any $i\Lambda:\mathbb{R}^{1,4}\to \frak{su}(2)$. Note that even with this condition in place, the global gauge orientation $g(t)$ remains intact as a physical degree of freedom \cite{Tong:2005un}. Writing $i\dot{\omega} = h^{-1} \dot{h}\in\frak{su}(2)$, (\ref{eq: transverse to gauge orbits}) is equivalent to
\begin{align}
  \hat{D}_i\hat{D}_i\dot{\omega} = -\hat{D}_i \left( \partial_0 \hat{A}_i \right)\ ,
\end{align}
where $\hat{D}_i$ denotes the gauge covariant derivative with respect to $\hat{A}_i$. Introducing the notation $z_i(t,x) = x_i - y_i(t)$ to denote the spatial displacement from the instanton centre at time $t$, we have the solution
\begin{align}
  \dot{\omega} = \frac{\rho^2}{z^2 \left( z^2 + \rho^2 \right)} \left( \eta_{ij}^\alpha \dot{y}_i z_j - \dot{u}^\alpha z^2 \right) g \sigma^\alpha g^{-1}\ ,
\end{align}
where we define $\dot{u}^\alpha$ by $g^{-1}\dot{g}=i\dot{u}^\alpha \sigma^\alpha$. This in turn determines $h$ for all time, given initial value $h_0= h(t_0)$ on some time slice $t=t_0$. Note, we could always shift $\dot{\omega}$ by some zero mode of $\hat{D}_i \hat{D}_i$, however, we shall see that such zero modes can be absorbed into the general solution for $A_0$.\\

We now turn to the other equations. Writing $X^I = h \hat{X}^I h^{-1}$, we find
\begin{align}
  E^{\left( X^I \right)}=0 \implies \hat{D}_i \hat{D}_i \hat{X}^I = 0 \ ,
\end{align}
while, noting that $D_i F_{0i} = D_i\left( \partial_0 A_i - D_i A_0 \right) = -D_i D_i A_0$, and writing $A_0 = h\hat{A}_0 h^{-1}$, we have
\begin{align}
  E^{\left( A_0 \right)}=0 \implies \hat{D}_i \hat{D}_i \hat{A}_0 = 0\ .
\end{align}
Hence, we can write our bosonic solution as
\begin{align}
  A_i &= h\hat{A}_i h^{-1} - i \left( \partial_i h \right)h^{-1} \ ,\nonumber \\
  A_0 &= h \hat{A}_0 h^{-1} \ ,\nonumber\\
  X^I &= h \hat{X}^I h^{-1}\ ,
\end{align}
where $\hat{X}^I$ and $\hat{A}_0$ are zero modes of the gauge covariant Laplacian in the background of the $k=-1$ instanton $\hat{A}_i$ in singular gauge. By performing a further \textit{time-independent} gauge transformation, it is easily seen that $h_0$ can be arbitrarily fixed by gauge transformations, and thus will not appear in any gauge invariant. Finally, we perform a gauge transformation to bring our bosonic solution to a more convenient form,
\begin{align}
  A_i &= \hat{A}_i \ ,\nonumber\\
  A_0 &= \hat{A}_0 - \dot{\omega} \ ,\nonumber\\
  X^I &= \hat{X}^I\ ,
  \label{eq: bosonic solution}
\end{align}
with
\begin{align}
  \hat{A}_i &= \frac{\rho^2}{z^2 \left( z^2 + \rho^2 \right)} \eta_{ij}^\alpha z_j g\sigma^\alpha g^{-1} \ ,\nonumber\\
  h^{-1} \dot{h} &= i \dot{\omega} =  i\frac{\rho^2}{z^2 \left( z^2 + \rho^2 \right)} \left( \eta_{ij}^\alpha \dot{y}_i z_j - \dot{u}^\alpha z^2 \right) g \sigma^\alpha g^{-1}\ .
\end{align}
We indeed see that any zero modes we could have added to $\dot{\omega}$ could be absorbed into $\hat{A}_0$.\\

With a bosonic solution in hand, we seek a general solution to the constraints (\ref{eq: SYM constraints}). In principle, starting from our bosonic solution, we can construct a solution with fermions turned on by performing a \textit{finite} supersymmetry transformation - sometimes referred to as the sweeping procedure \cite{Belitsky:2000ws}. Then, each field $\Phi$ is given by

\begin{align}
  \Phi = \left.\left( e^{\hat{\delta}}\Phi \right)\right|_{\text{bos. solution}} = \sum_{n=0}^\infty \frac{1}{n!} \left.\left( \hat{\delta}^n \Phi \right)\right|_{\text{bos. solution}}\ .
\end{align}
What's more, since $\hat{\delta}$ is parameterised by a finite number of Grassmann parameters, this series will necessarily terminate. However, this approach is cumbersome, and requires us to solve (\ref{eq: chi def}) for $\chi$ for generic configurations, and then calculate $\hat{\delta}\chi, \hat{\delta}^2\chi, \dots$.

Thankfully, there is a more straightforward approach we can take. First, we fix $A_i = \hat{A}_i$. Then, we note that to generate a non-trivial solution to $E^{\left( \Psi_+ \right)}=i\Gamma_i D_i \Psi_-=0$ we need only consider an \textit{infinitesimal} variation of the bosonic solution (\ref{eq: bosonic solution}). To see this, note that on $\Sigma$ we have $\hat{\delta} E^{\left( \Psi_+ \right)}=0$. In particular,
\begin{align}
  0 &= \left.\left( \hat{\delta} E^{\left( \Psi_+ \right)} \right)\right|_{\text{bos. solution}} \nonumber\\
  &= \left.\left( i\Gamma_i D_i \deltahat \Psi_- + \Gamma_i \left[ \deltahat A_i , \Psi_- \right] \right)\right|_{\text{bos. solution}} \nonumber\\
  &= i \Gamma_i D_i \left.\left( \deltahat \Psi_- \right)\right|_{\text{bos. solution}}\ .
\end{align}
And thus, $\deltahat \Psi_-$ evaluated on the bosonic solution provides a solution for $E^{\left( \Psi_+ \right)}=0$, for gauge field $A_i$ unchanged. Performing this calculation, we find the solution
\begin{align}
  \Psi_- = F_{ij} \Gamma_{ij} \left( \alpha + z_k \Gamma_k \beta \right) = F_{ij} \Gamma_{ij} \alpha + 4 F_{ij} z_j \Gamma_i \beta\ ,
\end{align}
where $\alpha,\beta$ are Majorana spinors of $\text{Spin}(1,10)$ with $\Gamma_{05}\alpha = \Gamma_{012345}\alpha = -\alpha$, $\Gamma_{05}\beta = -\Gamma_{012345}\beta = -\beta$. In particular, $\alpha$ and $\beta$ together make up $8|k|N=16$ Grassmann parameters, and thus we have the general solution for $\Psi_-$.\\

Solutions for the final two constraints for $X^I$ and $A_0$ cannot be generated in the same way, as they are both quadratic in $\Psi_+$. Instead, we solve them by inspection. It is convenient to write the solutions as
\begin{align}
  X^I &= \hat{X}^I + \tilde{X}^I \ ,\nonumber\\
  A_0 &= \hat{A}_0 - \dot{\omega} + \tilde{A}_0\ ,
\end{align}
where as before $D_i D_i \hat{X}^I=0$, $D_i D_i \hat{A}_0=0$, and
\begin{align}
  \tilde{X}^I &= \frac{\left( z^2 + \rho^2 \right)^2}{16\rho^2} \bar{\Psi}_- \Gamma_+\Gamma^I \Psi_- = -i F_{ij} \left( \bar{\alpha} \Gamma_+ \Gamma^I \Gamma_{ij} \alpha + 8 z_j \bar{\alpha} \Gamma_+ \Gamma^I \Gamma_i \beta + 4z_j z_k \bar{\beta} \Gamma_+ \Gamma^I \Gamma_{ik} \beta \right) \ ,\nonumber\\
  \tilde{A}_0 &= -\frac{\left( z^2 + \rho^2 \right)^2}{16\rho^2} \bar{\Psi}_- \Gamma_+ \Psi_- = i F_{ij} \left( \bar{\alpha} \Gamma_+ \Gamma_{ij} \alpha + 8 z_j \bar{\alpha} \Gamma_+  \Gamma_i \beta - 4z_j z_k \bar{\beta} \Gamma_+ \Gamma_{ik} \beta \right)\ .
  \label{eq: tilde defs}
\end{align}
In summary, we find the general solution to (\ref{eq: SYM constraints}) given by
\begin{align}
  A_i &= \frac{\rho^2}{z^2 (z^2 + \rho^2)} \eta^\alpha_{ij} z_j\,\, g \sigma^\alpha g^{-1} \ ,\nonumber\\
  \Psi_- &= F_{ij} \Gamma_{ij} \left( \alpha + z_k \Gamma_k \beta \right) \ ,\nonumber\\
  X^I &= \hat{X}^I + \frac{(z^2+\rho^2)^2}{16\rho^2}\bar{\Psi}_-\Gamma_+\Gamma^I \Psi_- \ ,\nonumber\\
  A_0 &= \hat{A}_0 + \frac{\rho^2}{z^2(z^2+\rho^2)}\left( z^2 \dot{u}^\alpha - \eta^\alpha_{ij} \dot{y}_i z_j \right) g \sigma^\alpha g^{-1} - \frac{(z^2+\rho^2)^2}{16\rho^2}\bar{\Psi}_-\Gamma_+ \Psi_-\ .
  \label{eq: general solution}
\end{align}
Finally, we need to solve for $\chi$ in order to determine the supersymmetry variations of the moduli parameterising the constraint space $\Sigma$. We can now split the terms in (\ref{eq: chi def}) into parts linear and cubic in $\Psi_-$, as
\begin{align}
  D_i D_i \chi &= -\Gamma_i\, [ \partial_0 A_i - D_i \hat{A}_0 + D_i \dot{\omega}, \Psi_- ] + \Gamma_i \Gamma^I [ D_i \hat{X}^I, \Psi_-] \nn\\
  &\qquad + \Gamma_i [ D_i \tilde{A}_0, \Psi_-] + \Gamma_i \Gamma^I [D_i \tilde{X}^I, \Psi_- ]\ ,
\end{align}
with $\tilde{X}^I$ and $\tilde{A}_0$ the fermion bilinears as defined in (\ref{eq: tilde defs}). We find the solution
\begin{align}
  \chi = &\,\,\hat{\chi} +\frac{\left( z^2+\rho^2 \right)^2}{16\rho^2}\Gamma_i\, [ \partial_0 A_i + D_i \dot{\omega}, \Psi_- ] \nonumber\\
  &-\frac{8\rho^2}{z^2\left( z^2 + \rho^2 \right)^3}\bigg( 4 \eta^\alpha_{il} z_l z_j z_k \Big( \left( \bar{\alpha}\Gamma_+ \Gamma_{ij} \alpha \right) \Gamma_k \alpha + \left( \bar{\alpha}\Gamma_+\Gamma^I \Gamma_{ij} \alpha \right) \Gamma^I\Gamma_k \alpha \Big) \nonumber\\
  &\hspace{32mm}+ 4\left( z^2 -\rho^2 \right)\eta^\alpha_{ik} z_k z_j \Big( \left( \bar{\alpha}\Gamma_+ \Gamma_{ij} \alpha \right) \beta + \left( \bar{\alpha}\Gamma_+\Gamma^I \Gamma_{ij} \alpha \right) \Gamma^I\beta \Big) \nonumber\\
  &\hspace{32mm}-4 z^2 \eta^\alpha_{ik}z_j z_l \Big( \left( \bar{\alpha}\Gamma_+ \Gamma_{ij} \alpha \right) \Gamma_{kl}\beta + \left( \bar{\alpha}\Gamma_+\Gamma^I \Gamma_{ij} \alpha \right) \Gamma^I\Gamma_{kl}\beta \Big) \nonumber\\
  &\hspace{32mm}+ 4 \eta^\alpha_{il} z_l z^2 \left( \rho^2 \delta_{jk} - z_j z_k \right) \Big( \left( \bar{\beta}\Gamma_+ \Gamma_{ij} \beta \right) \Gamma_{k}\alpha - \left( \bar{\beta}\Gamma_+\Gamma^I \Gamma_{ij} \beta \right) \Gamma^I\Gamma_{k}\alpha \Big) \nonumber\\
  &\hspace{32mm}+ \rho^2 z^4 \eta^\alpha_{ij} \Big( \left( \bar{\beta}\Gamma_+ \Gamma_{ij} \beta \right) \beta - \left( \bar{\beta}\Gamma_+\Gamma^I \Gamma_{ij} \beta \right) \Gamma^I\beta \Big) \bigg)\,\, g \sigma^\alpha g^{-1}\ ,
  \label{eq: chi solution}
\end{align}
where $\hat{\chi}$ is a solution to
\begin{align}
  D_i D_i \hat{\chi} = \Gamma_i\, [ D_i \hat{A}_0 , \Psi_- ] + \Gamma_i \Gamma^I [ D_i \hat{X}^I, \Psi_-]\ ,
  \label{eq: chi hat def}
\end{align}
for the so-far unspecified zero modes $\hat{X}^I$ and $\hat{A}_0$.

\subsubsection{The reduced theory}

Let $m^A$ formally denote the set of moduli $y_i,\rho,u^\alpha,\alpha,\beta$ combined with the time-dependant parameters describing the zero modes $\hat{X}^I$ and $\hat{A}_0$. We now want to determine the quantum mechanical action for the reduced theory, along with how supersymmetry acts on the $m^A$. The action $S_\Sigma$ is determined by simply evaluating our original action $S$ on the solutions (\ref{eq: general solution}).

To determine how supersymmetry acts on the $m^A$, we first need to augment the supersymmetry variation $\hat{\delta}$ by an infinitesimal gauge transformation with parameter $\tau$, such that the new variation lies transverse to gauge orbits. Indeed, we already did essentially the same thing in defining $\dot{\omega}$ to force time evolution to lie transverse to gauge orbits, and as such $\tau$ takes the same form. Then, letting $\delta_\Sigma$ denote the supersymmetry of the reduced theory, we have
\begin{align}
 \delta_\Sigma m^A \left( \frac{\partial A_i}{\partial m^A} \right) &= i\bar{\epsilon}_- \Gamma_i \Gamma_+ \Psi_- + D_i\tau  \ ,\nonumber \\
 \delta_\Sigma  m^A \left( \frac{\partial \Psi_-}{\partial m^A} \right) &= -F_{0i} \Gamma_i \epsilon_-  - \tfrac{1}{4} F_{ij} \Gamma_{ij} \Gamma_- \epsilon_+ + \left( D_i X^I \right) \Gamma_i \Gamma^I \epsilon_- + i [\tau,\Psi_-] \ ,\nonumber\\
  \delta_\Sigma  m^A \left( \frac{\partial X^I}{\partial m^A} \right) &= i\bar{\epsilon}_+ \Gamma^I \Psi_- + \bareps_- \Gamma_+ \Gamma^I \chi + \phi^I_{(X)} + i[\tau,X^I] \ ,\nonumber\\
    \delta_\Sigma  m^A \left( \frac{\partial A_0}{\partial m^A} \right) &= -i \bar{\epsilon}_+ \Psi_- - \bareps_-\Gamma_+ \chi + \phi_{(A)} +D_0 \tau \ ,
    \label{eq: reduced SUSY}
\end{align}
where
\begin{align}
  \tau = \hat{\tau}-\frac{\rho^2}{z^2 \left( z^2 + \rho^2 \right)}\Big( \eta^\alpha_{ij} \left( \delta_\Sigma y_i \right) z_j -z^2 \left( \delta_\Sigma  u^\alpha \right) \Big)\,\, g \sigma^\alpha g^{-1}\ ,
\end{align}
and $\hat{\tau}$ is any solution to $D_i D_i \hat{\tau}=0$, with different choices of $\hat{\tau}$ giving rise to different transformations of the $m^A$ under $\delta_\Sigma$. Choosing $\hat{\tau}=0$, first equation of (\ref{eq: reduced SUSY}) is solved to find
\begin{align}
  \delta_\Sigma y_i &= -4i \left( \bareps_- \Gamma_+ \Gamma_i \alpha \right) \ ,\nonumber\\
  \delta_\Sigma \rho &= -4i\rho \left( \bareps_- \Gamma_+ \beta \right) \ ,\nonumber\\
  \delta_\Sigma u^\alpha &= i \eta^\alpha_{ij} \left( \bareps_- \Gamma_+ \Gamma_{ij} \beta \right)\ .
\end{align}
In particular, we note that the 16 supercharges contained in $\epsilon_-$ descend to full super(conformal) symmetries of the reduce quantum mechanical model, while the final 8 supercharges in $\xi_+$ will appear as fermionic shift symmetries.

To proceed, we need to write down the general \textit{finite energy} configurations for the zero modes $\hat{X}^I, \hat{A}_0$. One can then solve (\ref{eq: chi hat def}) for $\hat{\chi}$, and then seek solutions for $\phi_{(X)}^I$ and $\phi_{(A)}$ such that $\hat{\delta} S$ as calculated in (\ref{eq: delta hat S}) vanishes for our solutions (\ref{eq: general solution}). Finally, $\delta_\Sigma m^A$ for the remaining moduli can be determined via (\ref{eq: reduced SUSY}), which will satisfy $\delta_\Sigma S_\Sigma=0$. For simplicity, we present a consistent truncation of the set of moduli $\{m^A\}$ to include only the spherically symmetric and regular modes of $\hat{X}^I$, $\hat{A}_0$; this amounts to having chosen spherically symmetric (but time-dependant) Dirichlet boundary conditions for $X^I$, $A_0$. We note, however, that due to the opposite sign gradient terms for $X^I$ and $A_0$ in $S$, we generically expect additional higher modes whose energies cancel, but which may nonetheless couple to fermions in the reduced theory. Leaving such analysis to future work, we write the spherically symmetric solutions as
\begin{align}
  \hat{X}^I = \frac{z^2}{z^2+\rho^2} v^{I,\alpha} g \sigma^\alpha g^{-1}\ ,\qquad \hat{A}_0 = \frac{z^2}{z^2+\rho^2} w^\alpha g \sigma^\alpha g^{-1}\ ,
  \label{eq: sph symm modes}
\end{align}
for some $v^{I,\alpha}(t), w^\alpha(t)$, as first considered in \cite{Lambert:2011gb}. It is easily checked that these solutions do indeed contribute finite energy for general $v^{I,\alpha}(t), w^\alpha(t)$. These choices admit the solution
\begin{align}
  \hat{\chi}= -\frac{\left( z^2+\rho^2 \right)^2}{16\rho^2} \left( \Gamma_i\, [ D_i \hat{A}_0 , \Psi_- ] + \Gamma_i \Gamma^I [ D_i \hat{X}^I, \Psi_-] \right)\ ,
\end{align}
for (\ref{eq: chi hat def}). Then, for these configurations (\ref{eq: sph symm modes}), we find
\begin{align}
  S_\Sigma = \frac{2\pi^2}{g^2} \int dt\,\, \bigg(& \dot{y}_i \dot{y}_i + 2 \dot{\rho}^2 + 2\rho^2 (\dot{u}^\alpha - w^\alpha)(\dot{u}^\alpha - w^\alpha) - 2\rho^2 v^{I,\alpha} v^{I,\alpha} \nonumber\\
  &-16 i \bar{\alpha} \Gamma_+ \dot{\alpha} - 32\rho^2 i \bar{\beta} \Gamma_+ \dot{\beta} \nonumber\\
  &- 8 \rho^2 i (\dot{u}^\alpha - w^\alpha)\eta^\alpha_{ij} \bar{\beta} \Gamma_+ \Gamma_{ij} \beta - 8 \rho^2 i v^{I,\alpha} \eta^\alpha_{ij} \bar{\beta} \Gamma_+\Gamma^I \Gamma_{ij} \beta \bigg)\ ,
  \label{eq: QM action}
\end{align}
while we find supersymmetries $\delta_\Sigma$ given by
{\allowdisplaybreaks
\begin{align}
    \delta_\Sigma  y_i &= -4i \left( \bareps_- \Gamma_+ \Gamma_i \alpha \right) \ ,\nonumber\\[0.3em]
  \delta_\Sigma  \rho &= -4i\rho \left( \bareps_- \Gamma_+ \beta \right) \ ,\nonumber\\[0.3em]
  \delta_\Sigma  u^\alpha &= i \eta^\alpha_{ij} \left( \bareps_- \Gamma_+ \Gamma_{ij} \beta \right) \ ,\nonumber\\[0.3em]
  \delta_\Sigma \alpha &= -\frac{1}{4} \dot{y}_i \Gamma_i \epsilon_- + \frac{1}{4} y_i \Gamma_i \Gamma_- \zeta_+ - \frac{1}{4} \Gamma_- \xi_+ \ ,\nonumber\\[0.3em]
  \delta_\Sigma \beta &= -\frac{\dot{\rho}}{4\rho}\epsilon_- - \frac{1}{16} \eta^\alpha_{ij} (\dot{u}^\alpha - w^\alpha) \Gamma_{ij} \epsilon_- + \frac{1}{16} \eta^\alpha_{ij} v^{I,\alpha} \Gamma^I \Gamma_{ij} \epsilon_- + \frac{1}{4}\Gamma_- \zeta_+ \nonumber\\*[0.3em]
  & \qquad - 8i \left( \bareps_- \Gamma_+ \beta \right) \beta + i \left( \bareps_- \Gamma_+ \Gamma_{ij} \beta \right) \Gamma_{ij}\beta \ ,\nonumber\\[1em]
  \delta_\Sigma  v^{I,\alpha} &= i\eta^\alpha_{ij} \left( \bareps_- \Gamma_+ \Gamma^I \Gamma_{ij} \dot{\beta} + \frac{2\dot{\rho}}{\rho} \bareps_- \Gamma_+ \Gamma^I \Gamma_{ij} \beta -2\bar{\zeta} \Gamma^I \Gamma_{ij} \beta  \right)\nonumber\\*[0.3em]
  &\qquad +i \varepsilon^{\alpha\beta\gamma}\eta^\beta_{ij} v^{J,\gamma} \left(\bareps_- \Gamma_+\Gamma^{IJ}\Gamma_{ij}\beta\right) + i \varepsilon^{\alpha\beta\gamma} \eta^\beta_{ij} \left( \dot{u}^\gamma - w^\gamma \right) \left(\bareps_- \Gamma_+\Gamma^{I}\Gamma_{ij}\beta\right) \nonumber\\*
  &\qquad + 8\eta^\alpha_{ij} \Big( 2 \left(\bar{\beta}\Gamma_+\Gamma_{ij}\beta\right)\left(\bareps_-\Gamma_+\Gamma^I\beta\right) - \left(\bar{\beta}\Gamma_+\Gamma_{ik}\beta\right)\left(\bareps_-\Gamma_+\Gamma^I\Gamma_{jk}\beta\right)  \Big) \ ,\nonumber\\
    \delta_\Sigma  w^\alpha &= -i\varepsilon^{\alpha\beta\gamma} \eta^\beta_{ij} v^{I,\gamma} \left(\bareps_- \Gamma_+\Gamma^I\Gamma_{ij}\beta\right)\nonumber\\*
    &\qquad - 8\eta^\alpha_{ij} \Big( 2\left(\bar{\beta}\Gamma_+\Gamma_{ik}\beta\right)\left(\bareps_-\Gamma_+\Gamma_{jk}\beta\right) - 2\left(\bar{\beta}\Gamma_+\Gamma^I\Gamma_{ij}\beta\right)\left(\bareps_-\Gamma_+\Gamma^I\beta\right) \nonumber\\*
    &\qquad\qquad\qquad   - \left(\bar{\beta}\Gamma_+\Gamma^I\Gamma_{ik}\beta\right)\left(\bareps_-\Gamma_+\Gamma^I\Gamma_{jk}\beta\right) \Big)\ ,
\end{align}
}%
with
\begin{align}
  \epsilon_- = \xi_- + t \Gamma_- \zeta_+\ .
\end{align}
Note, in verifying $\delta_\Sigma S_\Sigma=0$, it is helpful to use the Fierz relations
\begin{align}
  \BetaijBeta{i}{j}\BetaijBeta{i}{j}-\BetaIijBeta{I}{i}{j}\BetaIijBeta{I}{i}{j}&=0 \ ,\nonumber\\
  \BetaijBeta{i}{j}\BetaijBeta{i}{k}\EpsijBeta{j}{k}&=0\ .
\end{align}
We recognise $S_\Sigma$ as an extension of the standard $\mathcal{N}=(4,4)$ $\sigma$-model, with a flat metric on the target space $\mathbb{R}^4\times \mathbb{R}^4$ (here, $u^\alpha$ are the left-invariant $SU(2)$ forms of the unit $S^3$, which combined with radius $\rho$ gives a chart on $\mathbb{R}^4$). We note, however, that $g$ is indistinguishable from $-g$, and thus the actual target space is found by identifying $g\cong -g$ to find $\mathbb{R}^4 \times \left( \mathbb{R}^4/\mathbb{Z}_2 \right)$ which is indeed the hyper-K\"{a}hler moduli space of a single $SU(2)$ instanton. The model is extended by six $\frak{su}(2)$-valued time-dependant parameters $v^{I,\alpha}$, $w^\alpha$. We see that the full 8 rigid supersymmetries $\xi_-$ and 8 superconformal symmetries $\zeta_+$ now pair the bosonic and fermionic coordinates on $\Sigma$, while the rigid supersymmetry $\xi_+$ lives on as a shift symmetry for its Goldstino mode $\alpha$.\\

Naturally, the Lifshitz scaling symmetry (\ref{eq: scaling weights}) of the initial theory descends to the quantum mechanics, where it is more naturally viewed simply as a standard dilatation. In particular, in the notation of (\ref{eq: scaling weights}), we have
\begin{align}
  \left[ t \right] & = \left( -1 , -1 \right)\ , & \left[ \alpha \right] & = \left( -\tfrac{1}{2} , 0 \right) \ ,\nonumber\\
  \left[ y_i \right]  & = \left( -1 , -\tfrac{1}{2} \right)\ , & \left[ \beta \right] & = \left( \tfrac{1}{2} , \tfrac{1}{2} \right) \ ,\nonumber\\
  \left[ \rho \right] & = \left( -1 , -\tfrac{1}{2} \right) \ ,& \left[ \xi_- \right] & = \left( -\tfrac{1}{2} ,-\tfrac{1}{2} \right) \ ,\nonumber\\
  \left[ u^\alpha \right] & = \left( 0 , 0 \right) \ ,& \left[ \xi_+ \right] & = \left( -\tfrac{1}{2} , 0 \right) \ ,\nonumber\\
  \left[ v^{I,\alpha} \right]& = \left( 1 , 1 \right) \ , & \left[ \zeta_+ \right] & = \left( \tfrac{1}{2} , \tfrac{1}{2} \right)\ , \nonumber\\
  \left[ w^\alpha \right]& = \left( 1 , 1 \right)\ , & \left[ 1/g^2 \right]& = \left( 1 , 0 \right)  \ .
  \label{eq: QM scaling weights}
\end{align}
We also note that there is an additional class of symmetries acting on the parameters $v^{I,\alpha},w^\alpha$ corresponding to different choices for $\phi^I_{(X)}$ and $\phi_{(A)}$. These symmetries are trivially local, as $v^{I,\alpha},w^\alpha$ appear only algebraically in $S$. For instance, we have $\delta_\sigma S=0$ for
\begin{align}
  \delta_a v^{I,\alpha} &= \left( \bar{\sigma}_-\Gamma_+ \Gamma^{I} \Gamma_{ij} \beta \right)\Big( i\varepsilon^{\alpha\beta\gamma} \eta^\beta_{ij} \left( \dot{u}^\gamma - w^\gamma \right) - 8 \eta^\alpha_{ik} \BetaijBeta{j}{k} \Big)\ ,\nonumber\\
    \delta_a w^\alpha &= \left( \bar{\sigma}_-\Gamma_+ \Gamma^{I} \Gamma_{ij} \beta \right)\Big( i\varepsilon^{\alpha\beta\gamma} \eta^\beta_{ij} v^{I,\gamma} + 8 \eta^\alpha_{ik} \BetaIijBeta{I}{j}{k} \Big)\ ,
\end{align}
for spinor $\sigma_-(t)$ with $\Gamma_{012345}\sigma_-=-\Gamma_{05}\sigma_-=\sigma_-$. Such symmetries can be better understood by considering the transformations
\begin{align}
  \delta_\lambda v^{I,\alpha} &= \frac{v^{I,\alpha} - 2i\eta^\alpha_{ij} \left( \bar{\beta} \Gamma_+ \Gamma^I \Gamma_{ij} \beta \right)}{v^{J,b}v^{J,b} + 16 \left( \bar{\beta} \Gamma_+\Gamma_{kl} \beta \right)\left( \bar{\beta} \Gamma_+ \Gamma_{kl} \beta \right)} \lambda_v(t) \ ,\nonumber\\
  \delta_\lambda w^\alpha &= \frac{\left( \dot{u}^\alpha - w^\alpha \right) + 2i\eta^\alpha_{ij} \left( \bar{\beta} \Gamma_+ \Gamma_{ij} \beta \right)}{\left( \dot{u}^b - w^b \right)\left( \dot{u}^b - w^b \right) + 16 \left( \bar{\beta} \Gamma_+ \Gamma_{kl} \beta \right)\left( \bar{\beta} \Gamma_+  \Gamma_{kl} \beta \right)} \lambda_w(t)\ ,
\end{align}
where these expressions are understood as the corresponding (necessarily terminating) power series in $\beta$. Then, we have
\begin{align}
  \delta_\lambda S_\Sigma = \frac{2\pi^2}{g^2}\int dt\,\, \Big[ -4\rho^2 \left( \lambda_v + \lambda_w \right)  \Big]\ ,
\end{align}
and so in particular, $\delta_\lambda S_\Sigma=0$ for any $\lambda_w=-\lambda_v$. 


\chapter*{Discussion and further directions}
\addcontentsline{toc}{chapter}{Discussion and further directions}
\lhead{\textsl{\nouppercase{Part I, Discussion and further directions}}}


Let us summarise our findings in this Part of the thesis. In Chapter \ref{chap: construction from Lorentzian models} we described a limiting technique by which one can find non-Lorentzian supersymmetric field theories with Lifshitz scaling symmetry at particular fixed points of their Lorentzian cousins. We explored a few gauge theory examples relevant to M-theory, in particular recovering the theory on a stack of null-compactified M5-branes as well as their U-dual. In doing so, we elegantly recovered the DLCQ proposal for $(2,0)$ theory, seeing explicitly how dynamics is localised to instanton moduli space at the fixed point.

We then turned our attention to the explicit reduction of such fixed point theories to superconformal quantum mechanics on the moduli space of solitons in Chapter \ref{chap: reduction to soliton quantum mechanics}. We first got our bearings by considering a quite simple minimal kink model in $(1+1)$ dimensions. Interestingly, we demonstrated that the equivalence of the $\eta\to 0$ limit of the initial theory and the proposed fixed point theory persists at the quantum level, with 1-loop corrections to the path integral matching on either side. Finally, we returned to the five-dimensional Yang-Mills-like model for null-compactified M5-branes, and showed in principle (and explicitly for the single, $SU(2)$ anti-instanton) how the reduction to superconformal quantum mechanics works. \\


Let us finally discuss a few possible extensions and further directions. It would be interesting to consider non-Lorentzian RG flows and the subsequent soliton $\sigma$-models for other Yang-Mills solitons such as vortices and monopoles, especially when the parent theory includes coupling to matter. Here, one would expect to recover known results on the effective actions for these BPS solitons \cite{Gauntlett:1992yj,Gauntlett:1993sh,Cederwall:1995bc}, generically augmented by a set of couplings corresponding to matter fields of the parent theory.

It would also be interesting to generalise our analysis of the reduced theory in the SYM case, firstly to include all zero modes for $X^I$ and $A_0$, and then to use the ADHM formalism to investigate higher instanton number. Such analysis would in particular determine how the various modes of $X^I$ and $A_0$ appear in the reduced theory, and could thus shed light on how different M5-brane configurations are recovered in the $\sigma$-model. In doing such analysis, one may hope to recover results analogous to the D4-brane supertube \cite{Kim:2003gj} in the context of M5-branes. 

The quantum mechanical model (\ref{eq: QM action}), which constitutes an extension to the standard $\mathcal{N}=(4,4)$ $\sigma$-model with hyper-K\"{a}hler target space, bears a resemblance to models obtained by considering the low energy limit of $\mathcal{N}=4$ SYM in four dimensions, where one of the six scalar vevs is taken to be much larger than the rest\cite{Weinberg:2006rq}. In particular, such a model also possesses an adjoint valued $SO(5)$ vector, akin to our $v^I$. It would be interesting to make closer contact with this construction, especially as it may help in constraining the form of the aforementioned generalisations.

Finally, it would be of interest to revisit the non-Lorentzian theories found in Section (\ref{sec: M2s}) as fixed points of the $(1+2)$-dimensional BLG and ABJM/ABJ Chern-Simons-matter theories. The reduced theories would then be a $\sigma$-models on Hitchin moduli space \cite{Hitchin:1986vp}, again extended to include couplings to additional fields in the parent theory.


\authoredby{II}
\partfont{\color{IIcolor}}
\part{Non-Lorentzian descriptions of superconformal field theories}


\chapterfont{\color{IIcolor}}
\chapter*{Introduction}
\addcontentsline{toc}{chapter}{Introduction}
\lhead{\textsl{\nouppercase{Part II, Introduction}}}


One of the most interesting predictions to arise from string and M-theory is the existence of a rich spectrum of superconformal gauge field theories in dimensions greater than four. While the possibility of such theories first arose in the classification of Nahm \cite{Nahm:1977tg}, it was only much later that the importance of such models in the description of strongly coupled branes in string and M-theory was first appreciated \cite{Witten:1995zh,Strominger:1995ac,Seiberg:1996qx}. Key to a complete formulation of M-theory is an understanding of its fundamental objects, the M2- and M5-branes, whose low-energy dynamics are governed by the three-dimensional $\mathcal{N}=8$ and six-dimensional $(2,0)$ superconformal field theories, respectively.

While our handle on models of multiple M2-branes has progressed significantly through the use of Chern-Simons-matter theories \cite{Bagger:2006sk,Gustavsson:2007vu,Aharony:2008ug}, to date there is no satisfactory construction of the $(2,0)$ theory for M5-branes. We have already explored a number of proposals, but each is in its own way caveated, for instance in the requirement of a particular (often semi-compact) background geometry, with a non-compact picture recovered only at strong coupling.

The absense of such a complete description of the $(2,0)$ theory makes the computation of anything dynamical ({\it i.e.}\ unprotected) broadly inaccessible, at least from first principles. This challenge has in turn focussed attention on more indirect methods, with a great deal of progress made in computing correlations functions using a combination of conformal bootstrap and holographic methods. In particular 3-point correlators were first computed in \cite{Bastianelli:1999ab,Bastianelli:1999vm,Eden:2001wg} and 4-point correlators were computed in the supergravity approximation in \cite{Arutyunov:2002ff,Heslop:2004du,Beem:2015aoa,Rastelli:2017ymc}. The study of 4-point correlators beyond the supergravity approximation was initiated in \cite{Heslop:2017sco} and further developed in \cite{Chester:2018dga}, which fixed coefficients of higher-derivative corrections using a powerful chiral algebra conjecture formulated in \cite{Beem:2014kka}. Higher derivative corrections to eleven-dimensional supergravity were also studied from various other points of view in \cite{Chester:2018lbz,Chester:2018aca,Alday:2020tgi}.

The broad motivation for Part II of this thesis is to propose a Lagrangian description for the non-Abelian $(2,0)$ theory, that aims to add a more direct strategy to this toolbox of approaches towards explicit computation in the theory. The model and its analysis will evoke a number of familiar features of existing proposals for the M5-brane, but crucial differences suggest it offers a unique opportunity to better understand the $(2,0)$ on non-compact geometries.

What's more, in investigating the symmetries of the model and its implications for correlation functions, we will build the foundations of a far broader study of field theories in five dimensions with an $SU(1,3)$ spacetime symmetry, with or without supersymmetry. This work generalises the fundamental construction of conformal field theory to models with an inhomogeneous (i.e. Lifshitz) scaling symmetry, and thus forms the first step towards an expansion of modern conformal field theory techniques to such theories.


\subsection*{Outline of results}


Let us now outline the results of this Part of the thesis, in the order they are presented. First, in Chapter \ref{chap: SU(1,3) theories} we consider a class of field theories in five dimensions defined by their exotic $SU(1,3)$ spacetime symmetry group. In particular, the group is made up of the following transformations on $\mathbb{R}^5$, which we take to have coordinates $(x^-, x^i)$.
\begin{itemize}
  \item A non-Abelian subgroup of five translations, which acts transitively on $\mathbb{R}^5$\vspace{-0.8em}
  \item A $\frak{u}(1)\oplus \frak{su}(2)$ subgroup of four rotations in the $x^i$ directions\vspace{-0.8em}
  \item A Lifshitz scaling symmetry, under which $x^-$ scales twice as quickly as the $x^i$\vspace{-0.8em}
  \item Five `special' transformations, which in the representation theory play a role analogous to the special conformal transformations of the conformal group
\end{itemize}
The structure of the algebra as well as its representation in terms of vector fields is written in terms of a constant anti-self-dual matrix $\Omega_{ij}=-\Omega_{ji} = -\frac{1}{2}\epsilon_{ijkl}\Omega_{kl}$ that satisfies $\Omega_{ik}\Omega_{jk}=R^{-2}\delta_{ij}$ for a constant $R$.

We then build our analysis of these models from scratch, in analogy with the standard treatment of conformal field theories. We first build generic irreducible representations, generalising slightly to allow for operators with charge $p_+$ under an additional internal symmetry $P_+$. This additional generator together with the spacetime symmetries forms a (trivial) central extension to $\frak{su}(1,3)$ which we call $\frak{h}$, which after a suitable change of basis is identified simply as $\frak{h}=\frak{su}(1,3)\oplus \frak{u}(1)$. We then derive the corresponding Ward-Takahashi identities satisfied by generic correlation functions, which we solve explicitly for the case of scalar\footnote{A scalar operator is later defined within the broader representation theory of $\frak{h}$} operators. We find that the constraining power of the symmetry group lies somewhere between that of the Poincar\'{e} and conformal groups: the two-point function is fixed up to an overall normalisation, but the three-point function is only determined up to a function of a single invariant phase---itself a function of insertion points. More generally, the $N$-point function is determined up to a function of $N(N-3)/2$ cross-ratios and $(N-1)(N-2)/2$ phases.

It is at this point that we first encounter a curious complex structure that emerges most fundamentally from the $SU(1,3)$ symmetry. The correlation functions are written most naturally in terms of not the real coordinates $(x^-_1, x^i_1)$, $(x^-_2, x^i_2)$ of two operator insertions, but rather a particular complex distance $z_{12}$ that vanishes only when $(x^-_1, x^i_1)=(x^-_2, x^i_2)$. Indeed, in terms of these variables the correlation functions factorise naturally into holomorphic and anti-holomorphic parts, as is reminiscent of two-dimensional conformal field theory.\\


With the theory of models with $SU(1,3)$ spacetime symmetries safely in hand, in Chapter \ref{chap: the view from 6d CFT} we demonstrate precisely how such models are relevant in the study of any six-dimensional conformal field theory. We first detail a particular conformal compactification of six-dimensional Minkowski space; simply, a choice of coordinates $(x^+, x^-, x^i)$ such that $x^+$ runs over the finite inverval $x^+\in(-\pi R, \pi R)$ for constant $R$. Further, translations along the $x^+$ direction are conformal Killing vectors, with a conformal factor that can be made to vanish after a suitable Weyl rescaling, hence resulting in a simple isometry in the geometry. Thus, we are able to reduce an operator $\sOp$ in any six-dimensional conformal field theory into Fourier modes $\fOp_n$ on the $x^+$ interval, for $n\in\mathbb{Z}$. Each mode $\fOp_n$ is then in a representation of the maximal subalgebra of $\frak{so}(2,6)$ that commutes with $x^+$ translations; this is precisely $\frak{h}$, with the central element $P_+$ identified as translations along $x^+$. Thus, the Kaluza-Klein theory of these Fourier modes is precisely an $SU(1,3)$ theory, where $\fOp_n$ has charge $p_+=n/R$ under $P_+$.

Next, we consider the dimensional reduction of six-dimensional correlators to correlators of their modes. The point here is that six-dimensional correlators are constrained to solve the Ward-Takahashi identities of $\frak{so}(2,6)$, which are more constraining than those of $\frak{h}$. For instance, the 3-point function is completely determined. Thus, by dimensionally reducing six-dimensional correlators, we can further constrain the form of correlators in the five-dimensional $SU(1,3)$ theory. Conversely, we can see these results as necessary conditions on the correlators of a generic $SU(1,3)$ theory to admit a six-dimensional interpretation. 

For instance, consider a pair of operators $\sOp^{(1)},\sOp^{(2)}$ in six dimensions, whose 2-point function $\left\langle \sOp^{(1)} \sOp^{(2)} \right\rangle$ is determined up to an overall normalisation $C$. Conversely, the 2-point functions of their modes $\big\langle \fOp_n^{(1)} \fOp_{-n}^{(2)} \big\rangle$ are fixed by $\frak{h}$ only up to a set of normalisations $C_{n}$. By dimensional reduction, we are able to determine explicitly the required form for $C_n(C)$ such that the six-dimensional interpretation holds. Similarly, at 3-points we are able to fully fix the remaining freedom in our solution to the $\frak{h}$ Ward-Takahashi identities, up to the structure constants of the six-dimensional conformal field theory. \\


In Chapter \ref{chap: DLCQ}, we consider a particular, degenerate limit of our construction. We restrict the space of operators of the six-dimensional theory to only those that have period $2\pi R_+:=2\pi R/k$ on the $x^+$ interval for some $k\in\{1,2,\dots \}$, in effect considering a $\mathbb{Z}_k$ orbifold. This amounts to allowing only the modes $\fOp_{kn}$ in the Fourier decomposition.

In the limit that $R\to\infty$, the coordinates $(x^+,x^-,x^i)$ become regular lightcone coordinates on six-dimensional Minkowski space. Thus, by considering the limit $k,R\to\infty$ with $R_+=R/k$ fixed, we arrive in the limit at a six-dimensional CFT compactified along a null direction, with period $2\pi R_+$. 

This is the familiar geometric setup of Discrete Lightcone Quantisation (DLCQ), and as such we refer to this limit as the \textit{DLCQ limit}. The resulting Ward-Takahashi identities correspond to the $z=2$ Schr\"odinger algebra, and were first studied in \cite{Henkel:1993sg} in the context of condensed matter physics where they were solved at 2- and 3-points. The same structure became relevant to M-theory in the calculation of the 2-point function of protected operators of the $(2,0)$ theory through the corresponding DLCQ proposal \cite{Aharony:1997an,Aharony:1997th}. Here, we generalise these results to provide the general $N$-point solution to the Ward-Takahashi identities of the Schr\"odinger algebra.

With general solutions to the five-dimensional DLCQ Ward-Takahashi identities in hand, we take a step further and investigate the fate of the dimensionally reduced correlators of Chapter \ref{chap: the view from 6d CFT} as we approach the DLCQ limit. We demonstrate that dangerous divergences and associated ambiguities that arise upon a naive direct null reduction of correlation functions are regulated at finite $k,R$. Thus, by considering the relevant asymptotics, we provide a systematic way to extract the functional form of the DLCQ correlators. In this sense, the $\Omega$-deformed Lagrangian for conformally compactified M5-branes appears to be more fundamental than the original DLCQ proposal; indeed, the latter is shown to be an orbifold of the former. Further, by approaching the DLCQ limit from finite $R$ in this way,  we appear to provide clarity on the infamous zero mode problem \cite{Nakanishi:1976vf,Fitzpatrick:2018ttk}, showing for instance that the 2-point function of zero modes vanishes. \\

What we've done so far forms a detailed framework, through which one may in principle compute observables in a six-dimensional conformal field theory by instead performing calculations in a five-dimensional $SU(1,3)$ model, which are then Fourier resummed to six-dimensional results. The remaining Chapters of the thesis focus on a proposed realisation of this approach: an explicit five-dimensional $SU(1,3)$ Lagrangian, which we conjecture provides a description of the $(2,0)$ superconformal field theory on six-dimensional non-compact Minkowski space.\\

In Chapter \ref{chap: An explicit model and its symmetries}, we first present this model, which takes the form of a five-dimensional $SU(N_c)$ gauge theory coupled to five adjoint scalars and a single real 32-component adjoint spinor, with parameters $R,k$. The theory has a high degree of supersymmetry, with 16 real rigid supersymmetries, and a further 8 real conformal supersymmetries. The theory additionally has a Lagrange multiplier field, the effect of which is to localise the theory's dynamics to an $\Omega$-deformation of instanton moduli space. We review briefly the original holographic construction of the theory \cite{Lambert:2019jwi}, which hinges crucially on an application of the limiting technique defined and explored in Part I of this thesis.

Next, we consider the spacetime symmetries of the theory. This is most simply done by reformulating the theory as a particular gauge fixing of a six-dimensional diffeomorphism-invariant proxy theory, as explained in detail in Appendix \ref{app: proxy theory}. As a result, we find transformation rules for every field corresponding to spacetime transformations generated by each element of $\frak{su}(1,3)$. We find that if the gauge field is regular throughout $\mathbb{R}^5$, then the action has an $\frak{su}(1,3)$ spacetime symmetry.

Our claim, however, if that this theory encapsulates not just the zero modes, but all Kaluza-Klein modes of the conformal compactification of the $(2,0)$ theory. As such, we should be able to define operators in the theory for which the $\frak{su}(1,3)$ algebra closes only up to a non-zero central charge under the additional central element $P_+$, and thus the full symmetry algebra of the theory should be $\frak{h}$. The remainder of the Chapter shows that this is indeed the case.

The crucial step is to extend the configuration space of the theory by supposing that $F$ is regular only on $\mathbb{R}^5\setminus \{x_a\}$ for a number of isolated points $x_a$. Gauge bundles over this punctured space are then characterised by the integral of $\frac{1}{8\pi^2}\text{tr}\left(F\wedge F\right)$ on small $S^4$'s surrounding each of the points $x_a$, given by some integers $n_a$. Thus, the data of such a bundle is contained in the pairs $\{(x_a, n_a)\}$, which we call \textit{instanton insertions}. We consider the formulation of these instanton insertions in the path integral in terms of disorder operators known as \textit{instanton operators}, making contact with previous work \cite{Lambert:2014jna,Tachikawa:2015mha}. 

We show that when such sectors are included in the configuration space, the action is no longer invariant under $\frak{su}(1,3)$. Instead, under the `special' transformations, the action picks up an anomalous variation that is \textit{local} to the points $x_a$. Despite the non-invariance of the action, the theory is shown to still satisfy a set of deformed Ward-Takahashi identities. These are precisely those of a theory with spacetime symmetry algebra $\frak{h}$, with an operator inserted at $x_a$ having $P_+$ charge $p_+=kn_a/R$. Thus, for $k\in\mathbb{Z}$ we can identify such operators as the Fourier modes $\fOp_{kn_a}$ of some six-dimensional operator on the $x^+\in(-\pi R, \pi R)$ interval orbifolded by $\mathbb{Z}_k$. In particular, a choice of $k=\pm 1$ corresponds to non-compact six-dimensional Minkowski space. Hence, our theory passes a crucial test in its ability to encode six-dimensional physics.\\

At this point, there is some tension. On one hand, we have defined a five-dimensional gauge theory whose dynamics are constrained to an $\Omega$-deformation to instanton moduli space. In more detail, the constraint imposed by the Lagrange multiplier is $\mathcal{F}^+_{ij}:=\frac{1}{2}\left(\mathcal{F}_{ij}+\frac{1}{2}\varepsilon_{ijkl}\mathcal{F}_{kl}\right)=0$, where $\mathcal{F}=\frac{1}{2}\mathcal{F}_{ij} dx^i\wedge dx^j$ is an $\Omega$-deformation of the usual field strength given by $\mathcal{F}_{ij} = F_{ij}-\frac{1}{2}\Omega_{ik}x^k F_{-j}+\frac{1}{2}\Omega_{jk}x^k F_{-i}$. While when $R\to\infty$ we recover the usual anti-instanton equation in the $x^i$ directions, at finite $R$ the equation and resulting moduli space is suitably deformed.

On the other hand, we have shown that a six-dimensional interpretation of the theory necessarily requires an extended configuration space, allowing for non-zero instanton numbers $n_a$ on arbitrarily small surfaces surrounding a number of isolated points $x_a$.

It is a priori not clear that these two ideas are compatible: can we find configurations with non-zero $n_a$ that also lie on the constraint surface $\mathcal{F}^+=0$? The answer to this question is a resounding \textit{yes}, as is shown in Chapter \ref{chap: worldlines}. 

We explicitly construct gauge field configurations $A$ solving $\mathcal{F}^+=0$, with a number of interesting properties. Each solution has a number of special points $x_a$ joined by curves which must move only forward in the $x^-$ direction, but whose geometry is otherwise unconstrained. By considering the integral of $\frac{1}{8\pi^2}\text{tr}\left(F\wedge F\right)$ over general four-dimensional submanifolds, we demonstrate that these curves are precisely the worldlines of single anti-instantons, which may be created or annihilated at the points $x_a$. These points are then identified precisely as instanton insertions $\{(x_a, n_a)\}$, with the integer charge $n_a$ determined as the difference between the number of anti-instantons annihilated and created at the point. As such, we are able to write down configurations on the constraint surface with non-trivial instanton insertions $\{(x_a,n_a)\}$, and thus explicitly realise examples of the configurations shown in Chapter \ref{chap: An explicit model and its symmetries} to enhance the symmetries of the theory to see a sixth dimension.

We are finally able to explore a number of additional properties of the subspace of the constraint surface that has been explicitly constructed. In particular, we consider the DLCQ limit of such configurations, which beautifully recover the well-known 't Hooft solutions (\ref{eq: intro t Hooft solution}) with moduli able to depend on $x^-$. We also consider the dynamics on the constraint surface, in particular showing in principle how the full field content of the theory is reduced to a theory of interacting instanton-particle worldlines. 


\subsection*{Notation for transformations}


The focus of this Part of the thesis is spacetime symmetry, and as such we will often be talking about the transformations of coordinates and fields with respect to some Lie group $\mathcal{G}$, or infinitesimally its Lie algebra $\frak{g}$. This Lie group will in turn usually be the (conformal) isometry group of the geometry, and thus admit a representation in terms of (conformal) Killing vector fields under commutation. But before getting into more specific details, let us set up some generic notation.

First, for every $G\in\frak{g}$, denote by $G_\partial$ the representative of $G$ in the vector field representation. We can first define the transformation of coordinates. Given some point with coordinates $x^\alpha$ in our spacetime, and group element $g=e^{\epsilon G}$, we can define by $(xg)^\alpha$ the coordinates of the point $\epsilon$ distance along the integral curve of $G_\partial$ from $x^\alpha$. Equivalently, the functions $(xg)^\alpha(\epsilon)$ are uniquely determined by $\partial_\epsilon (xg)^\alpha = G^\alpha_\partial(xg)$ for all $\epsilon$, and $(xg)^\alpha|_{\epsilon=0}=x^\alpha$. Here, the vector field $G$ takes the form $G_\partial = G_\partial^\alpha(x) \partial_\alpha$ at each spacetime point $x$. We will often suppress the $\alpha$ index, and simply write $x$ and $xg$. Then, this defines a right group action of $\mathcal{G}$ on coordinates, i.e. $x(g_1 g_2)=(xg_1)g_2$.

With this notation in hand, we can next consider the transformation of some field $\Phi(x)$. We formulate a spacetime symmetry as an \textit{active} transformation throughout, so that under a transformation $g\in\mathcal{G}$, we have
\begin{align*}
  x		&\quad\longrightarrow\quad x		\ ,\nn\\
  \Phi(x) &\quad\longrightarrow\quad \Phi'(x) = g \Phi(x) :=  \mathcal{R}_g(xg^{-1})\Phi(xg^{-1})\ ,
\end{align*}
where $\mathcal{R}_g$ is some (generically spacetime-dependent) matrix acting on any indices of $\Phi$, and satisfying $\mathcal{R}_{g_2}(xg_1)\mathcal{R}_{g_1}(x) =\mathcal{R}_{g_1 g_2}(x)$. Taking $g$ then to act only on fields, so that for instance $g(\partial_i \Phi(x))=\partial_i \left( g\Phi (x) \right)$, we have that $(g_1 g_2)\Phi(x) = g_1\left( g_2\Phi(x) \right)$. Thus, we do indeed have that $\Phi$ falls into a representation of $\mathcal{G}$.

We can finally consider the corresponding infinitesimal transformations. Let us consider the above transformation for $g=e^{\epsilon G}$. Then, to leading order in $\epsilon$ we can write $\left(xg^{-1}\right)^\alpha = x^\alpha - \epsilon G_\partial^\alpha$ and $\mathcal{R}_g(x) = 1-\epsilon\, r_G(x)$ for some matrix $r_G(x)$ satisfying $[r_{G_1},r_{G_2}]+\left(G_1\right)_\partial r_{G_2} - \left(G_1\right)_\partial r_{G_1}=r_{[G_1,G_2]}$ for each $G_1, G_2\in\frak{g}$. Hence, we have
\begin{align*}
  \Phi'(x) = g\Phi(x) = \Phi(x) + \epsilon \delta_G \Phi(x),\quad \text{with}\quad \delta_G\Phi(x)=-G_\partial \Phi(x) - r_G(x) \Phi(x)\ .
\end{align*}
The $\delta_G$ form a representation of $\frak{g}$, i.e. $[\delta_{G_1},\delta_{G_2}]= \delta_{[G_1, G_2]}$, where we take the $\delta_G$ to act only on fields, so for instance $\delta_G\left(x^\alpha \partial_\alpha \Phi(x)\right) = x^\alpha \partial_\alpha \left(\delta_G \Phi(x)\right)$. Note, when $\Phi$ is tensorial, we have simply $\delta_G\Phi = - \mathcal{L}_{G_\partial} \Phi(x)$, i.e. (minus) the Lie derivative with respect to the vector field $G_\partial$. 


\subsection*{Some more conventions}


In this Part of the thesis, we will be using a number of different coordinate systems, and flittering quite frequently between fields and operators in both six-dimensional and five-dimensional theories. To help keep track, we stick with the following conventions throughout.

Let us first set a convention for indices. We fix the use of the following,
\begin{center}
\begin{tabular}{l|ll}
  	Index type & Use & Range \\
  	\hline
  	$\mu,\nu,\dots$ 			& Various six-dimensional coordinates		 & $\{+,-,1,2,3,4\}$				\\
  	$i,j,\dots$ 				& Various four-dimensional coordinates		 &	$\{1,2,3,4\}$				\\
  	$I,J,\dots$ 				& $SO(5)$ R-symmetry &	$\{6,7,8,9,10\}$				\\
  	$\alpha,\beta,\dots$ 		& $\frak{su}(2)$ basis and its representations &	$\{1,2,3\}$			\\
\end{tabular}
\end{center}
We will further use the indices $A,B,\dots$ to enumerate the anti-instanton worldlines of Chapter \ref{chap: worldlines}. Finally, the indices $a,b,\dots$ will be used a little more freely depending on context, such as for enumerating the operators in a correlation function in Chapter \ref{chap: SU(1,3) theories}, or to list instanton insertions in Chapter \ref{chap: An explicit model and its symmetries} (we will see that there's a good reason for this dual use).\\

At the heart of the main results of this Part is a relationship between the operators and correlators of six-dimensional conformal field theories, and those of five-dimensional $SU(1,3)$ field theories. As a general rule, we will always use $\sOp$ to denote a six-dimensional field or operator, and $\fOp$ to denote a five-dimensional field or operator.


\chapter{$SU(1,3)$ field theories and their correlators}\label{chap: SU(1,3) theories}
\lhead{\textsl{\nouppercase{Part II, \leftmark}}}


Our first aim is to build from the ground up the framework of theories in five dimensions with an exotic $SU(1,3)$ spacetime symmetry. We will follow an approach familiar from the foundations of conformal field theory, building generic irreducible representations of the symmetry algebra, and then finding and solving the resulting Ward-Takahashi identities constraining correlation functions.  


\section{An exotic spacetime symmetry}\label{sec: an exotic spacetime symmetry}


The spacetime symmetry algebra of the five-dimensional theories we are considering is a (necessarily trivial) central extension of $\frak{su}(1,3)$, which we call $\frak{h}$. The generators of $\frak{su}(1,3)$ are $\{P_-,P_i,B, C^\alpha,T,M_{i+},K_+\}$ with $i=1,\dots,4$, $\alpha=1,2,3$. The algebra is extended by a single central element $P_+$. A subset of the commutation relations of the algebra is  
\begin{align}
[M_{i+},P_j] \ &= \  -\delta_{ij} P_+ - \tfrac{1}{2}\Omega_{ij} T - \tfrac{2}{R} \delta_{ij} B + \Omega_{ik}\eta^\alpha_{jk} C^\alpha    \ ,	&	[T,P_-] \ 	&= \ -2P_-					\ ,\nn\\[-0.1em]
  [T,K_+] \	 &= \ 2K_+ \ ,					&		[P_-,P_i] 				\ &= \ 0		  				\ ,\nn\\[-0.1em]
  [K_+,P_-] \ &= \ -2T	\ ,					&		[P_-,M_{i+}] 		\ &= \ P_i   			\ ,\nn\\[-0.1em]
[M_{i+},M_{j+}] \ &= \ -\tfrac{1}{2} \Omega_{ij} K_+ \ , &										 		[K_+,P_i] 				\ &= \ -2M_{i+} 		\ ,\nn\\[-0.1em]
  [T,P_i] 	\ &= \ -P_i \ ,		&		[K_+,M_{i+}] 			\ &= \  0  						\ ,\nn\\[-0.1em]
  [T,M_{i+}] \ &= \ M_{i+} \ ,	&		[P_i,P_j] 		\ &= \ -\Omega_{ij} P_-  						\ ,
\label{eq: extended su(1,3) algebra}
\end{align}
where $\Omega_{ij}$ is anti-symmetric, anti-self-dual\footnote{Note, we could more generally consider $\Omega_{ij}$ not necessarily anti-delf-dual. However, we our main focus is on reductions of six-dimensional theories, for which an anti-self-dual $\Omega_{ij}$ plays a special role} and satisfies $\Omega_{ij}\Omega_{jk} = -R^{-2}\delta_{ik}$.
Here $R$ is a constant with dimensions of length. 
The rotations $B,C^\alpha$ form an $\frak{u}(1)\oplus \frak{su}(2)$ subalgebra;
\begin{align}
  [B,C^\alpha]	\ = \ 0\ ,\qquad [C^\alpha, C^\beta] \ = \ -\varepsilon^{\alpha\beta\gamma}C^\gamma \ .
\end{align} 
The remaining brackets are neatly summarised by noting that the `scalar' generators $S=P_-,T, K_+$ are inert under the rotation subgroup, i.e.\ $[S,B]=[S,C^\alpha]=0$, while the `one-form' generators $W_i=P_i,M_{i+}$ transform as
\begin{align}
  	[W_i,B] 	\ &= \ -\tfrac{1}{2} R\,\Omega_{ij} W_j 	\ ,	\qquad [W_i,C^\alpha] \	= \ \tfrac{1}{2}\eta^\alpha_{ij} W_j \ .
  	\label{eq: commutators with rotations}
\end{align}
The fact that $\frak{h}$ is a central extension to $\frak{su}(1,3)$ is far from obvious. In particular, setting $P_+=0$ in the above algebra, we have precisely that the $\{P_-,P_i,B, C^\alpha,T,M_{i+},K_+\}$ form a basis for $\frak{su}(1,3)$. This is most intuitively seen by a corresponding holographic construction, in which these generators are associated with the Killing vectors of the non-compact projective space $\tilde{\mathbb{CP}}^3$ \cite{Lambert:2019jwi}.

Indeed, we have that $\frak{h} = \frak{u}(1)\oplus \frak{su}(1,3)$, where the $\frak{u}(1)$ factor is spanned by $P_+$, while the $\frak{su}(1,3)$ factor has basis $\{P_-,P_i,\tilde{B}, C^\alpha,T,M_{i+},K_+\}$ with $\tilde{B} = B+ \frac{R}{2}P_+$. However, it will be more convenient for geometric reasons to continue to use $B$ rather than $\tilde{B}$, and thus refrain from making this direct sum decomposition of $\frak{h}$ manifest.

As with any spacetime symmetry, this algebra then admits a representation in terms of vector fields under commutation. Writing $(x^-, x^i)$ for the coordinates on our five-dimensional space, and $(\partial_-,\partial_i)$ for their derivatives, the vector field representation is then  \begin{align}
	\left( P_+ \right)_\partial 	\ &= \ 0 \ , 	\nn\\
	\left( P_- \right)_\partial \ 	&= \ \partial_-  \ ,  \nn\\
	\left( P_i \right)_\partial \ 	&= \ \tfrac{1}{2}\Omega_{ij} x^j \partial_- + \partial_i	\ , \nn\\
	\left( B \right)_\partial 		\ &= \ -\tfrac{1}{2}R\,\Omega_{ij}x^i\partial_j 		\ , \nn\\
	\left( C^\alpha \right)_\partial 	\ &= \ \tfrac{1}{2}\eta^\alpha_{ij}x^i\partial_j 								\ , \nn\\
	\left( T \right)_\partial 		\ &= \ 2x^- \partial_- + x^i \partial_i					\ , \nn\\
	\left( M_{i+} \right)_\partial 	\ &= \ \left( \tfrac{1}{2}\Omega_{ij} x^- x^j - \tfrac{1}{8}R^{-2} x^j x^j x^i \right)\partial_- + x^- \partial_i  +\tfrac{1}{4}( 2\Omega_{ik}x^k x^j + 2\Omega_{jk}x^k x^i - \Omega_{ij}x^k x^k )\partial_j	\ , \nn\\
	\left( K_{+} \right)_\partial 	 \ &= \ ( 2 ( x^- )^2 - \tfrac{1}{8} R^{-2} ( x^i x^i )^2 )\partial_- +( \tfrac{1}{2} \Omega_{ij} x^j x^k x^k + 2 x^- x^i )\partial_i	\ .
	\label{eq: 5d algebra vector field rep}
\end{align}
In particular, since $\left( P_+ \right)_\partial=0$, this is just a representation of $\frak{su}(1,3)$. Note, $\{\left(B\right)_\partial, \left(C^\alpha\right)_\partial\}$ span the $\frak{u}(1)\oplus \frak{su}(2)\subset \frak{su}(2)\oplus \frak{su}(2)$ subalgebra of rotations in the four-dimensional plane that commute with $\Omega_{ij}$, so that $\Omega_{ij}x^j$ transforms in the same way as $x^i$.

Note, given some point $x=(x^-, x^i)$, to compactify some more cumbersome equations we will sometimes use the index-free shorthand $\vex$ for the components $x^i$, and so $|\vex|^2=x^i x^i$.\\

We can finally build representations of the algebra (\ref{eq: extended su(1,3) algebra}). Although the algebra is not a conventional conformal algebra, it shares many properties with one. We still have a five-dimensional subalgebra of translations generated by $\{P_-, P_i\}$, although it is not Abelian, and a Lifshitz scaling $T$ which plays the role of the usual dilatation. Further, we still have pairs of ladder operators that raise and lower an operator or state's eigenvalues under $T$, except unlike usual conformal algebras there are two different gradations. The pair $(P_i, M_{i+})$  of raise and lower $T$ by one unit, while $(P_-, K_+)$ raise and lower by two units.

We therefore proceed in analogy with the familiar construction of conformal algebra representations. We first consider representations of the subalgebra that stabilises the origin $\{x^-=0, x^i=0\}$, which is generated by $\{B,C^\alpha,T,M_{i+},K_+\}$. Then, an operator $\fOp(0)$ at the origin transforms as
\begin{align}
  	\delta_B\fOp(0) \ 	&= \ -r_\fOp[B]\,\fOp(0)	\ ,\nn\\
  	\delta_{C^\alpha}\fOp(0) 	\ &= \ -r_\fOp[C^\alpha]\,\fOp(0)\ ,
\end{align}
for some representations $r_\fOp[B]$ and $r_\fOp[C^\alpha]$ of $\frak{u}(1)$ and $\frak{su}(2)$, respectively.

Further supposing that $r_\fOp[C^\alpha]$ is irreducible, by Schur's lemma we must have
\begin{align}
	\delta_{T}\fOp(0)			\ &= \ -\Delta \fOp(0)	 \ ,	\nn\\
	\delta_{M_{i+}} \fOp(0)	\ &= \ 0		\ ,					\nn\\
	\delta_{K_+}\fOp(0)		\ &= \ 0		 \ ,					\nn\\
	\delta_{P_+}\fOp(0)		\ &= \ ip_+\fOp(0) \ ,
	\label{eq: 5d alg at origin}
\end{align}
for some $p_+,\Delta\in\mathbb{C}$. In analogy with conventional conformal field theory, we regard $\fOp(0)$ as a \textit{primary field} at the origin. Each such primary is classified by the data \linebreak $\{r_\fOp[B], r_\fOp[C^\alpha], \Delta, p_+\}$. We can then act with $\delta_{P_-}$ and $\delta_{P_i}$ on $\fOp(0)$ to build \textit{descendants}. In particular, the form of the algebra dictates that acting with $\delta_{P_-}$ raises $\Delta$ by $2$, while acting with $\delta_{P_i}$ raises $\Delta$ by $1$. 

The corresponding primary field $\fOp(x)$ at a generic point $(x^-,x^i)$ is defined by
\begin{align}
  \fOp(x) \ = \ \exp\left( -x^- \delta_{P_-} - x^i \delta_{P_i} \right) \fOp(0)  \ .
\end{align}
Thus, the action of $\delta_{P_-}$ and $\delta_{P_i}$ on $\fOp(x)$ is determined by requiring
\begin{align}
  \fOp(x+\epsilon) - \fOp(x) \ = \ \epsilon^- \partial_- \fOp(x) + \epsilon^i \partial_i\fOp(x) \ ,
\end{align}
to leading order in $\epsilon^-,\epsilon^i$. Making use of the algebraic relation
\begin{align}
  &\exp\left( \left( x^-+\epsilon^- \right)P_- + \left( x^i + \epsilon^i \right)P_i \right) \nn\\
  &\qquad= \exp\left( x^- P_- + x^i P_i \right)\exp\left( \epsilon^- P_- + \epsilon^j P_j \right)\exp\left( -\tfrac{1}{2}\Omega_{kl}\epsilon^k x^l P_- \right)	\nn\\
  &\qquad=\exp\left( \tfrac{1}{2}\Omega_{kl}\epsilon^k x^l P_- \right)\exp\left( \epsilon^- P_- + \epsilon^j P_j \right)\exp\left( x^- P_- + x^i P_i \right)\ ,
\end{align}
which can be shown from the form of the algebra $\frak{h}$, we find
\begin{align}
  	\delta_{P_-}\fOp(x)		\	 &= \ -\partial_- \fOp(x)	\ = \ -\left( P_- \right)_\partial \fOp(x)	\ , \nn\\
  	\delta_{P_i}\fOp(x)	\ &= \ -\left( \partial_i + \tfrac{1}{2}\Omega_{ij} x^j \partial_- \right)\fOp(x) 	\ = \ -\left( P_i \right)_\partial \fOp(x) \ ,
\end{align}
as expected. Using (\ref{eq: extended su(1,3) algebra}), we determine the action of the whole algebra on the primary field $\fOp(x)$ to be
\begin{align}
  	\delta_{P_+}\fOp(x)		\ &=	 \ ip_+ \fOp(x)		\ ,												\nn\\
  	\delta_{P_-}\fOp(x)		\ &=	 \ -\left( P_- \right)_\partial \fOp(x)	\ ,							\nn\\
  	\delta_{P_i}\fOp(x)		\ &= \ -\left( P_i \right)_\partial \fOp(x)	\ ,							\nn\\
  	\delta_{B}\fOp(x)		\ &=	 \ -\left( B \right)_\partial \fOp(x) - r_\fOp[B]   \fOp(x)\ ,	\nn\\
  	\delta_{C^\alpha}\fOp(x)		\ &=	 \ -\left( C^\alpha \right)_\partial \fOp(x) - r_\fOp[C^\alpha] \fOp(x)	\ ,\nn\\
  	\delta_{T}\fOp(x)		\ &= \	-\left( T \right)_\partial \fOp(x) -  \Delta \fOp(x)	\ ,\nn\\
  	\delta_{M_{i+}}\fOp(x)	\ &= \	-\left( M_{i+} \right)_\partial \fOp(x) - \left(\tfrac{1}{2}\Delta \Omega_{ij} x^j - ip_+ x^i	 +\tfrac{2}{R}x^i r_\fOp[B] - \Omega_{ik} \eta^\alpha_{jk} x^j r_\fOp[C^\alpha]\right) \fOp(x)			\ ,\nn\\
  	\delta_{K_+}\fOp(x)			\ &=	 \ -\left( K_+ \right)_\partial \fOp(x)  - \left(2\Delta\, x^- - ip_+ x^i x^i +\tfrac{2}{R}x^i x^i r_\fOp[B] - x^i x^j \Omega_{ik}\eta^\alpha_{jk} r_\fOp[C^\alpha]   \right) \fOp(x) \ .
  	\label{eq: 5d algebra action on fields}
\end{align}
Once again in analogy with conventional conformal field theory, we will often refer to a primary field $\fOp$ with $r_\fOp[B]=r_\fOp[C^\alpha]=0$ as a \textit{scalar} primary, which is hence entirely characterised by $\{\Delta, p_+\}$.


\section{Constraining correlation functions}




The general Ward-Takahashi identity associated with symmetry generator $G$ acting on an $N$-point correlation function is
\begin{align}
	0 \ =& \ \sum_{a=1}^N \langle \fOp^{(1)}(x^-_1,x^i_1) \cdots \delta_G\fOp^{(a)}(x^-_a,x^i_a) \cdots \fOp^{(N)}(x^-_N,x^i_N) \rangle \ ,
\end{align}
for some operators $\fOp^{(a)}(x^-_a,x^i_a)$ inserted at spacetime points $(x^-_a,x^i_a)$. Let us further suppose that the $\fOp^{(a)}(x^-_a,x^i_a)$ are scalar primaries with scaling dimensions $\Delta_a$ and $P_+$ eigenvalues\footnote{We have suppressed the $+$ subscript on $p_+$ here and in the remaining subsections to improve readability. Moreover, $a,b$ indices are never subject to the Einstein summation convention.} $p_a$.

Then, the Ward-Takahashi identities for the generators $\{P_+,P_-,P_i,B, C^\alpha,T,M_{i+},K_+\}$ read %
{\allowdisplaybreaks
\begin{align}
	0 \ =& \ \sum_{a=1}^N \Bigg( - i p_a \Bigg) \langle \fOp^{(1)}(x^-_1,x^l_1) \cdots \fOp^{(N)}(x^-_N,x^l_N) \rangle \ , \label{WIpplus} \\
	0 \ =& \ \sum_{a=1}^N \Bigg(  \frac{\partial}{\partial x^-_a} \Bigg) \langle \fOp^{(1)}(x^-_1,x^l_1) \cdots \fOp^{(N)}(x^-_N,x^l_N) \rangle \ , \label{WIpminus} \\
	0 \ =& \ \sum_{a=1}^N \Bigg(  \frac{\partial}{\partial x^i_a} + \tfrac{1}{2}\Omega_{ij} x^j_a \frac{\partial}{\partial x^-_a} \Bigg) \langle \fOp^{(1)}(x^-_1,x^l_1) \cdots \fOp^{(N)}(x^-_N,x^l_N) \rangle \ , \label{WIpi} \\
	0 \ =& \ \sum_{a=1}^N  \Bigg(   - \tfrac{1}{2} R\,\Omega_{ij} x^i_a \frac{\partial}{\partial x^j_a} \Bigg) \langle \fOp^{(1)}(x^-_1,x^l_1) \cdots \fOp^{(N)}(x^-_N,x^l_N) \rangle \ , \label{WIB} \\
	0 \ =& \ \sum_{a=1}^N \Bigg(   \tfrac{1}{2}\eta^\alpha_{ij} x^i_a \frac{\partial}{\partial x^j_a} \Bigg) \langle \fOp^{(1)}(x^-_1,x^l_1) \cdots \fOp^{(N)}(x^-_N,x^l_N) \rangle \ , \label{WIC} \\
	0 \ =& \ \sum_{a=1}^N \Bigg(   2 x^-_a \frac{\partial}{\partial x^-_a} + x^i_a \frac{\partial}{\partial x^i_a} + \Delta_a \Bigg) \langle \fOp^{(1)}(x^-_1,x^l_1) \cdots \fOp^{(N)}(x^-_N,x^l_N) \rangle \ , \label{WIT} \\
	0 \ =& \ \sum_{a=1}^N \Bigg(   \big( \tfrac{1}{2}\Omega_{ij} x^-_a x^j_a - \tfrac{1}{8}R^{-2} x^j_a x^j_a x^i_a \big) \frac{\partial}{\partial x^-_a} + x^-_a \frac{\partial}{\partial x^i_a} + \tfrac{1}{2} \Delta_a \Omega_{ij} x^j_a - i p_a x^i_a \nn \\*
	&\quad + \tfrac{1}{4} \big( 2\Omega_{ik}x^k_a x^j_a + 2\Omega_{jk}x^k_a x^i_a - \Omega_{ij}x^k_a x^k_a \big) \frac{\partial}{\partial x^j_a} \Bigg) \langle \fOp^{(1)}(x^-_1,x^l_1) \cdots \fOp^{(N)}(x^-_N,x^l_N) \rangle \ , \label{WIMiplus}\\
	0 \ =& \ \sum_{a=1}^N \Bigg( \big( 2 ( x^-_a )^2 - \tfrac{1}{8} R^{-2} ( x^i_a x^i_a )^2 \big) \frac{\partial}{\partial x^-_a} + 2\Delta x^-_a - i p_a x^i_a x^i_a \nn \\*
	&\quad \qquad + \big( \tfrac{1}{2} \Omega_{ij} x^j_a x^k_a x^k_a + 2 x^-_a x^i_a \big) \frac{\partial}{\partial x^i_a} \Bigg) \langle \fOp^{(1)}(x^-_1,x^l_1) \cdots \fOp^{(N)}(x^-_N,x^l_N) \rangle \ .	 \label{WIKplus}
\end{align}
}%
There are some consequences of these equations which hold for any $N$-point function; the first equation immediately imposes
\begin{align}
	0 \ =& \ \sum_{a=1}^N  p_a \ , \label{momemtumconserv}
\end{align} 
whilst equations \eqref{WIpminus} through \eqref{WIC} force the correlation function to be a function of the variables
\begin{align}
	\tilde{x}_{ab} \ := & \ x^-_a - x^-_b + \tfrac{1}{2} \Omega_{ij} x^i_a x^j_b = -\tilde{x}_{ba} \ , \qquad  |\vex_{ab}|^2 \ = x_{ab}^ix_{ab}^i  \ .
	\label{eq: x tilde and x squared definitions}
\end{align}
where $x_{ab}^i = x_a^i - x_b^i$. It will also be convenient to define
\begin{equation}
\xi_{ab} \ \equiv \ \frac{x_{ab}^ix_{ab}^i}{\tilde{x}_{ab}} = - \xi_{ba}  \ .
\end{equation}
Finally, let us combine the two invariant distances (\ref{eq: x tilde and x squared definitions}) into a single complex object. In particular, define for any two points $x,y\in \mathbb{R}^5$ the complex distance
\begin{align}
z(x,y) = x^- - y^- + \tfrac{1}{2} \Omega_{ij} x^i y^j  + \frac{i}{4R} (x^i-y^i)(x^i-y^i) = - \bar{z}(y,x) \ . \label{Defzonetwo}
\end{align}
Then $z(x,y)$ defines a useful covariant distance due to its particularly simple transformations under $SU(1,3)$ transformations. Indeed, we see immediately that any correlator of scalar primaries is realised as a function only of the real and imaginary part of the $z(x_a, x_b)$. Further details on the finite transformations of $z(x,y)$ under $SU(1,3)$ can be found in Appendix \ref{app: finite transformations}. For our purposes here, we further use the shorthand $z_{ab}=z(x_a,x_b)$.\\

The further constraints that the remaining equations \eqref{WIT}--\eqref{WIKplus},  corresponding to the generators $T,M_{i+}$ and $K_+$, place on correlation functions are most easily considered on a case-by-case basis, and we turn to that question now.


\subsection{2-point Functions}\label{Sec2ptFn}


We begin with the simplest non-trivial correlation function of scalar operators and define
\begin{align}
	\langle \fOp^{(1)}(x^-_1,x^i_1) \fOp^{(2)}(x^-_2,x^i_2) \rangle \ = \ F( \tilde{x}_{12},|\vex_{12}|^2 ) \ ,
\end{align}
We make the redefinition
\begin{align}
	F \ = \ ( \tilde{x}_{12} )^{-\alpha} G( \tilde{x}_{12} , |\vex_{12}|^2 ) \ , \label{2ptFnAnsatz}
\end{align}
for some constant $\alpha$.
Then the Ward-Takahashi identity \eqref{WIT} associated with Lifshitz scaling is solved if
\begin{align}
	0 \ =& \ \alpha - \frac{1}{2} ( \Delta_1 + \Delta_2 ) \ ,\nonumber \\
	0 \ =& \ \bigg( \tilde{x}_{12} \frac{\partial}{\partial \tilde{x}_{12}} + |\vex_{12}|^2 \frac{\partial}{\partial |\vex_{12}|^2} \bigg) G( \tilde{x}_{12} , |\vex_{12}|^2 ) \ .
\end{align}
The second condition forces the functional form of $G$ to be
\begin{align}
	G( \tilde{x}_{12} , |\vex_{12}|^2 ) \ = \ G\bigg( \frac{|\vex_{12}|^2}{\tilde{x}_{12}} \bigg) \ = \ G( \xi_{12} ) \ .
\end{align}
Moving onto the Ward-Takahashi identity associated with the operator $M_{i+}$ we find using $F=( \tilde{x}_{12} )^{-\frac{1}{2} ( \Delta_1 + \Delta_2 )} G(\xi_{12})$ leads to
\begin{align}
	0 \ =& \ \frac{1}{4} ( \Delta_1 - \Delta_2 ) \Omega_{ij} x^j_{12} G(\xi_{12}) + x_{12}^i \bigg[ \Big( \frac{1}{8R^2} \xi_{12}^2 + 2 \Big) \frac{\mathrm{d}}{\mathrm{d} \xi_{12}} + \frac{1}{16R^2} ( \Delta_1 + \Delta_2 ) \xi_{12} - i p_1 \bigg] G(\xi_{12}) \ .
\end{align}
Since $x_{12}^i$ and $\Omega_{ij} x^j_{12}$ are independent for each value of $i$ and assuming a non-trivial $G(\xi_{12})$, we have two equations for this Ward-Takahashi identity to hold: 
\begin{align}
	0 \ =& \ \frac{1}{4} ( \Delta_1 - \Delta_2 ) \ ,\nonumber \\
	0 \ =& \ \bigg[ \Big( \frac{1}{8R^2} \xi_{12}^2 + 2 \Big) \frac{\mathrm{d}}{\mathrm{d} \xi_{12}} + \frac{1}{16R^2} ( \Delta_1 + \Delta_2 ) \xi_{12} - i p_1 \bigg] G(\xi_{12}) \ .
\end{align}
Solving we find $\Delta_1=\Delta_2$ and
\begin{align}
	G(\xi_{12}) \ = \ C_{\Delta_1,p_1} ( \xi_{12}^2 + 16 R^2 )^{-\frac{1}{2} \Delta_1} e^{\left( 2i p_1 R \arctan\big( \tfrac{\xi_{12}}{4R} \big) \right)} \ ,
\end{align}  
where $C_{\Delta_1,p_1}$ is a constant which may depend on $\Delta_1$ and $p_1$. The above formula can be written more explicitly using the identity
\begin{equation}
e^{2i\arctan (x)}=\frac{i-x}{i+x}\ .
\end{equation}
The final Ward-Takahashi identity \eqref{WIKplus} places no further constraints on $G(\xi_{12})$.
Therefore the 2-point correlation function is fully determined up to a constant by the symmetries, in direct analogy with the more familiar case of the $SO(p,q)$ conformal group. It is given by
\begin{align}
	\langle \fOp^{(1)}(x^-_1,x^i_1) \fOp^{(2)}(x^-_2,x^i_2) \rangle \ = \ \delta_{0,p_1+p_2} \delta_{\Delta_1,\Delta_2} \frac{C_{\Delta_1,p_1}}{ [ \tilde{x}_{12} ( \xi_{12}^2 + 16 R^2 )^{\frac{1}{2}} ]^{\Delta_1}} e^{ 2i R p_1  \arctan\big( \tfrac{\xi_{12}}{4R} \big) } \ .
\end{align}
Using the definition (\ref{Defzonetwo}), we can write this more neatly in terms of $z_{12}$. We find in particular a factorisation into  holomorphic and anti-holomorphic parts,
\begin{align}
\langle \fOp^{(1)}(x^-_1,x^i_1) \fOp^{(2)}(x^-_2,x^i_2) \rangle \ =& \ \delta_{0,p_1+p_2}\delta_{\Delta_1,\Delta_2}C_{\Delta_1,p_1}
\left(\frac{1}{z_{12}}\right)^{\Delta_1/2 -p_1 R} \left(\frac{1}{\bar z_{12}}\right)^{ \Delta_1/2 +p_1 R}  \ .
\label{eq: 2pt WI solution}
\end{align}
This resembles a 2-point function in a two dimensional CFT. From the Lagrangian gauge theory perspective it can be decomposed into a perturbative piece $(z_{12}\bar{z}_{12})^{-\Delta_1/2}$ with power-law decay times an oscillating non-perturbative piece $(z_{12}/\bar{z}_{12})^{p_1 R}$ associated  to a non-vanishing charge under $P_+$. 

It should be noted that the 2-point function (\ref{eq: 2pt WI solution})---and indeed all correlators in this thesis--- should be thought of as `bare' correlators, in the sense that they apply only at separated points. One generically expects short-distance singularities as insertion points are brought together; indeed, it is the regulation of such singularities through a UV regulator that underpins the relationship between such short-distance singularities and possible conformal anomalies in the theory \cite{Petkou:1999fv,Bzowski:2015pba}.


\subsection{3-point Functions}\label{subsec: 3pt from WIs}


We define
\begin{align}
	\langle \fOp^{(1)}(x^-_1,x^i_1) \fOp^{(2)}(x^-_2,x^i_2) \fOp^{(3)}(x^-_3,x^i_3) \rangle \ = \ F(\tilde{x}_{12},\tilde{x}_{23},\tilde{x}_{13},|\vex_{12}|^2,|\vex_{23}|^2,|\vex_{13}|^2 ) \ .
\end{align}
Note that the presence of $\Omega_{ij}$ in the definition of $\tilde{x}_{ab}$ means that they do not satisfy $\tilde{x}_{12}+\tilde{x}_{23}+\tilde{x}_{31}=0$.

To start we make the redefinition
\begin{align}
	F \ = \ ( \tilde{x}_{12} )^{-\alpha_{12}} ( \tilde{x}_{23} )^{-\alpha_{23}} ( \tilde{x}_{13} )^{-\alpha_{13}} G( \tilde{x}_{12},\tilde{x}_{23},\tilde{x}_{13},|\vex_{12}|^2,|\vex_{23}|^2,|\vex_{13}|^2 ) \ . \label{3ptFnAnsatz}
\end{align}
Then the Lifshitz Ward-Takahashi identity \eqref{WIT} is solved if
\begin{align}
	0 \ =& \ -\frac{1}{2} \Delta_\text{T}+\sum_{a<b}^3 \alpha_{ab}  \ , \label{3ptsumalphas} \\
	0 \ =& \ \bigg( \sum_{a < b}^3 \tilde{x}_{ab} \frac{\partial}{\partial \tilde{x}_{ab}} + |\vex_{ab}|^2 \frac{\partial}{\partial |\vex_{ab}|^2} \bigg) G( \tilde{x}_{12},\tilde{x}_{23},\tilde{x}_{13},|\vex_{12}|^2,|\vex_{23}|^2,|\vex_{13}|^2 ) \ . \label{3ptDiffDil}
\end{align}
where $\Delta_\text{T}$ denotes the total scaling dimension, $\Delta_\text{T} = \Delta_1+\Delta_2+\Delta_3$. Substituting the ansatz \eqref{3ptFnAnsatz} into the Ward-Takahashi identity \eqref{WIMiplus} associated with the generator $M_{i+}$ and using \eqref{3ptsumalphas} fixes
\begin{align}
	\alpha_{ab} =  \Delta_a + \Delta_b - \tfrac{1}{2} \Delta_\text{T}\ ,
	%
\end{align}
together with constraining the functional form of $G$ to be
\begin{align}
	G( \tilde{x}_{12},\tilde{x}_{23},\tilde{x}_{13},|\vex_{12}|^2,|\vex_{23}|^2,|\vex_{13}|^2 ) \ = \ G( \xi_{12},\xi_{23},\xi_{13} ) \ .
\end{align}
This functional form for $G$ also solves \eqref{3ptDiffDil}. Using all this information, the $M_{i+}$ Ward-Takahashi identity can be reduced to
\begin{align}
	0 \ =& \ \sum_{a < b}^3 x_{ab}^i \bigg[ \bigg( \frac{1}{8R^2} \xi_{ab}^2 + 2 \bigg) \frac{\partial}{\partial \xi_{ab}} + \frac{1}{8R^2} ( - \tfrac{1}{2} \Delta_\text{T} + \Delta_a + \Delta_b ) \xi_{ab} - \frac{i}{3} p_{ab} \bigg] G( \xi_{12},\xi_{23},\xi_{13} ) \ , 
\end{align}
where we have also defined\footnote{Here we have chosen not to impose total $p_+$ conservation given by \eqref{momemtumconserv} in order to display the most symmetric form for $G$.}
\begin{align}
	p_{ab} \ \equiv \ p_a-p_b \ .
\end{align}
A solution to these coupled partial differential equations is
\begin{align}
	G( \xi_{12},\xi_{23},\xi_{13} ) \ =& \ C_{123} \bigg[ \prod_{a<b}^3 ( \xi_{ab}^2 + 16 R^2 )^{\frac{1}{4} \Delta_\text{T} - \frac{1}{2} \Delta_a - \frac{1}{2} \Delta_b} e^{\left( \frac{2iR}{3} p_{ab} \arctan\big( \tfrac{\xi_{ab}}{4R} \big) \right)} \bigg] \nn \\
	&\qquad \times H\bigg( \sum_{a < b}^3 \arctan\big( \tfrac{\xi_{ab}}{4R} \big) \bigg) \ ,
\end{align}  
where $H$ is an arbitrary function of its argument and $C_{123}$ is a constant which may depend on the $\Delta$'s and $p$'s. As with 2-points, the remaining Ward-Takahashi identity \eqref{WIKplus} is satisfied automatically. The full 3-point function is
\begin{align}
	&\langle \fOp^{(1)}(x^-_1,x^i_1) \fOp^{(2)}(x^-_2,x^i_2) \fOp^{(3)}(x^-_3,x^i_3) \rangle \nn \\
	&\qquad = \ \delta_{0,p_1+p_2+p_3} C_{123} \bigg[ \prod_{a<b}^3  [ \tilde{x}_{ab} ( \xi_{ab}^2 + 16 R^2 )^{\frac{1}{2}} ]^{\frac{1}{2}\Delta_\text{T} - \Delta_a - \Delta_b} e^{\left( \frac{2iR}{3} p_{ab} \arctan\big( \tfrac{\xi_{ab}}{4R} \big) \right)} \bigg] \nn\\
	& \qquad \qquad \qquad \times H\bigg(  \sum_{a < b}^3 \arctan\big( \tfrac{\xi_{ab}}{4R} \big) \bigg) \ .
\end{align}
Using the identity
\begin{equation}
\arctan\alpha+\arctan\beta+\arctan\gamma=\arctan\left(\frac{\alpha+\beta+\gamma-\alpha\beta\gamma}{1-\alpha\beta-\beta\gamma-\gamma\alpha}\right) \ ,
\end{equation}
the undetermined function can also be written as
\begin{equation}
H=H\left(\frac{16R^2\left( \xi_{12}+\xi_{23}+\xi_{31} \right)-\xi_{12}\xi_{23}\xi_{31}}{16R^2-\xi_{12}\xi_{23}-\xi_{23}\xi_{31}-\xi_{31}\xi_{12}}\right) \ .
\end{equation}
Thus in theories with $SU(1,3)$ conformal symmetry, 3-point functions of scalar operators are determined only up to an arbitrary function of a single variable. This is in contrast to Lorentzian CFT's where 3-point functions are completely fixed by conformal symmetry.

Written in terms of the complex variables \eqref{Defzonetwo}, we again see factorisation into holomorphic and anti-holomorphic pieces times the undetermined function $H$:
\begin{align}
&\langle \fOp^{(1)}(x^-_1,x^i_1) \fOp^{(2)}(x^-_2,x^i_2) \fOp^{(3)}(x^-_3,x^i_3) \rangle\nonumber\\
&= \ \delta_{0,p_1+p_2+p_3}\, C_{123} \prod_{a<b}^3\left(\frac{1}{z_{ab}}\right)^{-\frac{1}{4}\Delta_\text{T} + \frac{1}{2}\Delta_a + \frac{1}{2}\Delta_b -\frac{R}{3} p_{ab}} 
\left(\frac{1}{\bar z_{ab}}\right)^{-\frac{1}{4}\Delta_\text{T} + \frac{1}{2}\Delta_a + \frac{1}{2}\Delta_b + \frac{R}{3} p_{ab}}H\left( \frac{z_{12}z_{23}z_{31}}{\bar{z}_{12}\bar{z}_{23}\bar{z}_{31}} \right)\label{eq: 3pt Ward-Takahashi identity solution}\ .
\end{align}
 
 
\subsection{4-point Functions}\label{Four-point Functions}


For 4-point and higher-point correlation functions we expect to see the appearance of analogues of cross-ratios which are annihilated by the $SU(1,3)$ generators given in \eqref{eq: 5d algebra vector field rep}. Inspection of the 2- and 3-point functions found in the previous subsections suggests that there are two types of variables to consider constructing invariants from. These are 
\begin{align}
	x'_{ab} \ = \ \tilde{x}_{ab} (\xi_{ab}^2 + 16R^2)^{\frac{1}{2}} \ , \qquad {\xi}'_{ab} \ = \ 4R \arctan\bigg( \frac{\xi_{ab}}{4R} \bigg) \ .
\end{align}
We find that cross ratios of the $x'$'s are indeed annihilated by all of the generators, and are therefore $SU(1,3)$ invariants. However, the infinitesimal transformation of the ${\xi}'_{ab}$ under $M_{i+}$ and $K_+$ is found to be
\begin{align}
	{\xi}'_{ab}e^{\epsilon^i M_{i+}} \ =& \ {\xi}'_{ab}+2\, \epsilon^ix_{ab}^i + \bigO\left(\epsilon^2\right) \ , \qquad {\xi}'_{ab}e^{\epsilon K_+} \ = \ {\xi}'_{ab} + 16\,\epsilon  \left( |\vex_a|^2 - |\vex_b|^2  \right) + \bigO\left(\epsilon^2\right) \ .
	\label{eq: xi' variation}
\end{align}
Hence, we see that it is linear combinations which will be invariant rather than cross ratios.

Specialising to the case of 4-points, there are two independent invariant cross ratios of the $x'$'s;
\begin{align}
	u \ =& \ \frac{{x}_{12}'{x}_{34}'}{{x}_{13}'{x}_{24}'} \ , \qquad  v \ = \ \frac{{x}_{14}'{x}_{23}'}{{x}_{13}'{x}_{24}'} \ ,
\end{align}
and three independent invariant combinations of the ${\xi}'$'s;
\begin{align}
	{\xi}'_{12} + {\xi}'_{24} + {\xi}'_{41} \ , \qquad {\xi}'_{13} + {\xi}'_{34} + {\xi}'_{41} \ , \qquad {\xi}'_{23} + {\xi}'_{34} + {\xi}'_{42} \ .
\end{align}
Hence at 4-points under the $SU(1,3)$ symmetry group there are five independent conformal invariants compared to the more familiar case of two cross ratios under $SO(p,q)$.

Starting from a generic function of these five variables, we may solve each of the Ward-Takahashi identities following the same steps as for 2- and 3-point functions. We eventually find\footnote{Once again, we have not imposed total $p_+$ conservation in order to present a more symmetrical result.}
\begin{align}
	\langle \fOp^{(1)}(x^-_1,x^i_1) &\fOp^{(2)}(x^-_2,x^i_2) \fOp^{(3)}(x^-_3,x^i_3) \fOp^{(4)}(x^-_4,x^i_4) \rangle \ \nn \\
	=& \ \delta_{0,p_1+p_2+p_3+p_4}  \, \bigg[ \prod_{a<b}^4 (x'_{ab})^{\frac{1}{6} \Delta_\text{T} - \frac{1}{2} ( \Delta_a + \Delta_b)}  e^{ i p_{ab} {\xi}'_{ab}/8 } \bigg] \nn \\
	&\qquad \times H\big( u , v , {\xi}'_{12} + {\xi}'_{24} + {\xi}'_{41} \ , {\xi}'_{13} + {\xi}'_{34} + {\xi}'_{41} \ , {\xi}'_{23} + {\xi}'_{34} + {\xi}'_{42} \big) \ ,
\end{align}
where $H$ is an undetermined function and now $\Delta_\text{T} = \Delta_1+\Delta_2+\Delta_3+\Delta_4$. 

It is also covenient to define an alternative basis for the space of conformal invariants involving the $\xi'_{ab}$, 
\begin{align}
\lambda_{1}\ &= \ {\xi}'_{12}+{\xi}'_{24}+{\xi}'_{43}+{\xi}'_{31} \ = \ +\left( {\xi}'_{12} + {\xi}'_{24} + {\xi}'_{41} \right)  - \left( {\xi}'_{13} + {\xi}'_{34} + {\xi}'_{41} \right)\ ,\nn \\
\lambda_{2} \ &= \ {\xi}'_{13}+{\xi}'_{32}+{\xi}'_{24}+{\xi}'_{41} \ = \ +\left( {\xi}'_{13} + {\xi}'_{34} + {\xi}'_{41} \right)  - \left( {\xi}'_{23} + {\xi}'_{34} + {\xi}'_{42} \right)\ ,\\
\lambda_{3}\ &= \ {\xi}'_{14}+{\xi}'_{43}+{\xi}'_{32}+{\xi}'_{21} \ = \ -\left( {\xi}'_{23} + {\xi}'_{34} + {\xi}'_{42} \right)  - \left( {\xi}'_{12} + {\xi}'_{24} + {\xi}'_{41} \right)\nn\ .
\end{align}
Then, for instance under the interchange $x_1\leftrightarrow x_2$ they transform as:
\begin{equation}
\left(u,v,\lambda_{1},\lambda_{2},\lambda_{3}\right)\ \longrightarrow \ \left(-u/v,1/v,\lambda_3,-\lambda_2,\lambda_1\right)\ ,
\end{equation}
with similar rules for the other five permutations of $S_3$.

Generically, we may then consider the resulting crossing equations constraining the function $H$. However, our main focus is a particular class of $SU(1,3)$ theories, explored in Chapter \ref{chap: the view from 6d CFT}, which admit a six-dimensional interpretation. For these theories, we will see that almost all crossing symmetry is broken by the parameters $p_a$ even if all the scaling dimensions are the same.\\ 

We will however later consider a maximally symmetric case, in which all scaling dimensions are equal, $\Delta_{a}=\Delta$, and all $p_a$ have equal magnitude $|p_a|=p> 0$. Writing $\fOp^{(a)}_{\pm}$ to denote such a primary with $p_a=\pm p$, we find after a convenient redefinition of $H$,
\begin{align}
  \langle \pm \pm \pm \pm  \rangle = \langle  \fOp^{(1)}_{\pm }\fOp^{(2)}_{\pm }\fOp^{(3)}_{\pm }\fOp^{(4)}_{\pm }\rangle = \left({x}_{13}'{x}_{24}'\right)^{-\Delta}\exp\left(\frac{i}{8}\sum_{a<b}p_{ab}{\xi}'_{ab}\right)G_{\pm\pm\pm\pm}\ ,
\label{eq: 4pt G}
\end{align}
where the $G_{\pm\pm\pm\pm}=G_{\pm\pm\pm\pm}(u,v,\lambda_{1},\lambda_{2},\lambda_{3})$. Clearly by the overall conservation of $P_+$ charge, only the combinations with two pluses and two minuses can be non-zero. We will in fact find that theories with a six-dimensional interpretation have only $G_{++--}$ and $G_{+-+-}$ non-zero, with all other orderings vanishing.

We may then consider the crossing relations of, say, $G_{++--}$ with itself. For example, the interchange $x_1\leftrightarrow x_2$ corresponds to
\begin{align}
  G_{++--}(u,v,\lambda_{1},\lambda_{2},\lambda_{3})  \,\,\longrightarrow\,\,  v^{-\Delta} G_{++--}(-u/v,1/v,\lambda_3,-\lambda_2,\lambda_1) \ .
  \label{eq: G++-- 12 crossing}
\end{align}

 
\subsection{Higher-point Functions}


At $\Npt$-points, the general solution to the conformal Ward-Takahashi identities for scalar operators is
\begin{align}
	&\langle \fOp^{(1)}(x^-_1,x^i_1) \dots \fOp^{(N)}(x^-_N,x^i_N) \rangle \ \nn \\
	&\qquad= \ \delta_{0,p_1+\dots+p_N}  \, \bigg[ \prod_{a<b}^N (x'_{ab})^{-\alpha_{ab}}  e^{i p_{ab} {\xi}'_{ab}/2N } \bigg]  \, H\!\left( \frac{x'_{ab}x'_{cd}}{x'_{ac}x'_{bd}},\xi'_{ab} +\xi'_{bc} + \xi'_{ca}  \right) \ ,
\end{align}
where the $\alpha_{ab}=\alpha_{ba}$ satisfy $\sum_{b\neq a}\alpha_{ab}=\Delta_a$ for each $a=1,\dots, N$. Here, $H$ is a generic function of $N^2-3N+1$ variables, which fall into two classes:
\begin{itemize}
  \item The $N(N-3)/2$ independent cross-ratios\footnote{Beyond 4-points, there are other invariant ratios of the $x_{ab}'$ we could consider. However, as is familiar from conventional conformal field theory, all such ratios can be written as a product of cross-ratios of the form (\ref{eq: x' cross ratios})} of the form
  \begin{align}
  \frac{x'_{ab}x'_{cd}}{x'_{ac}x'_{bd}}\ .
  \label{eq: x' cross ratios}
\end{align}
  \item The $(N-1)(N-2)/2$ independent triplets\footnote{One could also consider for instance combinations of the form $\xi'_{ab} +\xi'_{bc} + \xi'_{cd} +  \xi'_{da}$, which by (\ref{eq: xi' variation}) are invariant under $SU(1,3)$. However, since $\xi'_{ab}=-\xi'_{ba}$, this combination can be written in terms of triplets, as $\left(\xi'_{ab} +\xi'_{bc} + \xi'_{ca}\right) + \left(\xi'_{ac} + \xi'_{cd} +  \xi'_{da}\right) $. This generalises straightforwardly to show that any invariant linear combination is a sum of triplets of the form (\ref{eq: xi triplet})} of the form  
  \begin{align}
  \xi'_{ab} +\xi'_{bc} + \xi'_{ca}\ .
  \label{eq: xi triplet}
\end{align}\
\end{itemize}
To make explicit contact with our results at 3- and 4-points, note that one can always choose $H$ such that for $N\ge 3$ the $\alpha_{ab}$ can be taken as
\begin{align}
  \alpha_{ab} = \frac{1}{N-2}\left( \Delta_a+\Delta_b \right) - \frac{1}{(N-1)(N-2)}\Delta_\text{T} \ .
\end{align}
Where as before, $\Delta_\text{T}$ denotes the total scaling dimension, $\Delta_\text{T}=\sum_a \Delta_a$

As we are now used to, this $\Npt$-point function is more naturally written in terms of the complex variables $z_{ab}$. We find then that the \textit{general} $\Npt$-point function is written as
\begin{align}
	&\langle \fOp^{(1)}(x^-_1,x^i_1) \dots \fOp^{(N)}(x^-_N,x^i_N) \rangle \ \nn \\
	&\qquad= \ \delta_{0,p_1+\dots+p_N}  \, \left[ \prod_{a<b}^N (z_{ab} \bar{z}_{ab})^{-\alpha_{ab}/2}  \left( \frac{z_{ab}}{\bar{z}_{ab}} \right)^{p_{ab}R/N}\right]  H\left( \frac{|z_{ab}||z_{cd}|}{|z_{ac}||z_{bd}|}, \frac{z_{ab}z_{bc}z_{ca}}{\bar{z}_{ab}\bar{z}_{bc}\bar{z}_{ca}}  \right) \ .
	\label{eq: general N-point function}
\end{align}
The $SU(1,3)$ invariance of the $N(N-3)/2$ independent cross ratios of the form
\begin{align}
  \frac{|z_{ab}||z_{cd}|}{|z_{ac}||z_{bd}|}\ ,
  \label{eq: z cross ratios}
\end{align}
and the $(N-1)(N-2)/2$ independent phases of the form
\begin{align}
  \frac{z_{ab}z_{bc}z_{ca}}{\bar{z}_{ab}\bar{z}_{bc}\bar{z}_{ca}} = - \frac{z_{ab}z_{bc}z_{ca}}{z_{ac}z_{cb}z_{ba}}\ ,
  \label{eq: z phases}
\end{align}
is most easily seen from the $z_{ab}$'s finite transformation under $SU(1,3)$, as explored in Appendix \ref{app: finite transformations}. Indeed, we further show in this Appendix that these $N^2-3N+1$ combinations are the \textit{complete} set of independent $SU(1,3)$ invariants, and thus verify that (\ref{eq: general N-point function}) is indeed the general form of the scalar $N$-point function.

\addtocontents{toc}{\protect\pagebreak}
\chapter{The view from six-dimensional conformal field theory}\label{chap: the view from 6d CFT}


We have so far investigated an interesting albeit somewhat esoteric class of theories in five dimensions, admitting peculiar spacetime symmetries that permit an analysis of correlation functions comparable to standard conformal field theory. We now show that such theories naturally arise upon a particular conformal compactification of a generic six-dimensional conformal field theory.


\section{A conformal compactification}


We start with six-dimensional Minkowski spacetime in lightcone coordinates with metric 
\begin{align}
	ds^2_M = \hat g_{\mu\nu}d\hat x^\mu d\hat x^\nu =  -2d\hat x^+d\hat x^- + d\hat x^id \hat x^i\ ,
	\label{eq: 6d Minkowski metric}
\end{align}
where $\mu\in\{+,-,i\}$, and perform the coordinate transformation
\begin{align}\label{eq: coordinate transformation}
\hat x^+ &= 2R \tan (x^+/2R) \ , \nonumber\\ 
\hat x^- &= x^- +\frac{1}{4R}x^ix^i\tan (x^+/2R)\ ,\nonumber\\   
\hat x^i &=x^i - \tan(x^+/2R) R\Omega_{ij}x^j	\ .
\end{align}
Here $\Omega_{ij}$ is the same anti-self-dual constant spatial 2-form as appears in Chapter \ref{chap: SU(1,3) theories}. 

This transformation leads to the  metric
\begin{align}
ds^2_M = \frac{-2\, dx^+(dx^- - \frac12 \Omega_{ij}x^idx^j) + dx^idx^i}{\cos^2(x^+/2R)}\ .	
\label{5dmetric}
\end{align}
Following this we perform a Weyl transformation $ds^2_\Omega = {\cos^2(x^+/2R)}ds^2_M$ to find
\begin{align}
ds^2_\Omega = g_{\mu\nu}x^\mu x^\nu =  -2\, dx^+\left(dx^- - \frac12 \Omega_{ij}x^idx^j\right) + dx^idx^i \ .
\label{eq: conformally compactified metric}	
\end{align} 
Note the range of $x^+\in (-\pi R,\pi R)$ is finite. Thus we have conformally compactified the $x^+$ direction of six-dimensional Minkowski space. Note also that $\partial/\partial x^+$ is a Killing vector of $ds^2_\Omega$, while it was only a conformal Killing vector (with non-trivial conformal factor) of $ds^2_M$. This metric appears naturally as the conformal boundary to AdS$_7$ in the construction of \cite{Pope:1999xg}, arising from a $U(1)$ fibration of AdS$_7$ over a non-compact $\mathbb{C}P^3$.


\section{Mapping of symmetries and operators}


We first show how the five-dimensional symmetry algebra (\ref{eq: extended su(1,3) algebra}) is recovered from the full six-dimensional conformal algebra upon this null compactification. Let $\{P^\text{6d}_\mu,M^\text{6d}_{\mu\nu},D^\text{6d},K^\text{6d}_\mu\}$ denote the usual basis for $\frak{so}(6,2)$ in lightcone coordinates. These correspond to the conformal Killing vectors of the Minkowski metric (\ref{eq: 6d Minkowski metric}), which we take to be
\begin{align}
  \left( P^{\text{(6d)}}_\mu \right)_\partial &= \hat{\partial}_\mu \ , & \hat{\omega} &= 0  \ , \nn\\
\left( M^{\text{(6d)}}_{\mu\nu} \right)_\partial &= \hat{x}_\mu \hat{\partial}_\nu - \hat{x}_\nu \hat{\partial}_\mu \ , & \hat{\omega} &= 0 \ , \nn\\
\left( D^{\text{(6d)}} \right)_\partial &= \hat{x}^\mu \hat{\partial}_\mu \ , & \hat{\omega} &= 1 \ , \nn\\
\left( K^{\text{(6d)}}_\mu \right)_\partial &= \hat{x}_\nu \hat{x}^\nu \hat{\partial}_\mu - 2 \hat{x}_\mu \hat{x}^\nu \hat{\partial}_\nu \ , & \hat{\omega} &= -2 \hat{x}_\mu \ . 
\label{eq: 6d conformal Killing vectors}
\end{align}
Here, indices are raised and lowered with the metric $\hat{g}$ as in (\ref{eq: 6d Minkowski metric}).
These vector fields $V$ then satisfy $\mathcal{L}_V \hat{g} = 2\hat{\omega} \hat{g}$ for the conformal factor $\hat{\omega}$ as specified. Then, after the coordinate transformation (\ref{eq: coordinate transformation}) and Weyl transformation, these are still a basis for the space of conformal Killing vectors, just with shifted conformal factors; we have $\mathcal{L}_V g = 2\omega g$, where $\omega = \hat{\omega} - \tfrac{1}{2R}\tan\left( \frac{x^+}{2R} \right)V^+$ for each vector field $V$.

The metric $g_{\mu\nu}$ as in (\ref{eq: conformally compactified metric}) has a null isometry along the $x^+$ direction. In terms of the full algebra of conformal Killing vectors (\ref{eq: 6d conformal Killing vectors}), this is realised by the combination $P_+:=P^{\text{(6d)}}_++\tfrac{1}{4}\Omega_{ij} M^{\text{(6d)}}_{ij} + \tfrac{1}{8R^2}K^{\text{(6d)}}_-$, which has vector field representation $(P_+)_\partial=\partial_+$.

Our next step is to reduce the theory into modes along this $x^+$ direction. At the level of the symmetry algebra, this amounts to choosing a basis for the space of local operators which diagonalises $P_+$. Equivalently, we expand all six-dimensional local operators in a Fourier series in the $x^+$ direction, as will be explored in more detail below.

A mode in such a decomposition is an eigenvector of the symmetry variation $\delta_{P_+}$, and falls into a representation of the maximal subalgebra of $\frak{so}(6,2)$ that commutes with $P_+$. This subalgebra is precisely the algebra $\frak{h}=\frak{su}(1,3)\oplus \frak{u}(1)$ introduced in Chapter \ref{chap: SU(1,3) theories}, with basis identified in terms of the usual six-dimensional basis for $\frak{so}(2,6)$ as:
 \begin{align}
  	P_+ 	&= 	P^{\text{(6d)}}_+ + \tfrac{1}{4}\Omega_{ij} M^{\text{(6d)}}_{ij} + \tfrac{1}{8R^2} K^{\text{(6d)}}_-	\ , \nn\\
  	P_-		&=	P^{\text{(6d)}}_-	\ ,														\nn\\
  	P_i		&=	P^{\text{(6d)}}_i + \tfrac{1}{2}\Omega_{ij} M^{\text{(6d)}}_{j-}	 \ ,						\nn\\
  	B		&=	-\tfrac{1}{4}R\,\Omega_{ij} M^{\text{(6d)}}_{ij}			\ ,				\nn\\
  	C^\alpha		&=	\tfrac{1}{4}\eta^\alpha_{ij} M^{\text{(6d)}}_{ij}		\ ,						\nn\\
  	T		&=	D^{\text{(6d)}}-M^{\text{(6d)}}_{+-}		\ ,												\nn\\
  	M_{i+}	&=	M^{\text{(6d)}}_{i+}-\tfrac{1}{4}\Omega_{ij} K^{\text{(6d)}}_j	\ ,						\nn\\
  	K_+		&= 	K^{\text{(6d)}}_+ \ .
  	\label{eq: 5d alg in terms of 6d}
\end{align}
The corresponding six-dimensional algebra of vector fields in terms of the $x^\mu$ is then
\begin{align}
	\left( P_+ \right)_\partial 	&= \partial_+ 																								\ ,&\omega = 0	\ ,					\nn\\
	\left( P_- \right)_\partial 	&= \partial_- 																								\ ,&\omega = 0 \ ,					\nn\\
	\left( P_i \right)_\partial 	&= \tfrac{1}{2}\Omega_{ij} x^j \partial_- + \partial_i														\ ,&\omega = 0 \ ,					\nn\\
	\left( B \right)_\partial 		&= -\tfrac{1}{2}R\,\Omega_{ij}x^i\partial_j 																	\ ,&\omega = 0\ , 					\nn\\
	\left( C^\alpha \right)_\partial 	&= \tfrac{1}{2}\eta^\alpha_{ij}x^i\partial_j 																					\ ,&\omega = 0 \ ,					\nn\\
	\left( T \right)_\partial 		&= 2x^- \partial_- + x^i \partial_i																			\ ,&\omega = 1 \ ,					\nn\\
	\left( M_{i+} \right)_\partial 	&= x^i \partial_+ + \left( \tfrac{1}{2}\Omega_{ij} x^- x^j - \tfrac{1}{8}R^{-2} x^j x^j x^i \right)\partial_- + x^- \partial_i  				\nn\\
			&  \qquad +\tfrac{1}{4}( 2\Omega_{ik}x^k x^j + 2\Omega_{jk}x^k x^i - \Omega_{ij}x^k x^k )\partial_j	\ , &\omega = \tfrac{1}{2}\Omega_{ij}x^j \ , 	\nn\\
	\left( K_{+} \right)_\partial 	&= x^i x^i \partial_+ + ( 2 ( x^- )^2 - \tfrac{1}{8} R^{-2} ( x^i x^i )^2 )\partial_-															\nn\\			
			&  \qquad +( \tfrac{1}{2} \Omega_{ij} x^j x^k x^k + 2 x^- x^i )\partial_i							\ ,&\omega = 2x^- 	\ ,	
			\label{eq: 6d CKVs}	
\end{align}
which satisfy $\mathcal{L}_V g = 2\omega g$ with $\omega$ as given. The five-dimensional vector field representation (\ref{eq: 5d algebra vector field rep}) is then recovered from the push-forward of these vector fields with respect to the projection map $(x^+, x^-, x^i)\to (x^-, x^i)$. \\

Let us briefly comment on more generic choices of $\Omega_{ij}$. Note, the coordinate transformation (\ref{eq: coordinate transformation}) leading to metrics (\ref{5dmetric}) and (\ref{eq: conformally compactified metric}) goes through for any anti-symmetric $\Omega_{ij}$. The further requirement that $\Omega_{ij}$ should be anti-self-dual can be motivated by the bulk $S^1\hookrightarrow \text{AdS}_7 \to \tilde{\mathbb{CP}}^3$ geometry \cite{Lambert:2019jwi,Pope:1999xg}. But without recourse to holography, let us simply note that for generic $\Omega_{ij}$, the subalgebra of conformal isometries commuting with $P_+$ is smaller than $\frak{su}(1,3) \oplus \frak{u}(1)$. For instance, for generic $\Omega_{ij}$ only a subalgebra $\frak{u}(1)\oplus \frak{u}(1)\subset \frak{su}(2)\oplus \frak{su}(2) = \frak{so}(4)$ of rotations in the $x^i$ directions survives the reduction on the $x^+$ interval, which is enhanced to $\frak{su}(2) \oplus \frak{u}(1)$ when $\Omega_{ij}$ is anti-self-dual. 
Thus, by choosing $\Omega_{ij}$ to be anti-self-dual, we maximise the symmetries preserved under the dimensional reduction, which is similarly seen also at the level of supersymmetries \cite{Lambert:2019jwi}. \\
 
Next we consider the mapping of a local operator $\hat{\sOp}(\hat x^+,\hat x^-,\hat x^i)$ on six-dimensional Minkowski space.   Let us perform the coordinate transformation $\hat x^\mu\to x^\mu$ in (\ref{eq: coordinate transformation}) along with the Weyl transformation $d s^2 = \cos^2(x^+/2R) d\hat s^2$   which takes us to the $\Omega$-deformed space and maps $\hat{\sOp}(\hat x)$ to
\begin{align}
	 \sOp(x^+,x^-,x^i) = \cos^{-\Delta}(x^+/2R)\hat {\sOp}(\hat x^+(x),\hat x^-(x),\hat x^i(x))\ ,
	 \label{eq: 6d operator mapping}
\end{align}
where for simplicity we assumed that $\hat {\sOp}$ is a scalar operator with conformal dimension $\Delta$. 
Note that,  for operators that satisfy  $ \hat {\sOp}(\hat x)\to 0$ sufficiently quickly, {\it i.e.} faster than $1/|\hat x^+|^{\Delta}$,  as $\hat x^+\to \pm\infty$ then 
\begin{align}
	 \sOp(-\pi R,x^-,x^i)= \sOp(\pi R,x^-,x^i)= 0 \ . 
\end{align}
Such operators, as well as others,  can be expanded  in a Fourier series on $x^+\in (-\pi R,\pi R)$, as
\begin{align}
  \sOp(x^+,x^-,x^i) = \sum_{n\in\mathbb{Z}} e^{-inx^+/R}\fOp_n(x^-, x^i)\ .
  \label{eq: 6d op Fourier decomp}
\end{align}
In particular, the Fourier mode $\fOp_\n$ satisfies $\delta_{P_+}\fOp_\n(x^-, x^i)=ip_+\fOp_\n = i \tfrac{\n}{R}\fOp_\n$, {\it i.e.} $p_+=\tfrac{\n}{R}$, and is given by  
\begin{align}
	\fOp_\n(x^-,x^i) &= \frac{1}{2\pi R}\int_{-\pi R}^{\pi R}dx^+ e^{i\n x^+/R}\sOp(x^+,x^-,x^i)\nonumber\\
	& = \frac{1}{2\pi R}\int_{-\pi R}^{\pi R}dx^+ e^{i\n x^+/R}\cos^{-\Delta}(x^+/2R) \hat {\sOp}(2R\tan( x^+/2R), \hat x^-(x),\hat x^i(x)) \nonumber\\
	& =  \frac{(2R)^{1-\Delta}}{  \pi }\int_{-\infty}^{\infty} d \hat x^+  \frac{(2R+i\hat x^+)^{\n +\Delta/2-1}}{(2R-i \hat x^+)^{\n -\Delta/2+1}}  \hat {\sOp}( \hat x^+,\hat x^-(x),\hat x^i(x))\nonumber \\ 
	&= \frac{(-1)^n}{  \pi }\int_{-\infty}^{\infty} d u  \frac{(u-i)^{\n +\Delta/2-1}}{(u+i)^{\n -\Delta/2+1}}  \hat {\sOp}( 2Ru,\hat x^-(x),\hat x^i(x))\ ,
	\label{eq: Fourier expansion}
\end{align}
where we introduced  $u =\hat x^+/2R= \tan(x^+/2R)$ to simplify the integral.  This is still quite complicated since for instance $\hat{x}^-(x^-,x^i)=x^-+\frac{1}{4R}x^i x^i u$ depends explicitly on the integration variable $u$. However, the relation takes a simply form when translated to $\hat{x}^i=x^i=0$ and  $ \hat{x}^-=x^-=0$, so that 
\begin{align}
	\fOp_\n (0,0) &=\frac{(-1)^n}{  \pi }\int_{-\infty}^{\infty}  du \left(\frac{u-i}{u+i}\right)^{\n }\left({1+u^2}\right)^{\Delta/2-1}\hat {\sOp}(2Ru,0,0) \ ,
\end{align}
whose inverse is
\begin{align}
	 \hat {\sOp}( \hat x^+,0,0) &= \left(\frac{4R^2}{4R^2+( \hat x^+)^2}\right)^{\Delta/2}\sum_{\n \in{\mathbb Z}} \left(\frac{2R-i\hat x^+}{2R+i\hat x^+}\right)^{\n }{\fOp}_\n (0,0) \ .
\end{align}
Using this map between operators, we can determine the relationship between the quantum numbers of an operator $\hat {\sOp}(\hat {x})$ in six-dimensions and those of its Fourier modes $\fOp_\n (x)$. We specialise to scalar operators for simplicity so that $\hat {\sOp}({x})$ is wholly characterised by a scaling dimension $\Delta_\text{6d}$ such that $\delta_{D^\text{6d}}\hat {\sOp}(\hat {x})=-(\hat  {x}^\mu \hat {\partial}_\mu + \Delta_\text{6d} )\hat {\sOp}(\hat {x})$, and has no spin, {\it  i.e.} $\delta_{M^\text{6d}_{\mu\nu}}\hat{\sOp}(\hat {x})=-( \hat {x}_\mu \hat {\partial}_\nu - \hat {x}_\nu \hat {\partial}_\mu )\hat {\sOp}(\hat {x})$. Then, using the explicit forms (\ref{eq: 5d alg in terms of 6d}) we determine that the Fourier mode $\fOp_\n $ as defined in (\ref{eq: Fourier expansion}) is a scalar operator in a five-dimensional sense ($r_{\fOp_\n }[B]=r_{\fOp_\n }[C^\alpha]=0$), has scaling dimension $\Delta$ as defined in (\ref{eq: 5d alg at origin}) given simply by $\Delta=\Delta_\text{6d}$, and has Kaluza-Klein momentum $p_+=n/R$ so that $\delta_{P_+}\fOp_n = i \frac{n}{R}\fOp_n$.


\section{Further constraints on correlators}\label{sec: further constraints on correlators}


With such a six-dimensional picture in hand, we can consider the dimensional reduction of correlation functions from six dimensions down to five. This will in particular further constrain the necessary form of correlation functions in a five-dimensional $SU(1,3)$ theory if it is to admit a meaningful six-dimensional interpretation.

In much of the below, for a pair of points with $\hat{x}$ coordinates $\hat{x}_a,\hat{x}_b$, we write $\hat{x}_{ab}=\hat{x}_a-\hat{x}_b$. Similarly, for their $x$ coordinates $x_a,x_b$, we write $x_{ab}=x_a-x_b$.


\subsection{2-point Functions}\label{subsec: 2pt dim red}

Let us consider a 2-point function in the six-dimensional theory of the form
\begin{equation}
	\big\langle  \hat {\sOp}^{(1)}({\hat x^+_1,\hat x_1^-,\hat x^i_1})\hat {\sOp}^{(2)}({\hat x^+_2,\hat x_2^-,\hat x^i_2}) \big\rangle = \frac{\hat C_{12}}{(\hat x_{12} )^{2\Delta}}\ ,
	\label{eq: 6d 2pt}
\end{equation}
where $\hat{\sOp}^{(1,2)}$ both have scaling dimension $\Delta$, and $\hat{x}^2=\hat{x}^\mu \hat{x}_\mu=-2\hat{x}^+\hat{x}^-+\hat{x}^i \hat{x}^i$, and once again we specialise to separated points and thus remain blind to possible short-distance singularities. Using the reduction procedure described above, the five-dimensional 2-point function of the Fourier modes are given by
\begin{align}
	\left\langle \fOp^{(1)}_{n_1}(x_1)\fOp^{(2)}_{n_2}(x_2)\right\rangle &=\frac{\left( -1 \right)^{n_1+n_2}}{\pi^2}\int_{-\infty}^{\infty}\,d^2 u\prod_{a=1}^{2}{\left(u_{a}+i\right)^{-n_{a}+\Delta/2-1}\left(u_{a}-i\right)^{n_{a}+\Delta/2-1}}\nn\\
	&\hspace{40mm}\times \langle  \hat {\sOp}^{(1)}({2Ru_1,\hat x_1^-,\hat x^i_1})\hat {\sOp}^{(2)}({2Ru_2,\hat x_2^-,\hat x^i_2}) \rangle\nn\\[0.7em]
	&=\frac{\left( -1 \right)^{n_1+n_2}}{\pi^2}\hat C_{12}\int_{-\infty}^{\infty}\,d^2 u\,\frac{\Pi_{a=1}^{2}{\left(u_{a}+i\right)^{-n_{a}+\Delta/2-1}\left(u_{a}-i\right)^{n_{a}+\Delta/2-1}}}{(4R(u_2-u_1)(\hat x^-_1-\hat x^-_2)+(\hat x^i_1-\hat x_2^i)^2)^\Delta}\nn\\[0.7em]
&=\frac{\left( -1 \right)^{n_1+n_2}\hat C_{12}}{\pi^2(4 R)^{\Delta}}\frac{1}{\tilde{x}_{12}^{\Delta}}\int_{-\infty}^{\infty}\,\,d^2 u\,\frac{\Pi_{a=1}^{2}\left(u_{a}+i\right)^{-n_{a}+\Delta/2-1}\left(u_{a}-i\right)^{n_{a}+\Delta/2-1}}{\left(u_{2}-u_{1}+(1+u_{1}u_{2})\xi_{12}/4R\right)^{\Delta}}\ .	
	\label{eq: 2pt dim red integral}
\end{align}
We focus on protected operators in six-dimensions with $\Delta\in2\mathbb{Z}$, and thus the integrand is free of branch points. Nonetheless, the integral is ill-defined; it has a whole curve of poles in the $u_1-u_2$ plane, at
\begin{equation}
u_{2}=\frac{u_{1}-\xi_{12}/4R}{1+u_{1}\xi_{12}/4R}=v_{2}\ .
\end{equation}
This simply corresponds to the line of points at which the six-dimensional 2-point function diverges:\ when the two operators are light-like separated. Thankfully however the 2-point function $\langle \fOp^{(1)}_{n_1}\fOp^{(2)}_{n_2}\rangle$ is well-defined, and admits an integral representation given by a particular regularisation of (\ref{eq: 2pt dim red integral}). To see this, we note that the Lorentzian 2-point function (\ref{eq: 6d 2pt}) should more correctly be defined in terms of a Wick rotation of a six-dimensional Euclidean correlator. By considering this Wick rotation and its effect on the integral (\ref{eq: 2pt dim red integral}) more carefully, one determines this regularised integral, and hence a finite result for $\langle \fOp^{(1)}_{n_1}\fOp^{(2)}_{n_2}\rangle$.

The full details of this calculation can be found in Section \ref{app: 2pt}, while here we present a heuristic description of its mechanism, and the answer it produces. At the computational level, the Wick rotation from Euclidean signature is neatly encapsulated by an $i\epsilon$ prescription, which has the familiar effect of shifting poles in the variables $u_a$ off the real line and into the complex plane. For instance, in (\ref{eq: 2pt dim red integral}) the effect of this prescription is to shift the pole at $u_2=v_2$ infinitesimally into the lower-half-plane, hence rendering the $u_2$ integral well-defined, and easily computable by closing the contour in the upper-half-plane and thus taking only the residue at $u_2=i$. The $u_1$ integral is then also finite and easily computed, to find
\begin{align}
	\langle \fOp^{(1)}_{n_1}\fOp^{(2)}_{n_2}\rangle &= 
	\delta_{\n_1+\n_2,0}\,d(\Delta,\n_1)\hat{C}_{12}\frac{1}{\left(z_{12}\bar{z}_{12}\right)^{\Delta/2}}\left(\frac{z_{12}}{\bar{z}_{12}}\right)^{\n_1}\ ,
	\label{eq: dim red 2pt from heuristics}
\end{align}
while clearly if $\hat{\sOp}^{(1,2)}$ and hence the $\fOp^{(1,2)}$ had had differing scaling dimensions, we would have $\langle \fOp^{(1)}_{n_1}\fOp^{(2)}_{n_2}\rangle=0$. This result is indeed consistent with the general solution (\ref{eq: 2pt WI solution}) to the 2-point scalar Ward-Takahashi identities. The coefficient $d(\Delta,n)$ is found to be
\begin{align}
d(\Delta,n) = 	 \left( -2R\, i \right)^{-\Delta} {{\n+\tfrac{\Delta}{2}-1}\choose{\n-\tfrac{\Delta}{2}}}\ ,
\end{align} where we have  used the following convention for the generalised binomial coefficient here and throughout;
\begin{align}
  {\alpha\choose n} = \begin{cases}
 	\frac{\alpha(\alpha-1)\dots(\alpha-n+1)}{n!}	\qquad& n>0		\\
 	1	\qquad& n=0	\\
 	0	\qquad& n<0
 \end{cases}\ ,
\end{align}
for any $\alpha\in\mathbb{R}$, $n\in\mathbb{Z}$. Hence, the 2-point function in \eqref{eq: dim red 2pt from heuristics} vanishes unless $n_1 \geq \Delta/2$.


\subsubsection{Fourier Resummation}


To verify that our method of integral regularisation  is consistent we can perform the inverse procedure: resum the five-dimensional 2-point functions (\ref{eq: dim red 2pt from heuristics}) to get back to the six-dimensional 2-point function. So once again consider a scalar six-dimensional operator $\hat{\sOp}$ of scaling dimension $\Delta\in2\mathbb{Z}$. Then, we have
\begin{align}
  &\langle \hat{\sOp}^{(1)}(\hat{x}_1) \hat{\sOp}^{(2)}(\hat{x}_2) \rangle = \cos^\Delta\left( \tfrac{x_1^+}{2R} \right)\cos^\Delta\left( \tfrac{x_2^+}{2R} \right) \left\langle \sOp(x_1(\hat{x}_1)) \,\sOp(x_2(\hat{x}_2)) \right\rangle		\nn\\
  &\hspace{10mm}= \cos^\Delta\left( \tfrac{x_1^+}{2R} \right)\cos^\Delta\left( \tfrac{x_2^+}{2R} \right) \sum_{\n =\Delta/2}^\infty e^{-i\n x_{12}^+/R} \left\langle \fOp_{\n }(x_1^-, x_1^i) \fOp_{-\n }(x_2^-, x_2^i) \right\rangle		\nn\\[-0.5em]
  &\hspace{10mm}= \hat{C}_{12}(-2R\, i)^{-\Delta}\cos^\Delta\left( \tfrac{x_1^+}{2R} \right)\cos^\Delta\left( \tfrac{x_2^+}{2R} \right) \left( z_{12}\bar{z}_{12} \right)^{-\tfrac{\Delta}{2}} \sum_{\n =\Delta/2}^\infty e^{-i\n x_{12}^+/R}\,\, {{\n +\tfrac{\Delta}{2}-1}\choose{\n -\tfrac{\Delta}{2}}}\left( \frac{z_{12}}{\bar{z}_{12}} \right)^{\n }	\nn\\[-0.5em]
  &\hspace{10mm}= \hat{C}_{12} \left[ \frac{2R\, i \left(\bar{z}_{12} e^{ix_{12}^+/2R} - z_{12} e^{-i x_{12}^+/2R} \right)}{\,\cos\left( \tfrac{x_1^+}{2R} \right)\cos\left( \tfrac{x_2^+}{2R} \right)} \right]^{-\Delta} 	\nn\\
  &\hspace{10mm}= \hat{C}_{12} |-2 \hat{x}_{12}^+\hat{x}_{12}^- + \hat{x}_{12}^i \hat{x}_{12}^i|^{-\Delta}	\nn\\
  &\hspace{10mm}= \hat{C}_{12} (\hat{x}_{12})^{-2\Delta}\ ,
\end{align}
as required. Note, as it is written, the infinite sum in this calculation does not converge. However the resummation is made precise by recalling the definition of the Lorentzian correlator $\langle \hat{\sOp}^{(1)}(\hat{x}_1) \hat{\sOp}^{(2)}(\hat{x}_2) \rangle$ in terms of the $i\epsilon$ prescription (\ref{eq: i epsilon}), encoding its Wick rotation from Euclidean signature. This in particular replaces $e^{-i\n x_{12}^+/R}\to e^{-\left( \epsilon_1-\epsilon_2 \right)n/R}e^{-i\n x_{12}^+/R}$, and thus $\epsilon_1>\epsilon_2$ ensures the convergence of the sum.

 This demonstrates how in principle the correlation functions of the six-dimensional CFT can be computed from the five-dimensional theory.
 
 
\subsection{3-point Functions}\label{subsec: 3pt dim red}
 

We can now pursue the dimensionally-reduced 3-point function in a similar manner.  Let us start with the six-dimensional 3-point function 
\begin{align}
  \big\langle \hat{\sOp}^{(1)}(\hat{x}_1) \hat{\sOp}^{(2)}(\hat{x}_2) \hat{\sOp}^{(3)}(\hat{x}_3) \big\rangle &= \frac{\hat{C}_{123}}{\left( \hat{x}_{12}\right)^{2\alpha_{12}}\left( \hat{x}_{23} \right)^{2\alpha_{23}}\left( \hat{x}_{31} \right)^{2\alpha_{31}}} \ ,
  \label{eq: 6d 3pt}
\end{align}
determined up to structure constants $\hat{C}_{123}$, and as in section \ref{subsec: 3pt from WIs},
\begin{align}
  \alpha_{ab} =  \Delta_a + \Delta_b - \tfrac{1}{2} \Delta_\text{T}\ .
\end{align}
We can now dimensionally reduce this to verify that it agrees with the solution to the five-dimensional Ward-Takahashi identity (\ref{eq: 3pt Ward-Takahashi identity solution}), and in doing so determine the required form of the function $H$ such that the six-dimensional interpretation holds

Once again, one must appeal to a more detailed treatment of Lorentzian correlators to perform this calculation, which is given in full detail in Section \ref{app: 3pt}. For now, we simply state the result of this calculation,
\begin{align}
  &\big\langle \fOp^{(1)}_{n_1}(x_1) \fOp^{(2)}_{n_2}(x_2) \fOp^{(3)}_{n_3}(x_3) \big\rangle\nn\\
  &\quad = \delta_{\n_1+\n_2+\n_3,0}\hat{C}_{123} \left( -2R\, i \right)^{-\tfrac{1}{2}\left( \Delta_1+\Delta_2+\Delta_3 \right)} \left( z_{12}\bar{z}_{12} \right)^{-\tfrac{1}{2}\alpha_{12}}\left( z_{23}\bar{z}_{23} \right)^{-\tfrac{1}{2}\alpha_{23}}\left( z_{31}\bar{z}_{31} \right)^{-\tfrac{1}{2}\alpha_{31}}\nn\\
  &\qquad\quad \times\sum_{m=0}^\infty {-\n_3-\tfrac{\Delta_3}{2}+\alpha_{23} - m -1\choose -\n_3-\tfrac{\Delta_3}{2}-m}{\n_1-\tfrac{\Delta_1}{2}+\alpha_{12}-m-1 \choose \n_1-\tfrac{\Delta_1}{2}-m}{\alpha_{31}+m-1 \choose m} \nn\\
  &\qquad\hspace{20mm}\times \left( \frac{z_{12}}{\bar{z}_{12}} \right)^{\n_1-m-\tfrac{1}{2}\alpha_{31}}\left( \frac{z_{23}}{\bar{z}_{23}} \right)^{-\n_3-m-\tfrac{1}{2}\alpha_{31}}\left( \frac{z_{31}}{\bar{z}_{31}} \right)^{-m-\tfrac{1}{2}\alpha_{31}}\ ,
  \label{eq: 3pt dim red from heuristics}
\end{align}
where, given $z=r e^{i\theta}$, we've chosen the branches $\left( z\bar{z} \right)^{1/2}=r$ and $\left( \tfrac{z}{\bar{z}} \right)^{1/2}=e^{i\theta}$. Note that the sum terminates at $\min\left( \n_1-\tfrac{\Delta_1}{2},-\n_3-\tfrac{\Delta_3}{2} \right)$, and thus we have a finite, regularised result.

As with the 2-point function, the binomial coefficients encode the values of $\left( \Delta_a, \n_a \right)$ such that the 3-point function is non-vanishing. We immediately have that $\big\langle \fOp^{(1)}_{n_1} \fOp^{(2)}_{n_2} \fOp^{(3)}_{n_3} \big\rangle\neq 0$ requires $\n_1\ge \tfrac{\Delta_1}{2}$ and $\n_3\le -\tfrac{\Delta_3}{2}$. We find further constraints if either\footnote{At most one of the $\alpha_{ab}$'s can be non-positive.} $\alpha_{23}\le 0$ or $\alpha_{12}\le 0$. Writing these constraints back in terms of the conformal dimensions $\Delta_a$, we in particular have that if $\Delta_1\ge \Delta_2+\Delta_3$ we additionally need $\n_2=-\n_1-\n_3 \le -\tfrac{\Delta_2}{2}$, while if $\Delta_3\ge \Delta_2+\Delta_1$ we additionally need $\n_2=-\n_1-\n_3 \ge \tfrac{\Delta_2}{2}$.

We note also that the 3-point function admits a particularly compact form in terms of a contour integral of a generating function of two variables $w_1,w_2\in\mathbb{C}$,
\begin{align}
  &\hspace{-5mm}\big\langle \fOp^{(1)}_{n_1}(x_1) \fOp^{(2)}_{n_2}(x_2) \fOp^{(3)}_{n_3}(x_3) \big\rangle\label{eq: 3pt gen func}\\
   &= \delta_{\n_1+\n_2+\n_3,0}\hat{C}_{123} \left( -2R\, i \right)^{-\tfrac{1}{2}\left( \Delta_1+\Delta_2+\Delta_3 \right)}\nn\\
  &\quad \times \left( 2\pi i \right)^{-2}\oint_{|w_1|<1} dw_1 \oint_{|w_2|<1} dw_2\,\, \Big( w_1^{-n_1+\tfrac{\Delta_1}{2}-1}w_2^{n_3+\tfrac{\Delta_3}{2}-1} \left( \bar{z}_{23} - w_2 z_{23} \right)^{-\alpha_{23}}		\nn\\
  &\qquad \hspace{60mm} \times \left( \bar{z}_{12} - w_1 z_{12} \right)^{-\alpha_{12}}\left( z_{31} - w_1 w_2 \bar{z}_{31} \right)^{-\alpha_{31}} \Big)	\nn\\
  &= \delta_{\n_1+\n_2+\n_3,0}\hat{C}_{123} \left( -2R\, i \right)^{-\tfrac{1}{2}\left( \Delta_1+\Delta_2+\Delta_3 \right)}\nn\\
  &\quad \times \text{Res}_{\{w_1=0\}} \Big[ \text{Res}_{\{w_2=0\}}\,\, \Big[ w_1^{-n_1+\tfrac{\Delta_1}{2}-1}w_2^{n_3+\tfrac{\Delta_3}{2}-1} \left( \bar{z}_{23} - w_2 z_{23} \right)^{-\alpha_{23}}		\nn\\
  &\qquad \hspace{60mm} \times \left( \bar{z}_{12} - w_1 z_{12} \right)^{-\alpha_{12}}\left( z_{31} - w_1 w_2 \bar{z}_{31} \right)^{-\alpha_{31}} \Big]\Big]\ .\nn
\end{align}
Lastly we can compare this result with the general solution (\ref{eq: 3pt Ward-Takahashi identity solution}) to ensure the two are consistent. This is indeed the case and  the function $H$ is determined to be
\begin{align}
  &\hspace{-5mm}H = \hat{C}_{123} \left( -2R\, i \right)^{-\tfrac{1}{2}\left( \Delta_1+\Delta_2+\Delta_3 \right)}\nn\\
  &\quad \times\sum_{m=0}^\infty {-\n_3-\tfrac{\Delta_3}{2}+\alpha_{23} - m -1\choose -\n_3-\tfrac{\Delta_3}{2}-m}{\n_1-\tfrac{\Delta_1}{2}+\alpha_{12}-m-1 \choose \n_1-\tfrac{\Delta_1}{2}-m}{\alpha_{31}+m-1 \choose m} \nn\\
  &\hspace{20mm}\times \left( \frac{z_{12}z_{23}z_{31}}{\bar{z}_{12}\bar{z}_{23}\bar{z}_{31}} \right)^{-\tfrac{1}{2}\alpha_{31}+\tfrac{1}{3}(\n_1-\n_3)-m}\ ,
  \label{eq: H at 3pts, finite k}
\end{align}
where, without loss of generality, we use the overall factor of $\delta_{\n_1+\n_2+\n_3,0}$ in $\big\langle \fOp^{(1)}_{n_1} \fOp^{(2)}_{n_2} \fOp^{(3)}_{n_3} \big\rangle$ to write $H$ in terms of only $\n_1$ and $\n_3$. We indeed see that as required, $H$ depends only on the phase of $z_{12}z_{23}z_{31}$.


\subsubsection{Fourier resummation}


As we did for the 2-point function, we can verify the validity of our result (\ref{eq: 3pt dim red from heuristics}) by performing the inverse Fourier transform to recover the six-dimensional 3-point function (\ref{eq: 6d 3pt}). So, consider three six-dimensional operators $\hat{\sOp}^{(1)}, \hat{\sOp}^{(2)}, \hat{\sOp}^{(3)}$ with scaling dimensions $\Delta_1,\Delta_2,\Delta_3\in 2\mathbb{Z}$, respectively. Then, we have
\begin{align}
  &\langle \hat{\sOp}^{(1)}(\hat{x}_1) \hat{\sOp}^{(2)}(\hat{x}_2) \hat{\sOp}^{(3)}(\hat{x}_3) \rangle \nn\\
  &\hspace{10mm}= \hat{C}_{123} \left( -2R\, i \right)^{-\tfrac{1}{2}\left( \Delta_1+\Delta_2+\Delta_3 \right)} \left( \bar{z}_{12} \right)^{-\alpha_{12}}\left( \bar{z}_{23} \right)^{-\alpha_{23}}\left( z_{31} \right)^{-\alpha_{31}}\nn\\
  &\hspace{20mm}\times \cos^{\Delta_1}\left( \tfrac{x_1^+}{2R} \right)\cos^{\Delta_2}\left( \tfrac{x_2^+}{2R} \right) \cos^{\Delta_3}\left( \tfrac{x_3^+}{2R} \right)e^{-i\Delta_1 x_{12}^+/2R}e^{-i\Delta_3 x_{23}^+/2R} \mathcal{S}\ ,
\end{align}
where
\begin{align}
  \mathcal{S} &= \sum_{m=0}^\infty \sum_{\n_1=0}^\infty \sum_{\n_3=0}^\infty  e^{-i\n_1x_{12}^+/R}e^{-i\n_3x_{23}^+/R} \nn\\
  &\hspace{30mm}\times {\n_3+\alpha_{23} - m -1\choose \n_3-m}{\n_1+\alpha_{12}-m-1 \choose \n_1-m}{\alpha_{31}+m-1 \choose m}\nn\\
  &\hspace{30mm}\times \left( \frac{z_{12}}{\bar{z}_{12}} \right)^{\n_1-m}\left( \frac{z_{23}}{\bar{z}_{23}} \right)^{\n_3-m}\left( \frac{z_{31}}{\bar{z}_{31}} \right)^{-m}	\nn\\
  &= \left( 1-\frac{\bar{z}_{31}}{z_{31}}e^{ix_{31}^+/R} \right)^{-\alpha_{31}}\left( 1-\frac{z_{12}}{\bar{z}_{12}}e^{-ix_{12}^+/R} \right)^{-\alpha_{12}}\left( 1-\frac{z_{23}}{\bar{z}_{23}}e^{-ix_{23}^+/R} \right)^{-\alpha_{23}}\ .
\end{align}
Hence, we find\footnote{As we saw at 2-point, the precise way to perform this resummation is to once again shift $x_a^+\to x_a^+-i\epsilon_a$ as defined in by the $i\epsilon$ prescription (\ref{eq: i epsilon}). It is then straightforwardly seen that the strict ordering $\epsilon_1>\epsilon_2>\epsilon_3>0$ ensures that all three infinite sums converge.} 
\begin{align}
  \left\langle \hat{\sOp}^{(1)}(\hat{x}_1) \hat{\sOp}^{(2)}(\hat{x}_2) \hat{\sOp}^{(3)}(\hat{x}_3) \right\rangle &= \hat{C}_{123} \prod_{a<b}^3 \left[ \frac{2R\, i \left(\bar{z}_{ab} e^{ix_{ab}^+/2R} - z_{ab} e^{-i x_{ab}^+/2R} \right)}{\,\cos\left( \tfrac{x_a^+}{2R} \right)\cos\left( \tfrac{x_b^+}{2R} \right)} \right]^{-\alpha_{ab}}\nn\\[1em]
  &= \hat{C}_{123}(\hat{x}_{12})^{-2\alpha_{12}}(\hat{x}_{23})^{-2\alpha_{23}}(\hat{x}_{31})^{-2\alpha_{31}} \ .
\end{align}
as required.


\subsection{4-point Functions}


In the six-dimensional CFT, 4-point functions can be written in terms of general functions of two conformal cross ratios:
\begin{equation}
\hat{u}=\frac{\hat{x}_{12}^{2}\hat{x}_{34}^{2}}{\hat{x}_{13}^{2}\hat{x}_{24}^{2}},\,\,\,\hat{v}=\frac{\hat{x}_{14}^{2}\hat{x}_{23}^{2}}{\hat{x}_{13}^{2}\hat{x}_{24}^{2}}\ .
\end{equation}
Although 4-point functions are not fixed by conformal symmetry, they are heavily constrained by crossing symmetry. Under $1\leftrightarrow3$ exchange, they transform as follows:
\begin{equation}
(\hat{u},\hat{v})\rightarrow(\hat{v},\hat{u})\ .
\end{equation}
Under $1\leftrightarrow2$ exchange they transform as
\begin{equation}
(\hat{u},\hat{v})\rightarrow(\hat{u}/\hat{v},1/\hat{v})\ .
\end{equation}
For instance, consider a Lorentzian correlator of four scalar operators with identical scaling dimensions. In this case, the general solution to the conformal Ward-Takahashi identities is
\begin{equation}
\langle \hat{\sOp}^{(1)}(\hat{x}_1)\hat{\sOp}^{(2)}(\hat{x}_2)\hat{\sOp}^{(3)}(\hat{x}_3)\hat{\sOp}^{(4)}(\hat{x}_4)\rangle =\frac{G(\hat{u},\hat{v})}{\left(\hat{x}_{13}^{2}\hat{x}_{24}^{2}\right)^{\Delta}}\ ,
\end{equation}
where $G$ is an unspecified function. Invariance under crossing implies the following constraints on $G$:
\begin{align}
x_1\leftrightarrow x_3 &: \,\,\,G(\hat{u},\hat{v})=G(\hat{v},\hat{u})	\ ,\nn\\
x_1\leftrightarrow x_2 &: \,\,\,G(\hat{u}/\hat{v},1/\hat{v})=\hat{v}^{\Delta}G(\hat{u},\hat{v})\ .
\end{align}
Since 4-point functions are not fixed by symmetries, in order to proceed without making an assumption about dynamics we will consider disconnected correlators in a generalised free theory. Generalised free correlators are defined as those that decompose into products of 2-point correlators, and can be formally defined in theories like the six-dimensional $(2,0)$ theory \cite{Arutyunov:2002ff}. While such correlators constitute a rather severe assumption on the theory's dynamics, they provide a particularly simple setting through which to start to explore the $x^+$ decomposition of 4-point functions.

Given a four six-dimensional scalar operators $\hat{\sOp}^{(a)}(\hat{x}_a)$ with scaling dimensions $\Delta_a$, the disconnected free 4-point function is 
\begin{align}
  \big\langle  \hat{\sOp}^{(1)}\hat{\sOp}^{(2)}\hat{\sOp}^{(3)}\hat{\sOp}^{(4)} \big\rangle = \sum_{(ab,cd)\in I} \big\langle \hat{\sOp}^{(a)}\hat{\sOp}^{(b)} \big\rangle \big\langle \hat{\sOp}^{(c)}\hat{\sOp}^{(d)} \big\rangle\ ,
  \label{eq: 4pt gen free}
\end{align}
where we have suppressed the dependence on the $\hat{x}_a$. Here, we denote by $I$ the index set\footnote{Note that the ordering of operators in the 2-point functions is fixed by that of the operators in the 4-point function, which can be seen explicitly by defining Lorentzian correlators in terms of suitable $i\epsilon$-regulated Wick rotated Euclidean correlators as in (\ref{eq: i epsilon}).} $I=\{(13,24),(12,34),(14,23)\}$. 

We can proceed to then calculate the corresponding 4-point functions of the Fourier modes of the $\hat{\sOp}^{(a)}$. As we did at 2- and 3-points, we can determine $\big\langle \fOp^{(1)}_{n_1}\fOp^{(2)}_{n_2}\fOp^{(3)}_{n_3}\fOp^{(4)}_{n_4} \big\rangle$ as an integral over variables $u_a$ of the $i\epsilon$-regulated six-dimensional 4-point function. One finds however that the integrand and ordering of the $\epsilon_a$ are such that the expression factorises into pairs of the integral expression for the 2-point function (\ref{eq: deformed 2pt integral}). All of this is to say, the factorisation persists in the same form at the level of Fourier modes in the five-dimensional theory; we have
\begin{align}
  \left\langle \fOp^{(1)}_{n_1}\fOp^{(2)}_{n_2}\fOp^{(3)}_{n_3}\fOp^{(4)}_{n_4} \right\rangle  = \sum_{(ab,cd)\in I} \left\langle \fOp^{(a)}_{n_a}\fOp^{(b)}_{n_b} \right\rangle \left\langle \fOp^{(c)}_{n_c}\fOp^{(d)}_{n_d} \right\rangle \ ,
  \label{eq: dim red 4pt}
\end{align}
in terms of the 2-point functions (\ref{eq: dim red 2pt from heuristics}) we have already determined. This in particular vanishes unless there exists two pairs of the $n_a$, say $(n_a, n_b)$ and $(n_c, n_d)$, each obeying $P_+$ momentum conservation independently: $n_a+n_b=0, n_c+n_d=0$.

We note that this result is consistent with the general Ward-Takahashi identity solution (\ref{eq: general N-point function}) for the 4-point function, and hence determines the function $H$, but we omit details of this calculation.\\

For the sake of clarity, we finally consider a particular example of the generalised free 4-point function of Fourier modes, in which all of the scaling dimensions are equal, $\Delta_a=\Delta$, and all $P_+$ momenta $p_a=n_a/R$ are of equal magnitude, $|n_a|=n\ge \Delta/2$. Up to normalisation, this corresponds simply to
\begin{equation}\big\langle  \hat{\sOp}^{(1)}\hat{\sOp}^{(2)}\hat{\sOp}^{(3)}\hat{\sOp}^{(4)} \big\rangle=\frac{1}{\left(\hat{x}_{12}^{2}\hat{x}_{34}^{2}\right)^{\Delta}}+\frac{1}{\left(\hat{x}_{14}^{2}\hat{x}_{23}^{2}\right)^{\Delta}}+\frac{1}{\left(\hat{x}_{13}^{2}\hat{x}_{24}^{2}\right)^{\Delta}} \ .
\end{equation}
Using (\ref{eq: dim red 4pt}) as well as the vanishing conditions on 2-point functions, we find only two orderings of operators such that their 4-point functions are non-zero. Firstly, we have
\begin{align}
  \big\langle \fOp^{(1)}_{n}\fOp^{(2)}_{n}\fOp^{(3)}_{-n}\fOp^{(4)}_{-n} \big\rangle &= \big\langle \fOp^{(1)}_{n}\fOp^{(3)}_{-n} \big\rangle \big\langle \fOp^{(2)}_{n}\fOp^{(4)}_{-n} \big\rangle + \big\langle \fOp^{(1)}_{n}\fOp^{(4)}_{-n} \big\rangle \big\langle \fOp^{(2)}_{n}\fOp^{(3)}_{-n} \big\rangle\ ,
\end{align}
which has a manifest $\mathbb{Z}_2\times \mathbb{Z}_2$ crossing symmetry corresponding to $x_1\leftrightarrow x_2$ and $x_3\leftrightarrow x_4$. Secondly, we have
\begin{align}
  \big\langle \fOp^{(1)}_{n}\fOp^{(2)}_{-n}\fOp^{(3)}_{n}\fOp^{(4)}_{-n} \big\rangle &= \big\langle \fOp^{(1)}_{n}\fOp^{(2)}_{-n} \big\rangle \big\langle \fOp^{(3)}_{n}\fOp^{(4)}_{-n} \big\rangle 	\ ,
\end{align}
which has a manifest $\mathbb{Z}_2$ crossing symmetry corresponding to the simultaneous swap $(x_1,x_2)\leftrightarrow(x_3,x_4)$. We further that see that there is no crossing relation between these two results, which is evident from their representation in terms of 2-point functions.

Using the parameterisation in \eqref{eq: 4pt G}, these non-zero results correspond to
\begin{align}
  G_{++--} & = \  e^{i\n\lambda_2/4R}\left( 1+e^{-in\lambda_2/2R} v^{-\Delta} \right)	\ ,	\nn\\
  G_{+-+-} &= \ e^{-in\lambda_3/4R}u^{-\Delta}\ .
\end{align}
We indeed see that $G_{++--}$ is for instance invariant under the $x_1\leftrightarrow x_2$ crossing transformation \eqref{eq: G++-- 12 crossing}. 


\subsection{Higher point functions}\label{subsec: higher pt dim red}


Let us finally make some comments about $N$-point functions. The general $N$-point function of six-dimensional operators $\hat{\sOp}^{(a)}$ can be written
\begin{align}
   \big\langle \hat{\sOp}^{(1)}(\hat{x}_1) \dots \hat{\sOp}^{(N)}(\hat{x}_N) \big\rangle = F\left(\frac{\hat{x}_{ab}^2\hat{x}_{cd}^2}{\hat{x}_{ac}^2\hat{x}_{bd}^2}\right)\prod_{a< b} |\hat{x}_{ab}|^{-2\alpha_{ab}}\ ,
   \label{eq: general 6d Npt}
\end{align}
where the $\alpha_{ab}=\alpha_{ba}$ satisfy
\begin{align}
  \sum_{\substack{b=1\\ b\neq a}}^N \alpha_{ab} = \Delta_a\ ,
\end{align}
for each $a=1,\dots N$, and $F$ is an arbitrary function of the $N(N-3)/2$ independent cross ratios. Now, we can write
\begin{align}
  \hat{x}_{ab}^2 = -2\hat{x}^+_{ab}\hat{x}^-_{ab} + \hat{x}^i_{ab}\hat{x}^i_{ab} = \frac{2R\, i \left(\bar{z}_{ab} e^{ix_{ab}^+/2R} - z_{ab} e^{-i x_{ab}^+/2R} \right)}{\,\cos\left( \tfrac{x_a^+}{2R} \right)\cos\left( \tfrac{x_b^+}{2R} \right)}\ .
\end{align}
Although the $x_a^+$ run over the range $x^+_a\in(-\pi R, \pi R)$, we nonetheless note that the $|\hat{x}_{ab}|^2$ are periodic functions of both $x^+_a$ and $x^+_b$, with period $2\pi R$. Indeed, this follows from the definition (\ref{eq: coordinate transformation}) of the coordinates $(x^+, x^-, x^i)$. Thus, the $N$-point function (\ref{eq: general 6d Npt}) is a periodic function in the variables $x^+_a$ with period $2\pi R$.

In terms of the coordinates $x$, the cross ratios now take the form
\begin{align}
  \frac{\hat{x}_{ab}^2\hat{x}_{cd}^2}{\hat{x}_{ac}^2\hat{x}_{bd}^2} = \frac{\left(\bar{z}_{ab} e^{ix_{ab}^+/2R} - z_{ab} e^{-i x_{ab}^+/2R}\right)\left(\bar{z}_{cd} e^{ix_{cd}^+/2R} - z_{cd} e^{-i x_{cd}^+/2R}\right)}{\left(\bar{z}_{ac} e^{ix_{ac}^+/2R} - z_{ac} e^{-i x_{ac}^+/2R}\right)\left(\bar{z}_{bd} e^{ix_{bd}^+/2R} - z_{bd} e^{-i x_{bd}^+/2R}\right)}\ ,
  \label{eq: cross ratio relation}
\end{align}
which we note are translationally invariant in the $x^+$ direction, i.e. they are invariant under the simultaneous shift $x^+_a\to x^+_a + \epsilon$ for all $a\in\{1,\dots, N\}$. The $N$-point function (\ref{eq: general 6d Npt}) as a whole, however, is not due to the factors of $\cos(x^+_a/2R)$.\\

Now, at least when the $\hat{\sOp}^{(a)}$ are protected operators with $\Delta\in 2\mathbb{N}$, we don't have to worry about branch points and cuts and can write
\begin{align}
   &\big\langle \hat{\sOp}^{(1)}(\hat{x}_1) \dots \hat{\sOp}^{(N)}(\hat{x}_N) \big\rangle \nn\\
   &\qquad = \left(2R\, i\right)^{-\Delta_\text{T}/2}F\left(\frac{\hat{x}_{ab}^2\hat{x}_{cd}^2}{\hat{x}_{ac}^2\hat{x}_{bd}^2}\right)\left(\prod_a \cos^{\Delta_a}\!\left(\frac{x^+_a}{2R}\right)\right)\prod_{a< b} \left(\bar{z}_{ab} e^{ix_{ab}^+/2R} - z_{ab} e^{-i x_{ab}^+/2R} \right)^{-\alpha_{ab}}\ ,
\end{align}
where $\Delta_\text{T}=\sum_a \Delta_a$. Thus, for the Weyl rescaled operators $\sOp^{(a)}$ as defined by (\ref{eq: 6d operator mapping}) we have simply
\begin{align}
   \left\langle \sOp^{(1)}(\hat{x}_1) \dots \sOp^{(N)}(\hat{x}_N) \right\rangle = \left(2R\, i\right)^{-\Delta_\text{T}/2}F\left(\frac{\hat{x}_{ab}^2\hat{x}_{cd}^2}{\hat{x}_{ac}^2\hat{x}_{bd}^2}\right)\prod_{a< b} \left(\bar{z}_{ab} e^{ix_{ab}^+/2R} - z_{ab} e^{-i x_{ab}^+/2R} \right)^{-\alpha_{ab}}\ .
\end{align}
Hence, we see that in contrast to (\ref{eq: general 6d Npt}), the $N$-point function of the Weyl rescaled operators $\sOp^{(a)}$ has $x^+$ translational symmetry. Indeed, we already knew must be the case since $\partial_+$ is an isometry of the Weyl-rescaled metric (\ref{eq: conformally compactified metric}).

Then, if the function $F$ is known, one can follow similar steps as before to dimensionally reduce this correlator to five-dimensional correlation functions for the Fourier modes of the $\sOp^{(a)}$.\\

Finally, note that the restriction to protected operators ensures that the $x^+_a$ periodicity of the $N$-point function is preserved under Weyl rescaling, essentially because $\cos^{\Delta_a}(x_a^+/2R)$ has period $2\pi R$ when $\Delta_a\in 2\mathbb{N}$. Explicitly, under $x_a^+\to x_a^+\to x_a^++2\pi R$ for a \textit{single} $a\in\{1,\dots, N\}$, we have
\begin{align}
   \left\langle \sOp^{(1)}(\hat{x}_1) \dots \sOp^{(N)}(\hat{x}_N) \right\rangle \quad\longrightarrow\quad &\left(-1\right)^{-\sum_{b\neq a}\alpha_{ab}}  \left\langle \sOp^{(1)}(\hat{x}_1) \dots \sOp^{(N)}(\hat{x}_N) \right\rangle		\nn\\
   =&\left(-1\right)^{-\Delta_a}  \left\langle \sOp^{(1)}(\hat{x}_1) \dots \sOp^{(N)}(\hat{x}_N) \right\rangle		\nn\\
   =&\left\langle \sOp^{(1)}(\hat{x}_1) \dots \sOp^{(N)}(\hat{x}_N) \right\rangle\ ,
\end{align}
and so the $N$-point function is indeed a periodic function of the $x^+_a$ with period $2\pi R$.


\chapter{Recovering the DLCQ description}\label{chap: DLCQ}


We now investigate an interesting limit of our construction, in which our operators and correlators become those of six-dimensional Minkowski space compactified on a null direction.

It is helpful to first reiterate our geometric set-up. We have coordinates $(x^+, x^-, x^i)$ which cover six-dimensional Minkowski space $\mathbb{R}^{1,5}$, as defined in (\ref{eq: coordinate transformation}). While $x^-,x^i$ are non-compact coordinates, $x^+$ runs over a finite interval $x^+\in(-\pi R, \pi R)$. In the limit $R\to\infty$, the transformation (\ref{eq: coordinate transformation}) degenerates and $(x^+, x^-, x^i)$ become normal lightcone coordinates on $\mathbb{R}^{1,5}$.
\begin{center}
\begin{minipage}{\textwidth}
\centering
\captionof{figure}{The $\mathbb{Z}_k$ orbifold of the $x^+\in(-\pi R,\pi R)$ interval}\label{fig: orbifold}
\hspace{-8mm}\includegraphics[width=102mm]{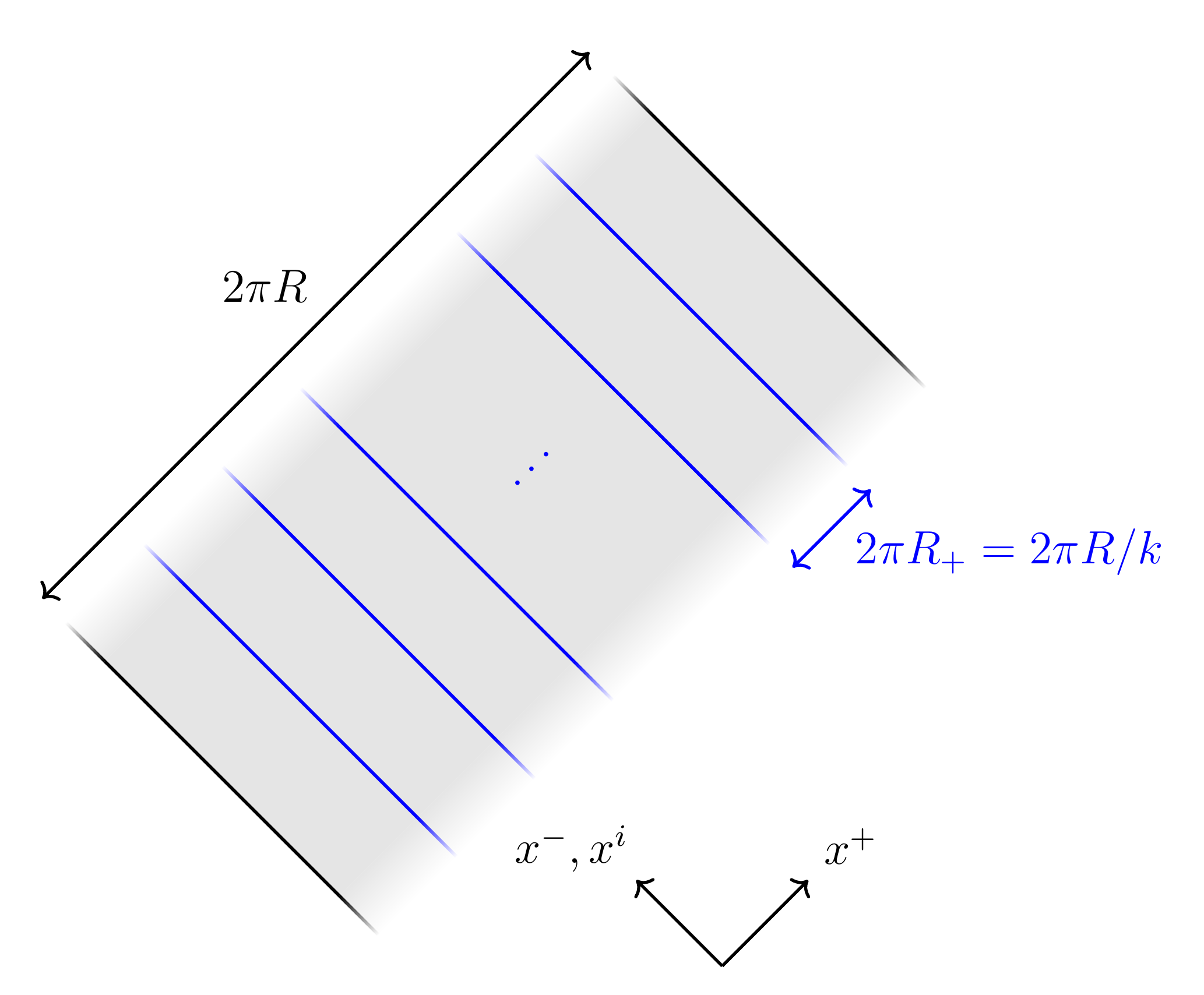}	
\end{minipage}
\end{center}
We now consider splitting the interval $x^+\in(-\pi R, \pi R)$ into $k\in\mathbb{N}$ subintervals each of length $2\pi R/k = 2\pi R_+$ where $R_+=R/k$, as depicted in Figure \ref{fig: orbifold}. We can then reduce our space of operators to only those which repeat on each subinterval; in other words, those satisfying $\sOp(x^++2\pi R_+)=\sOp(x^+)$. This defines a $\mathbb{Z}_k$ orbifold of our geometry, and the theory living on it. \\

Looking ahead a little, let us briefly comment on the holographic interpretation of this orbifold, in the case that the six-dimensional CFT is the $(2,0)$ superconformal theory dual to M-theory on $\text{AdS}_7\times S^4$. Viewing $\text{AdS}_7$ as a circle fibration over a non-compact $\mathbb{C}P^3$, one identifies $x^+$ as the coordinate along the fibre \cite{Lambert:2019jwi}. Hence, the $\mathbb{Z}_k$ orbifold we have described, in which the space of operators is restricted to those with periodicity $2\pi R_+$, corresponds to the same $\mathbb{Z}_k$ orbifold of this fibration. This is how the holograhic duality is defined for $k>1$. We note that this is analogous to what was done in the context of M2-brane theories \cite{Aharony:1997an}, where the dual geometry is $\text{AdS}_4 \times S^7$ and the $S^7$ is thought of as a circle fibration of a compact $\mathbb{C}P^3$, which is then subjected to a $\mathbb{Z}_k$ orbifold.\\

Given this $\mathbb{Z}_k$ orbifold, we now consider the combined limit in which we take $R\to\infty$ and $k\to\infty$ while holding $R_+=R/k$ fixed. In this limit, $\left( x^+, x^-, x^i \right)$ become standard lightcone coordinates on $\mathbb{R}^{1,5}$, but all operators must be periodic with period $2\pi R_+$ along the null direction $x^+$. In other words, we arrive at a null compactification $x^+\sim x^+ + 2\pi R_+$ of Minkowski space, a background first considered for the M5-brane in the DLCQ proposal of \cite{Aharony:1997an,Aharony:1997th}. For this reason, we will refer to this combined limit as the \textit{DLCQ limit}.

It is worth briefly noting that for our purposes, this should be taken as \textit{defining} what we mean by the DLCQ of the M5-brane, and of six-dimensional conformal field theory more generally. One may then question to what extent such a limit coincides with simply periodically identifying a null direction. Indeed, this question arises naturally in the context of Matrix theory---the DLCQ description of M-theory---where one seeks to define this DLCQ as a limit of space-like compactifications \cite{Seiberg:1997ad,Sen:1997we}, and in doing so provide more control than simply periodically identifying a null direction. One indeed finds that the two descriptions differ subtly \cite{Maldacena:2008wh}, as evidenced by supergravity computations \cite{Dine:1997sz}, with the former (sometimes dubbed the `$L^3$ limit' \cite{Hellerman:1997yu}) deemed correct for the eleven-dimensional picture to hold. Analogous considerations in AdS$_7$ \cite{Maldacena:2008wh} are particularly relevant for the present work if one would like to understand the DLCQ limit, as presented here, in the bulk AdS$_7$ geometry.  \\


In the remainder of this Chapter, we will investigate the behaviour of correlators in the DLCQ limit. Mirroring our analysis of the finite $R$ theory, we will first explore the constraining power of spacetime symmetries on the correlators of Fourier modes on a null compactification, recovering and extending the results of \cite{Henkel:1993sg,Aharony:1997an}.

We will then seek the precise form of these five-dimensional Fourier mode correlators by dimensionally reducing known six-dimensional correlators. Equivalently, this will determine necessary conditions on a five-dimensional theory to admit a six-dimensional interpretation, which necessarily go beyond the constraints of five-dimensional symmetries. Although this calculation can in principle be performed in a way analogous to the dimensional reduction performed at finite $R$ in Chapter \ref{chap: the view from 6d CFT}, in practice one encounters divergences essentially due to the infinite range of $x^+$ as we approach $R\to\infty$. We will therefore pursue the five-dimensional DLCQ correlators by considering the DLCQ limit of our results at finite $R$, determining the leading order asymptotics of the 2-point, 3-point and some special 4-point functions.
  


\section{Ward-Takahashi identities of the Schr\"odinger group}


We first suppose that we lie exactly at the DLCQ limit. Then, $\left( x^+, x^-, x^i \right)$ are standard lightcone coordinates on $\mathbb{R}^{1,5}$, but with periodic null coordinate $x^+ \sim x^+ + 2\pi R_+$. We first determine and solve the Ward-Takahashi identities which constrain the correlators of Fourier modes on this compact null direction $x^+$. At the level of symmetries, this once again simply corresponds to choosing a basis of operators which diagonalise the translation $\left( P_+ \right)_\partial = \partial_+$. Such operators then fall into representations of the maximal subalgebra of $\frak{so}(2,6)$ that commutes with $P_+$.

This subalgebra is guaranteed to include 16 generators obtained by simply taking the $R\to\infty$ of our $\frak{su}(1,3)$ generators at finite $R$. In fact  the algebra is enhanced as $R\to\infty$ by an additional 2 generators, corresponding to the subalgebra $\frak{u}(1)\oplus \frak{su}(2) = \text{span}\left\{ B, C^\alpha \right\}$ of spatial rotations which preserve $\Omega_{ij}$ becoming enhanced to the full $\frak{su}(2)\oplus \frak{su}(2)\cong \frak{so}(4)$. This 18-dimensional symmetry algebra has the vector field representation
\begin{align}
	\left( P_+ \right)_\partial 	&= \partial_+ 																								\ ,					\nn\\
	\left( P_- \right)_\partial 	&= \partial_- 			 																					\ ,					\nn\\
	\left( P_i \right)_\partial 	&= \partial_i							\  ,					\nn\\
	\left( M_{ij} \right)_\partial 		&= x^i \partial_j -x^j \partial_i	\ ,\nn\\
	\left( T \right)_\partial 		&= 2x^- \partial_- + x^i \partial_i																			\  ,					\nn\\
	\left( M_{i+} \right)_\partial 	&= x^i \partial_+ + x^- \partial_i  				\ ,\nn\\
	\left( K_{+} \right)_\partial 	&= x^i x^i \partial_+ +  2 ( x^- )^2 \partial_-															 + 2 x^- x^i \partial_i							 	\ . 		
\end{align}
which is identified as precisely the $z=2$ Schr\"odinger group in 4 spatial dimensions, as is standard upon any null compactification \cite{Taylor:2015glc}. Indeed, in the $R\to\infty$ limit the embedding (\ref{eq: 5d alg in terms of 6d}) of these generators inside $\frak{so}(2,6)$ degenerates to the usual embedding of the Schr\"odinger algebra in $\frak{so}(2,6)$ \cite{Maldacena:2008wh}, with the two additional rotations included by defining simply $M_{ij}=M_{ij}^{(\sixd)}$.

Hence, the correlators of Fourier modes on the compact null direction $x^+$ are constrained by the Ward-Takahashi identities of the Schr\"odinger group. We now present solutions to these Ward-Takahashi identities.\\

An operator $\sOp^\text{DCLQ}$ on our null compactified geometry can be written as a sum of Fourier modes $\fOp^\text{DCLQ}_n$, with $P_+$ eigenvalues $\delta_{P_+}\fOp^\text{DCLQ}_n=i\left( n/R_+ \right)\fOp^\text{DCLQ}_n$. The 2-point function of such operators is then completely fixed by the Ward-Takahashi identities to be
\begin{align}
  \left\langle \fOp^{(1),\text{DCLQ}}_{\n_1}\left( x_1^-, x_1^i \right) \fOp^{(2),\text{DCLQ}}_{\n_2}\left( x_2^-, x_2^i \right) \right\rangle \,\,\propto\,\, \delta_{n_1+n_2,0} \delta_{\Delta_1,\Delta_2} \left( x_{12}^- \right)^{-\Delta_1} \exp\left( \frac{i\n_1}{2R_+}\frac{|\vex_{12}|^2}{x_{12}^-} \right)\ ,
  \label{eq: 2pt DLCQ WI solution}
\end{align}
as was first found in \cite{Henkel:1993sg} and later stated in \cite{Aharony:1997an}. We note in particular that in contrast to the result at finite $R$, there is no longer any spatial power-law decay.

At 3-points, the general solution is
\begin{align}
  &\left\langle \fOp^{(1),\text{DCLQ}}_{\n_1}\left( x_1^-, x_1^i \right) \fOp^{(2),\text{DCLQ}}_{\n_2}\left( x_2^-, x_2^i \right) \fOp^{(3),\text{DCLQ}}_{\n_3}\left( x_3^-, x_3^i \right) \right\rangle\nn\\ &\quad\propto  \left[ \prod_{a<b}^3 \left( x^-_{ab} \right)^{\tfrac{1}{2}\Delta_{\text{T}} -\Delta_a-\Delta_b} e^{i\left( n_a-n_b \right)\xi_{ab}/6R_+} \right] H\left( \xi_{12}+\xi_{23}+\xi_{31} \right)\ ,
  \label{eq: 3pt DLCQ WI solution}
\end{align}
where $H$ is a general function of one variable, and we abuse notation slightly by using $\xi_{ab}$ to denote the DLCQ limit of $\xi_{ab}$ at finite $R$,
\begin{align}
  \xi_{ab} = \lim_{R\to\infty} \left( \frac{|\vex_{ab}|^2}{x_{ab}^-+\tfrac{1}{2}\Omega_{ij} x_a^i x_b^j} \right) =  \frac{|\vex_{ab}|^2}{x_{ab}^-}\ .
  \label{eq: xi DLCQ limit}
\end{align}
More generally, at $\Npt$-points we have
\begin{align}
  &\left\langle \fOp^{(1),\text{DCLQ}}_{\n_1}\left( x_1^-, x_1^i \right) \dots \fOp^{(N),\text{DCLQ}}_{\n_N}\left( x_N^-, x_N^i \right) \right\rangle\nn\\ &\quad\propto  \left[ \prod_{a<b}^N \left( x^-_{ab} \right)^{-\alpha_{ab}} e^{i\left( n_a-n_b \right)\xi_{ab}/2NR_+} \right] H\left( \frac{x_{ab}^- x_{cd}^-}{x_{ac}^- x_{bd}^-},\xi_{ab}+\xi_{bc}+\xi_{ca} \right)\ ,
  \label{eq: DLCQ Npt WI sol}
\end{align}
where the $\alpha_{ab}=\alpha_{ba}$ satisfy $\sum_{b\neq a}\alpha_{ab} = \Delta_a$ for each $a=1,\dots, N$.  


\section{Correlators from six dimensions}


We now  repeat our analysis from Chapter \ref{chap: the view from 6d CFT} and consider the dimensional reduction of known six-dimensional correlators to determine the correlators of Fourier modes on a null compactification exactly.

Suppose first that we lie exactly at the DLCQ point $k,R\to\infty$. Given an operator $\sOp$ on the non-compact spacetime, the naive construction for an operator $\sOp^\text{DLCQ}$ on the null-compactified space $x^+\sim x^+ + 2\pi R_+$ is then a sum over images,
\begin{align}
  \sOp^\text{DLCQ}\left( x^+, x^-, x^i \right) = \sum_{\s\in\mathbb{Z}}  \sOp \left( x^+ + 2\pi R_+ \s, x^-, x^i \right)\ .
\end{align}
The utility of this formulation is that we know the correlation functions of the operators $\sOp$. However, we soon run into issues if we try to use them to write down the correlation functions of our new compactified operators. In particular, for a generic $\Npt$-point function we have
\begin{align}
  &\left\langle\sOp^{(1),\text{DLCQ}} \left( x_1^+\right) \sOp^{(2),\text{DLCQ}} \left( x_2^+ \right) \dots \sOp^{(\Npt),\text{DLCQ}} \left( x_\Npt^+ \right)		\right\rangle		\nn\\
  & \quad= \sum_{\s_a\in \mathbb{Z}}  \left\langle\sOp^{(1) } \left( x_1^+ +2\pi R_+ \s_1 \right) \sOp^{(2) } \left( x_2^++2\pi R_+ \s_2\right) \dots \sOp^{(\Npt)} \left( x_\Npt^++2\pi R_+ \s_N\right)		\right\rangle		\nn\\
  & \quad= \sum_{\tilde{\s}_a\in \mathbb{Z}}  \left\langle\sOp^{(1) } \left( x_1^+ \right) \sOp^{(2) } \left( x_2^++2\pi R_+ \tilde{\s}_2\right) \dots \sOp^{(\Npt)} \left( x_\Npt^++2\pi R_+ \tilde{\s}_N\right)		\right\rangle\ ,
 \label{eq: DLCQ Npt divergent}
\end{align}
where $a=1,\dots,\Npt$, and we have suppressed dependence on the coordinates $x_a^-, x_a^i$. Here, we have used the translational symmetry of the six-dimensional correlators in the $x^+$ direction to move the first operators to $x_1^+$ for every term in the multiple sums. Hence, the terms being summed over have no dependence on $\tilde{s}_1$, yet we still sum over all $\tilde{s}_1$, introducing a divergence. Therefore, the decomposition is ill-defined, and requires regularisation. Crucially, we do \textit{not} encounter this divergence in the finite $k,R$ theory, simply due to the finite range of $x^+\in(-\pi R, \pi R)$. We can therefore regularise the DLCQ $\Npt$-point function by approaching from finite $k,R$ more carefully. \\

So consider the behaviour of correlators at finite $k,R$ as we approach the DLCQ limit. We begin with the theory at finite $R$, and perform the $\mathbb{Z}_k$ orbifold as defined at the beginning of Chapter \ref{chap: DLCQ}. This amounts to restricting to operators $\sOp^\text{orb}$ on our interval $x^+\in (-\pi R, \pi R)$ which have periodicity $2\pi R_+=2\pi R/k$. Such an operator can be written in terms of an operator $\sOp$ on the un-orbifolded spacetime by once again summing over images; but now, only $k$ images need to be summed over. As a formal device to simplify notation, we let $x^+$ be a periodic coordinate $x^+\sim x^+ + 2\pi R$. Then, this sum of images is written simply as
\begin{align}
  \sOp^\text{orb}\left( x^+, x^-, x^i \right) = \sum_{s=0}^{k-1} \sOp\! \left( x^++2\pi R_+ s, x^-, x^i \right)\ .
  \label{eq: orbifolded operator}
\end{align}
We can now once again seek the correlators of these compactified operators in terms of the known correlations of the $\sOp$. We have in particular,
\begin{align}
  &\left\langle\sOp^{(1),\text{orb}} \left( x_1^+\right) \sOp^{(2),\text{orb}} \left( x_2^+ \right) \dots \sOp^{(\Npt),\text{orb}} \left( x_\Npt^+ \right)		\right\rangle		\nn\\
  & \quad= \sum_{\s_a=0}^{k-1}  \left\langle\sOp^{(1) } \left( x_1^+ +2\pi R_+ \s_1 \right) \sOp^{(2) } \left( x_2^++2\pi R_+ \s_2\right) \dots \sOp^{(\Npt) } \left( x_\Npt^++2\pi R_+ \s_N\right)		\right\rangle		\nn\\
  & \quad= \sum_{\tilde{\s}_a=0}^{k-1}  \left\langle\sOp^{(1) } \left( x_1^+ \right) \sOp^{(2) } \left( x_2^++2\pi R_+ \tilde{\s}_2\right) \dots \sOp^{(\Npt) } \left( x_\Npt^++2\pi R_+ \tilde{\s}_N\right)		\right\rangle		\nn\\
  & \quad= k\sum_{\substack{\tilde{\s}_a=0\\a\neq 1}}^{k-1}  \left\langle\sOp^{(1) } \left( x_1^+ \right) \sOp^{(2) } \left( x_2^++2\pi R_+ \tilde{\s}_2\right) \dots \sOp^{(\Npt) } \left( x_\Npt^++2\pi R_+ \tilde{\s}_N\right)		\right\rangle	\ .
  \label{eq: DLCQ reg sum}
\end{align}
We have  utilised translational invariance, so that the summation variable that drops out contributes a multiplicative factor of $k$. Note, this use of translational invariance, in particular the $x^+$ translational invariance of the $\Npt$-point function of the $\sOp$'s, is slightly subtle. While $\partial/\partial x^+$ is not an isometry of six-dimensional Minkowski space (it is a conformal Killing vector field with \textit{non-zero} conformal factor), it \textit{is} an isometry of the Weyl rescaled metric (\ref{eq: conformally compactified metric}) on which the $\sOp$ are defined. Hence, their correlation functions are invariant under $x^+$ translations, as indeed we saw explicitly in Section \ref{subsec: higher pt dim red}. In fact, we need a little more than this; the calculation (\ref{eq: DLCQ reg sum}) is valid only if the $N$-point function of the $\sOp^{(a)}$ is \textit{periodic} in each of the variables $x^+_a$ with period $2\pi R$. This is indeed true when $\Delta\in 2\mathbb{N}$, as was shown in Section \ref{subsec: higher pt dim red}.\\

Unsurprisingly, the $\Npt$-point function diverges at large $k$ due to the overall factor of $k$. In this way, $k$ provides a regulator for the divergence encountered in (\ref{eq: DLCQ Npt divergent}). It does however make sense to consider an asymptotic expansion of the $N$-point function of orbifolded operators as $k\to\infty$. In particular, we find that the 2-point function has leading order behaviour
\begin{align}
  \left\langle\sOp^{(1),\text{orb}} \left( x_1^+\right) \sOp^{(2),\text{orb}} \left( x_2^+ \right) 	\right\rangle \sim k \sum_{\s\in\mathbb{Z}} \left\langle\sOp^{(1) } \left( x_1^+ \right) \sOp^{(2) } \left( x_2^++2\pi R_+ \s\right)\right\rangle\ ,
  \label{eq: 2pt leading k}
\end{align}
where the coefficient of this leading order term is straightforwardly seen to converge by the known functional form of the 2-point function. Similarly, at 3-points we have
\begin{align}
  &\left\langle\sOp^{(1),\text{orb}} \left( x_1^+\right) \sOp^{(2),\text{orb}} \left( x_2^+ \right)\sOp^{(3),\text{orb}} \left( x_3^+ \right) 	\right\rangle \nn\\
  &\qquad \sim k \sum_{\s_a\in\mathbb{Z}} \left\langle\sOp^{(1) } \left( x_1^+ \right) \sOp^{(2) } \left( x_2^++2\pi R_+ \s_1\right)\sOp^{(3) } \left( x_3^++2\pi R_+ \s_2\right)\right\rangle\ ,
  \label{eq: 3pt leading k}
\end{align}
where once again, the function form of the 3-point function guarantees convergence of this leading order coefficient.

At higher points, we generically encounter yet more divergent behaviour, corresponding to the orbifolding of correlators with additional degrees of translational symmetry. This is explored in more detail in section \ref{subsubsec: DLCQ 4pt and beyond}. \\

We can finally consider the behaviour of the five-dimensional correlators of Fourier modes as we approach the DLCQ limit. Using the definition (\ref{eq: orbifolded operator}) of $\sOp^\text{orb}$ in terms of images of the operator $\sOp$ along with the Fourier series decomposition (\ref{eq: 6d op Fourier decomp}), we find the expansion
\begin{align}
  \sOp^\text{orb}\left( x^+, x^-, x^i \right) = \sum_\n e^{-inx^+/R_+} \,\fOp^\text{orb}_{n}( x^-, x^i ) = \sum_\n e^{-inx^+/R_+}\Big( k\,\fOp_{kn}( x^-, x^i )\Big)\ ,
\end{align}
in terms of the Fourier modes of $\sOp$, i.e. $\fOp^\text{orb}_{n}=k\fOp_{kn}$. Then, the six-dimensional orbifolded $\Npt$-point function is reconstructed from Fourier mode correlators by
\begin{align}
  &\left\langle\sOp^{(1),\text{orb}} \left( x_1^+\right) \sOp^{(2),\text{orb}} \left( x_2^+ \right) \dots \sOp^{(\Npt),\text{orb}} \left( x_\Npt^+ \right)		\right\rangle		\nn\\
  & \quad= \sum_{n_a\in\mathbb{Z}} \exp\left( -\tfrac{i}{R_+} \textstyle{\sum_a} n_a x_a^+ \right)  \left\langle \fOp^{(1),\text{orb}}_{n_1}\fOp^{(2),\text{orb}}_{n_2}\dots \fOp^{(N),\text{orb}}_{n_N} \right\rangle 	\nn\\
  & \quad= \sum_{n_a\in\mathbb{Z}} \exp\left( -\tfrac{i}{R_+} \textstyle{\sum_a} n_a x_a^+ \right) \Big( k^N \left\langle \fOp^{(1)}_{kn_1}\fOp^{(2)}_{kn_2}\dots \fOp^{(N)}_{kn_N} \right\rangle \Big)\ ,
  \label{eq: orbifolded Npt decomposition}
\end{align}
where the sum over Fourier modes is regulated by the $i\epsilon$ prescription (\ref{eq: i epsilon}). We therefore have a formula by which to reconstruct six-dimensional correlators on the orbifolded spacetime from the five-dimensional correlators of the $(kn)^\text{th}$ Fourier modes of $\sOp$, which we have already calculated at 2-points (\ref{eq: dim red 2pt from heuristics}), 3-points (\ref{eq: 3pt dim red from heuristics}), and for a particular example at 4-points (\ref{eq: dim red 4pt}). We now present the leading order asymptotics of these results at large $k$, which can be used to \textit{define} the DLCQ limit of the theory.


\subsection{2-point Functions}


The six-dimensional 2-point function $\left\langle\sOp^{(1),\text{orb}} \sOp^{(2),\text{orb}} \right\rangle$ is decomposed into a sum over Fourier mode correlators $\big\langle\fOp^{(1),\text{orb}}_{n_1} \fOp^{(2),\text{orb}}_{n_2} \big\rangle$, which are then in turn determined in terms of Fourier mode correlators in the un-orbifolded theory. We are concerned with the leading order asymptotics of these correlators at large $k$, which from (\ref{eq: 2pt leading k}) are determined to appear at order $k$.

We find that $\big\langle\fOp^{(1),\text{orb}}_{n_1} \fOp^{(2),\text{orb}}_{n_2} \big\rangle$ vanishes, unless $\Delta_1=\Delta_2=\Delta$ and $n_1=-n_2=n>0$. In this case, we find the leading order asymptotics
\begin{align}
  \big\langle\fOp^{(1),\text{orb}}_{n} \fOp^{(2),\text{orb}}_{-n} \big\rangle &=k^2\big\langle \fOp^{(1)}_{\k \n}(x_1) \fOp^{(2)}_{-\k \n}(x_2)  \big\rangle \nn\\
 	  &\quad\sim k\,\frac{\hat{C}_\Delta \left( 2R_+ i \right)^{-\Delta} \n^{\Delta-1}}{\Gamma(\Delta)} \left( x_{12}^- \right)^{-\Delta} \exp \left( \frac{i\n}{2R_+} \frac{|\vex_{12}|^2}{x_{12}^-} \right) \ ,
 \label{eq: DLCQ 2pt from limit}
\end{align}
which is consistent with the general solution (\ref{eq: 2pt DLCQ WI solution}) to the DLCQ Ward-Takahashi identities.


\subsection{3-point Functions}


The result at 3-points is most straightforwardly approached from the generating function representation (\ref{eq: 3pt gen func}) for the 3-point function at finite $k$. The resulting 3-point functions $\langle \fOp^{(1),\text{orb}}_{\n_1}(x_1) \fOp^{(2),\text{orb}}_{\n_2}(x_2) \fOp^{(3),\text{orb}}_{\n_3}(x_3)  \rangle=\k^3\langle \fOp^{(1)}_{\k\n_1}(x_1) \fOp^{(2)}_{\k\n_2}(x_2) \fOp^{(3)}_{\k\n_3}(x_3)  \rangle$ vanish, unless $n_1+n_2+n_3=0$, $n_1>0$ and $n_3<0$. In this case, we find the large $k$ asymptotics,
\begin{align}
  &\langle \fOp^{(1),\text{orb}}_{\n_1}(x_1) \fOp^{(2)}_{-\n_1-\n_3}(x_2) \fOp^{(3)}_{\n_3}(x_3)  \rangle\label{eq: DLCQ 3pt from limit}\\
  &\,  \sim k\, \hat{C}_{123} \left( 2R_+ i \right)^{-\tfrac{1}{2}\left( \Delta_1+\Delta_2+\Delta_3 \right)} \left( x_{12}^- \right)^{-\alpha_{12}}\left( x_{23}^- \right)^{-\alpha_{23}}\left( x_{31}^- \right)^{-\alpha_{31}} \exp \left( \tfrac{in_1}{2R_+}\xi_{12}-\tfrac{in_3}{2R_+}\xi_{23} \right)	\nn\\
  &\,\hspace{6mm} \times \text{Res}_{\{w_1=0\}}\text{Res}_{\{w_2=0\}}\Bigg[ \left( e^{n_3w_2-n_1w_1} w_1^{-\alpha_{12}}w_2^{-\alpha_{23}}\left( w_1+w_2 - \tfrac{i}{2R_+}\left( \xi_{12}+\xi_{23}+\xi_{31} \right) \right)^{-\alpha_{31}} \right)	\nn\\
  &\, \hspace{44mm} +\left\{\begin{aligned}
	\exp\left( \tfrac{in_3}{2R_+} \left( \xi_{12}+\xi_{23}+\xi_{31} \right)\right) e^{n_3w_2-\left( n_1+n_3 \right)w_1} w_1^{-\alpha_{12}} w_2^{-\alpha_{31}}\qquad\\
	\times\left( w_2-w_1 + \tfrac{i}{2R_+}\left( \xi_{12}+\xi_{23}+\xi_{31} \right) \right)^{-\alpha_{23}} \text{ if } n_1+n_3\ge 0 \\[2em]
	\exp\left( -\tfrac{in_1}{2R_+} \left( \xi_{12}+\xi_{23}+\xi_{31} \right)\right) e^{\left( n_1+n_3 \right)w_1 -n_1 w_2} w_1^{-\alpha_{23}} w_2^{-\alpha_{31}}\qquad\\
	\times\left( w_2-w_1 + \tfrac{i}{2R_+}\left( \xi_{12}+\xi_{23}+\xi_{31} \right) \right)^{-\alpha_{12}} \text{ if } n_1+n_3< 0\nn
\end{aligned}\,\right]\ ,
\end{align}
where once again the $\bigO(k)$ behaviour is as expected from (\ref{eq: 3pt leading k}). This result is easily shown to be consistent with the general solution (\ref{eq: 3pt DLCQ WI solution}) for the 3-point function to the DLCQ Ward-Takahashi identities, and hence determines the function $H$, although we omit the details of this calculation.

As we saw at finite $k,R$, the 3-point function takes on special forms if any of the $\alpha_a\le 0$. In this DLCQ limit, we see this simply as the generating function reducing to a single term. Further, it is straightforward to see that if $\Delta_1\ge \Delta_2+\Delta_3$, then a non-zero 3-point function requires $n_2=-n_1-n_3<0$, while if $\Delta_3\ge \Delta_2+\Delta_1$, then a non-zero 3-point function requires $n_2=-n_1-n_3>0$. Otherwise, the 3-point function is non-zero for all $n_1>0, n_3<0$.


\subsection{4-point and Higher-point Functions}\label{subsubsec: DLCQ 4pt and beyond}


We can finally investigate the behaviour of the generalised free 4-point function (\ref{eq: 4pt gen free}) as we approach the DLCQ limit of large $k$. The form of the 4-point function, and in particular it's representation in terms of disconnected diagrams, leads to an overall $\bigO(k^2)$ asymptotic behaviour at large $k$, rather than the $\bigO(k)$ behaviour encountered at 2- and 3-points.

To see this, consider the 4-point function of orbifolded operators,
\begin{align}
  &\left\langle\sOp^{(1),\text{orb}} \left( x_1^+\right) \sOp^{(2),\text{orb}} \left( x_2^+ \right)\sOp^{(3),\text{orb}} \left( x_3^+ \right)\sOp^{(4),\text{orb}} \left( x_4^+ \right) 	\right\rangle \nn\\
  &\qquad = k \sum_{\substack{\s_a=0\\ a=1,2,3}}^{k-1} \left\langle\sOp^{(1) } \left( x_1^+ \right) \sOp^{(2) } \left( x_2^++2\pi R_+ \s_1\right)\sOp^{(3) } \left( x_3^++2\pi R_+ \s_2\right)\sOp^{(4) } \left( x_4^++2\pi R_+ \s_3\right)\right\rangle\ .
\end{align}
However, the form of $ \big\langle  \hat{\sOp}^{(1)}\hat{\sOp}^{(2)}\hat{\sOp}^{(3)}\hat{\sOp}^{(4)} \big\rangle$ implies the same factorisation for the (coordinate and Weyl) transformed operators $\sOp^{(a)}$, namely
\begin{align}
  \big\langle  \sOp^{(1)}\sOp^{(2)}\sOp^{(3)}\sOp^{(4)} \big\rangle = \sum_{(ab,cd)\in I} \big\langle \sOp^{(a)}\sOp^{(b)} \big\rangle \big\langle \sOp^{(c)}\sOp^{(d)} \big\rangle\ ,
\end{align}
which also persists for the orbifolded operators,
\begin{align}
  &\left\langle\sOp^{(1),\text{orb}} \left( x_1^+\right) \sOp^{(2),\text{orb}} \left( x_2^+ \right)\sOp^{(3),\text{orb}} \left( x_3^+ \right)\sOp^{(4),\text{orb}} \left( x_4^+ \right) 	\right\rangle \nn\\
  &\qquad = k^2 \sum_{(ab,cd)\in I} \sum_{\substack{\s_a=0\\ a=1,2}}^{k-1} \left\langle\sOp^{(a) } \left( x_a^+ \right) \sOp^{(b) } \left( x_b^++2\pi R_+ \s_1\right)\right\rangle \left\langle\sOp^{(c) } \left( x_c^+\right)\sOp^{(d) } \left( x_d^++2\pi R_+ \s_2\right)\right\rangle			\nn\\
  &\qquad= \sum_{(ab,cd)\in I} \left\langle\sOp^{(a),\text{orb}} \left( x_a^+\right) \sOp^{(b),\text{orb}} \left( x_b^+ \right)\right\rangle \left\langle\sOp^{(c),\text{orb}} \left( x_c^+ \right)\sOp^{(d),\text{orb}} \left( x_d^+ \right) 	\right\rangle\ .
  \label{eq: extra divergence for 4pt}
  \end{align}
Since we have $\left\langle\sOp^{(a),\text{orb}} \left( x_a^+\right) \sOp^{(b),\text{orb}} \left( x_b^+ \right)\right\rangle = \bigO(k)$ as $k\to \infty$, we see that the 4-point function goes as $k^2$. In other words, the additional degree of $x^+$ translational symmetry enjoyed by the terms of the generalised free correlator gives rise to an additional degree of divergence as we approach the DLCQ limit $k\to \infty$.\\

We can similarly consider the large $k$ asymptotics of the Fourier modes of the 4-point function, as determined in (\ref{eq: orbifolded Npt decomposition}). Making use of the factorisation (\ref{eq: dim red 4pt}) of the Fourier mode 4-point function at finite $k$, have simply
\begin{align}
  \big\langle \fOp^{(1),\text{orb}}_{\n_1} \fOp^{(2),\text{orb}}_{ \n_2} \fOp^{(3),\text{orb}}_{\n_3}\fOp^{(4),\text{orb}}_{\n_4}  \big\rangle &= k^4\big\langle \fOp^{(1)}_{\k \n_1} \fOp^{(2)}_{\k \n_2} \fOp^{(3)}_{\k \n_3}\fOp^{(4)}_{\k \n_4}  \big\rangle \nn\\
 	  &= \sum_{(ab,cd)\in I} \Big( k^2\big\langle \fOp^{(a)}_{\k\n_a}\fOp^{(b)}_{\k\n_b} \big\rangle \Big) \Big( k^2\big\langle \fOp^{(c)}_{\k\n_c}\fOp^{(d)}_{\k\n_d}\big\rangle  \Big)\ ,
 	  \label{eq: 4pt DLCQ final}
\end{align}
where the large $k$ asymptotics of the right-hand-side are given by (\ref{eq: DLCQ 2pt from limit}). One can further check that this result is consistent with the general solution (\ref{eq: DLCQ Npt WI sol}) to the DLCQ Ward-Takahashi identities, and hence determines the function $H$, although we omit the details of this calculation.\\

We finally make some comments on the situation at higher points. Following (\ref{eq: DLCQ reg sum}), we see that the general $\Npt$-point function goes at least as $k$ in the DLCQ limit. But in a generalisation of our analysis of the generalised free 4-point function, we find that if a $\Npt$-point function factorises into $A$ connected sub-correlators, then the correlator grows at least as $k^A$ at large $k$. Indeed, we find the curious result that in the DLCQ limit, correlators are entirely \textit{dominated} by contributions that are maximally disconnected.


\chapter{An explicit model and its symmetries}\label{chap: An explicit model and its symmetries}


We have now built the framework through which we can build six-dimensional conformal field theories using $SU(1,3)$ theories in five dimensions. In more detail, we considered a particular choice of coordinates on six-dimensional Minkowski space admitting a conformal Killing vector $\partial_+$ along an interval $x^+\in(-\pi R, \pi R)$. We were thus able to reduce a generic six-dimensional conformal field theory along this direction, and consider the correlation functions of the resulting modes. We found in particular that these modes fall into representations of $\frak{h}\subset \frak{so}(2,6)$, which is $\frak{su}(1,3)$ extended by a single central element $P_+$ encoding momentum in the $x^+$ direction. 

This chapter, and indeed the rest of this thesis, is concerned with using this framework to propose a Lagrangian description of the six-dimensional non-Abelian $(2,0)$ theory, describing multiple M5-branes in M-theory. 


\section{Five-dimensional $\Omega$-deformed super-Yang-Mills}\label{sec: five-dimensional Omega-deformed super-Yang-Mills}

Our starting point is a non-Abelian gauge theory in five-dimensions, with gauge group $SU(N_c)$. As above, we use the coordinates $(x^-, x^i)$ on $\mathbb{R}^5$, with $i,j,\dots=1,\dots,4$. In addition to its gauge field $A=(A_-, A_i)$, the theory has five scalar fields $X^I$, where $I,J,\dots=6,\dots,10$ and a real 32-component spinor $\Psi$ of $\text{Spin(1,10)}$. Finally, we also have a field $G_{ij}=-G_{ji}$ which is self-dual, $G_{ij}=\frac{1}{2}\epsilon_{ijkl} G_{kl}$. All of the fields $X^I, \Psi$ and $G_{ij}$ transform in the adjoint of $SU(N_c)$.

We choose a $32\times 32$ real representation $\{\Gamma_0,\Gamma_1,\dots, \Gamma_{10}\}$ of the $(1+10)$-dimensional Clifford algebra with signature $(-,+,\dots, +)$, and additionally define the combinations $\Gamma_\pm = (\Gamma_0 \pm \Gamma_5)/\sqrt{2}$ which project onto spinors of definite chirality under $\Gamma_{05}$. The fermion $\Psi$ then satisfies $\Gamma_{012345}\Psi = - \Psi$.

The action of the theory is then $S=\int dx^- d^4 x \mathcal{L}$, where
\begin{align}
  \mathcal{L} = \frac{k}{4\pi^2R} \text{tr} \bigg\{ & \,\frac{1}{2}F_{-i}F_{-i} - \frac{1}{2}\hat{D}_i X^I \hat{D}_i X^I + \frac{1}{2} \mathcal{F}_{ij} G_{ij}\nn\\
  &\,\,-\frac{i}{2} \bar{\Psi} \Gamma_+ D_- \Psi + \frac{i}{2} \bar{\Psi} \Gamma_i \hat{D}_i \Psi - \frac{1}{2}\bar{\Psi} \Gamma_+ \Gamma^I [X^I, \Psi] \bigg\}\ ,
  \label{eq: Lagrangian}
\end{align}
where $F=dA-iA\wedge A$ is the field strength of $A$, and $D_-, D_i$ are adjoint gauge covariant derivatives for the gauge field $A_-, A_i$, {\it i.e.}\ $D_-=\partial_- - i [A_-,\,\cdot\,]$ and $D_i=\partial_i - i [A_i,\,\cdot\,]$. In terms of these more conventional objects, we have used the corresponding $\Omega$-deformed objects,
\begin{align}
\hat D_i &= D_i - \tfrac12 \Omega_{ij}x^jD_- \ , \nn\\
  \mathcal{F}_{ij}  &= F_{ij}	 - \frac12\Omega_{ik}x^kF_{-j}+\frac12\Omega_{jk}x^kF_{-i}\ ,
\end{align}
where as before, $\Omega_{ij}$ is anti-self-dual and normalised as $\Omega_{ik}\Omega_{jk}=R^{-2}\delta_{ij}$. We also define $\hat{\partial}_i=\partial_- - \tfrac{1}{2}\Omega_{ij} x^j \partial_-$ for later use.

We see then that $G_{ij}$ acts as a Lagrange multiplier, imposing the constraint that $\mathcal{F}_{ij}$ is anti-self-dual, i.e. $\mathcal{F}^+ =0$, where $\mathcal{F}^+_{ij}=\frac{1}{2}\left(\mathcal{F}_{ij}+\frac{1}{2}\epsilon_{ijkl} \mathcal{F}_{kl}\right)$.

Note, we can define the usual notion of length, with $x^-$ and the $x^i$ each having length dimension 1. It follows that the constant $R$ also has dimensions of length, while $k$ is dimensionless.\\

Let us comment on the origin of this theory \cite{Lambert:2019jwi}. The AdS/CFT correspondence tells us that the worldvolume theory for a stack of M5-branes is dual to M-theory on an $\text{AdS}_7\times S^4$ background. In analogy with the ABJM construction \cite{Aharony:2008ug} and following the geometric considerations of \cite{Pope:1999xg}, one first considers $\text{AdS}_7$ as a timelike circle fibration $S^1\hookrightarrow \text{AdS}_7\to \tilde{\mathbb{CP}}^3$ over the non-compact complex projective space $\tilde{\mathbb{CP}}^3$. One can then write down a non-Abelian action describing the reduction along the fibre of a stack of M5-branes at fixed $\tilde{\mathbb{CP}}^3$ radius. The geometry suggests such a theory should possess eight real supercharges, and it does. Finally, one takes the embedding radius to infinity, effectively sending the stack of M5-branes to the boundary of $\text{AdS}_7$. This boundary is described by the metric (\ref{eq: conformally compactified metric}), with $x^+$ identified as the coordinate along the fibre along which we have reduced\footnote{In order to make this holographic picture precise, one should treat $x^+$ as a \textit{periodic} coordinate $x^+\sim x^+ + 2\pi R$ \cite{Pope:1999xg}, and thus restrict also to periodic operators in the six-dimensional theory}.

However, as we take this limit, certain terms in the action diverge. These terms are precisely of the form (\ref{eq: simple S_{-1}}), and thus we are still able to propose an action lying at the fixed point. This computation is precisely an $\Omega$-deformed generalisation of the null limit of five-dimensional maximal super-Yang-Mills as in Section \ref{sec: M5s}, and results in the Lagrangian (\ref{eq: Lagrangian}). Indeed, we see that in the limit $k,R\to\infty$ with $k/R$ fixed, the Lagrangian (\ref{eq: Lagrangian}) degenerates to precisely the theory (\ref{eq: final M5 action}), with $x^0$ identified with $x^-$. 

 Once again, we can look at the geometry of the M5-brane embeddings to predict the number of supercharges observed by the worldvolume theory. In particular, once the branes reach the boundary, the circle reduction breaks only one quarter of the supersymmetry, and so we can expect the theory to have 24 real supercharges. This is indeed the case, with 8 realised as rigid supersymmetries, and the remaining 16 as conformal supersymmetries \cite{Lambert:2019jwi}. Note, by considering the corresponding sub-superalgebra of the $(2,0)$ algebra \cite{Lipstein:Yangian}, one can see explicitly that in the limit $k,R\to\infty$ with $k/R$ fixed, 8 of the conformal supersymmetries degenerate to rigid ones, thus arriving precisely at the supersymmetries (\ref{eq: 5d fixed point theory SUSYs}).

\section{Spacetime symmetries of the action}\label{sec: spacetime symmetries of the action}


As well as its high degree of supersymmetry, the action has a manifest $SO(5)$ R-symmetry, rotating the scalars $X^I$. Our primary interest now however will be the spacetime symmetries of $S$.

There is a little nuance here regarding what we should expect. We interpret $S$ as describing $N$ M5-branes that have been conformally compactified, and their worldvolume theory therefore reduced along the direction $x^+\in(-\pi R, \pi R)$. Let us first suppose, like in a standard Kaluza-Klein reduction, that in doing this reduction we have truncated the spectrum of the theory maximally; in other words, the theory $S$ describes only the zero modes on the $x^+$ interval. We know then that such modes will fall into representation of $\frak{h}$ with vanishing charge under $P_+$; in other words, representations of $\frak{su}(1,3)$. Thus, we expect $S$ to admit an $\frak{su}(1,3)$ spacetime symmetry.

Conversely, just as five-dimensional maximal super-Yang-Mills is conjectured to in fact describe \textit{all} modes of a spatial compactification of M5-branes through the inclusion of local operators with non-zero instanton charge, we also propose that our action $S$ should describe all modes of the conformal compactification. Modes with non-zero charge under $P_+$ are expected to be realised only when the configuration space is extended to allow for isolated singular points, around which one measures non-zero instanton number.

What we will show first is that if we disallow such configurations, then the theory does indeed admit an $\frak{su}(1,3)$ spacetime symmetry. It will already be clear however at this point that something goes wrong when the configuration space is extended. We will indeed show in Section \ref{subsec: instantons and symmetry breaking} that in this case we precisely recover modes with charge $p_+=kn/R$ under $P_+$, and thus the spacetime symmetry algebra is extended to $\frak{h}$.

\subsection{Field variations}

It is the norm in Lorentzian theories for the action to be written in a manifestly Lorentz-invariant way, with the transformations of fields straightforward to write down. For our theory and its $\frak{su}(1,3)$ spacetime symmetry, we do not have this luxury. The transformations of the fields of the theory can in principle be derived by trial and error. However, there turns out to be an elegant and useful way to derive them from a diffeomorphism-invariant six-dimensional theory, to which the theory is subtly related. The full details of this construction can be found in Appendix \ref{app: proxy theory}. Here we state the results.

We consider a transformation generated by some $G\in\frak{su}(1,3)$. Then, the components of the gauge field transform in a standard way,
\begin{align}
  	\delta_G A_- &= - G_\partial A_- - \big(\partial_- G_\partial^-\big) A_- - \big(\partial_- G_\partial^i\big) A_i	\ ,\nn\\
  	\delta_G A_i &= - G_\partial A_i - \big(\partial_i G_\partial^-\big) A_- - \big(\partial_i G_\partial^j\big) A_j\ ,
\end{align}
i.e. as (minus) the Lie derivative along the vector field $G_\partial$.

The scalar fields $X^I$ also transform under the usual Lie derivative for scalars, except that they are also subject to a compensating Weyl rescaling for $G\in\{T, M_{i+}, K_+\}$. This Weyl factor is given by
\begin{align}
  \omega:= \frac{1}{4}\hat{\partial}_i G_\partial^i\ ,
\end{align}
which takes the values
\begin{align}
  G &= \,\makebox[10mm][l]{$T$}  \longrightarrow \quad \omega = 1		\ ,\nn\\
  G &= \,\makebox[10mm][l]{$M_{i+}$}  \longrightarrow\quad  \omega = \tfrac{1}{2}\Omega_{ij} x^j	\ ,	\nn\\
  G &= \,\makebox[10mm][l]{$K_+$}  \longrightarrow\quad \omega = 2x^-	\ ,
\end{align}
while vanishing for the remaining generators. Then, we have
\begin{align}
  \delta_G X^I = - G_\partial X^I - 2\omega X^I\ .
  \label{eq: X transformation}
\end{align}
This is indeed entirely analogous to the familiar interpretation of usual conformal field theory as a gauge fixing of a theory with both diffeomorphism and Weyl invariance. There, like here, it is a coordinated combination of a diffeomorphism and Weyl rescaling which leaves the metric invariant, and thus forms a symmetry of the gauge fixed theory.

For the fermions, we find
\begin{align}
  \delta_G \Psi = - G_\partial \Psi - \frac{1}{2}\omega\left( 5+\Gamma_{-+} \right) \Psi + \Omega_{ij}( \hat{\partial}_j \omega )\Gamma_+\Gamma_i\Psi + \frac{1}{4}\lambda^{ij}\Gamma_{ij} \Psi\ ,
\end{align}
where
\begin{align}
  \lambda^{ij} = (\hat{\partial}_j G_\partial^i) - \omega \delta_{ij} = - \lambda^{ji}\ .
\end{align}
Explicitly, we find that $\lambda^{ij}=0$ for $G\in\{P_-, P_i, T\}$, while for the remaining generators,
\begin{align}
  G &= B  &\longrightarrow \quad \lambda^{ij}&=	\tfrac{1}{2}R\,\Omega_{ij}		\ ,\nn\\
  G &= u^\alpha C^\alpha &\longrightarrow \quad \lambda^{ij}&=	u^\alpha \left( - \tfrac{1}{2}\eta^\alpha_{ij}\right)		\ ,\nn\\
  G &= v^i M_{i+} &\longrightarrow \quad \lambda^{jk}&=	\tfrac{1}{2}v^i\left(\Omega_{jk} x^i+\Omega_{ik}x^j - \Omega_{ij} x^k + \delta_{ik}\Omega_{jl} x^l - \delta_{ij} \Omega_{kl} x^l\right)		\ ,\nn\\
  G &= K_+  &\longrightarrow \quad \lambda^{ij}&= \tfrac{1}{2}\Omega_{ij} x^k x^k + \Omega_{ik} x^k x^j - \Omega_{jk} x^k x^i		\ .
\end{align}
for some constant directions $u^\alpha,v^i$. Then, when acting on the fields $A,X^I, \Psi$, we find that the $\{\delta_G\}$ with $G\in \mathcal{B}=\{P_-, P_i, B, C^\alpha, T, M_{i+}, K_+\}$ are precisely the generators of a representation of $\frak{su}(1,3)$. It is notationally convenient to introduce a trivial variation $\delta_{P_+}$, acting as $\delta_{P_+}X^I=0$, $\delta_{P_+}\Psi=0$ and $\delta_{P_+}A=0$. We further write $\mathcal{B}_+ = \mathcal{B} \cup \{P_+\}$. Then, the $\{\delta_G\}_{G\in \mathcal{B}_+}$ generate a representation of $\frak{h}$ with $P_+$ trivially represented, and we have $[\delta_{G_1},\delta_{G_2}]=\delta_{[G_1,G_2]}$ for all $G_1,G_2\in\frak{h}$, with brackets as given in (\ref{eq: extended su(1,3) algebra})--(\ref{eq: commutators with rotations}).\\

We finally come to the Lagrange multiplier field $G_{ij}$, which arises in a more complicated fashion from the six-dimensional proxy theory. We find
\begin{align}
  \delta_G G_{ij} &= - G_\partial G_{ij} - 4\omega G_{ij} - \left( \lambda^{mi}G_{mj} - \lambda^{mj} G_{mi}  \right)			\nn\\
  &\qquad + 2R^2\left( \Omega_{im}(\hat{\partial}_m \omega) F_{-j} - \Omega_{jm}(\hat{\partial}_m \omega) F_{-i}+ \varepsilon_{ijkl} \Omega_{km}(\hat{\partial}_m \omega) F_{-l}  \right)\ ,
  \label{eq: Gij infinitesimal variation}
\end{align}
as well as $\delta_{P_+}G_{ij}=0$. Note that the right hand-side is indeed self-dual on $\{i,j\}$. We note in particular that, in contrast the other fields, the algebra only closes on $G_{ij}$ on the constraint surface $\mathcal{F}^+=0$. In particular, for each $G_1,G_2\in\mathcal{B}_+$ we have $[\delta_{G_1},\delta_{G_2}]G_{ij} = \delta_{[G_1,G_2]}G_{ij}$ except for
\begin{align}
  [\delta_{M_{i+}}, \delta_{M_{j+}}]G_{kl} &= \delta_{[M_{i+}, M_{j+}]}G_{kl} + \bar{\delta}_{ij}G_{kl}		\nn\\
  [\delta_{M_{i+}}, \delta_{K_+}]G_{jk} &= \delta_{[M_{i+}, K_+]}G_{jk} + 2 x^l\bar{\delta}_{il}G_{jk}\ ,
  \label{eq: Gij algebra extension}
\end{align}
where
\begin{align}
  \bar{\delta}_{ij} G_{kl}			&=2\left(\delta_{ik} \mathcal{F}^+_{jl}-\delta_{il} \mathcal{F}^+_{jk}-\delta_{jk} \mathcal{F}^+_{il}+\delta_{jl} \mathcal{F}^+_{ik}\right)\ .
\end{align}
A discussion of the origin of this extension to the algebra can be found in Appendix \ref{app: proxy theory}.

Note that we have $\mathcal{F}_{kl}\bar{\delta}_{ij}G_{kl}=0$, which ensures that $\bar{\delta}$ is a symmetry of the Lagrangian. Indeed, since $G_{ij}$ appears only algebraically in $\mathcal{L}$, we have local symmetries $\epsilon(x)\bar{\delta}$ for any function $\epsilon(x)$, and thus we should think of $\bar{\delta}$ as generating an auxiliary gauge symmetry which become trivial on the constraint surface.\\

Finally, note that under Lifshitz scalings as generated by $T$, we have
\begin{align}
  	X^I(x^-, x^i)\quad &\longrightarrow\quad \lambda^{-2}  X^I(\lambda^{-2}x^-, \lambda^{-1}x^i)		\ ,\nn\\
  	A_-(x^-, x^i)\quad &\longrightarrow\quad \lambda^{-2}  A_-(\lambda^{-2}x^-, \lambda^{-1}x^i)		\ ,\nn\\
  	A_i(x^-, x^i)\quad &\longrightarrow\quad \lambda^{-1}  X^I(\lambda^{-2}x^-, \lambda^{-1}x^i)		\ ,\nn\\
  	G_{ij}(x^-, x^i)\quad &\longrightarrow\quad \lambda^{-4}  G_{ij}(\lambda^{-2}x^-, \lambda^{-1}x^i)	\ ,\nn\\
  	\Psi_+(x^-, x^i)\quad &\longrightarrow\quad \lambda^{-3}  \Psi_+(\lambda^{-2}x^-, \lambda^{-1}x^i)	\ ,\nn\\
  	\Psi_-(x^-, x^i)\quad &\longrightarrow\quad \lambda^{-2}  \Psi_-(\lambda^{-2}x^-, \lambda^{-1}x^i)	\ ,
  	\label{eq: Lifshitz scalings of fields}
\end{align}
where we denote by $\Psi_\pm$ the components of $\Psi$ with definite chirality under $\Gamma_{-+}=\Gamma_{05}$, so that $\Gamma_{-+}\Psi_\pm = \pm \Psi_\pm$.

\subsection{Repackaging as primaries}

At this stage, we seem to have two conflicting perspectives. On one hand, in Section \ref{sec: an exotic spacetime symmetry}, we built general representations of $\frak{h}$ by defining the notion of primaries and descendants, in analogy with the representation theory of the conformal algebra. On the other, we found the transformation properties of the fields in our explicit theory (\ref{eq: Lagrangian}) in a somewhat ad hoc fashion, with results that do not immediately appear compatible.

However, with the exception of $G_{ij}$, the fields of our theory fall into off-shell representations of $\frak{h}$. As such, they must be expressible in terms of primary fields (with $p_+=0$) and their descendants. Let us now present this reorganisation of fields, with a focus on the bosonic field content.

Firstly, by comparison between (\ref{eq: X transformation}) and the transformations (\ref{eq: 5d algebra action on fields}) of a generic primary field, we identify each of the components of $X^I$ simply as a scalar primary, with $\Delta=2$ and $p_+=0$.

The situation is less straightforward with the gauge field $A$, which is not a primary field. Instead, the degrees of freedom contained in $A$ are reorganised into a 4-component field $\hat{A}_i$ and a singlet $\hat{A}_-$. These are each $p_+=0$ primary fields, with the following scaling dimensions and rotation representations,
\begin{align}
	&\hat{A}_i 	= A_i - \tfrac{1}{2}\Omega_{ij}x^j A_-&	 &\hspace{-25mm}\implies
  		\begin{cases}
			\Delta				&=	1		\\
			r[B]\hat{A}_i		&= 	-\tfrac{1}{2}R\,\Omega_{ij} \hat{A}_j		\\
			r[C^\alpha]\hat{A}_i		&=  \tfrac{1}{2}\eta^\alpha_{ij}\hat{A}_j
		\end{cases}	\label{eq: Ai hat def}\ ,\\
	&\hat{A}_- 	= A_- - \tfrac{1}{2} R^2 \Omega_{ij} \hat{\partial}_i\hat{A}_j& 	 &\hspace{-25mm}\implies
  		\begin{cases}
			\Delta				&=	2		\\
			r[B]\hat{A}_-		&= 	0		\\
			r[C^\alpha]\hat{A}_-		&=  0
		\end{cases}\ .
\label{eq: A hat def}
\end{align}
Note finally that the combination $\Omega_{ij}[\hat{A}_i, \hat{A}_j]$ is a scalar primary of dimension $\Delta=2$, and so we can add any multiple of it to $\hat{A}_-$ while preserving its representation under $\frak{h}$.

\subsection{Variation of the Lagrangian}

We have shown that the full field content of the theory falls into representations of $\frak{h}$ (at least on the constraint surface, in the case of $G_{ij}$) under the variations $\delta_G$. We reiterate, these are indeed representations of $\frak{su}(1,3)\subset \frak{h}$, as $P_+$ is trivially represented: $\delta_{P_+}\fields=0$ on all fields $\fields=A, X^I, \Psi, G_{ij}$.

Further, we showed that the anomalous extension to the symmetry algebra (\ref{eq: Gij algebra extension}) when acting on $G_{ij}$ is parameterised by an additional variation $\bar{\delta}_{ij}$, which explicitly annihilates the Lagrangian. Thus, the Lagrangian (\ref{eq: Lagrangian}) transforms in a representation of $\frak{su}(1,3)\subset\frak{h}$.

So let us state the variation of the Lagrangian $\mathcal{L}$. In addition to the trivial $\delta_{P_+}\mathcal{L}=0$, for $G\in\{P_-, P_i, B, C^\alpha, T\}$ we find
\begin{align}
  -\delta_{P_-}\mathcal{L}	&= \partial_- \mathcal{L}		\ ,\nn\\
  -\delta_{P_i}\mathcal{L}	&= \partial_- \left( \tfrac{1}{2}\Omega_{ij} x^j \mathcal{L} \right) + \partial_i \mathcal{L} 	\ ,\nn\\
  -\delta_{B}\mathcal{L}		&= \partial_i \left( \tfrac{1}{2} \Omega_{ij} x^j\mathcal{L} \right) 			\ ,\nn\\
  -\delta_{C^\alpha}\mathcal{L}	&= \partial_i \left( -\tfrac{1}{2} \eta^\alpha_{ij} x^j\mathcal{L} \right) 				\ ,\nn\\
  -\delta_{T}\mathcal{L}		&= \partial_- \left( 2x^-\mathcal{L} \right) + \partial_i \left( x^i\mathcal{L} \right) 	\ ,
\end{align}
and hence with suitable boundary conditions on the 4-sphere $S^4_\infty$ at infinity, we have $\delta_G S=0$. More care must be taken, however, in the case of $G\in\{M_{i+}, K_+\}$. We find\footnote{Here and throughout, we take $\star$ to denote the Hodge star with respect to the Euclidean metric on $\mathbb{R}^5$, which satisfies $\star^2=1$ on all forms, and $\left( \star d \star \omega \right)_{\alpha_1\dots \alpha_{p-1}}=\partial^\beta \omega_{\alpha_1\dots \alpha_{p-1} \beta}$ for generic $p$-form $\omega$}
\begin{align}
  -\delta_{M_{i+}} \mathcal{L} &= \star \left( dx^i \wedge \left( \frac{k}{8\pi^2R}\tr\left( F\wedge F \right) \right) \right)		\nn\\
  &\qquad+   \partial_- \left[ \left( \frac{1}{2}\Omega_{ij}x^- x^j - \frac{1}{8}x^j x^j x^i \right)\mathcal{L} -\frac{k}{16\pi^2R} x^i \,\text{tr}\big( X^I X^I \big) \right] 	 				\nn\\
  &\qquad+ \partial_j \left[ \frac{1}{4}\left( 2\Omega_{ik} x^k x^j + 2\Omega_{jk} x^k x^i - \Omega_{ij} x^k x^k +4x^- \delta_{ij} \right)\mathcal{L} -\frac{k}{8\pi^2R} \Omega_{ij} \,\text{tr}\big( X^I X^I \big) \right]	\ ,\nn\\
  -\delta_{K_+} \mathcal{L} &= \star \left( d\left( x^i x^i \right) \wedge \left( \frac{k}{8\pi^2R}\tr\left( F\wedge F \right) \right) \right)		\nn\\
  &\qquad+   \partial_- \left[ \left( 2\left( x^- \right)^2-\frac{1}{8}(x^i x^i)^2 \right)\mathcal{L} -\frac{k}{8\pi^2R} x^ix^i \,\text{tr}\big( X^I X^I \big) \right] 	 				\nn\\
  &\qquad+ \partial_i \left[ \left( \frac{1}{2}\Omega_{ij} x^j x^k x^k+2 x^- x^i \right)\mathcal{L} +\frac{k}{4\pi^2R} \Omega_{ij}x^j \,\text{tr}\big( X^I X^I \big) \right]\ .
\end{align}
If we require that the gauge field $A$ is globally defined and regular everywhere, then we can write
\begin{align}
  dx^i \wedge \left( \frac{k}{8\pi^2R}\tr\left( F\wedge F \right) \right) &= d\left( \frac{k}{8\pi^2R} x^i \tr\left( F\wedge F \right) \right)	\ ,\nn\\
  d\left( x^i x^i \right) \wedge \left( \frac{k}{8\pi^2R}\tr\left( F\wedge F \right) \right) &= d\left( \frac{k}{8\pi^2R} x^i x^i \tr\left( F\wedge F \right) \right)\ ,
\end{align}
and hence, in both cases, $\delta_G\mathcal{L}$ is a total derivative, and for suitable boundary conditions on $S^4_\infty$ we have $\delta_G S=0$.


\section{Instantons and symmetry breaking}\label{subsec: instantons and symmetry breaking}


We instead allow the configuration space of our theory to be broader. It is clear that all $SU(N_c)$-principal bundles $P\to \mathbb{R}^5$ are trivialisable. Consider instead however removing a set of points $\{x_a\}_{a=1}^M$ and considering $SU(N_c)$-principal bundles over $M_5=\mathbb{R}^5\setminus\{x_a\}_{a=1}^M$. Such bundles are then characterised by the integral of the second Chern class over small 4-spheres surrounding each of the $x_a$, which are quantised as
\begin{align}
  n_a = \frac{1}{8\pi^2} \int_{S^4_a} \text{tr}\left( F\wedge F \right) \in\mathbb{Z}\ ,
  \label{eq: flux quantisation}
\end{align}
with $S^4_a$ denoting a small 4-sphere surrounding the puncture at $x_a$. We then call each pair $(x_a, n_a)$ an \textit{instanton insertion}, with $x_a\in\mathbb{R}^5$ the instanton insertions's \textit{position}, and $n_a\in\mathbb{Z}$ its \textit{charge}. We could also in principle consider allowing for non-zero instanton number on $S^4_\infty$, but we instead consider only configurations with
\begin{align}
  \frac{1}{8\pi^2}\int_{S^4_\infty} \tr \left( F\wedge F \right) = 0\ .
  \label{eq: vanishing flux at infinity}
\end{align}
Since the finite $SU(1,3)$ transformations generated by $M_{i+}$ and $K_+$ move the point at infinity, this is chosen as a convenience, rather than a restriction.

Note then that since $d\,\tr\left( F\wedge F \right)=0$ throughout $M_5$, we have
\begin{align}
  0 = \frac{1}{8\pi^2}\int_{S^4_\infty} \tr \left( F\wedge F \right) = \sum_{a=1}^M \frac{1}{8\pi^2} \int_{S^4_a} \text{tr}\left( F\wedge F \right) = \sum_{a=1}^M n_a\ .
  \label{eq: sum of charges is zero}
\end{align}
Thus, the data of the bundle is contained within the set of instanton insertions $\{(x_a, n_a)\}_{a=1}^M$, with the $x_a$ distinct and the $n_a$ summing to zero. Necessarily, $M_5$ must now be covered in a number of patches, on each of which $A$ is defined. One can however consider a limit of such an open cover, such that $A$ is now globally defined and regular except along 1-dimensional strings where it is singular. These strings, which are analogous to the Dirac string, extend between the insertions $x_a$. Then, the integral of the Chern-Simons 3-form on any $S^3$ through which such a string is threaded is quantised, ensuring that (\ref{eq: flux quantisation}) is satisfied. We will however not need the details of this picture here, but will later in Chapter \ref{chap: worldlines} explore gauge field configurations with precisely this form.\\

We are now able to extend our field content back to the whole of $\mathbb{R}^5$, so long as we allow for particular singular behaviour of the field strength $F$. We in particular have
\begin{align}
  d\left( \frac{1}{8\pi^2}\tr\left( F\wedge F \right) \right) = d^5 x\,\sum_{a=1}^M n_a \delta^{(5)}(x-x_a)\ .
\end{align}
Such configurations with maximal symmetry about the points $x_a$ will behave as
\begin{align}
  \frac{1}{8\pi^2}\tr\left( F\wedge F \right) \sim  \frac{n_a}{\pi^2}\star d\left( \frac{1}{|x-x_a|^3} \right)\ ,
  \label{eq: sph symmetric instanton}
\end{align}
as we approach $|x-x_a|\to 0$, where here $|x|^2=(x^-)^2+|\vex|^2$. However, we more generally only expect the pullback to the $S^4$ surrounding $x_a$ to behave as
  \begin{align}
  \left.\frac{1}{8\pi^2}\tr\left( F\wedge F \right)\,\right|_{S^4} \sim n_a\Omega_4\ ,
\end{align}
as we approach $|x-x_a|\to 0$, where $\Omega_4$ encodes angular dependence, and satisfies $\int_{S^4}\Omega_4=1$. Thus, the components of $\left.\frac{1}{8\pi^2}\tr\left( F\wedge F \right)\,\right|_{S^4}$ in Cartesian coordinates on $\mathbb{R}^5$ go as $|x-x_a|^{-4}$ as we approach $|x-x_a|\to 0$.

Explicit examples of such configurations on $S^4$ can be constructed by suitable stereographic projection from corresponding configurations on $\mathbb{R}^4$. The minimal such construction \cite{Bergman:2016avc}, in which the $SU(2)$ BPST instanton of size $\rho$ is mapped to $S^4$, corresponds to $n_a=\pm 1$, with $\rho=1$ producing the spherically symmetric result (\ref{eq: sph symmetric instanton})

Later in Chapter \ref{chap: worldlines}, we will construct explicit configurations on $\mathbb{R}^5$ which feature an arbitrary number of instanton insertions at points $x_a$, as well as vanishing flux on $S^4_\infty$ as in (\ref{eq: vanishing flux at infinity}). The details of such configurations, which additionally have finite action and satisfy the constraint $\mathcal{F}^+=0$ imposed by $G_{ij}$, will not be required here.\\

So, we now take the configuration space of our theory to be extended to the disjoint union of subspaces, on each of which we specify instanton insertions $\{(x_a,n_a)\}$. Note, for the sake of later notational convenience, we allow for any of the $n_a$ to be zero, in which case $F$ can be smoothly extended to $x_a$. 

Importantly, $SU(1,3)$ still admits a natural action on this extended configuration space. In particular, the form of $gA$ ensures that
\begin{align}
  d\left( \frac{1}{8\pi^2}\tr\left( F[A]\wedge F[A] \right) \right) = d^5 x\,\sum_{a=1}^M n_a \delta^{(5)}(x-x_a) \nn\\ 
  \implies \quad  d\left( \frac{1}{8\pi^2}\tr\left( F[gA]\wedge F[gA] \right) \right) &= d^5 ( x g^{-1} )\,\sum_{a=1}^M n_a \delta^{(5)}(xg^{-1}-x_a)		\nn\\
  &= d^5 x\sum_{a=1}^M n_a \delta^{(5)}(x-x_ag)\ ,
  \label{eq: moving of instanton insertions}
\end{align}
and hence if $A$ has instanton insertions $\{(x_a, n_a)\}_{a=1}^M$, the transformed field $gA$ has instanton insertions $\{(x_ag, n_a)\}_{a=1}^M$.\\

Let us now return to the $\su(1,3)$ variation of the Lagrangian. We find now that in the presence of instanton insertions, the variation of $\mathcal{L}$ under $M_{i+}, K_+$ is no longer a total derivative, and the action is no longer invariant. We find
\begin{align}
  	\delta_{M_{i+}}\mathcal{L} &= \frac{k}{R}\sum_{a=1}^M n_a x_a^i \delta^{(5)}(x-x_a) + \star\, d \left( \dots \right)		\ ,\nn\\
  	\delta_{K_+}\mathcal{L} &= \frac{k}{R}\sum_{a=1}^M n_a x_a^i x_a^i \delta^{(5)}(x-x_a) + \star\, d \left( \dots \right)		\ ,\nn\\
\end{align}
and hence, for suitable boundary conditions on $S^4_\infty$, we have
\begin{align}
  	\delta_{M_{i+}}S &= \frac{k}{R}\sum_{a=1}^M n_a x_a^i		\ ,\nn\\
  	\delta_{K_+}S &= \frac{k}{R}\sum_{a=1}^M n_a x_a^i x_a^i		\ .
  	\label{eq: M and K action variation}
\end{align}
Thus, we find that the classical action is no longer invariant under $SU(1,3)$. However, we note that the variation of the action is \textit{local} to the punctures $\{x_a\}$. It is precisely this fact that allows for a recasting of the classical non-invariance of $S$ as a symmetry \textit{deformation} in the quantum theory.

Alternatively, one could explore introducing in the action a coupling to massive point particles localised to the points $\{x_a\}$, so as to try to cancel the non-invariance (\ref{eq: M and K action variation}), although we do not explore this possibility here. \\

However, before exploring this we finally note the transformation of the action under the finite transformations generated by $M_{i+}$ and $K_+$, which are found by exponentiating the infinitesimal results (\ref{eq: M and K action variation}).

Let $\fields=A, X^I, \Psi, G_{ij}$ be shorthand for the set of fields of the theory, and suppose that the gauge field $A$ has insertions $\{(x_a, n_a)\}_{a=1}^M$. Then, we find
\begin{align}
  S[e^{\epsilon^i M_{i+}}\fields] = S[\fields] - ik \sum_{a=1}^M n_a \log \left( \frac{\overline{\mathcal{M}_\epsilon(x_a)}}{\mathcal{M}_\epsilon(x_a)} \right)\ ,
\end{align}
where given some 4-vector $\alpha^i$, we define
\begin{align}
  \mathcal{M}_\alpha(x) &= \left( 1-\tfrac{1}{2}\Omega_{ij}\alpha^i x^j+\tfrac{1}{16R^2}|\vec{\alpha}|^2|\vex|^2 \right) - \frac{i}{4R}\left( |\vec{\alpha}|^2 x^- + 2\alpha^i x^i \right)\nn\\
  &= 1+z(x,(0,\alpha^i))-z(x,0) - z(0,(0,\alpha^i))-\frac{i}{4R}|\vec{\alpha}|^2 z(x,0)\ ,
  \label{eq: curly M definition}
\end{align}
and as we first saw in Chapter \ref{chap: SU(1,3) theories}, we have the complex distance
\begin{align}
  z(x_1,x_2)=\left( x_1^--x_2^-+\frac{1}{2}\Omega_{ij} x_1^i x_2^j \right) + \frac{i}{4R}(x_1^i-x_2^i)(x_1^i-x_2^i) = -\bar{z}(x_2,x_1)\ .
\end{align}
Equivalently, we can write
\begin{align}
  \exp\left( {iS[e^{\epsilon^i M_{i+}}\fields]} \right) = e^{iS[\fields]} \prod_{a=1}^M \left( \frac{\overline{\mathcal{M}_\epsilon(x_a)}}{\mathcal{M}_\epsilon(x_a)} \right)^{kn_a}\ .
  \label{eq: finite action variation M}
\end{align}
Similarly, we find
\begin{align}
  S[e^{\epsilon K_+}\fields] = S[\fields] - ik\sum_{a=1}^M n_a\log \left( \frac{1-2\epsilon \bar{z}(x_a,0)}{1-2\epsilon z(x_a,0)} \right)\ ,
\end{align}
or equivalently,
\begin{align}
  \exp\left( {iS[e^{\epsilon K_+}\fields]}  \right)= e^{iS[\fields]} \prod_{a=1}^M \left( \frac{1-2\epsilon \bar{z}(x_a,0)}{1-2\epsilon z(x_a,0)} \right)^{kn_a}\ .
  \label{eq: finite action variation K}
\end{align}
Let us briefly note that the multiplicative factors appearing in the finite variations (\ref{eq: finite action variation M}) and (\ref{eq: finite action variation K}) have branch cuts for generic $k$. This is suggestive of the possible existence if closed loops in configuration space around which $e^{iS[\fields]}$ picks up a non-trivial phase, indicating a lack of single-valuedness of $e^{iS[\fields]}$ on configuration space. One can indeed explicitly construct such loops, implying that the parameter $k$ can only take discrete values $k\in \frac{1}{2}\mathbb{Z}$ when we allow for non-trivial instanton insertions \cite{Lambert:SymmEnhance}.

In Section \ref{subsec: extended symmetry} we will require $k\in\mathbb{Z}$ in order that the theory admits a six-dimensional interpretation. It is unclear whether any such novel interpretation holds when $k\in \{\frac{1}{2},\frac{3}{2},\dots\}$ or whether such cases can be rules out by other means. 


\section{Quantum recovery}\label{subsec: quantum recovery}


We now consider the fate of our $\frak{su}(1,3)$ symmetry in the corresponding quantum theory. Despite the non-invariance of the action, we find a set of Ward-Takahashi identities satisfied by all correlation functions of the theory.

Such identities are of the usual form, in particular involving the divergence of some vector current; the Noether current for the respective symmetry. The derivation of such \textit{local} Ward-Takahashi identities and corresponding currents is left until Section \ref{subsec: local WTIs}. We first derive the corresponding \textit{global} identities---also obtainable by integrating their local counterparts over $\mathbb{R}^5$---directly, so as to elucidate the quantum recovery of the theory's symmetries most straightforwardly.\\

 First suppose we forbid instanton insertions, and define the configuration space of the theory to have globally regular $F$.  We can then formally define correlation functions of operators $\fOp^{(1)},\dots,\fOp^{(N)}$ by the path integral\footnote{We are here a little cavalier about operator ordering, which is discussed in more detail in Section \ref{sec: relation to correlation functions}}
\begin{align}
  \left\langle \fOp^{(1)}(x_1)\dots \fOp^{(N)}(x_N) \right\rangle = \int D\fields\, \fOp^{(1)}(x_1)\dots \fOp^{(N)}(x_N) e^{iS[\fields]}\ ,
  \label{eq: basic path integral}
\end{align}
where we use $\fields$ to denote the fields $X^I, A, \Psi, G_{ij}$ of the theory, and the $\fOp^{(a)}$ are generically composite functions of $X^I, \Psi, A$ and their derivatives. The partition function is $\mathcal{Z}=\langle\mathds{1}\rangle$.

Symmetries are then realised by Ward-Takahashi identities for correlations functions. Under some $SU(1,3)$ transformation $g$, we have transformed fields $\fields'=g\fields$. Making use of the fact that $S[\fields']=S[\fields]$, and assuming $D\fields' = D\fields$, we have
\begin{align}
  \langle \fOp^{(1)'}(x_1)\dots \fOp^{(N)'}(x_N) \rangle &= \int D\fields\, \fOp^{(1)'}(x_1)\dots \fOp^{(N)'}(x_N) e^{iS[\fields]}		\nn\\
  &= \int D\fields'\, \fOp^{(1)'}(x_1)\dots \fOp^{(N)'}(x_N) e^{iS[\fields']}		\nn\\
  &= \int D\fields\, \fOp^{(1)}(x_1)\dots \fOp^{(N)}(x_N) e^{iS[\fields]}		\nn\\
  &=  \langle \fOp^{(1)}(x_1)\dots \fOp^{(N)}(x_N) \rangle\ ,
  \label{eq: global WTI, no instantons}
\end{align}
where viewing $\fOp[\fields]$ as a composite function of the fields $\fields$, we have $\fOp'=\fOp[\fields']$. This is the global Ward-Takahashi identity for the symmetry $g$. We can equivalently write the infinitesimal form,
\begin{align}
  \sum_{a=1}^A \left\langle \fOp^{(1)}(x_1)\dots \delta_{G}\fOp^{(a)}(x_a)\dots \fOp^{(N)}(x_N) \right\rangle = 0\ ,
  \label{eq: usual Ward-Takahashi identities}
\end{align}
for each $G\in \frak{su}(1,3)$.\\

Let us now consider what changes when we allow for instanton insertions. The configuration space of the theory is now the disjoint union of subspaces on which we specify instanton insertions $\{(x_a,n_a)\}$. Hence, in calculating the correlation function of a set of operators $\fOp_a$, we must also specify which of these subspaces we perform the path integral over. We are lead then to define
\begin{align}
  \left\langle \fOp^{(1)}(x_1)\dots \fOp^{(N)}(x_N) \right\rangle_{\{(x_a,n_a)\}} := \int_{\{(x_a,n_a)\}} D\fields\, \fOp^{(1)}(x_1)\dots \fOp^{(N)}(x_N) e^{iS[\fields]}\ ,
\end{align}
where the path integral is performed only over configurations with instanton insertions $\{(x_a,n_a)\}_{a=1}^N$. Note, the operator insertion points are the same as the instanton insertion points, denoted $x_a$. This is done without loss of generality, since we allow for any of the operators $\fOp^{(a)}$ to be the identity operator $\mathds{1}$, and we allow any of the $n_a=0$.

Next, consider some $SU(1,3)$ transformation $g$, with corresponding transformed fields $\fields'(x) = g\fields(x)$. Again assuming no non-trivial Jacobian factor, we then have
\begin{align}
  \int_{\{( x_ag^{-1}, n_a)\}} D\fields = \int_{\{(x_a,n_a)\}} D\fields'\ ,
  \label{eq: change of integration variable}
\end{align}
since if $\fields$ has instanton insertions $\{(x_a g^{-1}, n_a)\}$, then $\fields'$ has instanton insertions $\{(x_a, n_a)\}$. Then, consider in particular $g$ lying in the subgroup of $SU(1,3)$ generated by $\{P_-, P_i, B, C^\alpha, T\}$, for which we additionally have $S[\fields']=S[\fields]$. Then, we find
\begin{align}
  \langle \fOp^{(1)'}(x_1)\dots \fOp^{(N)'}(x_N) \rangle_{\{( x_a g^{-1}, n_a)\}} &= \int_{\{(x_ag^{-1}, n_a)\}} D\fields\, \fOp^{(1)'}(x_1)\dots \fOp^{(N)'}(x_N) e^{iS[\fields]}		\nn\\
  &= \int_{\{(x_a,n_a)\}} D\fields'\, \fOp^{(1)'}(x_1)\dots \fOp^{(N)'}(x_N) e^{iS[\fields]}		\nn\\
  &= \int_{\{(x_a,n_a)\}} D\fields'\, \fOp^{(1)'}(x_1)\dots \fOp^{(N)'}(x_N) e^{iS[\fields']}		\nn\\
  &= \int_{\{(x_a,n_a)\}} D\fields\, \fOp^{(1)}(x_1)\dots \fOp^{(N)}(x_N) e^{iS[\fields]}		\nn\\
  &= \langle \fOp^{(1)}(x_1)\dots \fOp^{(N)}(x_N) \rangle_{\{(x_a, n_a)\}}\ .
  \label{eq: WT identity nice generators}
\end{align}
We now consider the rest of $SU(1,3)$. The only difference here is that we no longer necessarily have $S[\fields']=S[\fields]$. First consider $g=\exp\left( \epsilon^i M_{i+} \right)$. Then, we have
\begin{align}
  \langle \fOp^{(1)'}(x_1)\dots \fOp^{(N)'}(x_N) \rangle_{\{( x_a g^{-1}, n_a)\}} &= \int_{\{(x_ag^{-1}, n_a)\}} D\fields\, \fOp^{(1)'}(x_1)\dots \fOp^{(N)'}(x_N) e^{iS[\fields]}		\nn\\
  &= \int_{\{(x_a,n_a)\}} D\fields'\, \fOp^{(1)'}(x_1)\dots \fOp^{(N)'}(x_N) e^{iS[\fields]}		\nn\\
  &= \prod_{a=1}^N \left( \frac{\overline{\mathcal{M}_{-\epsilon}(x_a)}}{\mathcal{M}_{-\epsilon}(x_a)} \right)^{kn_a}\langle \fOp^{(1)}(x_1)\dots \fOp^{(N)}(x_N) \rangle_{\{(x_a, n_a)\}}\ .
  \label{eq: WT identity M}
\end{align}
Following the same steps, for $g=\exp\left( \epsilon K_+ \right)$ we have
\begin{align}
  &\langle \fOp^{(1)'}(x_1)\dots \fOp^{(N)'}(x_N) \rangle_{\{( x_a g^{-1}, n_a)\}} = \prod_{a=1}^N \left( \frac{1+2\epsilon \bar{z}(x_a,0)}{1+2\epsilon z(x_a,0)} \right)^{kn_a}\langle \fOp^{(1)}(x_1)\dots \fOp^{(N)}(x_N) \rangle_{\{(x_a, n_a)\}}\ .
  \label{eq: WT identity K}
\end{align}
Hence, through (\ref{eq: WT identity M}) and (\ref{eq: WT identity K}) we find that in the quantum theory, we still have global Ward-Takahashi identities corresponding to $M_{i+}, K_+$. But these identities are deformed from the naive result (\ref{eq: global WTI, no instantons}), which holds only in an absence of instanton insertions.\\

Before moving on to find the more general local counterparts to these Ward-Takahashi identities, let us describe an equivalent but nonetheless useful formulation of the quantum theory. This formulation, in terms of \textit{instanton operators}, will in particular allow for an infinitesimal form of (\ref{eq: WT identity nice generators})--(\ref{eq: WT identity K}), while also making contact with previous work in Lorentzian Yang-Mills theories in five dimensions \cite{Lambert:2014jna,Tachikawa:2015mha,Bergman:2016avc}.

We introduce a set of local disorder operators $\{\I_n(x)\}_{n\in\mathbb{Z}}$, known as instanton operators. If there are no instanton operators present in a correlation function, then we perform the path integral assuming no instanton insertions. Otherwise, the insertion of $\I_n(x)$ instructs us to perform the path integral with an instanton insertion at $x$, with charge $n$. In this way, we have
\begin{align}
  \int D\fields\,\, \I_{n_1}(x_1)\I_{n_2}(x_2)\dots \I_{n_N}(x_N)\Big( \dots \Big) = \int_{\{(x_a, n_a)\}} D\fields \Big(\dots \Big)\ .
\end{align}
We further take the $\I_n(x)$ at each point $x$ to form an Abelian group isomorphic to $(\mathbb{Z},+)$, i.e. $\I_n(x)\I_m(x)=\I_m(x) \I_n(x) = \I_{n+m}(x)$.

We can next consider how each $\I_n$ transforms under $SU(1,3)$. We denote by $\I_n'(x)$ the finite transformation under $\delta_G$ of $\I_n(x)$, which we take to match that of a local scalar primary of $\frak{su}(1,3)$ with Lifshitz scaling dimension $0$. In other words, we have simply $\I_n'(x) = g\I_n(x) = \I_n(xg^{-1})$, or infinitessimally $\I_n(x) = \I_n(x) + \delta_G\I_n(x)$ with $\delta_G\I_n(x) = -G_\partial \I_n(x)$ for each $G\in\su(1,3)$. This ensures that for any operator $\fOp(x)$, $\I_n(x)\fOp(x)$ transforms in the same representation as $\fOp(x)$.

Then, the change of path integration field (\ref{eq: change of integration variable}) is recast simply as
\begin{align}
  \int D\fields\,\, \I'_{n_1}(x_1)\I'_{n_2}(x_2)\dots \I'_{n_N}(x_N)\Big( \dots \Big) = \int D\fields'\,\, \I_{n_1}(x_1)\I_{n_2}(x_2)\dots \I_{n_N}(x_N)\Big( \dots \Big)\ .
\end{align}
Written in terms of the $\I_n(x)$, (\ref{eq: WT identity nice generators}) becomes
\begin{align}
  \big\langle \I'_{n_1}(x_1)\fOp^{(1)'}(x_1)\dots \I'_{n_N}(x_N)\fOp^{(N)'}(x_N) \big\rangle = \big\langle \I_{n_1}(x_1)\fOp^{(1)}(x_1)\dots \I_{n_N}(x_N)\fOp^{(N)}(x_N) \big\rangle\ ,
\end{align}
which holds for $G\in\text{span}\{P_-, P_i, B, C^\alpha, T\}$. The infinitesimal form of this identity is then
\begin{align}
  \sum_{a=1}^N \left\langle \delta_{G}\big( \I_{n_a}(x_a) \fOp^{(a)}(x_a) \big)\prod_{b\neq a} \I_{n_b}(x_b)\fOp^{(b)}(x_b) \right\rangle = 0\ .
  \label{eq: infinitesimal undeformed WTIs}
\end{align}
We can similarly write (\ref{eq: WT identity M}) and (\ref{eq: WT identity K}) in terms of the $\I_{n_a}(x_a)$, which the give rise to the infinitesimal forms
\begin{align}
  \sum_{a=1}^A \left\langle \Big( \delta_{M_{i+}} + \tfrac{ik}{R}n_a x_a^i  \Big)\big( \I_{n_a}(x_a)\fOp^{(a)}(x_a) \big)\prod_{b\neq a} \I_{n_b}(x_b)\fOp^{(b)}(x_b) \right\rangle &= 0		\ ,\nn\\
  \sum_{a=1}^A \left\langle  \Big( \delta_{K_+} + \tfrac{ik}{R}n_a x_a^i x_a^i \Big)\big( \I_{n_a}(x_a)\fOp^{(a)}(x_a) \big)\prod_{b\neq a} \I_{n_b}(x_b)\fOp^{(b)}(x_b) \right\rangle &= 0		\ ,
  \label{eq: infinitesimal deformed WTIs}
\end{align}
respectively.


\section{Local Ward-Takahashi identities and Noether currents}\label{subsec: local WTIs}


Having now seen that symmetry is restored in the quantum theory, in which Ward-Takahashi identities are deformed in the presence of instanton operators, let us know present the much more general \textit{local} Ward-Takahashi identities. These will in particular determine the corresponding Noether currents.

We derive the identities following the standard procedure. We consider the variation of correlation functions under a broader class of transformations, in which the $\su(1,3)$ variations are allowed to vary locally according to some function $\epsilon(x)$. Note however that $\epsilon(x)$ must still be approximately constant in a neighbourhood of the points $x_a$, to ensure that the resulting transformations still map into the extended configuration space. Then, taking the functional derivative with respect to $\epsilon(x)$ of the resulting expression, for each $G\in\su(1,3)$ we arrive at
\begin{align}
  &-i\,\left\langle \mathcal{W}_G(x) \prod_{a=1}^N \I_{n_a}(x_a) \fOp^{(a)}(x_a) \right\rangle \nn\\
  &\hspace{30mm}= \star\sum_{a=1}^N \delta^{(5)}(x-x_a) \left\langle \delta_G\big( \I_{n_a}(x_a)\fOp^{(a)}(x_a) \big)\prod_{b\neq a}\I_{n_b}(x_b)\fOp^{(b)}(x_b) \right\rangle\ ,
  \label{eq: general local WTI}
\end{align}
Note, we have for simplicity restricted to operators $\fOp^{(a)}$ that depend only on the fields $A, X^I, \Psi$ and not their derivatives. More generally, one would find additional terms one the right-hand side of the form $\partial_\mu \left( \text{contact term} \right)^\mu$.

For $G\in\{P_-, P_i, B, C^\alpha, T\}$, the top forms $\mathcal{W}_G$ are given by
\begin{align}
  \mathcal{W}_G =  d \star J_G\ ,
\end{align}
for Noether currents $J_G$. Once again, the story is different for $M_{i+}, K_+$, for which we find
\begin{align}
  \mathcal{W}_{M_{i+}} 	&= d \star J_{M_{i+}} + x^i\, d\left( \frac{k}{8\pi^2R}  \text{tr}\left( F\wedge F \right) \right)		\nn\\
  						&= d \star J_{M_{i+}} + \star \frac{k}{R}\sum_{a=1}^N n_a x_a^i \delta^{(5)}(x-x_a)\ ,
\end{align}
and
\begin{align}
  \mathcal{W}_{K_+} 	&= d \star J_{K_+} + x^i x^i\, d\left( \frac{k}{8\pi^2}  \text{tr}\left( F\wedge F \right) \right)		\nn\\
  						&= d \star J_{K_+} + \star \frac{k}{R}\sum_{a=1}^N n_a x_a^i x_a^i \delta^{(5)}(x-x_a)\ .
\end{align}
The explicit forms of the Noether currents $J_G$ can be found in Appendix \ref{app: Noether currents}.\\

It is natural then to reorganise terms in (\ref{eq: general local WTI}) for $G=M_{i+}, K_+$, to bring the set of Ward-Takahashi identities to a more familiar form. We have
\begin{align}
  &-i\,d\star\left\langle J_G(x) \prod_{a=1}^N \I_{n_a}(x_a) \fOp^{(a)}(x_a) \right\rangle\nn\\
  &\hspace{30mm} = \star\sum_{a=1}^N \delta^{(5)}(x-x_a) \left\langle \tilde{\delta}_G\big( \I_{n_a}(x_a)\fOp^{(a)}(x_a) \big)\prod_{b\neq a}\I_{n_b}(x_b)\fOp^{(b)}(x_b) \right\rangle\ ,
  \label{eq: general local WTI with tildes}
\end{align}
where the new variations $\tilde{\delta}_G$ act as $\tilde{\delta}_G \fOp(x)= \delta_G \fOp(x)$ on normal local operators of the theory, but as
\begin{align}
  \tilde{\delta}_G \I_n(x) 			&=		\delta_G\I_n(x)\quad \text{for } G\in\{P_-, P_i, B, C^\alpha, T\}		\ ,\nn\\
  \tilde{\delta}_{M_{i+}} \I_n(x)	&= \delta_{M_{i+}} \I_n(x) + \tfrac{ik}{R}nx^i \I_n(x)									\ ,\nn\\
  \tilde{\delta}_{K_+} \I_n(x)	&= \delta_{K_+} \I_n(x) + \tfrac{ik}{R}nx^ix^i \I_n(x)	\ ,
  \label{eq: tilde variations of I}
\end{align}
on instanton operators. Then, by integrating (\ref{eq: general local WTI with tildes}) over $\mathbb{R}^5$ and taking suitable boundary conditions on $S^4_\infty$, we recover the global Ward-Takahashi identities (\ref{eq: infinitesimal undeformed WTIs})--(\ref{eq: infinitesimal deformed WTIs}), written compactly in terms of the $\tilde{\delta}_G$ as
\begin{align}
  \sum_{a=1}^N \left\langle \tilde{\delta}_{G}\big( \I_{n_a}(x_a) \fOp^{(a)}(x_a) \big)\prod_{b\neq a} \I_{n_b}(x_b)\fOp^{(b)}(x_b) \right\rangle = 0\ .
  \label{eq: infinitesimal WTIs with tilde}
\end{align}

\section{Extended symmetry, and the sixth dimension}\label{subsec: extended symmetry}

Let us summarise our findings so far. The classical theory admitted an $SU(1,3)$ spacetime symmetry in the absence of instanton insertions. The corresponding infinitesimal variation of fields is denoted $\delta_G$ for each $G\in\su(1,3)$, which form a representation of $\su(1,3)$ when acting on the gauge field $A$ and matter fields $X^I, \Psi$. We extended this to include a variation $\delta_{P_+}$ that acts trivially on all fields $\delta_{P_+}\fields=0$, and in this way realised the $\{\delta_G\}$ as a representation of $\frak{h}$, with brackets  $[\delta_{G_1},\delta_{G_2}]=\delta_{[G_1,G_2]}$ as in Chapter \ref{chap: SU(1,3) theories}.

We found that this symmetry was broken in the classical theory in the presence of instanton operators. However, this breaking is local to the instanton insertion points $x_a$, and thus the resulting Ward-Takahashi identities in the quantum theory could nonetheless be written in the standard form (\ref{eq: general local WTI with tildes}) in terms of Noether currents $J_G$. Integrating these local identities over $\mathbb{R}^5$, we recovered the infinitesimal form of the global Ward-Takahashi identities (\ref{eq: infinitesimal undeformed WTIs})--(\ref{eq: infinitesimal deformed WTIs}).\\

The Ward-Takahashi identities (\ref{eq: general local WTI with tildes}) are written not in terms of our original variations $\delta_G$, but instead in terms of variations $\tilde{\delta}_G$, which we have defined for each $G\in \mathcal{B}=\{P_-, P_i, B, C^\alpha, T, M_{i+}, K_+\}$. In particular, they differ from the $\delta_G$ for $G=M_{i+}, K_+$ when acting on instanton operators, as in (\ref{eq: tilde variations of I}). 

We are then lead to ask: are the $\{\tilde{\delta}_G\}_{G\in \mathcal{B}}$ the generators of a representation of $\frak{su}(1,3)$ under commutation, like the $\delta_G$ are? The answer is in fact \textit{no}. In particular, we find a \textit{single} commutator that does not close on $\frak{su}(1,3)$, which is
\begin{align}
  [\tilde{\delta}_{M_{i+}},\tilde{\delta}_{P_j}] \big(\I_n(x)\fOp(x)\big)  &= \left( -\tfrac{1}{2}\Omega_{ij} \tilde{\delta}_T - \tfrac{2}{R}\delta_{ij} \tilde{\delta}_B+\Omega_{ik} \eta^\alpha_{jk} \tilde{\delta}_{C^\alpha} - \tfrac{ik}{R}\, \delta_{ij}  n \right)\big(\I_n(x)\fOp(x)\big)	\ .
  \label{eq: deformed tilde bracket}
\end{align}
Suppose however that we define a new variation $\tilde{\delta}_{P_+}$ that annihilates all standard operators $\fOp(x)$, but acts as
\begin{align}
  \tilde{\delta}_{P_+}\I_n(x) = \frac{ik}{R}n\,\I_n(x)\ ,
\end{align}
on instanton operators, and hence
\begin{align}
  [\tilde{\delta}_{M_{i+}},\tilde{\delta}_{P_j}] \big(\I_n(x)\fOp(x)\big)  &= \left( -\delta_{ij} \tilde{\delta}_{P_+}-\tfrac{1}{2}\Omega_{ij} \tilde{\delta}_T - \tfrac{2}{R}\delta_{ij} \tilde{\delta}_B+\Omega_{ik} \eta^\alpha_{jk} \tilde{\delta}_{C^\alpha}\right)\big(\I_n(x)\fOp(x)\big)	\ .
\end{align}
Then, by direct comparison with the algebra (\ref{eq: extended su(1,3) algebra}), we find quite remarkably that the full set of variations $\{\tilde{\delta}_G\}_{\mathcal{B}_+}$ \textit{do} generate a representation of $\frak{h}$, with the instanton operator $\I_n$ now carrying charge $p_+=kn/R$ under $\tilde{\delta}_{P_+}$.

In particular, suppose the composite field $\fOp$ is, under the $\delta_G$, an $\frak{su}(1,3)$ primary with scaling dimension $\Delta$ and rotation representations $r_\fOp[B], r_\fOp[C^\alpha]$. Then, under the $\{\tilde{\delta}_G\}_{G\in\frak{h}}$, the operator $\I_n \fOp$ is precisely an $\frak{h}$ primary field with $P_+$ charge $p_+=kn/R$, as well as scaling dimension $\Delta$ and rotation representations $r_\fOp[B], r_\fOp[C^\alpha]$. Indeed, one can identify explicitly the $p_+$ dependence of the variations (\ref{eq: 5d algebra action on fields}) as arising from the action of $\tilde{\delta}_{M_{i+}},\tilde{\delta}_{K_+}$ and $\tilde{\delta}_{P_+}$ on $\I_n$! \\

We can similarly now extend the local Ward-Takahashi identity to read once again
\begin{align}
  &-i\,d\star\left\langle J_G(x) \prod_{a=1}^N \I_{n_a}(x_a) \fOp^{(a)}(x_a) \right\rangle \nn\\
  &\hspace{30mm} = \star\sum_{a=1}^N \delta^{(5)}(x-x_a) \left\langle \tilde{\delta}_G\big( \I_{n_a}(x_a)\fOp^{(a)}(x_a) \big)\prod_{b\neq a}\I_{n_b}(x_b)\fOp^{(b)}(x_b) \right\rangle\ ,
  \label{eq: final local WTIs}
\end{align}
which now holds for all $G\in\mathcal{B}_+$, where we define
\begin{align}
  J_{P_+} = -\frac{k}{8\pi^2 R}\star \tr\left( F\wedge F \right)\ .
  \label{eq: P_+ Noether current}
\end{align}
It is indeed straightforward to see that for $G=P_+$, (\ref{eq: final local WTIs}) is satisfied trivially. Further, (\ref{eq: final local WTIs}) holds for all $G\in\frak{h}$ with $J_{G_1+G_2}=J_{G_1}+J_{G_2}$ and $J_{[G_1,G_2]} = \tilde{\delta}_{G_1} J_{G_2} - \tilde{\delta}_{G_2} J_{G_1}$ for all $G_1,G_2\in\frak{h}$. Integrating over $\mathbb{R}^5$, we once again arrive at the global identities
\begin{align}
  \sum_{a=1}^N \left\langle \tilde{\delta}_{G}\big( \I_{n_a}(x_a) \fOp^{(a)}(x_a) \big)\prod_{b\neq a} \I_{n_b}(x_b)\fOp^{(b)}(x_b) \right\rangle = 0\ ,
  \label{eq: final global infinitesimal WTIs}
\end{align}
which hold for all $G\in\frak{h}$. We can equivalently write this in its finite form, as
\begin{align}
 	 \big\langle \I'_{n_1}(x_1)\fOp^{(1)'}(x_1)\dots \I'_{n_N}(x_N)\fOp^{(N)'}(x_N) \big\rangle = \left\langle \I_{n_1}(x_1)\fOp^{(1)}(x_1)\dots \I_{n_N}(x_N)\fOp^{(N)}(x_N) \right\rangle\ ,
 	 \label{eq: final global finite WTIs}
\end{align}
where in this expression, $\I'(x) = \exp(\tilde{\delta}_G)\I(x)$ and $\fOp'(x) = \exp(\tilde{\delta}_G)\fOp(x) = \exp(\delta_G)\fOp(x)$. The explicit forms of these finitely-transformed operators can be found in Appendix \ref{app: finite transformations}. It is in particular straightforward to then see that (\ref{eq: final global finite WTIs}) reproduces (\ref{eq: WT identity nice generators})--(\ref{eq: WT identity K}). \\

The algebra $\frak{h}$, its representations, and the solutions of the resulting Ward-Takahashi identities (\ref{eq: final global infinitesimal WTIs}) for scalar primaries have already been studied extensively in the previous Chapters. For instance, if we consider a pair of scalar primaries $\fOp^{(1)},\fOp^{(2)}$ of the theory with scaling dimensions $\Delta_1,\Delta_2$, respectively, and dress each of them with an instanton operator of generic charge, the resulting 2-point function satisfies the Ward-Takahashi identities (\ref{eq: final global finite WTIs}). These in turn have a unique solution up to an overall constant, given by
\begin{align}
  \big\langle \I_{n_1}(x_1)\fOp^{(1)}(x_1) \I_{n_2}(x_2)\fOp^{(2)}(x_2) \big\rangle = \delta_{\Delta_1,\Delta_2}\delta_{0,n_1+n_2}d(\Delta_1,n_1)\frac{1}{(z_{12}\bar{z}_{12})^{\Delta_1/2}}\left(\frac{z_{12}}{\bar{z}_{12}}\right)^{n_1}\ ,
\end{align}
for some constant $d(\Delta_1,n_1)$. In particular, taking for each of $\fOp^{(1)},\fOp^{(2)}$ the unit operator, we have
\begin{align}
  \big\langle \I_{n_1}(x_1) \I_{n_2}(x_2) \big\rangle \propto \delta_{0,n_1+n_2}\left(\frac{z_{12}}{\bar{z}_{12}}\right)^{n_1}\ .
\end{align}
%
So, we learn that we can interpret our theory with extended configuration space as a theory of Fourier modes of a conformally compactified six-dimensional CFT. Owing to the $2\pi$ interval over which $x^+$ runs, such an interpretation requires the eigenvalues of $P_+$ to be valued in $i\mathbb{Z}$. This is indeed the case, so long as $k\in\mathbb{Z}$. Then, $\tilde{\delta}_{P_+}\left( \I_n(x)\fOp(x) \right)=\tfrac{ik}{R}n\,\I_n(x)\fOp(x)$, which precisely identifies $\I_n(x)\fOp(x)$ as the Fourier mode $\fOp_{kn}$ of some six-dimensional operator as per the expansion (\ref{eq: 6d op Fourier decomp}). In particular, a choice of $k=\pm 1$ allows for the realisation of the full spectrum of Fourier modes on the conformal compactification, while higher $k$ corresponds to a $\mathbb{Z}_{|k|}$ orbifold thereof. \\

One of the virtues of introducing instanton operators $\I_n(x)$ is that we can use a single path integral formulation to encapsulate not just one but all topological sectors of the theory. In particular, we can form a coherent state of Fourier modes, and so define the notion of a six-dimensional operator in our theory.

In more detail, given a local operator $\fOp(x)$, we are lead to define six-dimensional operator
\begin{align}
  \sOp(x^+, x^-, x^i):= \sum_{n\in\mathbb{Z}} e^{-iknx^+/R}\I_n(x^-, x^i) \fOp(x^-, x^i)\ ,
  \label{eq: six d operators definition}
\end{align}
for some new coordinate $x^+$. Then, we have 
\begin{align}
  \tilde{\delta}_{P_+} \sOp (x^+, x^-, x^i) = -\frac{\partial}{\partial x^+}\sOp (x^+, x^-, x^i)\ ,
\end{align}
and so $P_+$ is identified as translations along an emergent sixth dimension\footnote{Note, the sign here is consistent with the convention used to define the $\tilde{\delta}_G$, i.e. with $\tilde{\delta}_{P_-}\sOp(x)=-\partial_-\sOp(x)$}.

Indeed, it is straightforward to go a step further, and show that for generic $G\in\frak{h}$, we have $\tilde{\delta}_G\sOp(x) = - G _\partial^\text{6d} \sOp(x) - r^\text{6d}(x)\sOp(x)$, where as usual $r^\text{6d}$ acts on any indices of $\sOp$, while the six-dimensional vector fields $G_\partial^\text{6d}$ form precisely the algebra of conformal Killing vector fields of six-dimensional Minkowski space which commute with $(P_+^\text{6d})_\partial=\partial_+$. These are given explicitly in (\ref{eq: 6d CKVs}).


\chapter{Instanton worldlines on the constraint surface}\label{chap: worldlines}


The theory (\ref{eq: Lagrangian}) exhibits a number of very interesting features. We have seen in particular that by extending the configuration space to allow for instanton insertions, encoded in the path integral by instanton operators, we can realise non-trivial Kaluza-Klein modes in the sixth dimension in a very concrete way.

However, we also have that the dynamics of the theory are constrained by the Lagrange multiplier $G_{ij}$ to live only on the surface $\mathcal{F}^+=0$, a particular $\Omega$-deformation of instanton moduli space.

From the outset, it is not at all clear that these two notions are compatible. It could very possibly be the case that the additional topological sectors of the theory required to realise Kaluza-Klein modes are entirely disjoint from the constraint surface $\mathcal{F}^+$, and thus inaccessible.\\

In this final Chapter in which we construct and analyse explicit solutions to $\mathcal{F}^+=0$. Motivated by the surprising breadth of solutions to the normal instanton equations (\ref{eq: intro instanton equations}) captured by the 't Hooft solutions (\ref{eq: intro t Hooft solution}), we pursue solutions via an $\Omega$-deformation of the ansatz (\ref{eq: tHooft ansatz}).

What we find is a broad class of solutions that we interpret as describing the position of and backreaction due to an arbitrary number of anti-instanton worldlines. In particular, points at which these worldlines begin or end, where an anti-instanton is created or annihilated, are necessarily singular points in the field strength $F$. We show in particular that, in the language of the Chapter \ref{chap: An explicit model and its symmetries}, a point $y$ at which $m$ worldlines end and $n$ worldlines begin is precisely identified as the position of an instanton insertion $(y,m-n)$.

We finally consider some aspects of the resulting dynamics by exploring solutions to the remaining equations of motion on these non-trivial instanton backgrounds. We also investigate the DLCQ limit once again, recovering the usual 't Hooft solutions (\ref{eq: intro t Hooft solution}) with generic time-dependant moduli, while also constructing an interesting deformation thereof.

\section{Generalised 't Hooft ansatz}


So, we wish to solve $\mathcal{F}^+=0$. It turns out that this equation is usefully written in terms of the derivatives $\hat{\partial}_i = \partial_i - \frac{1}{2}\Omega_{ij} x^j \partial_-$ and primary field $\hat{A}_i = A_i - \tfrac{1}{2}\Omega_{ij} x^j A_-$ as introduced in Chapter \ref{chap: An explicit model and its symmetries}. Then, we have
\begin{align}
  \mathcal{F}_{ij} &= \hat{F}_{ij} - \Omega_{ij} A_-	\ ,	\nn\\
  \hat{F}_{ij} &= \hat{\partial}_i \hat{A}_{j} - \hat{\partial}_j \hat{A}_{i} - i[\hat{A}_i, \hat{A}_j]\ .
\end{align}
Thus, since $\Omega_{ij}$ is manifestly anti-self-dual, the constraint can be written
\begin{align}
  \hat{F}_{ij}^+=\frac{1}{2}\left(\hat{F}_{ij}+\tfrac{1}{2}\varepsilon_{ijkl} \hat{F}_{kl}\right)=0\ .
\end{align}
Note that for $x^-$-independent solutions, $\hat A_i$ can be viewed as an ordinary anti-self-dual gauge field and  $\hat{F}_{ij}$ its field strength. Thus every $x^-$-independent solution to the familiar anti-self-dual gauge field condition, given by the ADHM construction in terms of moduli, is also a solution to the  $\Omega$-deformed anti-self-dual gauge condition. 

However,   when there is a non-trivial dependence on $x^-$, the $\Omega$-deformed anti-self-duality condition is more restrictive since it constrains the dependence on $x^-$, whereas in the undeformed ($\Omega_{ij}=0$) theory there are no constraints on the $x^-$-dependence of the moduli.\\

It would be interesting to obtain an ADHM-like construction for the general solution to  $\hat{F}_{ij}^+=0$. In lieu of this we restrict attention to an $SU(2)$ gauge group and make the 't Hooft-like ansatz
\begin{align}
\hat A_i = \eta^\alpha_{ij}\partial_j B\sigma^\alpha + \Omega_{ij}\eta^\alpha_{jk} C_k\sigma^\alpha\ ,
 \end{align}
 for Pauli matrices $\sigma^\alpha$, and $\eta^\alpha_{ij}$ the `t Hooft symbols as discussed in the Introduction.
 
With this ansatz we find that we have $\hat{F}_{ij}^+=0$ precisely if
\begin{align}
  -2\partial_i\partial_i B + 4 \partial_i B \partial_i B + \Omega_{ij} x^j \partial_-\partial_i B + R^{-2} x^i \partial_- C_i + 8 \Omega_{ij} \partial_i B\, C_j + 4R^{-2} C_i C_i - 2\Omega_{ij} \partial_i C_j &= 0 \ , \nonumber\\
\eta^\alpha_{ki}\Omega_{kj}\left( -x^j \partial_-\partial_i B + \Omega_{li} x^j \partial_- C_l - 2\partial_i C_j \right) &= 0\ .
\end{align} 
A solution for $C_i$ in the second equation is 
\begin{align}
C_i = -\frac 12 x^i \partial_- B	\ .
\end{align}
Then, the vanishing of the first equation further requires
\begin{align}
  -2\left( \partial_i \partial_i B - 2\partial_i B \partial_i B \right) + 2\Omega_{ij} x^j \left( \partial_i \partial_- B - 2 \partial_- B \partial_i B  \right) - \frac{|\vex|^2 }{2R^2} \left( \partial_- \partial_- B - 2\partial_- B \partial_- B \right)=0\ ,
\end{align}
Then, introducing $B=-\frac{1}{2}\log\tHooft$, we have
\begin{align}
  \partial_i \partial_i\tHooft - \Omega_{ij} x^j \partial_-\partial_i \tHooft + \frac{1}{4}R^{-2} |\vex|^2 \partial_-\partial_- \tHooft = 0 \quad\Longleftrightarrow\quad \hat\partial_i\hat\partial_i \tHooft= 0\ ,
\label{eq: linearised eqn}
\end{align}
and
\begin{align}
\hat A_i = 
-\frac12\eta^\alpha_{ij}\sigma^\alpha\hat\partial_j \log \tHooft \ .
\label{eq: 't Hooft form for A}
 \end{align} 
Note that it follows from this that $\hat \partial_i\hat A_i=\hat D_i\hat A_i=0$ as $\Omega_{ij}\eta^\alpha_{ij}=0$. One can also compute 
\begin{align}
	\hat{ F}_{ij} = \frac12\tHooft \hat \partial_{i}\hat \partial_{k} \tHooft^{-1}\eta^\alpha_{jk}\sigma^\alpha-\frac12\tHooft \hat \partial_{j}\hat \partial_{k} \tHooft^{-1}\eta^\alpha_{ik}\sigma^\alpha+ \frac12\tHooft^2\hat\partial_k\tHooft^{-1} \hat\partial_k\tHooft^{-1} \eta^\alpha_{ij}\sigma^\alpha\ .
\end{align}
Thus we find that using a 't Hooft ansatz the self-duality condition reduces to a linear second order differential equation for $\tHooft$. This suggests that a more general ADHM construction could also be obtained. 


\subsection{Spherically symmetric solutions}


Let us start by looking for spherically symmetric solutions where $\tHooft=\tHooft\left( x^-, |\vex|^2 \right)$.  Then we find
\begin{align}
  4|\vex|^2\frac{\partial^2\tHooft}{\partial\left( |\vex|^2 \right)^2} + 8\frac{\partial\tHooft}{\partial|\vex|^2} + \frac{ |\vex|^2}{4R^2} \frac{\partial^2\tHooft}{\partial({x^-})^2}  = 0\ .
\end{align}
This can be written in a nicer form in terms of the complex variable
\begin{align}
 z=z(x,0) = x^- + \frac{i}{4R} |\vex|^2. 
\end{align}
Then we have
\begin{align}
  \left( z-\bar{z} \right)\partial \bar{\partial} \tHooft - \left( \partial - \bar{\partial} \right) \tHooft = \partial \bar{\partial}\big((z-\bar{z})\tHooft\big) = 0\ .
\end{align}
Hence, the \textit{general} spherically symmetric solution of $\tHooft$ is given by
\begin{align}
  \tHooft = \frac{1}{z-\bar{z}} \Big(\, \varphi_+\left( z \right) + \varphi_-\left( \bar{z} \right)\Big)\ .
\end{align}
However Hermiticity of $\hat A_i$ requires that   $\tHooft$ is real and hence imposes  $\varphi_-= -\bar{\varphi}_+$ and so 
\begin{align}
  \tHooft = \frac{1}{z-\bar{z}} \Big(\, \varphi_+\left( z \right) - \bar \varphi_+\left( \bar{z} \right)\Big)\ .
\end{align}
Furthermore to avoid serious singularities in $\hat A_i$  we also require that $\tHooft >0$. Since the imaginary part of $z$ is positive definite this requires that ${\rm Im}(\varphi_+)>0 $ on the upper half-plane. According to \cite{Aronszajn:1956} the general solution can be written as  \begin{align}
\varphi_+(z) &=\alpha +   \gamma^2(z-{\rm Re}(z_0)) 	+\int^\infty_{-\infty}\left(\frac{1}{\tau-z} - \frac{\tau -{\rm Re}(z_0)}{|\tau - z_0|^2} \right)\mu(\tau)d\tau\ ,
\end{align}
where $z_0$ lies in the complex upper half-plane but $\alpha,\beta,\gamma$ and $\tau$ are real and $\mu(\tau)\ge 0$ is arbitrary so long as the integral exists. Note that, assuming $\gamma\ne 0$, we can rescale $\tHooft$ to set $\gamma=1$ without altering the gauge field $\hat A_i$. Thus we see that $\tHooft$ takes the form
\begin{align}
 \tHooft & =  1  + \int^\infty_{-\infty} \frac{\mu(\tau)}{|\tau-z|^2}  d\tau\ .
 \label{eq: Phi spherically symmetric solution}
\end{align}
In particular, $\tHooft$ is regular except for at the line of points at $x^i=0$ which additionally have $\mu(x^-)>0$. As we will see in more detail in Section \ref{sec: gauge field topology}, these lines can be seen as the worldlines of single anti-instantons ({\it i.e.}\ instantons with charge $-1$), with points at which $\mu(x^-)$ transitions between $\mu(x^-)=0$ and $\mu(x^-)>0$ interpreted as their creation or annihilation. Such transition points are precisely the instanton insertions of Chapter \ref{chap: An explicit model and its symmetries}, with creation points having charge $n=-1$ and annihilation points $n=+1$.\\

Finally, it is worth briefly noting that the inclusion of $\Omega_{ij}\neq 0$ introduces a preferred duality relation, which in particular breaks the straightforward symmetry between solutions of $\hat{F}^+=0$ and $\hat{F}^-=0$ that is present when $\Omega_{ij}=0$, where here $\hat{F}_{ij}^\pm = \frac{1}{2}\left(\hat{F}_{ij}\pm\tfrac{1}{2}\varepsilon_{ijkl} \hat{F}_{kl}\right)$. To see this, let us firstly briefly review the case $\Omega_{ij}=0$, where $\hat{F}_{ij}=F_{ij}$ is the usual field strength of the gauge field $A_i$. Solutions of $\hat{F}^+=0$ with generic instanton number $k<0$ are found in singular gauge as\footnote{One could alternatively try this ansatz with the anti-self-dual 't Hooft matrices $\bar{\eta}^\alpha_{ij}$, but would only find the $k=-1$ instanton in regular gauge.} $A_i=-\frac12\eta^\alpha_{ij}\sigma^\alpha\partial_j \log \tHooft$ for harmonic $\tHooft$. To find solutions to the opposite equation $\hat{F}^-$ with $k>0$, one simply swaps $\eta^\alpha_{ij} \to \bar{\eta}^\alpha_{ij}$.

Let us now go back to $\Omega_{ij}\neq 0$, and take $\Omega_{ij}$ anti-self-dual as we do throughout this thesis. In solving $\hat{F}^\pm=0$, we may in principle consider two different ans\"{a}tze: $\hat{A}_i=-\frac12\eta^\alpha_{ij}\sigma^\alpha\hat\partial_j \log \tHooft$, or $\hat{A}_i=-\frac12\bar{\eta}^\alpha_{ij}\sigma^\alpha\hat\partial_j \log \tHooft$. As we have seen, it is the former ansatz involving $\eta^\alpha_{ij}$ that proves fruitful in solving $\hat{F}^+=0$. One might then hope that the latter ansatz involving $\bar{\eta}^\alpha_{ij}$ will be similar useful in solving $\hat{F}^-=0$. This is however not the case. In particular, considering the following two parameterisations,
\begin{align}
\tHooft=\frac{g_{A}(z,\bar{z})}{z-\bar{z}}=\left(z-\bar{z}+g_{B}(z,\bar{z})\right)^{-1},
\end{align}
we find the following constraints depending on which equation we are trying to solve, and which ansatz we are using\footnote{Note, if we had instead chosen $\Omega_{ij}$ self-dual, we need simply to swap the rows and columns of this table, so that each entry is exchanged with its diagonal opposite.}:
\[
\begin{array}{l|cc}
 & \eta & \bar{\eta}\\
 \hline\hat{F}^+=0   & \partial\bar{\partial}g_{A}=0 & \partial\bar{\partial}g_{B}=\partial^{2}g_{B}=\bar{\partial}^{2}g_{B}=0\\
\hat{F}^-=0   & \partial\bar{\partial}g_{B}=\partial^{2}g_{B}=\bar{\partial}^{2}g_{B}=0 & \partial\bar{\partial}g_{A}=\partial_{-}g_{A}=0.
\end{array}
\]
The important take away is that the top left and bottom right entries are qualitatively different, and in particular the $\bar{\eta}$ ansatz gives only static solutions to $\hat{F}^-=0$. In this way, we see that the anti-self-duality of $\Omega_{ij}$ breaks the symmetry between $\hat{F}^+=0$ and $\hat{F}^-=0$.


\subsection{Simple examples}


Before we continue let us first look at some simple forms for $\tHooft$. If $\mu = \rho^2/4\pi R$ is constant, then we find from \eqref{eq: Phi spherically symmetric solution}
\begin{align}\label{S'tH}
\tHooft = 1 + \frac{\rho^2}{|\vex|^2}	\ .
\end{align}
This gives back the usual static instanton located at $x^i=0$ and with size $\rho$. A more interesting example is
\begin{align}
\mu(\tau) = \frac{\rho^2_0l^2/4\pi R}{(\tau-\tau_0)^2+l^2}	\ ,
\end{align}
which leads to
\begin{align}\label{ex2}
	\tHooft = 1 + \frac{\rho^2_0}{|\vex|^2}\frac{l(l+\tfrac{1}{4R}|\vex|^2)}{(x^--\tau_0)^2 + (l+\tfrac{1}{4R}|\vex|^2)^2}\ .
\end{align}
Here the small $|\vex|$ behaviour is unchanged except that the instanton size grows and then decays in $x^-$. Note that  $\tHooft \sim 1/|\vex|^4$ as $|\vex|\to\infty$. Taking the limit $l\to\infty$ leads to the static instanton.  On the other hand taking the limit $l\to 0$ with $\rho^2=\rho^2_0 l/4R $ fixed gives
\begin{align}\label{ex3}
	\tHooft = 1 + \frac{\rho^2} {(x^--\tau_0)^2 + \tfrac{1}{16R^2}|\vex|^4}\ .
\end{align}
Which corresponds to  $\mu(\tau) = \rho^2\delta(\tau-\tau_0)$ and does not lead to an instanton  as $\tHooft$ is smooth as a function of $x^i$ except at $x^-=\tau_0$ where it produces a singular gauge field. 

We can also consider a simple oscillating  anti-instanton by taking $\mu(\tau) = A-B\cos(\tau)$ with $A\ge B>0$:
\begin{align}
	\tHooft = 1 +  {4\pi R }\frac{A-Be^{-{|\vex|^2/4R}}\cos(x^-)}{|\vex|^2}.
\end{align}
Note that taking $A=B$ the small $|\vex|$ limit gives
\begin{align}\label{wavey}
	\tHooft = 1 +  {4\pi A R }\frac{1- \cos(x^-)}{|\vex|^2}+\ldots\ ,
\end{align}
corresponding to instantons that shrink to zero size and then grow again.

Another $x^-$-dependent example is simply  a step function
\begin{align}
\mu(\tau) = \begin{cases}
0 & \tau <\tau_1 \\
\rho^2/4\pi R &  \tau_1\le \tau\le \tau_2\\
0 &  \tau_1 <\tau 	\\
 \end{cases}\ ,
\end{align}
 so that
\begin{align}
\tHooft  =&\  1 +\frac{\rho^2}{ |\vex|^2}\frac{1}{2\pi i}\left[\ln \left(\frac{\tau_2 - z}{\tau_2 - \bar z}\right)-\ln \left(\frac{\tau_1 - z}{\tau_1 - \bar z}\right)\right]	 \ .\end{align} 
Note we must choose a branch of the logarithm such that for $z$ in the upper half-plane, $\ln (z/\bar{z})=2i\arg(z)$, with $0\le \arg(z) \le \pi$. The logarithms are bounded and regular everywhere except for $(x^-,x^i)=(\tau_1,0)$ and  $(x^-,x^i)=(\tau_2,0)$. Then, for $x^i\to0$ we find 
\begin{align}
	\ln \left(\frac{\tau-z}{\tau - \bar z}\right)\to \begin{cases}
2\pi i   & x^-<\tau  \\ 
0 &x^->\tau	
 \end{cases}\ ,
\end{align}
but note that if $x^->\tau_2$ then  $x^->\tau_1$ and if $x^-<\tau_1$ then  $x^-<\tau_2$ hence, as $|\vex|^2\to0$,
\begin{align}
	\ln \left(\frac{\tau_2 - z}{\tau_2 - \bar z}\right)-\ln \left(\frac{\tau_1 - z}{\tau_1 - \bar z}\right)\to \begin{cases}
0  & x^->\tau_{2}  \\ 
2\pi i  &\tau_1<x^-<\tau_{2}\\
	0  & x^-<\tau_{1}
 \end{cases}\ .
\end{align}
Thus we   create an instanton centred at the origin at $x^-=\tau_1$ and destroy it at $x^-=\tau_2$.  
Taking $\tau_1\to-\infty$ and $\tau_2\to\infty$ we recover the static solution.  
However this solution has infinite action (at least when $A_-=0$) arising from the jump discontinuities in $\mu(\tau)$. 

A smoother, continuous, example is
  ($\tau_4>\tau_3>\tau_2>\tau_1$)
 \begin{align}
  \mu(\tau)=\left\{\begin{aligned}
  &\,\, 0 &\tau < \tau_1		\nn\\
  &\,\,\frac{\rho^2}{4\pi R }\left( \frac{\tau-\tau_1 }{\tau_2-\tau_1}\right)\qquad	&\tau_1 \le  \tau \le  \tau_2\nn\\
  &\,\,\frac{\rho^2} {4 \pi R}	\qquad	&\tau_2<\tau<\tau_3	\nn\\
    &\,\,  \frac{\rho^2}{  4\pi R }\left( \frac{\tau_4-\tau }{\tau_4-\tau_3}\right)	\qquad	&\tau_3 \le \tau \le \tau_4	\nn\\
  &\,\,0 &\tau >    \tau_4		
\end{aligned}\right.\ .
\end{align}
The resulting $\tHooft$ takes the rather ugly form
\begin{align}
\tHooft  = 1 &+\frac{\rho^2}{|\vex|^2} \frac{ 1}{2\pi i} \frac{1}{\tau_2-\tau_1}\left[ z\ln \left(\frac{\tau_2-z}{\tau_1  -  z}\right) -\bar z\ln\left(\frac{\tau_2-\bar z}{\tau_1 - \bar z}\right) \right] \nonumber\\
&-\frac{\rho^2}{|\vex|^2} \frac{ 1}{2\pi i} \frac{\tau_1}{\tau_2-\tau_1}\left[ \ln\left(\frac{\tau_2 - z}{\tau_2- \bar  z}\right) - \ln\left(\frac{\tau_1 - z}{\tau_1 -\bar  z}\right)\right]\nonumber\\
&+\frac{\rho^2}{|\vex|^2} \frac{ 1}{2\pi i}\left[   \ln\left(\frac{\tau_3- z}{\tau_3 -  \bar z}\right)-\ln\left(\frac{\tau_2- z}{\tau_2 -  \bar z}\right) \right]\nonumber\\
&+\frac{\rho^2}{|\vex|^2} \frac{ 1}{2\pi i}\frac{\tau_4}{\tau_4-\tau_3}\left[   \ln\left(\frac{\tau_4 - z}{\tau_4- \bar  z}\right) - \ln\left(\frac{\tau_3 - z}{\tau_3 -\bar  z}\right)\right]\nonumber\\
&-\frac{\rho^2}{|\vex|^2} \frac{ 1}{2\pi i}\frac{1}{\tau_4-\tau_3}\left[ z\ln \left(\frac{\tau_4-z}{\tau_3  -  z}\right) -  \bar z\ln \left(\frac{\tau_4-\bar z}{\tau_3  - \bar  z}\right)\right]	\ .
\end{align}
However one can see that,  since $\bar z = z - \frac{i}{2R}|\vex|^2$,  the first and last lines are finite as $|\vex|\to 0$, whereas the middle lines behave similarly to the previous case. Thus we find a continuous solution to the constraint  that represents the creation and then annihilation of an instanton at $x^i=0$.


\subsection{Allowing general worldlines}


We can significantly generalise the spherically symmetric solution (\ref{eq: Phi spherically symmetric solution}). Since the equation for $\tHooft$ is linear we can obtain new solutions by summing over existing solutions. However any sum over spherically symmetric solutions remains spherically symmetric and hence just changes  the form of the function $\mu$. To find more solutions we can leverage the $SU(1,3)$ spacetime symmetry enjoyed by the theory. 
As we have seen, the behaviour of the action under these symmetries is subtle in the presence of instantons. However 
it turns out that  the constraint equation $\mathcal{F}^+_{ij}=0$ is manifestly $SU(1,3)$ invariant. Further, the translations $\{P_-, P_i\}$, as well as the Lifshitz scaling $T$, preserve the 't Hooft form (\ref{eq: 't Hooft form for A}), in the sense that under their action, $\hat{A}_i$ in 't Hooft form is sent to a different $\hat{A}_i$ that is \textit{still} in this form, just for a different $\tHooft$. 

So let us consider a translation. These take the form
\begin{align}
	x'^- = x^- - y^- + \frac12 \Omega_{ij}x^i y^j \ , \qquad x'^i = x^i - y^i\ .
	\label{eq: translated coordinates}
\end{align}  If we let
\begin{align}
\tHooft'(x^-,x^i) 
	= \tHooft(x'^-,x'^i)\ ,
\end{align}
then we find that
\begin{align}
\hat  \partial_i \hat \partial_i\tHooft'&=  \partial_i \partial_i\tHooft' - \Omega_{ij} x^j \partial_-\partial_i \tHooft' + \frac{1}{4R^2}  |\vex|^2 \partial_-^2 \tHooft'\nonumber\\ & =   
   \partial'_i \partial'_i\tHooft - \Omega_{ij} x'^j\partial'_-\partial'_i \tHooft + \frac{1}{4R^2} |\vex'|^2 {\partial'}_-^2\tHooft\nonumber \\
   &= \hat  \partial'_i \hat \partial'_i\tHooft'\ .
  \end{align}
Hence if $\tHooft$ satisfies (\ref{eq: linearised eqn}) then so does $\tHooft'$. Let us use these translations to derive a significant generalisation of (\ref{eq: Phi spherically symmetric solution}).

We note that the solution (\ref{eq: Phi spherically symmetric solution}) is a continuous linear sum over solutions of the form $(z-\tau)^{-1}(\bar{z}-\tau)^{-1}$, for any $\tau\in\mathbb{R}$. By acting with the above translation on such a solution, we find the more general solution
\begin{align}
  \hat{\partial}_i\hat{\partial}_i \left(\frac{1}{z(x,y)\bar{z}(x,y)}\right)=0\quad \text{ for all } x\neq y\ ,
\end{align}
for any point $y=(y^-, y^i)$. 

%
Then, in order that $\tHooft$ continues to describe a particle-like configuration, we integrate over a one-parameter family of the translated solutions, each centred at some point $y(\tau)=(y^-(\tau), y^i(\tau))$. The result is the solution
\begin{align}
  \tHooft = 1+ \int d\tau \frac{\mu(\tau)}{z(x,y(\tau))\bar{z}(x,y(\tau))}\ ,
  \label{eq: Phi general single particle}
\end{align}
where $\mu(\tau) \ge 0$.

Let us now interpret this solution. We once again find this solution describes a particle-like ({\it i.e.}\ co-dimension four) object. This is seen by noting that $\tHooft$ and hence $\hat{A}_i$ is singular precisely at any point $x$ such that there exists some $\tau$ with $x=y(\tau)$ and $\mu(\tau)>0$. We see that the spacetime curve $y(\tau)=(y^-(\tau), y^i(\tau))$ is precisely the worldline of this particle, with $\tau$ providing a local parameterisation along it.
We  recover the spherically symmetric solution (\ref{eq: Phi spherically symmetric solution}) by considering the case $y^i(\tau)=0$ and  $y^-(\tau)=\tau$. 

Note that we are free to sum up $N$ disjoint  particle-like configurations for $\tHooft$. Thus a  yet more general solution is then given by
\begin{align}
  \tHooft = 1+ \sum_{A=1}^N\int d\tau_A \frac{\mu_A(\tau)}{z(x,y_A(\tau_A))\bar{z}(x,y_A(\tau))}\ ,
  \label{eq: Phi general solution}
\end{align}
for any $N$. Such a solution then describes not one but a  swarm   of $N$ instanton particles. Note that such solutions can nonetheless be brought back to the form (\ref{eq: Phi general single particle}) by connecting each worldline end-to-end with new segments along which $\mu(\tau)=0$. As such, the form (\ref{eq: Phi general solution}) is an equivalent rather than generalised form for $\tHooft$, which is nonetheless a useful representation of the solution.\\

Let us briefly comment on the gauge group embedding. If we take a solution
to the anti-self-duality constraint $\hat{F}^+_{ij}=0$ and do the following
\begin{align}
\hat{A}_{i}\rightarrow U^{-1}\hat{A}_{i}U\ ,
\end{align}
where $U(x^{-})\in SU(2)$, then the field strength becomes
\begin{align}
\hat{F}_{ij}\rightarrow U^{-1}\left(\hat{F}_{ij}+\Lambda_{ij}\right)U\ ,
\end{align}
where
\begin{align}
\Lambda_{ij}=\Omega_{[i|k|}x^{k}\left[\hat{A}_{j]},U\partial_{-}U^{-1}\right]\ .
\end{align}
The anti-self-duality constraint is then preserved if
\begin{align}
\Lambda_{ij}+\frac{1}{2}\varepsilon_{ijkl}\Lambda_{kl}=0\ .
\end{align}
It is unclear if this constraint has any nontrivial solutions. It would therefore be interesting to see if the gauge group embedding can be implemented more naturally by generalising the worldline representation in \eqref{eq: Phi general solution}.\\

Finally, as a simple example, let us consider the boosted version of the static
solution. In particular, take $\mu$ to be constant
\begin{align}
\mu(\tau)=\frac{\rho^{2}}{4\pi R} \ , 
\end{align}
but allow it to move in the $x_{4}$ direction with velocity $v$: 
\begin{align}
y^{-}(\tau)=\tau \ , \ y_{1}=y_{2}=y_{3}=0 \ , \ y_{4}=v\tau \ .
\end{align}
Further choosing $\Omega_{ij}=-R^{-1}\bar{\eta}^2_{ij}$ for concreteness, we find the covariant distance
\begin{align}
z(x,y)=x^- - \tau +  \frac{i}{4R}|\vex |^2 + \frac{1}{2R}\left(x_{2}-ix_{4}\right)v\tau+\frac{i}{4R}v^{2}\tau^{2}\ ,
\end{align}
and the integral in \eqref{eq: Phi general single particle} gives 
\begin{align}
\tHooft=1+\frac{\rho^{2}R\left(2\left(2R-vx_{2}\right)+2{\rm Re}\left(\frac{\left(2R-vx_{2}\right)\left(2R-vx_{2}+ivx_{4}\right)-2iRv^{2}x^{-}}{\sqrt{4R^{2}-4iRv\left(vx^{-}-x_{4}-ix_{2}\right)+v^{2}\left(\left(x_{2}-ix_{4}\right)^{2}+\left|\vex\right|^{2}\right)}}\right)\right)}{4Rv\left(Rvx^{-}+vx_{2}x_{4}-2Rx_{4}\right)+\left(vx_{2}-2R\right)^{2}\left|\vex\right|^{2}} \ .
\end{align}
This solution looks complicated, but in the limit where the velocity
goes to zero we recover the usual static solution:
\begin{align}
\lim_{v\rightarrow0}\tHooft=1+\frac{\rho^{2}}{\left|\vex\right|^{2}} \ .
\end{align}
Moreover, in the $R\to\infty$ ({\it i.e.}\ $\Omega_{ij}\to 0$) limit we recover a boosted version of the above
solution: 
\begin{align}
\lim_{R\rightarrow\infty}\tHooft=1+\frac{\rho^{2}}{x_{1}^{2}+x_{2}^{2}+x_{3}^{2}+\left(x_{4}-vx^{-}\right)^{2}} \ .
\end{align}
Indeed, we later show in Section \ref{subsec: DLCQ} that we generally reproduce the usual 't Hooft form solutions in the $R\to\infty$ limit, with moduli that are allowed to vary arbitrarily with $x^-$.

\section{Gauge topology and instantons}\label{sec: gauge field topology}

We have already begun thinking of the `t Hooft-like solutions for $\hat{A}_i$ corresponding to $\tHooft$ of the form (\ref{eq: Phi general solution}) as anti-instanton particles of the gauge field $A=(A_-, A_i)$. Let us now justify this.

We first return to the case of a single instanton centred at the origin, as given by $\tHooft$ in (\ref{eq: Phi spherically symmetric solution}). Here, we will recover almost all of the important qualitative properties of the much more general solution (\ref{eq: Phi general solution}), while avoiding many of the more technical details.

We then generalise our analysis, and show that the solution (\ref{eq: Phi general solution}) describes an arbitrary number of anti-instanton particles, generically travelling between points at which they are created and annihilated.

\subsection{The single, spherically symmetric anti-instanton}

Let us consider again the spherically-symmetric solution for $\hat{A}_i$ given by
\begin{align}
  \tHooft[\mu] = 1+ \int_{-\infty}^\infty \frac{\mu(\tau)}{|\tau-z|^2} d\tau = 1+\frac{1}{z-\bar{z}} \int_{-\infty}^\infty d\tau\, \mu(\tau) \left(\frac{1}{\tau-z}-\frac{1}{\tau-\bar{z}}\right) \ ,
  \label{eq: Phi for single instanton at origin}
\end{align}
where recall the shorthand $z=z(x,0)=x^- + \frac{i}{4R}|\vec  x|^2$. This corresponds to a choice of $y^-(\tau)=\tau$ and $y^i(\tau)=0$ in the more general solution (\ref{eq: Phi general single particle}).

It is clear then that for suitable behaviour of $\mu(\tau)$ at large $\tau$ ({\it i.e.}\ that it is bounded), the integral converges when $|\vec  x|>0$, and so $\tHooft$ is regular away from the origin. Indeed, the solution we will be most interested in are those for which $\mu(\tau)$ has compact support. Conversely, $\tHooft$ and hence $\hat{A}_i$ is singular at all points such that $x^i=0$ and $\mu(x^-)>0$.

Now let $\mathbb{R}^4_*\cong \mathbb{R}^4$ denote the spatial slice defined by fixing some $x^- = x^-_\ast$. We can then consider the total instanton flux through this slice, defined by
\begin{align}
  Q_* = \frac{1}{8\pi^2}\int_{\mathbb{R}^4_*} \text{tr}\left(F\wedge F\right)\ .
  \label{eq: Q star definition}
\end{align}
In our conventions, $F=dA-i A\wedge A$, and so we have locally $\tr(F\wedge F)=d\omega_3(A)$, where $\omega_3(A)=\text{tr}\left(A\wedge dA -\frac{2i}{3}A\wedge A\wedge A\right)$ is the Chern-Simons 3-form.

It may be that the gauge field $A$ is regular throughout $\mathbb{R}^4_*$.
Then, $Q_*$ reduces to an integral of $\omega_3 (A)$ over the 3-sphere at spatial infinity. However more generically the 't Hooft ansatz produces  
solutions where $A$ is singular at $x^i=0$, if $\mu(x^-_*)>0$. Hence, we generally have that $Q_*$ reduces to
\begin{align}
  Q_*= \frac{1}{8\pi^2}\int_{S^3_\infty} \omega_3(A) - \frac{1}{8\pi^2}\int_{S^3_0} \omega_3(A)\ ,
\end{align}
where $S^3_0$ and $S^3_\infty$ denotes 3-spheres around the origin and at spatial infinity, respectively. 

Our key result is that, for suitable boundary conditions on $A_-$, we have $Q_*=0$ if $\mu(x^-_*)=\dot{\mu}(x^-_*)=0$, and $Q_*=-1$ if $\mu(x^-_*)>0$. In more detail, we find that when $\mu(x^-_*)>0$, the leading order behaviour of $A_i$ near the origin precisely matches that of a single $SU(2)$ anti-instanton in singular gauge centred at the origin. As such, the integral of $(1/8\pi^2)\,\omega_3(A)$ over $S^3_0$ is quantised in the integers; it is indeed simply equal to $1$. If instead $\mu(x^-_*)=\dot{\mu}(x^-_*)=0$, then $A_i$ is regular at the origin and so the contribution to $Q_*$ from $S^3_0$ vanishes. Conversely, we show that under reasonable assumptions on $\mu(\tau)$ as $\tau\to\pm\infty$, the contribution from the integral at $S^3_\infty$ vanishes, for any $\mu(x^-_*)\ge 0$.

There are a number of steps required to arrive at these results. Firstly, we consider how the asymptotic behaviour of $\hat{A}_i$ both near the origin and at infinity is dictated by that of $\tHooft$. Secondly, we must translate these asymptotics to those of the gauge field $A_i$ rather than $\hat A_i$, for which we must additionally consider the asymptotic behaviour of $A_-$.\\

We note that the form of $\hat{A}_i$ involves both $\tHooft$ and $\partial_- \tHooft$. This however does not pose much of a computational complication, since provided $\mu(\tau)$ is bounded as $\tau\to\pm\infty$, we have\footnote{Note, this relation holds only for solutions with $y^-=\tau$ and $y^i$ constant.} $\partial_-\tHooft[\mu]=\tHooft[ \dot \mu ]-1$, where $\dot \mu = d\mu/d\tau$. Thus, we have
\begin{align}
  \hat{A}_i (x) = -\eta^\alpha_{ij} \sigma^\alpha x^j \tHooft[\mu]^{-1}  \frac{\partial\tHooft[\mu]}{\partial|\vex|^2} + \frac{1}{4} \eta^\alpha_{ij} \sigma^\alpha \Omega_{jk} x^k \tHooft[\mu]^{-1} \left(\tHooft[\dot{\mu}]-1\right)\ .
\end{align}
Let us now consider the behaviour of $\tHooft[\mu]$, both as $|\vex|\to\infty$ and $|\vex|\to 0$. Firstly, it is immediate that as $|\vex|\to \infty$, we have $\tHooft[\mu]=1+\bigO(|\vex|^{-2})$. However, the corresponding leading order behaviour of $\hat{A}_i$ requires that we know the next-to-leading-order behaviour of $\tHooft$, which in turn depends subtly on the global properties of the function $\mu(\tau)$. 

First suppose that the integral of $\mu(\tau)$ over $\tau\in\mathbb{R}$ converges, so that in particular $\mu(\pm\infty)=0$.\footnote{Note that this is a sufficient but not necessary condition for $\tHooft$ to be finite away from the worldline, which requires only that $\mu(\tau)$ is bounded as $\tau\to \pm\infty$.} Then, in the limit $|\vex|^2\to \infty$, we have
\begin{align}\label{muint}
  \tHooft[\mu] = 1 + \frac{16R^2}{|\vex|^4} \left(\int_{-\infty}^\infty d\tau\, \mu(\tau)\right) + \bigO\left(|\vex|^{-6}\right)\ .
\end{align}
More generally however we may consider profiles for $\mu(\tau)$ such that the limits $\mu(\pm\infty)=\lim_{\tau\to\pm\infty}\mu(\tau)$ exist but may be non-zero. Such choices will still give rise to finite $\tHooft$ away from worldlines, but now the behaviour as $|\vex|\to\infty$ is adjusted. We find\footnote{This can be seen by writing $\mu = \frac12(\mu(\infty)+\mu(-\infty))+\tfrac12(\mu(\infty)-\mu(-\infty))\tanh(\tau-x^-)+\mu_0(\tau)$  where $\mu_0$ has a finite integral, performing the integral of the first two terms explicitly, and then using (\ref{muint}) for the $\mu_0$ contribution. Note, $\int d\tau \tanh \left(\tau - x^-\right)/|\tau - z|^2=0$ as the integrand is odd about the point $\tau=x^-$.}
\begin{align}
  \tHooft[\mu] = 1 + \frac{2\pi R}{|\vex|^2}\left(\mu(\infty)+\mu(-\infty)\right) + \bigO\left(|\vex|^{-4}\right)\ ,
  \label{eq: large x Upsilon}
\end{align}
provided that $\mu$ converges to $\mu(\pm \infty)$ at least as quickly as $\tau^{-2}$ as $\tau\to\pm \infty$.

Next, we can investigate the behaviour of $\tHooft[\mu]$ as $|\vex|\to 0$. This is most easily seen by first considering the Fourier transform of the function $\mu(\tau)$,
\begin{align}
  \mu(\tau) &= \int_{-\infty}^\infty d\omega\, e^{i\omega\tau} \tilde{\mu}(\omega)	\ , 	\nonumber\\
  \tilde{\mu}(\omega) &= \frac{1}{2\pi}\int_{-\infty}^\infty d\tau\, e^{-i\omega\tau} \mu(\tau)	\ ,
\end{align}
with reality of $\mu(\tau)$ implying $\overline{\tilde{\mu}(\omega)}=\tilde{\mu}(-\omega)$. Note, non-zero $\mu(\pm\infty)$ corresponds to allowing for singular profiles for $\tilde{\mu}(\omega)$.

We then have for all $|\vex|\neq 0$,
\begin{align}
  \int_{-\infty}^\infty d\tau\, \frac{\mu(\tau)}{\tau-z} = \int_{-\infty}^0 d\omega \,\tilde{\mu}(\omega) \int_{-\infty}^{\infty} d\tau\, \frac{e^{i\omega \tau}}{\tau-z} + \int_0^{\infty} d\omega \,\tilde{\mu}(\omega) \int_{-\infty}^{\infty} d\tau\, \frac{e^{i\omega \tau}}{\tau-z}\ .
\end{align}
Both integrals over $\tau$ can then be computed by a corresponding contour integral, with the contour closed in the lower half-plane for the former, and upper half-plane for the latter. Since $z$ lies in the upper half-plane, only the latter integral survives, and we have
\begin{align}
  \int_{-\infty}^\infty d\tau\, \frac{\mu(\tau)}{\tau-z} = 2\pi i \int_0^\infty d\omega\,\tilde{\mu}(\omega)  e^{i\omega z}\ ,
\end{align}
and hence,
\begin{align}
  \tHooft[\mu] 	&= 1+\frac{4\pi R}{|\vex|^2} \int_0^\infty d\omega \left(\tilde{\mu}(\omega)e^{i\omega z}+\overline{\tilde{\mu}(\omega)}e^{-i\omega \bar{z}}\right)		\nn\\
  				&= 1+ \frac{4\pi R}{|\vex|^2}\left[ \int_{-\infty}^0 d\omega\, \tilde{\mu}(\omega) e^{i\omega \bar{z}} + \int^{\infty}_0 d\omega\, \tilde{\mu}(\omega) e^{i\omega z} \right]\ .
\end{align}
Therefore, we find that as we approach $|\vex|\to 0$,
\begin{align}\label{smallx}
  \tHooft[\mu] = \frac{4\pi R}{|\vex|^2} \mu(x^-) + \bigO(1)\ .
\end{align}
With these results in hand, we are ready to write down the asymptotic behaviour of $\hat{A}_i$. Firstly, as $|\vex|\to\infty$ we have
\begin{align}
  \hat{A}_i = \bigO\left( |\vex|^{-3} \right) \ , 
\end{align}
assuming that $\lim_{\tau\to\pm\infty}\dot{\mu}(\tau)=0$. Note that this holds even if we allow $\mu(\pm\infty)$ to be non-zero.

Next, consider the limit $|\vex|\to 0$. Then, if $\mu(x^-_*)>0$, we find
\begin{align}
  \hat{A}_i = \frac{1}{|\vex|^2}\eta^\alpha_{ij} x^j\sigma^\alpha +\bigO\left(|\vex|\right)\ .
  \label{eq: A hat i leading order}
\end{align}
In contrast, if $\mu(x^-_*)=0$ and $\dot{\mu}(x^-_*)=0$, then we have
\begin{align}
  \hat{A}_i = \bigO\left(|\vex|\right)\ .
\end{align}
Before we can say anything about $Q_*$, we must finally determine the corresponding asymptotic behaviour of $A_i$, determined in terms of $\hat{A}_i$ and $A_-$ by
\begin{align}
  A_i = \hat{A}_i +\frac{1}{2}\Omega_{ij} x^j A_-\ .
\end{align}
Then, if $A_-$ dies away at least as quickly as $|\vex|^{-3}$ as $|\vex|\to\infty$, and is no more singular than $|\vex|^{-1}$ as $|\vex|\to 0$, then the leading order behaviour of $A_i$ in these limits if $\mu(x^-_*)>0$ is given by
\begin{align}
  	A_i &= \bigO\left( |\vex|^{-2} \right) &&\hspace{-20mm} \text{as }|\vex|\to \infty 	\ ,	\nn\\
  	A_i &= \frac{1}{|\vex|^2}\eta^\alpha_{ij} x^j\sigma^\alpha +\bigO\left(1\right) \qquad  &&\hspace{-20mm}\text{as }|\vex|\to 0 \ .
\end{align}
Hence, the contribution to $Q_*$ from the integral over $S^3_\infty$ vanishes. Conversely, the behaviour of $A_i$ near the origin is precisely that of a single $SU(2)$ anti-instanton in singular gauge, centred at the origin, and thus the resulting contribution to $Q_*$ is quantised in the integers. Indeed, we have that the Chern-Simons 3-form $\omega_3$ pulled back to $S^3_0$ is given by $\omega_3|_{S^3_0}=\left(4+\bigO(|\vex|^{-2})\right)d\Omega_3$, where $d\Omega_3$ is the standard volume form on $S^3$. Hence, the contribution to $Q_*$ from the integral over $S^3_0$ is precisely $-1$, and so we find $Q_*=-1$.

Conversely, for the same boundary conditions on $A_-$ we find that if $\mu(x^-_*)=\dot{\mu}(x^-_*)=0$ then we have the asymptotic behaviour
\begin{align}
  	A_i &= \bigO\left( |\vex|^{-2} \right)  &&\hspace{-20mm} \text{as }|\vex|\to \infty 		\ ,\nn\\
  	A_i &= \bigO\left(1\right)\qquad  &&\hspace{-20mm}\text{as }|\vex|\to 0 \ .
\end{align}
and hence, the contribution to $Q_*$ on both $S^3_\infty$ and $S^3_0$ vanishes, and we have $Q_*=0$.

We are of course free to make such a choice of boundary condition for $A_-$, in effect defining some refined subspace of the total configuration space in which we require $A_-$ sits. However, it is a priori not clear that this subspace intersects with the subspace of solutions to the classical equations of motion, and thus such a boundary condition may violate any straightforward variational principle in the theory. However we will see below that there are solutions for $A_-$ which leave the instanton number  of $A_i$ intact and curiously that there are also solutions which precisely cancel the divergent behaviour of $\hat{A}_i$ near the worldline, and hence have $Q_*=0$ for all $x^-_*$.\\

For clarity, let us summarise our findings so far. We have determined how the value of $Q_*$ as defined at some $x^-_*$ in (\ref{eq: Q star definition}) depends on the function $\mu(\tau)$ appearing in $\hat{A}_i$ and thus $A_i$ through (\ref{eq: Phi for single instanton at origin}). Provided that $\mu$ converges to some asymptotic values $\mu(\pm \infty)$ at least as quickly as $\tau^{-2}$ as $\mu\to\pm \infty$, and $A_-$ has suitable asymptotic behaviour, we find the following. If $\mu(x^-_*) = \dot\mu(x^-_*) = 0$, then $Q_*=0$, while if $\mu(x^-_*)>0$ then $Q_*=-1$. In most of what follows




\subsection{Creation and annihilation}\label{subsubsec: creation and annihilation}


We have found that the instanton flux $Q_*$ on a slice $\mathbb{R}^4_*$ of constant $x^-=x^-_*$ depends in a crucial way on whether $\mu(x^-_*)>0$ or $\mu(x^-_*)=0$. This is indicative of a singularity not just in the gauge field but in the field strength $F$ itself, located at points at the spatial origin at which $\mu(x^-)$ transitions from a zero to non-zero value.

We can understand this as follows. Suppose $\mu(\tau)>0$ on $\tau\in(\tau_1, \tau_2)$, $\tau_1<\tau_2$, and identically zero otherwise. We have then that the total instanton flux over a constant $x^-$ slice is $-1$ if $x^-\in (\tau_1, \tau_2)$, while it is zero for $x^-\in (-\infty,\tau_1)\cup (\tau_2,\infty)$. We can thus interpret the point $x_1=(\tau_1,\vec 0)$ as the location at which an anti-instanton is created, and $x_2=(\tau_2,\vec 0)$ as the point at which it is annihilated.

We can gain further insight into the behaviour of the gauge field at the transition points $x_1, x_2$ by considering the instanton charge over more general four-dimensional submanifolds. Define $\Omega_4 = (1/8\pi^2)\,\text{tr}\left(F\wedge F\right)$, and write $Q(S)=\int_S \Omega_4$ for some submanifold $S$, so that for instance $Q_*=Q(\mathbb{R}^4_*)$. Note, it is clear that away from $x^i= 0$, we have $d\Omega_4=0$. We in fact have that $d\Omega_4=0$ everywhere except for at the transition points. This is seen by considering the integral of $\Omega_4$ over generic Gaussian pillboxes. Consider in particular $Q(P)$ where $P$ is a cylinder whose top and bottom lie transverse to the line $x^i= 0$. If such a cylinder does not intersect the line $x^i= 0$, then $A$ is defined globally over $P$ and hence $Q(P)=0$. Suppose instead that $P$ does intersect the spatial origin, but that it does not contain a transition point (see Figure \ref{fig: trivial pillbox}).
\begin{center}
\begin{minipage}{\textwidth}
\centering
\captionof{figure}{Pillbox integral with vanishing $Q$.}\label{fig: trivial pillbox}
\hspace{-8mm}\includegraphics[width=70mm]{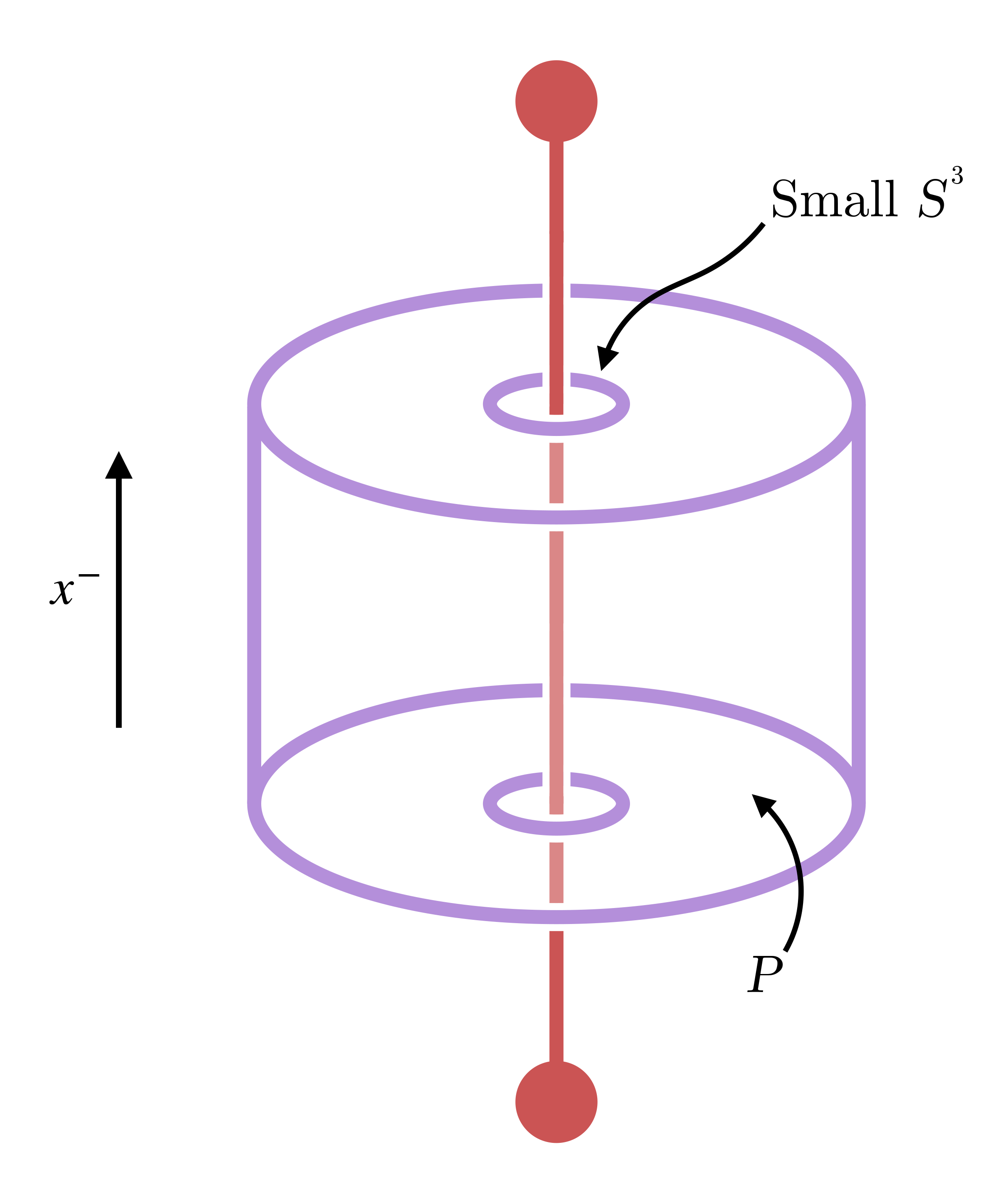}	
\end{minipage}
\end{center}
 Then, $Q(P)$ reduces to a pair of integrals of $\omega_3(A)$ on the small 3-spheres surrounding the two points at which the origin intersects $P$, with a relative minus sign due to orientation. But these two contributions are equal, and thus $Q(P)=0$. This is then sufficient to ensure that $d\Omega_4=0$ everywhere away from transition points.\\
 
We can next consider $Q(S)$ for a generic submanifold $S$. The fact that $\Omega_4$ is closed away from the two transition points implies that $Q(S)$ is topological; we can smoothly deform $S$ without changing $Q(S)$, provided that such a deformation does not drag $S$ through a transition point. In particular, if $S$ doesn't contain either of $x_1, x_2$, then $S$ can be shrunk to a point and $Q(S)=0$. Suppose instead that $S$ contains $x_1$ (but not $x_2$). We can then smoothly deform $S$ to a cylinder of the type described previously (see Figure \ref{fig: creation pillbox}). It is clear then that $Q(S)$ receives a contribution of $-1$ from the top of the cylinder, but zero from the bottom, and hence $Q(S)=-1$. Similarly, for $S$ containing $x_2$, but not $x_1$, we have $Q(S)=+1$.
\begin{center}
\begin{minipage}{0.8\textwidth}
\centering
\vspace{1em}\captionof{figure}{Instanton charge around a creation point.}\label{fig: creation pillbox}\vspace{-1em}
\includegraphics[width=140mm]{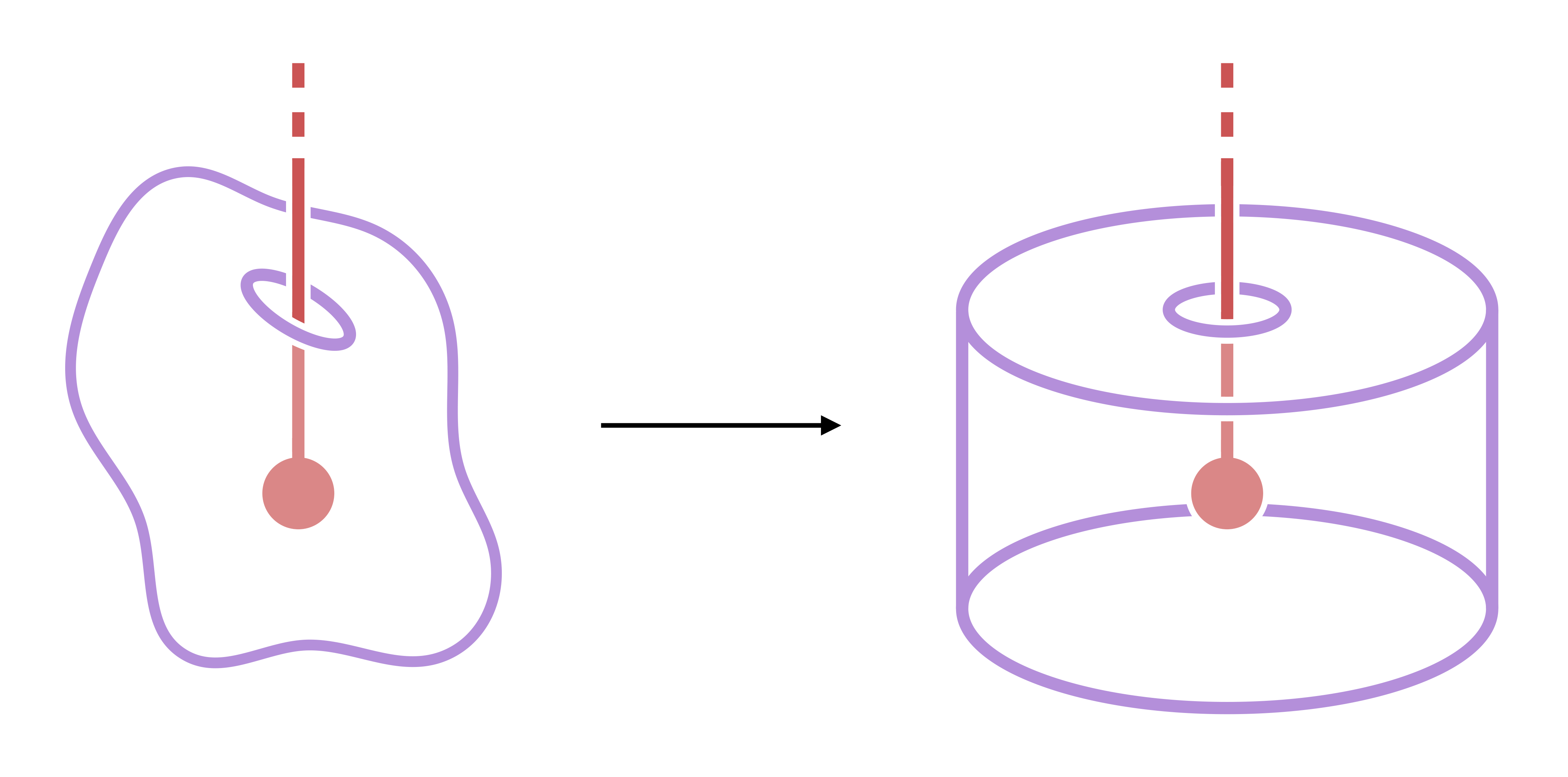}	
\end{minipage}
\end{center}
Indeed, we can consider $S$ to be some arbitrarily small 4-sphere about either a creation or annihilation point, for which we will still have $Q(S)=-1$ or $Q(S)=+1$, respectively. Thus, we learn that $x_1$ and $x_2$ are precisely the positions of \textit{instanton insertions} $(x_1,-1)$ and $(x_2,+1)$ as defined in Chapter \ref{chap: An explicit model and its symmetries}. The data of the corresponding bundle is precisely $\{(x_1,-1),(x_2,+1)\}$.

Note finally that this analysis generalises trivially to the case that $\mu$ varies from zero to non-zero and back not once but a number of times. Such a configuration describes an anti-instanton being created then annihilated, followed by another being created then annihilated, and so on as in example (\ref{wavey}).

\subsection{Moving away from the origin}

This analysis generalises easily to describe a single, static anti-instanton sitting not necessarily at $x^i=0$ but at a generic constant worldline $x^i=y^i$. This is not immediate, due to the unconventional translational symmetries generated by $\{P_-, P_i\}$, but is nonetheless not much more work to show. The $\tHooft$ giving rise to such an anti-instanton can be written as
\begin{align}
  \tHooft = 1+ \int d\tau \frac{\mu(\tau)}{z(x,y(\tau))\bar{z}(x,y(\tau))}\ ,
\end{align}
for $y(\tau)=(\tau, y^i)$ with constant $y^i\in\mathbb{R}$. More explicitly, we have
\begin{align}
  \tHooft = 1+ \int_{-\infty}^\infty \frac{\mu(\tau)}{|\tau-z|^2} d\tau = 1+\frac{1}{z-\bar{z}} \int_{-\infty}^\infty d\tau\, \mu(\tau) \left(\frac{1}{\tau-z}-\frac{1}{\tau-\bar{z}}\right) \ ,
\end{align}
as in (\ref{eq: Phi for single instanton at origin}), except now $z=z(x,(0,y^i))=x^- +\frac{1}{2}\Omega_{ij} x^i y^j +  \frac{i}{4R}(x^i-y^i) (x^i-y^i)$. The important point, however, is that since the instanton is still static, $z$ is independent of $\tau$, and hence we can proceed identically as before.

We are again interested in $Q_*$, the instanton charge on a slice of constant $x^-_*$. In particular, the large $|\vex - \vey|\sim |\vex|$ behaviour is such that $Q_*$ receives no contribution from spatial infinity, for suitable behaviour of $A_-$. Conversely, we can assess the behaviour near the wordline $x^i\to y^i$ by Fourier transform of $\mu(\tau)$, which gives us
\begin{align}
  \tHooft	&= 1+ \frac{4\pi R}{|\vex-\vey|^2}\left[ \int_{-\infty}^0 d\omega\, \tilde{\mu}(\omega) e^{i\omega \bar{z}} + \int^{\infty}_0 d\omega\, \tilde{\mu}(\omega) e^{i\omega z} \right]\ ,
\end{align}
and hence as $|\vex - \vey|\to 0$,
\begin{align}
  \tHooft &=\frac{4\pi R}{|\vex - \vey|^2}\mu\hspace{-0.5mm}\left(x^-+\tfrac{1}{2}\Omega_{ij} x^i y^j\right) + \bigO(1) 	\nn\\
  &=\frac{4\pi R}{|\vex - \vey|^2}\mu\hspace{-0.5mm}\left(x^-\right) + \bigO(|\vex - \vey|^{-1}) \ .
  \label{eq: single, static asymptotics away from origin}	
\end{align}
This is then enough to ensure that $\hat{A}_i$ and, for suitable boundary conditions for $A_-$, the gauge field $A_i$ behaves near the worldline precisely like a single anti-instanton in singular gauge, provided that $
\mu(x^-_*)>0$. If this is indeed the case, then $Q_*=-1$. If however $\mu(x^-_*)=0$ and $\dot{\mu}(x^-_*)=0$, we have $Q_*=0$. Indeed, the interpretation of transition points between these regions as creation and annihilation points, each carrying non-zero instanton charge on surrounding 4-spheres, generalises in the obvious way.


\subsection{General worldlines}


We have seen that in the case that $y^i(\tau)$ is constant and $A_-$ well-behaved, the resulting gauge field $A_i$ describes an anti-instanton sitting at $x^i = y^i$, that is created whenever $\mu$ transitions from a zero to non-zero value, and then annihilated when it returns to zero. These transition points are then special points in the spacetime, carrying non-zero instanton charge, which are precisely identified as the instanton insertions required to define non-trivial six-dimensional Kaluza-Klein modes in the theory,

This interpretation extends in the natural way to the more general form of $\tHooft$,
\begin{align}
  \tHooft(x) = 1+ \sum_{A=1}^N\int d\tau_A \frac{\mu_A(\tau_A)}{z(x,y_A(\tau_A))\bar{z}(x,y_A(\tau_A))}\ ,
\end{align}
where we can assume without loss of generality that each of the $\mu_A(\tau_A)$ is strictly non-zero on some open interval $(\tau_{A,1}, \tau_{A,2})\subseteq \mathbb{R}$, and otherwise identically zero. Then, $\tHooft(x)$ is regular throughout $\mathbb{R}^5$, except along curves defined by $y_A(\tau)=(y^-_A(\tau),y^i_A(\tau))$ for $\tau\in (\tau_{A,1}, \tau_{A,2})$, at which it is singular. If we further suppose that each of these curves extends in the $x^-$ direction without turning---more precisely that each of the $y^-_A(\tau)$ is a strictly monotonic function, which we are free to take as strictly increasing\footnote{This is because if we have a worldline with $y^-$ strictly decreasing, we can simply reparameterise $\tau\to-\tau$.}---then the resulting gauge field $A$ describes $N$ anti-instantons. Each is created at $y_A(\tau_{A,1})$, follows the worldline $y_A(\tau_A)$, and then is annihilated at $y_A(\tau_{A,2})$. 

To see this, let us first for simplicity of notation restrict our attention to the case of a single monotonic worldline and, as we did in the spherically symmetric case, consider the asymptotic behaviour of $\tHooft$ at fixed $x^-$ as we approach the worldline. We are once again really interested in the resulting asymptotics of $\hat{A}_i$ which, for suitable boundary conditions on $A_-$, dictate the instanton charge $Q_*$ as measured over the slice $\mathbb{R}^4_*$ at constant $x^- = x^-_*$. We have
\begin{align}
  \tHooft(x) = 1+ \int d\tau \frac{\mu(\tau)}{z(x,y(\tau))\bar{z}(x,y(\tau))}\ ,
\end{align}
where we have implicitly used the monotonicity of $y^-$ to reparameterise the worldline such that $y(\tau)=(\tau, y^i(\tau))$, in doing so redefining $\mu(\tau)$ appropriately. The function $\mu(\tau)$ is strictly non-zero on $(\tau_1, \tau_2)\subseteq \mathbb{R}$, and identically zero otherwise.

Let us first fix $x^-=x^-_*$ such that $\mathbb{R}^4_*$ does not intersect the worldline, {\it i.e.}\ $x^-_*\notin [\tau_{1},\tau_{2}]$. Then, $\tHooft$ is perfectly regular throughout $\mathbb{R}^4_*$, and dies away sufficiently fast as $|\vex|\to \infty$ to ensure that for suitable behaviour of $A_-$, we have $Q_*=0$.

Suppose instead that $\mathbb{R}^4_*$ cuts through the interior of the worldline, that is $x^-_*\in (\tau_1,\tau_2)$. Then, as we approach $x^i \to y^i(x^-_*)$, the integral becomes increasingly divergent, with the dominant contribution from a neighbourhood of $\tau=x^-_*$. In this neighbourhood, we can write $y^i(\tau) = y^i(x^-_*)+\bigO\left(\tau- x^-_*\right)$, and thus as $|\vex - \vey(x^-_*)|\to 0$,
\begin{align}
  \tHooft(x) &\sim 1+ \int d\tau \frac{\mu(\tau)}{z(x,(\tau,y^i(x^-_*)))\bar{z}(x,(\tau,y^i(x^-_*)))} \nn\\
  &\sim \frac{4\pi R}{|\vex - \vey (x^-_*) |^2} \mu(x^-_*)\ ,
\end{align}
which follows from (\ref{eq: single, static asymptotics away from origin}). Hence, noting the sufficiently small behaviour as $|\vex|\to\infty$, and for suitable behaviour of $A_-$, we find $Q_*=-1$.

These asymptotics then generalise to the case of $N$ monotonic worldlines, provided they do not intersect. Indeed, by generalising the arguments of Section \ref{subsubsec: creation and annihilation}, we can learn how to read off the value of $Q(S)$ for $S$ a 4-dimensional submanifold that does not pass through a transition point.

This can be summarised as follows. Suppose we have $N$ monotonic, disjoint worldlines, and let $S$ be some 4-dimensional submanifold that does not pass through any of the creation or annihilation points. Each anti-instanton is created at a point $x=y_A(\tau_{A,1})$ and annihilated at a point $x=y_A(\tau_{A,2})$, with $\tau_{A,2}> \tau_{A,1}$ and hence $y^-(\tau_{A,2})>y^-(\tau_{A,1})$. Thus, the $x^-$ direction defines an intrinsic direction of each worldline. Then, each time a worldline passes in this direction through\footnote{We assume that the intersection of $S$ and the set of all worldlines is a set of disjoint points in $\mathbb{R}^5$.} $S$ `upwards' in a right-handed sense, $Q(S)$ receives a contribution of $-1$, while each time it passes through `downwards', we pick up a $+1$. See Figure \ref{fig: multi disjoint} for an illustrative example. In particular, for $S$ a small 4-sphere surrounding a creation point, $Q(S)=-1$, while around an annihilation point, $Q(S)=+1$.

\begin{center}
\begin{minipage}{\textwidth}
\centering
\begin{minipage}{0.8\textwidth}
\captionof{figure}{The instanton charge $Q(S)=-1$ for a closed 4-dimensional surface $S$, which intersects some disjoint worldlines. Let $M$ be the region in $\mathbb{R}^5$ enclosed by $S$. Then, the two leftmost worldlines are those of anti-instantons created in $M$, and annihilated outside. The third describes an anti-instanton created outside $M$, and annihilated inside. The final, rightmost anti-instanton is both created \textit{and} annihilated outside $M$, but has worldline that nonetheless passes through $M$.}\label{fig: multi disjoint}
\end{minipage}
\includegraphics[width=140mm]{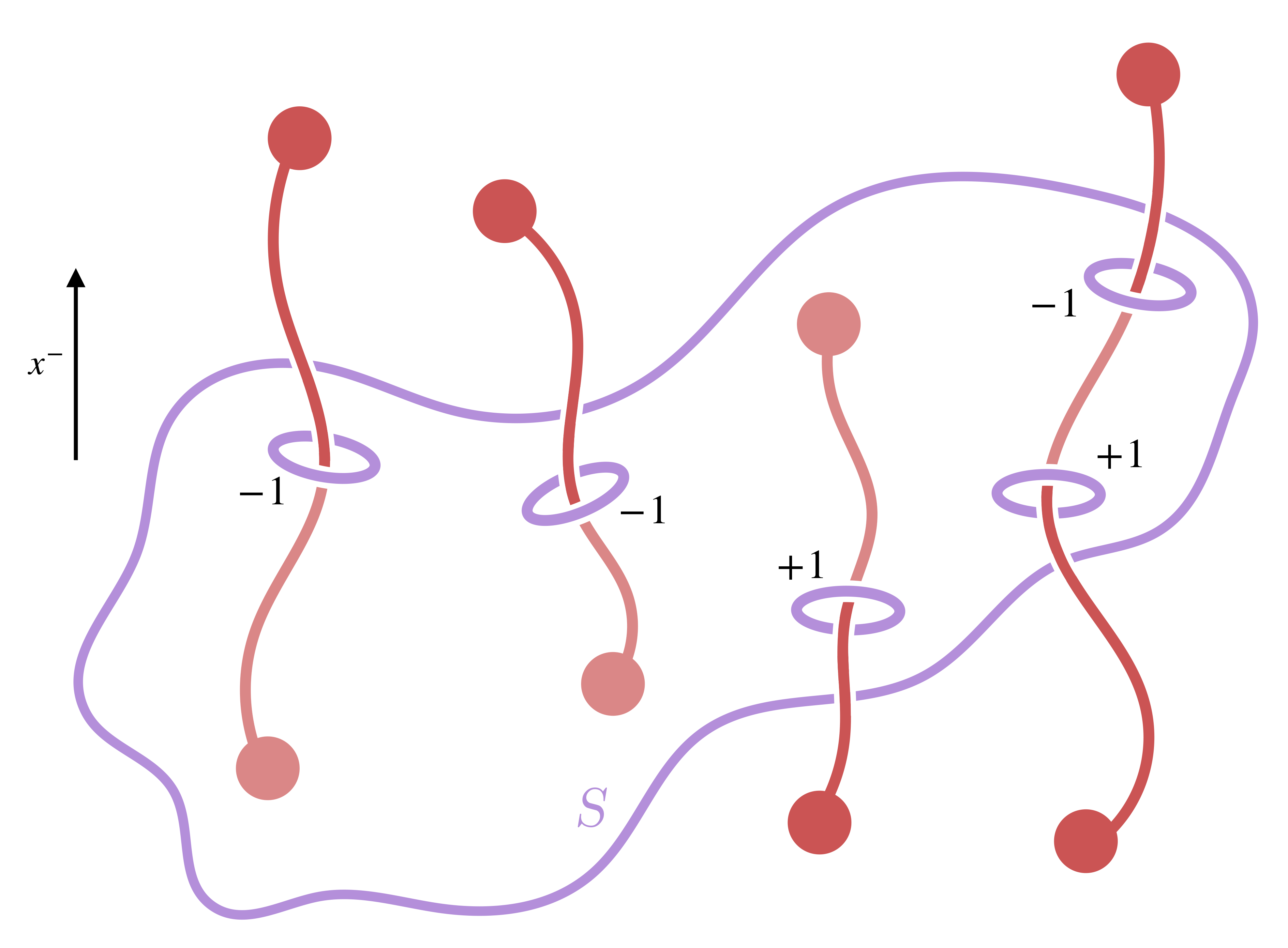}	
\end{minipage}
\end{center}


\subsection{Intersections, turning points, and graphs}


Once we allow for general worldlines, there are a number of interesting additional features our worldlines may have that were not present in the static case. These give rise to an extended space of possible worldline configurations. Then, using the asymptotics we've already found, we can show that such configurations include transition points with not only $Q=\pm 1$, but generic $Q\in\mathbb{Z}$, and thus recover instanton insertions of general charge.

So suppose we once again start with the $N$-instanton solution,
\begin{align}
  \tHooft(x) = 1+ \sum_{A=1}^N\int d\tau_A \frac{\mu_A(\tau_A)}{z(x,y_A(\tau_A))\bar{z}(x,y_A(\tau_A))}\ ,
\end{align}
but let us go beyond the choice of monotonic, disjoint worldlines. First, we can consider what happens when a pair of worldlines intersects at one or more isolated points. By suitably splitting up any worldlines that intersect in their interiors into smaller worldlines joined end-to-end, we can reformulate this configuration as a set of worldlines that are disjoint in their interiors, but may share creation and annihilation points.

We could also suppose that one or more worldline has a turning point: a point at which the corresponding $\dot{y}^-(\tau)$ flips sign, and the worldline turns around. However, so long as we restrict our focus to worldlines for which $\dot{y}^-=0$ only at isolated points, we can once again split up such worldlines into into smaller sections, on each of which $y^-$ is monotonic. For example, a worldline with $y^-(\tau)=\tau(\tau-1)(\tau+1)$ and $\mu(\tau)$ non-zero for $\tau\in(-2,2)$ is split into three monotonic worldlines, forming a graph between four transition points (see Figure \ref{fig: turning points}). It is important here to remember that, due to our freedom to reparameterise, the only sense of `direction' for a worldline is that which corresponds to increasing $x^-$. Thus, there is no sense in which such a split-up worldline `remembers' it was once a single worldline with turning points.
\begin{center}
\begin{minipage}{0.8\textwidth}
\centering
\captionof{figure}{Splitting of a turning worldline into several monotonic pieces}\label{fig: turning points}
\includegraphics[width=80mm]{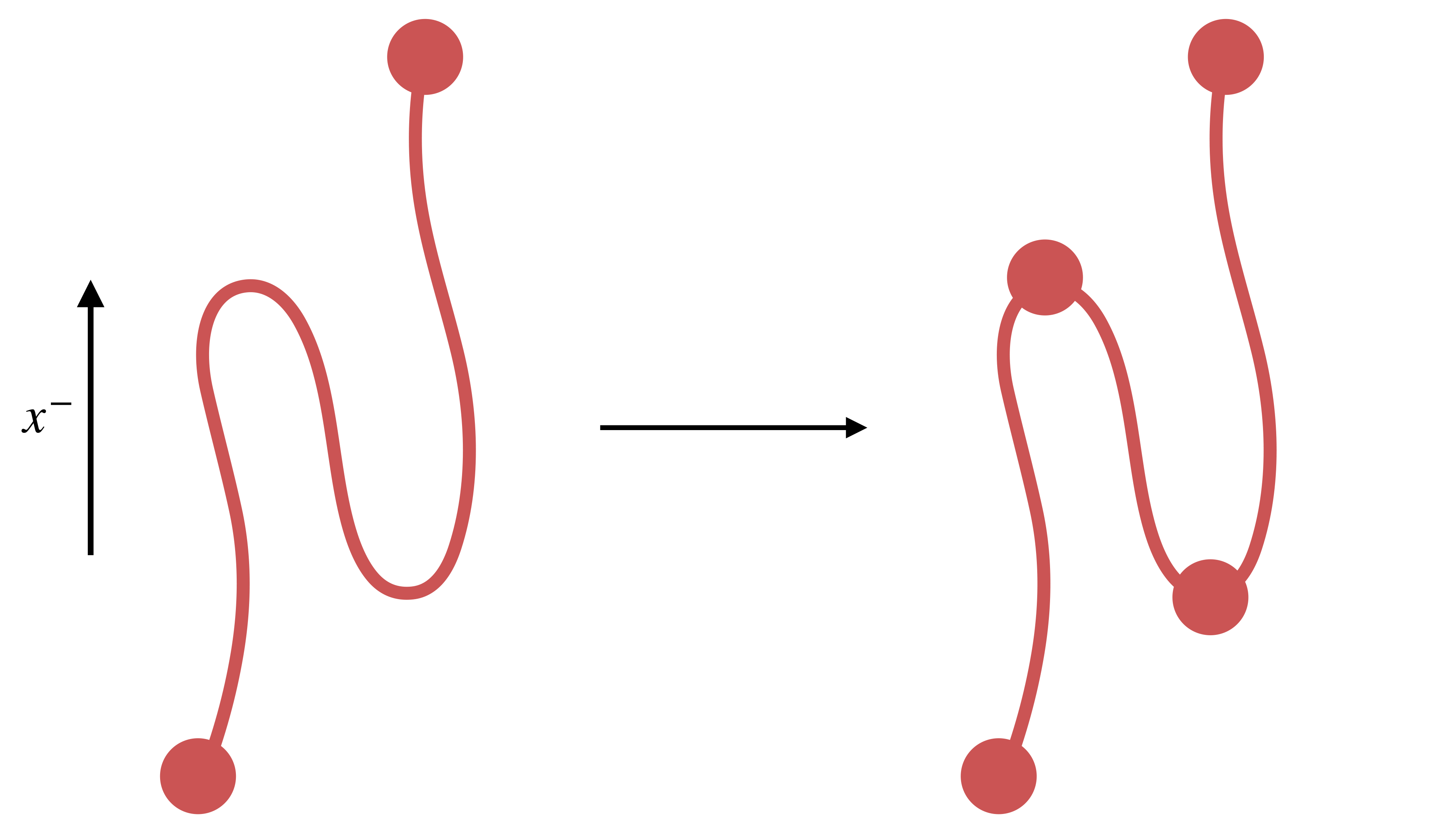}	
\end{minipage}
\end{center}
So, we are lead to a more general set-up:\ we still have $N$ monotonic worldlines, but now they are allowed to share beginnings and ends. A general worldline configuration is then a graph, whose nodes are a set of transition points, and whose edges are a set of monotonic worldlines. However, despite this generalisation, our rules for computing $Q(S)$ for some generic 4-dimensional surface $S$ carry over straightforwardly, as they care only about the asymptotic behaviour of the gauge field in a neighbourhood of the point at which a worldline intersects $S$. In particular, each time a worldline passes through $S$ `upwards' in a right-handed sense, $Q(S)$ receives a contribution of $-1$, while each time it passes through `downwards', we pick up a $+1$. See Figure \ref{fig: general graph} for an illustrative example.
 
\begin{center}
\begin{minipage}{0.8\textwidth}
\centering
\captionof{figure}{The instanton charge $Q(S)=+1$ on a closed surface $S$, amongst transition points joined with monotonic worldlines. The middle two transition points lie in the region enclosed by $S$, while the other two are outside.}\label{fig: general graph}\vspace{-1em}
\includegraphics[width=110mm]{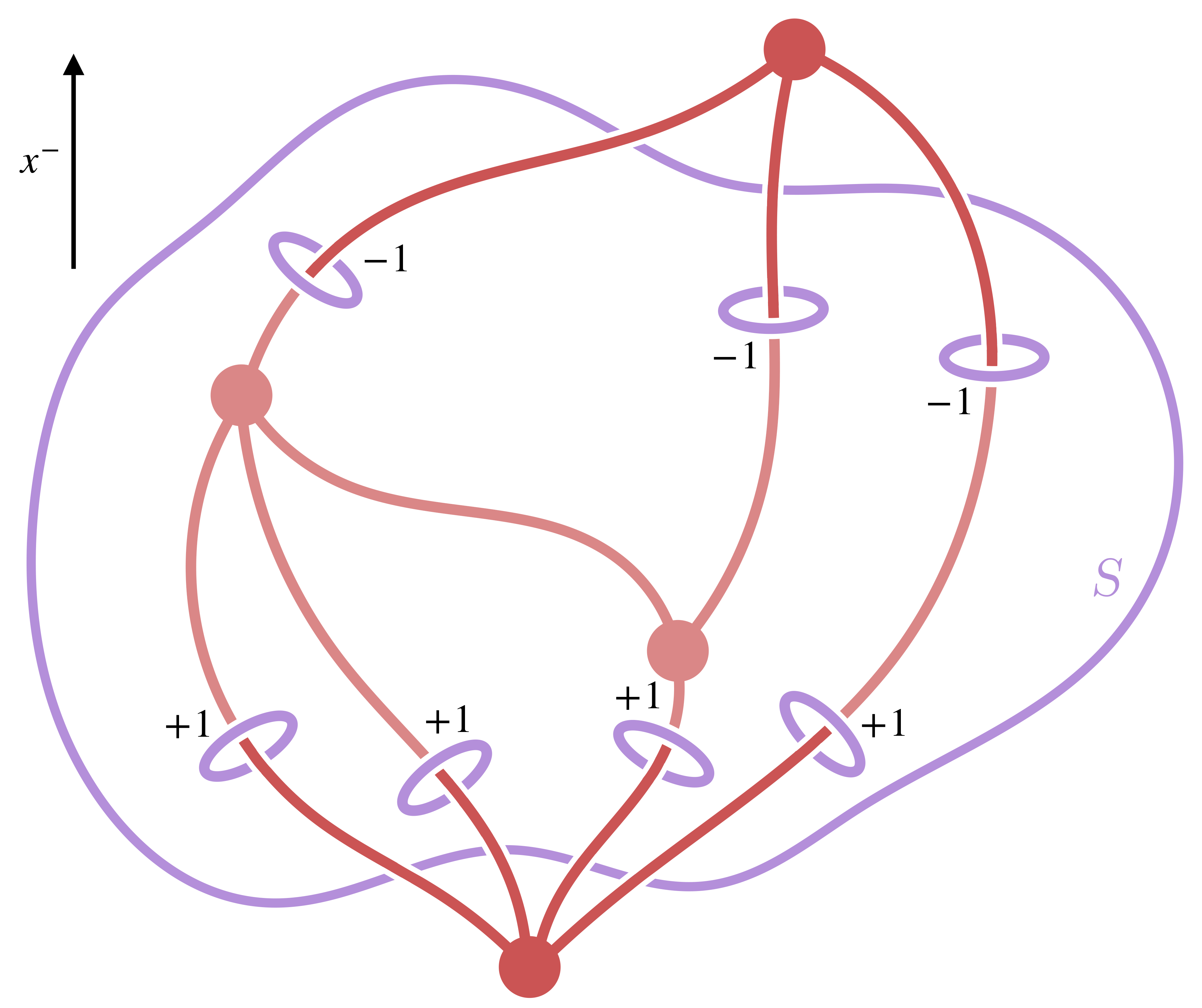}	
\end{minipage}
\end{center}
So, we can now ask what $Q(S)$ is when $S$ is a small 4-sphere surrounding some transition point $y\in\mathbb{R}^5$. It is given simply by the number $m$ of anti-instantons annihilated at $y$, minus the number $n$ created at $y$ (see Figure \ref{fig: general transition point}). Thus, the point $y$ is precisely the position of an instanton insertion $(y,m-n)$.
\begin{center}
\begin{minipage}{0.8\textwidth}
\centering
\captionof{figure}{A general transition point, with $Q(S)=m-n$ on a small 4-sphere $S$ surrounding it.}\label{fig: general transition point}
\includegraphics[width=90mm]{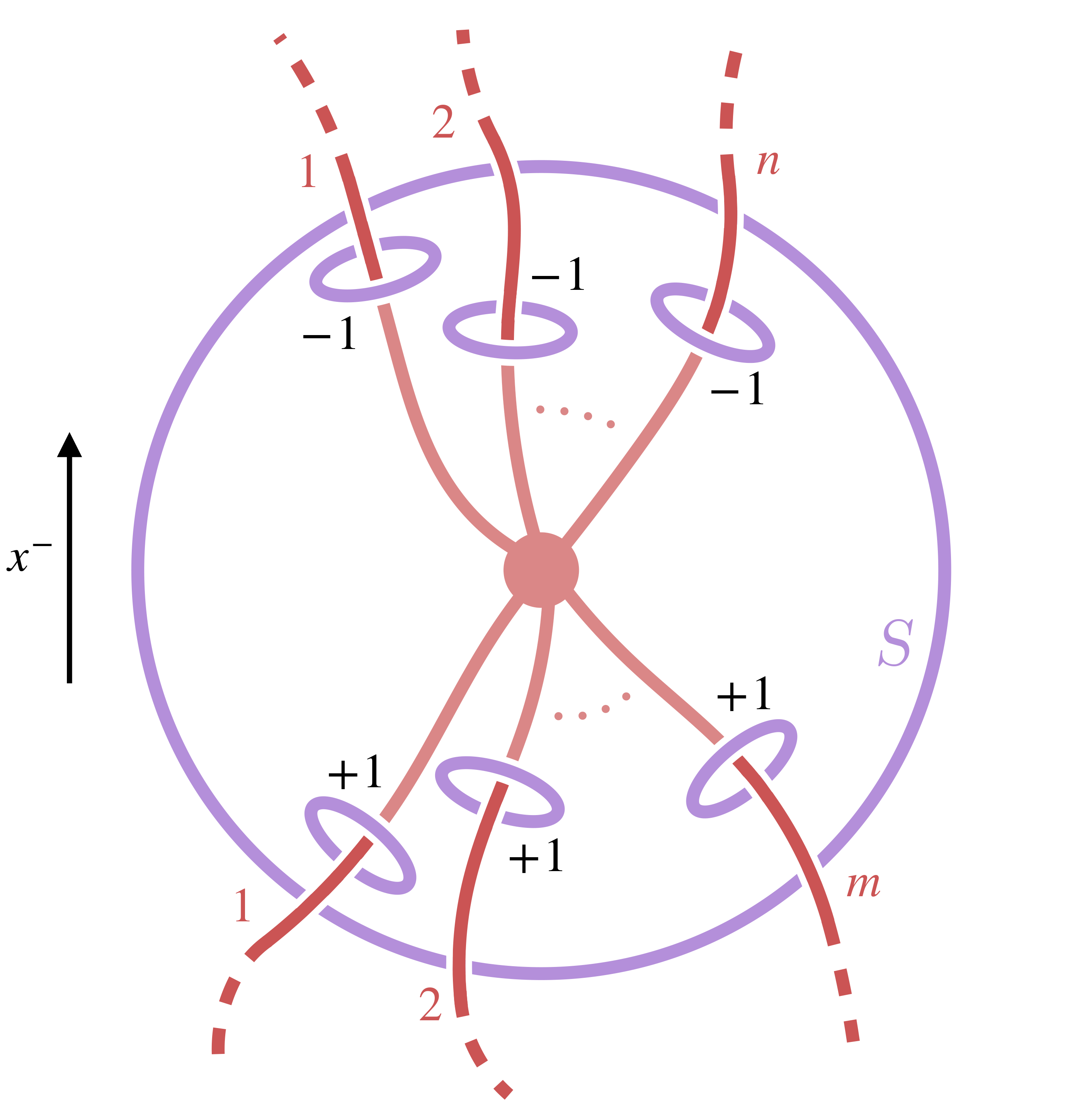}	
\end{minipage}
\end{center}
Let us finally make some comments. Firstly it might appear as though there is a little ambiguity in our analysis: given any graph of transition points and monotonic worldlines, we can always split any worldline into a pair of monotonic worldlines, in effect introducing a new transition point. However, such a point has $Q=0$ on a small 4-sphere surrounding it, and so therefore need not give rise to a singularity in $F$.

Secondly, note that the reinterpretation of turning points as transition points---{\it i.e.}\ points at which $\text{tr}\left(F\wedge F\right)$ diverges---is forced upon us. This is seen most simply by considering $Q(S)$ for $S$ a small 4-sphere surrounding such a point, which following our discussion gives $Q(S)=-2$ at a local minimum (the creation of two anti-instantons), or $Q(S)=+2$ at a local maximum (the annihilation of two anti-instantons). It is nonetheless instructive to look at a simple example. We can consider a worldline with $y^-(\tau)=\tau^2$ and $y^i(0)=0$, which has a local minimum when $\tau=0$ at the spacetime origin $x=(0,\vec{0})$. Suppose we attempt to calculate $Q(S)$ for $S\cong \mathbb{R}^4$ the spatial slice at constant $x^-=0$, which includes the turning point. As we approach $|\vex|\to 0$, we have
  \begin{align}
	\tHooft &\sim 1 +\int_{-\infty}^\infty d\tau  \frac{\mu(\tau) }{|x^--\tau^2+\tfrac{i}{4R}|\vex|^2|^2} \nonumber\\
	& = 1 +\int_0^\infty d\tau'\frac{1}{2\sqrt{\tau'}}  \frac{\mu(\sqrt{\tau'})+\mu(-\sqrt{\tau'}) }{|x^--\tau'+\tfrac{i}{4R}|\vex|^2|^2}		\nn\\
	&\sim \frac{2\pi R}{|\vex|^2} \lim_{x^- \to 0}\left(\frac{\mu(\sqrt{x^-})+\mu(-\sqrt{x^-})}{\sqrt{x^-}}\right) \ .
 \end{align}
If $\mu(0)\neq 0$, this limit does not exist and $\tHooft$ is too divergent at the origin. Hence, $Q(S)$ is ill-defined, and $x=(0,\vec{0})$ is a transition point. Conversely, if $\mu(0)= 0$, then $x=(0,\vec{0})$ is by definition the start of two distinct worldlines. In either case, we find that the turning point is indeed a transition point.

Looking a little ahead, let us finally make a brief comment on the analytic properties of the functions $\mu_A(\tau_A)$. Our analysis of the asymptotics of the gauge field, giving rise to contributions $Q$, required only that $\mu_A$ was such that $\tHooft$ is finite away from worldlines---i.e. that the integrals in (\ref{eq: Phi general solution}) exist when we are sat away from all worldlines---and that $\mu_A$ admits a well-defined Fourier transform. Such choices could, for instance, include jump discontinuities. However, as we will see in Section \ref{sec: dynamics}, a sufficient condition for $\hat{A}_i$ to give rise to a finite action is that the functions $\mu_A$ are sufficiently smooth. In particular, for a monotonic worldline, $\mu$ must tend sufficiently smoothly to zero at the creation and annihilation points. Note further that for a worldline with turning points, the splitting into a set of shorter monotonic worldlines is then consistent only if $\mu_A\to 0$ as one approaches such turning points. Thus,  we learn that a sufficient condition for a worldline with turning points to give rise to a finite action is that the size $\mu$ shrinks to zero sufficiently smoothly at the turning points.

 
\subsection{Constraints on instanton charges}


Despite our freedom in defining worldline configurations with instanton insertions $(x_a, n_a)$ of generic charges $n_a\in\mathbb{Z}$, the global structure of such configurations nonetheless give rise to interesting constraints on what the $n_a$ can be.

Suppose we have a worldline configuration with transition points $x_a$, $a=1,\dots, N$, with $x_a^-<x_{a+1}^-$ for each $a=1,\dots, N-1$, and suppose further that all worldlines are created or annihilated at one of these transition points, as opposed to any escaping to or from infinity. The corresponding bundle is then defined by the data $\{(x_a, n_a)\}_{a=1}^N$, where recall for each $a$ we have charge $n_a=Q(S^4_a)\in\mathbb{Z}$, where $S^4_a$ is a small 4-sphere surrounding $x_a$.

By considering $Q(S)$ for some $S$ surrounding every $x_a$, we have $\sum_a n_a=0$. Indeed, it is clear from (\ref{eq: sum of charges is zero}) that this is true more generally in the entire extended configuration space, rather than just on the constraint surface.

We can also consider $\sum_{a=1}^r n_a$ for some $r=1,\dots, N-1$. Then, we have,
\begin{align}
  \sum_{a=1}^r n_a = \sum_{a=1}^r Q(S^4_a) = Q\left(\cup_{a=1}^r S^4_a\right) = Q\left(M_r\right) \ .  
  \label{eq: multi point deformation}
\end{align}
Here, we smoothly deformed the disjoint union of small 4-spheres into some closed 4-manifold $M_r$  that encloses the $x_1,\dots, x_r$, but not the $x_{r+1},\dots, x_N$. Crucially, this deformation can always be done without the 4-manifold passing through any transition points, and as such, $Q$ is invariant under the deformation.

Then, we have $Q(S_r)\le 0$. To see this, note that we can further deform $M_r$ to a cylinder $P_r$, with top and bottom at $x^-=y^-_1, y^-_2$ respectively, again without passing through any transition points. We have in particular that the top of $P_r$ lies somewhere between $x_r$ and $r_{r+1}$ ({\it i.e.}\ $x^-_r<y^-_2<x^-_{r+1}$), while its bottom lies below all transition points ({\it i.e.}\ $y^-_1<x^-_1$). We can further take the radius of the cylinder to be sufficiently large that the only points at which a worldline passes through $P_r$ is on its top. Hence, following the rules of the previous section, and taking note of the orientation of $P_r$ as inherited from that of $M_r$, we have $Q(M_r)=Q(P_r)\le 0$.

Figure \ref{fig: multi point deformation} provides a schematic of this calculation for $N=4$ and $r=2$, in particular demonstrating the continuous deformation of $S^4_1\cup S^4_2$ into $M_2$ and then into $P_2$.

Therefore, using $Q(S_r)\le 0$, we find
\begin{align}
  \sum_{a=1}^r n_a\le 0 \quad \text{for all }r=1,\dots N \ , 
  \label{eq: graph rule}
\end{align}
or equivalently, $ \sum_{a=r}^N n_a\ge 0$ for all $r=1,\dots, N$. 

Let us finally suppose further that the graph of transition points and worldlines is \textit{connected}. In the above discussion, this then implies that $Q(S_r)=Q(P)<0$ for all $r=1,\dots, N-1$, {\it i.e.}\ that the bound (\ref{eq: graph rule}) is saturated only for $r=N$. Therefore, we have the strengthened statement,
\begin{align}
	\sum_{a=1}^r n_a &< 0\qquad \text{for all }r=1,\dots, N-1	\ ,
  \label{eq: connected graph rule}
\end{align}
or equivalently, $\sum_{a=r}^N n_a > 0$ for all $r=2,\dots, N$.

\begin{center}
\begin{minipage}{0.8\textwidth}
\centering
\captionof{figure}{A schematic showing the continuous deformation of a pair of small 4-spheres $S^4_1\cup S^4_2$, to a single surface $M_2$ that encloses $x_1$ and $x_2$, and finally to a cylinder $P_2$ that is only pierced by worldlines on its top. The arrows represent orientation. Indeed, we can read off $Q(S^4_1)+Q(S^4_2)=-3+1=-2=Q(M_2)=Q(P_2)$.}\label{fig: multi point deformation}
\includegraphics[width=100mm]{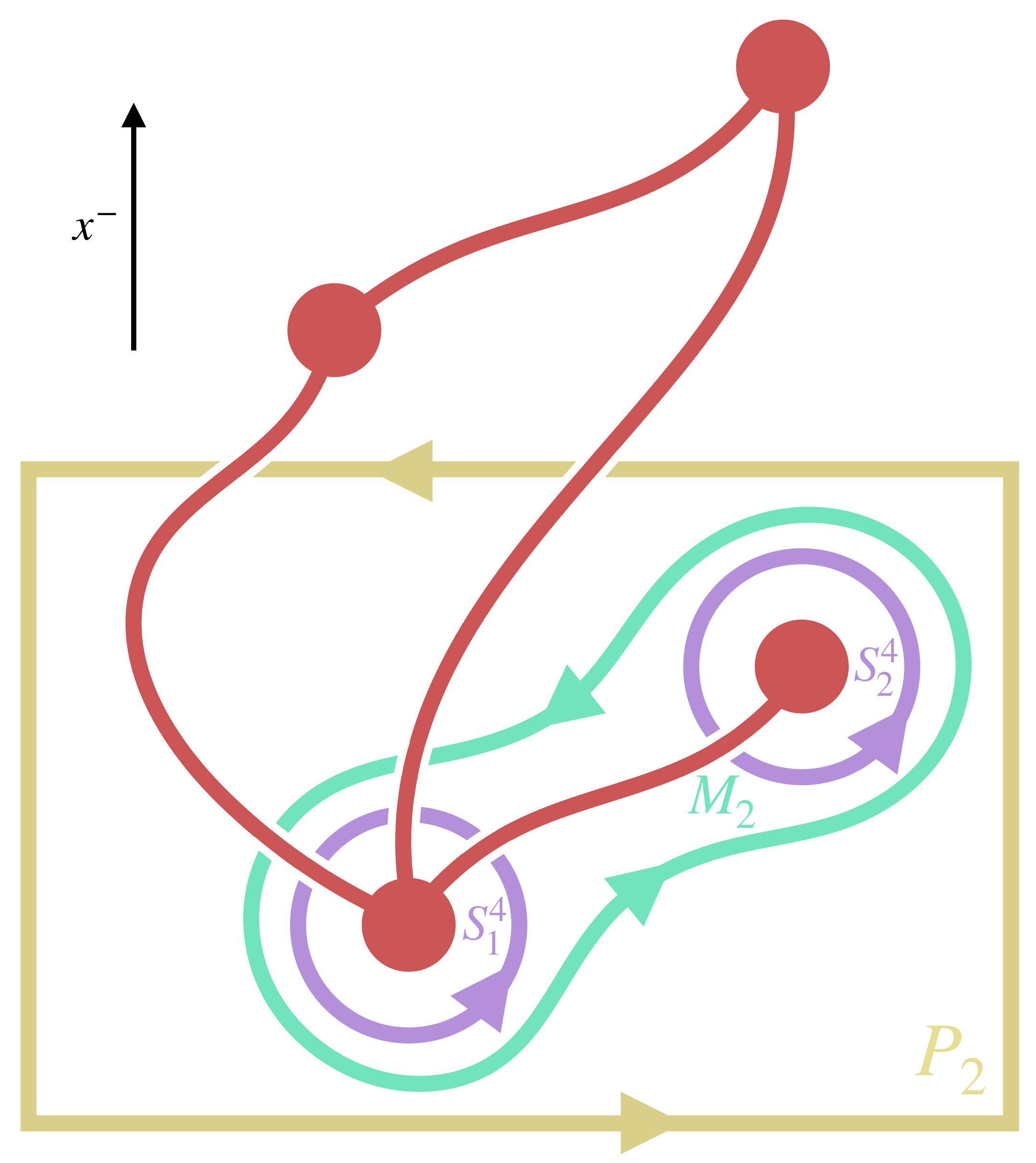}	
\end{minipage}
\end{center}
Each of these results (\ref{eq: graph rule}), (\ref{eq: connected graph rule}) is a straightforward application of the fact that on any spatial surface we will always have $Q({\mathbb R}^4_{y^-})\le 0$ for any $y^-$, provided there are no transition points lying precisely on ${\mathbb R}^4_{y^-}=\{x^-=y^-\}$. This is strengthened to $Q({\mathbb R}^4_{y^-})< 0$ in the case that the worldline graph is connected, and $x_1^-<y^-<x_m^-$.

\section{Relation to correlation functions}\label{sec: relation to correlation functions}

So let us now relate these results to our results on the correlation functions. Recall, at the level of five-dimensional symmetries, we were able to constrain the form of correlation functions to the form (\ref{eq: general N-point function}). However, the requirement that these correlation functions could be resummed to produce solutions to the six-dimensional conformal Ward-Takahashi identities put significant further constraints on correlation functions. In more detail, let us suppose that given some local scalar operator $\fOp$ in the five-dimensional theory, we can identify\footnote{Alternative choices of Fourier resummation are discussed later in Discussion and Further Directions} the $\fOp_n(x)=\I_n(x)\fOp(x)$ as the Fourier modes of some six-dimensional scalar primary operator. Then, the 2-point functions $\left\langle \fOp_{n_1}(x_1)\fOp_{n_2}(x_2) \right\rangle$ should be completely fixed by (\ref{eq: dim red 2pt from heuristics}) up to a single overall coefficient, which is the normalisation of the corresponding six-dimensional 2-point function (\ref{eq: 6d 2pt}). Similarly, the 3-point functions $\left\langle \fOp_{n_1}(x_1) \fOp_{n_2}(x_2) \fOp_{n_3}(x_3) \right\rangle$ should be completely fixed by (\ref{eq: 3pt dim red from heuristics}) up to an overall coefficient, which is the structure constant of the corresponding six-dimensional 3-point function (\ref{eq: 6d 3pt}).

Thus, the question we are lead to ask is: does our theory produce precisely these correlation functions? Does the six-dimensional interpretation of the theory actually hold? We believe that the solutions to the constraint $\mathcal{F}^+=0$ explored in this Chapter provide compelling evidence that this is indeed the case.\\

Let us first make a comment about operator ordering. While in a Euclidean theory operators commute within correlation functions, the same is not true in a Lorentzian theory where order \textit{does} matter. The ordering of operators can be encoded by the $N!$ essentially different ways of Wick rotating an $N$-point function from Euclidean to Lorentzian signature. Indeed, it is this perspective and its representation in terms of the so-called $i\epsilon$ prescription that provided a crucial regulator in our dimensional reduction of Lorenztain correlators in Appendix \ref{app: dimensional reduction of correlators}. A comprehensive review of this formalism can be found in \cite{Hartman:2015lfa}.

The general solutions (\ref{eq: general N-point function}) to our five-dimensional Ward-Takahashi identities depend on the ordering of operators only through the unfixed function $H$, and thus at this level, we could very well have a theory whose correlators are inert under reordering, much like a Euclidean theory. However, the correlation functions at 2-, 3- and 4-points found by dimensional reduction of their six-dimensional counterparts in Chapter \ref{chap: the view from 6d CFT} tell a different story. Here, it is very clear that for our theory to really admit a six-dimensional interpretation, operator ordering \textit{must} matter. For instance, we found that the lift to six dimensions requires that we find the 2-point functions 
\begin{align}
	\langle \fOp_n(x_1)\fOp_{-n}(x_2)\rangle &=  {{\n+\tfrac{\Delta}{2}-1}\choose{\n-\tfrac{\Delta}{2}}}\frac{\left( -2R\, i \right)^{-\Delta}\hat{C}_{12}}{\left(z_{12}\bar{z}_{12}\right)^{\Delta/2}}\left(\frac{z_{12}}{\bar{z}_{12}}\right)^{\n}\ .
	\label{eq: 2pt function worldlines}
\end{align}
If $\sOp$ is the six-dimensional operator with modes $\fOp_n$, then the 2-point functions (\ref{eq: 2pt function worldlines}) resum to produce the six-dimensional Lorentzian correlator $\left\langle \sOp(x^+_1, x^-_1, x^i_1) \sOp(x^+_2, x^-_2, x^i_2) \right\rangle $ for any $x^+_1,x^+_2$, rather than $\left\langle  \sOp(x^+_2, x^-_2, x^i_2) \sOp(x^+_1, x^-_1, x^i_1) \right\rangle $. Each of these two distinct Lorentzian correlators is in turn defined relative to their mutual Euclidean counterpart by the somewhat unconventional $i\epsilon$ prescription (\ref{eq: i epsilon}).

Note, while correlators in a Lorentzian theory depend on ordering rather subtly through a suitable $i\epsilon$ prescription, the 2-point function (\ref{eq: 2pt function worldlines}) depends on ordering in a very clear way. We see that it is not symmetric under $n\to -n$, and it indeed \textit{vanishes} unless $n\ge \Delta/2$. \\

Now, there is an important nuance that has been swept under the rug in supposing that we may recover the normalisation (\ref{eq: 2pt function worldlines}) directly in the five-dimensional theory. In order to actually show that the operator $\I_n \fOp$ can be identified as the non-trivial Fourier mode $\fOp_n$ of a six-dimensional operator, we had to appeal to a path integral formulation of the theory, which computes correlation functions as in (\ref{eq: basic path integral}). It was at this point that we made a somewhat grievous (albeit common) abuse of notation. On one hand, we have been using $\left\langle \dots \right\rangle$ to denote the Wightman functions of some abstract quantisation of the theory, within which operators are generically non-commuting. On the other hand, in expressions such as (\ref{eq: basic path integral}) we have used $\left\langle \dots \right\rangle$ to denote an expectation value computed in a path integral. However, we are always free to commute (bosonic) fields under a path integral sign, corresponding to the familiar fact that the path integral is only good for computing a particular subset of the theory's correlation functions, defined by some \textit{fixed} ordering prescription. Indeed, in a Lorentzian theory one usually defines the path integral such that it computes precisely the \textit{time ordered} correlators of the full theory.

Thus, as with any path integral formulation, the two uses of $\left\langle \dots \right\rangle$ coincide only for operators that are ordered according to some fixed prescription. At least formally we are free to choose any prescription; this comes along with the definition of the path integral. However, we find that there is a single, simple prescription that appears most convenient in realising the theory's six-dimensional interpretation. For some operators $\fOp^{(1)}\dots,\fOp^{(N)}$, we suppose that the path integral computes
\begin{align}
  \left\langle \mathcal{T}\left[\fOp^{(1)}(x_1)\dots \fOp^{(N)}(x_N)\right] \right\rangle = \int D\fields\, \fOp^{(1)}(x_1)\dots \fOp^{(N)}(x_N) e^{iS[\fields]}\ ,
  \label{eq: path integral with order}
\end{align}
where $\mathcal{T}$ denotes ordering with respect to the coordinate $x^-$, so that for example
\begin{align}
  \mathcal{T}\left[ \fOp^{(1)}(x_1) \fOp^{(2)}(x_2) \right] = \begin{cases}
 	\fOp^{(1)}(x_1) \fOp^{(2)}(x_2)	\quad \text{if } x_1^-> x_2^-	\\
 	\fOp^{(2)}(x_2) \fOp^{(1)}(x_1)	\quad \text{if } x_2^-> x_1^-
 \end{cases}\ .
\end{align}
This corresponds to prescribing in the path integral boundary conditions on the $\mathbb{R}^4$'s at $x^-\to\pm \infty$; these are the `in' and `out' configurations between which the path integral integrates. We take these boundary conditions to be such that no worldlines escape to $x^-\to\pm \infty$, and thus all anti-instantons must be created and annihilated at finite points\footnote{This interpretation technically holds only for our solutions of the form (\ref{eq: Phi general solution}). More generally, we simply require that there exists some $X\in \mathbb{R}_{\ge 0}$ such that for all $|x^-|> X$, the instanton charge $Q(\mathbb{R}^4_{x^-})$ vanishes.}.\\

Let us now explain why this prescription appears to be the right one for us to have a chance of realising the full six-dimensional theory. For some scalar primary $\fOp$ in the five-dimensional theory and points $x_1,x_2$ with $x_1^->x_2^-$, consider the 2-point function $F_n=\left\langle \I_n(x_1) \fOp(x_1) \I_{-n}(x_2)\fOp(x_2) \right\rangle$, which is computed by the path integral (\ref{eq: path integral with order}). Then, the lift to six-dimensions \textit{requires} that $F_n$ is given precisely by (\ref{eq: 2pt function worldlines}), and so in particular vanishes for $n< 0$ (and indeed for all $n<\Delta/2$).

The path integral is performed in a single topological sector, defined by the data\linebreak $\{(x_1,n),(x_2,-n)\}$. Within this sector, the path integral is further localised to the subspace of finite action solutions to $\mathcal{F}^+=0$, which we denote $\Sigma^n$. Let us further denote by $\Sigma^n_\tHooft\subset \Sigma^n$ the subspace of $\Sigma^n$ captured by solutions of the form (\ref{eq: Phi general solution}), which we will soon see do indeed have finite action.

At this point we understand why the ordering prescription (\ref{eq: path integral with order}) is crucial. Since $x_1^->x_2^-$, we see that $x_2$ is a creation point, and $x_1$ is an annihilation point. Thus, it is immediate that $\Sigma^n_\tHooft=\emptyset$ for all $n<0$, since we are only able to create \textit{anti}-instantons. Then, if we further suppose that this extends to the entire constraint surface so that $\Sigma^n=\emptyset$ for all $n<0$, we see that the path integral vanishes and so we do indeed have $F_n=0$ for all $n<0$. Conversely, we have $\Sigma^n\supset \Sigma^n_\tHooft\neq \emptyset$ for all $n\ge \Delta/2$, and so it is reasonable to expect $F_{n}\neq 0$ in such cases.

The situation becomes somewhat cloudier, however, for $n$ in the range $0\le n<\Delta/2$. One one hand, we have $\Sigma^n\supset\Sigma^n_\tHooft\neq \emptyset$ for all $n$ in this range, and yet we are supposed to find that we still have $F_n=0$. We can get a little closer by supposing that the path integral should be performed only over \textit{connected} worldline diagrams, giving us $\Sigma^0_\tHooft=\emptyset$ and hence suggesting $F_0=0$. One can indeed go further a propose additional restrictions on the domain of the path integral to argue that $F_n=0$ for all $n$ in this range \cite{Lambert:SymmEnhance}.

One can then play similar games at 3-points. Indeed, it is straightforward to see at the level of worldlines why $\left\langle \I_{n_1}(x_1)\fOp(x_1)\I_{n_2}(x_2)\fOp(x_2)\I_{n_3}(x_3)\fOp(x_3)  \right\rangle$ should indeed vanish whenever $n_3>0$ or $n_1<0$, as we saw in our result (\ref{eq: 3pt dim red from heuristics}) from dimensional reduction. As with the 2-point function, one finds a rich and encouraging relationship between dimensionally reduced correlators and worldline diagrams, suggesting that six-dimensional physics is indeed encoded within the correlation functions of this five-dimensional theory \cite{Lambert:SymmEnhance}.\\

Our prescription (\ref{eq: path integral with order}) constitutes identifying the $x^-$ direction as a preferred direction, with respect to which the path integral naturally orders. To proceed to compute Wightman functions of operators in generic orders then requires a suitable adaptation of the Schwinger-Keldysh formalism \cite{Schwinger:1960qe,Bakshi:1962dv,Bakshi:1963bn,Keldysh:1964ud}, helpfully reviewed in \cite{Skenderis:2008dg}, in which the various orderings of insertions correspond to evolution along different choices of contours in the complex time plane.


\section{Scalar field solutions}


Let us now examine the behaviour of the scalar fields on the constraint surface. In the theory (\ref{eq: Lagrangian}), we have five scalar fields $X^I$ transforming in the adjoint of the gauge group. However, it turns out that this model is in fact a special case of a more general class of theories, which generically have half the supersymmetry and thus form proposals for five-dimensional realisations of the six-dimensional $(1,0)$ theories \cite{Lambert:2020jjm}. We will not need to know much about such theories here, but note simply that in general there is only one scalar $\lambda$ in the adjoint representation that comes from the tensor multiplet in six dimensions.

There is then additionally any number of complex scalar fields contained in the fields $X^z{}_m=\left(X_z{}^m\right)^\dagger$, where the $m$ index runs over the number of hypermultiplets from which these scalars descend, while $z=1,2$. Any such a scalar, which we denote by  $X$, appears in the action  through
\begin{align}
S_\text{scalar} =-\frac{k}{4\pi^2 R}\int dx^-d^4x \ \hat D_i X^\dag \hat D_i X\ .
\end{align}
Here $X$ is taken to be an any unitary representation $R$ of the gauge group (which we take to be $SU(2)$). In particular, let us write write the gauge field as $A=A^\alpha \nu^\alpha$ where we take the standard basis $i\nu^\alpha = \frac{i}{2}\sigma^\alpha$ for $\frak{su}(2)$. Then, we have
\begin{align}
\hat D_i X = \hat\partial_i X - i\hat A^\alpha_iT_\alpha(X)\ ,	
\end{align}
where the $T_\alpha$ span the representation $R$. First look at the classical equation of motion
\begin{align}
	\hat D_i\hat D _i X =0\ .
\end{align}
Smooth solutions to this equation are unique up to their behaviour at the boundary by a variation of the usual argument (note we    need the full five-dimensional integral here):
\begin{align}
 \int dx^-d^4x \hat D_i X^\dag \hat D _i X& = \int\hat dx^-d^4x\, \partial_i( X^\dag\hat D_i  X)-\int dx^-d^4x X^\dag \hat D_i\hat D _i X    \nonumber\\
 & = \int dx^-d^4x\,\hat \partial_i( X^\dag\hat D_i  X)\ .
\end{align}
Thus if $X$ vanishes on the boundary then \begin{align}
\hat D_iX	=0\ ,
\end{align}
everywhere and so $X=0$. In addition if $X$ is the difference   between two solutions which agree on the boundary then $X$ is also a solution but since it vanishes on the boundary the  two solutions must be equal everywhere.

Let us look more carefully at the boundary term. Since $\hat \partial_i$ contains derivatives in $x^-$ and $x^i$ we find
\begin{align}
 \int dx^-d^4x\,\hat{\partial}_i( X^\dag\hat D_i  X) & =    \int_{|\vex|\to\infty} dx^-d\Omega_3 \, 	|\vex|^2 x^i X^\dag \hat D_i X \nonumber\\
 &\qquad -\frac12\ \Omega_{ij} \int_{x^-\to\infty} d^4x \, x^jX^\dag \hat D_i X +\frac12\ \Omega_{ij} \int_{x^-\to-\infty} d^4x \, x^jX^\dag \hat D_i X \ ,
 \label{eq: boundary term shorthand}
 \end{align}
where  $d\Omega_3$ is the volume element on a unit 3-sphere. Thus specifying the behaviour on the boundary means that we must specify the spatial behaviour in the form $X = X_{0} + X_1/|\vex|^2+\dots $ but also the early and late values of $  X$ over all of ${\mathbb R^4}$. We will examine these terms in greater detail below.

In the case of the 't Hooft ansatz we can be quite explicit and 
compute
\begin{align}\label{DDX}
\hat D_i\hat D _i X =	\hat\partial_i\hat\partial_i X -\frac{C_R}{\tHooft^2}\hat\partial_i\tHooft \hat\partial_i\tHooft X  + \frac{2i}{\tHooft}\eta^\alpha_{ij} \hat\partial_i\tHooft T_a(\hat\partial_j X)\ ,
\end{align}
where we introduced $C_R$ as the quadratic Casimir of the representation:
\begin{align}
\sum_\alpha T_\alpha T_\alpha = C_R{\mathbb I}	\ .
\end{align}
For example  in the adjoint representation $(T_\alpha)_{\beta\gamma}=-i\varepsilon_{\alpha\beta\gamma}$ and hence $C_{adj}=2$, whereas for the fundamental representation $T_\alpha=\tfrac12 \sigma^\alpha$ and hence $C_{fund}=3/4$. More generally $C_R=s(s+1)$ with $s=0,\tfrac12,1,...$.
To solve this equation we impose the ansatz $X=X(\tHooft(x^-,\vex))$   so that the last term in (\ref{DDX}) vanishes. In this case we find
\begin{align}
\hat D_i\hat D _i X = \hat\partial_i\tHooft \hat\partial_i\tHooft \left(X'' - \frac{C_R}{\tHooft^2} X\right) =0 	\ ,
\end{align}
where a prime denotes a derivative with respect to $\tHooft$. Thus the  solutions to this take the form
\begin{align}
X = X_{0}\tHooft^{\frac{1-\sqrt{1+4C_R}}{2}} + X'_{0}\tHooft^{\frac{1+\sqrt{1+4C_R}}{2}}\ ,
\end{align}
for constant vectors $X_{0},X'_{0}$. 
However we want well behaved solutions at the poles of $\tHooft$ and hence we find
\begin{align}
X &= X_{0}{\tHooft^{-{\frac{\sqrt{1+4C_R}-1}{2}}}}\nonumber\\
&= X_{0}{\tHooft^{-s}} \ .\end{align}


\section{Dynamics}\label{sec: dynamics}


In this section we will describe how to solve the equations of motion on the constraint surface and evaluate the action. To this end write the Bosonic part of the action as \cite{Lambert:2020jjm}
\begin{align}
S = \frac{k}{4\pi^2 R} \int dx^- d^4x\, \Big\{ \frac12 \tr \big(F_{-i}F_{-i})  + \frac{1}{2} \tr\big(\hat {F}_{ij}G^+_{ij} \big ) -\frac{1}{2} \tr \big(\hat D_{i} \lambda\hat D_{i} \lambda \big) -   \hat D_i {X}_{z}{}^{m}\hat D_i X^{z}{}_{m} \Big\}\ .
\end{align}
Varying with respect to $G^+_{ij},  A_i,  A_-, \lambda$ and $X^z{}_m$ respectively we find the equations of motion
\begin{align}
\hat{ F}_{ij}^+ &= 0 \ , \nonumber\\
\hat D_j G^+_{ij}&= D_-F_{-i}  - i[\lambda, \hat D_i\lambda] - i[[{X}_{z}{}^{m},\hat D_i X^{z}{}_{m}]] \ , \nonumber\\
\frac12 \Omega_{ik}x^k\hat D_j G^+_{ij}&=D_iF_{-i}- \frac{i}{2} \Omega_{ik}x^k[\lambda, \hat D_i\lambda] - \frac{i}{2} \Omega_{ik}x^k[[{X}_{z}{}^{m},\hat D_i X^{z}{}_{m}]] \ , \nonumber\\
0&= \hat D_i \hat D_i \lambda \ , \nonumber\\
0&= \hat D_i \hat D_i X^z{}_m\ .
\end{align}
Here \begin{align}
 [[X_z{}^m,\hat D_i X^z{}_m]]	=\sum_\alpha X_{z }{}^mT_\alpha(\hat D_i X^{z }{}_m) \nu^\alpha\ ,
 \end{align}
  where $T_a$ are the $SU(2)$ generators for  the representation that $X^\alpha{}_m$ belongs to, and as above $\nu^\alpha = \frac{1}{2}\sigma^\alpha$. 

We view the first equation as restricting the dynamics to the constraint surface defined by $\hat{ F}_{ij}^+=0$.
We can view the second equation as determining $G^+_{ij}$. However there is no need to explicitly solve for $G^+_{ij}$ as its contribution  to the action   will vanish on the constraint surface. 
Combining the second and third equations we simply find 
\begin{align}\label{A-eq}
\hat D_i F_{-i} =\hat D_i \partial_-\hat A_i	- \hat D_i \hat D_i A_-	=0\ .
\end{align}
Here we find a scalar Laplace equation for $  A_-$ but now with a source. We can therefore find a unique solution for $  A_-$ for a given choice of boundary condition. Let us decompose 
\begin{align}
A_- = a_- + A'_-	 \ ,
\end{align}
where 
\begin{align}\label{A'-}
	\hat D_i\hat D_i a_-=\hat D_i\partial_- \hat A_i\ ,\qquad \hat D_i\hat D_i  A'_-=0\ ,
\end{align}
In particular we choose $a_-$ such that $a_-=0$ when $\partial_-\hat A_i=0$. Furthermore since $\partial_-\hat A_i\to 0$ on the boundaries we expect $a_-\to 0$ there  whereas  $\hat A'_-$ can be non-vanishing. Note that under a gauge transformation we require
 \begin{align}
 a_- &\to ig\partial_-g^{-1} + g a_- g^{-1}\ , \nonumber\\
 A'_- &\to g A'_- g^{-1}	\ ,
 \end{align}
 so we can think of $A_-'$ as an adjoint valued scalar which satisfies the same equation of motion as the scalar $\lambda$, although it will have a different interpretation. Clearly if we start from a static ansatz with $a_-=A'_-=0$ can make $a_-$ non-zero by considering $x^-$-dependent gauge transformation while maintaining $A'_-=0$. In this sense we can think of $\hat D_i\hat D_i a_-=\hat D_i\partial_- \hat A_i$ as a gauge fixing condition.
   Thus we expect to find unique solutions for $a_-$ as well as $A'_-$, $\lambda$ and $X^z{}_m$ given their boundary values. 
   
We can now evaluate the action on the constraint surface to be 
 %
\begin{align}\label{S}
S  
&= \frac{k}{4\pi^2 R} \int dx^- d^4x  \ \frac12 {\rm tr}(  \partial_- \hat A_i -  \hat D_i a_-\ -\hat D_i  A'_-)^2     -\frac{1}{2} \tr (\hat D_{i} \lambda\hat D_{i} \lambda \big ) -   \hat D_i {X}_{z}{}^{m}\hat D_i X^{z}{}_{m}\nonumber\\
 &= \frac{k}{4\pi^2 R} \int dx^-d^4x   \frac12 {\rm tr}(  \partial_- \hat A_i -  \hat D_i a_-)^2 \nn\\
 &\qquad- \frac{k}{4\pi^2 R}\int dx^-d^4x\, 	\hat{\partial}_i\Big[  \tr \big( (\partial_- \hat A_i -  \hat D_i a_-)   A'_-  \big) -   \frac12 \tr ( A'_- \hat D_i A'_- )
	\Big]
\nonumber\\ &\qquad - \frac{k}{4\pi^2 R}\int dx^-d^4x\,  	\hat{\partial}_i\Big[  \frac12\tr ( \lambda  \hat D_i \lambda \big ) +   {X}_{z}{}^{m}\hat D_i   X^{z}{}_{m} \Big] \ .
\end{align}
The first term gives an action for the gauge field $\hat A_i$. All the remaining terms in (\ref{S}) are boundary contributions and as such depend on the choice of the asymptotic values of $A_-',\lambda$ and $X^z{}_m$ which are not fixed.\\

 Let us now discuss what this action looks like when we take the gauge field to be of the form (\ref{eq: 't Hooft form for A}), with $\tHooft$ of the generic form (\ref{eq: Phi general solution}).
 As seen above, for the scalars,  we take
\begin{align}\label{scalars}
 A'_- = A_{0-}\tHooft^{-1}\ ,\qquad \lambda = \lambda_0\tHooft^{-1}\ ,\qquad X^\alpha{}_m = X^\alpha_{0m}\tHooft^{-s_m} \ ,
 \end{align}
from some constants $A_{0-},\lambda_0\in \frak{su}(2)$ and $X^z_{0m}$ in the $SU(2)$ representation space of $X^{z}{}_m$ with spin $s_m$.  These choices correspond to a specific set of boundary conditions where the fields approach constant values as $|\vex|\to\infty$ whereas the  $x^-=\pm \infty$ behaviour is determined by $\tHooft$.

However we need to determine $a_-$. This was required to solve $\hat D_i\hat D_i a_- = \hat D_i\partial_-\hat A_i$ such that it vanishes when $\partial_-\hat A_i=0$. This seems too complicated to do in general. However it is important to look at the solution near the instanton worldlines to check that they do not affect the original $A_i$ gauge field topology, as discussed in Section \ref{sec: gauge field topology}. For simplicity we can consider a static worldline  at $ x^i=0$ and take the small $|\vex|$ expansion (\ref{smallx}). We find that, to lowest order in $x^i$, the solution is
\begin{align}
	a_- = \frac{1}{24}\Omega_{ik}\eta^\alpha_{kj}\sigma^\alpha x^ix^j\left(\mu(x^-)\partial_-^2 \mu(x^-) - (\partial_- \mu(x^-))^2\right)+\ldots \ , 
\end{align}
where the ellipsis denotes higher order powers of $x^i$. We assume that there are solutions which remain suitably bounded at $|\vex|\to\infty$.
In particular $a_-$ is finite and does not affect the  singular nature of the $A_i$ gauge field at $|\vex|\to 0$. 
For a moving instanton we expect a solution similar to that found in \cite{Lambert:2011gb} which behaves as 
$a_- \sim \eta^\alpha_{ij}x^i\partial_-y^j/|\vex|^2$ near worldlines and leads to a finite contribution to $A_i$. We also note that  $a_-=0$ at transition points where $\mu=\dot\mu=0$.

\subsubsection{Gauge field action}

  The first term in the action can then in principle be evaluated to give an expression involving $\tHooft$ and $\partial_-\tHooft$ given in terms of multiple integrals of $\mu_A(\tau)$   over the instanton  worldlines. We leave this as an exercise to the enthusiastic reader, but let us argue that for suitably well-behaved $\mu_A(\tau)$, the contribution from this term is finite. To see this, for simplicity let us take the case of a single worldline at the spatial origin. Then, provided that $\mu(\tau)$ and $\dot\mu(\tau)$ are bounded at $\tau\to \pm \infty$, we have
\begin{align}
  \text{tr}\left(\partial_- \hat{A}_i \partial_- \hat{A}_i\right) &= 6 |\vex|^2 \left[\tHooft[\mu]^{-2}\left(\tHooft[\dot{\mu}]-1\right)\frac{\partial \tHooft[\mu]}{\partial|\vex|^2} - \tHooft[\mu]^{-1}\frac{\partial \tHooft[\dot{\mu}]}{\partial|\vex|^2}\right]^2	\nn\\
  &\qquad  + \frac{3}{8R^2}|\vex|^2 \bigg[ \tHooft[\mu]^{-1} \left(\tHooft[\ddot{\mu}]-1\right) - \tHooft[\mu]^{-2} \left(\tHooft[\dot{\mu}]-1\right)^2 \bigg]^2		
    \label{eq: gauge part of action expanded}
\end{align}
Our aim is to determine sufficient conditions on $\mu$ such that the integral of $\text{tr}\left(\partial_- \hat{A}_i \partial_- \hat{A}_i\right)$ over $\mathbb{R}^5$ converges. Provided that $\dot{\mu}$ and $\ddot{\mu}$ exist, and that all of $\mu,\dot{\mu}$ and $\ddot{\mu}$ are bounded for all $\tau\in (-\infty,\infty)$, we have that $\tHooft[\mu,\dot{\mu},\ddot{\mu}]$ are all regular at all points except along the line at $\vex = \vec{0}$. Thus, we need to look at the behaviour of (\ref{eq: gauge part of action expanded}) in the neighbourhood of the worldline $\vex=\vec{0}$, and also as $(x^-,\vex)\to\infty$ where we could in principle encounter a large volume divergence.

Near to the worldline, the behaviour of $\tHooft$ is determined in (\ref{smallx}). It is then straightforward to show that the expression (\ref{eq: gauge part of action expanded}) is perfectly finite as we approach $|\vex|\to 0$. This can be traced back to the fact that the divergent part of $\hat{A}_i$ is independent of generic infinitesimal variations of $\mu$, as is seen in (\ref{eq: A hat i leading order}). However, note that this leading-order behaviour differs depending on whether or not $\mu$ vanishes.  It is important to note that while the leading order behaviour of $\hat{A}_i$ is discontinuous as the value of $x^-$ passes through an instanton insertion, at which $\mu$ transitions between a zero and non-zero value, the integrand (\ref{eq: gauge part of action expanded}) is well-behaved so long as $\mu$ itself is suitably well behaved---i.e. satisfying the conditions above---at the transition point.

Next we consider the large $(x^-, \vex)$ region. Let us write $\chi^2=(x^-)^2+\tfrac{1}{16R^2}|\vex|^4$. Then, as $\chi\to\infty$ we find 

\begin{align}
  \tHooft[\mu] = 1+ \chi^{-2} \int_{-\infty}^\infty d\tau\, \mu(\tau) + \mathcal{O}\left(\chi^{-3}\right)
\end{align}
provided that the integral exists; this is indeed ensured if $\mu(\tau)$ vanishes at least as quickly as $\tau^{-2}$ as $\tau\to\pm\infty$. Thus, to leading order we can take $\tHooft[\mu]$ to depend on coordinates only through $\chi$, with angular dependence coming at lower orders in $\chi$. Then, at leading order we can perform a trivial integral over angles to arrive at $\int dx^- d^4x \left(\dots\right) = V \int \chi^2 d\chi \left(\dots\right)$ for some constant $V$. Further noting that $\tHooft[\dot{\mu},\ddot{\mu}]=1+ \mathcal{O}\left(\chi^{-3}\right)$ provided $\dot{\mu}$ and $\ddot{\mu}$ vanish as $\tau\to\pm\infty$, we find that as $\chi\to\infty$,
\begin{align}
  \int dx^- d^4 x\, \text{tr}\left(\partial_- \hat{A}_i \partial_- \hat{A}_i\right) \propto \int \chi^{-3} d\chi 
\end{align}
and hence the integral over $\mathbb{R}^5$ converges.

If we finally assume that this convergence is not spoiled by the gauge fixing potential $a_-$, then we find that the gauge field part of the on-shell action (\ref{S}) can be made finite for a broad class of functions $\mu(\tau)$.

\subsubsection{Scalar field contributions}

Next we look at the remaining terms in the action (\ref{S}) which are all boundary terms arising from scalar fields which are given by (\ref{scalars}). For simplicity we set $A_{0-}=0$.  For a  generic scalar solution of the Laplacian, which we simply denote by $X$, in a spin $s$ representation of the  $SU(2)$ gauge group, we have
\begin{align}\label{expansion}
X^\dag \hat D_i X = -sX_{0}^\dag X_{0} \frac{\hat \partial_i\tHooft}{\tHooft^{2s+1}}	 + i\eta^\alpha_{ik}X_{0}^\dag T_\alpha(X_{0})\frac{\hat \partial_k\tHooft}{\tHooft^{2s+1}}\ .
\end{align}
First we consider the component of the boundary at $|\vex|\to\infty$. As $|\vex|\to \infty$, the generalisation of (\ref{eq: large x Upsilon}) to an arbitrary number of monotonic worldlines is 
\begin{align}
	\tHooft =  1 + \frac{2\pi R}{|\vex|^2}\sum_A(\mu_A(\infty)+\mu_A(-\infty))+\mathcal{O}\left(|\vec x|^{-3}\right)\ ,
	\label{eq: Upsilon at large vec x}
\end{align}
where each worldline is parameterised such that $y_A^-(\tau)=\tau$. Note, to arrive at this expression, we further require that none of the worldlines fly off to spatial infinity; that is, $|\vec{y}_A(\tau)|$ is bounded for all $\tau\in (-\infty,\infty)$.
 
Then, the first term in (\ref{expansion}) leads to a contribution 
\begin{align}\label{Vis}
 -sX_{0}^\dag X_{0}\int_{|\vex|\to\infty} dx^-d\Omega_3  	|\vex|^2 x^i \frac{\hat \partial_i\tHooft}{\tHooft^{2s+1}}& =  -sX_{0}^\dag X_{0}\int_{|\vex|\to\infty} dx^-d\Omega_3  	|\vex|^2 \frac{ x^i\partial_i\tHooft}{\tHooft^{2s+1}}\nonumber\\ 
 &=  {8\pi^3 R s} X_{0}^\dag X_{0} \int dx^- 	\sum_A ( \mu_A(\infty)+\mu_A(-\infty)) \ .
\end{align}
Thus to obtain a finite action we require $X_0=0$ or $\mu_A(\pm\infty)=0$. This latter condition can be thought of as the requirement that there are no instantons present at $x^-\to\pm\infty$ (although there can be solutions where there are instantons at any finite value of $x^-$, just with a size that shrinks to zero as in (\ref{ex2})).
 
 From the second term in  (\ref{expansion}) we find
 \begin{align} 	
 &i\eta^\alpha_{ik}X_{0}^\dag T_\alpha(X_{0})\int_{|\vex|\to\infty} dx^-d\Omega_3  	|\vex|^2 x^i \frac{\hat \partial_k\tHooft}{\tHooft^{2s+1}}	 =i\eta^\alpha_{ik}X_{0}^\dag T_\alpha(X_{0})\int_{|\vex|\to\infty} dx^-d\Omega_3  	|\vex|^2  \frac{x^i }{\tHooft^{2s+1}}\nonumber\\ 
 &\hspace{120mm}\times \left(\partial_k \tHooft - \frac12 \Omega_{kj}x^j\partial_-\tHooft\right) \nonumber\\
 &\hspace{65mm}=-\frac{i}{2}\eta^\alpha_{ik}\Omega_{kj}X_{0}^\dag T_\alpha(X_{0})\int_{|\vex|\to\infty} dx^-d\Omega_3  	|\vex|^2 x^i x^j \frac{\partial_-\tHooft }{\tHooft^{2s+1}} \ ,
  \end{align}
where the $\partial_k\tHooft$ contribution vanishes since $\partial_k\tHooft \sim x_k/|\vex|^4+\ldots$. 
Thus so long as $\partial_-\tHooft\to 0$ faster than $1/|\vex|^{4}$ this contribution will vanish.\\

 Next we consider 
the contributions from the $x^-\to\pm\infty$ boundary pieces. There are in principle two ways in which divergences may occur: from points at which worldlines escaping to infinity pierce the boundaries at $x^-\to\pm\infty$, and large volume diverges due to insufficient fall-off at large $|\vex|$. However, we find that in the vicinity of some worldline at $\vec{y}(\pm \infty)$, the integrand goes as $d^4 x|\vec{x}-\vec{y}|^{4s}$, and so is perfectly regular for all $s=0,1/2,\dots$. Thus, we need only look at the $|\vex|\to\infty$ region.

To begin with we  have contributions from the first term in (\ref{expansion}):
\begin{align}
\frac{s}2 X^\dag_0X_0\Omega_{ij}\int_{x^-\to \pm\infty} d^4 x x^j	\frac{\hat \partial_i\tHooft}{\tHooft^{2s+1}} &= \frac{s}2 X^\dag_0X_0\Omega_{ij}\int_{x^-\to \pm\infty} d^4 x x^j	\frac{\partial_i\tHooft}{\tHooft^{2s+1}}\nonumber\\
&\hskip 2cm - \frac{s}{4R^2} X^\dag_0X_0 \int_{x^-\to\pm\infty} d^4 x |\vex|^2 	\frac{\partial_-\tHooft}{\tHooft^{2s+1}}\ .
\label{eq: misc eq 1}
\end{align}
For spherically symmetric solutions the first term vanishes. To deal with the more general case, note that the first two terms in a large $|\vec x|$ expansion of $\Upsilon[\mu]$ are $x^-$ independent, and are found in (\ref{eq: Upsilon at large vec x}). In particular, we note that the term at order $|\vex|^{-2}$ is spherically symmetric, and thus does not contribute to the integral. The first potentially non-spherically symmetric term is at order $|\vec x|^{-3}$, providing a finite contribution to the integral. 
	 The second term in (\ref{eq: misc eq 1}) will be convergent if  $\partial_-\tHooft\to 0$ faster than $1/|\vex|^{6}$. 
	 
	 Lastly we  have the contributions from the second term in (\ref{expansion}):
\begin{align}
\frac{i}{2} X^\dag_0T_\alpha(X_0)\Omega_{ij}\int_{x^-\to \pm\infty} d^4 x x^j\eta_{ik}^\alpha	\frac{\hat \partial_k\tHooft}{\tHooft^{2s+1}} &= \frac{i}{2} X^\dag_0T_\alpha(X_0)\Omega_{ij}\eta^\alpha_{ik}\int_{x^-\to \pm\infty} d^4 x x^j	\frac{\partial_k\tHooft}{\tHooft^{2s+1}}\ .
\end{align}
Again given the form  (\ref{eq: Upsilon at large vec x}), the spherical symmetry of the order $|\vex|^{-2}$ piece ensures that at leading order we encounter integrals of the form
\begin{align}\label{symint}
	\int d^4x x^jx^k F(|\vex|^2) = \frac14\delta^{jk} \int d^4x |\vex|^2F(|\vex|^2)\ ,\end{align}
for a suitable choice of $F$,  whose contribution will therefore vanish as $\eta^\alpha_{ij}\Omega_{ij}=0$. Therefore, the leading non-vanishing contribution comes from the term at order $|\vex|^{-3}$, which is finite.\\

We have found that the scalar field contributions to the on-shell action arising form boundary integrals at $x^-\to\pm\infty$ are finite provided that the $|\vec{y}_A(\tau)|$ are bounded for all $\tau$, and that $\partial_- \Upsilon$ vanishes faster than $1/|\vex|^{6}$ as $|\vex|\to\infty$ (at any value of $x^-$). In order than the contributions from the boundaries at $|\vex|\to\infty$ were finite (and indeed vanishing), we further required that $\mu_A(\pm\infty)=0$.
In particular these conditions are satisfied by the solution (\ref{ex2}) as $\mu(\pm\infty)=0$  and $\partial_-\tHooft \sim 1/|\vex|^8$ as $|\vex|\to\infty$.

To understand the constraint on $\partial_- \Upsilon$ better, it is instructive to  consider a single instanton fixed at a fixed spatial point, for which we have $\partial_- \Upsilon[\mu] = \Upsilon[\dot{\mu}]-1$. Provided then that $\dot\mu(\pm\infty)=0$, which for well-behaved $\mu$ simply follows from $\mu(\pm \infty)=0$, we see from (\ref{muint}) that as $|\vex|\to\infty$, we have
\begin{align}
  \partial_- \Upsilon [\mu] 		&= \frac{16R^2}{|\vex|^4} \int_{-\infty}^\infty d\tau \dot{\mu}(\tau) + \mathcal{O} \left(|\vex|^{-6}\right)	\nn\\
  								&= \frac{16R^2}{|\vex|^4} \left(\mu(\infty)-\mu(-\infty)\right) + \mathcal{O} \left(|\vex|^{-6}\right)		\nn\\
  								&= \mathcal{O} \left(|\vex|^{-6}\right)
\end{align}
Thus, we find that generally $\partial_- \Upsilon[\mu]$ vanishes only \textit{as fast} as $1/|\vex|^6$ as $|\vex|\to\infty$. However, one can manipulate further to obtain
\begin{align}
  \partial_- \Upsilon[\mu] = -\frac{256 R^4}{|\vex|^8} \int_{-\infty}^\infty d\tau \, \dot{\mu}(\tau) \left(\tau - x^-\right)^2 + \mathcal{O}\left(|\vex|^{-10}\right)
\end{align}
provided that the integral exists. In particular, we require that $\dot{\mu}(\tau)$ approaches zero as $\tau\to\pm\infty$ at least as fast as $\tau^{-4}$.

Thus, at least in this simple case, we find that the boundary conditions on $\mu(\tau)$ at $\tau=\pm \infty$ in fact \textit{imply} the correct boundary conditions on $\partial_- \Upsilon$. It is not unreasonable to expect that this is true in full generality.

\subsubsection{In summary}

A number of results in this Section have relied on quite particular conditions on the solution (\ref{eq: Phi general solution}) and the functions $\mu_A$ therein, so let us briefly collect the salient points.

We considered evaluating the on-shell action (\ref{S}) with the constraint on $\hat{A}_i$ solved by the particular solution (\ref{eq: 't Hooft form for A}), with function $\Upsilon$ given most generally in (\ref{eq: Phi general solution}), and the scalar Laplace equation solved as in (\ref{scalars}). We argued that under some assumptions on the functions $\mu_A(\tau)$, which roughly speaking parameterise the size of the anti-instantons that $\hat{A}_i$ describes, the resulting action is finite. These assumptions fall into two categories:
\begin{itemize}
  \item The $\mu_A$ are well-behaved: in order to show that the gauge field contribution to the action was finite, it was sufficient that both $\dot{\mu}_A(\tau)$ and $\ddot{\mu}_A(\tau)$ exist and are bounded for all $\tau$. In practise, one is safe if they just take the $\mu_A$ to be smooth
  \item The $\mu_A$ vanish as $\mu\to\pm\infty$: in order to demonstrate finiteness of various terms in the action, we required that $\mu_A$ and $\dot{\mu}_A$ vanish sufficiently quickly as $\tau\to\pm\infty$. Each of these conditions is satisfied if $\mu_A$ goes as $\tau^{-3}$ as $\tau\to\pm \infty$. In practise, we are most interested in choices of $\mu_A$ with compact support, describing anti-instantons that are created and annihilated at finite points
\end{itemize}
It appears natural, therefore, to consider standard bump function profiles for the $\mu_A$, which are both smooth and compactly supported, to describe anti-instantons created and annihilated at finite points.


\subsection{A curious exact solution}


Remarkably, assuming the 't Hooft ansatz, we can find an exact form for $A_-$ that solves (\ref{A-eq}):
 \begin{align}\label{omegasol}
 A_- &= -\frac{1}{4}R^2\Omega_{ij}{\hat F}_{ij}- \frac{i}{2}R^2\Omega_{ij}[\hat A_i,\hat A_j] 
 \nonumber\\
 & = \frac{1}{4}R^2\Omega_{ik}\eta^\alpha_{kj}\sigma^\alpha\tHooft^{-1}\hat \partial_i\hat\partial_j \tHooft	 \ .
 \end{align}
 In addition we have the option to add zero-modes such as $u\tHooft^2+v\tHooft^{-1}$
 where $u,v$ are  constant $\frak{su}(2)$ matrices. However a non-zero $u$ leads to singular configurations whereas  solutions  with $v$ non-zero do not  change our discussion below.  This solution is notable as it means that we have explicitly solved all the dynamical field equations in terms of the function $\tHooft$.  It would be interesting to know if a similar solution exists more generally, beyond the 't Hooft ansatz.
 
 Furthermore we find that near a worldline, which we take to be at $x^i=0$,
 \begin{align}
 	A_i = \eta^\alpha_{ij}\sigma^\alpha\frac{x^j}{|\vex|^2} - R^2\Omega_{il}\eta^\alpha_{km}\sigma^\alpha\Omega_{kn}\frac{x^lx^mx^n}{|\vex|^4}+\ldots \ .
 \end{align} 
 The extra contribution to the singularity in the gauge field actually cancels the instanton number arising from the first term! More precisely, we find that as we approach $|\vex|\to 0$, the Chern-Simons 3-form goes as $\omega_3|_{S^3_0}=\bigO(|\vex|^{-2})d\Omega_3$, in contrast to the finite behaviour found previously. Noting further that this solution for $A_-$ dies away as $|\vex|\to\infty$ sufficiently fast to not affect the contribution to $Q$ from the integral at $S^3_\infty$, we find $Q=0$. Note that the second term on its own does not define a gauge field with instanton number, but adding it to the anti-instanton removes the instanton.  Thus we find exact solutions given by $\tHooft$ but all with vanishing instanton number for the original gauge field strength $F_{ij}$. These solutions presumably still can be interpreted as some sort of worldline as the energy density is peaked along a curve $(x^-(\tau),x^i(\tau))$.
 
 Note that in this case $A_-$ does not vanish if $\partial_-\hat A_i =0$. As such it doesn't represent a solution for $a_-$ that was introduced in (\ref{A'-}). Rather it must be identified with $a_-+ A'_-$ for the  $a_-$  as defined and some $A_-'$. On the other hand we argued above that we also expect there to exist classical solutions where $a_-$ does not affect the gauge field instanton number. For example if we consider the static case then we see that there are at least two acceptable solutions for $A_-$ (and again we could include the zero-modes). One is simply $A_-=0$ in which case $A_i=\hat A_i$ and we indeed find the $A_i$ has a non vanishing instanton number. However we can also take $A_-$ to be given by (\ref{omegasol}) in which case $A_i$ does not carry any instanton number, although $\hat A_i$ remains the same. For non-static solutions  (\ref{omegasol}) is a valid solution again leading to a vanishing instanton number for the gauge field $A_i$. However we have argued that in this case there also exists a solution for $a_-$ such that 
 (\ref{A-eq}) is solved and the instanton number of $A_i$ is non-vanishing.


\subsection{The DLCQ limit}\label{subsec: DLCQ}


Lastly let us consider the DLCQ limit $R\to \infty$, and so $\Omega_{ij}\to 0$. Here the  constraint simply states that $A_i$ has an anti-self-dual field strength on ${\mathbb R}^4$. As such it is determined in complete generality by the ADHM construction as an explicit function of $\vex$ as well as a finite set of moduli $m^I$, where here $I$ runs over some unspecified range. As far as the constraint is concerned these moduli can depend arbitrarily on $x^-$. The action and equations of motion take the same form but now $\hat A_i=A_i$ and $\hat D_i=D_i$. We can then write
 \begin{align}
 \partial_- A_i = \partial_I A_i \partial_- m^I	\ ,
 \end{align}
and we expand
\begin{align}
	a_- = \omega_I\partial_-m^I\ ,
\end{align}
so
that (\ref{A'-}) becomes 
\begin{align}
 D_iD_i \omega_I = D_i\partial_I A_i\ .
\end{align}
The interpretation is that  $a_-$ acts as a compensating gauge transformation which ensures that
\begin{align}
\delta A_i =( \partial_IA_i - D_i\omega_I)\delta m^I	 \ , \qquad \delta m^I = \partial_-m^I\delta x^-\ ,
\end{align}
is orthogonal to a gauge transformation in the sense that:
\begin{align}
D_i\delta A_i =0	\ .
\end{align}
 In this way $\delta A_i$ can be viewed as a tangent vector to the moduli space of anti-self-dual gauge fields (see \cite{Tong:2005un}).  We are not aware of any closed form expression for $a_-$ in the $\Omega_{ij}=0$ case\footnote{The solution (\ref{omegasol}) diverges in the $\Omega_{ij}\to0$ limit.}. 
 
 If we now evaluate the action (\ref{S}) we find (still assuming $A'_-=0$)
 \begin{align}
S = \frac{k}{4\pi^2 R}\int dx^- \Big[  \frac12 g_{IJ}\partial_- m^I \partial_- m^J - V\Big] \ ,	
\label{eq: reduced action worldlines}
\end{align}
where the moduli space metric is defined by 
\begin{align}\label{metric}
	  g_{IJ} = \int d^4x\ {\rm tr} \left((  \partial_I A_i - D_i\omega_I)(  \partial_J A_i - D_i\omega_J)\right)\ ,
\end{align} 
and now we find a potential for the moduli $V$ that comes from the scalar field boundary terms (\ref{Vis})
\begin{align}
V &= \sum_{scalars}\int d\Omega_3 |\vex| x^i X^\dag D_i X	\nonumber\\
& = 4\pi^2 \left(\sum_{\text{scalars}} s_X X^\dag_0X_0\right) \sum_A \rho^2_A\ ,
\end{align}
where in the second line we evaluated the integral using the standard 't Hooft ansatz obtained by taking linear combinations of the solution (\ref{S'tH}) but translated to have poles at points $\vey_A \in {\mathbb R}^4$. In this case only the boundary components at $|\vex|\to \infty$ arise. Furthermore the $\rho_A$ are the size moduli. These, along with  $\vey_A$ and the gauge embedding moduli, are allowed to be arbitrary functions of $x^-$. Thus we recover the  conventional description of dynamics on the moduli instanton space. Indeed, in the single $SU(2)$ anti-instanton case (with five adjoint scalars) the action (\ref{eq: reduced action worldlines}) is realised explicitly by the (bosonic sector of) the action (\ref{eq: QM action}), with the $w^\alpha$ recovered by turning on the zero mode $A'_-$.  

Thus for  $\Omega_{ij}= 0$  the dynamics takes place on the moduli space of anti-self-dual gauge fields, as proposed in \cite{Aharony:1997th,Aharony:1997pm,Aharony:1997an}. This space is a disconnected sum with each component labelled by  instanton number and parameterised by a set of positions, sizes and gauge embedding moduli, which we have denoted by $m^I$, all of which are dynamical. Finite action configurations consist of fluctuations of all the moduli in each connected component subjected to a potential for their size when the scalars have a vacuum expectation value.\\

We know that in the $R\to\infty$ limit, the constraint equation reduces to the usual spatial anti-instanton equation $F^+_{ij}=0$. It is worth therefore asking which solutions to these equations are found in the $R\to\infty$ limit of our 't Hooft solutions at finite $R$.

So consider the $N$ worldline solution (\ref{eq: Phi general solution}). As discussed in Section \ref{sec: gauge field topology}, we can without loss of generality take each of these worldlines to be monotonic, and disjoint except for possibly at their endpoints. Then let us consider the $R\to\infty$ limit. Much of our work is already done, since the $R\to\infty$ limit is very similar to the limit in which we approach the worldline, $|\vex-\vec{y}|\to 0$, where the integrals over the $\tau_A$ localise. In particular, one can take the limit explicitly by making use of the Fourier techniques in Section \ref{sec: gauge field topology}. Then, noting that in the limit $\hat{A}_i = A_i$ and $\hat{\partial}_i=\partial_i$, and further normalising as $\mu_A(\tau)=(\rho_A(\tau))^2/4\pi R$ for some functions $\rho_A(\tau)$, we find that as $R\to\infty$,
\begin{align}
  A_i = \hat{A}_i = -\frac{1}{2}\eta^\alpha_{ij} \sigma^\alpha \partial_j \log \tHooft,\qquad \tHooft(x) = 1+\sum_{A=1}^N \frac{(\rho_A(x^-))^2}{|\vex - \vey_A(x^-)|^2} \ .
  \label{eq: DLCQ limit of t Hooft solution}
\end{align}
Thus, we precisely recover the usual 't Hooft ansatz, in which the size modulus $\rho(x^-)$ and position moduli $y^i(x^-)$ are allowed to vary arbitrarily with time. The solution therefore describes a set of $N$ anti-instantons moving arbitrarily. Interestingly, we can still take any of the $\rho_A$ to have compact support, resulting in anti-instantons that are created and annihilated. Indeed, all of the analysis of Section \ref{sec: gauge field topology} persists in the $R\to\infty$ limit, and is indeed much more immediate due to local form of (\ref{eq: DLCQ limit of t Hooft solution}).

It is interesting that at finite $R$, the value of the gauge field at some point $x=(x^-, \vex)$ away from worldlines is determined by the moduli $\mu_A(\tau), \vec{y}_A(\tau)$ at \textit{every} point along every worldline, while in the $R\to \infty$ limit the solution localises, in the sense that the gauge field now depends only on the $\rho_A(\tau),\vec{y}_A(\tau)$ at $\tau=x^-$.

Finally, let us consider the fate of our solution (\ref{omegasol}) for $A_-$ in the $R\to\infty$ ($\Omega_{ij}\to 0$) limit. In fact, the solution (\ref{omegasol}) diverges, however $R^{-1}A_-$ is finite and becomes harmonic: $\partial_i\partial_i (R^{-1}A_-)\to 0$. Therefore  the combination
\begin{align}
A_i &= \hat A_i + \frac12 \Omega_{ij}x^j A_- \nonumber\\
&=  	 \hat A_i + \frac12 R\Omega_{ij}x^j \frac{1}{R}A_-\ , 
\end{align}
will remain finite and in the limit becomes 
\begin{align}
A_i	= -\frac{1}{2\tHooft}\eta^\alpha_{ij}\sigma^\alpha  \partial_j \tHooft+ \frac{1}{8\tHooft}\Omega'_{ij}x^j \Omega'_{mk}\eta^\alpha_{kn}\sigma^\alpha  \partial_m \partial_n \tHooft	\ ,
\end{align}
where $\Omega' = R\Omega_{ij}$ is a constant non-degenerate anti-self-dual two-form on ${\mathbb R}^4$ and $\tHooft$ is harmonic: $\partial_i\partial_i\tHooft=0$.  The first term is the usual 't Hooft ansatz solution and carries an instanton number of minus one from each of the poles in $\tHooft$.	But as we have seen the two terms together give a  solution which does not carry any instanton number. This suggests that even in the ordinary instanton case there is an $\Omega'$-deformation which removes the instanton singularities   by shifting $A_i$
 \begin{align}
 A_i\to A_i +\frac12 \Omega'_{ij}x^jA_-''	\ ,
 \end{align}
 where 
 \begin{align}
A_-'' = \frac{1}{4\tHooft} \Omega'_{mk}\eta^\alpha_{kn}\sigma^\alpha  \partial_m \partial_n \tHooft\ ,	
\end{align}
 is a harmonic function which is not spherically symmetric, even if $\tHooft$ is.
 The only difference is that there is no preferred choice for $ \Omega'_{ij}$ but rather a three-parameter family of choices.


\chapter*{Discussion and further directions}
\addcontentsline{toc}{chapter}{Discussion and further directions}
\newcounter{StarEquations}
\stepcounter{StarEquations}
\lhead{\textsl{\nouppercase{Part II, Discussion and further directions}}}


Let us now take stock. Everything we have done in Part II of the thesis has been motivated by a singular aim: to build a five-dimensional Lagrangian field theory capable of describing the $(2,0)$ superconformal field theory on six-dimensional Minkowski space. However, at many points along the way we have sought to generalise our analysis and broaden our horizons, resulting in a programme of work that takes many new steps towards a broader understanding of $SU(1,3)$ field theories, with or without supersymmetry.\\

We first built up the theory of five-dimensional theories with a curious $SU(1,3)$ spacetime symmetry from scratch, following the basic steps of conformal field theory. From here, we were able to derive and solve the Ward-Takahashi identities, thus determining the general form of correlators in such theories.

Next came our embedding of such theories within the more familiar setting of six-dimensional conformal field theory. We explicitly constructed a conformal compactification of six-dimensional Minkowski space, and showed that the dynamics of its modes are described by an $SU(1,3)$ theory. More precisely, we showed that for such a six-dimensional interpretation to hold, the symmetry algebra of this $SU(1,3)$ theory must extend to a central extension $\frak{h}$, with non-zero Fourier modes charged under the single central element $P_+$.\\

If that was the prelude, next came the whole point, at least as far as M5-branes are concerned: we exhibited an explicit Lagrangian gauge theory in five dimensions, and showed it had all the properties one would expect of a five-dimensional $SU(1,3)$ description of the non-Abelian $(2,0)$ theory. We first saw that, naively, the theory has precisely an $\frak{su}(1,3)$ spacetime symmetry. From here, one may suppose that the model describes only the zero modes of the reduction from six dimensions.

However, salvation is found---and higher Kaluza-Klein modes uncovered---by extending the theory's configuration space to allow for isolated singularities at points $x_a$. Such singular points are the positions of instanton insertions $(x_a,n_a)$, which each carry an instanton charge $n_a\in\mathbb{Z}$. Any local operator inserted at $x_a$ is now found to be charged under the central element $P_+$. This charge is $kn_a/R$, and thus for $k\in\mathbb{Z}$ we find perfect agreement with our previous construction: this five-dimensional theory has precisely the properties required to describe the Fourier modes of a six-dimensional CFT on a $\mathbb{Z}_k$ orbifold of six-dimensional Minkowski space. In particular, for $k=1$, we have simply non-compact Minkowski space.

Additionally, the theory possesses a high degree of super(conformal) symmetry; indeed, precisely the right amount one would expect from such a reduction of the $(2,0)$ theory. Thus, we are lead to conjecture that this theory provides a Lagrangian for the non-Abelian $(2,0)$ theory on non-compact Minkowski space.\\

We finally employed an $\Omega$-deformation of the standard ansatz (\ref{eq: tHooft ansatz}) to explore a broad set of solutions to the constraint $\mathcal{F}^+=0$. These solutions describe a set of anti-instanton worldlines, generically created and annihilated at some points $x_a$, as shown for instance in Figure \ref{fig: general graph}. Then, crucially, as a corollary to a more general set of rules for calculating the instanton charge $Q(S)$ of any 4-surface $S$, we showed that a point $x_a$ at which $m$ anti-instantons are annihilated and a further $n$ are created is precisely identified as an instanton insertion $(x_a,m-n)$. 

Thus, we exhibited explicit gauge field configurations on the constraint surface $\mathcal{F}^+$ featuring points with non-zero instanton charge, and hence showed that the theory on the constraint surface is indeed a theory of a full tower of Kaluza-Klein modes of a six-dimensional conformal field theory.  \\


So let us now discuss a number of interesting aspects of this work, and in doing so detail potential further directions.

\subsection*{Parameters and regimes}

Both our purely geometric conformal compactification of six-dimensional Minkowski space, and our five-dimensional Lagrangian theory feature two parameters: $R$ and $k$. Indeed, the central result of Part II of the thesis is the non-trivial identification of the $R$ and $k$ of the former with that of the latter. Let us briefly review how this works.

First, consider the geometry. We first defined coordinates $(x^+, x^-, x^i)$ on six-dimensional Minkowski space by (\ref{eq: coordinate transformation}), where in particular $x^+$ has finite range, $x^+\in(-\pi R, \pi R)$. Thus, we see that from a six-dimensional perspective, $R$ has dimensions of length. After considering a theory reduced into modes $\fOp_n$ along this $x^+$ interval, in Chapter \ref{chap: DLCQ} we went a step further and considered restricting only to modes $\fOp_{kn}$, for some $k\in\mathbb{Z}$. From the perspective of the geometry, this corresponds simply to performing a $\mathbb{Z}_k$ orbifold of the $x^+$ interval, on which operators must have period $2\pi R/k = 2\pi R_+$.

Now let's look at the Lagrangian theory (\ref{eq: Lagrangian}), for a moment forgetting about any relation to six dimensions. Firstly, we see that $R$ is omnipresent; as well as appearing explicitly in front of the action, it appears also in each instance of $\Omega_{ij}\sim R^{-1}$. Additionally, if we define a notion of length, with $x^-, x^i$ both having dimensions of length, then once again, $R$ has dimensions of length. In contrast, $k$ appears simply as an overall normalisation of the action, and is a priori just some real number. However, we found that the finite variation of the action under $K_+, M_{i+}$ is single-valued only for integer $k$, suggesting that we should in fact quantise as $k\in\mathbb{Z}$, so let us do so.

Let us then recall how these two initially disparate setups converge. By considering a broader configuration space for the Lagrangian theory, we found the $\frak{su}(1,3)$ symmetry enhanced to a central extension $\frak{h}$, identified as the subalgebra of the six-dimensional conformal algebra $\frak{so}(2,6)$ preserved under the above compactification. In particular, we found operators $\fOp_{kn}$ with momentum $kn/R$ in the $x^+$ direction, and thus, the $R$ and $k$ of the Lagrangian theory are identified as those of the geometrical setup.\\

What can we say about different regimes of the parameters $R$ and $k$? Guided by conventional Yang-Mills theory, let us define $g^2=4\pi^2 R/k$, and now treat $g$ and $R$ as the two parameters of the theory. Then, after the standard field redefinitions $A\to g A$, $X^I\to g X^I$, $\Psi\to g \Psi$, we see that all interaction terms come with some positive power of $g$. In this sense we can think of $g$ as the coupling of the theory; in particular, since $g$ is dimensionful, we see that the theory becomes strongly coupled at distances shorter than $\sim R/k$. Note then that if we fix $R$ and increase $k$, we push the strongly coupled regime to shorter and shorter distances, $R/k<< R$, and as such can think of this as a sort of weak coupling limit. Conversely, again keeping $R$ fixed, we identify $k=1$ as the `strongly coupled' limit, as $k\in\mathbb{Z}$. 

Let us contrast this with analogous statements about the standard, Lorentzian $\mathcal{N}=2$ (maximal) super-Yang-Mills theory in five dimensions. There we have a Yang-Mills coupling $g^2=4\pi^2 R_5$, where $R_5$ is identified as the radius of the M-theory circle on which the M5-brane is wrapped. Again, instanton particles are conjectured to provide the Kaluza-Klein modes of this compactification. But to describe a non-compact stack of M5-branes, one must go to infinite strong coupling, $R_5\to\infty$, where the Lagrangian theory itself breaks down. For our theory here, precisely \textit{because} we have performed merely a \textit{conformal} compactification, we realise the non-compact geometry at $k=1$ and so $g^2=4\pi^2R/k=4\pi^2R$ for \textit{any} $R$, with orbifolds thereof realised at higher $k$.\\

So we know how the regard $k$ while we keep $R$ fixed. Then, how do we treat $R$? If the six-dimensional interpretation of the theory holds, then $R$ can be thought of as an artefact: although it will appear in any object computed in the five-dimensional theory, it will drop out upon Fourier resummation to six dimensions, as indeed we saw explicitly for correlation functions in Section \ref{sec: further constraints on correlators}. In fact, $R$ can be completely absorbed into coordinates and fields. This can be seen first at the level of the six-dimensional geometry, which in terms of coordinates $(\tilde{x}^+,\tilde{x}^-,\tilde{x}^i)=(R^{-1}x^+, Rx^-, x^i)$ has metric
\begin{align*}\tag{$*.\theStarEquations$}
  ds^2 = - 2\, d\tilde{x}^+ \left(d\tilde{x}^- - \frac{1}{2}R\,\Omega_{ij} \tilde{x}^i d\tilde{x}^j\right) + d\tilde{x}^i d\tilde{x}^i\ ,
\end{align*}\stepcounter{StarEquations}%
which is independent of $R$. Note, $\tilde{x}^+$ is dimensionless from the perspective of the original coordinates, and has range $\tilde{x}^+\in(-\pi, \pi)$, while $\tilde{x}^-$ has length dimension 2.

We can then continue to eliminate $R$ from the action, by writing it in terms of $\tilde{x}^-, \tilde{x}^i$ and the rescaled fields $(\tilde{A}_-, \tilde{A}_i)=(R^{-1}A_-, A_i)$, $\tilde{X}^I = R^{-1} X^I$, $\tilde{G}_{ij} = R^{-2}G_{ij}$, $\tilde{\Psi}_+ = R^{-3/2}\Psi_+$ and $\tilde{\Psi}_- = R^{-1/2}\Psi_-$. Practically speaking, the effect on all formulae is to simply put a tilde on everything, and set $R=1$.

The fact that such a rescaling exists is no surprise, and in fact follows from what we already knew. There are two notions of dimensionality we can assign to parameters, coordinates and fields: that of length dimension as ordained by the coordinates $(x^+, x^-, x^i)$, and that of Lifshitz scaling dimension, as determined by the transformations (\ref{eq: Lifshitz scalings of fields}). The action $S$ is invariant under rescaling either of these dimensions, but only the Lifshitz scaling is a true \textit{symmetry}, since only coordinates and fields (but not parameters) scale. Then, these two notions of dimensionality coincide for the coordinates $\tilde{x}^-, \tilde{x}^i$ and fields $\tilde{X}^I, \tilde{A}_-, \dots$. Thus, writing $S$ in terms of these coordinates and fields, it follows that $S$ \textit{cannot} depend on $R$, which has dimensions of length, but zero Lifshitz dimension.

So now we sit in the rescaled theory, with only a single parameter $k$. The length dimension of coordinates and fields is dictated by their Lifshitz dimension, and so in particular $x^-$ has length dimension 2, while the $x^i$ have length dimension 1. The sole parameter $k$ is then dimensionless, and indeed we take it to be integer.

We can once again identify a coupling, which is now the dimensionless $g^2=4\pi^2/k$. By the same field redefinition as above, we indeed see that all interactions come with positive powers of $g$. The theory is strongly coupled at $k=1$, where it should describe M5-branes on non-compact Minkowski space, and weakly coupled as $k>>1$, where the six dimensional geometry is orbifolded by $\mathbb{Z}_k$\\

What we learn is that, as far as the qualitative analysis of the five-dimensional theory (\ref{eq: Lagrangian}) is concerned, the parameter $R$ plays no role.  We can happily use the rescaled theory, with its single, discrete parameter $k$, effectively choosing units such that $R=1$. In contrast, $k$ has profound impact on the six-dimensional interpretation of the theory. Indeed, it is only at $k=1$ that we expect addition, special enhancements to occur to the theory, as we discuss next.

\subsection*{Back to (M theory) basics}


It is worth at this point to briefly remember our roots, and ground a number of objects defined in this Part within the broader M-theory picture. In analogy with the open strings ending on D-branes in string theory, it is natural to think of the dynamics of M5-branes as arising from that of M2-branes ending on them \cite{Strominger:1995ac,Witten:1995zh}. From the perspective of the M5-brane worldvolume theory, the ends of such M2-branes are realised as self-dual string-like solitons \cite{Howe:1997ue}.

Upon compactification to a five-dimensional gauge theory on D4-branes, these string-like solitons give rise to two kinds of objects, related by electric-magnetic duality. If the string extends along the compact direction, we end up with a particle-like state in five-dimensions, while otherwise we find a string-like state, corresponding to a D2-D4-brane intersection.\\

Now, if the string in six dimensions carries no momentum on the M-theory circle, the resulting particle-like states lie within the perturbative spectrum of the five-dimensional gauge theory, with the corresponding string-like states their electro-magnetic duals. As we have seen at many points throughout this thesis, however, non-perturbative particle-like configurations carrying instanton charge are naturally thought of as higher Kaluza-Klein modes on the M-theory circle. More precisely, we identify an instanton-particle of instanton charge $n$ in the five-dimensional theory as a self-dual string in six dimensions, both wrapping and carrying\footnote{In the conventions used in this thesis} $\left(-n\right)$ units of momentum around the M-theory circle.

In the theory (\ref{eq: Lagrangian}), with respect to which $x^+$ provides the M-theory direction, we find that the operator $\I_{-n}(x^-, x^i)$ creates such a state of Kaluza-Klein momentum $n$, while $\I_n(y^-, y^i)$ annihilates it at some $y^-> x^-$. As previously discussed, the definite sign of the momentum of allowed created states is natural in any null reduction.

With this picture in mind, one can consider extending the instanton operator formalism explored in Chapter \ref{chap: An explicit model and its symmetries} to include creation and annihilation operators for the non-perturbative string-like states corresponding to unwrapped self-dual strings with non-zero momentum on $x^+$ \cite{Lambert:2010iw}. This would amount to defining a second set of disorder operators $\hat\I_n (\Gamma)$ defined not at a single point but rather a line $\Gamma\subset \mathbb{R}^{1,4}$. Then, in analogy with the $\I_n$, one would take $\hat\I_{-n}$ to create a string-like state of $x^+$ momentum $n$, while $\hat\I_n$ then annihilates it.\\

There is a broad literature\footnote{See for instance \cite{Berman:2007bv} for a useful review} concerning the computation of the dynamics of M5-branes, many applying only in a particular background, or along a particular branch of moduli space corresponding to separated branes. It is important therefore to appreciate that if the model (\ref{eq: Lagrangian}), suitably quantised with care taken to include non-trivial topological sectors as explored in Chapter \ref{chap: An explicit model and its symmetries}, is to truly provide a Lagrangian description for M5-branes, then each of these previous results must in turn be recovered.

One such area of study is that of the dynamics and scattering behaviour of the aforementioned self-dual strings within the M5-brane worldvolume. In particular one can propose a worldvolume action for such self-dual strings, at least away from the origin of moduli space (the conformal point), so that such a string has finite tension \cite{Gustavsson:2001uw,Arvidsson:2002tr,Arvidsson:2003ya,Arvidsson:2003pc,Arvidsson:2004xa}. Such a perspective then proves useful, for instance in the computation of the scattering amplitude of waves off the self-dual string.

But perhaps most relevant to our discussion here is the proposed worldvolume Lagrangian itself. In particular, it would be interesting to assess whether such an explicit model can be derived from the Lagrangian (\ref{eq: Lagrangian}). Once again, the way in which this might be expected to occur differs depending on whether the self-dual string in question does or does not extend along the $x^+$ direction. To arrive at the $(1+1)$-dimensional $\sigma$-model of the unwrapped string, one should evaluate the action on string-like solitonic configurations and then seek a decoupling limit which one should hope to match with the proposed $\sigma$-model. Conversely, the action evaluated on the particle-like instanton-solitons, as given in (\ref{S}), should describe a self-dual string extended and carrying fixed momentum along the $x^+$ direction, and thus should at least in some decoupling limit produce a quantum mechanics identified as this very same $\sigma$-model evaluated on a sector of fixed $x^+$ momentum. One particularly attractive prospect of such a study is the identification of the M5-brane worldvolume two-form, whose coupling to the self-dual string is prescribed in \cite{Arvidsson:2003pc}.


\subsection*{Symmetry enhancement at strong coupling}


Let us revisit the question that started all of this: what symmetries do we expect the theory (\ref{eq: Lagrangian}) to have? We know the symmetries of the M5-brane, given by the $(2,0)$ superconformal algebra\footnote{Here and throughout we are a little cavalier which real form of the underlying complexified superalgebra we mean, here for instance really meaning $\frak{osp}(2,6|4)$} $\frak{osp}(8|4)$. This consists of a spacetime symmetry $\frak{so}(2,6)$ and R-symmetry $\frak{so}(5)$, along with 16 real rigid supercharges and 16 real conformal supercharges.

At generic $k$, our theory should describe M5-branes on a $\mathbb{Z}_k$ orbifold of Minkowski space. Of the bosonic symmetries, this geometry preserves all of the R-symmetry $\frak{so}(5)$, but only a subalgebra $\frak{h}=\frak{u}(1)\oplus \frak{su}(1,3)\subset \frak{so}(2,6)$ of the spacetime symmetries. As we have shown in this thesis, these are indeed the bosonic symmetries of the theory. Further, the geometry breaks precisely one quarter of the supersymmetries, leaving only 24 supercharges. Again, these are indeed the supersymmetries realised by the theory \cite{Lambert:2019jwi}, which combined with the spacetime and R-symmetry can be shown to form $\frak{osp}(6|4)\oplus \frak{u}(1)$ \cite{Lipstein:Yangian}.\\

However, at strong coupling ($k=1$), there is no orbifold, and so we should in fact find the full $\frak{osp}(8|4)$. Specifically, we should see that the spacetime symmetry enhanced from $\frak{h}$ to the full conformal algebra $\frak{so}(2,6)$, and 8 additional supercharges should also emerge, topping us up to 32 in total. To realise these enhanced symmetries at strong coupling is a crucial next step in the analysis of this theory.

Let us speculate on how this might work. Our approach to symmetries has been very much guided by standard conformal field theory approaches; namely, deriving (and solving) Ward-Takahashi identities for correlators. We can proceed in a similar vein \cite{Lambert:SymmEnhance}. Consider for instance the two-point function $\left\langle \I_n(x_1)\fOp(x_1) \I_{-n}(x_2) \fOp(x_2)\right\rangle$ for some $\frak{h}$ scalar primary $\fOp$ of scaling dimension $\Delta$ in the five-dimensional theory. Then, the 2-point function is constrained by the $\frak{h}$ Ward-Takahashi identities to be precisely of the form
\begin{align}\tag{$*.\theStarEquations$}
  \left\langle \I_n(x_1)\fOp(x_1) \I_{-n}(x_2) \fOp(x_2)\right\rangle = C(\Delta,n) \frac{1}{\left(z\bar{z}\right)^{\Delta/2}}\left(\frac{z}{\bar{z}}\right)^n\ ,
  \label{eq: solution to h WTIs}
\end{align}\stepcounter{StarEquations}%
where $z=z(x_1,x_2)$, and the $C(\Delta,n)$ are some constants. Now, at the level of solving the $\frak{h}$ Ward-Takahashi identities, the constants $C(\Delta,n)$ and, say, $C(\Delta,n+1)$ know nothing about each other. This is because these identities (\ref{WIpplus})--(\ref{WIKplus}) are partial differential equations just for each of the $ \left\langle \I_n(x_1)\fOp(x_1) \I_{-n}(x_2) \fOp(x_2)\right\rangle$ independently, rather than one that mixes them. This is in turn precisely because $[\frak{h},P_+]=0$. In the language of Chapter \ref{chap: An explicit model and its symmetries}, we might say that these Ward-Takahashi identities live in a single, fixed topological sector.

Further, we saw how to translate these identities into Ward-Takahashi identities for the six-dimensional operators (\ref{eq: six d operators definition}), written in terms of the six-dimensional vector field representation of $\frak{h}$ as in (\ref{eq: 6d CKVs}). If the theory's symmetries are indeed enhanced at strong coupling, then these Ward-Takahashi identities must only be a subset of a whole $\frak{so}(2,6)$-worth (and indeed $\frak{osp}(8|4)$-worth) of identities that are obeyed. So a natural next step towards showing this is simply to consider the form of this broader class of six-dimensional Ward-Takahashi identities, and understand how they descend to identities for the correlators of Fourier modes.\\

For example, let $\fOp_n$ denote the modes of some six-dimensional scalar primary $\sOp$, as per (\ref{eq: 6d op Fourier decomp}). Then, we find that the six-dimensional scalar 2-point function is invariant under the six-dimensional dilatation $D^{(\sixd)}$, which has $[D^{(\sixd)},P_+]\neq 0$ and thus does not lie in $\frak{h}$, only if the 2-point functions of Fourier modes obey
\begin{align}
  &\Big(-\tfrac{1}{2}\left(n+1\right)+\tfrac{1}{4}\Delta + \tfrac{i}{8}|\vex_1|^2\partial^1_-+\tfrac{1}{4}x_1^i \partial^1_i-\tfrac{i}{4}\Omega_{ij} x_1^j \partial^1_i\Big)\left\langle \fOp_{n+1}(x_1) \fOp_{-n-1}(x_2)\right\rangle		\nn\\
  &\qquad\qquad+ \Big(-\tfrac{1}{2}n+\tfrac{1}{4}\Delta + \tfrac{i}{8}|\vex_2|^2\partial^2_-+\tfrac{1}{4}x_2^i \partial^2_i-\tfrac{i}{4}\Omega_{ij} x_2^j \partial^2_i\Big)\left\langle \fOp_{n}(x_1) \fOp_{-n}(x_2)\right\rangle = 0\ ,
  \tag{$*.\theStarEquations$}\label{eq: TWTI}
\end{align}\stepcounter{StarEquations}%
for every $n\in\mathbb{Z}$. Here, $\partial^{1,2}_- = \partial/\partial x_{1,2}^-$ and $\partial^{1,2}_i = \partial/\partial x_{1,2}^i$. Let us stare at this for a moment. It certainly takes the usual form of an infinitesimal Ward-Takahashi identity, but crucially it is a partial differential equation that mixes correlators computed in \textit{different} topological sectors. Indeed, plugging the solution (\ref{eq: solution to h WTIs}) into (\ref{eq: TWTI}), we find that the identification $\fOp_n=\I_n\fOp$ requires a non-trivial relation between $C(\Delta,n)$ and $C(\Delta,n+1)$,
\begin{align}\tag{$*.\theStarEquations$}
   \left(n+\frac{\Delta}{2}\right)C(\Delta,n) - \left(n-\frac{\Delta}{2}+1\right)C(\Delta,n+1)=0\ ,
   \label{eq: recursion relation}
\end{align}\stepcounter{StarEquations}%
to be satisfied. As a consistency check, note that this identity is indeed satisfied by the coefficients $C(\Delta, n)=d(\Delta,n)$, found in Section \ref{subsec: 2pt dim red} from the dimensional reduction of the 2-point function of a protected six-dimensional scalar operator with $\Delta\in 2\mathbb{N}$.\\

This property---a non-trivial mixing of topological sectors---is characteristic of any constraint on our five-dimensional correlators that descends from a Ward-Takahashi identity in six dimensions for some $G\in\frak{so}(2,6)$ with $[G,P_+]\neq 0$. It is straightforward to derive these constraints at $N$-points. In general, we refer to these new constraints as \textit{topological} Ward-Takahashi identities, since from the perspective of our five-dimensional theory they do indeed mix topological sectors.

Thus, we have refined our aim: in the topological Ward-Takahashi identites, we find necessary and sufficient conditions for the full six-dimensional conformal algebra $\frak{so}(2,6)$ to be realised as a symmetry at the level of correlation functions. The next step is to show, just as we did for the more conventional $\frak{h}$ Ward-Takahashi identities, that these identities are indeed satisfied by generic correlators of the theory.\\

One may hope to refine these necessary and sufficient conditions yet further, by leveraging the worldline interpretation of the non-trivial sectors of the theory as explored in Chapter \ref{chap: worldlines}. Given some correlator $\left\langle \I_{n_1}\fOp^{(1)} \dots \I_{n_N}\fOp^{(N)} \right\rangle$, it is straightforward to draw all possible worldline diagrams as in Figure \ref{fig: general graph} that are compatible with the ordering of the $x_1^-,\dots, x_N^-$ and integer charges $n_1,\dots, n_N$. One can then imagine an associated calculus, through which they can read off the value of at least some part of the correlator from these worldline diagrams. Crucially, such a procedure would define non-trivial relations between correlators as the charges $n_1,\dots, n_N$ are varied. The aim then in designing such an algorithm would be for it to always produce solution to the topological Ward-Takahashi identities of the form (\ref{eq: TWTI}). Following from the foundations laid out in Section \ref{sec: relation to correlation functions}, it appears that such a calculus may exist, and if it will provide a refined and perhaps more fundamental target in showing completely the $\frak{so}(2,6)$ symmetry enhancement of the theory \cite{Lambert:SymmEnhance}.\\

One may further hope to probe the theory's ability to compute six-dimensional observables through the analysis of conformal anomalies. In particular, as noted in Chapter \ref{chap: SU(1,3) theories}, one can examine a theory's conformal anomalies through analysis of short-distance singularities appearing in correlation functions \cite{Petkou:1999fv,Bzowski:2015pba}. Then, just as in Chapter \ref{chap: the view from 6d CFT} we dimensionally reduced the bare correlators at separated points to arrive at required forms for the corresponding bare correlators in the five-dimensional theory, one may aim to leverage the extensive literature on conformal anomalies in six dimensions with $(2,0)$ superconformal symmetry (see for instance \cite{Bonora:1985cq,Deser:1993yx,Henningson:1998gx,Bastianelli:2000hi}) to extend this analysis, and provide criteria on short-distance singularities in the five-dimensional theory such that the correct six-dimensional anomalies are reproduced. In doing so, we would acquire yet further criteria that the five-dimensional model must satisfy in order for its six-dimensional interpretation to hold. Such criteria would in particular encode the famous $N^3$ behaviour of the six-dimensional conformal anomaly \cite{Henningson:1998gx} within the short-distance behaviour of five-dimensional correlators; indeed, it is a key challenge in the full realisation of six-dimensional physics to obtain this characteristic growth from the model (\ref{eq: Lagrangian}). \\  

Let us make one final comment on this topic. What we have shown is that given any local operator $\fOp$ in the five-dimensional theory, the operator $\I_n\fOp$ can be interpreted as the $n^\text{th}$ Kaluza-Klein mode of some six-dimensional operator. This is seen by noting that such operators satisfy precisely the correct Ward-Takahashi identities corresponding to the algebra $\frak{h}$. However, there is a great deal of ambiguity in precisely how, given some six-dimensional operator whose correlators we wish to compute, we should identify the modes $\fOp_n$ as a function of these five-dimensional operators of the form $\I_n \fOp$.

Recall, we generically decompose a six-dimensional operators $\sOp$ as
\begin{align}\tag{$*.\theStarEquations$}
  \sOp(x^+,x^-,x^i) = \sum_{n\in\mathbb{Z}} e^{-inx^+/R}\fOp_n(x^-, x^i)\ .
  \label{eq: conclusion fourier resum}
\end{align}\stepcounter{StarEquations}%
 We have often proposed that we should then choose $\fOp_n=\I_n\fOp$ for some local operator $\fOp$ in the five-dimensional theory. The simplest generalisation, however, is that we identify $\fOp_n = f(n) \I_n \fOp$ for some function $f(n)$, effectively introducing a non-trivial normalisation for the instanton operators $\I_n$. This is perfectly valid since, even after introducing $f(n)$, the correlators of the $\fOp_n$ continue to satisfy the $\frak{h}$ Ward-Takahashi identities. Indeed, it is in many ways \textit{unnatural} to set $f(n)=1$. It is then straightforward to see that we only expect at most some special choices of the $f(n)$ to give rise to a six-dimensional operator $\sOp$ that is local in the $x^+$ direction. Conversely, other choices of $f(n)$ may form natural tools in the construction of defect operators in the six-dimensional theory extended along the $x^+$ direction
 
Most generally, we can consider a single six-dimensional operator $\sOp$ as being built from some set $\{\fOp^{(a)}\}_{a\in\mathbb{Z}}$ of five-dimensional operators, so that its Fourier modes are identified as $\fOp_n = \I_n \fOp^{(n)}$. Indeed, the above generalisation involving $f(n)$ is precisely recovered when the $\fOp^{(a)}$ are each multiples of one another.

 
Therefore, an important step in the realisation of six-dimensional physics will be a classification of these different choices of Fourier resummation, and in particular an understanding of which give rise to local operators in six dimensions. Further it would be interesting to understand exactly how the $\frak{so}(2,6)$ representations of these local operators are built from those of $\frak{h}$.

For instance, consider a scalar primary of $\frak{so}(2,6)$. We know then that its Fourier modes are necessarily scalar primaries of $\frak{h}$. But can we make stronger converse statements? For example, given a single scalar primary $\fOp$ of $\frak{h}$, do the $\fOp_n = f(n)\I_n \fOp$ always resum to produce a scalar primary $\sOp$ of $\frak{so}(2,6)$ for some $f(n)$?

It is reasonable to expect that a crucial tool in the resolution of the above issues are the topological Ward-Takahashi identities. In particular, let us suppose that the theory does indeed realise the full six-dimensional conformal group $SO(2,6)$ as a symmetry at strong coupling $(k=1)$. The content of this statement is that the correlators of the modes $\fOp_n$ of some operator must satisfy the full set of TWTIs, and thus realise the symmetry enhancement $SU(1,3)\times U(1)\to SO(2,6)$, again at strong coupling. Thus, we arrive at concrete equations that must be satisfied by such five-dimensional correlators, that should in principle constrain the form of the $\fOp_n$ as functions of the $\I_n \fOp$.


\subsection*{The view from the bulk}


As we've discussed, the M5-brane worldvolume theory enjoys a rich holographic duality through the $\text{AdS}_7/\text{CFT}_6$ correspondence. In particular, our model at strong coupling ($k=1$) can be seen as a \textit{definition} M-theory in the bulk $\text{AdS}_7\times S^4$ geometry.

As we then increase $k$, this is realised in the dual geometry as a $\mathbb{Z}_k$ orbifold of the fibration $S^1\hookrightarrow \text{AdS}_7\to \tilde{\mathbb{CP}}^3$. In particular, as we go to $k>>1$, the fibre shrinks and we approach the regime of M-theory that is well described by Type IIA string theory on $\tilde{\mathbb{CP}}^3\times S^4$. As discussed previously, in this string theory picture the Kaluza-Klein modes on the $S^1$ fibre are realised as D0-brane bound states. Going back the other way, as we take $k$ small again, these D0-brane bound states become light and the string theory picture breaks down.\\

It would be interesting to pursue opportunities to match results from our model to objects like Witten diagrams on the gravity side. Such matching is likely to be approachable only in the supergravity approximation, corresponding in the gauge theory to large $N$. To this end, it would be interesting to explore our theory at large $N$, which will amount to deriving a suitable $\Omega$-deformation to the now well-understood large $N$ properties of the usual ADHM moduli space \cite{Dorey:1999pd}. Further, it may be interesting to consider some combined $k,N\to\infty$ limit, in analogy with the planar limit of the ABJM model, which we discuss next.


\subsection*{Comparison with M2-brane models}


As we have seen, we expect our theory's symmetries to be enhanced at strong coupling. In fact, from a group-theoretic perspective, this is identical to the symmetry enhancement experienced by the ABJM model for a stack of M2-branes. Further, the specific details of this enhancement show that the two scenarios share a number of analogous features. Let us explore these details now, which are also demonstrated in Table \ref{tab: M-brane symmetries}. 
\begin{center}
\begin{minipage}{0.8\textwidth}
\centering
\captionof{table}{The symmetries realised by Lagrangian M-brane models, and their necessary enhancement at strong coupling}\label{tab: M-brane symmetries}
\includegraphics[width=130mm]{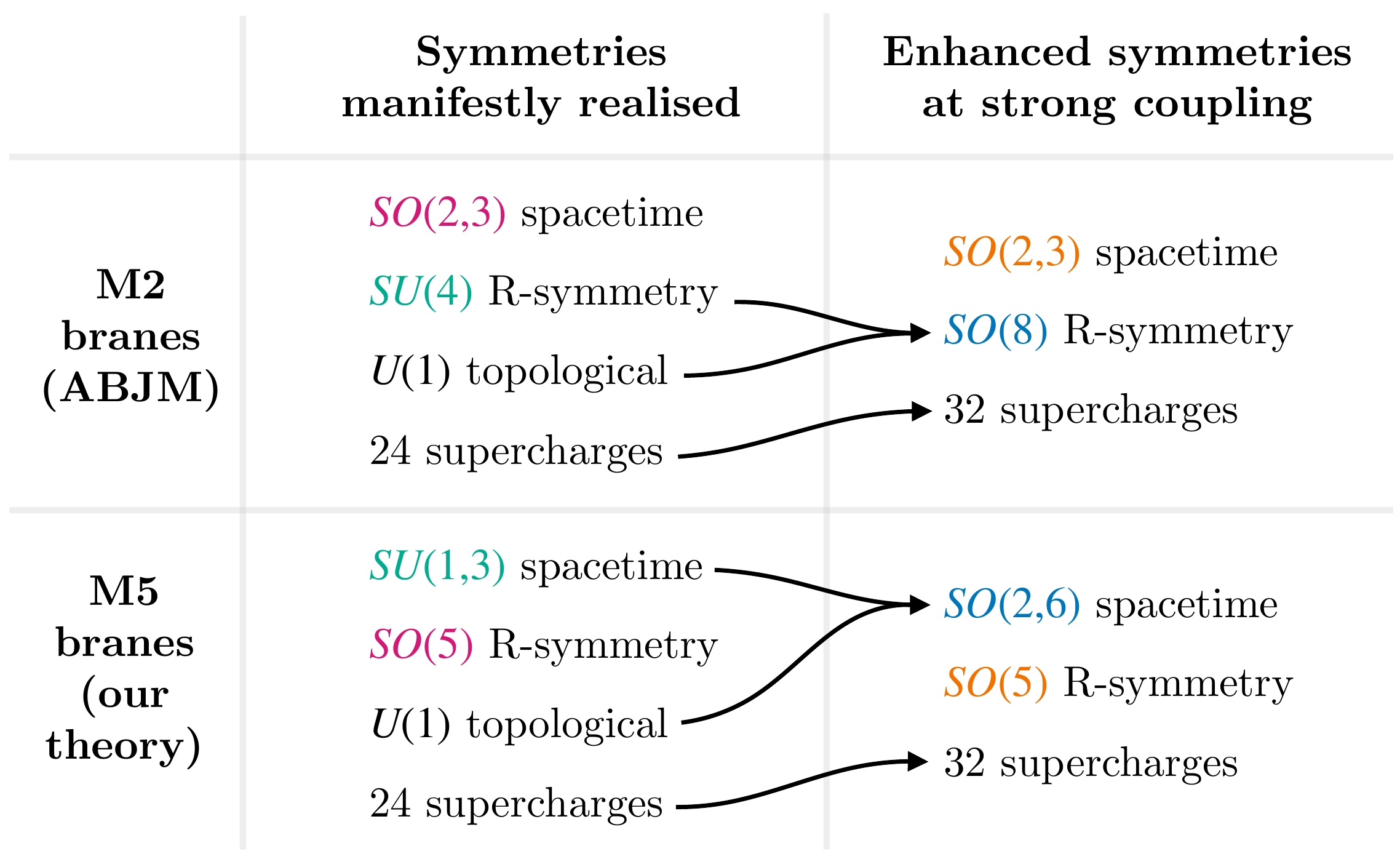}	
\end{minipage}
\end{center}
As discussed previously, we can describe a generic number of M2-branes using Chern-Simons-matter theories. More precisely, recall that the $U(N)_k \times U(N)_{-k}$ ABJM model describes $N$ M2-branes on a transverse geometry $\mathbb{C}^4/\mathbb{Z}_k$. At $k=1$ this is simply flat space, and so the dual geometry in the bulk is $\text{AdS}_4\times S^7$. More generally, the $\mathbb{Z}_k$ orbifold is realised in the bulk as a $\mathbb{Z}_k$ orbifold of the fibration $S^1\hookrightarrow S^7\to \mathbb{CP}^3$.

At general $k$, the symmetries realised by the theory are precisely those expected from the geometry. The orbifold leaves intact the worldvolume geometry, and thus the spacetime symmetry is unbroken: the theory is a three-dimensional conformal field theory, i.e. it has the spacetime symmetry $\frak{so}(2,3)$. Conversely, the orbifold's action on the transverse geometry generically breaks the R-symmetry to $U(1)\times SU(4)$, with these factors corresponding in the dual geometry to isometries on the $S^1$ fibre and the $\mathbb{CP}^3$ base, respectively. While the $SU(4)$ R-symmetry is realised in a conventional way (i.e. an internal symmetry acting on fields), the $U(1)$ symmetry\footnote{Technically there are \textit{two} topological $U(1)$ currents, corresponding to each of the components of the $U(N)\times U(N)$ gauge group, but only a symmetric combination acts non-trivially on the physical spectrum \cite{Bergman:2020ifi}} is generated by a topological current $J\sim \star \text{tr}(F)$. Finally, the orbifold implies the theory should only realise 24 of the usual 32 supercharges, which is indeed the case. Altogether, these symmetries make up the superalgebra $\frak{osp}(6|4)\oplus \frak{u}(1)$, where the $\frak{so}(2,3)$ spacetime symmetry and $\frak{su}(4)$ R-symmetry have been combined with the 24 supercharges into the superalgebra $\frak{osp}(6|4)$, while the topological $\frak{u}(1)$ sits alone.

Then, at strong coupling ($k=1$) there is no orbifold, and so the symmetries are enhanced to $\frak{osp}(8|4)$. In detail, one includes in the theory a class of disorder operators called \textit{monopole operators}, which instruct the path integral to run over configurations with singular points $x_a$, on a small 2-sphere around which one measures a non-zero monopole number $\frac{1}{2\pi}\int_{S^2} F=n_a$. By including such operators, one explicitly sees the R-symmetry enhancement $U(1)\times SU(4)\to SO(8)$ as well as the addition of 8 new supercharges.\\

Let us play the same game for a stack of M5-branes. We claim that our model at rank $N$ and level $k$ describes $N$ M5-branes on a $\mathbb{Z}_k$ orbifold of six-dimensional Minkowski space. At level $k=1$ this is simply flat space, and so the dual geometry in the bulk
 is $\text{AdS}_7 \times S^4$. More generally, the $\mathbb{Z}_k$ orbifold is realised in the bulk as a $\mathbb{Z}_k$ orbifold of the fibration $S^1\hookrightarrow \text{AdS}_7 \to \tilde{\mathbb{CP}}^3$.

At general $k$, the symmetries realised by the theory are those expected from the geometry. Now, the orbifold acts on the worldvolume geometry, and so generically breaks the full six-dimensional conformal group from $SO(2,6)$ down to\footnote{Recall, we had $\frak{h}= \frak{u}(1)\oplus  \frak{su}(1,3)$} $U(1)\times SU(1,3)$. These factors correspond in the dual geometry to isometries of the $S^1$ fibre and the $\tilde{\mathbb{CP}}^3$ base, respectively. While the $SU(1,3)$ symmetry is realised in a conventional way (i.e. a spacetime symmetry acting on fields and coordinates), the $U(1)$ symmetry is generated by a topological current $J\sim \star \text{tr}\left(F\wedge F\right)$. Conversely, the orbifold leaves the transverse geometry untouched, and thus the full $SO(5)$ R-symmetry is manifestly realised. Finally, the orbifold implies the theory should only realise 24 of the usual 32 supercharges, which is indeed the case. Altogether, these symmetries make up the superalgebra $\frak{osp}(6|4)\oplus \frak{u}(1)$, matching identically (again, up to the choice of real form) with the ABJM model. The point here is that, up to the choice of real form, the spacetime symmetries of our model are the R-symmetries of the ABJM model, and vice-versa. This is again very natural in the bulk picture, where for the M2-brane the compact space $S^7$ is fibred over, while for the M5-brane it is the $\text{AdS}_7$ that is fibred over.

Once again, at strong coupling ($k=1$) there is no orbifold, and so we predict that the symmetry is enhanced to the full $\frak{osp}(8|4)$. As we have seen, this enhancement will depend crucially on the inclusion of a class of disorder operators called \textit{instanton operators}, which instruct the path integral to run over configurations with singular points $x_a$, on a small 4-sphere around which one measures a non-zero instanton number $\frac{1}{8\pi^2}\int_{S^4}\text{tr}(F\wedge F)=n_a$. We in particular expect the spacetime symmetry to be enhanced as $U(1)\times SU(1,3)\to SO(2,6)$, and also to see the addition of 8 new supercharges.\\


This close analogy with the ABJM model lends support to the proposed symmetry enhancements at strong coupling in our theory. What's more, the identification of the superalgebras of the two theories, at least up to choice of real form, raises the possibility that the rich literature of results in the ABJM model found through integrability may carry over in some form to our model \cite{Minahan:2008hf}. In particular, $\frak{osp}(6|4)$ admits an infinite-dimensional extension known as Yangian symmetry (the superconformal algebra of $\mathcal{N}=4$ super-Yang-Mills also exhibits such an extension, see \cite{Beisert:2010jr} for a review). Originally seen at the level of scattering amplitudes \cite{Bargheer:2010hn}, a proposal for seeing Yangian symmetry at the Lagrangian level was recently described in \cite{Beisert:2018zxs}. It would be interesting to understand whether similar analysis will uncover such extended symmetry at the Lagrangian level in our theory \cite{Lipstein:Yangian}.


\subsection*{Direct computations in five dimensions}


If the theory (\ref{eq: Lagrangian}), with a suitably extended configuration space as explored in Chapter \ref{chap: An explicit model and its symmetries}, provides an action principle for M5-branes on non-compact Minkowski space, then the ultimate goal is to utilise it for explicit computation of correlators and other observables that probe the dynamics of the $(2,0)$ theory.

To this end, it would be interesting to understand the reduction from six dimensions down to five, and conversely the resummation from five dimensions up to six, for a broader class of operators and correlation functions. Firstly, it would be interesting to investigate the dimensional reduction of 4-point functions of protected ($\Delta\in 2\mathbb{Z}$) scalar operators in the $(2,0)$ theory, which can be computed in the large-$N$ expansion \cite{Arutyunov:2002ff,Heslop:2004du,Rastelli:2017ymc,Heslop:2017sco,Chester:2018dga,Abl:2019jhh,Alday:2020lbp,Alday:2020tgi}.

 It would also be an important step to understand the five/six-dimensional relationship for the correlators of operators with more general $\Delta\notin 2\mathbb{N}$ corresponding to unprotected operators in six dimensions. On one hand, the topological Ward-Takahashi identities of the form (\ref{eq: TWTI}) will admit solutions for $\Delta\in\mathbb{C}$, although as we have seen there are interesting discontinuities in the solution space as we approach $\Delta\in 2\mathbb{N}$. On the other hand, much of the relative simplicity of our dimensional reduction of correlation functions arose due to the meromorphicity of the integrand of integrals like (\ref{eq: 2pt dim red integral}), and more generally those of the form (\ref{eq: N point integral}). This is clearly broken when $\Delta\notin 2\mathbb{N}$, and one must think more carefully about branch points and cuts. It would be useful therefore to understand how this continuation to $\Delta\notin 2\mathbb{N}$ works, and in particular seek a consistent analytic continuation of results in the five-dimensional theory that behave correctly as we resum back to six dimensions. It is this last requirement in particular that is key in order for the five-dimensional theory to be truly useful for dynamical computations in the $(2,0)$ theory.\\

We should again pay attention to different regimes of the theory. Indeed, it is reasonable to expect that many of the possible routes forward detailed below will be tractable only at weak coupling ($k>>1$). As mentioned previously, it would also be very interesting to probe the large $N$ theory, and indeed some combined large $k,N$ limit, in which one may hope to define a limit analogous to the planar limit of the ABJM model.

However, regardless of the regime we are in, the theory is far from conventional, and indeed the results we expect may be at odds with our usual quantum field theory intuition. Consider for instance (at $k=1$) the 2-point functions $\left\langle \I_{n}(x_1)\fOp(x_1) \I_{-n}(x_2)\fOp(x_2) \right\rangle$ for scalar operator $\fOp$ of dimension $\Delta$. Suppose further that we can identify the $\I_n\fOp$ as the modes of a protected scalar operator of the $(2,0)$ theory with $\Delta\in 2\mathbb{Z}$, whose 2-point functions we derived explicitly by dimensional reduction from six dimensions in Section \ref{subsec: 2pt dim red}. We then expect any correct procedure for direct computation in the theory to find $\left\langle \I_{n}(x_1)\fOp(x_1) \I_{-n}(x_2)\fOp(x_2) \right\rangle$ given precisely by (\ref{eq: dim red 2pt from heuristics}), in particular \textit{vanishing} in any sector with $n<\Delta/2$, including the trivial sector $n=0$! Indeed, even as we go to weak coupling and thus consider only the modes $\I_{kn}\fOp$ with $k>>1$, the trivial sector $n=0$ must still be projected out at 2-points. Let us caveat this, however, with the above observation that much of the nuance here may be contained within the correct identification of the Fourier modes of a given protected scalar operator in six dimensions.\\

Let us speculate on possible routes towards explicit computations in the five-dimensional theory. Continuing from Section \ref{sec: dynamics}, an obvious next step would be to further study the model obtained by the reduction of the Lagrangian (\ref{eq: Lagrangian}) to the subspace of the total constraint surface $\mathcal{F}^+=0$ captured by the ansatz (\ref{eq: 't Hooft form for A}). Such an investigation would constitute an $\Omega$-deformation of the results of Section \ref{sec: Five-dimensional super-Yang Mills and the M5-brane}. The key difference, however, is that the resulting model will generically be a non-local quantum mechanics for the worldline positions $y_A(\tau)$ and sizes $\mu_A(\tau)$. It would therefore be interesting to consider limits, such as that of close worldlines, in which this non-locality may be sub-leading and the theory more tractable. One could also consider a formal limit in the parameter $(1/R)$, which may form a natural stratification of non-local effects.

It is not unlikely, however, that access to quantitative results as well as potentially more tractable regimes such as large $N$ will require a more complete understanding of the constraint surface $\mathcal{F}^+=0$. Although the solutions explored in Chapter \ref{chap: worldlines} are constrained to be of the 't Hooft form (\ref{eq: 't Hooft form for A}), it is non unreasonable to expect that their interpretation as encapsulating both the presence of and backreaction due to anti-instanton worldlines persists on the entire constrain surface $\mathcal{F}^+=0$, at least when worldlines are well-separated. Indeed, the success of a 't Hooft-like ansatz for the $\Omega$-deformed instanton equation offers hope that a suitably $\Omega$-deformed ADHM construction may be fruitful in exploring the total constrain surface.\\

Finally, with an action in hand there are a number of other directions one may consider. A Lagrangian such as (\ref{eq: Lagrangian})---one in which a Lagrange multiplier imposes a constraint on the dynamics---is naturally quantised canonically in Dirac's formalism for constrained systems \cite{Dirac:1964,Hanson:1976cn}. Indeed, the monotonicity of the worldline solutions of Chapter \ref{chap: worldlines} is very suggestive that a canonical approach with Hilbert space defined on constant $x^-$ slices may be fruitful, especially considering the considerations of operator ordering in Section \ref{sec: relation to correlation functions}. In such a formulation, one would at least formally expect a raising operator on the Hilbert space to create a worldline at some `time' $x^-_1$, only for a lowering operator at some later time $x^-_2>x^-_1$ to annihilate it.

A more modern approach to direct computation in supersymmetric field theories is that of supersymmetric localisation \cite{Pestun:2016zxk}. Our theory's high degree of supersymmetry may make it a suitable candidate for such techniques. It should be noted that this approach would likely involve the compactification of theory; indeed, it is interesting in its own right to consider enacting both five-dimensional and six-dimensional compactifications in the theory.


\subsection*{Structure and implications of supersymmetry}


Much of our analysis of the theory (\ref{eq: Lagrangian}) has revolved around its spacetime symmetries. Indeed, as we saw in Chapter \ref{chap: An explicit model and its symmetries}, the classical breaking of the symmetries $M_{i+},K_+$ in the presence of instanton operators was sufficient to reveal the extension of the $\frak{su}(1,3)$ spacetime symmetry algebra  to $\frak{h} = \frak{su}(1,3) \oplus \frak{u}(1)$ in the quantum theory, with operators dressed with non-trivial instanton operators having charge under the central element $P_+$.

However, one may now seek a more complete treatment to verify that the quantum theory's superalgebra is indeed $\frak{osp}(6|4)\oplus \frak{u}(1)$. By considering the sub-superalgebra within $\frak{osp}(8|4)$ that commutes with $P_+$, one finds that 16 of the 24 supersymmetries of the five-dimensional theory (\ref{eq: Lagrangian}) should be broken is the presence of instanton operators but recovered in the quantum theory, in a way conceptually identically to what we saw for the generators $M_{i+}, K_+$.\\

It would be interesting to then consider the implications of these supersymmetries for correlators in the five-dimensional theory. Indeed, the resulting supersymmetric Ward-Takahashi identities in the DLCQ limit have been studied previously \cite{Henkel:2005dj}, giving a useful result against which to match. Further, it would be interesting to explore more abstractly the representation theory of the superalgebra, including the notions of short multiplets and thus protected operators. Such analysis will be closely related to comparable statements in the ABJM theory, but with differences attributable to the swapping of the spacetime and R-symmetry groups between the two theories.


\subsection*{The DLCQ limit, altogether now}


Throughout this thesis, we have grounded many of our results on M5-brane models in terms of the older DLCQ proposal for the $(2,0)$ theory \cite{Aharony:1997an,Aharony:1997th}. With the benefit of hindsight, let us draw these connections together into a single story.

We showed in Chapter \ref{chap: DLCQ} that the DLCQ setup---a simple compactification of a null coordinate---could be recovered in a particular limit of the construction at finite $R$. This limit required that we took $R\to\infty$, where the coordinate transformation (\ref{eq: coordinate transformation}) degenerates such that $(x^+, x^-, x^i)$ are simple lightcone coordinates on $\mathbb{R}^{1,5}$. But we also had to introduce the orbifold parameter $k$, and thus restrict to modes of period $2\pi R/k = 2\pi R_+$ on the $x^+$ compactification. Then, taking simultaneously $k,R\to\infty$ but with the ratio $R_+=R/k$ fixed, we arrived at the DLCQ picture \cite{Maskawa:1975ky}: a compactification on a null circle, $x^+\sim x^++2\pi R_+$.\\

We were then able to consider the fate of correlation functions in this DLCQ limit. We first considered the limit of the solution (\ref{eq: general N-point function}) to the $\frak{h}$ Ward-Takahashi identities at $N$-points, leading to the general form of the $N$-point function in the DLCQ limit (\ref{eq: DLCQ Npt WI sol}) and thus extending significantly the results of \cite{Aharony:1997an,Henkel:1993sg}. Interestingly, while at finite $R$ the generic $N$-point function falls away polynomially at both large $x^-$ and $x^i$ separation, correlators in the DLCQ limit generically retain only the decay with $x^-$ separation.

 More interesting still was the DLCQ limit of the 2-, 3- and special 4-point functions found at finite $R$ from dimensional reduction, which by construction satisfy the $\frak{so}(2,6)$ Ward-Takahashi identities. We saw in particular that as we approach the limit, correlation functions necessarily diverge at least as badly as $\sim k$, and indeed we saw that the degree of this divergence was subtly related to the extent to which the initial six-dimensional correlator factorised. Indeed, the Fourier mode expansion of correlators degenerates and becomes ill-defined in the DLCQ limit, and thus the finite $R$ theory acts as a useful and consistent regulator.
 
 In a similar vein, the finite $R$ theory appears to provide clarity on the infamous zero mode problem that is ubiquitous to DLCQ formulations \cite{Nakanishi:1976vf,Fitzpatrick:2018ttk} (see also \cite{Hellerman:1997yu,Bilal:1998ys} for discussions relevant to M-theory). Roughly speaking, a central tenet of any DLCQ description is that (with sign depending on convention), processes with momentum transfer $n>0$ on the compact null direction are expressly projected out, while those with $n<0$ survive. The issue of how best to deal with the $n=0$ processes, however, is the source of over five decades of debate, starting with \cite{Chang:1968bh,Yan:1973qg}, and tackled much more recently in a number of works, including \cite{Collins:2018aqt,Fitzpatrick:2018ttk}. To understand the relevance here, consider once again the dimensionally-reduced 2-point function $\left\langle \fOp_{n}\fOp_{-n} \right\rangle$ at finite $R$ (\ref{eq: dim red 2pt from heuristics}), which vanishes unless\footnote{Recall, we have dimensionally reduced the 2-point function of protected scalar operators with $\Delta\in 2\mathbb{Z}$} $n\ge \Delta/2>0$, assuming $\Delta>0$. Thus, we see that the non-zero 2-point functions are in a sense gapped, with the sector of vanishing momentum transfer strictly off-limits! Now, in the DLCQ limit we write $n\to kn$, and consider the limit of large $k$. Thus, we find that for all $n>0$ and $k$ sufficiently large, $kn\ge \Delta/2$ and thus the 2-point function is non-zero. Conversely, for all $n\le 0$, we will always have $kn<\Delta/2$, no matter how large we make $k$. Thus, we see in the DLCQ limit, the only remaining 2-point functions are those with strictly negative momentum transfer.\\

Finally, we can consider the DLCQ limit of the finite $k,R$ theory (\ref{eq: Lagrangian}). But this is nothing other than the action (\ref{eq: final M5 action}), with $x^0\to x^-$ and coupling $g^2=4\pi^2 R_+$. In particular, we see that in the DLCQ limit, the Lagrange multiplier constraint is simply $F_{ij}+\frac{1}{2}\varepsilon_{ijkl}F_{kl}=0$, the usual anti-instanton equation. We thus neatly recover the standard DLCQ proposal, in which dynamics is reduced to instanton moduli space as we explored in explicit detail in Section \ref{sec: Five-dimensional super-Yang Mills and the M5-brane}. To say it another way, we have simply added the final strand in a web of relations. On one hand, we can straight off the bat consider M5-branes on a null compactification, where their equations of motion arise from the action (\ref{eq: final M5 action}) \cite{Lambert:2018lgt,Lambert:2010wm,Lambert:2011gb}. On the other, we can first consider a conformal compactification of M5-branes, described by action (\ref{eq: Lagrangian}), and then consider a limit in which this compactification becomes a simple null compactification, again arriving at (\ref{eq: final M5 action}).


\subsection*{Other topics}


Let us now discuss a final few aspects of this work, and some related future directions. First, note that in addition to the $(2,0)$ theory of the M5-brane, there is a family of superconformal field theories in six dimensions with half as much supersymmetry that allow for more general matter content. These $(1,0)$ theories are realised by various constructions in M-theory and massive Type IIA string theory \cite{Ganor:1996pc,Bah:2017wxp}. In line with this, one finds  that the theory (\ref{eq: Lagrangian}) is in fact just a special case of a far broader class of five-dimensional $SU(1,3)$ theories with 12 supercharges, which are similarly proposed as Lagrangian descriptions of $(1,0)$ theories \cite{Lambert:2020jjm}. Indeed, we considered the bosonic part of such theories in Section \ref{sec: dynamics}. Further, the results of Chapter \ref{chap: An explicit model and its symmetries} depend only on the gauge sector of the theory, and thus can be straightforwardly shown to apply more generally to the broader class of 12 supercharge theories. It is therefore reasonable to expect that the description of six-dimension superconformal field theories proposed in this thesis apply for these less supersymmetric examples, too.\\

It would be interesting to explore the inclusion of defects in our construction. From the perspective of the five-dimensional theory, one could look for solutions to the Ward-Takahashi identities on semi-infinite space, as was done in the DLCQ limit in \cite{Henkel:1993sg}. Conversely, it is clear that defects extended along the $x^+$ direction will arise from different linear combinations of the Fourier modes as discussed below (\ref{eq: conclusion fourier resum}). From the perspective of the M5-brane theory, one could aim to recover five-dimensional manifestations of the $(2,0)$ defects considered in \cite{Drukker:2020atp,Drukker:2020swu}.\\

More generally, one could explore theories with an $SU(1,d/2)$ spacetime symmetry, obtained from an analogous conformal compactification of conformal field theories of dimension $d=2,4$ \cite{Lambert:Frontiers}. The origin of this symmetry group can be understood holographically by considering $\text{AdS}_{d+1}$ as a circle fibration over a non-compact $\mathbb{CP}^{d/2}$ \cite{Pope:1999xg}, making manifest the subgroup $U(1)\times SU(1,d/2)\subset SO(2,d)$. It would be interesting to derive the conformal blocks for these symmetry groups and develop the corresponding (non-relativistic) conformal bootstrap, which may have applications to condensed matter physics \cite{Chen:2020vvn,Seiberg:2020bhn,Moroz:2019qdw,Orlando:2020idm,Hellerman:2020eff}. Further, one may hope to make contact with known sectors of the $\mathcal{N}=4$ super-Yang-Mills theory that exhibit an $SU(1,1)$ and $SU(1,2)$ symmetries \cite{Baiguera:2020jgy,Baiguera:2020mgk}.


\authoredby{final}
\partfont{\color{black}}
\chapterfont{\color{black}}
\chapter*{Some final words}
\addcontentsline{toc}{part}{Some final words}
\lhead{\textsl{\nouppercase{\leftmark}}}


So there we are. This thesis has approached the topic of non-Lorentzian theories with an inhomogeneous scaling symmetry from a number of different angles. In doing so, we hope it provides a valuable contribution to the growing literature on non-Lorentzian models in string theory, M-theory and gravity.

The construction of supersymmetric theories with an inhomogeneous scaling symmetry in Part I touched upon a number of topics, both old and new. Aside from providing a concrete route to construct such models from their Lorentzian cousins, this work shed light on the dynamics of slowly-moving supersymmetric solitons.

The foundational results on $SU(1,3)$ theories in Part II closely followed the modern construction of conformal field theories, with a number of close analogies made, but also several interesting differences noted. It seems only natural to suppose, then, that such models are ripe to be studied in the same forensic detail as conformal field theories have in recent years.

Finally, we hope we have provided compelling evidence that non-Lorentzian models provide a encouraging path towards dynamical computations in six-dimensional superconformal field theories. In doing so, we add a new chapter to a long-standing story, of the role that non-Lorentzian models have to play in the construction of M-theory and its branes.


\appendix
\addtocontents{toc}{\protect\setcounter{tocdepth}{0}}
\part*{Appendices}
\addcontentsline{toc}{part}{Appendices}

\addtocontents{toc}{
\setlength{\cftbeforechapskip}{0.3\cftbeforechapskip}
\setlength{\cftchapindent}{\cftsecindent}
\protect\renewcommand{\cftchapfont}{\cftsecfont}
\protect\renewcommand{\protect\cftchapdotsep}{\cftsecdotsep}
}
\chapter{Finite $SU(1,3)\times U(1)$ transformations of coordinates and fields}\label{app: finite transformations}


In Part II of this thesis, we were concerned with field theories with a curious spacetime symmetry algebra $\frak{h}=\frak{su}(1,3)\oplus \frak{u}(1)\subset \frak{so}(2,6)$, which has basis $\{P_+,P_-, P_i, B, C^\alpha, M_{i+},K_+\}$ and brackets as in Chapter \ref{chap: SU(1,3) theories}. In the DLCQ limit ($R\to \infty$) this algebra degenerates and, with the additional of the two missing rotations in the $x^i$ directions, becomes simply the maximal subalgebra of the six-dimensional conformal group that commutes with a lightcone translation $P_+$. This is the well-studied Schr\"odinger algebra.

Let us however consider the symmetries at finite $R$. In Chapter \ref{chap: SU(1,3) theories}, we stated the representation of $\frak{h}$ in terms of vector fields on $\mathbb{R}^5$ (\ref{eq: 5d algebra vector field rep}), which were later realised as the push-forward of a set of six-dimensional conformal Killing vector fields (\ref{eq: 6d CKVs}) with respect to the simple projection map $(x^+, x^-, x^i)\to (x^-, x^i)$. We also defined primary operators of $\frak{h}$, which we found to transform as in (\ref{eq: 5d algebra action on fields}).

Working in the explicit theory introduced in Chapter \ref{chap: An explicit model and its symmetries}, we saw that the fields $A, X^I, \Psi$ were naively inert under $P_+$, and thus built of primaries and descendants with $p_+=0$. However, when inserted at points with non-zero instanton charge, we saw how in the quantum theory, we could build operators with non-zero charge under $P_+$. This was encoded in the language on instanton operators $\I_n(x)$.\\

Almost everything in the body of this thesis with regards to symmetries has been \textit{infinitesimal}. This was sufficient to derive infinitesimal Ward-Takahashi identities, and thus constrain correlation functions. The purpose of this appendix is to provide the corresponding finite transformations, both of coordinates and fields, found by exponentiating the infinitesimal results.

\section{Coordinates}

Let us first consider the finite transformations of the coordinates $(x^+,x^-,x^i)$ under $SU(1,3)\times U(1)$, as generated by the vector fields (\ref{eq: 6d CKVs}). These are most straightforwardly determined by considering the well-known finite transformations of the coordinates $(\hat{x}^+, \hat{x}^-, \hat{x}^-)$ under the conformal group, and using then using the coordinate relations (\ref{eq: coordinate transformation}) along with the embedding (\ref{eq: 5d alg in terms of 6d}) of $\frak{su}(1,3)\oplus \frak{u}(1)\subset \frak{so}(2,6)$ to determine the corresponding transformations of the coordinates $(x^+, x^-, x^i)$. We then find
{\allowdisplaybreaks
\begin{align}
  g=e^{ \epsilon P_+ }
  \, &\longrightarrow \, \left\{\,\,\begin{aligned}
 	 (x g)^+  &= 	x^+ + \epsilon  		\\
 	 (x g)^-  &= 	x^- 		\\
 	 (x g)^i &= x^i
 \end{aligned}\right.\ ,\nn\\
  g=e^{ \epsilon P_- }
  \, &\longrightarrow \, \left\{\,\,\begin{aligned}
  	 (x g)^+  &= 	x^+   \\
 	 (x g)^-  &= 	x^- + \epsilon		\\
 	 (x g)^i &= x^i
 \end{aligned}\right.\ ,\nn\\
 g=e^{ \epsilon^i P_i }
  \, &\longrightarrow \, \left\{\,\,\begin{aligned}
    (x g)^+  &= 	x^+   \\
 	(x g)^- &= x^- +\tfrac{1}{2} \Omega_{ij} \epsilon^i x^j			\\
 	(x g)^i &= x^i + \epsilon^i
 \end{aligned}\right.\ ,\nn\\
 g=e^{ \epsilon B }
  \, &\longrightarrow \, \left\{\,\,\begin{aligned}
    (x g)^+  &= 	x^+   \\
 	(x g)^- &= x^-			\\
 	(x g)^i &= \cos\left( \tfrac{\epsilon}{2} \right)x^i + \sin\left( \tfrac{\epsilon}{2} \right) R\,\Omega_{ij} x^j
 \end{aligned}\right.\ ,\nn\\
 g=e^{ \epsilon^\alpha C^\alpha }
  \, &\longrightarrow \, \left\{\,\,\begin{aligned}
    (x g)^+  &= 	x^+   \\
 	(x g)^- &= x^-			\\
 	(x g)^i &= \cos\left( \tfrac{|\epsilon^\alpha|}{2} \right)x^i - \sin\left( \tfrac{|\epsilon^\alpha|}{2} \right) \tfrac{\epsilon^\alpha}{|\epsilon^\alpha|} \eta^\alpha_{ij} x^j
 \end{aligned}\right.\ ,\nn\\
 g=e^{ \epsilon T }
  \, &\longrightarrow \, \left\{\,\,\begin{aligned}
    (x g)^+  &= 	x^+   \\
 	(x g)^- &= e^{2\epsilon} x^-			\\
 	(x g)^i &= e^{\epsilon} x^i
 \end{aligned}\right.\ ,\nn\\
 g=e^{ \epsilon^i M_{i+}}
  \, &\longrightarrow \, \left\{\,\,\begin{aligned}
    (x g)^+  &= 	x^+ +2R  \arctan \left(\frac{1}{4R}\frac{|\vec{\epsilon}|^2 x^- + 2\epsilon^i x^i}{1-\tfrac{1}{2}\Omega_{ij}\epsilon^i x^j+\tfrac{1}{16R^2} |\vec{\epsilon}|^2 |\vex|^2}\right)   \\
 	(x g)^- &= \left( x^--\tfrac{1}{2}x^-\Omega_{ij}\epsilon^i x^j - \tfrac{1}{8R^2}|\vex|^2\epsilon^i x^i \right) 			\\
 	&\qquad \times \Big( \left( 1-\tfrac{1}{2}\Omega_{ij}\epsilon^i x^j+\tfrac{1}{16R^2} |\vec{\epsilon}|^2 |\vex|^2 \right)^2+\tfrac{1}{16R^2}\left( |\vec{\epsilon}|^2 x^- + 2\epsilon^i x^i \right)^2		\Big)^{-1}		\\
 	(x g)^i &= \Big( x^i + \epsilon^i x^- + \tfrac{1}{4} \Omega_{ij} \epsilon^j |\vex|^2 +\tfrac{1}{2}\Omega_{ij}x^j x^k \epsilon^k +\tfrac{1}{2}\Omega_{kl}x^i x^k \epsilon^l			\\
 	&\qquad\hspace{2mm} -\tfrac{1}{8R^2}|\vex|^2 \epsilon^i \epsilon^j x^j + \tfrac{1}{16R^2}|\vex|^2 x^i |\vec{\epsilon}|^2 - \tfrac{1}{8}\Omega_{ij} \Omega_{kl} \epsilon^j \epsilon^k x^l |\vex|^2			\\
 	&\qquad\hspace{2mm} + \tfrac{1}{4} x^- \Omega_{ij}x^j |\vec{\epsilon}|^2 + \tfrac{1}{2}x^- \Omega_{ij} \epsilon^j \epsilon^k x^k -\tfrac{1}{2}x^- \Omega_{kl} \epsilon^i \epsilon^k x^l			\\
 	&\qquad\hspace{2mm} +\tfrac{1}{64R^2}\Omega_{ij} \epsilon^j |\vec{\epsilon}|^2 |\vex|^4 + \tfrac{1}{4}\left( x^- \right)^2 \Omega_{ij} \epsilon^j |\vec{\epsilon}|^2	\Big)	\\
 	&\qquad \times \Big( \left( 1-\tfrac{1}{2}\Omega_{ij}\epsilon^i x^j+\tfrac{1}{16R^2} |\vec{\epsilon}|^2 |\vex|^2 \right)^2 +\tfrac{1}{16R^2}\left( |\vec{\epsilon}|^2 x^- + 2\epsilon^i x^i \right)^2		\Big)^{-1}
 \end{aligned}\right.\ ,\nn\\
 g=e^{ \epsilon K_+ }
  \, &\longrightarrow \, \left\{\,\,\begin{aligned}
    (x g)^+  &= 	x^+ +2R  \arctan \left(\frac{\epsilon}{2R}\frac{|\vex|^2}{1-2\epsilon x^-}\right)  \\
 	(x g)^- &= \frac{x^-\left( 1-2\epsilon x^- \right)-\tfrac{1}{8R^2}\epsilon |\vex|^4}{\left( 1-2\epsilon x^- \right)^2 + \tfrac{1}{4R^2}\epsilon^2 |\vex|^4}			\\
 	(x g)^i &= \frac{x^i\left( 1-2\epsilon x^- \right)+\tfrac{1}{2}\epsilon\, \Omega_{ij} x^j |\vex|^2}{\left( 1-2\epsilon x^- \right)^2 + \tfrac{1}{4R^2}\epsilon^2 |\vex|^4}	
 \end{aligned}\right.\ ,
 \label{eq: finite coordinate transformations}
\end{align}}for constants $\epsilon, \epsilon^i$ and $\epsilon^\alpha$. Note, the global action of these transformations---in particular the correct branch of the arctangent for the transformation of $x^+$---can be fixed by transformation back to the coordinates $(\hat{x}^+, \hat{x}^-, \hat{x}^i)$ and considering the standard action of the $\frak{so}(2,6)$ generators as in (\ref{eq: 5d alg in terms of 6d}) on the one-point compactification of $\mathbb{R}^{1,5}$, although we omit details here.

Recall that given any $x_1,x_2\in\mathbb{R}^5$, we can define
\begin{align}
  z(x_1,x_2)=\left( x_1^--x_2^- + \frac{1}{2}\Omega_{ij}x_1^i x_2^j \right) +\frac{i}{4R} \left( x_1^i-x_2^i \right)\left( x_1^i-x_2^i \right) = -\bar{z}(x_2,x_1)\ .
\end{align}
This object arose naturally in our study of correlation function in Chapter \ref{chap: SU(1,3) theories}, as its real and imaginary parts are the unique combinations invariant under translations and rotations. Further, it played a crucial role in the constraint surface analysis of Chapter \ref{chap: worldlines}.

Clearly $z(x_1,x_2)$ is invariant under $P_+$. Thus, we can consider the action of $SU(1,3)\subset SU(1,3)\times U(1)$ on $z$, defining $z(x_1,x_2)g=z(x_1g,x_2g)$ for each $g\in SU(1,3)$. Additionally writing $z_{12}=z(x_1,x_2)$ for brevity, we find
\begin{align}
    g=\exp\left( \epsilon P_- \right)
  \quad &\longrightarrow \quad z_{12}g=z_{12}\ ,\nn\\
 g=\exp\left(\epsilon^i P_i \right)
  \quad &\longrightarrow \quad z_{12}g=z_{12}\ ,\nn\\
 g=\exp\left(\epsilon B \right)
  \quad &\longrightarrow \quad z_{12}g=z_{12}\ ,\nn\\
 g=\exp\left(\epsilon^\alpha C^\alpha \right)
  \quad &\longrightarrow \quad z_{12}g=z_{12}\ ,\nn\\
 g=\exp\left(\epsilon T \right)
  \quad &\longrightarrow \quad z_{12}g=e^{2\epsilon}z_{12}\ ,\nn\\
 g=\exp\left(\epsilon^i M_{i+} \right)
  \quad &\longrightarrow \quad z_{12}g=\frac{z_{12}}{\mathcal{M}_\epsilon(x_1)\overline{\mathcal{M}_\epsilon(x_2)}}\ ,\nn\\
 g=\exp\left(\epsilon K_+ \right)
  \quad &\longrightarrow \quad z_{12}g= \frac{z_{12}}{\big( 1-2\epsilon z(x_1,0) \big)\big( 1-2\epsilon \bar{z}(x_2,0) \big) }\ ,
  \label{eq: z transformation}
\end{align}
with $\mathcal{M}_\epsilon(x)$ as defined in (\ref{eq: curly M definition}).\\

With this handle on finite transformations, let us revisit a question answered in Chapter \ref{chap: SU(1,3) theories}, and verify the result. We now consider $N$ points $x_a=(x_a^-, x_a^i)$, $a=1,\dots, N$, and also write $z_{ab}=z(x_a, x_b)$. Let us now ask once again: what are the independent $SU(1,3)$ invariants we can construct?

The only combinations that are invariant under the translations $\{P_-, P_i\}$ are
\begin{align}
  x_a^- - x_b^- + \frac{1}{2}\Omega_{ij} x_a^i x_b^j\qquad \text{and}\qquad x_a^i - x_b^i~ .
\end{align}
The first of these is then also invariant under the rotations $\{B, C^\alpha\}$. From the spatial distances $\left( x_a^i - x_b^i\right)$, we can form two types of rotationally-invariant object: $\left( x_a^i - x_b^i\right)\left( x_c^i - x_d^i\right)$ and $\left( x_a^i - x_b^i\right)\Omega_{ij}\!\left( x_c^j - x_d^j\right)$. Here, we do not necessarily take the $\{a,b\}$ as disjoint from the $\{c,d\}$.

However, not all of these objects are independent. Firstly, it is clear that any of the $\left( x_a^i - x_b^i\right)\left( x_c^i - x_d^i\right)$ can be written as some linear combination of the $\left( x_a^i - x_b^i\right)\left( x_a^i - x_b^i\right)$. Secondly, we have
\begin{align}
  \tfrac{1}{2}\left( x_a^i - x_b^i\right)\Omega_{ij}\!\left( x_c^j - x_d^j\right) &= \left(x_a^- - x_c^- + \tfrac{1}{2}\Omega_{ij} x_a^i x_c^j\right) - \left(x_b^- - x_c^- + \tfrac{1}{2}\Omega_{ij} x_b^i x_c^j\right)			\nn\\
  &\qquad - \left(x_a^- - x_d^- + \tfrac{1}{2}\Omega_{ij} x_a^i x_d^j\right) + \left(x_b^- - x_d^- + \tfrac{1}{2}\Omega_{ij} x_b^i x_d^j\right)\ .
\end{align}
Hence, we learn that the independent objects invariant under translations and rotations are precisely the real and imaginary parts of the $z_{ab}$. Noting that $\bar{z}_{ba}=-z_{ab}$, we see that we can write \textit{any} $SU(1,3)$ invariant as some holomorphic function of the $z_{ab}$.

So we are finally left to determine which functions of the $z_{ab}$ are invariant under the three remaining transformations generated by $\{T, M_{i+}, K_+\}$. This in fact follows rather straightforwardly from (\ref{eq: z transformation}). We see in particular that the general $SU(1,3)$ invariant is
\begin{align}
  \frac{z_{a_1 b_1}z_{a_2 b_2}\dots z_{a_m b_m}}{z_{a_1 b_{\sigma(1)}}z_{a_2 b_{\sigma(2)}}\dots z_{a_m b_{\sigma(m)}}}\ ,
\end{align}
for any permutation $\sigma\in S_m$.

There is however a large amount of degeneracy here. We would like to understand what the \textit{independent} invariants are. To this end, let us take as our building blocks
\begin{align}
  u_{ab}:=|z_{ab}|^2\ ,\quad v_{ab} = \frac{z_{ab}}{\bar{z}_{ab}}\ .
\end{align}
It is clear that any $SU(1,3)$ invariant can be written as a function of the $u_{ab}$ and $v_{ab}$. Indeed, it is evident from the transformation rules of the $u_{ab}$ and $v_{ab}$ that there are no non-trivial invariants made up of some combination of $u$'s and $v$'s. Thus, we are lead to find invariant combinations of the $u$'s, and invariant combinations of the $v$'s.

The only way to make an invariant out of the $u_{ab}$ is to take some ratio of them. Indeed, by precisely the same logic as in conventional conformal field theory \cite{Ginsparg:1988ui}, we find that \textit{any} such invariant ratio can be written as a function of the cross-ratios
\begin{align}
  \frac{u_{ab}u_{cd}}{u_{ac}u_{bd}}~,
\end{align}
of which $N(N-3)/2$ are independent. These ratios (or rather, their square root) are precisely the variables (\ref{eq: z cross ratios}).

Again making use of the rules (\ref{eq: z transformation}), we find that any invariant made up of the $v_{ab}$ can be written in terms of combinations of the form
\begin{align}
  v_{a_1 a_2}v_{a_2 a_3}\dots v_{a_{m-1} a_m}v_{a_m a_1}\ .
\end{align}
However, by inserting factors of $1=v_{ab}v_{ba}$, we see that any such combination can be further reduced to a function of the invariants
\begin{align}
  v_{ab}v_{bc}v_{ca}\ ,
\end{align}
of which $(N-1)(N-2)/2$ are independent. These are precisely the phases (\ref{eq: z cross ratios}).

Thus, we have successfully shown from scratch that the cross-ratios (\ref{eq: z cross ratios}) and phases (\ref{eq: z phases}) form a \textit{complete} set of $SU(1,3)$ invariants, and hence the expression (\ref{eq: general N-point function}) does indeed provide the \textit{general} $N$-point function of scalar primaries in an $SU(1,3)$ theory.

\section{Fields under $SU(1,3)$}

We have defined the notion of primary fields in the five-dimensional theory, transforming infinitesimally as in (\ref{eq: 5d algebra action on fields}). Let us focus on the sector with vanishing $P_+$ charge, $p_+=0$. 

First consider a scalar primary $\fOp$, i.e. a primary field with $r_\fOp[B]=r_\fOp[C^\alpha]=0$, which is defined by a single scaling dimension $\Delta$. Then, under a finite $SU(1,3)$ transformation we have
\begin{align}
  g\fOp(x) = \det\left(\hat{\partial}_i (xg^{-1})^j\right)^{\Delta/4} \fOp(xg^{-1})\ .
  \label{eq: scalar primary finite transformation}
\end{align}
Explicitly, we have that $\det\left(\hat{\partial}_i (xg^{-1})^j\right)=1$ for $G\in\{P_-, P_i, B, C^\alpha\}$. For the remaining generators, we have
\begin{align}
  	g=e^{\epsilon T} \quad &\longrightarrow \quad \det\left(\hat{\partial}_i (xg^{-1})^j\right)^{1/4} = e^{-\epsilon}		\ ,\nn\\
  	g=e^{\epsilon^i M_{i+}} \quad &\longrightarrow \quad \det\left(\hat{\partial}_i (xg^{-1})^j\right)^{1/4} = \frac{1}{\left|\mathcal{M}_{-\epsilon} (x)\right|}{}		\ ,\nn\\
  	g=e^{\epsilon K_+} \quad &\longrightarrow \quad \det\left(\hat{\partial}_i (xg^{-1})^j\right)^{1/4} = \frac{1}{\left|1+2\epsilon z(x,0)\right|}{}\ ,
\end{align}
where $\mathcal{M}_\alpha(x)$ is as defined in (\ref{eq: curly M definition}).

Indeed, the bosonic fields of the theory (\ref{eq: Lagrangian}) could be recast in terms of the primaries $X^I, \hat{A}_-, \hat{A}_i$. In particular, $X^I$ and $\hat{A}_-$ are both scalar primaries with $\Delta=2$, and thus their $SU(1,3)$ transformation as generated by the $\delta_G$ are given by (\ref{eq: scalar primary finite transformation}). Further, note that the instanton operators $\I_n(x)$ transform under $\delta_G$ as scalar primaries with $\Delta=0$.\\

The field $\hat{A}_i$, on the other hand, is not a scalar but instead has `spin' under $B$ and $C^\alpha$ as given in (\ref{eq: Ai hat def}). We find then the finite transformation
\begin{align}
  g\hat{A}_i(x) = e^{\delta_G}\hat{A}_i(x) = \hat{\partial}_i(xg^{-1})^j \hat{A}_j (xg^{-1})\ .
  \label{eq: A_i hat finite transformation}
\end{align}
Let us finally comment on the behaviour of the hatted derivative $\hat{\partial}_i$, which features centrally in the finite transformations of fields (\ref{eq: scalar primary finite transformation}), (\ref{eq: A_i hat finite transformation}). Let us define a generalised object,
\begin{align}
  \hat{\partial}_i^y = \frac{\partial}{\partial y^i} - \frac{1}{2}\Omega_{ij} y^j \frac{\partial}{\partial y^-}\ ,
\end{align}
where $ y^\alpha(x)$ is some function on $\mathbb{R}^5$. We in particular recover $\hat{\partial}_i = \hat{\partial}_i^x$. Then, we find that $\hat{\partial}_i^y$ satisfies a novel chain rule,
\begin{align}
  \hat{\partial}_i^{xg} = \big( \hat{\partial}_i^{xg} x^j \big)\hat{\partial}_j^x\ ,
\end{align}
for any $g\in SU(1,3)$. It is precisely this fact that ensures that the finite transformations (\ref{eq: scalar primary finite transformation}), (\ref{eq: A_i hat finite transformation}) do indeed form representations of $SU(1,3)$, i.e. $(g_1 g_2)\fOp(x) = g_1(g_2 \fOp(x))$ and $(g_1 g_2)\hat{A}_i(x) = g_1 (g_2 \hat{A}_i(x))$, respectively.\vspace{-1em}

\section{Including $P_+$ charge}

We now understand fully the transformation properties of the fields in the theory (\ref{eq: Lagrangian}) under the variations $\delta_G$. However, we later saw in Section \ref{subsec: quantum recovery} that the action is not invariant under these transformations in the presence of instanton operators, but that the variation of the action nonetheless lead to Ward-Takahashi identities of the usual form (\ref{eq: final local WTIs}) in terms of deformed variations $\tilde{\delta}_G$ which form a representation of $\frak{h}$. We now present the finite field transformations as generated by $\tilde{\delta}_G$.

For normal operators $\fOp(x)$, we have $\tilde{\delta}_G\fOp(x) = \delta_G \fOp(x)$, and so we're done. In contrast, while instanton operators simply transform under $\delta_G$ as $\Delta=0$ scalar primaries, their behaviour under $\tilde{\delta}_G$ is deformed as in (\ref{eq: tilde variations of I}). Then, we find
\begin{align}
  \exp(\tilde{\delta}_G)\,\I_n(x) = \exp(\delta_G)\,\I_n(x) = \I_n(xg^{-1}) \quad \text{for } G\in\text{span}\left( \{P_-, P_i, B, C^\alpha, T\} \right)\ ,
\end{align}
and otherwise,
\begin{align}
  	\exp\left( \epsilon\, \tilde{\delta}_{P_+} \right)\I_n(x) &= e^{i\epsilon k n}\I_n(x)			\ ,\nn\\
  	\exp\left( \epsilon^i\, \tilde{\delta}_{M_{i+}} \right)\I_n(x) &= \left( \frac{\mathcal{M}_{-\epsilon}(x)}{\overline{\mathcal{M}_{-\epsilon}(x)}} \right)^{kn}\I_n(xg^{-1})			\ ,\nn\\
  	\exp\left( \epsilon\, \tilde{\delta}_{K_+} \right)\I_n(x) &= \left( \frac{1+2\epsilon z(x,0)}{1+2\epsilon \bar{z}(x,0)} \right)^{kn}\I_n(xg^{-1})		\ .
  	\label{eq: finite tilde transformation I}	
\end{align}
These forms indeed follow straightforwardly from the finite variation of the action (\ref{eq: finite action variation M})--(\ref{eq: finite action variation K}).


\chapter{Derivation of dimensionally reduced correlators}\label{app: dimensional reduction of correlators}


\section{2-point Functions}\label{app: 2pt} 
Let us now derive the result of Section \ref{subsec: 2pt dim red}, and in particular show how it arises from a six-dimensional $i\epsilon$ prescription. We follow the familiar routine of defining Lorentzian correlation functions by Wick rotating their Euclidean counterparts. In doing so, one encounters ambiguities corresponding to how the branch points in the complex time plane are navigated. The resulting family of Lorentzian results are naturally captured through a Wightman function, inside which operators no longer commute. In this way, the Wick rotation induces a natural operator ordering, which is efficiently captured by a so-called $i\epsilon$ prescription usually taken with respect to the time direction.

A useful review of this perspective on Lorentzian quantum field theory can be found in \cite{Hartman:2015lfa}, while the grounding of such a viewpoint in a path integral framework is achieved through the Schwinger-Keldysh formalism \cite{Schwinger:1960qe,Bakshi:1962dv,Bakshi:1963bn,Keldysh:1964ud}, helpfully reviewed in \cite{Skenderis:2008dg,Weinberg:1995mt}.

 We instead choose our prescription with respect to the coordinate $x^+$, which can be seen as a deformation of the more familiar ordering by the regular lightcone coordinate $\hat{x}^+$ \cite{Fitzpatrick:2018ttk}. This then defines the Lorentzian $n$-point function to be
\begin{align}
  \left\langle \hat{\sOp}^{(1)}(\hat{x}_1^+,\hat{x}_1^-,\hat{x}_1^i)\dots \hat{\sOp}^{(N)}(\hat{x}_N^+,\hat{x}_N^-,\hat{x}_N^i) \right\rangle &=\nn\\ 
  &\hspace{-60mm}\lim_{\substack{\epsilon_a\to 0 \\ \epsilon_1>\dots >\epsilon_N>0}} \left\langle \hat{\sOp}^{(1)}(\hat{x}_1^+(\epsilon_1),\hat{x}_1^-(\epsilon_1),\hat{x}_1^i(\epsilon_1))\dots \hat{\sOp}^{(N)}(\hat{x}_N^+(\epsilon_N),\hat{x}_N^-(\epsilon_N),\hat{x}_N^i(\epsilon_N)) \right\rangle_\text{Wick}\ ,
  \label{eq: i epsilon}
\end{align}
where the correlation function on the right hand side is the naive result of Wick rotating the Euclidean correlation function, and we define
\begin{align}
  	\hat{x}^+(\epsilon) &= \hat{x}^+\left(x^+-i\epsilon,x^-, x^i\right) = 2R\left( \frac{\left( \tfrac{\hat{x}^+}{2R} \right)-i\tanh\left( \tfrac{\epsilon}{2R} \right)}{1+i\tanh\left( \tfrac{\epsilon}{2R} \right)\left( \tfrac{\hat{x}^+}{2R} \right)} \right)  \ ,	\nn\\
  	\hat{x}^-(\epsilon)	&= \hat{x}^-\left(x^+-i\epsilon,x^-, x^i\right) 	\ ,\nn\\
  	\hat{x}^i(\epsilon)	&= \hat{x}^i\left(x^+-i\epsilon,x^-, x^i\right) \ .
\end{align}
It is important here to keep track of this deformation for \textit{finite} $\epsilon$, since the action of the prescription in $\hat{x}^+$ space is inhomogeneous. Note, the effect of this $i\epsilon$ prescription on the integral (\ref{eq: 2pt dim red integral}) is to shift any potential poles in $\hat{x}_1^+$ and $\hat{x}_2^+$ on the real line infinitesimally up or down into the complex plane, thus regularising the integral.

Using this, we dimensionally reduce the six-dimensional 2-point function (\ref{eq: 6d 2pt}), finding five-dimensional 2-point function
\begin{align}
  F_2:&= \left\langle \fOp^{(1)}_{n_1}(x_1)\fOp^{(2)}_{n_2}(x_2)\right\rangle\nn\\
   &= \frac{\hat{C}_\Delta}{\pi^2}\left( 4R \right)^{-\Delta}(-1)^{\n_1+\n_2} \nn\\
  &\qquad\times\lim_{\substack{\epsilon_a\to 0\\ \epsilon_1>\epsilon_2>0}}\int_{-\infty}^{\infty} du_1\int_{-\infty}^{\infty} du_2\,\, \prod_{a=1}^2 \left( u_a+i \right)^{-\n_a-1}\left( u_a-i \right)^{\n_a-1} \left( u^\epsilon_a+i \right)^{\frac{\Delta}{2}}\left( u^\epsilon_a-i \right)^{\frac{\Delta}{2}}		\nn\\
  &\hspace{58.5mm}\times \Big( \tilde{x}_{12} \left( u^\epsilon_2-u^\epsilon_1 +\left( 1+ u^\epsilon_1 u^\epsilon_2 \right)\tfrac{\xi_{12}}{4R} \right) \Big)^{-\Delta} \ ,
  \label{eq: deformed 2pt integral}
\end{align}
where $a=1,2$, and we define
\begin{align}
  u^\epsilon_a = \frac{u_a-i\epsilon_a}{1+i\epsilon_a u_a} \ .
  \label{eq: u epsilon def}
\end{align}
We have importantly made sure to apply the $i\epsilon$ prescription also to the Weyl rescaling factor (\ref{eq: 6d operator mapping}). Also, in arriving at this expression we have exploited the strict monotonicity of $\tanh$ to replace $\tanh\left( \tfrac{\epsilon_a}{2R} \right)$ with simply $\epsilon_a$ while preserving the ordering of the $\epsilon_a$. In doing this, we have also shortened the range of $\epsilon_a$ to $\epsilon_a\in(0,1)$.

We now find that the integrand is strictly non-zero for all $u_1,u_2\in\mathbb{R}$, and goes like $u_a^{-2}$ as $|u_a|\to\infty$. Hence, it is convergent.

Although the $\epsilon_a$ enter in $u_a^\epsilon$ in a somewhat complicated way, their effect on the total integrand is simplified  by noting the identity
\begin{align}
  \tilde{x}_{12} \Big( u^\epsilon_2-u^\epsilon_1 +\left( 1+ u^\epsilon_1 u^\epsilon_2 \right)\tfrac{\xi_{12}}{4R} \Big) &= \frac{i}{2}\Big[ \left( u_1^\epsilon - i \right)\left( u_2^\epsilon+i \right)\bar{z}_{12} -  \left( u_1^\epsilon + i \right)\left( u_2^\epsilon-i \right)z_{12} \Big]\ ,
\end{align}
with $z_{12}=z(x_1,x_2)$, and hence
\begin{align}
  &\frac{\left( \tilde{x}_{12} \left( u^\epsilon_2-u^\epsilon_1 +\left( 1+ u^\epsilon_1 u^\epsilon_2 \right)\tfrac{\xi_{12}}{4R} \right) \right)^2}{\left( u_1^\epsilon +i \right)\left( u_1^\epsilon -i \right)\left( u_2^\epsilon +i \right)\left( u_2^\epsilon -i \right)} \nn\\
  &\qquad= -\frac{1}{4} \frac{\Big[ \left( 1+\epsilon_1 \right)\left( 1-\epsilon_2 \right)\left( u_1 - i \right)\left( u_2+i \right)\bar{z}_{12} -  \left( 1-\epsilon_1 \right)\left( 1+\epsilon_2 \right)\left( u_1 + i \right)\left( u_2-i \right)z_{12} \Big]^2}{\left( 1-\epsilon_1^2 \right)\left( 1-\epsilon_2^2 \right)\left( u_1 +i \right)\left( u_1 -i \right)\left( u_2 +i \right)\left( u_2 -i \right)}\ .
  \label{eq: fraction identity}
\end{align}
We can therefore write
\begin{align}
  F_2 &= \frac{\hat{C}_\Delta}{\pi^2}\left( 2R\, i \right)^{-\Delta}(-1)^{\n_1+\n_2}\label{eq: deformed 2pt integral 2}\\
  &\times\lim_{\substack{\epsilon_a\to 0\\ \epsilon_1>\epsilon_2>0}}\left( 1-\epsilon_1^2 \right)^{\frac{\Delta}{2}}\left( 1-\epsilon_2^2 \right)^{\frac{\Delta}{2}}\int_{-\infty}^{\infty} du_1\int_{-\infty}^{\infty} du_2\,\, \prod_{a=1}^2 \left( u_a+i \right)^{-\n_a+\frac{\Delta}{2}-1}\left( u_a-i \right)^{\n_a+\frac{\Delta}{2}-1} 	\nn\\
  &\times \Big( \left( 1+\epsilon_1 \right)\left( 1-\epsilon_2 \right)\left( u_1 - i \right)\left( u_2+i \right)\bar{z}_{12} -  \left( 1-\epsilon_1 \right)\left( 1+\epsilon_2 \right)\left( u_1 + i \right)\left( u_2-i \right)z_{12} \Big)^{-\Delta} \nn \ .
\end{align}
There are now many ways to proceed to calculate $F_{12}^{(\text{5d})}$ explicitly. Here, we follow a particularly streamlined approach. For a more general discussion of the evaluation of integrals of this type, including an alternative contour derivation of their explicit values, see appendix \ref{app: contours}.\\

To proceed, note that we can write the final part of the integrand as
\begin{align}
  &\Big( \left( 1+\epsilon_1 \right)\left( 1-\epsilon_2 \right)\left( u_1 - i \right)\left( u_2+i \right)\bar{z}_{12} -  \left( 1-\epsilon_1 \right)\left( 1+\epsilon_2 \right)\left( u_1 + i \right)\left( u_2-i \right)z_{12} \Big)^{-\Delta}\nn\\
  &\hspace{5mm}= \left( \left( 1+\epsilon_1 \right)\left( 1-\epsilon_2 \right)\left( u_1 - i \right)\left( u_2+i \right)\bar{z}_{12} \right)^{-\Delta}\left( 1 -  \frac{\left( 1-\epsilon_1 \right)\left( 1+\epsilon_2 \right)\left( u_1 + i \right)\left( u_2-i \right)z_{12}}{\left( 1+\epsilon_1 \right)\left( 1-\epsilon_2 \right)\left( u_1 - i \right)\left( u_2+i \right)\bar{z}_{12}} \right)^{-\Delta}\ .
  \label{eq: Taylor expn 1}
\end{align}
Then, we have
\begin{align}
  \left|\frac{\left( 1-\epsilon_1 \right)\left( 1+\epsilon_2 \right)\left( u_1 + i \right)\left( u_2-i \right)z_{12}}{\left( 1+\epsilon_1 \right)\left( 1-\epsilon_2 \right)\left( u_1 - i \right)\left( u_2+i \right)\bar{z}_{12}}\right| = \frac{\left( 1-\epsilon_1 \right)\left( 1+\epsilon_2 \right)}{\left( 1+\epsilon_1 \right)\left( 1-\epsilon_2 \right)}<1\quad \text{ since }\quad \epsilon_1>\epsilon_2\ ,
\end{align}
and hence we can use the Taylor expansion for $(1-w)^{-\Delta}$, as the argument falls just within the radius of convergence. Indeed, the partial sums converge uniformly to functions both of $u_1$ and $u_2$, allowing us to integrate term-wise. Hence, we see in this way how our integral is regularised: we have a convergent series expansion which would have otherwise been indeterminate. Further, if we had instead  $\epsilon_1<\epsilon_2$, we would have instead written $(1-w)^{-\Delta}=\left( -w \right)^{-\Delta}\left( 1-w^{-1} \right)^{\Delta}$ and used the series expansion for the latter factor; in this way, we can see how the ordering prescription manifests in our calculation.

So we now simply substitute this series expansion into (\ref{eq: deformed 2pt integral}) to find
\begin{align}
  F_2 &= \frac{\hat{C}_\Delta}{\pi^2}\left( -2R\, i \right)^{-\Delta}(-1)^{\n_1+\n_2}   \nn\\
  &\qquad\times\lim_{\substack{\epsilon_a\to 0\\ \epsilon_1>\epsilon_2>0}}\sum_{m=0}^\infty \left( \tfrac{\left( 1-\epsilon_1 \right)\left( 1+\epsilon_2 \right)}{\left( 1+\epsilon_1 \right)\left( 1-\epsilon_2 \right)} \right)^{\tfrac{\Delta}{2}+m}{\Delta+m-1\choose m}\left( z_{12} \right)^m \left( \bar{z}_{12} \right)^{-\Delta-m}\nn\\
  &\qquad\hspace{30mm}\times\left( \int_{-\infty}^{\infty} \frac{du_1}{1+u_1^2} \left( u_1+i \right)^{-\n_1+\tfrac{\Delta}{2}+m}\left( u_1-i \right)^{\n_1-\tfrac{\Delta}{2}-m}	 \right)	\nn\\
  &\qquad\hspace{30mm}\times \left( \int_{-\infty}^{\infty} \frac{du_2}{1+u_2^2} \left( u_2+i \right)^{-\n_2-\tfrac{\Delta}{2}-m}\left( u_2-i \right)^{\n_2+\tfrac{\Delta}{2}+m} \right)	\ .
  \label{eq: 2pt integral expanded}
\end{align}
We can finally perform the integrals explicitly, using the identity
\begin{align}
  \int_{-\infty}^\infty \frac{du}{1+u^2} \left( u+i \right)^m\left( u-i \right)^{-m} = \pi \delta_{m,0} \ .
  \label{eq: orthogonality relation}
\end{align}
We in particular find that every term in the sum vanishes, \textit{unless} $\n_1=-\n_2=\tfrac{\Delta}{2}+m$ for some $m\in\{0,1,\dots\}$. If this is the case, then there is a single non-zero term in the sum. Computing this term and then safely taking the limits $\epsilon_a\to 0$, we finally arrive at the dimensionally reduced 2-point function,
\begin{align}
  \left\langle \fOp^{(1)}_{\n_1}(x_1^-, x_1^i) \fOp^{(2)}_{\n_2}(x_2^-,x_2^i) \right\rangle=
	\delta_{\n_1+\n_2,0}\hat{C}_\Delta \left( -2R\, i \right)^{-\Delta} {{\n_1+\tfrac{\Delta}{2}-1}\choose{\n_1-\tfrac{\Delta}{2}}} \left( z_{12}\bar{z}_{12} \right)^{-\tfrac{\Delta}{2}} \left( \frac{z_{12}}{\bar{z}_{12}} \right)^{\n_1} 
	\label{eq: dim red 2pt final} \ .
\end{align}

\pagebreak
\section{3-point Functions}\label{app: 3pt}

We also present the full derivation of the dimensionally reduced 3-point function of Section \ref{subsec: 3pt dim red}. Using the $i\epsilon$ prescription (\ref{eq: i epsilon}), we find that the Fourier modes of the 3-point function (\ref{eq: 6d 3pt}) are given by the regularised integral
\begin{align}
  F_3:&=\left\langle \fOp^{(1)}_{\n_1}(x_1) \fOp^{(2)}_{\n_2}(x_2)\fOp^{(3)}_{\n_3}(x_3) \right\rangle 	
\nn\\
& =\lim_{\substack{\epsilon_a\to 0\\ \epsilon_1>\epsilon_2>\epsilon_3>0}}\int_{-\infty}^{\infty} d^3 u\, \frac{\hat{C}_{123}}{\pi^3} \left( 4R \right)^{-\frac{1}{2}\left( \Delta_1+\Delta_2+\Delta_3 \right)}\left( -1 \right)^{\n_1+\n_2+\n_3} \nn\\
  &\qquad\times\prod_{a=1}^3 \frac{ \left( u_a-i \right)^{\n_a-1} }{\left( u_a+i \right)^{\n_a+1}} \left( \left( u^\epsilon_a+i \right)\left( u^\epsilon_a-i \right) \right)^{\frac{\Delta_a}{2}}\prod_{a<b}^3\Big( \left( \tilde{x}_{ab} \right)\left( -u^\epsilon_a+u^\epsilon_b +\left( 1+ u^\epsilon_a u^\epsilon_b \right)\tfrac{\xi_{ab}}{4R} \right) \Big)^{-\alpha_{ab}}\ ,
\end{align}
The $u^\epsilon_a$ are as defined in (\ref{eq: u epsilon def}), and we have again assumed $\Delta_1,\Delta_2,\Delta_3\in 2\mathbb{Z}$ to avoid the issue of branch points.

As we saw for the 2-point function, the role of the $i\epsilon$ prescription is made clearer by rewriting this as
\begin{align}
  F_3 &= \frac{\hat{C}_{123}}{\pi^3} \left( 2R\, i \right)^{-\tfrac{1}{2}\left( \Delta_1+\Delta_2+\Delta_3 \right)}\left( -1 \right)^{\n_1+\n_2+\n_3} \nn\\
  &\quad\times\lim_{\substack{\epsilon_a\to 0\\ \epsilon_1>\epsilon_2>\epsilon_3>0}} \prod_{a=1}^3 \left( 1-\epsilon_a^2 \right)^{\frac{\Delta_a}{2}}\int_{-\infty}^{\infty} d^3 u \, \prod_{a=1}^3 \frac{ \left( u_a-i \right)^{\n_a+\frac{\Delta_a}{2}-1} }{ \left( u_a+i \right)^{\n_a-\frac{\Delta_a}{2}+1} } \nn\\
  &\quad\times \prod_{a<b}^3 \Big( \left( 1+\epsilon_a \right)\left( 1-\epsilon_b \right)\left( u_a - i \right)\left( u_b+i \right)\bar{z}_{ab} -  \left( 1-\epsilon_a \right)\left( 1+\epsilon_b \right)\left( u_a + i \right)\left( u_b-i \right)z_{ab} \Big)^{-\alpha_{ab}}	\ .
  \label{eq: 3pt ready for series expn}
\end{align}
with $z_{ab}=z(x_a, x_b)$.

To proceed to calculate $F_{123}^{(\text{5d})}$ explicitly, we follow the same procedure as we did at 2-points. For a more general discussion of the evaluation of integrals of this type, including an alternative contour derivation of their explicit values, see Appendix \ref{app: contours}.\\

As we saw at 2-points, the final three terms can be expanded in convergent series expansions, making use of $\epsilon_1>\epsilon_2$, $\epsilon_2>\epsilon_3$ and $\epsilon_1>\epsilon_3$ respectively. Doing so, we arrive at
\begin{align}
  F_3&= \frac{\hat{C}_{123}}{\pi^3} \left( -2R\, i \right)^{-\tfrac{1}{2}\left( \Delta_1+\Delta_2+\Delta_3 \right)}\left( -1 \right)^{\n_1+\n_2+\n_3}\nn\\
  &\quad \times\sum_{m_1,m_2,m_3=0}^\infty {\alpha_{23}+m_1-1 \choose m_1}{\alpha_{31}+m_2-1 \choose m_2}{\alpha_{12}+m_3-1 \choose m_3} \nn\\
  &\hspace{20mm}\times \left( z_{12} \right)^{m_3}\left( \bar{z}_{12} \right)^{-\alpha_{12}-m_3}\left( z_{23} \right)^{m_1}\left( \bar{z}_{23} \right)^{-\alpha_{23}-m_1}\left( z_{31} \right)^{-\alpha_{31}-m_2}\left( \bar{z}_{31} \right)^{m_2}\nn\\
  &\hspace{20mm}\times\left( \int_{-\infty}^\infty \frac{du_1}{1+u_1^2}\,\, \left( u_1+i \right)^{-\n_1+\tfrac{\Delta_1}{2}+m_2+m_3}\left( u_1-i \right)^{\n_1-\tfrac{\Delta_1}{2}-m_2-m_3} \right)		\nn\\
  &\hspace{20mm}\times\left( \int_{-\infty}^\infty \frac{du_2}{1+u_2^2}\,\, \left( u_2+i \right)^{-\n_2+\tfrac{1}{2}\left( \Delta_3-\Delta_1 \right)+m_1-m_3}\left( u_2-i \right)^{\n_2-\tfrac{1}{2}\left( \Delta_3-\Delta_1 \right)-m_1+m_3} \right)		\nn\\
  &\hspace{20mm}\times\left( \int_{-\infty}^\infty \frac{du_3}{1+u_3^2}\,\, \left( u_3+i \right)^{-\n_3-\tfrac{\Delta_3}{2}-m_1-m_2}\left( u_3-i \right)^{\n_3+\tfrac{\Delta_3}{2}+m_1+m_2} \right)		\ ,
\end{align}
which again is finite due to the identity (\ref{eq: orthogonality relation}). Hence, we finally find the dimensionally reduced 3-point function
\begin{align}
  F_3= \delta_{\n_1+\n_2+\n_3,0}&\hat{C}_{123} \left( -2R\, i \right)^{-\tfrac{1}{2}\left( \Delta_1+\Delta_2+\Delta_3 \right)} \left( z_{12}\bar{z}_{12} \right)^{-\tfrac{1}{2}\alpha_{12}}\left( z_{23}\bar{z}_{23} \right)^{-\tfrac{1}{2}\alpha_{23}}\left( z_{31}\bar{z}_{31} \right)^{-\tfrac{1}{2}\alpha_{31}}\nn\\
  \qquad \times\sum_{m=0}^\infty &{-\n_3-\tfrac{\Delta_3}{2}+\alpha_{23} - m -1\choose -\n_3-\tfrac{\Delta_3}{2}-m}{\n_1-\tfrac{\Delta_1}{2}+\alpha_{12}-m-1 \choose \n_1-\tfrac{\Delta_1}{2}-m}{\alpha_{31}+m-1 \choose m} \nn\\
  &\times \left( \frac{z_{12}}{\bar{z}_{12}} \right)^{\n_1-m-\tfrac{1}{2}\alpha_{31}}\left( \frac{z_{23}}{\bar{z}_{23}} \right)^{-\n_3-m-\tfrac{1}{2}\alpha_{31}}\left( \frac{z_{31}}{\bar{z}_{31}} \right)^{-m-\tfrac{1}{2}\alpha_{31}}\ ,
  \label{eq: final dim red 3pt}
\end{align}
where, given $z=r e^{i\theta}$, we've chosen the branches $\left( z\bar{z} \right)^{1/2}=r$ and $\left( \tfrac{z}{\bar{z}} \right)^{1/2}=e^{i\theta}$. We note that the sum terminates at $\min\left( \n_1-\tfrac{\Delta_1}{2},-\n_3-\tfrac{\Delta_3}{2} \right)$, and thus as promised we have a finite, regularised result.


\chapter{$N$-point integrals and their residue representation}\label{app: contours}


In Section \ref{sec: further constraints on correlators}, we performed the dimensional reduction of six-dimensional 2-point, 3-point, and some special 4-point functions to five-dimensions. Here, we present a more general discussion of integrals that would appear at $\Npt$-points. This will in particular include an alternative route to explicitly calculating their values, {\it via} contour integrals.\\

In dimensionally reducing an $\Npt$-point function in six dimensions, we will encounter integrals of the form
\begin{align}
  \mathcal{I}^{(\Npt)} := \prod_{a=1}^\Npt\int_{-\infty}^{\infty}  du_a \left( u_a + i \right)^{-\n_a+\tfrac{\Delta_a}{2}-1}\left( u_a - i \right)^{\n_a+\tfrac{\Delta_a}{2}-1} \prod_{b<c} \left( \tilde{x}_{bc}\left( u_c-u_b +\left( 1+ u_b u_c \right)\tfrac{\xi_{bc}}{4R} \right) \right)^{-\alpha_{bc}}\ ,
  \label{eq: N point integral}
\end{align}
where the $\alpha_{ab}=\alpha_{ba}$ are integers, which from six-dimensional scale invariance satisfy $\sum_{b\neq a}\alpha_{ab}=\Delta_a$. These integrals are however generically ill-defined; although the integrand has integrable behaviour at large $|u_a|$ ({\it i.e.} $u_a\sim u_a^{-2}$ as $|u_a|\to\infty$), there are poles at finite points which render it divergent.

To see this more explicitly, consider trying to perform the $u_\Npt$ integral. Then, the integrand generically has $(\Npt-1)$ poles at the points
\begin{align}
  u_\Npt = v_{\Npt,a}:=\frac{4Ru_a-\xi_{a\Npt}}{4R+\xi_{a\Npt}u_a},\quad\text{ for each }\quad a\neq \Npt\ ,
\end{align}
and hence the integral is not well defined. \\

We can regularise $\mathcal{I}^{(\Npt)}$ by redefining it as the limit of a well-defined $\epsilon$-deformed integral,
\begin{align}
  \mathcal{I}^{(\Npt)} := \lim_{\substack{\epsilon_a\to 0\\ \epsilon_1>\dots\epsilon_\Npt>0}}\prod_{a=1}^\Npt\int_{-\infty}^{\infty}  du_a \left( u_a + i \right)^{-\n_a-1}\left( u_a - i \right)^{\n_a-1}\left( u^\epsilon_a + i \right)^{\tfrac{\Delta_a}{2}}\left( u^\epsilon_a - i \right)^{\tfrac{\Delta_a}{2}} \nn\\
  \times\prod_{b<c} \left( \tilde{x}_{bc}\left( u^\epsilon_c-u^\epsilon_b +\left( 1+ u^\epsilon_b u^\epsilon_c \right)\tfrac{\xi_{bc}}{4R} \right) \right)^{-\alpha_{bc}}\ .
\end{align}
Utilising a generalisation of the identity (\ref{eq: fraction identity}), we can rewrite this as
\begin{align}
  \mathcal{I}^{(\Npt)} = \lim_{\substack{\epsilon_a\to 0\\ \epsilon_1>\dots\epsilon_\Npt>0}}\prod_{a=1}^\Npt\left( 2i\left( \epsilon_a^2-1 \right) \right)^{\tfrac{\Delta_a}{2}}\int_{-\infty}^{\infty}  du_a \left( u_a + i \right)^{-\n_a+\tfrac{\Delta_a}{2}-1}\left( u_a - i \right)^{\n_a+\tfrac{\Delta_a}{2}-1} \hspace{25mm} \nn\\
  \times\prod_{b<c} \Big( \left( 1+\epsilon_b \right)\left( 1-\epsilon_c \right)\left( u_b-i \right)\left( u_c+i \right) \bar{z}_{bc} - \left( 1-\epsilon_b \right)\left( 1+\epsilon_c \right)\left( u_b+i \right)\left( u_c-i \right) z_{bc} \Big)^{-\alpha_{bc}}\ ,
\end{align}
which is indeed a generalisation of (\ref{eq: deformed 2pt integral 2}) at 2-points and (\ref{eq: 3pt ready for series expn}) at 3-points. We could then proceed to calculate this explicitly by series expanding the factors involving the $z_{bc}$, and using the relation (\ref{eq: orthogonality relation}) to pick out the non-zero terms in the resulting sums.\\

We will now  explore an alternative way in which we could proceed, which will in particular demonstrate that the action of $i\epsilon$ prescription is to shift the integrand's poles off the real line and into the complex $u_\Npt$-plane. Further, this occurs in a controlled way, which allows for relatively simple expression for $\mathcal{I}^{(\Npt)}$ in terms of iterated residues. 

First, we perform the $u_\Npt$ integral by continuing $u_\Npt$ to $\mathbb{C}$ and seeking a related contour integral. In addition to the integrand's obvious potential poles at $u_\Npt=\pm i$, we have up to $(\Npt-1)$ additional poles at
\begin{align}
  u_\Npt = v_{\Npt,a}^\epsilon &:=  -i\, \frac{\left( 1+\epsilon_a \right)\left( 1-\epsilon_\Npt \right)\left( u_a-i \right)\bar{z}_{a\Npt}+\left( 1-\epsilon_a \right)\left( 1+\epsilon_\Npt \right)\left( u_a+i \right)z_{a\Npt}}{\left( 1+\epsilon_a \right)\left( 1-\epsilon_\Npt \right)\left( u_a-i \right)\bar{z}_{a\Npt}-\left( 1-\epsilon_a \right)\left( 1+\epsilon_\Npt \right)\left( u_a+i \right)z_{a\Npt}}\nn\\
  &\phantom{:}= v_{n,i} + \bigO\left( \epsilon_1,\epsilon_2 \right)\ .
\end{align}
We find in particular the imaginary part,
\begin{align}
  \Im\left( v_{n,i}^\epsilon \right) = -\frac{\left( 1+u_a^2 \right)\left( 16R^2+\xi_{a\Npt}^2 \right)\left( 1-\epsilon_a\epsilon_\Npt \right)\left( \epsilon_a-\epsilon_\Npt \right)}{\left( \epsilon_a-\epsilon_\Npt \right)^2\left( 4R u_a-\xi_{a\Npt} \right)^2+\left( 1-\epsilon_a\epsilon_\Npt \right)^2\left( 4R+u_a\xi_{a\Npt} \right)^2}\ ,
  \label{eq: imaginary part}
\end{align}
and hence because we have $\epsilon_a>\epsilon_\Npt$ for all $a\neq \Npt$, we see that all $(\Npt-1)$ of the $v_{\Npt,a}^\epsilon$ lie strictly in the lower-half-plane. Hence, we choose to complete our contour with a large semi-circle in the upper-half-plane, as shown in Figure \ref{fig: n-pt contours}.
\begin{center}
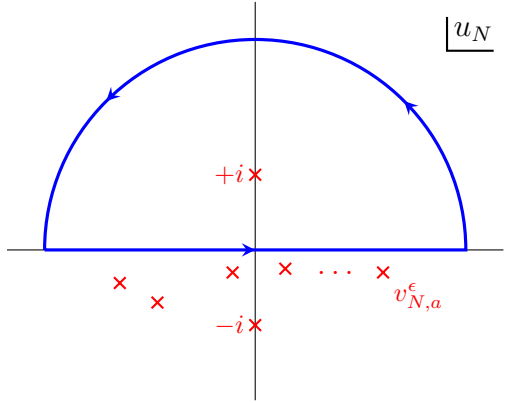
\captionof{figure}{The $u_n$ contour integral, with $(N-1)$ poles shifted into the lower-half-plane}\vspace{1em}\label{fig: n-pt contours}.
\begin{tikzpicture}
\draw  (-8,0) -- (-1.4,0) ;

\draw  (-4.7,-2) -- (-4.7,3.3) ;

\node  at (-1.8,2.9) {$u_N$};
\draw [thick] (-2.15,3.1) -- (-2.15,2.65) -- (-1.53,2.65) ;

\node [cross=3,thick,red ] at (-4.7,1) {};
\node [left,red] at (-4.7,1) {\footnotesize\,$+i$};

\node [cross=3,thick,red ] at (-4.7,-1) {};
\node [left,red] at (-4.7,-1) {\footnotesize\,$-i$};

\node [cross=3,thick,red ] at (-3,-0.3) {};
\node [cross=3,thick,red ] at (-4.3,-0.25) {};
\node [cross=3,thick,red ] at (-6.5,-0.44) {};
\node [cross=3,thick,red ] at (-5,-0.3) {};
\node [cross=3,thick,red ] at (-6,-0.7) {};
\node [red] at (-3.6,-0.3) {$\dots$};
\node [below right, red] at (-3,-0.3) {\footnotesize$v_{\Npt,a}^\epsilon$};

\path [draw=blue,very thick,postaction={on each segment={mid arrow=blue}}] (-7.5,0) -- (-1.9,0) arc(0:180:2.8) ;

\end{tikzpicture}
\end{center}

We omit the explicit form of this residue, but note that it is a meromorphic function of the remaining variables $u_1,\dots, u_{\Npt-1}$ with possible poles only at $u_a= \pm i$. This is seen by noting

\begin{align}
&\Big( \left( 1+\epsilon_a \right)\left( 1-\epsilon_N \right)\left( u_a-i \right)\left( u_N+i \right)\bar{z}_{aN}\nn\\
&\qquad -\left( 1-\epsilon_a \right)\left( 1+\epsilon_N \right)\left( u_a+i \right)\left( u_N-i \right)z_{aN} \Big)\Big|_{u_N=i} = 2i\left( 1+\epsilon_a \right)\left( 1-\epsilon_N \right)\left( u_a-i \right)\bar{z}_{aN}\ .
\end{align}
Thus, its only effect of evaluating the residue at $u_N=i$ on the singularity structure of the resulting integrand is to shift the degree of the poles at $u_a= i$.

So we now do the $u_{\Npt-1}$ integral. There are now $(\Npt-2)$ potential poles other than $u_{\Npt-1}=\pm i$ in the integrand, but since $\epsilon_a>\epsilon_{\Npt-1}$ for all $a<\Npt-1$, once again we see that all of these poles lie strictly in the lower-half-plane. Hence, the integral is once again given entirely by the residue at $u_{\Npt-1}=i$. Iterating this procedure, we finally find
\begin{align}
 \mathcal{I}^{(\Npt)} =(2\pi i)^N\,\text{Res}_{\{u_1=i\}}\bigg[ \,\text{Res}_{\{u_2=i\}}\bigg[ \, \dots \, \text{Res}_{\{u_\Npt=i\}}\bigg[ \hspace{70mm}\nn\\
  \prod_{a=1}^\Npt \left( u_a + i \right)^{-\n_a+\tfrac{\Delta_a}{2}-1}\left( u_a - i \right)^{\n_a+\tfrac{\Delta_a}{2}-1} \prod_{b<c} \left( \tilde{x}_{bc}\left( u_c-u_b +\left( 1+ u_b u_c \right)\tfrac{\xi_{bc}}{4R} \right) \right)^{-\alpha_{bc}}\bigg]\dots \bigg] \bigg]\ .
\end{align}
Computing this directly for the 2-point and 3-point functions, we indeed recover the results (\ref{eq: dim red 2pt from heuristics}) and (\ref{eq: 3pt dim red from heuristics}), respectively.


\chapter{Symmetries from a six-dimensional proxy theory}\label{app: proxy theory}


In this appendix we  provide a six-dimensional origin for the spacetime symmetries of the action $S=\int dx^- d^4 x \mathcal{L}$, with $\mathcal{L}$ as in (\ref{eq: Lagrangian}). Of course the main problem is that there is no known Lagrangian for the $(2,0)$ theory in six dimensions, nor is there expected to be one. However    let us consider the following action
\begin{align}
  S_{6D} = \frac{k}{8\pi^3 \om^2 } \text{tr} \int d^6 x\,\, \sqrt{-g} \bigg\{ &-\frac{1}{12} H_{\mu\nu\rho}H^{\mu\nu\rho} - \frac{1}{2} g^{\mu\nu} D_\mu X^I D_\nu X^I \nonumber\\
  & +\frac{\ii}{2} \bar{\Psi} \Gamma^\mu {\cal D}_\mu \Psi -\frac{1}{2} V^\mu \bar{\Psi} \Gamma_\mu \Gamma^I \left[ X^I, \Psi \right]  \bigg\}\ .
\end{align}
Note that in this appendix we use $\Gamma_\mu$ to denote six-dimensional curved space $\Gamma$-matrices. In the body of this thesis all $\Gamma$-matrices are those of Minkowski space and as such can be identified with the tangent frame $\Gamma$-matrices that appear  in this Appendix. Furthermore
here $\mu=\{+,-,i\}$ and we have introduced a three-form $H_{\mu\nu\rho}$ and vector field $V^\mu$. 

We emphasise that we are not proposing $S_{6D}$ as a candidate for the $(2,0)$ theory. Rather we merely wish to use it to motivate the symmetries of the theory (\ref{eq: Lagrangian}). In particular we will use two features of $S_{6D}$: it has six-dimensional diffeomorphism invariance and, using a suitable ansatz,  it can be dimensionally reduced to $S$, up to a single topological term whose variation is a total derivative. We will then see that the somewhat unusual transformations we used above have a more standard interpretation  within the context of $S_{6D}$.

We have a vielbein $e^{\underline{\mu}}{}_\mu$ satisfying $e^{\underline{\mu}}{}_\mu \eta_{\underline{\mu}\underline{\nu}}e^{\underline{\nu}}{}_\nu=g_{\mu\nu}$, where we choose lightcone coordinates in the tangent frame, {\it i.e.}\ $\eta_{\underline{+}\underline{-}}=-1, \eta_{\underline{+}\underline{+}}=\eta_{\underline{-}\underline{-}}=0, \eta_{\underline{i}\underline{j}}=\delta_{ij}$. Then, we have $\Gamma^\mu = e^{\mu}{}_{\underline{\mu}}\Gamma^{\underline{\mu}}$ and $\Gamma^I = \delta^I_{\und{I}} \Gamma^{\und{I}}$, where $\left\{ \Gamma^{\underline{\mu}}, \Gamma^{\underline{I}} \right\}$ form a (real) basis for the eleven-dimensional Clifford algebra.  The gauge covariant derivative   is $ D_\mu  = \partial_\mu  - \ii \left[ A_\mu, \,. \,\, \right]$, while on $\Psi$  we have
\begin{align}
  {\cal D}_\mu \Psi = \left( D_\mu + \frac{1}{4}  \omega_\mu^{\underline{\mu\nu} }\Gamma_{\underline{\mu\nu}} \right) \Psi \ .
\end{align}
By construction $S_{6D}$ is invariant under six-dimensional diffeomorphisms. In particular
given a vector field ${k}^\mu$, the infinitesimal diffeomorphism generated by $k^\mu$ is given by
\begin{align}
  \delta_d x^\mu &= {k}^\mu \ ,\nonumber\\
  \delta_d \tensor{T}{^{\mu_1}^{\dots}^{\mu_r}_{\nu_1}_{\dots}_{\nu_s}} &=   \left( \partial_\rho {k}^{\mu_1} \right) \tensor{T}{^{\rho}^{\mu_2}^{\dots}^{\mu_r}_{\nu_1}_{\dots}_{\nu_s}} + \dots \nonumber\\
  &\hspace{12mm} - \left( \partial_{\nu_1} {k}^{\rho}  \right) \tensor{T}{^{\mu_1}^{\dots}^{\mu_r}_{\rho}_{\nu_2}_{\dots}_{\nu_s}} - \dots \nonumber\\
  &= - \tensor{\left(\mathcal{L}_{{k}} T \right)}{^{\mu_1}^{\dots}^{\mu_r}_{\nu_1}_{\dots}_{\nu_s}} + {k}^\rho \partial_\rho \tensor{T}{^{\mu_1}^{\dots}^{\mu_r}_{\nu_1}_{\dots}_{\nu_s}} \ ,\nonumber\\
  \delta_d \Psi &= \frac{1}{4} {\lambda}^{\und{\mu\nu}} \Gamma_{\und{\mu\nu}} \Psi \ ,\nonumber\\
  \delta_d e^{\und{\mu}}_{\mu} &= - \left( \partial_\mu {k}^\rho \right) e^{\und{\mu}}_{\rho} +  \tensor{{\lambda}}{^{\und{\mu}}_{\und{\nu}}} e^{\und{\nu}}_{\mu} \ ,\nonumber\\
  \delta_d \tensor{\omega}{_\mu^{\und{\mu\nu}}} &= -\left( \partial_\mu {k}^\rho \right) \tensor{\omega}{_\rho^{\und{\mu\nu}}} + \tensor{{\lambda}}{^{\und{\mu}}_{\und{\rho}}}\, \tensor{\omega}{_\mu^{\und{\rho\nu}}} + \tensor{{\lambda}}{^{\und{\nu}}_{\und{\rho}}}\, \tensor{\omega}{_\mu^{\und{\mu\rho}}} - \partial_\mu {\lambda}^{\und{\mu\nu}}\ ,
\end{align}
where ${T}^{\mu_1 \dots \mu_r}_{\nu_1 \dots \nu_s}$ is a general $(r,s)$-tensor, and we've allowed for a local infinitesimal Lorentz transformation $\tensor{{\lambda}}{^{\und{\mu}}_{\und{\nu}}}$ in the tangent frame. We are   assuming  here that the components of ${k}^\mu$ in a given coordinate frame are small so that we need only consider the first order terms. Note also that we are here regarding the diffeomorphism as a \textit{passive} transformation.

Next we want to write  $S_{6D}$ explicitly in a coordinate frame in which the metric is given by (\ref{eq: conformally compactified metric}	). This metric admits the choice of vielbein $e^{\und{+}}{}_{+} = 1, e^{\und{-}}{}_{-} = 1,  e^{\und{-}}{}_{i}=\frac{1}{2} \Omega_{ij}x^j$ and $e^{\und{i}}{}_{j} = \delta_{ij}$, with all other components vanishing. We suppose that the vector $V^\mu$ takes the form $V^+=1$ with all other components vanishing. Furthermore we choose to turn off any $x^+$ dependance of the fields, and set $A_+=0$, in turn implying $F_{+\mu}=0$. We can then make the identification
\begin{align}\label{FH}
F_{\mu\nu} = H_{\mu\nu +}	\ ,
\end{align}
while we also define
\begin{align}\label{GH}
G_{ij}=H_{-ij}+\frac{1}{2} \varepsilon_{ijkl}H_{-kl} \ .
\end{align}
After performing the trivial $x^+$ integral, we find that $S$  agrees with the reduced $S_{6D}$   up to two additional terms:
\begin{align}\label{SS}
  S &= \frac{k}{4\pi^2 \om}\text{tr} \int d^5x\, \bigg\{ \frac12 F_{-i}F_{-i} - \frac12  {\nabla}_i X^I {\nabla}_i X^I + \frac12 {\cal F}_{ij}G_{ij} \nonumber\\
  &\hspace{33mm}  -\frac{\ii}{2}\bar\Psi\Gamma_{\und{+}}D_-\Psi + \frac{\ii}{2}\bar\Psi\Gamma_{\und{i}} {\nabla}_i\Psi - \frac{1}{2}\bar\Psi\Gamma_{\und{+}}\Gamma^{I}[X^I,\Psi] \bigg\}\nonumber\\
  &= S_{6D} + \frac{k}{4\pi^2 \om} \text{tr} \int d^5x\, \bigg\{ \frac{1}{4} \varepsilon_{ijkl} \mathcal{F}_{ij} H_{-kl} \nonumber\\
  &\hspace{45mm}+ \frac{1}{12} \left( H_{ijk} + \frac{3}{2} \Omega_{l[i|}H_{-|jk]} x^l \right) \Big( H_{ijk} + \frac{3}{2} \Omega_{m[i|}H_{-|jk]} x^m \Big)  \bigg\} \ ,\end{align}
where, as above,  $ {\nabla}_i = D_i -\frac12\Omega_{ij}x^jD_-$ and ${\cal F}_{ij}  = F_{ij} - \frac12\Omega_{ik}x^kF_{-j}+ \frac12 \Omega_{jk}x^kF_{-i}$. 
Lastly we can impose the relation
\begin{align}\label{Hijk}
H_{ijk}= - \frac{3}{2} \Omega_{l[i|}H_{-|jk]} x^l\ . 
\end{align}
This ensures that the second line in  (\ref{SS}) vanishes and as such we   have  
\begin{align}
  S  =S_{6D} +  \frac{k}{4\pi^2\om}   \text{tr}  \int d^5x\, & \frac{1}{4} \varepsilon_{ijkl} \mathcal{F}_{ij} H_{-kl} \ .
  \end{align}
Note that (\ref{Hijk}) differs from that used in the construction of \cite{Lambert:2019jwi}. However we emphasise again that $S_{6D}$ should not be taken literally as an action for the $(2,0)$ theory. In particular with the ansatz here $H_{\mu\nu\lambda}$ is not self-dual. 

We now wish to construct a bosonic symmetry $\delta$ for $S_\Omega$ that descends from the diffeomorphisms for $S_{6D}$. In particular we start with a natural guess $\delta_\text{trial}$ that comes from diffeomorphisms which we then need to slightly correct using the scaling symmetry to find the total variation $\delta_{}$. For a generic object $\Phi$, we are free to replace $\Phi$ in $S_{6D}$ with an explicit expression $\Phi(x)$ in some coordinate frame and preserve a passive diffeomorphism ${k}^\mu$ only if we have
\begin{align}
  \hat{\delta} \Phi := {k}^\rho \partial_\rho \Phi - \delta_d \Phi = 0\ .
  \label{eq: tilde hat def}
\end{align}
In other words the transformation of $\Phi$, as induced by its dependence on $x^\mu$, must match its transformation under $\delta_d$. For a tensor field $T$, we have $\hat{\delta} T = \mathcal{L}_{{k}} T$, and so for ${k}^\mu$ Killing, we have $\hat{\delta}g_{\mu\nu} = 0$. We will consider instead the more general space of conformal Killing vectors with $\partial_+ {k}^\mu=0$, which has basis as listed in (\ref{eq: 6d CKVs}). In particular, if
\begin{align}
  k^\mu &= a_1 \!\left(P_+\right)_\partial^\mu + a_2 \!\left(P_+\right)_\partial^\mu  + a_3^i \!\left(P_i\right)_\partial^\mu + a_4 \!\left(B\right)_\partial^\mu + a_5^\alpha \!\left(C^\alpha\right)_\partial^\mu  \nn\\
  &\qquad+ \omega_1 \!\left(T\right)_\partial^\mu + v_i\!\left(M_{i+}\right)_\partial^\mu + \omega_2 \!\left(K_+\right)_\partial^\mu
  \label{eq: k explicit form}
\end{align}
then $k^\mu$ satisfies $\mathcal{L}_k g_{\mu\nu} = 2\omega g_{\mu\nu}$, with $\omega= \omega_1 + \tfrac{1}{2}\Omega_{ij} v_i x^j + 2\omega_2 x^-$.

So we choose to replace $\{g_{\mu\nu},e^{\und{\mu}}{}_{\mu},\tensor{\omega}{_\mu^{\und{\mu \nu}}}, V^\mu \}$ with their coordinate expressions. Then, $\delta_\text{trial}$ is defined to act as $k^\rho\partial_\rho$ on these fields, and as $\delta_d$ on everything else. Equivalently, we have $\delta_\text{trial}=\delta_d + \hat{\delta}$, where $\hat{\delta}$ as defined in (\ref{eq: tilde hat def}) acts only on $\{g_{\mu\nu},e^{\und{\mu}}{}_{\mu},\tensor{\omega}{_\mu^{\und{\mu  \nu}}}, V^\mu \}$.

As we've already seen, we have $\hat{\delta}g_{\mu\nu}=2\omega g_{\mu\nu}$. Next, we note that the conformal Killing equation implies that
\begin{align}
  k^\rho \partial_\rho e^{\und{\mu}}{}_{\mu} + \left( \partial_\mu k^\rho \right) e^{\und{\mu}}{}_{\rho} = \tensor{\lambda}{^{\und{\mu}}_{\und{\nu}}} e^{\und{\nu}}{}_{\mu} +  \omega e^{\und{\mu}}{}_{\mu}\ ,
\end{align}
for local Lorentz transformation $\tensor{\lambda}{^{\und{\mu}}_{\und{\nu}}}$ given by
\begin{align}
  \tensor{\lambda}{^{\und{\mu}}_{\und{\nu}}} = \left( \partial_\mu k^\nu \right) e_{\und{\nu}}{}^{\mu} e^{\und{\mu}}{}_{\nu} + k^\nu e_{\und{\nu}}{}^{\mu} \partial_\nu e^{\und{\mu}}{}_{\mu} -  \omega \delta^{\und{\mu}}_{\und{\nu}}\ .
\end{align}
One can show using the conformal Killing equation that this does indeed satisfy $\tensor{\lambda}{_{\und{\mu \nu}}}+\tensor{\lambda}{_{\und{\nu \mu}}}=0$. Then, choosing this $\tensor{\lambda}{_{\und{\mu \nu}}}$ for the diffeomorphism $\delta_d$, we have $\hat{\delta}e^{\und{\mu}}{}_{\mu}= \omega e^{\und{\mu}}{}_{\mu}$. Next we find that for the spin connection term we have
\begin{align}
  \hat{\delta}\left( \frac{1}{4} \bar{\Psi} \Gamma^{{\mu}}\, \tensor{\omega}{_\mu^{\und{\nu\rho}}} \Gamma_{\und{\nu\rho}}\Psi \right) = 0\ .
\end{align}
Finally, we simply have $\hat{\delta}V^\mu = 0$.

To continue we observe that
\begin{align}\label{dSS}
  \delta_\text{trial} S  = \hat{\delta} S_{6D}
    + \frac{k}{4\pi^2\om} \delta_\text{trial}\Bigg[ \text{tr}  \int d^5x\, & \frac{1}{4} \varepsilon_{ijkl} \mathcal{F}_{ij} H_{-kl}   \Bigg]\ , \nonumber
\end{align}
where we have used $\delta_d S_{6D}=0$.   Note that once we impose (\ref{Hijk}) it is not necessary to also require that 
\begin{align}
	\delta_\text{trial}\left[H_{ijk}+ \frac{3}{2} \Omega_{l[i|}H_{-|jk]} x^l\right]=0\ ,
\end{align} 
to ensure that the variation of the  second line in (\ref{SS}) vanishes since the right hand side is quadratic in $H_{ijk}+ \tfrac{3}{2} \Omega_{l[i|}H_{-|jk]} x^l$. We also do not need to worry about the relation (\ref{GH}) as this defines $G_{ij}$ and hence will define its variation.

However  we do require that the identification (\ref{FH}) is consistent with the diffeomorphism. Under a general diffeomorphism ${k}^\mu$   we have
\begin{align}
  \delta_\text{trial} H_{+\mu\nu} &= -( \partial_\mu k^\lambda ) H_{+\lambda\nu} - ( \partial_\nu k^\lambda ) H_{+\mu\lambda} -(\partial_+ {k}^\lambda )H_{\lambda\mu\nu}\nonumber\\
  \delta_\text{trial} F_{\mu\nu} &= -( \partial_\mu k^\lambda ) F_{\lambda\nu} - ( \partial_\nu k^\lambda ) F_{\mu\lambda}\ .
\end{align}
We see that $\delta_\text{trial} F_{\mu\nu} = \delta_\text{trial} H_{+\mu\nu}$ only if $\partial_+ {k}^\mu=0$ and so (\ref{FH})  is invariant under this restricted set of diffeomorphisms. Unsurprisingly this breaks the space of symmetries to those ${k}^\mu$ and $\tensor{{\lambda}}{^{\und{\mu}}_{\und{\nu}}}$ that are independent of $x^+$.

Thus we are led to the $\mathcal{F}_{ij}H_{-kl}$ term. We find
  \begin{align}
  \delta_\text{trial} \mathcal{F}_{ij} &= - 2\omega \mathcal{F}_{ij} - \left( \delta_i^{\und{i}} \tensor{\lambda}{^{\und{k}}_{\und{i}}}\delta^k_{\und{k}} \right) \mathcal{F}_{kj} + \left( \delta_j^{\und{j}} \tensor{\lambda}{^{\und{k}}_{\und{j}}}\delta^k_{\und{k}} \right) \mathcal{F}_{ki} \nonumber\\
  \delta_\text{trial} H_{-kl} &= -4\omega H_{-kl} - \left( \delta_k^{\und{k}} \tensor{\lambda}{^{\und{i}}_{\und{k}}}\delta^i_{\und{i}} \right) H_{-il} + \left( \delta_l^{\und{l}} \tensor{\lambda}{^{\und{i}}_{\und{l}}}\delta^i_{\und{i}} \right) H_{-ik} \nonumber \\
  & \qquad +\left(  v_k + 2\omega_2 x^k \right)F_{-l} - \left(  v_l + 2\omega_2 x^l \right)F_{-k}\ .
\end{align}
Indeed, these forms follow almost immediately when one notes the forms of $\mathcal{F}_{ij}$ and $H_{-kl}$ in terms of tangent frame fields; $\mathcal{F}_{ij} = H_{\und{+ij}}$, $H_{-kl}=H_{\und{-kl}}$. Then, noting that $\delta_\text{trial}\left( d^5 x \right)=3\omega d^5 x$ and that the local Lorentz pieces exactly vanish, we find
\begin{align}
 \frac{k}{4\pi^2\om}  \delta_\text{trial}\Bigg[ \text{tr} \int d^5x\, \bigg( \frac{1}{4} \varepsilon_{ijkl} \mathcal{F}_{ij} H_{-kl} \bigg)  \Bigg] =  \frac{k}{4\pi^2\om} \text{tr} \int d^5x\, \frac{1}{2}\varepsilon_{ijkl} \left( v_k + 2 \omega_2 x^k  \right) F_{ij} F_{-l}\ .
\end{align}
This term is essentially $d k^+\wedge \text{tr} \left( F\wedge F \right)$, and so is a total derivative. In particular, we have
\begin{align}
  \varepsilon_{ijkl} \left( v_k + 2 \omega_2 x^k  \right) \text{tr} \left( F_{ij} F_{-l} \right) =& - \partial_- \Big(  k^+  \varepsilon_{ijkl} \text{tr} \left( F_{ij} F_{kl} \right) \Big) \nonumber\\
  &+4\partial_i \Big(  k^+   \varepsilon_{ijkl} \text{tr} \left( F_{-j} F_{kl} \right) \Big)\ .
\end{align}
Hence, for suitable boundary conditions on $S^4_\infty$, and regular behaviour of $F$ throughout $\mathbb{R}^5$, the resulting boundary terms vanish. However, it is precisely the relaxation of this latter condition that gives rise to the action non-invariance explored in Section \ref{subsec: instantons and symmetry breaking}.

For now, let us assume regular behaviour of $F$, and so we are left with
\begin{align}
  \delta_\text{trial} S = \hat{\delta} S_{6D} = \frac{k}{4\pi^2\om} \text{tr} \int d^5 x\, \bigg\{ &4\omega \left( -\frac{1}{2} {\nabla}_i X^I {\nabla}_i X^I \right) \nonumber \\
  &+ 5\omega \left( -\frac{\ii}{2} \bar{\Psi} \Gamma_{\und{+}}D_- \Psi + \frac{\ii}{2} \bar{\Psi} \Gamma_{\und{i}}  {\nabla}_i \Psi \right) \nonumber\\
  &+ 7 \omega \left( -\frac{1}{2} \bar{\Psi} \Gamma_{\und{+}} \Gamma^{I} \left[ X^I, \Psi\right] \right) \bigg\}\ .
  \label{eq: S variation before scaling}
\end{align}
Lastly if we  augment $\delta_\text{trial}$ by a simple scaling by $\omega$
\begin{align}
  \delta' X^I &= - 2\omega X^I \ ,\nonumber\\
  \delta' \Psi &= -\frac{5}{2} \omega \Psi\ ,
 \end{align}
then for $\delta_{}= \delta_\text{trial} + \delta'$, we have $\delta_{}S=0$.

In summary  we have
\begin{align}
  \delta_{} x^\mu &= k^\mu \ ,\nonumber\\
  \delta_{} A_- &= -\left( \partial_- k^- \right) A_- - \left( \partial_- k^i \right) A_i \ ,\nonumber\\
  \delta_{} A_i &= -\left( \partial_i k^- \right) A_- - \left( \partial_i k^j \right) A_j \ ,\nonumber\\
  \delta_{} X^I &= -2\omega X^I \ ,\nonumber\\
 \delta_{} G_{ij} &= -4\omega G_{ij} - \frac{1}{2} \left( \delta_i^{\und{i}} \tensor{\lambda}{^{\und{k}}_{\und{i}}}\delta^k_{\und{k}} \right) G_{kj} + \frac{1}{2} \left( \delta_j^{\und{j}} \tensor{\lambda}{^{\und{k}}_{\und{j}}}\delta^k_{\und{k}} \right) G_{ki} - \frac{1}{2} \varepsilon_{ijkl} \left( \delta_k^{\und{k}} \tensor{\lambda}{^{\und{m}}_{\und{k}}}\delta^m_{\und{m}} \right) G_{ml} \nonumber \\
  &\quad +  \left( v_i + 2\omega_2 x^i \right) F_{-j} - \left( v_j + 2\omega_2 x^j \right) F_{-i} + \varepsilon_{ijkl}\left( v_k + 2\omega_2 x^k \right) F_{-l}\ ,\nonumber\\ 
  \delta_{} \Psi &= -\frac{5}{2} \omega \Psi +  \frac{1}{4} \lambda^{\und{\mu \nu}} \Gamma_{\und{\mu \nu}} \Psi \ ,
  \label{eq: fields under passive transformation}
\end{align}
where $\lambda^{\und{\mu \nu}}=-\lambda^{\und{\nu \mu}}$ and
\begin{align}
  \lambda^{\und{-+}} &= -\omega \ ,\nonumber\\
  \lambda^{\und{-i}} &= 0 \ ,\nonumber\\
  \lambda^{\und{+i}} &= v_i + 2 \omega_2 x^i \nonumber\\
  \lambda^{\und{ij}} &= M_{ij} + \frac{1}{2}\left( \Omega_{ij} v_k x^k + \Omega_{ik} v_k x^j - \Omega_{jk} v_k x^i + \Omega_{ik} v_j x^k - \Omega_{jk} v_i x^k \right) \ ,\nonumber\\
  &\qquad + \frac{1}{2} \omega_2 \left( \Omega_{ij} |\vex|^2 + 2 \left( \Omega_{ik} x^k x^j - \Omega_{jk} x^k x^i \right)\right)\ . 
\end{align}
Given the  form for $k^\mu$ specified in (\ref{eq: k explicit form}) we can compute an explicit expression for $\delta_{}$. 
From the point view of the five-dimensional field theory one can decompose  $\delta_{}$ into a diffeomorphism contribution, a scale transformation, and a tensor-like transformation that mixes the various components of the fields. To arrive at the field transformations listed in Chapter \ref{chap: An explicit model and its symmetries}, we simply change our perspective of the transformation $\delta$ from the passive to the active picture.\\

Let us finally make a brief note about representations. As diffeomorphisms in the proxy theory $S_\text{6D}$, the variations $\delta_\text{trial}$ for the general set of six-dimensional conformal Killing vectors form a natural representation of the six-dimensional conformal algebra $\frak{so}(2,6)$. Then restricting to only those conformal Killing vectors that commute with $\partial_+$, we find a representation of $\frak{h}=\frak{u}(1)\oplus \frak{su}(1,3)$. One can further verify that this is \textit{not} spoiled by the shifts $\delta'$, and so the variations $\delta$ form a representation of $\frak{h}$ in the six-dimensional theory.

However, once we have reduced to the five-dimensional theory $S$, the $\delta$ not longer form a representation of $\frak{h}$ when acting on the Lagrange multiplier field $G_{ij}$, as shown explicitly in Section \ref{sec: spacetime symmetries of the action}. What's gone wrong? The answer is quite subtle. Recall, we imposed the relation $\mathcal{H}_{ijk}:=\left(H_{ijk}+ \frac{3}{2} \Omega_{l[i|}H_{-|jk]} x^l\right)=0$ in order to massage $S_\text{6D}$ to look (almost) like $S$. It was crucial that this combination appeared only quadratically in $S_\text{6D}$, and thus in order to preserve the invariance of $S_\text{6D}$, we were \textit{not} required to ensure that $\delta_\text{trial}\mathcal{H}_{ijk}$ also vanished. Indeed, one finds that $\delta_\text{trial}\mathcal{H}_{ijk}\neq 0$ under transformations generated by $M_{i+}$ and $K_+$.

However, we find that the variation $\delta_\text{trial}G_{ij}$ involves $\mathcal{H}_{ijk}$ \textit{linearly}, in the case of a transformation generated by $M_{i+}$ or $K_+$. To arrive at the expression for $\delta G_{ij}$ in (\ref{eq: fields under passive transformation}), we simply threw away this term. But in doing so, we deformed the symmetry algebra at \textit{second order}, as $\delta_\text{trial}\mathcal{H}_{ijk}\neq 0$ for some transformations. We see in particular that we generically expect the brackets $[\delta_{M_{i+}}, \delta_{M_{j+}}]$ and $[\delta_{M_{i+}}, \delta_{K_+}]$ to be deformed, as indeed is precisely the case as shown in (\ref{eq: Gij algebra extension}).


\chapter{Noether currents}\label{app: Noether currents}


We state here the explicit $\frak{su}(1,3)$ Noether currents of the theory introduced in Chapter \ref{chap: An explicit model and its symmetries}, which feature in the local infinitesimal Ward-Takahashi identities (\ref{eq: general local WTI with tildes}). Note that we use $J_G$ to denote a vector field and 1-form interchangeably, as the musical isomorphism with respect to the Euclidean metric on $\mathbb{R}^5$ that relates them is trivial. The expressions below are written in terms of the Lagrangian (\ref{eq: Lagrangian}) and the vector fields (\ref{eq: 5d algebra vector field rep}).

\subsection*{$G\in\{P_-, P_i, B, C^\alpha, T\}$}
\begin{align}
  	J_{G}^- 		&= 	-\left( G_\partial \right)^- \mathcal{L}  +\frac{k}{4\pi^2R}\text{tr}\,\bigg[\,\,  -\left( F_{-i} + \tfrac{1}{2}\Omega_{jk} x^k G_{ij} \right)\, \delta_G A_i  -\tfrac{1}{2}\Omega_{ij} x^j \big( \hat{D}_i X^I \big) \delta_G X^I \nonumber\\
  &\hspace{95mm} +\tfrac{i}{2} \bar{\Psi} \left( \Gamma_+ + \tfrac{1}{2} \Omega_{ij} x^j \Gamma_i \right)\delta_G\Psi \,\,\bigg]	\ ,\nn\\
  	J_{G}^i 		&= -(G_\partial)^i \mathcal{L} +\frac{k}{4\pi^2R}\text{tr}\,\bigg[\,\, \left( F_{-i} + \tfrac{1}{2}\Omega_{jk} x^k G_{ij} \right)\, \delta_G A_-  - G_{ij}\, \delta_G A_j  + \big( \hat{D}_i X^I \big)\, \delta_G X^I \nonumber\\
  &\hspace{122mm} -\tfrac{i}{2}\bar{\Psi} \Gamma_i \delta_G \Psi  \,\,\bigg]	\ .	
\end{align}

\subsection*{$M_{i+}$}
\begin{align}
  	J_{M_{i+}}^- 		&= 	\left( -\frac{k}{8\pi^2R}x^i \star\text{tr}\left( F\wedge F \right) \right)^-	-\left( M_{i+} \right)_\partial^- \mathcal{L}\nn\\
  	&\qquad  +\frac{k}{4\pi^2R}\text{tr}\,\bigg[\,\,  \tfrac{1}{4}x^i X^I X^I-\left( F_{-i} + \tfrac{1}{2}\Omega_{jk} x^k G_{ij} \right)\, \delta_G A_i   \nonumber\\
  &\hspace{45mm}-\tfrac{1}{2}\Omega_{ij} x^j \big( \hat{D}_i X^I \big) \delta_G X^I +\tfrac{i}{2} \bar{\Psi} \left( \Gamma_+ + \tfrac{1}{2} \Omega_{ij} x^j \Gamma_i \right)\delta_G\Psi \,\,\bigg]\ ,	\nn\\
  	J_{M_{i+}}^j 		&= 	\left( -\frac{k}{8\pi^2R}x^i \star\text{tr}\left( F\wedge F \right) \right)^j	-\left( M_{i+} \right)_\partial^j \mathcal{L}	\nn\\
  	&\qquad+\frac{k}{4\pi^2R}\text{tr}\,\bigg[\,\, \tfrac{1}{2}\Omega_{ij} X^I X^I +\left( F_{-i} + \tfrac{1}{2}\Omega_{jk} x^k G_{ij} \right)\, \delta_G A_-   \nonumber\\
  &\hspace{30mm}- G_{ij}\, \delta_G A_j  + \big( \hat{D}_i X^I \big)\, \delta_G X^I -\tfrac{i}{2}\bar{\Psi} \Gamma_i \delta_G \Psi  \,\,\bigg]	\ .
\end{align}

\subsection*{$K_+$}
\begin{align}
  	J_{K_+}^- 		&= 	\left( -\frac{k}{8\pi^2R}x^i x^i \star\text{tr}\left( F\wedge F \right) \right)^-	-\left( K_+ \right)_\partial^- \mathcal{L}\nn\\
  	&\qquad  +\frac{k}{4\pi^2R}\text{tr}\,\bigg[\,\,  \tfrac{1}{2}x^i x^i X^I X^I-\left( F_{-i} + \tfrac{1}{2}\Omega_{jk} x^k G_{ij} \right)\, \delta_G A_i   \nonumber\\
  &\hspace{45mm}-\tfrac{1}{2}\Omega_{ij} x^j \big( \hat{D}_i X^I \big) \delta_G X^I +\tfrac{i}{2} \bar{\Psi} \left( \Gamma_+ + \tfrac{1}{2} \Omega_{ij} x^j \Gamma_i \right)\delta_G\Psi \,\,\bigg]\ ,	\nn\\
  	J_{K_+}^i 		&= 	\left( -\frac{k}{8\pi^2R}x^j x^j \star\text{tr}\left( F\wedge F \right) \right)^i	-\left( K_+ \right)_\partial^i \mathcal{L}	\nn\\
  	&\qquad+\frac{k}{4\pi^2R}\text{tr}\,\bigg[\,\, \Omega_{ij} x^j X^I X^I +\left( F_{-i} + \tfrac{1}{2}\Omega_{jk} x^k G_{ij} \right)\, \delta_G A_-   \nonumber\\
  &\hspace{30mm}- G_{ij}\, \delta_G A_j  + \big( \hat{D}_i X^I \big)\, \delta_G X^I -\tfrac{i}{2}\bar{\Psi} \Gamma_i \delta_G \Psi  \,\,\bigg]	\ .
\end{align}


\bibliography{thesis}
\bibliographystyle{JHEP}

\end{document}